\documentclass{elsart}

\usepackage{graphicx}
\usepackage{pxfonts}
\usepackage{lineno}
\usepackage{appendix}

\usepackage{amssymb}
\usepackage{footnote}
\usepackage{multirow}
\usepackage{varwidth}
\usepackage{array}
\usepackage{epstopdf}
\usepackage{url}

\journal{}

\begin{document}

\newpage

\begin{frontmatter}



\title{SPECIAL RELATIVITY}
\title{Applications to astronomy and the accelerator physics}


\author{Evgeny Saldin}
\newpage
\address[]{ Final draft }

If you have any comments, please send an email to evgueni.saldin@desy.de.

\end{frontmatter}


\newpage
*Preface

Many books have been written on the classical subject of special relativity. However, after years of experience in both
relativistic engineering and research, I have come to believe that there is room for a new perspective. 
This book is not quite like the others, as it aims to shed light on often-overlooked aspects of the subject. 

To clarify, this is not a textbook on relativity theory. Instead, it focuses on the nature of special relativistic kinematics, its connection to space and time, and the operational interpretation of coordinate transformations.  
Every theory contains a number of quantities that can be measured by experiment and expressions that cannot possibly be observed. Whenever we have a theory containing an arbitrary convention, we should examine what parts of the theory depend on the choice of that convention and what parts do not. Unfortunately,  
this distinction is often overlooked, leading some authors to mistakenly classify certain data as observable when, in reality, they rely on arbitrary choices rather than physical  experiments. This oversight results in inconsistencies and paradoxes that should be avoided.

The practical approach adopted in the book should appeal to astronomers, space engineers, accelerator engineers, and more broadly, relativistic engineers. This approach, unusual in the relativistic literature, may be clarified by quoting one of the problems discussed in the text: the new light beam kinematics for rotating frames of references. Since we live on such a rotating (Earth-based) frame of reference, the difference in relativistic kinematics between rotating and non-rotating frames of reference is of great practical as well as theoretical significance. 
A correct solution of this problem requires the use of relativistic principles even at low velocities since the first-order terms in $(v/c)$, where $v$ is the orbital velocity, play a fundamental role in the non-inertial relativistic kinematics of light propagation.

All the results presented here are derived from the first principles, with every step  
involving physical reasoning explicitly detailed. To maintain a self-consistent style, auxiliary derivations
are provided in the appendices. To encourage readers to develop their own perspectives, each chapter 
includes a suggested bibliography with relevant remarks. The references list only papers I personally consulted, and while many valuable works remain unmentioned, I apologize for any omission.

I am  deeply grateful to my longtime friends Gianluca Geloni and Vitaly Kocharyan for years of discussions on much of the material  covered in this book. I also extend my sincere thanks to DESY (Deutsches  Electronen-Synchrotron) for the opportunity to work in this fascinating field.

\newpage

\tableofcontents

\newpage

\section{Introduction}

This book begins with a critical examination of current approaches to special relativity. Traditionally, the theory is framed around Einstein’s two postulates: the principle of relativity and the constancy of the speed of light. However, in a more general geometric approach, the principle of relativity is not a fundamental axiom but rather a consequence of space-time geometry.

We emphasize that the core of special relativity lies in the following fundamental postulate: all physical processes occur in four-dimensional space-time, whose geometry is pseudo-Euclidean. This perspective shapes the presentation of the subject in this book, where the four-dimensional geometric formulation takes precedence over conventional approaches.

The space-time geometric approach accommodates all possible choices of coordinates for reference frames, making the second Einstein postulate—the constancy of the coordinate speed of light—unnecessary in this broader framework. The coordinate speed of light remains isotropic and constant only in Lorentz coordinates, where Einstein’s synchronization of distant clocks and Cartesian space coordinates are applied. Thus, the basic elements of the space-time geometric formulation of the special relativity  and the usual Einstein's formulation, are quite different.  

It is important to emphasize that, in practical applications, there are two useful choices for clock synchronization conventions:

(a) Einstein's convention, which leads to the Lorentz transformations between reference frames.

(b) The absolute time convention, which leads to the Galilean transformations between reference frames.

Absolute time (or simultaneity) can be introduced in special relativity without altering either the logical structure or the convention-independent predictions of the theory. While this choice may seem unconventional within relativity, it is often the most practical when connecting theoretical results to laboratory reality.

It is widely believed that only philosophers of physics discuss the issue of distant clock synchronization. Indeed, a typical physics laboratory does not contain a predefined space-time grid. It is important to recognize that a rule-clock structure exists only in our minds, and the use of hypothetical clocks in special relativity is a necessary prerequisite for applying dynamics and electrodynamics in a coordinate-based framework.
This situation often leads physicists to assume that the theory of relativity can be applied to physical processes without a detailed understanding of the clock synchronization procedure. However, many problems in special relativity can only be adequately addressed using a non-standard absolute time synchronization approach.
A distinctive feature of this book is its exploration of absolute time coordinatization—specifically, the use of Galilean transformations—within the framework of special relativity.

Chapter 3 presents an "operational interpretation" of Lorentz and absolute time coordinatizations, making it arguably the most important chapter of this book.
Today, asserting the validity of Galilean transformations is considered a "shocking heresy," conflicting with the prevailing relativistic intuition and the widely accepted interpretation of special relativity among physicists.
The distinction between absolute time synchronization and Einstein’s time synchronization, from an operational perspective, will be a significant revelation for any expert in special relativity. To our knowledge, neither the operational interpretation of absolute time synchronization nor this key distinction has been explored elsewhere in the literature.

We begin by examining the phenomenon of the aberration of light.
Light, as a special case of electromagnetic waves, is governed by electrodynamics. It is well known that electrodynamics fully complies with the principles of relativity and, therefore, must accurately describe the properties of light—an inherently relativistic entity.

In Chapters 4 and 5, we undertake a critical reexamination of the existing theory of light aberration. Even at first order in  $v/c$, the phenomenon remains nontrivial, and the literature contains numerous incorrect or incomplete results.

To analyze optical phenomena involving the relative motion of two (or more) light sources, it is essential to formulate electrodynamics within an absolute-time framework. When the sources are independent, the introduction of Lorentz coordinates presents no difficulty: each source may be assigned its own coordinate system with an independent set of clocks. The situation changes fundamentally, however, when a secondary source interacts with radiation emitted by a primary source.

From the standpoint of special relativity, a key difficulty arises in such cases: a single inertial frame cannot provide a common Lorentz coordinatization for both sources. This fundamental limitation necessitates a deeper examination of how electrodynamics should be applied when light emitted by one source is subsequently processed by another moving system.

It is widely assumed in textbooks that when a mirror moves tangentially to its surface, the law of reflection remains identical to that of a stationary mirror. According to this view, a plane wave incident normally on a moving mirror is reflected as an oblique beam. 
This assumption, however, is incorrect. 

The  fundamental error in standard textbook treatments is conceptual. The light beam emitted by a stationary source is described using the Minkowski metric, and it is implicitly assumed that the same metric applies to a moving mirror. In other words, it is assumed that the energy transport velocity coincides with the phase velocity. This assumption is often treated as self-evident.

It is worth noting that using the conventional approach to calculate reflection from a moving mirror does not necessarily lead to errors.
When analyzing the reflection of light from a mirror moving normally to its surface, the standard textbook reasoning correctly predicts the variation in light frequency upon reflection.
The key to understanding this discrepancy lies in the nature of the motion. In the case of a normally moving mirror, the velocity is perpendicular to the plane of oscillating elementary sources (dipoles). As a result, the plane of simultaneity in absolute time coordinates retains the same orientation in Lorentz coordinates.


There is an intuitively plausible reason why textbook authors conclude that light reflected from a tangentially moving mirror experiences no aberration. A comparison between conventional aberration theory for light and sound theory can provide insight.
In Chapter 4, we will consider a sound emitter at rest in the atmospheric frame and examine a mirror moving tangentially along its surface. According to sound theory, the reflected beam’s energy transport does not deviate because the group velocity equals the phase velocity. When a plane wave falls the mirror normally, it produces an oblique sound beam. The diagonal wave equation for the moving mirror in the atmospheric frame is identical to that of an emitter at rest.  However, in sound theory, we operate within Newtonian space and time rather than the pseudo-Euclidean geometry of space-time. This distinction is crucial: the standard analysis of light aberration within a single inertial frame fails to account for the fundamental difference between the velocity of light and that of sound.

In the electrodynamics case,  the mirror undergoes acceleration along the radiation wavefront (i.e. the initial plane of phase simultaneity). Since the orientation of the radiation wavefront is no longer absolute, it depends on the chosen clock synchronization convention.

To illustrate the physical consequences, consider a point source and an observer, together with their respective measuring devices, all at rest in an inertial laboratory frame. We show that when a finite-aperture mirror moves tangentially at a constant velocity and interacts with radiation in the far field—where the point source effectively produces a plane wave at the mirror—the energy transport of the reflected light exhibits a measurable deviation. This effect has important practical implications. According to Babinet’s principle, our theory of light aberration predicts an analogous deviation for light transmitted through a hole in a moving opaque screen, or equivalently, through the open end of a moving telescope barrel.

The agreement between the predictions of this theoretical framework and empirical observations lends strong support to our revised formulation of special relativity incorporating absolute simultaneity (see Chapter 4 for further discussion).

Questions related to transmission through a transversely moving pupil detection system (e.g., the moving end of a telescope barrel) often lead to serious misunderstandings. These arise from an inadequate grasp of several complex aspects of statistical optics.
Modern explanations suggest that the phenomenon of aberration of light can be interpreted using a ray model. A particularly interesting case is the transmission of light through a moving telescope tube. The rays of light from a star impinge on the telescope tube, and according to the prevailing literature, they do not interact with its sides.

One might naively assume—following textbook reasoning—that the domain of ray optics applies to all spatially incoherent radiation. However, this is a misconception. A completely incoherent spatial source (such as an incandescent lamp or a star) consists of elementary, statistically independent point sources with different offsets. The characteristic dimension of each elementary statistically independent source is approximately $\lambda$, where $\lambda$ is the wavelength of visible radiation.
The radiation field generated by such a source can be considered a linear superposition of fields from individually incoherent sources. Each elementary source effectively produces a plane wave in front of the pupil detection system. Moreover, any measuring instrument inevitably influences the detected radiation due to the unavoidable diffraction of a plane wave by an aperture. It is important to note that a linear superposition of radiation fields from elementary point sources preserves fundamental single-point source characteristics, including their independence from source motion.

Chapter 6 explores astronomical applications.
Stellar aberration is often considered one of the simplest phenomena in astronomical observations. However, despite its apparent simplicity, it remains one of the most intricate effects of special relativity.
A common belief is that stellar aberration depends on the relative velocity between the source (star) and the observer. However, observations clearly indicate that aberration is independent of the relative motion between stars and telescopes on Earth. 
The lack of symmetry, between the cases when either the source or telescope is moving is shown clearly on the basis of the separation of binary stars. 
The relative motion of these stars with respect to each other (and hence, with respect to the Earth) is never followed by any aberration, although the motion of these stars is, sometimes, much faster than that of the Earth around the Sun. 
It should be stressed that it is the telescope and not the star that must change its velocity (relative to the fixed stars) to cause aberration. 

The contradiction is so evident that some physicists argue stellar aberration conflicts with the special theory of relativity.
Currently, no explanation accounts for why observational data on stellar aberration align with a moving Earth, yet the symmetric case—where the star has relative transverse motion—fails to produce compatible predictions.
We show that the absence of widely separated binary stars does not necessitate a fundamental revision of physical principles. Instead, it highlights the need to interpret stellar aberration within a space-time geometric framework.

The phenomenon of the aberration of light is often interpreted, within the corpuscular model of light, as analogous to the oblique fall of raindrops observed by a moving observer. This classical kinematic approach has been employed in astronomy for nearly three centuries to compute stellar aberration.
This book presents a relativistic theory of the aberration of light in rotating frames. To explore the effects of relativistic modifications to stellar aberration, we begin with the classical aberration increment, given by  $\theta_a = v/V$ , where $v$ is the observer's velocity and $V$ is the velocity of the particles. It is assumed that $v \ll V$.
According to the conventional approach, by neglecting terms of order $v^2/c^2$ in comparison to 1, the stellar aberration increment simplifies to the elementary formula $\theta_a = v/c$. This result is particularly notable, as the study of stellar aberration traditionally relies on classical kinematics, specifically the Galilean vectorial law of velocity addition.
We conclude that the standard analysis of stellar aberration fails to account for the fundamental distinction between the velocity of light and the velocity of raindrops.

A satisfactory treatment of relativistic modifications should be based on two key relativistic parameters.
For an observer on Earth, the velocity relative to the solar reference frame is approximately 30 km/s, corresponding to Earth's motion around the Sun. In the theory of stellar aberration, we typically consider the small expansion parameter $v/c \simeq 10^{-4}$, neglecting terms of the order of  $v^2/c^2$.
However, in addition to this parameter, we must also account for the relativistic quantity  $V/c$. For light, $V/c = 1$, and according to the special theory of relativity, stellar aberration cannot be treated purely within the framework of classical mechanics. Light is always an "ultra-relativistic" phenomenon, regardless of how small the ratio $v/c$ may be.
Since physical processes occur within the metric structure of space-time, as dictated by the special theory of relativity, the geodesic interval must be used to accurately describe the phenomenon of stellar aberration.

The challenges associated with Earth-based measurements are resolved by recognizing a fundamental asymmetry between Earth-based and Sun-based observers—specifically, the acceleration of the traveling Earth-based observer relative to the fixed stars.
We demonstrate that explaining the aberration of light in a rotating frame of reference does not require modifying special relativity or invoking general relativity. Instead, a rigorous application of special relativity is sufficient.

All phenomena in non-inertial reference frames should be analyzed within the space-time geometric framework using the metric tensor. Employing the Langevin metric in a rotating frame of reference allows us to account for all Earth-based experiments. A correct treatment of this problem in a rotating frame necessitates the use of the metric tensor, even for first-order effects in $v/c$. This is because the cross term in the Langevin metric—which represents the first-order deviation of the metric tensor from its Minkowski form—plays a crucial role in the non-inertial kinematics of light beam propagation. The historical progression of studies on optical effects in rotating frames is quite unusual. The Sagnac effect was discovered in 1913 and later described by Langevin in 1921. It is intriguing that the Langevin metric has never been previously applied to stellar aberration theory.

We demonstrate that the aberration of light is a complex phenomenon that must be categorized into various types based on their origin. These categories depend on the cause of the aberration—whether it arises from a change in the observer’s velocity (relative to the fixed stars) or the velocity of the light source. Furthermore, aberration can be further subdivided based on the physical influence of the optical instrument on the measurement.

A theory of stellar aberration within the Earth-based frame of reference must account for two key observations: (1) the annual apparent motion of fixed stars relative to their positions, and (2) the absence of apparent aberration in rotating binary systems (they exhibit aberrations not different from other stars).
We present a theory that addresses both of these phenomena while also providing new insights. All Earth-based experiments can be explained by considering the effect of the measuring instrument (i.e., the inevitable physical influence of the telescope on the measurement) and the acceleration of the Earth-bound observer relative to the fixed stars.

In Chapter 11, we analyze the potential for using Earth-based sources to confirm predictions of the relativistic aberration of light theory. We introduce a simple scaling model for stellar aberration, emphasizing that the motion of stars relative to Earth is not accompanied by aberration. Notably, an aberration shift occurs even when a star moves at the same velocity as Earth.
We derive a condition for optical similarity between the aberration of light from a distant star moving with Earth's velocity and that from an Earth-based incoherent source. The proposed method for measuring the aberration angle leverages Earth-based sources and offers a significant advantage: Earth's rotation should induce a corresponding shift in the observed image. This aberration shift depends solely on $v_{\perp}$, the component of Earth's orbital velocity perpendicular to its rotation axis. The apparent position of the source image is thus always a little displaced in the direction of the Earth's motion around the Sun at that moment, and hence describes a small elliptical figure during the annular revolution of the Earth around its axis. In principle, observations could be recorded within a single day.

Special relativity is generally considered a reciprocal theory. However, the aberration of light in an accelerated frame reveals a fundamental asymmetry between accelerated and inertial observers. Notably, without referencing anything external to the Earth-based accelerated frame, one could determine Earth's velocity relative to the Sun-based frame by measuring the aberration of an Earth-based incoherent light source. While no such experiment has been conducted, astronomical observations confirm what the outcome would be.

Many people who learn special relativity in the usual way find this argument unsettling. The reasoning behind the claim that aberration shift must be symmetrical typically proceeds as follows:

1. A fundamental principle of special relativity states that the metric contains all the necessary information about the physics of a given situation as described in the chosen coordinates.

2. The direction of an observer's acceleration is encoded in the cross term of the Langevin metric.

3. It is always possible to choose a coordinate system in which the metric of an accelerated frame becomes diagonal. This follows from the pseudo-Euclidean geometry of space-time.

4. Maxwell’s equations hold in all inertial frames, ensuring that light propagates at the same velocity $c$ in each of them. This eliminates any privileged frame, implying that the concept of absolute motion has no physical meaning.

In the argument above, steps (2) and (3) are correct, but step (1), and consequently step (4), are incorrect. Step (3) asserts that all inertial frames are equivalent with respect to physical laws, but not necessarily with respect to physical facts. At first glance, the diagonalization of the Langevin metric might seem to establish symmetry between inertial frames, but where does the asymmetry arise?

The electrodynamics equations alone do not provide a complete description of the physical situation. To solve them, one must also specify the initial conditions. The time after diagonalization
is readily obtained by introducing the time offset factor. This time shift introduces a rotation of the plane of simultaneity, which in turn leads to a rotation of the plane wavefront in the accelerated frame. Consequently, after metric diagonalization, the information about the observer’s acceleration is no longer contained in the metric but is instead encoded in the initial conditions—specifically, in the orientation of the radiation wavefront.

One can conclude that not everything is relative in relativity, as the theory also encompasses certain absolute features. Many people assume that, because time and distance in the Lorentz coordinatization have direct physical significance, there must be an underlying physical (dynamical) reason for wavefront rotation.
Some believe that the dynamics governed by physical fields are merely concealed within the language of pseudo-Euclidean geometry. However, the paradoxical asymmetry between inertial and accelerated reference frames is best explained through a dynamical perspective. The resolution of this asymmetry paradox identifies the inertial (pseudo-gravitational) force within the accelerated system as the cause of the asymmetry.
By applying the principle of equivalence, one can address non-inertial kinematic problems using dynamical methods. In this context, wavefront rotation, which arises when transforming from an inertial frame to an accelerated frame (relative to the fixed stars), can be understood as the effect of a pseudo-gravitational potential gradient during the acceleration process.

Several key points arise from the above results. One interesting question is why we are not discussing the aberration of light emitted by a laser.
When considering light aberration, it is essential to distinguish between incoherent and laser sources. Naively, one might expect Earth's motion around the Sun to introduce an aberration significant enough to affect precise laser-based observations. However, it is important to emphasize that laser light aberration cannot be measured.
Inside a laser resonator, the electromagnetic wave travels back and forth, reflecting between mirrors. This repeated reflection effectively cancels out any asymmetry over successive round trips, preventing the manifestation of aberration.
A crucial distinction exists between laser and incoherent sources. The radiation from an incoherent source depends on initial conditions, influenced by a mix of spatial and temporal factors—an effect akin to pseudo-gravity experienced by an accelerated observer. In contrast, in an optical resonator, the (3-space) boundary conditions dictate the direction of energy propagation.

Many books claim that no experiment can determine which observer remains "at rest" during acceleration because, according to the principle of relativity, only relative motion has physical significance.
From this, they conclude that any observed asymmetry would contradict the principle of relativity. However, this argument is incorrect.
There is no conflict between the fundamental structure of special relativity and the aberration of light phenomena. The principle of special relativity, as stated in Einstein’s writings and every textbook on the subject, asserts that the laws of physics are the same in all inertial frames. The principle of special relativity states that velocity is irrelevant to physical laws, but not to everything. \footnote{As Ferraro \cite{FER} note:  ``In 1905, Einstein published the article entitled “On the electrodynamics of
moving bodies” (1905), where he reformulated the notions of space and time starting from two postulates: 

\textit{ ... the same laws of electrodynamics and optics will be valid for all frames
of reference for which the equations of mechanics hold good. }  (Principle of
relativity) 

\textit{...  light is always propagated in empty space with a definite velocity c
which is independent of the state of motion of the emitting body.}

The postulates refer to the validity of Maxwell’s laws in all inertial frames.''  Minkowski, in 1908, recognized that these postulates could be reformulated as a single geometric axiom: spacetime possesses a pseudo-Euclidean geometry.  }

It is widely assumed that covariant equations must yield covariant solutions. This notion, commonly found in textbooks, is incorrect. For example, from a mathematical perspective, the Lorentz covariance of electrodynamics equations does not necessarily ensure covariant solutions. Here, we emphasize a crucial point: the principle of special relativity, which states that all physical equations must remain invariant under Lorentz transformations (i.e., treating the pseudo-Euclidean geometry of space-time as an axiom of the theory), does not inherently impose reciprocal symmetry in nature.

Textbook authors mistakenly assume that all inertial frames are equivalent. Relativistic (or reciprocal) symmetry is commonly associated with symmetry under Lorentz boosts. However, it is well known that Lorentz boosts alone do not form a group—the composition of two boosts in different directions results in a combination of a pure boost and a spatial (Wigner) rotation.
From a mathematical perspective, symmetry between inertial frames requires that the relevant transformations form a group. Since Lorentz boosts alone do not satisfy this criterion, true symmetry between inertial frames is absent. Inertial frames are defined by motion at a constant velocity relative to the fixed stars.
A fundamental distinction exists between an accelerated inertial frame (relative to the fixed stars) and one without an acceleration history. In the case of the aberration of light, this difference is closely linked to the Wigner rotation effect. \footnote{A deeper tension emerges between the principle of relativity—which asserts the equivalence of all inertial frames—and the geometric structure of Minkowski spacetime. In fact, only one of these principles can hold unconditionally.
Given the strength of experimental evidence supporting Minkowski geometry of spacetime, we adopt Lorentz covariance as the foundational principle of the theory.}

It is important to note that all well-known tests of special relativity rely on second-order optical (interference) measurements, with the Michelson-Morley experiment being the most prominent example. In such experiments, light from a source is split into two beams, which later recombine to produce an interference pattern. 
It is not possible to detect the orbital velocity using second-order (interference) experiments.

Phase, as a 4D invariant, is merely a number—it remains unchanged regardless of the chosen inertial frame or coordinate system. This invariance stems from the fundamental geometry of space-time. Consequently, in interference experiments, neither the Doppler effect nor the aberration of light exist as independent, well-defined physical phenomena; only phase has physical significance.

This reasoning explains why special relativity predicts a null fringe shift in Earth-based frames. A careful analysis of all second-order optical experiments within an inertial frame confirms that observed phenomena remain unaffected by uniform motion relative to the fixed stars. While the standard formulation of relativity incorporates the principle of velocity irrelevance, this principle strictly applies in the limiting case of optical second-order measurements.

Beyond interference-based experiments, several other second-order tests exist that do not involve light interference. This raises the question of whether it is possible to determine the motion of an initial inertial frame from within an accelerated frame using time dilation effects. We demonstrate that the slowing of a physical clock accelerated relative to the fixed stars is independent of the reference frame in which this effect is measured. In other words, time dilation is an absolute effect, not a relative one—a fundamental property of pseudo-Euclidean space-time.

Now, let us return to the topic of metric diagonalization. In textbooks, time dilation is typically derived from the diagonal (Minkowski) metric. After diagonalization, these textbooks assert that coordinates in inertial frames should be related through the Lorentz transformation.
At first glance, the problem appears completely symmetrical, suggesting that time dilation is a relative effect. However, something is evidently amiss. In Chapter 13, we explore the connection between time dilation and the Langevin metric.

Perhaps the only way to demonstrate that time dilation is absolute is through a mathematical perspective. Mathematically, the argument for the absoluteness of time dilation follows this reasoning:
The metric tensor must be a continuous quantity. The Langevin metric is obtained by smoothly transitioning between the metric tensors of an accelerated frame and an inertial frame. To achieve this smooth transition, it is sufficient to relate the coordinates and time of the accelerated observer to those of the inertial observer using a Galilean boost. Under these conditions, the inertial frame’s metric naturally evolves into the Langevin metric of the accelerated frame.

However, after diagonalization, the metric tensor in the accelerated frame must abruptly shift from the Langevin metric to the Minkowski (diagonal) form. This discontinuity leads to a critical conclusion: when applying the Lorentz transformation, if the transformation fails to preserve the continuity of the metric tensor, then symmetry between inertial frames does not hold.
Thus, any change in velocity or acceleration—relative to the fixed stars—has an absolute significance.

The formalism of relativistic physics is fundamentally tied to the concept of an absolute space-time structure. This structure is inherently linked to an initial inertial frame, which, in turn, serves as an absolute rest frame. Consequently, uniform motion is not relative, and a particle's absolute velocity can be determined within this initial frame. The key distinction between the initial inertial frame and Newton's concept of absolute space lies in the lessons of special relativity, which introduced a unified space-time model. In this framework, Newton’s separate notions of absolute space and absolute time are merged into a single, cohesive space-time continuum.

The proper time interval can be associated with various physical phenomena, including particle lifetimes, atomic transition periods, and nuclear half-life.
Until now, all tests of time dilation using atomic clocks in an Earth-based frame have involved circular path geometries. For instance, atomic clocks flown on commercial jets were compared with a reference clock upon returning to the laboratory, ensuring that both were at rest relative to the Earth.

According to special relativity, any time dilation experiment investigating the effects of orbital velocity should use a linear (one-way) path geometry. This is typically observed in experiments involving the decay of rapidly moving unstable particles. However, current experimental techniques lack the sensitivity required to detect the influence of orbital velocity on the average distance traveled by a decaying particle.


In Chapter 17, we describe the particle dynamics method in the Lorentz lab frame, using lab time $t$  as the evolution parameter. The study of relativistic particle motion in a constant magnetic field, as traditionally approached in accelerator engineering, remains closely tied to classical (Newtonian) kinematics. Specifically, it relies on the Galilean vectorial law for velocity addition. This approach, which is non-covariant, implicitly assumes an absolute time coordinatization—though this remains hidden within the formalism.

The absolute time synchronization convention is self-evident, which is why it is rarely discussed in accelerator physics. Standard textbooks suggest that when only a single frame is considered, a detailed understanding of relativity is unnecessary. In this conventional approach, the only modification to classical mechanics is the introduction of relativistic mass, thereby maintaining a distinction between time and space.

From the lab frame perspective, conventional particle tracking treats a trajectory as the result of successive Galilean boosts that follow the motion of an accelerated particle. The preference for a non-covariant approach within the framework of dynamics stems from its simplicity and practicality. By avoiding the intermixing of spatial coordinates and time, the (3+1)-dimensional non-covariant tracking method remains intuitive, transparent, and well-suited to laboratory conditions.

It can be demonstrated that this approach poses no fundamental difficulties in mechanics or electrodynamics. It is both effective and consistent, as the choice of transformation does not alter the underlying physical reality. What matters is that once a particular convention is adopted, it must be applied consistently across both dynamics and electrodynamics.

A common mistake in accelerator physics arises from an incorrect algorithm for solving electromagnetic field equations. When using Maxwell's equations, only the covariant formulation of the dynamics equations (i.e., in Lorentz coordinates) ensures the correct coupling between Maxwell’s equations and particle trajectories in the lab frame.

Interestingly, employing the conventional coupling of Maxwell's equations with a corrected Newtonian equation to calculate radiation from a moving source does not always lead to errors. In cases where the source moves rectilinearly along with the emitted light beam, both covariant and non-covariant approaches yield identical trajectories, making Maxwell’s equations consistent with conventional particle tracking. However, this method was mistakenly extended to non-collinear geometries, where it breaks down.

To better understand this issue, we first examine the reasoning presented in textbooks. It is generally assumed that, in Lorentz coordinates, the magnetic field can only alter an electron’s direction of motion, not its speed. However, there is a counterargument to this widely accepted derivation of velocity composition.

Consider the composition of two velocities that do not lie along a straight line. Conventional particle tracking, following Newtonian mechanics, treats velocity as an ordinary vector and combines them geometrically using the parallelogram method. However, this approach fails to account for relativistic effects and leads to inconsistencies in non-collinear motion.

Solving problems related to covariant particle tracking presents two key challenges. First, conventional particle tracking within a single inertial frame is typically based on classical Newtonian mechanics, which does not incorporate Lorentz transformations. Second, there is considerable confusion regarding the combination of non-parallel velocities in special relativity. The standard textbook treatment of relativistic velocity transformation implicitly assumes that the $(x',y',z')$ axes of a moving observer remain parallel to the $(x,y,z)$ axes of the lab frame. In other words, it presumes that both observers share a common three-dimensional space—a misconception.
Textbooks often overlook the fact that in Lorentz coordinatization, two observers following different trajectories experience different three-spaces. Special relativity lacks an absolute notion of simultaneity, leading to a fundamental mixing of spatial and temporal coordinates. For one observer, spatial measurements inherently include a contribution from the time coordinate as perceived by another observer. Consequently, there is no well-defined concept of an instantaneous three-space.

At first glance, textbooks on special relativity typically derive the relativistic velocity addition formula analytically from the combination of Lorentz boosts. Many authors argue that the principle of relativity requires the velocity addition law to obey a group composition structure. However, the non-associativity of Einstein’s velocity addition for non-collinear velocities is often overlooked, despite being demonstrable through straightforward algebra.
Consequently, constructing a fully covariant trajectory using only Einstein’s velocity addition is impossible—additional structure is needed.

We emphasize that the addition of non-collinear velocities in Lorentz-coordinatized spacetime is governed by the Wigner rotation. A key consequence of the non-commutativity of non-collinear Lorentz boosts is the lack of a shared, global "ordinary" space. It is crucial to recognize that Einstein’s velocity addition does not directly follow from the full structure of Lorentz transformations. Textbook derivations of transverse (perpendicular) velocity addition often invoke the relation  $dt' = dt/\gamma$, yet fail to account for the Wigner rotation, which introduces a coupling between transverse position and time.

Only by solving the equations of motion within a fully covariant framework—one that includes the Wigner rotation in the treatment of non-collinear velocities—can the coupling between Maxwell’s equations and particle trajectories in the laboratory frame be properly described.

Our analysis highlights the distinction between the concepts of path and trajectory.
A path is a purely geometric notion, representing the spatial course of a particle without reference to time. In contrast, a trajectory provides more information, as it specifies not only the particle’s spatial position but also the corresponding time instant.

The path has a precise, objective meaning. For example, in a magnetic field, the curvature radius of the path—and consequently, the three-momentum—is convention-invariant and thus objectively defined. However, the trajectory 
$\vec{x}(t)$ is convention-dependent, reflecting the inherent conventionality of velocity, and therefore lacks an exact objective meaning. While dynamical theory involves the concept of a particle’s trajectory, direct verification of it is unnecessary. Instead, the trajectory serves as a theoretical tool in the analysis of electrodynamics problems.

In Chapter 17, we derived specific quantitative results for relativistic velocity addition using only elementary principles of special relativity. However, readers seeking a more mathematically rigorous treatment may prefer Chapter 18, where we present a comprehensive account of the mathematical structure underlying the composition of Lorentz boosts.

In Chapter 20, we present a critical reexamination of radiation theory. To evaluate radiation fields generated by external sources, we must determine the particle velocity  $\vec{v}$
and position $\vec{x}$ as functions of the lab-frame time $t$. As discussed earlier, Maxwell’s equations should be solved in the lab frame, incorporating the current and charge density produced by particles moving along their covariant trajectories. For arbitrary values of $v/c$, covariant calculations of radiation processes become highly complex. However, in certain cases, significant simplifications are possible. One such case is a non-relativistic radiation setup.
The non-relativistic assumption enables the dipole approximation, which is of great practical importance. When considering only the dipole component of radiation, we disregard all details of the electron trajectory. Consequently, dipole radiation is insensitive to differences between covariant and non-covariant particle trajectories. However, this is merely the first and most practically relevant term in the expansion.
To compute higher-order corrections beyond the dipole approximation, detailed knowledge of the electron trajectory is required. In such cases, relying on the non-covariant approach is insufficient, and the covariant trajectory must be used instead.

It is generally assumed that Maxwell’s equations and the corrected form of Newton’s second law can explain all experiments conducted within a single inertial (laboratory) frame. According to standard textbooks, if only one frame is involved, knowledge of relativity is deemed unnecessary. Many physicists, having learned special relativity from textbooks, might say, “For those who want just enough understanding to solve problems, special relativity merely modifies Newton’s laws by introducing a correction factor to mass.”
However, this is a misconception. The velocity of light transforms like that of a particle, and one cannot simply apply classical kinematics to mechanics while using relativistic kinematics for electrodynamics. We highlight errors in the standard approach to coupling fields and particles through a relatively simple example, ensuring that the core physical principles remain clear and unobscured by unnecessary mathematical complexity.
We hope this example will draw greater attention to the crucial role of Wigner rotation in the transformation of non-collinear velocities.

Accelerator physicists studying the application of relativity theory to synchrotron radiation often face a fundamental issue: the distinction between covariant and non-covariant particle trajectories has never been fully understood. Consequently, the contribution of relativistic kinematic effects to synchrotron radiation has been overlooked. Traditionally, accelerator physics has relied on Newtonian kinematics, which is incompatible with Maxwell’s equations. This raises an important question: if storage rings are designed without accounting for relativistic kinematics, how do they function? In reality, electron dynamics in storage rings are significantly influenced by radiation emission.
We address this question in detail in Chapter 21, focusing on spontaneous synchrotron radiation from bending magnets and undulators.

Just as the non-relativistic asymptote simplifies calculations, the ultrarelativistic asymptote does the same for covariant analysis, as it naturally leads to the paraxial approximation. The motion of electrons in a bending magnet exhibits cylindrical symmetry, which gives rise to several remarkable effects. We demonstrate that covariant particle tracking is unnecessary for deriving bending magnet radiation. However, there is one case where the conventional approach fails: the covariant framework predicts a nonzero redshift of the critical frequency due to perturbations of electron motion along the magnetic field.

A direct laboratory test of synchrotron radiation theory could highlight the incompatibility between standard relativistic electrodynamics—based on Maxwell’s equations—and non-covariant particle tracking. Despite decades of measurements, synchrotron radiation theory remains experimentally unverified in certain aspects. We explore the potential of synchrotron radiation sources to test the predictions of a revised synchrotron radiation theory, proposing experiments that have never been conducted at existing facilities.

The theory of relativity demonstrates that the relativity of simultaneity—a fundamental effect of relativistic kinematics—is closely linked to extended relativistic objects. Until the 21st century, no macroscopic objects were known to travel at relativistic speeds, and it was widely believed that only microscopic particles in experiments could reach velocities close to the speed of light. However, the 2010s saw rapid advancements in laser light sources in the X-ray wavelength range. An X-ray free electron laser (XFEL) is an example where improvements in accelerator technology make it possible to develop ultrarelativistic macroscopic objects with an internal structure (modulated electron bunches), and the relativistic kinematics plays an essential role in their description.

In Chapter 22, we critically reexamine existing XFEL theory, particularly for readers with limited knowledge of accelerator physics. We focus on the production of coherent undulator radiation by a modulated ultrarelativistic electron beam.
Fortunately, understanding this process does not require prior knowledge of undulator radiation theory from Chapter 21, as it can be explained in a straightforward manner. Relativistic kinematics plays a fundamental role in XFEL physics, particularly through the rotation of the modulation wavefront, which, in the ultrarelativistic approximation, is closely linked to the relativity of simultaneity.
When particle trajectories are analyzed in a Lorentz reference frame—an inertial frame where time coordinates are assigned using the Einstein synchronization procedure—relativistic kinematic effects, such as the relativity of simultaneity, must be considered. 
In the ultrarelativistic limit, the orientation of the modulation wavefront, i.e. the orientation of the plane of simultaneity, is always perpendicular to the electron beam velocity when the evolution of the beam's evolution is described using Lorentz coordinates.

It is important to recognize that Maxwell's equations are valid only within Lorentz reference frames. Consequently, Einstein's concept of time ordering must be applied consistently in both dynamics and electrodynamics. According to Maxwell's equations, the wavefront of a laser beam is always orthogonal to its direction of propagation. Notably, in the framework of special relativity, a modulated electron beam in the ultrarelativistic (asymptotic) limit exhibits the same kinematic behavior as a laser beam when described in Lorentz coordinates. Experimental evidence confirms the accuracy of this prediction.

The theory of relativity, formulated within a space-time framework based on pseudo-Euclidean geometry, has been developed for over a century. Recently, it has experienced a rapid expansion in practical applications, particularly in XFEL physics.

\newpage

\section{A Critical Survey of Present Approaches to Special Relativity}

\subsection{What is Special Relativity?}

Special relativity is founded on the principle that the laws of physics are invariant under Lorentz transformations. This principle is inherently restrictive: it does not dictate the specific form of the dynamics involved, but rather constrains any physical theory to be consistent with Lorentz symmetry.

Understanding the postulates of special relativity is conceptually similar to understanding the principle of energy conservation. Initially, we encounter energy conservation as a general rule. Later, we seek deeper, microscopic theories—such as classical mechanics or quantum field theory—that not only comply with this principle but also explain the underlying mechanisms. These deeper theories must, of course, recover energy conservation as a consequence of their more fundamental laws.

The utility of the energy conservation principle lies in its ability to guide analysis, even when the full fundamental theory is not yet known. A similar methodological perspective applies to the postulates of special relativity. For example, suppose we do not fully understand the microscopic mechanism behind muon decay. However, if we know the decay law in the muon's rest frame, we can treat this as a phenomenological law. By applying the principles of special relativity, we can generalize this decay law to any Lorentz frame and make testable predictions.

In particular, if we transform the decay law to the laboratory frame using a Lorentz boost, we find that the population of muons is reduced to half its original value after they travel a distance of $\gamma v\tau_0$, where $\gamma$ is the Lorentz factor and  $\tau_0$ is the muon’s proper lifetime. This implies that, from the lab’s perspective, the muon’s lifetime appears to be extended to $\gamma\tau_0$.
Such predictions are direct consequences of the relativistic transformation of time intervals.

Despite its successes, special relativity is not a complete theory. It is a restrictive framework—more a set of symmetry constraints than a constructive (microscopic) explanation. Constructive theories, such as electrodynamics or quantum field theory, provide a more detailed understanding of physical phenomena. Special relativity, in this context, serves primarily as a kinematical interpretation of how physical processes appear from different inertial frames.

Importantly, all relativistic effects can, in principle, be derived directly from underlying physical laws—without invoking relativity as an independent framework—provided those laws are Lorentz covariant. For instance, muons in motion exhibit relativistic behavior because the quantum field equations governing their interactions and decay are Lorentz invariant.

In the "microscopic" view of relativistic phenomena, the Lorentz covariance of the fundamental laws of physics remains, much like energy conservation, an empirical and unexplained fact—but ultimately, all explanations must terminate at some foundational level.

\subsection{Different Approaches to Special Relativity}

In literature, three main approaches to special relativity are discussed: Einstein's approach, the standard covariant approach, and the space-time geometric approach.

Einstein's formulation is based on two fundamental postulates: the principle of relativity and the constancy of the speed of light.

The standard covariant formulation describes special relativity in terms of pseudo-Euclidean space-time geometry and the invariance of the space-time interval $ds$. However, this interpretation is often limited to cases where the metric is strictly diagonal. Assuming a diagonal metric implicitly imposes Lorentz coordinates, ensuring that different inertial frames are related by Lorentz transformations.

The space-time geometric approach, on the other hand, places primary emphasis on the geometry of space-time. In this framework, space-time is assumed to have a pseudo-Euclidean structure, and only four-tensor quantities are considered physically meaningful. Unlike Einstein’s formulation, where the principle of relativity is postulated, in this approach, it naturally follows from the geometric properties of space-time.

Since the space-time geometric approach accommodates all possible coordinate choices within a given reference frame, Einstein’s second postulate—asserting the constancy of the coordinate speed of light—does not generally hold. The coordinate speed of light remains isotropic and constant only in Lorentz coordinates, where Einstein’s synchronization of distant clocks and Cartesian space coordinates are used.

\subsubsection{The Usual Einstein's Approach}

Traditionally, the special theory of relativity is founded on two key postulates:

1. Principle of relativity. The laws of nature are the same  (or take the same form) in all inertial frames

2. Constancy of the speed of light. Light propagates with constant velocity $c$ independently of the direction of propagation, and of the velocity of its source.

However, the constancy of the speed of light in all inertial frames is not a fundamental statement of relativity. The core principle of special relativity is the Lorentz covariance of all fundamental physical laws. Contrary to common textbook presentations, the second postulate is not an independent physical assumption but rather a convention that cannot be experimentally tested.

By assuming the constancy of the speed of light in all inertial frames, we implicitly adopt Lorentz coordinates and assume that different inertial frames are related by Lorentz transformations. In this restrictive interpretation of relativity, only Lorentz transformations are used to map event coordinates between inertial observers.

\subsubsection{The Usual Covariant Approach}

In the standard covariant approach, special relativity is formulated as a theory of space-time with pseudo-Euclidean geometry. Physical quantities are represented by tensors in four-dimensional space-time, making them covariant, and the laws of physics are expressed as four-tensor equations to ensure manifest covariance.

To develop space-time geometry, a metric or measure $ds$ of space-time intervals must be introduced. The choice of measure determines the nature of the geometry. In this framework, an event is mathematically represented by a point in space-time, called a world-point, while the evolution of a particle is depicted as a curve in space-time, known as a world-line. If $ds$ represents the infinitesimal displacement along a particle's world-line, then

\begin{eqnarray}
&& ds^2 =  c^2 dT^2 - dX^2 - dY^2 - dZ^2~ ,\label{MM1}
\end{eqnarray}
where we have selected a special type of coordinate system (a Lorentz coordinate system),  defined by the requirement that Eq. (\ref{MM1}) holds.

To simplify our writing we will use, instead of variables $T, X, Y, Z$,  variables $X^{0} = cT, ~ X^{1} = X,~ X^{2} = Y,~ X^{3} = Z$. Then, by adopting the tensor notation, Eq. (\ref{MM1}) becomes $ds^2 = \eta_{ij}dX^{i}dX^{j}$, where Einstein summation is understood. Here $\eta_{ij}$ are the Cartesian components of the metric tensor and by definition, in any Lorentz system, they are given by $\eta_{ij} = \mathrm{diag}[1,-1,-1,-1]$, which is the metric canonical, diagonal form. As a consequence of the space-time geometry, Lorentz coordinate systems are connected by Lorentz transformations.

Physical quantities are represented by space-time geometric (tensor) entities. When a basis is introduced, a tensor as a geometric object consists of both its components and the basis itself.
In the conventional covariant approach, one typically considers only the components of tensors in a Lorentz coordinate system—i.e., under the assumption that the basis four-vectors are orthogonal. Consequently, physical laws are expressed exclusively through four-tensor equations written in component form.

However, in this approach, the definition of a tensor relies on the transformation properties of its components. For instance, in standard covariant formalism, the electromagnetic "tensor" $F^{\mu\nu}$ is not actually a full tensor; rather, it consists only of component values implicitly taken in an orthogonal basis. These components are coordinate-dependent quantities and do not fully describe the physical entity, as they lack explicit reference to the space-time basis itself. This limitation is not an issue when restricting transformations to those between orthogonal bases, such as Lorentz transformations, which are assumed to correctly map event coordinates. However, according to the conventional covariant approach, transformations from a standard orthogonal basis to a non-orthogonal one—such as Galilean transformations—are considered "incorrect."

\subsubsection{The Space-Time Geometric Approach}

We emphasize the significant freedom in choosing a coordinate system for Minkowski space-time. The space-time continuum, defined by the interval in Eq. (\ref{MM1}), can be described using arbitrary coordinates, not just Lorentz coordinates. When transitioning to an arbitrary coordinate system, the geometry of four-dimensional space-time remains unchanged. In special relativity, there are no restrictions on the choice of coordinate system: the spatial coordinates $x^1, x^2, x^3$ can be any parameters defining the position of particles, while the time coordinate  $x^0$ can correspond to any arbitrary clock.

The components of the metric tensor in the coordinate system $x^i$ can be determined by performing the transformation from the Lorentz coordinates  $X^{i}$ to the arbitrary variables $x^{j}$, which are fixed as $X^{i} = f^{i}(x^{j})$. One then obtains

\begin{eqnarray}
&& ds^2 = \eta_{ij}dX^{i}dX^{j} = \eta_{ij}\frac{\partial X^{i}}{\partial x^{k}}\frac{\partial X^{j}}{\partial x^{m}}dx^{k}dx^{m} = g_{km}dx^{k}dx^{m} ~ ,\label{MM3}
\end{eqnarray}

In textbooks and monographs, special relativity is typically presented in terms of an interval $ds$ expressed in the Minkowski form of Eq. (\ref{MM1}), whereas Eq. (\ref{MM3}) is often associated with general relativity. However, in the space-time geometric approach, special relativity is fundamentally a theory of four-dimensional space-time with pseudo-Euclidean geometry. In this formulation, the space-time continuum can be described equivalently in any coordinate system, provided that $ds$ remains invariant. In contrast, the conventional presentation of special relativity also relies on the invariance of $ds$ but interprets it in a more restricted sense, assuming a strictly diagonal metric.

Textbook presentations of the special theory of relativity typically follow Einstein's approach or, more generally, the standard covariant approach, which, as discussed above, deals only with the components of 4-tensors in specific (orthogonal) Lorentz bases. However, these presentations often overlook the fact that transitioning to arbitrary coordinates does not alter the underlying geometry of space-time.
As a result, a common misconception among experts is that transforming from an orthogonal Lorentz basis to a nonorthogonal one is incorrect, while a Lorentz transformation—between two orthogonal Lorentz bases—is valid. This belief is unfounded. Physics can be described in any coordinate system; changes in coordinates merely affect how the components of 4-tensors are expressed, without altering the tensors themselves.
Although Einstein synchronization (i.e., the choice of Lorentz coordinates) is preferred by physicists for its simplicity and symmetry, it is no more "physical" than any other coordinate system. In this book, we will explore an especially unconventional choice: the absolute time coordinate system.

\subsection{The Myth About the Incorrectness of Galilean Transformations}

The role of Galilean transformations within the framework of relativity requires careful reconsideration. Many physicists continue to view Galilean transformations—specifically, transformations from an orthogonal Lorentz basis to a non-orthogonal one—as outdated and incorrect when applied to spatial coordinates and time. A common argument against their validity is that they fail to preserve the form-invariance of Maxwell’s equations under a change of inertial frame. This reasoning is frequently found in well-known textbooks and monographs.
\footnote{It is widely believed that Galilean transformations, being pre-relativistic, are incompatible with the special theory of relativity. For instance, Bohm \cite{Bohm} states: “… the Galilean law of addition of velocities implies that the speed of light should vary with the speed of the observing equipment. Since this predicted variation is contrary to fact, the Galilean transformations evidently cannot be correct.” Similar assertions appear in more recent pedagogical literature. Drake and Purvis \cite{DR}, for example, write: “One of the great insights of relativity theory was the realization that Galilean transformations are incorrect. The proper way to translate space-time measurements between inertial frames is through Lorentz transformations.”}

However, the assertion that Galilean transformations are inherently incorrect is a misconception. Special relativity is fundamentally a theory of four-dimensional space-time with a pseudo-Euclidean geometry. From this perspective, the principle of relativity naturally emerges as a geometric property of space-time, which can be described using arbitrary coordinate systems. 
Thus, contrary to the standard presentation in most textbooks, Galilean transformations are in fact compatible with the principle of relativity, even though they modify the explicit form of Maxwell’s equations.

The mathematical claim, commonly found in standard textbooks, that a transition to arbitrary coordinates changes the geometry of space-time is erroneous. As L. Landau and E. Lifshitz state \cite{LL}:

``This formula is called the Galilean transformation. It is easy to verify that this transformation, as was to be expected, does not satisfy the requirements of the theory of relativity; it does not leave the interval between events invariant.''

This conclusion results from a failure to distinguish between the convention-dependent components of the metric tensor and the convention-invariant geometry of space-time. In pseudo-Euclidean geometry, the space-time interval between two events remains invariant under an arbitrary coordinate transformation, although the components of the metric tensor generally change.

We illustrate this point with a simple example.

A comparison with ordinary three-dimensional Euclidean space is particularly instructive. It is well known that Euclidean space may be described equally well using Cartesian coordinates $(x,y,z)$, cylindrical coordinates $(r,\phi,z)$, spherical coordinates $(\rho,\theta,\phi)$, or any other admissible coordinate system. Depending on the choice of coordinates, the line element takes different forms:

\[
ds^2 = dx^2 + dy^2 + dz^2,
\]

\[
ds^2 = dr^2 + r^2 d\phi^2 + dz^2,
\]

and

\[
ds^2 = d\rho^2 + \rho^2 d\theta^2   + \rho^2 \sin^2\theta d\phi^2.
\]

Clearly, the geometry of Euclidean space has not changed. What changes is merely the representation of the metric in terms of the chosen coordinates. In general, the line element is written as

\[
ds^2 = g_{ik} dx^i dx^k.
\]

In Cartesian coordinates, the components of the metric tensor are constant:

\[
g_{ij} = \mathrm{diag}(1,1,1).
\]

In cylindrical or spherical coordinates, the components of the metric tensor are different and depend on the coordinates themselves. Nevertheless, all these coordinate systems describe precisely the same Euclidean geometry.

The same distinction must be made in Minkowski space-time. Lorentz transformations relating inertial frames equipped with Einstein-synchronized coordinates leave the components of the metric tensor unchanged. However, a more general coordinate transformation need not preserve the numerical form of these components. The metric components may change, and off-diagonal terms may appear, but this does not imply any change in the underlying pseudo-Euclidean geometry. The invariant space-time interval remains the same geometrical quantity, independently of the coordinates used to describe it.

\subsection{The Non-Relativistic Limit of Lorentz Transformations}

It is commonly asserted in textbooks that the Lorentz transformation reduces to the Galilean transformation in the non-relativistic limit. We argue that this widely accepted claim is both incorrect and misleading. Kinematics is inherently a comparative study of two coordinate systems and therefore requires the assignment of time coordinates to both systems. These time coordinates depend on the adopted clock-synchronization convention, and different synchronization schemes lead to different time assignments.

In Minkowski space-time, the choice of a synchronization convention is equivalent to selecting a particular coordinate system within a given inertial frame of reference. Practical considerations may favor one coordinatization over another, depending on the physical context. In relativistic engineering, for example, one commonly works either with an absolute-time coordinatization or with Lorentz time. Importantly, the space-time continuum can be described consistently in either framework. This implies that, even within the theory of relativity and for arbitrary particle velocities, Galilean coordinate transformations may effectively characterize the change of reference frame from a laboratory inertial observer to a comoving inertial observer, provided a suitable synchronization convention is adopted.

A review of the literature reveals that the treatment of the non-relativistic limit of Lorentz transformations is surprisingly confused. In particular, Ungar’s works—written at a high scientific level—and his recently published monographs (\cite{UNG1,UNG2}), which summarize his long-standing research on special relativity, devote significant attention to the relationship between Galilean and Lorentz transformations.

As an illustration, consider the following statement by Ungar from a 2011 paper \cite{UN}:

``Hence, we emphasize that in the Newtonian limit of large vacuum speed of light $c, ~ c \to \infty$ , the Lorentz boost $L(\vec{v})$, reduces to the Galilean boost $G(\vec{v})$ .... As we see from (3.2) - (3.3),   our space time coordinates are $(t, \vec{x})$ . In contrast, some authors present spacetime coordinates as $(ct, \vec{x})$, resulting in a symmetric Lorentz boost matrix presentation.  Since in our approach to special relativity analogies with classical results form the right tool, the representation of spacetime coordinates as $(t, \vec{x})$ is more advantages than the representation as $(ct, \vec{x})$. Indeed, unlike the latter, the former allows one to recover the Galilei boost by taking the Newtonean limit of large speed of light $c$, as shown in (3.3).  ''  \footnote{Similar statement about   Newtonian limit of large vacuum speed of light  is found in  the world-famous course of theoretical physics. For instant,  Landau and Lifshitz state \cite{LL}: ``It is easy to see from (4.3) that on making the transition to the limit $c \to \infty$, the formula for the Lorentz transformation actually goes into the Galileo transformation.''}

Several comments are in order. It is puzzling that the analysis is carried out using space and time variables with different physical dimensions, while treating the dimensional constant  $c$ as a parameter whose numerical value depends on the chosen time units. From a mathematical standpoint, asymptotic analysis requires that variables be expressed in consistent units. When time and space are measured in the same units, the Lorentz transformation depends only on the single dimensionless parameter $v/c$. A dimensionless parameter can be meaningfully regarded as small or large only in comparison with unity.

Let us therefore consider the non-relativistic limit  $v/c \ll 1$ that for speeds small compare to $c$.
For $v/c$ small enough that terms of order $v^2/c^2$ can be neglected, the Lorentz transformation simplifies to $x' = x - x_0v/c$, $x_0' = x_0 - xv/c$, where $x_0 = ct$ 
is the time coordinate measured in units of length. This infinitesimal Lorentz transformation differs from the infinitesimal Galilean transformation $x' = x - x_0(v/c)$, $x_0' = x_0$. The difference lies in the additional term $xv/c$ in the Lorentz transformation for time, which is a first-order correction.

We emphasize the key point: at first order in  $v/c$, an infinitesimal Lorentz transformation differs from the Galilean transformation solely through the relativity of simultaneity. This is the only genuinely relativistic effect appearing at this order. Other effects—such as length contraction and time dilation—arise from higher-order terms and emerge mathematically only through the iteration of infinitesimal Lorentz transformations. \footnote{There is a common misconception that the Lorentz transformation reduces to the Galilean transformation in the non-relativistic limit. For example, French \cite{FR} states, "The reduction of $t'= \gamma(t - vx/c^2)$ to the Galilean relation $t' = t$ requires $x \ll ct$ as well as $v/c \ll 1$".  Similar statements are found in recent textbooks, such as Rafelski \cite{RAFELSKI}, who asserts: "A wealth of daily experience shows that the Galilean coordinate transformation (GT) is correct in the nonrelativistic limit in which the speed of light is so large that it plays no physical role. Any coordinate transformation replacing the GT must also agree with this experience, and thus must contain the GT in the nonrelativistic limit." This conclusion is mathematically flawed. As Baierlein \cite{BA} points out: "If the Lorentz transformation for infinitesimal $v/c$ were to reduce to the Galilean transformation, then the iterative process could never generate a finite Lorentz transformation that is radically different from the Galilean transformation."}

The fundamental distinction between Lorentz coordinatization and absolute-time coordinatization lies in the transformation laws relating space and time between relatively moving systems. It is therefore incorrect to claim, as is often done, that the expression  $t'= \gamma(t - vx/c^2)$ reduces to the Galilean relation  $t' = t$
 in the non-relativistic limit. Such a claim would imply that infinitesimal Lorentz transformations coincide with infinitesimal Galilean transformations, which they do not. The essence of Lorentz (and Galilean) transformations is already contained in their infinitesimal form, and genuinely relativistic kinematic effects cannot be generated by iterating infinitesimal Galilean transformations alone.

\subsection{The Myth About the Constancy of the Speed of Light}

It is widely believed that experiments demonstrate the independence of the speed of light in a vacuum from both the source and the observer \footnote{Many textbook authors still treat conventional quantities as if they are empirically measurable. For instance, Cristodoulides \cite{Cr} states: "The fact that the Galilean transformation does not preserve Maxwell's equations has already been mentioned [...] On the other hand, experiments show that the speed of light in a vacuum is independent of the source or observer."}. This claim is commonly presented in textbooks but is, in fact, incorrect. The constancy of the speed of light is not an empirical fact but rather a consequence of the chosen synchronization convention, making it impossible to test experimentally \footnote{Since we can empirically access only the round-trip average speed of light, any claims about the magnitude and isotropy of the one-way speed of light depend on the chosen time coordinate system. Such values vary with the synchronization scheme used. As Anderson, Vetharaniam, and Stedman \cite{AV} state: "No experiment, then, is a 'one-way' experiment. An empirical test of any property of the one-way speed of light is not possible. Such quantities as the one-way speed of light are irreducibly conventional in nature, and recognizing this aspect is to recognize a profound feature of nature."}.

To measure the one-way speed of light, one must first synchronize an infinite set of clocks distributed across space to enable time measurements. However, an inherent circularity arises when clocks are synchronized under the assumption that the one-way speed of light is $c$. If the synchronization process is based on this assumption, then any measurement using these clocks will necessarily yield $c$, as the measurement framework has been defined accordingly.

Thus, the one-way speed of light is a matter of convention rather than a physical quantity with empirical meaning. In contrast, the two-way (or round-trip) speed of light has physical significance because it is directly measurable without reliance on clock synchronization. Round-trip experiments involve observing simultaneity (or lack thereof) at a single spatial location, making them independent of any synchronization convention. All well-established methods for measuring the speed of light are, in fact, round-trip measurements. A prime example is the Michelson-Morley experiment, which employs an interferometer to compare light beams in a two-way manner.

Apparent paradoxes emerge when Galilean transformations are applied within electrodynamics. Many sources claim that Galilean velocity transformations are inconsistent with the electron-theoretic explanations of optical phenomena such as refraction and reflection. This notion is widespread in standard literature. Pauli, for instance, states \cite{PA}: "It is essential that the spherical waves emitted by the dipoles in the body interfere with the incident wave. If we consider the body at rest while the light source moves relative to it, the waves emitted by the dipoles will have a velocity different from that of the incident wave, making interference impossible."

However, such claims stem from a misunderstanding. It is often asserted that when a light source moves relative to a stationary medium (like glass), the resulting wave emitted by the oscillating electrons in the glass cannot interfere properly with the incident wave due to differing velocities. This is incorrect. Regardless of the light's velocity, the frequency of the incident wave determines the oscillation frequency of the electrons in the medium. These electrons then re-emit light of the same frequency. Therefore, both the incident and re-emitted waves maintain the same frequency and can interfere. The velocity difference merely results in a spatially varying phase difference, not a breakdown in interference.

\subsection{Convention-Dependent  Aspects of the Theory}

Consider the motion of a charged particle in a given magnetic field. According to the theory of relativity, the trajectory $\vec{x}(t)$ of the particle in the lab frame depends on the chosen synchronization convention for clocks in that frame. Whenever a theory involves an arbitrary convention, it is essential to examine which aspects of the theory depend on that choice and which do not. The former are convention-dependent, while the latter are convention-invariant. Clearly, physically meaningful measurement results must be convention-invariant.

Now, consider the motion of two charged particles in a given magnetic field, which controls their trajectories. Suppose there are two apertures located at points $A$ and $A'$. From the solution of the equations of motion, we conclude that the first particle passes through aperture $A$ while the second passes through aperture $A'$ simultaneously. The occurrence of these two events—particles passing through $A$ and $A'$—has an exact objective meaning, making them convention-invariant. However, the simultaneity of these events is convention-dependent and lacks an exact objective meaning. It is important to emphasize that, consistent with the conventionality of simultaneity, the value of the particle’s speed is also a matter of convention and does not possess a definite objective meaning.

To determine which aspects of dynamical theory depend on a chosen convention and which do not, we examine the distinction between the notions of path and trajectory.
Consider the motion of a particle in three-dimensional space, described by the vector-valued function $\vec{x}(t)$
The particle follows a prescribed curve, or path, as it moves. This motion along the path is parameterized by $l(t)$, where $l$ represents a certain parameter—specifically, in our case of interest, the arc length.

Now, if we take the origin of a Cartesian coordinate system and connect it to a point on the path, the resulting vector is the position vector $\vec{x}(l)$. \footnote{For a more detailed discussion on the distinction between path and trajectory, see \cite{JA}.} The key difference between trajectory $\vec{x}(t)$ and path $\vec{x}(l)$
is fundamental: the path has an exact, objective meaning—it is independent of any convention. In contrast, and consistent with the inherent conventionality of velocity, the trajectory $\vec{x}(t)$ is convention-dependent and lacks an exact objective interpretation.

To illustrate this distinction, consider experiments in accelerator physics. Suppose we wish to measure a particle’s momentum. A uniform magnetic field can be used to construct a momentum analyzer for high-energy charged particles. Importantly, this method of determining momentum is convention-independent. The curvature radius of the path within the magnetic field—and consequently, the three-momentum—has an objective, convention-invariant meaning. While dynamical theory includes the concept of a particle’s trajectory, this quantity is not directly measured but instead serves as a tool in the analysis of electrodynamics problems.

\subsection{Relativistic Time Dilation and  Length Contraction}

Experts in the theory of relativity often mistakenly conflate the fundamental properties of Minkowski space-time with the specific forms that certain convention-dependent quantities take under standard Lorentz coordinatization. These quantities, commonly referred to as "relativistic kinematic effects," include time dilation, length contraction, and Einstein’s velocity addition. There is a widespread misconception that such effects have direct physical significance, rather than being artifacts of the chosen synchronization convention.

Relativistic kinematic effects are inherently coordinate-dependent and lack absolute physical meaning. As noted by Leubner, Auflinger, and Krumm \cite{LAK}, many physicists incorrectly equate the standard formulations of these effects with relativity itself. Their study demonstrates that adopting a different synchronization convention can lead to significant changes in the appearance of these quantities.

For instance, under Lorentz coordinatization, one observes time dilation. However, in a framework based on absolute time coordinatization, relativistic kinematic effects disappear—no time dilation is observed. Despite such variations, all coordinate-independent quantities, such as the particle path $\vec{x}(l)$ and momentum  $\vec{p}$, remain unaffected by changes in clock synchronization. 

It is a misconception to believe that physics must rely solely on concepts that are directly measurable. Special relativity, when formulated in Lorentz coordinates, includes constructs such as time dilation and length contraction that are not directly observable. Nevertheless, these constructs are essential for making theoretical predictions. The strength of a theory lies in its predictive power, not in the observability of every individual element.

This idea can be further illustrated. Both the velocity of an electron and the speed of light are convention-dependent quantities. In absolute-time coordinatization, the speed of light differs from the electrodynamic constant $c$, as Maxwell’s equations are not invariant under Galilean transformations. Conversely, in Lorentz coordinatization, the speed of light equals $c$, while the electron’s velocity is changed due to relativistic effects. Yet the dimensionless ratio of the electron's velocity to the speed of light remains convention-independent. This ratio, unaffected by changes in clock synchronization or rhythm, forms the basis for physically meaningful predictions in electrodynamics—for example, in the analysis of synchrotron radiation emitted by bending magnets.

\subsection{Relativistic Particle Dynamics}

Problems in physics can be separated into two classes: 

- solution of dynamical problem, i.e. finding the motion of the particles under the action of the given electromagnetic fields;

- solution of the electrodynamic problem, i.e. finding the electromagnetic fields generated by a given distribution of charges and currents.  \footnote{The problem of electromagnetic wave amplification in free electron lasers  (FEL) refers to a class of self-consistent problems. In this case, the field equations and equations of motion should be solved simultaneously. The motion of the particles is determined not only by the external electromagnetic fields, but also by the fields produced by the electron beam itself.}   

The accelerated motion of a relativistic charged particle is governed by a covariant equation of motion under the influence of the four-force in the Lorentz lab frame. The particle’s trajectory, $\vec{x}_{cov}(t)$, as observed in this frame, results from a sequence of infinitesimal Lorentz transformations. Importantly, the lab frame time $t$ in the equation of motion cannot be treated as independent of the spatial coordinates. This dependence arises because Lorentz transformations inherently mix space and time coordinates, with relativistic kinematic effects manifesting as a consequence of the relativity of simultaneity.

Let us consider the conventional approach to particle tracking. It is generally accepted that, to describe the dynamics of relativistic particles in an inertial laboratory reference frame, one only needs to account for the relativistic dependence of momentum on velocity. The treatment of relativistic particle dynamics is based on a modified form of Newton's second law.
In a given lab frame, there is an electric field $\vec{E}$ and magnetic field $\vec{B}$. They push on a particle in accordance with

\begin{eqnarray}
&& \frac{d\vec{p}}{dt} = e\left(\vec{E} + \frac{\vec{v}}{c}\times \vec{B}\right) ~,\cr &&
\vec{p} = m\vec{v}\left(1 - \frac{v^2}{c^2}\right)^{-1/2}~ ,\label{N1}
\end{eqnarray}

where here the particle's mass, charge, and velocity are denoted by $m$, $e$, and $\vec{v}$  respectively.
The Lorentz force law, together with measurements of the acceleration components of test particles, serves to define the components of the electric and magnetic fields. Once these field components are determined from the acceleration of test particles, they can be used to predict the accelerations of other particles.

This solution to the dynamics problem in the lab frame does not rely on Lorentz transformations. Conventional particle tracking treats space-time in a non-relativistic manner, as a (3+1) manifold.
In this approach, the only modification to classical mechanics is the introduction of relativistic mass, while time remains distinct from space. There is no mixing of spatial and temporal coordinates.
\footnote{Clarification of the true content of the non-covariant theory can be found in various advanced textbooks. 
To quote e.g. Ferrarese and Bini \cite{RCM}: "... within a single inertial frame, the time is an absolute quantity in special relativity also. As a consequence, \textit{if no more than one frame is involved}, one would not expect differences between classical and relativistic kinematics. But in the relativistic context, there are differences in the transformation laws of the various relative quantities (of kinematics or dynamics), when passing from one reference frame to another." These authors emphasize the unique role of a "single inertial frame." This term suggests that a defining characteristic of non-covariant theory is the absence of relativistic kinematics in describing particle motion.
A similar viewpoint is expressed by Friedman \cite{Frid}: "Within any single inertial frame, things look precisely the same as in Newtonian kinematics: there is an enduring Euclidean three-space, a global time $t$, and the law of motion."}

Many fascinating phenomena arise when charges move under the influence of electromagnetic fields, often in highly complex situations. However, here we focus on a simpler case: the accelerated motion of particles in a constant magnetic field. In a non-covariant treatment, a magnetic field can change only the direction of a particle’s motion, not its speed (or mass). The study of relativistic particle motion in a constant magnetic field—common in accelerator physics—appears similar to its non-relativistic counterpart in Newtonian dynamics and kinematics. The trajectory of a particle,  $\vec{x}(t)$, derived from the corrected form of Newton’s second law, does not account for relativistic kinematic effects, as it relies on the Galilean vector addition of velocities and disregards the relativity of simultaneity.

Now, let us address the important problem of velocity addition in relativity. It is evident that the velocity of light transforms just as a particle’s velocity does. We cannot use one set of kinematics for a particle’s velocity and another for the velocity of light. Consider a scenario where a light source in the lab frame is accelerated from rest to a velocity $u$. Suppose that, in the Lorentz inertial frame where the light source is at rest, an observer sees light propagating with velocity  $v = c$. How will this appear to an observer in the lab frame? According to the relativistic velocity addition formula, the answer remains $v = c$.
Maxwell’s equations retain their form under Lorentz transformations. However, these transformations lead to non-Galilean velocity transformation rules. Consequently, special relativity dictates that if Maxwell’s equations are to hold in the lab frame, particle trajectories must incorporate relativistic kinematic effects. That is, Maxwell’s equations can be consistently applied in the lab frame only if particle trajectories are described as the result of successive infinitesimal Lorentz transformations.

The absence of relativistic kinematic effects is a fundamental flaw in conventional (3+1) non-covariant particle tracking, making it clearly inconsistent with Maxwell's electrodynamics. The relativity of simultaneity—i.e., the interdependence of spatial positions and time—is entirely absent from this framework. Consequently, the approach of coupling fields and particles within a "single inertial frame" is not just flawed but fundamentally and profoundly incorrect.

\subsection{Commonly Used Method of Coupling Fields and Particles}

The electrodynamics problem is generally considered solvable within a "single inertial frame" framework, without requiring Lorentz transformations. Standard derivations assume that Maxwell's equations, along with a corrected version of Newton's second law, can fully account for all experiments conducted within a single inertial frame, such as the laboratory reference frame.

Going to the electrodynamics problem, the differential form of  Maxwell's equations  describing electromagnetic phenomena in the same inertial lab frame (in cgs units)  is given by the following expressions:

\begin{eqnarray}
&& \vec{\nabla}\cdot \vec{E} = 4 \pi \rho~, \cr && \vec{\nabla}\cdot
\vec{B} = 0~ , \cr && \vec{\nabla}\times \vec{E} =
-\frac{1}{c}\frac{\partial \vec{B}}{ \partial t}~,\cr &&
\vec{\nabla}\times \vec{B} = \frac{4\pi}{c}
\vec{j}+\frac{1}{c}\frac{\partial \vec{E}}{\partial t}~. \label{CD11}
\end{eqnarray}

Here the charge density $\rho$ and current density $\vec{j}$ are written as

$ $

\begin{eqnarray}
&&\rho(\vec{x}, t) = \sum_{n} e_n\delta(\vec{x} - \vec{x}_n(t)) ~,\cr &&
\vec{j}(\vec{x},t) = \sum_{n} e_n\vec{v}_n(t)\delta(\vec{x} - \vec{x}_n(t))~, \label{CD}
\end{eqnarray}

where $\delta(\vec{x} - \vec{x}_n(t))$ is three-dimensional delta function, $m_n, e_n, \vec{x}_n(t)$, and $\vec{v}_n =   d\vec{x}_n(t)/dt$ denote the mass, charge, position, and the velocity of the $n$th particle, respectively.
To evaluate radiation fields arising from external sources Eq. (\ref{CD11}) we need the velocity $\vec{v}_n$ and the position $\vec{x}_n$ as a function of lab frame time $t$. It is generally accepted by the physics community that the equation of motion, which describes how coordinates of the particle carrying the charge change with time $t$, is described by corrected Newton's second law Eq. (\ref{N1}).

This coupling of Maxwell's equations with the corrected Newtonian equation is widely recognized as a valuable approach in accelerator physics, particularly in analytical and numerical studies of radiation properties. This method of treating relativistic dynamics and electrodynamics often leads accelerator physicists to assume that designing particle accelerators is possible without a deep understanding of the theory of relativity.

However, a common misconception in accelerator physics arises from the distinction between the trajectories $\vec{x}(t)$ and $\vec{x}_{cov}(t)$. To examine this issue from the perspective of electrodynamics of relativistically moving charges, consider the evaluation of fields generated by external sources. This requires knowledge of their velocity and position as functions of the lab-frame time $t$.
Now, suppose one aims to compute the properties of radiation. Given our previous discussion, an important question emerges: Should one solve Maxwell’s equations in the lab frame using the current and charge density associated with a particle moving along the non-covariant trajectory $\vec{x}(t)$? We argue that the answer is no.

The conventional approach—solving Maxwell’s equations in the lab frame with currents and charge densities based on non-covariant trajectories—is widely regarded as relativistically correct. However, this method contradicts the fundamental principles of special relativity. Remarkably, this critical issue has largely gone unnoticed in the physics community. The only correct way to ensure consistency between Maxwell’s equations and particle trajectories in the lab frame is to solve the equations of motion in fully covariant form.

\subsection{Geometry of Spacetime and the Equivalence of Inertial Frames}

It is important to distinguish between observer-independence and Lorentz covariance. The Principle of Relativity is often interpreted as implying the complete physical equivalence of all inertial frames. Strictly speaking, however, Lorentz covariance establishes only that the fundamental laws of nature have the same mathematical form in every inertial frame.

According to the covariance principle, the equations expressing the laws of physics are invariant under Lorentz transformations. Consequently, once these laws have been formulated in one inertial frame, they can be applied without modification in any other inertial frame. Expressed in the coordinates and time variables of different inertial observers, the equations retain the same mathematical structure.

Taylor and Wheeler \cite{TW}, for example, write:

``Einstein's Principle of Relativity says that once the laws of physics have been established in one inertial frame, they can be applied without modification in any other inertial frame. Both the mathematical form of the laws of physics and the numerical values of the basic physical constants that these laws contain are the same in every inertial frame. So far as concerns the laws of physics, all inertial frames are equivalent.''

Notice, however, what the Principle of Relativity does \emph{not} imply. It does not require that the physical circumstances—or the solutions describing a particular physical situation—be identical in different inertial frames. The geometry of Minkowski spacetime guarantees Lorentz covariance of the fundamental equations, but it does not by itself guarantee that every physical solution possesses the same symmetry.

This distinction is well illustrated by the role of initial conditions. Although Maxwell's equations are Lorentz covariant, they do not uniquely determine a physical solution. A complete description requires both the field equations and appropriate initial or boundary conditions. These additional conditions may themselves break symmetries that are present in the equations.

A familiar example is time-reversal symmetry. Maxwell's equations admit both retarded (outgoing) and advanced (incoming) radiation solutions. In practice, however, electromagnetic radiation emitted by physical sources is described by the retarded solution. The preference for outgoing waves is not dictated by Maxwell's equations alone, but by the physical boundary conditions established by experiment.

A second example concerns the homogeneous solution of Maxwell's equations. In conventional classical electrodynamics it is usually assumed that, in the absence of sources, both the electric and magnetic fields vanish identically. This assumption is not required by Maxwell's equations themselves; it is an additional physical postulate specifying the initial conditions. Alternative choices are possible. For example, one may postulate the existence of a source-free fluctuating electromagnetic field, as in stochastic electrodynamics (see the footnote below). The field equations remain unchanged—the difference lies entirely in the adopted initial conditions. \footnote{One may instead postulate the existence of classical zero-point electromagnetic fields in empty space. In this approach, the vacuum possesses an energy $\hbar\omega/2$ per normal mode. Since these fields satisfy Maxwell's equations, such an assumption is mathematically consistent. Here the Planck constant merely specifies the experimentally required magnitude of the vacuum fluctuations and does not alter the classical character of the theory. This extension, commonly known as stochastic electrodynamics, reproduces a number of quantum phenomena within a classical framework \cite{PWM}.}

The significance of boundary conditions extends beyond electrodynamics. As discussed in Section~2.1, many relativistic effects can be derived directly from the underlying dynamical laws without introducing relativistic kinematics as an independent postulate. In particular, the aberration of particles observed when transforming from an inertial frame to an accelerated frame may be interpreted as the dynamical consequence of forces acting on the system.

This raises a natural question. Why does an accelerated frame differ physically from an inertial frame with no acceleration history? The answer developed in this book is based on the possibility of source-free electromagnetic fields in classical relativistic electrodynamics.

To illustrate the idea, consider an electron gun initially at rest in an inertial frame and emitting relativistic electrons along the $z$-axis. Suppose that the entire apparatus is subsequently accelerated to velocity $\vec{v}$ in the $x$-direction. After the acceleration, both the source and the observer are at rest in the new frame. The electromagnetic forces governing the emitted electron beam are modified during the acceleration process, producing a change in the beam trajectory. As shown in Chapter~9, this change can be explained entirely within classical electrodynamics.

The electromagnetic fields in the accelerated frame may be obtained by requiring continuity of the metric tensor during the transformation. If the original inertial frame contains only an electric field,
\[
\vec{B}=0,
\]
then, to first order in $v/c$, the transformed fields become
\[
\vec{E}_n=\vec{E},
\qquad
\vec{B}_n=-\frac{\vec{v}\times\vec{E}}{c}.
\]

Within this description, particle aberration appears as the action of the Lorentz force produced jointly by the electric field $\vec{E}_n$ and the induced magnetic field $\vec{B}_n$.

At first sight this result appears paradoxical. Both fields are static, yet the magnetic field exists in the absence of electric currents. According to the standard textbook picture, static magnetic fields originate from moving charges. Why, then, does a magnetic field appear here?

The resolution lies in the relativistic transformation properties of measuring devices. A conventional magnetometer is itself affected by the transformation and therefore cannot detect the induced field independently.

To illustrate this point, consider a proton-resonance magnetometer based on nuclear magnetic resonance. In relativistic notation, the magnetic dipole moment is represented by the antisymmetric tensor
\[
m_{\mu\nu}=(\vec{m},\vec{d}),
\]
where $\vec{m}$ and $\vec{d}$ denote the magnetic and electric dipole moments, respectively. If the electric dipole moment initially vanishes, the Lorentz transformation gives
\[
\vec{m}_n=\vec{m},
\qquad
\vec{d}_n=-\frac{\vec{v}\times\vec{m}}{c}.
\]

The interaction energy in the accelerated frame is therefore
\[
U_p=-\vec{m}\cdot\vec{B}_n-\vec{d}_n\cdot\vec{E}_n.
\]

The two contributions cancel exactly, so that the induced magnetic field produces no observable signal in the magnetometer. This conclusion is not restricted to NMR devices; any conventional magnetometer based on static electromagnetic interactions yields the same result.

A relativistic test particle, however, behaves differently. Its motion is influenced simultaneously by $\vec{E}_n$ and $\vec{B}_n$, and the resulting transverse momentum change is precisely the relativistic aberration of the particle trajectory. Consequently, in the relativistic domain the induced magnetic field cannot be measured independently of the accompanying aberration effect. The observable phenomenon is the combined action of both transformed fields.

Special relativity often challenges physical intuition because Lorentz transformations mix spatial and temporal quantities. Many apparent paradoxes associated with source-free fields arise from interpreting spacetime phenomena in purely three-dimensional terms. Once the full four-dimensional geometry of Minkowski spacetime is taken into account, these paradoxes disappear and the theory remains internally consistent.

\subsection{The Minkowski Geometry  and 3-Vector Velocity between the Frames}

It is commonly assumed that a Lorentz transformation relating two inertial observers is parametrized by a reciprocal relative velocity.
For example, Landau and Lifshitz \cite{LL} write:

``We try to find the formula of transformation from an inertial reference frame $K$ to system $K'$ moving relative to the $K$ with velocity $V$ along the $x$ axis. ... The inverse formulas, expressing $x', y', z', t'$ are most easily obtained by changing $V$ to $-V$ (since the $K$ system moves with velocity $-V$ relative to the $K'$ system). The same formulas can be obtained directly by solving equations (4.3) for $x', y', z', t'.$''

Thus, the reciprocity relation

\[
\vec v^{-1}=-\vec v
\]

is usually regarded as self-evident. Implicitly, this assumes that the inverse operation in special relativity has exactly the same geometric meaning as in Galilean--Newtonian mechanics.

At first sight, this reciprocity relation appears to involve only ordinary three-dimensional vectors. Within the framework of special relativity, however, such an interpretation is far from obvious. Since spacetime possesses a pseudo-Euclidean geometry rather than a universal Euclidean space, one must first ask where the vectors $\vec v$ and $\vec v^{-1}$ are actually defined. Are they vectors in spacetime itself, or do they belong to different three-dimensional spaces associated with different observers?

In Newtonian mechanics, the components of the velocity vector transform under Galilean transformations in exactly the same way as the components of the position vector. This correspondence exists because all inertial observers share a common three-dimensional Euclidean space, while time is absolute. Consequently, both position and velocity vectors may be regarded as belonging to the same universal spatial manifold.

In special relativity, the situation is fundamentally different. There is no universal instantaneous three-dimensional space shared by all inertial observers. Instead, each observer possesses a private three-space defined by the hypersurface orthogonal to its timelike four-velocity. This distinction between Newtonian geometry and the pseudo-Euclidean geometry of Minkowski spacetime is often left implicit in standard textbook treatments.\footnote{Most textbooks do not discuss explicitly whether the velocity vector $\vec v$ should be regarded as a vector in spacetime or as a vector belonging to an observer's three-space. For example, Bohm \cite{Bohm} writes: ``Note that the inverse [Lorentz] transformation follows by changing $\vec v$ to $-\vec v$. This should be expected, because if the velocity of frame B relative to frame A is $\vec v$, then the velocity of frame A relative to frame B is $-\vec v$. Therefore, in passing from B to A we use the same functions of the velocity (taking the sign into account) as in passing from A to B.''}

The transformation properties of four-vectors present no such ambiguity. The four-velocity

\[
u^\mu=\frac{dx^\mu}{ds}
\]

transforms between Lorentz frames in exactly the same way as the spacetime position four-vector

\[
x^\mu=(ct,\vec x).
\]

The subtleties arise only when one returns to the familiar three-velocity,

\[
\vec v=\frac{d\vec x}{dt},
\]

whose definition requires a particular decomposition of spacetime into space and time. These subtleties become especially important for non-collinear motion.

Indeed, the reciprocity relation $\vec v^{-1}=-\vec v$ is not automatically compatible with the relativity of simultaneity. A relative velocity is always tangent to the simultaneity hypersurface of the observer who measures it. Since simultaneity is observer-dependent, different observers possess different instantaneous three-spaces. Consequently, the vectors $\vec v$ and $\vec v^{-1}$ belong to different spatial hypersurfaces, which in general cannot be identified with one another.

The history of an inertial reference system is represented in spacetime by a timelike worldline together with its unit four-velocity. The corresponding three-velocity is a spacelike vector belonging to the observer's private three-space. Different choices of the timelike four-velocity therefore generate different $(1+3)$ decompositions of spacetime, and these decompositions are not canonically equivalent because they correspond to different families of simultaneity hypersurfaces.

From this viewpoint, a unique relative three-velocity exists only after a preferred $(1+3)$ decomposition has been specified. Once a preferred reference system is chosen, every material reference system is assigned a unique velocity within the common three-space associated with that observer. In the conventional matrix representation of Lorentz transformations, this preferred decomposition is introduced implicitly.

It is therefore important to distinguish between two different concepts. The vector $-\vec v$ appearing in the inverse Lorentz transformation is not the velocity directly measured by the observer in $K'$, which itself moves relative to the preferred frame $K$ with velocity $\vec{v}$.
Rather, $-\vec{v}$ is the velocity of $K$ relative to $K'$, as reconstructed by the preferred observer $K$.

The standard representation of Lorentz transformations employs $4\times4$ matrices acting on spacetime four-vectors. Every Lorentz transformation may be decomposed into a spatial rotation and a boost. This decomposition, however, requires the selection of a distinguished timelike basis vector,

\[
e_0 = (1,0,0,0),
\]

which is the unit four-velocity of the chosen reference observer. Spatial rotations act within the three-space orthogonal to $e_0$, while boosts are parametrized by velocity vectors lying in that same three-space. Consequently, although the Lorentz group satisfies

\[
(L_{\vec v})^{-1}=L_{-\vec v},
\]

the explicit matrix representation of the transformation, as well as the meaning of the velocity parameter $\vec v$, depends on the chosen $(1+3)$ decomposition of spacetime.

As discussed in the previous section, additional mathematical considerations suggest that the pseudo-Euclidean geometry of spacetime does not by itself single out a unique physical decomposition into space and time. Nevertheless, there is no conflict between the mathematical structure of special relativity and the introduction of a preferred reference system. The Lorentz covariance of the fundamental laws of physics remains intact. What changes is not the form of the dynamical equations, but the specification of the initial conditions. The principle of relativity requires the laws of physics to be Lorentz invariant; it does not require all physically relevant initial conditions to be independent of the choice of inertial frame.

Within the framework adopted in this book, the mathematical formalism of special relativity is therefore developed with respect to an initial inertial frame that has not undergone prior acceleration relative to the distant stars. This frame serves as the preferred reference system.

On the basis of the available experimental evidence, we identify the Sun-centered inertial frame as this preferred system (see Chapter~16). Once such a frame has been specified, every pair of inertial reference systems possesses a unique relative three-velocity defined within the preferred three-space.

For terrestrial applications, the Earth's orbital velocity,

\[
v_{\rm orb}\simeq30~\mathrm{km/s},
\]

is sufficiently small that present experimental techniques are generally unable to detect its influence on relativistic particle dynamics. Consequently, for most problems encountered in accelerator physics, the Earth-based laboratory frame may be regarded, to excellent approximation, as the preferred inertial frame.

The reader may wonder why these conceptual issues are introduced at this stage, before the necessary relativistic formalism has been fully developed. The reason is that the underlying geometry is subtle and is best appreciated gradually. Our purpose here is to establish the geometric framework and to motivate the questions that naturally arise. A detailed quantitative treatment will be presented in Chapter~18.

\newpage

\section{Space-Time and Its Coordinatization}

\subsection{Choosing a Coordinate System in an Inertial Frame}

Let us explore an "operational interpretation" of Lorentz and absolute time coordinatizations.
Every physical phenomenon occurs in both space and time, and a concrete means of representing these dimensions is a frame of reference. A single space-time continuum can be described using different coordinate-time grids, corresponding to various frames of reference. Even the simplest space-time coordinate systems require careful definition.

Clocks play a crucial role in tracking a particle’s motion through a coordinate-time grid. The general approach to determining a particle's motion is as follows: at any instant, a particle possesses a well-defined velocity, $\vec{v}$, as measured in a laboratory frame of reference. But how is this velocity determined? It is found once coordinates are assigned in the lab frame and then measured over appropriate time intervals along the particle’s trajectory. However, measuring a time interval between events occurring at different spatial points poses a fundamental challenge. To do so—and thereby measure a particle’s velocity within a single inertial lab frame—it is first necessary to synchronize distant clocks. The concept of synchronization is central to understanding special relativity.

There are multiple methods to synchronize distant clocks.\footnote{According to the thesis of the conventionality of simultaneity \cite{Frid, M, RE, LOG, BROWN}, simultaneity between distant events is a matter of convention, as it can be legitimately defined in different ways within any given inertial reference frame. Moeller \cite{M}, for example, states: "All methods for the regulation of clocks meet with the same fundamental
difficulty. The concept of simultaneity between two events in different places has
no exact objective meaning at all since we cannot specify any experimental method
by which this simultaneity could be ascertained. The same is therefore true also for
the concept of velocity."} The choice of a synchronization convention is ultimately a matter of selecting a coordinate system within an inertial frame of reference in Minkowski space-time.

The space-time continuum can be described using arbitrary coordinates. However, altering these coordinates does not change the underlying geometry of four-dimensional space-time. In special relativity, we are therefore unrestricted in our choice of a coordinate system. By leveraging the geometric structure of Minkowski space-time, one can define a class of inertial frames and adopt a Lorentz frame with orthonormal basis vectors. Within this chosen Lorentz frame, Einstein’s synchronization procedure—which relies on the constancy of the speed of light in all inertial frames—is applied, along with Cartesian spatial coordinates.

\subsection{The Inertial Frame Where a Light Source is at Rest}

Let us provide an operational interpretation of Lorentz coordinatization. The fundamental laws of electrodynamics are governed by Maxwell’s equations, which state that light propagates at a constant velocity $c$ in all directions. This isotropy arises naturally from the structure of Maxwell’s theory, which lacks any intrinsic directional dependence. It is often asserted that Maxwell’s equations, in their original form, are strictly valid within inertial frames. However, these equations can only be expressed in coordinate form once a space-time coordinate system has been properly defined.

Assigning Lorentz coordinates to the laboratory frame when a light source is in motion presents a complex challenge. To approach this problem systematically, we begin with the simpler case: assigning space-time coordinates to an inertial frame in which the light source remains at rest.

Our goal is to provide a practical and operational procedure for this assignment. The most natural synchronization method involves first bringing all ideal clocks to the same spatial location, where they can be synchronized. Once synchronized, the clocks are then slowly transported to their designated positions—a process known as slow clock transport. \footnote{Eddington, in a text first published in 1923, was apparently the first to describe a synchronization procedure based on the slow transport of clocks \cite{EDD}. Further details can be found in the review \cite{AV} (see also \cite{CLOC}).}

Maxwell’s equations in their usual form are valid in any inertial frame where the sources are at rest, and the time coordinate is assigned using the slow clock transport procedure. The same principles apply when charged particles move in a non-relativistic manner. In particular, oscillating charged particles emit radiation. In the non-relativistic case, where the velocities of the oscillating charges are much smaller than $c$, the emitted radiation is predominantly dipole in nature and is well-described by Maxwell’s equations in their standard form.

Let us now examine in more detail how the dipole radiation term arises. The retardation time in the integrands of the radiation field amplitude expression can be neglected when the charge's trajectory changes insignificantly during this time. The conditions for this approximation can be determined as follows. Denote by $a$ the characteristic size of the system. The retardation time is then of the order $a/c$. To ensure that the charge distribution remains nearly unchanged during this period, it is required that $a \ll \lambda$, where $\lambda$  is the radiation wavelength. This condition implies that the system’s dimensions must be small relative to the radiation wavelength. Expressing this constraint differently, we obtain  $v \ll c$, where $v$  represents the characteristic velocity of the charges.
Since dipole radiation theory accounts only for the dipole component of the emitted radiation, it disregards detailed information about the electron trajectory. Consequently, it is not surprising that the fields predicted by dipole radiation theory closely resemble those obtained from instantaneous approximations.

The theory of relativity provides an alternative method for clock synchronization, based on the fundamental principle that the speed of light remains constant in all inertial frames. While this is often regarded as a postulate, as we have seen, it is actually a matter of convention. The synchronization process that follows adheres to the standard Einstein synchronization procedure.
Consider a dipole radiation source. When the dipole source is at rest, the governing field equations are given by Maxwell’s equations. In dipole radiation theory, we introduce a small expansion parameter, $v/c \ll 1$,  and neglect terms of order $v/c$. This effectively means that the theory employs a zero-order nonrelativistic approximation.
Einstein synchronization is defined using light signals emitted by the dipole source at rest, under the assumption that light propagates isotropically with velocity $c$ in all directions. Applying the Einstein synchronization procedure in the rest frame of the dipole source allows us to establish a Lorentz coordinate system.

Slow transport synchronization is equivalent to Einstein's synchronization in an inertial system where the dipole light source is at rest. \footnote{Many textbook authors still attribute a reality status to the one-way speed of light. To quote Hrasko \cite{HA}: "It is sometimes claimed that Einstein's synchronization of distant clocks $A$ and $B$ is circular. ... This argument, however, is fallacious. It is true that a one-way measurement of light velocity can be performed only if clocks at the endpoints are correctly synchronized. However, since they need not show the correct coordinate time, they can be synchronized without light signals by transporting them symmetrically from a common site. ... As we see, the described thought experiment is capable of proving the constancy of the speed of light if it is true or disproving it if it is false. It therefore provides a solid logical foundation for Einstein's synchronization prescription." This reasoning, however, is incorrect. Slow clock transport synchronization is equivalent to Einstein's synchronization only in an inertial system where the light source is at rest.}
In other words, suppose we have two sets of synchronized clocks spaced along the $x$-axis. Suppose that one set of clocks is synchronized by using the slow clock transport procedure and the other by light signals. If we would ride together with any clock in either set, we could see that it has the same time as the adjacent clocks, with which its reading is compared. This is because in our case of interest when the light source is at rest, field equations are the usual Maxwell's equations and Einstein synchronization is defined in terms of light signals emitted by a source at rest assuming that light propagates with the same velocity $c$ in all directions.  
Applying either synchronization method in the rest frame establishes a Lorentz coordinate system. In this system, the metric of the light source takes the Minkowski form.
We now turn to the topic of electromagnetic waves. In free space, the electric field $\vec{E}$ of an electromagnetic wave emitted by a stationary dipole source satisfies the wave equation:
$\Box^2\vec{E} =  \nabla^2\vec{E} - \partial^2\vec{E}/\partial(ct)^2  = 0$. This result is closely related to the Minkowski metric  $ds^2 =  c^2 dt^2 - dx^2 - dy^2 - dz^2$.

\subsection{Motion of a Light Source Relative to an Inertial Frame}

We now consider the case in which a light source, initially at rest in the lab frame, is accelerated to a velocity $v$ along the $x$-axis. In other words, we examine an active (physical) boost.
When we say that the source undergoes acceleration while the inertial observer (equipped with measuring devices) does not, we mean acceleration relative to the fixed stars. Any acceleration relative to the fixed stars—that is, any active boost of velocity—has an absolute meaning.

In the case of an active boost, we analyze the motion of the same physical system over time, as observed from a single reference frame. A key question is whether the lab’s clock synchronization method depends on the motion of the light source. The answer to this is a matter of convention. The simplest synchronization method involves maintaining the same set of uniformly synchronized clocks used when the light source was at rest—essentially preserving clock transport synchronization (or Einstein synchronization, which is defined using light signals from a stationary dipole source).
This choice is the most practical from a laboratory perspective, as it preserves simultaneity and follows the absolute time (or absolute simultaneity) convention.

Absolute simultaneity can be incorporated into special relativity without altering either the logical structure or the (convention-independent) predictions of the theory.
We begin with the metric as the true measure of space-time intervals for an inertial observer with coordinates $(t,x,y,z)$.  Applying a Galilean boost, we transform the spatial coordinate as $x \to x-vt$ while keeping time unchanged. Substituting this into the Minkowski metric, $ds^2 = c^2 dt^2 - d x^2 - dy^2 - dz^2$ yields

\begin{eqnarray}
&& ds^2 = c^2(1-v^2/c^2)dt^2 + 2vdxdt - dx^2 - dy^2 -dz^2 ~ .\label{GGG11}
\end{eqnarray}

This metric describes the electrodynamics of the moving light source from the viewpoint of the inertial observer measurements.
From the structure of the equation, it is evident that the measurement of the electromagnetic field configuration—expressed in terms of the coordinates $x$ and $t$ of a measuring device at rest in the inertial frame—yields the same result as that obtained at the spatial position  $x - vt$ at time $t = 0$. 

To our knowledge, the metric in Eq. 6 is presented here for the first time.
This result has not been previously reported in the literature. The agreement
between this theoretical framework and empirical observations lends sup-
port to our revised formulation of special relativity incorporating absolute
simultaneity (see Section 4.14 for further details).
By contrast, existing literature often assumes—incorrectly—that an inertial
observer and an accelerated light source share the same Minkowski metric,
treating this assumption as self-evident.

To understand the physical interpretation of Eq. \ref{GGG11}, it is important to recognize that space-time points can be labeled in infinitely many ways using different coordinate systems. These systems are related through coordinate transformations. A new frame of reference can always be introduced by relabeling coordinates and describing physical phenomena using the new labels—a process known as a passive (coordinate) transformation. 
For instance, in a laboratory frame, one can define a comoving coordinate system to analyze radiation from a moving source in terms of the new (comoving) coordinates. In this system, fields are expressed as functions of the independent variables 
$x', y', z',t'$,  which are related to the original coordinates  $x, y, z, t$  via a passive Galilean transformation. 
Specifically, after the passive transformation, the Cartesian coordinates of the source transform as $x' = x-vt, ~ y' = y, ~ z' = z$. This transformation completes with the invariance of simultaneity, $t' = t$. The transformation of time and spatial coordinates of any event has the form of a Galilean transformation.  One must take into account that $(t,x,y,z)$ are coordinates of the rule clock structure at rest in the inertial frame.

The equivalence of the active and passive perspectives arises because shifting a system in one direction is equivalent to shifting the coordinate system in the opposite direction by the same amount. However, there is no a priori guarantee that the physical world must behave this way. For example, Newtonian mechanics is not invariant under passive rotations, as demonstrated by Newton’s bucket experiment.
The empirical principle states that within a single inertial frame, every passive Galilean boost is postulated to correspond to an active Galilean boost.

According to this equivalence of passive and active boosts, Maxwell’s equations remain valid in the comoving frame. In this frame, the fields are expressed as functions of the independent variables $x', y', z'$, and $t'$. 
The electric field $\vec{E}'$ of an electromagnetic wave satisfies the equation $\Box'^2\vec{E}' =  \nabla'^2\vec{E}' - \partial^2\vec{E}'/\partial(ct')^2  = 0$. 
However, the variables $x', y', z', t'$ can be expressed in terms of the independent variables $x, y, z, t$ using a Galilean transformation, so that fields can be written in terms of  $x, y, z, t$. From the Galilean transformation $x' = x - vt, ~ y' = y, ~ z' = z, ~ t' = t $, after partial differentiation, one obtains $\partial/{\partial t} = \partial/{\partial t'} - v\partial/{\partial x'}$, $\partial/{\partial x} = \partial/{\partial x'}$.
Hence the wave equation transforms into

\begin{eqnarray}
&& \Box^2\vec{E} = \left(1-\frac{v^2}{c^2}\right)\frac{\partial^2\vec{E}}{\partial x^2}  - 2\left(\frac{v}{c}\right)\frac{\partial^2\vec{E}}{\partial ct\partial x}
+ \frac{\partial^2\vec{E}}{\partial y^2} + \frac{\partial^2\vec{E}}{\partial z^2}
- \frac{1}{c^2}\frac{\partial^2\vec{E}}{\partial t^2} = 0 ~ , \label{GGT2}
\end{eqnarray}
This last result is closely related to the metric equation Eq. \ref{GGG11}.

The solution of this equation $F[x - (c +v)t] + G[x + (-c +v)t]$ is the sum of two arbitrary functions, one of argument $x - (c +v)t$ and the other of argument $x + (-c +v)t$.
Here we obtained the solution for waves that move in the $x$ direction by supposing that the field does not depend on $y$ and $z$. The first term represents a wave traveling forward in the positive $x$ direction, and the second term represents a wave traveling backward in the negative direction.

We conclude that the speed of light emitted by a moving source, as measured in the laboratory frame  $(t, x, y, z)$, depends on the relative velocity $v$ between the source and the observer. In other words, in this scenario, the speed of light follows the Galilean law of velocity addition. Specifically, the coordinate velocity of light along the $x$-axis is given by $dx/dt = c + v$ in the positive direction and $dx/dt = -c + v$ in the negative direction.
This apparent deviation from the electrodynamics constant $c$ arises because the clocks are synchronized according to an absolute time convention. Once such a convention is chosen, it must be consistently applied to both mechanics and electrodynamics. Under this absolute time coordinatization, the velocity of light transforms analogously to that of a particle, as expected.

\subsection{Discussion}

So far, we have examined the Galilean transformation of the electrodynamics equations. Now, we highlight a crucial distinction between fundamental and phenomenological theories.
For instance, Newton’s equation in a co-moving frame can be regarded as a phenomenological law. It does not provide a microscopic interpretation of a particle’s inertial mass; instead, rest mass is introduced in an ad hoc manner. This suggests that the particle may possess internal variables that remain unknown. Any phenomenological law valid in the Lorentz rest frame can be embedded in four-dimensional space-time only through Lorentz coordinatization.
Now, let us turn our attention to electrodynamics. It is essential to emphasize that electrodynamic laws do not rely on any hidden internal variables or underlying mechanisms. Electromagnetic fields are fundamental, and electrodynamics fully complies with the principles of relativity. As a result, any synchronization scheme can be used to describe electromagnetic field dynamics.

The flexibility in describing physical systems allows us to choose the most suitable representation for each problem. A new frame of reference can always be introduced by relabeling coordinates, enabling the discussion of physical phenomena through passive transformations. In contrast, an active transformation involves the actual motion of a physical system—where its characteristics change due to internal or external interactions. Here, we are concerned with the evolution of the same physical system over time, analyzed from the perspective of a single reference frame.

Let us now explore the correspondence between passive and active transformations, drawing a comparison between special relativity and Newtonian mechanics.
In Newton’s non-relativistic theory, the formalism is grounded in an absolute space structure. However, there is relativity concerning uniform motion, meaning we cannot determine a particle's absolute velocity.
Passive Galilean boosts are dynamical symmetries of Newtonian mechanics. The equivalence between active and passive Galilean boosts arises from the fact that moving a system in one direction is equivalent to shifting the coordinate system in the opposite direction by the same amount, as illustrated by Galileo's famous ship experiment. This symmetry, known as the Galilean group, exists because the equations of motion in Newtonian mechanics do not explicitly depend on velocity. The internal dynamics of the system remain unchanged under a Galilean boost.
However, Newtonian mechanics is not invariant under passive rotations. The form of the equations of motion changes due to the introduction of pseudo-forces. As a general discussion of the correspondence between passive and active transformations, we refer the reader to the paper \cite{APT}. Newton’s bucket argument contrasts two situations: one where a vessel of water is at rest, and another where the vessel is gradually rotated to a state of uniform angular velocity. The argument against the equivalence of active and passive rotations is based on the observation that, while there is no relative motion between the water and the bucket in both cases, the shape of the water differs.
There is no relativity of accelerations in Newtonian mechanics, and we can determine the absolute acceleration of a particle in an inertial frame. Absolute acceleration refers to acceleration relative to the fixed stars.

Let us now examine the correspondence between active and passive transformations in special relativity.
Within a single inertial frame, kinematical symmetry corresponds to dynamical symmetry, and the symmetry principle (i.e., the equivalence of the active and passive Galilean boosts) holds. It is important to note that a passive Galilean boost, when considered within a single inertial frame, is merely a different parametrization of the observations made by the inertial observer.
Under the passive transformation ($x \to x - vt$, $t \to t$), the motion of fixed stars relative to the observer and their devices remains unchanged. Note that $(t,x,y,z)$ represent the coordinates of the clock structure at rest in the inertial frame.

\subsection{A Clock Re-Synchronization  Procedure}

Most existing studies assume that an observer at rest and a moving source share the same Minkowski metric, thereby neglecting the need for any clock re-synchronization procedure. However, this standard approach is flawed, as it fails to account for the principle of equivalence between active and passive boosts within a single inertial frame.

Our analysis shows that after applying a Galilean boost under an absolute time coordinatization, the homogeneous wave equation for the field in the lab frame appears nearly—but not exactly—in its standard form (valid in the absence of acceleration). The key deviation lies in the presence of a mixed derivative term, $\partial^2/\partial t,\partial x$, which complicates the equation’s solution.

To resolve this, we note that a simplification is always possible by introducing a suitable change of the time variable. If we limit our analysis to first-order terms in $v/c$, a time transformation of the form  $t \to t + x v/c^2$
can be applied. This time shift effectively eliminates the mixed derivative term, reducing the equation to the standard wave equation form.

Let us now consider the case of an arbitrary velocity along the $x$-axis.
We have, then, a general method for finding a solution of the electrodynamics problem in the case of absolute time coordinatization. The new independent variables $(x_L,t_L)$ can be expressed in terms of the old independent variables
$(x,t)$: 

\begin{eqnarray}
&& ct_L = t\sqrt{1 - v^2/c^2}  +  vx/[c^2\sqrt{1 - v^2/c^2}] ~, ~ x_L =  x/\sqrt{1 - v^2/c^2}  ~ , \label{GGT3}
\end{eqnarray}

Using the transformation Eq. (\ref{GGT3}),
the velocity transformation becomes 

\[
dx_L/dt_L = (dx/dt)/[1 - v^2/c^2 + vdx/dt)]   . 
\]

Substituting
$dx/dt = v$ gives  $dx_L/dt_L = v$.  

Since the change variables 
completed by the Galilean transformation
is mathematically equivalent to the Lorentz transformation, 
it follows that transforming to new variables $x_L, t_L$ leads to the usual Maxwell's equations.
In particular, the  wave equation Eq. (\ref{GGT2})  transforms into

\begin{eqnarray}
\Box_L^2E = \nabla_L^2E - \partial^2E/\partial(ct_L)^2 = 0  ~.
\end{eqnarray}

In the new variables, the velocity of light emitted by a moving source is constant in all directions, and equal to the electrodynamics constant $c$. 

In the new variables, inertial observer  has the spacetime interval given by

\begin{equation}
ds^2 = (dt_L - vdx_L/c^2)^2 /(1 - v^2/c^2)  - (1 - v^2/c^2)dx_L^2    .
\label{STI}
\end{equation}

The combined effect of the Galilean boost and variable changes results in the Lorentz transformation under absolute time coordinatization in the lab frame. However, in this context, the transformation should be regarded merely as a mathematical tool that simplifies solving the electrodynamics problem while maintaining absolute time synchronization.

This raises an interesting question: Is it necessary to transform the solution back into the original variables? We argue that the variable changes introduced above have no intrinsic significance; their meaning is purely conventional. Notably, the time shift $t + xv/c^2$ is directly related to the issue of clock synchronization. Moreover, the final rescaling of time and spatial coordinates remains physically unobservable. Crucially, the convention-independent results remain unchanged in the new variables. Consequently, there is no need to revert the solution to its original form.

Consider two independent light sources 
The question now arises of how to assign a time coordinate to the lab reference frame. We have two choices: an absolute time coordinate or a Lorentz time coordinate. The most natural choice, in terms of connecting to laboratory reality, is absolute time synchronization. In this framework, simultaneity is absolute, meaning that a single set of synchronized clocks is sufficient for both sources in the lab frame. However, Maxwell’s equations are not form-invariant under Galilean transformations, meaning their mathematical form differs in the lab frame. Using absolute time synchronization leads to more complex field equations, which also vary for each source. To introduce Lorentz coordinates, we must adopt a different approach. The only viable method in this scenario is to assign an individual coordinate system—i.e., an independent set of synchronized clocks—to each source.

The situation becomes more complex when a secondary source interacts with light emitted by a primary source. A fundamental challenge in special relativity is that both sources cannot, in general, be described using a single shared Lorentz frame within a common inertial reference system. This issue, and its implications for electrodynamics, is discussed in more detail in the next chapter.

\subsection{A Moving Light Source: Peculiarities of Collinear Geometry}

So far, we have considered both covariant and non-covariant approaches to solving the problem of radiation emitted by a moving source in an inertial frame. Now, let us examine the acceleration of a dipole source in the laboratory inertial frame, where it accelerates to a velocity $v$ along the $x$-axis. This raises the question of how to assign synchronization in the lab frame after the source has undergone acceleration.
Before acceleration, we adopted a Lorentz coordinate system. However, if synchronization in the lab frame remains unchanged after the acceleration, the electrodynamics of moving charges becomes significantly more complex. In this case, the transformation of time and spatial coordinates does not retain the standard Lorentz form. Maxwell’s equations remain valid in the lab inertial frame only when Lorentz coordinates are properly assigned.
To preserve Lorentz coordinatization after acceleration, it is necessary to redefine the space-time coordinate system by introducing new variables $(x_L,y_L,z_L,t_L)$, as given in Eq. (\ref{GGT3}). In this new coordinate system, the velocity of light emitted by the moving source remains equal to the electrodynamic constant $c$ in all directions.

The aberration of light and the Doppler effect serve as practical examples to illustrate the distinction between covariant and non-covariant approaches.
The most effective way to calculate radiation is by applying the Lorentz transformation formulas between inertial reference frames. Specifically, consider a radiation source at rest in the frame $S'$, emitting dipole radiation with frequency $\omega_0$. In the laboratory frame $S$, where the source moves with velocity $v$ and the radiation propagates along the direction of motion, we observe the so-called radial Doppler effect. This effect is governed by the well-known formula: $\omega = \omega_0\sqrt{1 - v^2/c^2}/(1 - v/c)$. The effect of the factor  $\sqrt{1 - v^2/c^2}$ can be summarized in the following statement: on the moving object, time is flowing slower than expected (time dilation).  

It is worth noting that using the conventional coupling of Maxwell’s equations and the corrected Newton’s equation to calculate radiation from a moving source does not necessarily lead to errors.
For rectilinear motion of both the source and the emitted light beam, non-covariant and covariant approaches yield identical trajectories, ensuring compatibility between Maxwell’s equations and conventional particle tracking. However, this method was incorrectly extended to non-collinear geometries.

This raises a reasonable question: why does the same method produce incorrect results when the source moves perpendicular to the direction of its emitted radiation?
The key distinction in collinear geometry—where the source moves along the same line as the radiated beam—is that the velocity is perpendicular to the plane of the radiation wavefront (i.e., the plane of simultaneity). Consequently, for collinear motion, the plane of simultaneity in absolute time coordinatization retains the same orientation in Lorentz coordinatization.

According to the theory of special relativity, the standard description of optical phenomena related to the motion of a light source remains valid when radiation propagates along the direction of velocity.
For instance, in a collinear geometry, the conventional coupling of fields and particles can provide a useful framework for analyzing the Doppler effect.

To understand the relativistic redshift of a light source in a moving system, we must analyze the mechanism of the source and observe its behavior while in motion. Since this can be quite challenging, we will consider a simplified source that captures the fundamental principles.
In a vacuum, an electron emits radiation only when it undergoes acceleration. In the non-relativistic case, this radiation exhibits a dipole pattern. When a non-relativistic electron moves within a magnetic field, the emitted radiation is commonly known as cyclotron radiation. The frequency of the cyclotron radiation (the dipole radiation) is of course equal to the frequency of electron rotation in the magnetic field $ \vec{H} = H\vec{e}_x$, i.e. $\omega_{L0} = eH/(mc)$. In the case of circular motion (with the velocity component parallel to the field $v_x = 0$) the radius of the orbit is 

\[
r_L = v_c/\omega_{L0} = [mc^2/(eH)]v_c/c = \lambdabar_{L0} v_c/c       , 
\]

where $\lambda_{L0}$    is the wavelength of the cyclotron radiation and $v_c$ is the electron velocity. When $v_c/c \ll 1 $ we clearly always have $r_{L}/\lambdabar_{L0}   \ll 1$, which means that the dipole approximation is applicable.

The relativistic motion of an electron is described by  $d\vec{p}/dt = e\vec{v}\times \vec{H}/c$ with $\vec{p} = m\vec{v}/\sqrt{1 - v^2/c^2}$. This equation can be written as $d\vec{v}/dt = \vec{v}\times \vec{\omega}_L$ where 

\[
\vec{\omega}_L = e\vec{H}\sqrt{1 - v^2/c^2}/( mc) = \vec{\omega}_{L0}\sqrt{1 - v^2/c^2}  
\]

is the relativistic Larmor frequency.   
When the source is accelerated, the speed of electrons is increased, and therefore the mass is also increased and the electron is heavier.          
The emissivity presents a spectrum in which the frequency is given by

\[
\omega = \omega_{L0}\sqrt{1 - v^2/c^2}/(1 - v/c)        .
\]

The frequency is multiplied by the relativistic Larmor frequency accounting at the same time through the denominator for the Doppler effect caused by the motion parallel to the magnetic field. 

We can thus conclude that the frequency of an electron's oscillations decreases when the source moves with velocity $v$.
At first glance, the mechanism of the source described may not seem to involve relativistic time dilation. However, it is implicitly present in the assumption that the mass of a moving object corresponds to its relativistic mass, given by
$m_r = m/\sqrt{1 - v^2/c^2}$.

According to the covariant approach, various relativistic kinematic effects relevant to the dipole radiation setup emerge in successive orders of approximation.
The first-order kinematic term $v/c$ plays a crucial role only in the description of dipole radiation in the perpendicular geometry. In the case of collinear geometry, however, the motion of the dipole source affects only the second-order kinematic terms  $(v/c)^2$, in accordance with the theory of relativity.

In the non-covariant approach, solving the dynamics problem in the lab frame does not involve Lorentz transformations. This means that, within the lab frame, particle motion appears exactly as predicted by classical mechanics, which assumes an absolute time framework. This absolute-time approach is well-suited for explaining experimental outcomes in collinear geometry. However, we argue that this method of solving Maxwell’s equations in the lab frame is not applicable in the transverse case.

In this case, the source undergoes acceleration along the radiation wavefront (i.e., the initial plane of phase simultaneity).
Since the orientation of the radiation wavefront is no longer absolute, it depends on the chosen clock synchronization convention.

To analyze this, let us first consider a Lorentz transformation. Specifically, we use a Lorentz boost to describe the uniform translational motion of the light source in the lab frame.
It is evident that wavefront phases, which are simultaneous in $S'$ but occur at different $x'$-locations, are not simultaneous in $S$. When applying a Lorentz boost, a time transformation is introduced: $t' = t - xv/c^2$. 
The effect of this transformation is equivalent to a rotation of the radiation wavefront in the lab frame by an angle of approximately $v/c$ in the first-order approximation.

According to Maxwell's electrodynamics, radiation is always emitted in a direction normal to the radiation wavefront. Consequently, the radiated light beam propagates at an angle $v/c$, leading to the phenomenon known as the aberration of light. This remarkable effect will be the focus of our discussion in the next chapter.

In the conventional (3+1) approach, simultaneity is considered absolute, and there is no mixing of spatial coordinates and time when sources change their velocities in an inertial frame.
Thus, it appears that the conventional (3+1) approach cannot account for the geometric phenomenon of aberration. By applying Maxwell's equations alongside an absolute time transformation, neither the wavefront plane nor the direction of energy transport undergoes deflection.

The conventional accelerator theory  typically relies on a hybrid approach: Newtonian kinematics—corrected for relativistic mass—is applied to particle dynamics, while Einsteinian principles underpin electrodynamics. In practice, the relativistic treatment of particle motion often reduces to a modified form of Newton’s second law, with velocity composition still grounded in Galilean transformations.

For rectilinear motion of the ultrarelativistic modulated electron beam, both covariant and non-covariant approaches yield identical particle trajectories. Specifically, in this case, the modulation wavefront -defined as plane of simultaneity - is always perpendicular to the velocity of the electron beam.
Consequently, Maxwell’s equations remain consistent with conventional particle tracking in such cases. 
However, the relativity of simultaneity—which involves the intermixing of spatial and temporal coordinates—leads to fundamental differences between covariant and non-covariant descriptions when the modulated beam follows a curved path. In such cases, the relativistically correct coupling between fields and particles predicts outcomes that sharply contrast with those of conventional treatments.

The use of electrodynamics in absolute-time coordinatization becomes essential when adopting a non-covariant (3+1)-dimensional approach (i.e., "old kinematics") to relativistic particle dynamics. Conventional particle tracking relies on the lab time $t$ as the independent variable, thereby excluding relativistic kinematic effects from the description. This approach is based on the assumption of absolute time synchronization in the lab frame. If one intends to use corrected Newtonian equations—where time serves as the independent variable—the electrodynamic equations must be formulated in a non-covariant manner. Only in this framework does the coupling between electrodynamic equations and particle trajectories remain consistent within a single reference frame.

\newpage

\section{Aberration of Light in an Inertial Frame of Reference}

In this chapter, we explore several practical applications of the concepts introduced earlier, using the spacetime metric given in Eq.~(\ref{GGG11}). Our first topic addresses the phenomenon of light aberration caused by a light source moving perpendicular to the direction of radiation.

Light aberration in an inertial frame is generally understood as an apparent shift in the direction of light propagation due to the motion of the source. While often treated as a first-order effect in $v/c$, a rigorous description is not trivial—even at this level of approximation. Many standard textbooks contain inaccurate or oversimplified accounts, underscoring the subtlety of the underlying physics. To keep the analysis manageable, we will limit our discussion to first-order effects in $v/c$.

Our focus will be on the physical influence of optical components on the observed aberration of light. We begin by examining the case of a finite-aperture mirror moving parallel to its surface, with an incident plane wave striking it normally. We will show that the reflected light exhibits aberration, observable as a deviation in the direction of energy transport.

We then turn to a more practically significant scenario: the transmission of light through an aperture in a moving opaque screen, invoking Babinet’s principle. A particularly important application is the passage of light through the moving end of a telescope barrel—a configuration we will analyze in detail.

In this chapter, the binary star paradox is resolved by recognizing that when light passes through the end of a telescope barrel, diffraction perturbs its fields. As a result, the light beam no longer carries information about the star’s motion relative to the fixed stars. 
It is crucial to emphasize that aberration is caused by changes in the telescope's velocity, not the star's.
Thus, planetary acceleration plays a key role in the phenomenon. The asymmetry paradox is resolved by recognizing acceleration as defining factor.
In the next chapter, we derive stellar aberration using the Langevin metric in special relativity.

\subsection{The "Plane Wave" Emitter}

To build intuitive insight into the phenomenon of aberration, we begin by analyzing a simple, idealized case: a single "plane-wave" emitter.

We model this emitter as a two-dimensional array of identical, coherent elementary sources (dipoles), uniformly distributed over a fixed ($x$–$y$) plane. These sources are assumed to begin radiating simultaneously in the lab frame, where the plane is at rest. All dipoles oscillate with the same frequency $\omega$, amplitude, and phase, and their motion is confined to the plane.

By decreasing the spacing between adjacent sources to a value much smaller than the radiation wavelength, $\lambda = 2\pi c/\omega$, this arrangement effectively approximates an ideal plane-wave emitter. The resulting wavefronts are planar, and the emitted radiation resembles an ideal plane wave.

While the concept of an (infinitely extended) plane wave is a useful mathematical abstraction—owing to its simplicity and exact correspondence with solutions to Maxwell’s equations—it is not physically realizable, as such a wave would carry infinite energy. In any real scenario, the emitter has a finite aperture.

It is also assumed that any detector used to measure the radiation direction is sensitive to energy flow (rather than to phase fronts, for instance) and has an aperture large enough compared to the beam size to capture the overall direction of energy propagation. What is commonly described as aberration, in this context, is in fact a shift in the direction of energy transport.

\subsection{A  Moving Emitter: Galilean Transformations-Based Explanation}

Let us consider the case of an emitter lying in the $(x-y)$ plane of an inertial laboratory frame, which is accelerated from rest to a velocity $v$ along the $x$-axis. Suppose an observer at rest in this inertial frame measures the direction of energy transport.

There are two possible approaches to analyzing the aberration of light emitted by a single moving source. The first is the covariant approach, commonly used in the literature, which explains the aberration effect by applying a Lorentz boost to determine how the direction of a light beam depends on the velocity of the source relative to the inertial frame. The second is a non-covariant approach, which employs an old kinematics description  without explicit reference to Lorentz transformations.

Both approaches, whether formulated using Einstein's synchronization convention or an absolute time synchronization scheme, yield the same result in the case of a single moving emitter. Thus, the choice between them is a matter of pragmatism rather than fundamental necessity.

\begin{figure}
	\centering
	\includegraphics[width=0.8\textwidth]{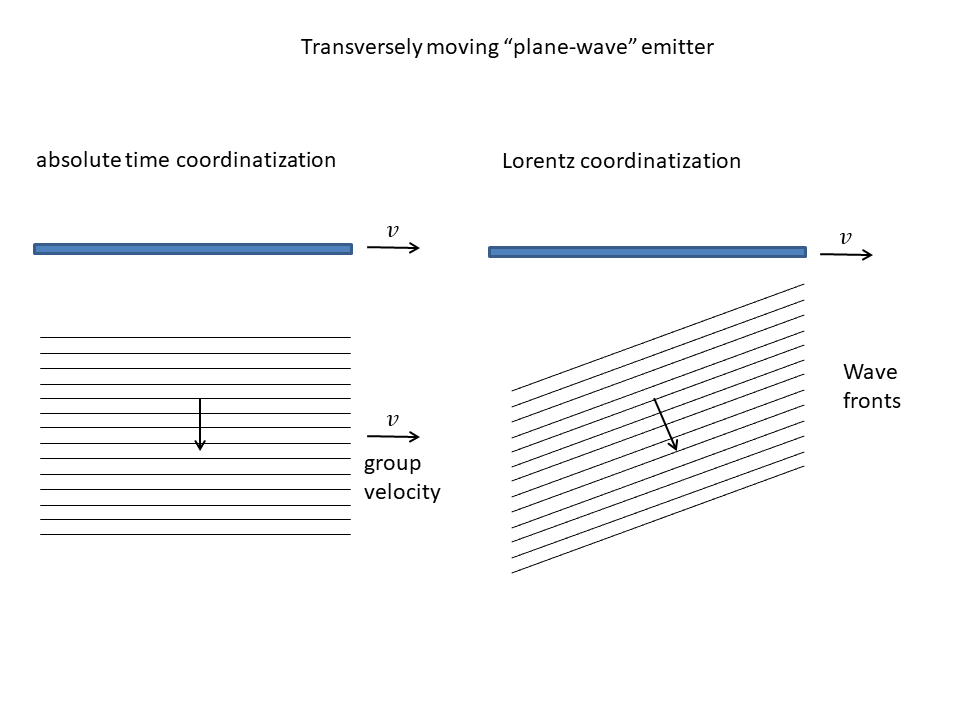}
	\caption{Illustration of a plane wave emitter, initially at rest in an inertial frame, accelerating to velocity $v$ along the $x$-axis. The effect of aberration is analyzed using a Galilean boost (left) and a Lorentz boost (right).}
	\label{B103}
\end{figure}

In this chapter, we present both approaches, beginning with the non-covariant one. It is important to emphasize that the (3+1) non-covariant formulation in relativistic electrodynamics is entirely valid and provides a consistent description. The aberration of light can be treated within a single-inertial-frame framework, without invoking Lorentz transformations. In this context, time remains an absolute quantity in special relativity. But what does "absolute time" mean? It signifies that simultaneity is preserved within the inertial frame, and there is no intermixing of spatial positions and time coordinates when sources change velocity within this frame.
A distinctive feature of this non-covariant approach is the absence of relativistic kinematics in the description of light aberration. For a tangentially moving emitter (relative to its surface), the direction of energy transport of the emitted light undergoes a deviation. Within the "single-frame" approach, this effect arises due to the Doppler effect, which induces angular frequency dispersion of the emitted light waves (see Fig. \ref{B103}, left).

The current approach to the moving emitter problem employs the Fourier transform method. In the analysis of linear systems, it is often useful to break down a complex input into simpler components, determine the system's response to each elementary function, and then superimpose these responses to obtain the total outcome. Fourier analysis serves as a fundamental tool for this decomposition. \footnote{For a broader discussion on Fourier transform methods in spatial filtering theory or Abbe diffraction theory, we recommend referring to \cite{AB}.}

Consider the inverse transform relationship 

\[
g(x) = \int_{-\infty}^{\infty}G(K)\exp(iKx)dK
\]

which express the profile function in terms of its wavenumber spectrum. We may regard this expression as a decomposition of the function $g(x)$ into a linear combination (in our case into an integral) of elementary functions, each with specific form $\exp(iKx)$. From this, it is clear that the number $G(K)$ is simply a weighting factor that must be applied to the elementary function of wavenumber $K$ to synthesize the desired $g(x)$.

An emitter with a finite aperture is a kind of active medium that breaks up the radiated beam into a number of diffracted beams of plane waves. Each of these beams corresponds to one of the Fourier components into which an active medium can be resolved.  Let us assume that the current density of the elementary dipole sources varies according to the law  

\[
j_{pol} = g(K_{\perp})\sin{({K}_{\perp}x)}   . 
\]

From this, we conclude that the active medium of the emitter is sinusoidally space-modulated.

Let us examine the electrodynamics of a moving source.
To explain the phenomenon of radiation in our case of interest, we apply an (active) Galilean boost to describe the uniform translational motion of the source in the inertial lab frame.
According to equivalence of passive and active boosts, Maxwell's equations remains valid in the comoving frame.
The measurement of the electromagnetic field configuration - expressed in terms of coordinates $x$ and $t$ of a measuring device at rest - yields the same result that obtained at the spatial position $x-vt$ at time $t = 0$. 
Applying a Galilean boost, we transform the spatial coordinate as $x \to x - vt$, while keeping time unchanged. 
Since Maxwell's equations are not invariant under Galilean transformations, their form is not preserved.

Starting from the Galilean transformation and applying partial differentiation, we obtain the wave equation   Eq.(\ref{GGT2}). Our goal is to demonstrate how the additional terms introduced in the field equations due to the Galilean transformation lead to the prediction of the Doppler effect.
Consider as a possible solution a radiated plane wave $\exp(i\vec{k}\cdot\vec{r} - i\omega t)$. 
With a plane wave $\exp(i\vec{k}\cdot\vec{r} - i\omega t)$ with the  wavenumber vector $\vec{k}$ and the frequency $\omega$ equation Eq.(\ref{GGT2}) becomes:

\begin{eqnarray}
&&(1-v^2/c^2)k_x^2 + 2vk_x \omega/c^2 + k_z^2 -\omega^2/c^2 = 0 ~  .  \label{CD2}
\end{eqnarray}

The wavevector $\vec{k}$ is determined by the initial conditions $\vec{k}' = (k'_x, k'_ z)$ before the acceleration and subsequent Galilean transformation. Specifically, we set $k'_x = K_{\perp}$, where $K_{\perp}$ corresponds to the wavenumber of the sinusoidally space-modulated dipole density, and 

\[
(k_z')^2  = \omega_i^2/c^2 - K_{\perp}^2   , 
\]

with $\omega_i$  denoting the frequency of the emitted radiation prior to acceleration.

The phase of the wave at the world point $(\vec{r},t)$ is invariant at a change of the reference system (when the field is zero, everyone measures the field as zero).
This invariance holds regardless of the specific coordinate transformation used to describe the change in the reference system.
Therefore, the phase $\vec{k}\cdot\vec{r} - \omega t$ must be invariant of the Galilean transformation. 
Consequently, 

\[
\vec{k}'\cdot\vec{r}' - \omega' t' = \vec{k}\cdot\vec{r} - \omega t   .
\]

Substituting the Galilean transformation formulae $x' = x - vt, t' = t$ into the phase equality formula we obtain  $\omega = \omega_i + K_{\perp}v$. This transformation completes with the invariance of wavenumber vector 

\[
k_x = K_{\perp}    ,       \qquad                    k_z^2 =  \omega_i^2/c^2 - K_{\perp}^2   .
\]

Substituting these into dispersion relation confirms that it remains satisfied, as expected.

Let us first remind the reader that the velocity of waves is typically defined in terms of the phase difference between oscillations observed at two different points in a free plane wave. This concept is primarily used to compute interference fringes, which visually highlight phase differences. In a plane wave, the phase velocity is given by the ratio  $\omega/k$. In addition to phase velocity, another type of velocity—known as the group velocity or energy propagation velocity—can be defined by considering the propagation of a disturbance, such as a change in amplitude, superimposed on a wave train. 
A simple example of a wave group is formed by the superposition of two waves with frequencies and wave numbers

\[
\omega_1 = \omega + \Delta\omega, ~ k_1 = k + \Delta k    ,   \qquad        \omega_2 = \omega - \Delta\omega, ~ k_2 = k - \Delta k   .
\]

This combination represents a carrier wave with frequency  $\omega$ and a modulation with frequency $\Delta\omega$. The resulting wave can be described as a succession of moving beats (or groups), where the carrier wave travels with velocity $\omega/k$, and the group velocity is given by  

\[
v_g = \Delta\omega/\Delta k   ,
\]

which, in the limit, becomes   $v_g = d\omega/dk$.

Many textbooks on electromagnetic theory discuss the phenomenon of light aberration in the context of plane waves. However, using a plane wave model to explain light aberration leads to an incorrect understanding. When an infinite sinusoidal wave propagates, it maintains a uniform average energy density throughout space. The question then arises: does this energy remain stationary, or does it propagate through space? It is impossible to determine this with certainty based on the plane wave model.
All experimental methods used to measure the aberration of light rely on light signals, which measure not the phase velocity, but the signal velocity. In our case of interest, this signal velocity coincides with the group velocity.

In this example, plane waves with different wavenumber vectors propagate from the moving emitter, each with its own frequency. The relation $\Delta\omega/\Delta k_x =  v$ holds for each radiated wave, regardless of the sign or magnitude of the radiated angle. Specifically, in this case, $\Delta\omega$ represents the Doppler shift, given by 

\[
\Delta\omega = \vec{K}_{\perp}\cdot\vec{v}   ,
\]

and $\Delta k_x$ is simply the transverse component of the radiated wavenumber vector, $\Delta k_x = K_{\perp}$. These equations indicate that a light beam with a finite transverse size propagates along the $x$-direction with a group velocity $d\omega/dk_x = v$.

In our discussion, we assume that the aberration angle is significantly larger than the divergence of the emitted beam. Mathematically, this is expressed as  
 $\lambdabar/D_e \ll v/c$ (i.e. $cK_{\perp}/\omega_i \ll v/c$), where $D_e$ is the characteristic size of the emitter and the  $\lambdabar = c/\omega_i$ is the reduced wavelength.

\subsection{A  Moving Emitter: Lorentz Transformations-Based Explanation}

An alternative and insightful approach to understanding the aberration of light from a moving source is through Lorentz coordinatization. This method provides a relativistically consistent explanation of the phenomenon within the framework of Maxwell's equations.

This explanation utilizes a clock-re-synchronization procedure. The time coordinate $t_L$ in the lab frame, under Lorentz coordinatization, is obtained by introducing a time offset of the form given in Eq. (\ref{GGT3}).  

Although Maxwell’s equations are invariant under Lorentz transformations, the Lorentz boost introduces a transformation of the time coordinate of the form  $t' = t - xv/c^2$. This spatially varying time shift introduces a delay in the emission of radiation across the source plane, effectively tilting the phase front of the emitted wave. To first order in $v/c$, this corresponds to a rotation of the phase front by an angle approximately equal to  $v/c$.

This rotation manifests physically in the fact that features of the wave, such as the electric field maximum, reach different spatial positions at different times in the lab frame. For instance, the field maximum at position  $x$ along the $x$-axis is delayed relative to the field maximum at $x = 0$, with a time shift given by $\Delta t = xv/c^2$.

In accordance with Maxwell’s electrodynamics, coherent radiation is always emitted perpendicular to the phase front. Since Maxwell’s equations contain no intrinsic directional bias, the tilt in the wavefront—induced by the Lorentz time transformation—results in a corresponding shift in the direction of energy propagation. Within this relativistic description, the moving emitter exhibits a phase chirp across its aperture, given by $d\phi/dx = k_x = v\omega_i/c^2$,
which means that the individual elementary sources now oscillate with position-dependent phase. As a result, the wavefront undergoes a rotation, and the emitted radiation propagates at an angle $\theta_a = v/c$, capturing the essence of the aberration effect (see Fig.~\ref{B103}, right).

We now examine the group velocity of the emitted beam. For a plane wave described by $\exp(i\vec{k}\cdot\vec{r} - i\omega t)$, the dispersion relation in vacuum, derived from Maxwell’s equations, is

\[
 k_z^2 + k_x^2 - \omega^2/c^2 = 0   .
\]

In special relativity, the four quantities $(\omega,k_x,k_y,k_z)$ form a four-vector and transform accordingly under Lorentz transformations. 
We will restrict our analysis to terms up to first order in $v/c$. Using the initial conditions and the Lorentz transformation, we obtain  the following expression: 

\[
\omega = (\omega_i + v K_{\perp})    ,      \qquad           k_z = \omega_i/c   ,    \qquad     k_x = (v\omega_i/c^2 + K_{\perp})   ,
\]

where $K_{\perp}$ is the wavenumber of sinusoidally space-modulated current density, $\omega_i$ is the frequency of the emitter radiation before the acceleration. Substituting these into the dispersion relation confirms that it remains satisfied, as expected.

As a consequence of the Doppler effect in Lorentz coordinatization, we observe an angular frequency dispersion in the light waves emitted from a moving source with a finite aperture.
The Doppler shift, $\Delta\omega$, of a radiated light wave (in the first order approximation) is given by  $\Delta\omega = \vec{K}_{\perp}\cdot\vec{v}$   ,
where $K_{\perp}$ is the transverse component of the radiated wavenumber vector. This equation implies that a light beam with a finite transverse extent propagating along the $x$-direction has a group velocity given by $d\omega/dk_x = v$.

It is important to recognize the significance of the fact that both covariant (Lorentz-based) and non-covariant (absolute time-based) approaches yield the same group velocity. This agreement underscores the physical objectivity of the result: group velocity is a measurable, convention-independent quantity, and must therefore remain invariant regardless of the chosen coordinatization.

The distinction between absolute time coordinatization and Lorentz coordinatization, as discussed in Chapter 3, is particularly illuminating. For non-relativistic velocities, the transition between these two frameworks can be understood as a simple redefinition of the time variable: $t \to t + xv/c^2$.

When combined with the Galilean transformation, this leads naturally to the Lorentz transformation, even within the context of an absolute-time formalism. However, this change of variables has no intrinsic physical significance—it represents merely a reparametrization of time, not a real physical effect.

This coordinate transformation is closely tied to the issue of clock synchronization. Crucially, the physical content of electrodynamics remains unchanged under such transformations. As a result, once a solution is obtained in the transformed variables, there is no need to revert to the original (3+1) variables; the results remain valid and physically meaningful.

It is commonly assumed that the theory of relativity can be applied to physical processes without a detailed understanding of the clock synchronization procedure. Most textbooks suggest that an operational interpretation of Lorentz coordinatization within a single inertial frame is unnecessary.

Nevertheless, distinguishing between absolute-time and Lorentz coordinatization (from an operational standpoint) deepens conceptual understanding—even in the simple case of a single moving source. While some physicists have expressed concern about the lack of a "dynamical" explanation for wavefront rotation in Lorentz coordinates, Chapter 3 offers a compelling perspective: the rotation arises naturally from the transformation of the time coordinate, which alters the simultaneity structure of spacetime events. Within the absolute-time framework, this provides an intuitive—though coordinate-dependent—explanation of the aberration phenomenon.

\subsection{Reflection from a Mirror Moving Parallel to Its Surface}

Until now, the previous two sections have not introduced any fundamentally new results. As previously discussed, when the sources are independent, adopting Lorentz coordinates presents no difficulty—each source can be assigned its own coordinate system, along with an independent set of clocks. The situation becomes more complex, however, when a secondary source interacts with light emitted by a primary source within a single inertial frame. This challenge in special relativity can be effectively addressed using an electrodynamics-based approach framed within an absolute time perspective using metric Eq. (\ref{GGG11}).

As noted earlier, the metric in Eq. (\ref{GGG11}) has not been previously reported in the literature. Existing studies generally assume—incorrectly—that both sources share the same Minkowski metric and are governed by Maxwell's equations within a single inertial (laboratory) frame. This assumption is often treated as self-evident.
In effect, the conventional approach relies on a hidden premise: that the principle of equivalence between active and passive boosts does not holds within a single inertial frame. Our formulation challenges this assumption.

The agreement between our theoretical framework and experimental observations lends strong support to our revised formulation of special relativity, which incorporates the metric of Eq. (\ref{GGG11}) (see Section 4.14 for further discussion).

With the results of Chapter 3 now established, we are prepared to examine the intriguing phenomenon of radiation from moving sources.

\begin{figure}
	\centering
	\includegraphics[width=0.6\textwidth]{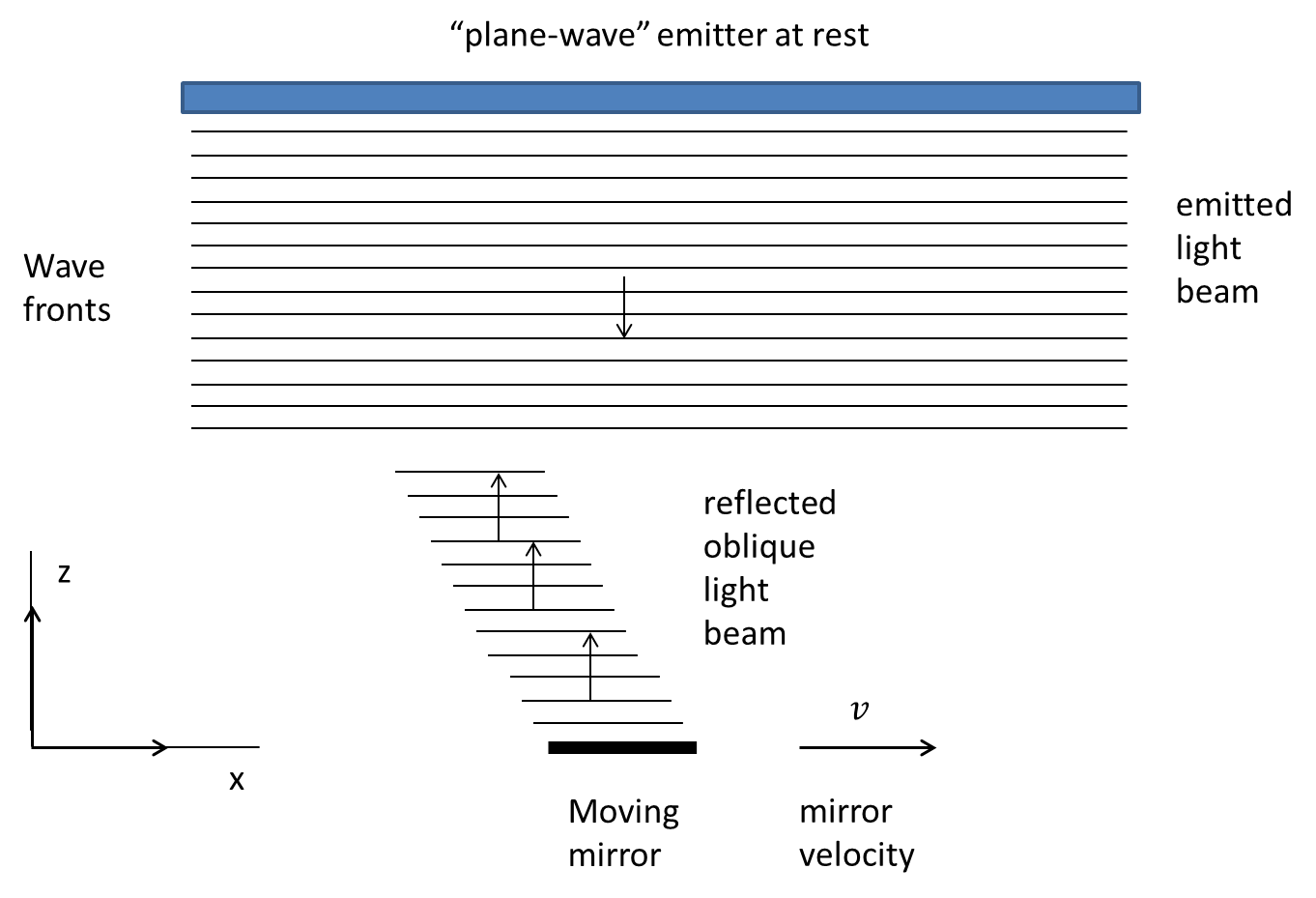}
	\caption{A mirror moving parallel to itself. 
		According to standard textbooks treatments, the energy transport of the reflected light beam remains undeviated.}
	\label{B33}
\end{figure}

We begin by analyzing the intensity of light reflected when a stationary source illuminates a moving mirror. It is commonly assumed that if the mirror moves tangentially to its surface, the law of reflection remains unaffected—just as it does for a stationary mirror (see Fig. \ref{B33}). According to this view, a monochromatic plane wave incident normally on a small aperture of a moving mirror is reflected as an oblique beam. In other words, it is assumed that the energy transport velocity coincides with the phase velocity. This assumption, frequently found in textbooks (see, for example, \cite{PA} and \cite{S}), is, however, incorrect.

\begin{figure}
	\centering
	\includegraphics[width=1.0\textwidth]{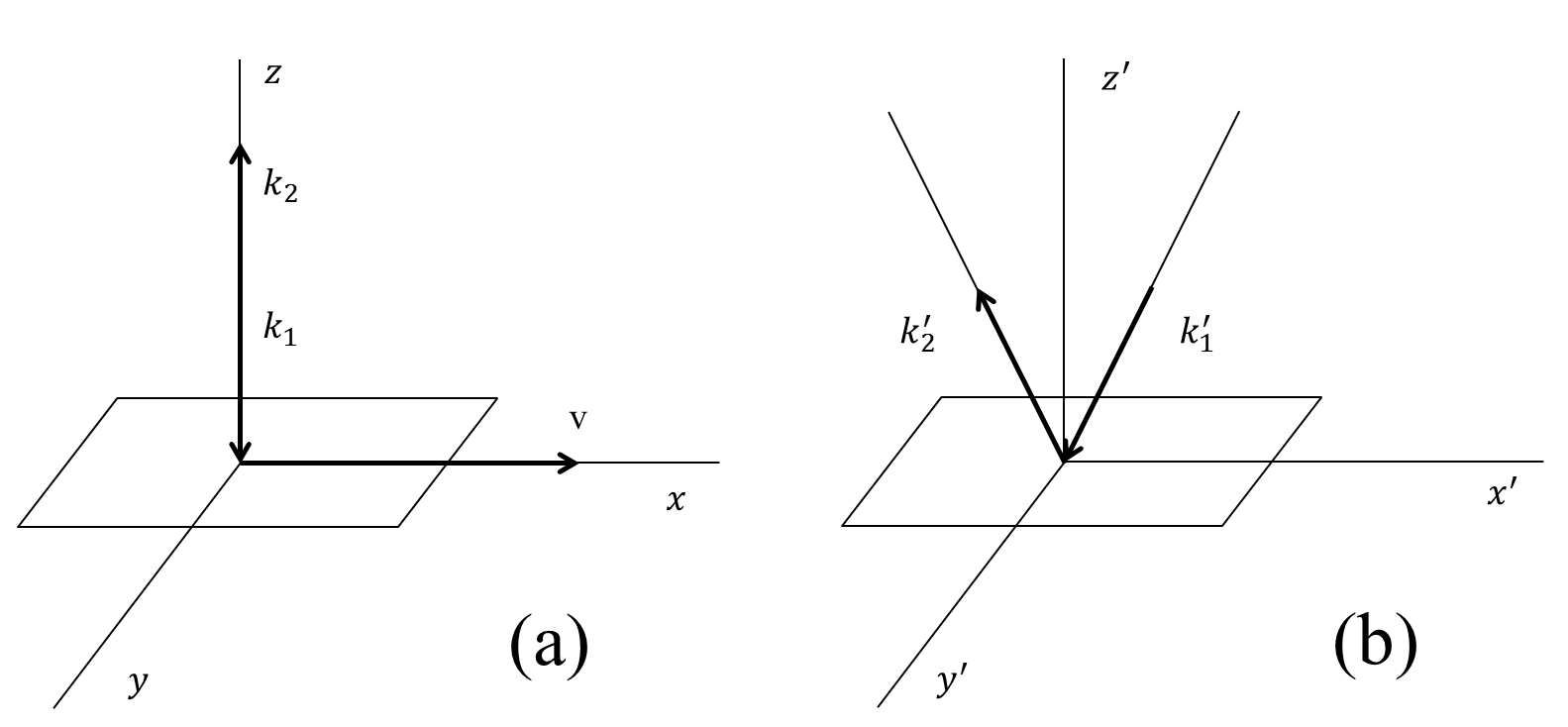}
	\caption{The aberration of light explained using relativistic kinematics, formulated in terms of the wavenumber four-vector. Reflection geometry as observed from (a) the laboratory frame and (b) an inertial frame co-moving with the mirror.}
	\label{conventional}
\end{figure}

To examine this claim, we analyze the reasoning found in standard texts. \footnote{
Many sources claim that no aberration occurs when light reflects off a mirror moving parallel to its surface. For example, Sommerfeld \cite{S} states: "For a mirror moving tangentially to its surface, the law of reflection that holds for a stationary mirror is preserved."}
Textbooks typically analyze reflection using two Lorentz reference frames. The fixed (lab) frame $S$ is at rest relative to the plane-wave emitter, while the moving frame $S'$ travels at velocity $v$, in which the mirror is at rest. Both frames use a Cartesian coordinate system with the $x-y$ plane tangent to the mirror’s surface. The $x$-axis aligns with the mirror's velocity $v$. For simplicity, we consider light incident from the $z$-direction in the lab frame. The incident wave is described by its four-dimensional wave vector, where the time-like component represents the angular frequency $\omega$, and the space-like components define the propagation direction. In the lab frame  $(t,x,y,z)$, the wave vector has components  

\[
k_1 = (\omega, 0, 0, -\omega/c)   ,
\]

with the negative sign indicating propagation toward the mirror (see Fig. \ref{conventional}a). Our goal is to determine the wave vector of the reflected beam.

The argument that light reflected from a tangentially moving mirror experiences no aberration proceeds as follows. It is most straightforward to analyze the reflection in the mirror's rest frame (Fig. \ref{conventional}b), where the surface is stationary and the standard laws of optical reflection apply. We consider aberration effects only up to the first order in $v/c$. In this frame, an observer at rest with respect to the mirror perceives the incoming wave vector as

\[
k_1' = (\omega, -v\omega/c^2, 0, -\omega/c)   . 
\]

The effect of reflection is to reverse the sign of the $z'$ component of the wave vector, 

\[
k'_2 = (\omega, -v\omega/c^2, 0, \omega/c)   .
\]

We now obtain the reflected wave vector in the lab frame by applying the inverse Lorentz transformation: 

\[
k_2 = (\omega, 0,0,\omega/c)   .
\]

This vector represents a light beam traveling away from the mirror, having the same frequency as the incoming beam. This confirms that the reflection follows the usual laws of geometrical optics and that the beam undergoes no aberration. 

There are two fundamental errors in solving problems related to the reflection of light from a mirror moving tangentially to its surface.
The first error arises from applying the concept of a plane wave and an infinite plane mirror to the case of tangential motion. This assumption leads to an inherent contradiction: if the mirror were truly infinite, the problem would not be time-dependent. This hidden assumption—namely, the absence of time dependence—eliminates aberration, which is unsurprising. However, only the motion of a finite mirror is physically meaningful.
From an electrodynamics perspective, only the velocity of a finite mirror has physical significance. Textbook treatments often overlook the crucial interaction between light and the moving edges of the mirror. A well-defined energy transport problem requires specifying both the source and mirror apertures—an essential detail frequently neglected in conventional explanations.

We consider the case of a finite-aperture mirror moving tangentially to its surface. For simplicity, we assume that the mirror's size is much smaller than that of the "plane-wave" emitter, as illustrated in Fig. \ref{B33}. Notably, we analyze a scenario where the aberration angle is significantly larger than the divergence of the reflected radiation. 
Mathematically, this is expressed as  $\lambdabar/D_e \ll \lambdabar/D_m \ll v/c$, where $D_e$ and  $D_m$  are the sizes of the emitter and mirror, respectively.
Classical optics textbooks state that energy transport in the reflected light beam remains unchanged.

The second error in the textbook argument is conceptual. Consider an observer at rest in the inertial frame who measures the direction of energy transport. This observer describes the light beam emitted from a stationary source using the Minkowski metric. According to textbooks, the metric of the moving mirror is assumed to be the same Minkowski metric as that of the stationary emitter. \footnote{We should make one additional remark regarding two common errors when solving problems related to reflection from a mirror moving tangentially to its surface. These errors are not independent. Textbooks typically assume that the metric of the moving mirror is diagonal, similar to the metric of a 'plane wave' emitter at rest, and that the group velocity equals the phase velocity. As a result, regardless of the mirror's size, there is no deviation in the energy transport of the reflected light beam.} 


The textbook authors incorrectly assume that a common Lorentz time coordinate axis can be assigned to both the emitter and the mirror. This is a misconception. A fundamental question arises: how should synchronization be established in the laboratory frame after the mirror has undergone acceleration?

A distinctive feature of this problem in the context of relativistic kinematics is that the emitter (along with the observer and measuring devices) remains at rest in the lab's inertial frame, while the mirror moves at a constant velocity relative to the lab frame, interacting with the radiated light beam.
How can we address the issue of the emitter-mirror relative velocity?
One approach is to assign a Lorentz time coordinate to describe the emitter’s radiation. However, for the moving mirror, this time coordinate would act as an absolute time reference, and its motion would be described by a Galilean transformation in this framework. Without altering synchronization in the lab frame, a common set of synchronized clocks for both the mirror and the emitter can only be established under absolute time coordinatization—that is, when simultaneity is absolute.

Suppose we re-synchronize the clocks in the lab frame to define the Lorentz time coordinate for the boosted mirror. In this framework, the reflection of light is described using Maxwell's equations. However, this time transformation merely results in a rotation of the radiation phase front of the incoming plane wave by an angle $v/c$. Consequently, in this new coordinatization, Maxwell's equations no longer directly apply to describing the emitter’s radiation.
Notably, a key feature of this coordinatization is that the energy transport velocity differs from the phase velocity of the incoming light beam.

This problem in special relativity can be effectively analyzed using an absolute time coordinatization approach. We demonstrate that when a mirror with a finite aperture moves tangentially while a plane wave of light falls normally onto it, the energy transport of the reflected light deviates. In this case, the aberration of light arises directly from the time dependence of the mirror’s position, necessitated by its finite aperture.

Notably, in absolute time coordinatization, the wave equation explicitly depends on the velocity vector. Consequently, the solution naturally involves light beams of different frequencies, incorporating the Doppler effect. For a tangentially moving mirror with a finite aperture, angular frequency dispersion—an effect of first order in $v/c$—cannot be neglected.

We have already emphasized that Chapter 3 is arguably the most important part of this book. To our knowledge, the operational interpretation of absolute time coordinatization, as well as the distinction between absolute time synchronization and Einstein’s time synchronization from an operational perspective, has not been explicitly addressed in the literature.
Applying the Galilean boost, we substitute $x \to x-vt$ while keeping time unchanged in the Minkowski metric, $ds^2 = c^2 dt^2 - d x^2 - dy^2 - dz^2$, leading to the metric given in Eq. \ref{GGG11}. This metric characterizes the electrodynamics of the moving mirror as observed from an inertial reference frame, representing the measurements from the perspective of an inertial observer.

\subsection{Solving the Emitter-Mirror Problem in (3+1) Space-Time}

The aberration of light problem can be addressed within a "single inertial frame" framework, without invoking Lorentz transformations.
When an object is illuminated by a monochromatic, spatially coherent source, a particularly simple method exists for calculating the reflected intensity. This approach, rooted in Fourier transform techniques from spatial filtering theory—specifically, Abbe diffraction theory—offers conceptual elegance.

In this context, Abbe's method treats the mirror as a diffraction grating that decomposes an incident plane wave into multiple diffracted beams, each representing a Fourier component of the reflected power. Since a finite-aperture mirror is a non-periodic object, it produces an infinite continuum of diffracted beams.

\begin{figure}
	\centering
	\includegraphics[width=0.4\textwidth]{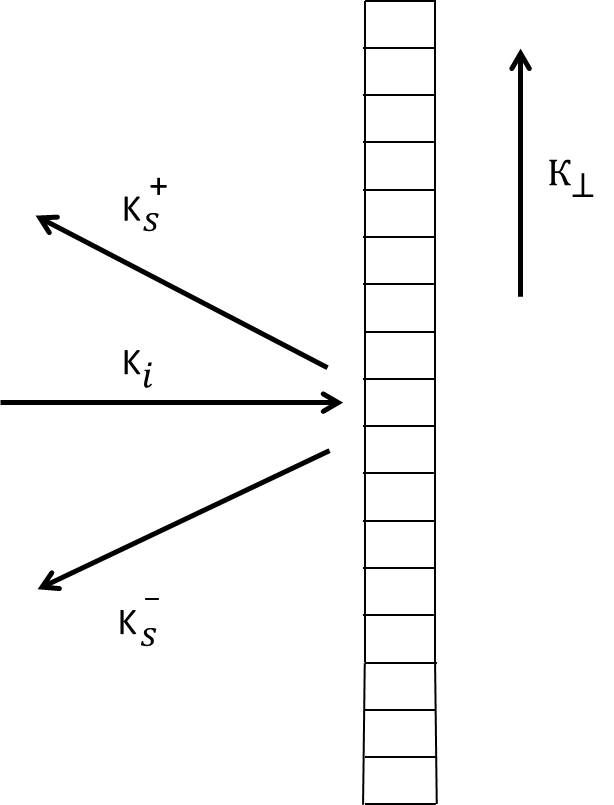}
	\caption{Bragg diffraction grating at normal incidence. The reflectance is sinusoidally space-modulated. }
	\label{G1}
\end{figure}

A basic example of a diffraction grating is illustrated in Fig. \ref{G1}.
Assuming the grating’s reflectance follows the function

\[
R = g(K_{\perp})\sin{(\vec{K}_{\perp}\cdot\vec{r})}     ,
\]

the reflectance is sinusoidally modulated in space. It is important to note that the permanent reflectance distribution grating discussed here is only our mathematical model and we do not need to discuss how it can be created.

The wave vectors $\vec{k}_i$  shown in Fig.\ref{G1} represent the propagation vector of the incident plane wave, which is assumed to be directed perpendicularly to the surface. The vectors $\vec{k_s}^{(+)}$ and $\vec{k_s}^{(-)}$ indicate the scattered light.
The Bragg condition $\vec{k}_s = \vec{k}_i \pm \vec{K}_{\perp}$,  describes the relationship of the incident and scattered waves. Since the scattering occurs from a sinusoidal grating rather than a set of discrete planes (grooves), the first-order maximum dominates.

We assume that the grating vector $\vec{K}_{\perp}$ lies parallel to the surface of the grating, while the incident wave is normal to the surface, as illustrated in Fig.~\ref{G1}. Under the Bragg condition for small angles, the scattering angle is approximately given by $\theta \approx K_{\perp} / k_i$.

Let us analyze the reflection of light from a mirror moving tangentially (i.e., parallel to its surface), relying solely on relativistic kinematics. To describe this scenario, we adopt the framework of absolute time coordinatization and apply a Galilean boost. In this context, we utilize the metric given by Eq. \ref{GGG11}, which characterizes the electrodynamics of a moving light source as observed from an inertial frame.

From the structure of Eq. \ref{GGG11}, it becomes evident that the electromagnetic field configuration—expressed in terms of the coordinates $x$ and $t$ of a detector at rest in the inertial frame—coincides with the field configuration at position $x - vt$ at time $t = 0$.

Using a Fourier analysis, we show that the problem can be reduced to that of reflection from a tangentially moving diffraction grating. Consider a plane wave of the form  $\exp(i\vec{k}\cdot\vec{r} - i\omega t)$ where $\vec{k}$ is the wavevector and  $\omega$ the angular frequency. In the absolute time coordinatization, the dispersion relation takes the form Eq. \ref{CD2}.

Within a single inertial frame, the equivalence between active and passive Galilean boosts implies that translating the system in one direction is indistinguishable from shifting the coordinate system in the opposite direction by the same amount. This equivalence ensures that Maxwell’s equations retain their form in the comoving coordinate system.

In the comoving frame, the wavevector of the diffracted wave is determined by the initial conditions. Specifically, we have:

\[
(k_z')^2 = \omega_i^2/c^2 - (k_x')^2   ,   \quad   k_x' = K_{\perp}     , 
\]

where $K_{\perp}$ denotes the spatial frequency  of the sinusoidally space-modulated reflectance. Transforming back to the lab frame using a Galilean transformation, we find that the frequency shifts according to: $\omega = \omega_i + K_{\perp}v$, while the components of the wavevector remain unchanged.

Substituting these expressions into the dispersion relation  Eq. \ref{CD2} yields: 

\[
(1-v^2/c^2)K_{\perp}^2 + 2vK_{\perp} (\omega_i + K_{\perp}v)/c^2 + \omega_i^2/c^2 - K_{\perp}^2   -(\omega_i + K_{\perp}v)^2/c^2 = 0   ,
\]

which is readily verified to be satisfied, as expected.

This analysis leads to the conclusion that the frequency shift $\Delta \omega$ and the spatial frequency $K_{\perp}$
are related by: $\Delta \omega = K_{\perp}v$. This relation is recognized as a manifestation of the Doppler effect, arising from reflection off a moving (as opposed to stationary) surface.

A crucial aspect of describing a "single inertial frame" cannot be overstated. If the light source is at rest while the mirror is in motion, it follows that the electrodynamic equations must remain identical for all electromagnetic waves. In other words, the dispersion equation in absolute time coordinatization must be consistently applied to both incoming and scattered waves (Fig. \ref{G1}).
In our previous discussion on absolute time coordinatization, we established that an emitter at rest must still be governed by Maxwell's electrodynamics. In this framework, the dispersion equation simplifies to  $k_i^2 -\omega_i^2 = 0$. From the initial conditions, we derive  $\vec{k}_i = \vec{e}_z k_z$ and $\omega_i = ck_z$. However, an apparent contradiction arises—one that is resolved through a geometric analysis of light reflection.
A key feature of this geometry is that even after applying a Galilean transformation along the $x$-axis, the dispersion equation in absolute time coordinatization retains its diagonal form: $k_z^2 - \omega^2 = 0$ for the incident wave. To clarify the fundamental physical principles, we examined a simple case.

One of the consequences of the Doppler effect is the angular frequency dispersion of light waves reflected from a moving mirror with a finite aperture.
If $\vec{n} = \vec{k}/|\vec{k}|$ denotes a unit vector in the direction of the wave normal, and $\vec{v}$ is the mirror velocity vector relative to the lab frame, we get the equation 

\[
\omega_s = \omega_i(1 + \vec{n}\cdot\vec{v}/c) = \omega_i  + (\omega_i v/c)\cos \theta   . 
\]

The Doppler effect is responsible for angular frequency dispersion to the first order of $v/c$ even when  $\vec{n}\cdot\vec{v} = 0$ (i.e when $\cos\theta = 0$). In fact, 

\[
\d\omega_s/d\theta = - (\omega_iv/c)\sin\theta = -\omega_iv/c
\]

at $\theta = \pi/2$.  We can rewrite this equation in a different way. The differential of the scattered angle is given by $d\theta =  - dk_x/k_i$. With the help of this relation and account for that $k_i = \omega_i/c$ we have $d\omega_s/dk_x = v$.

\begin{figure}
	\centering
	\includegraphics[width=0.7\textwidth]{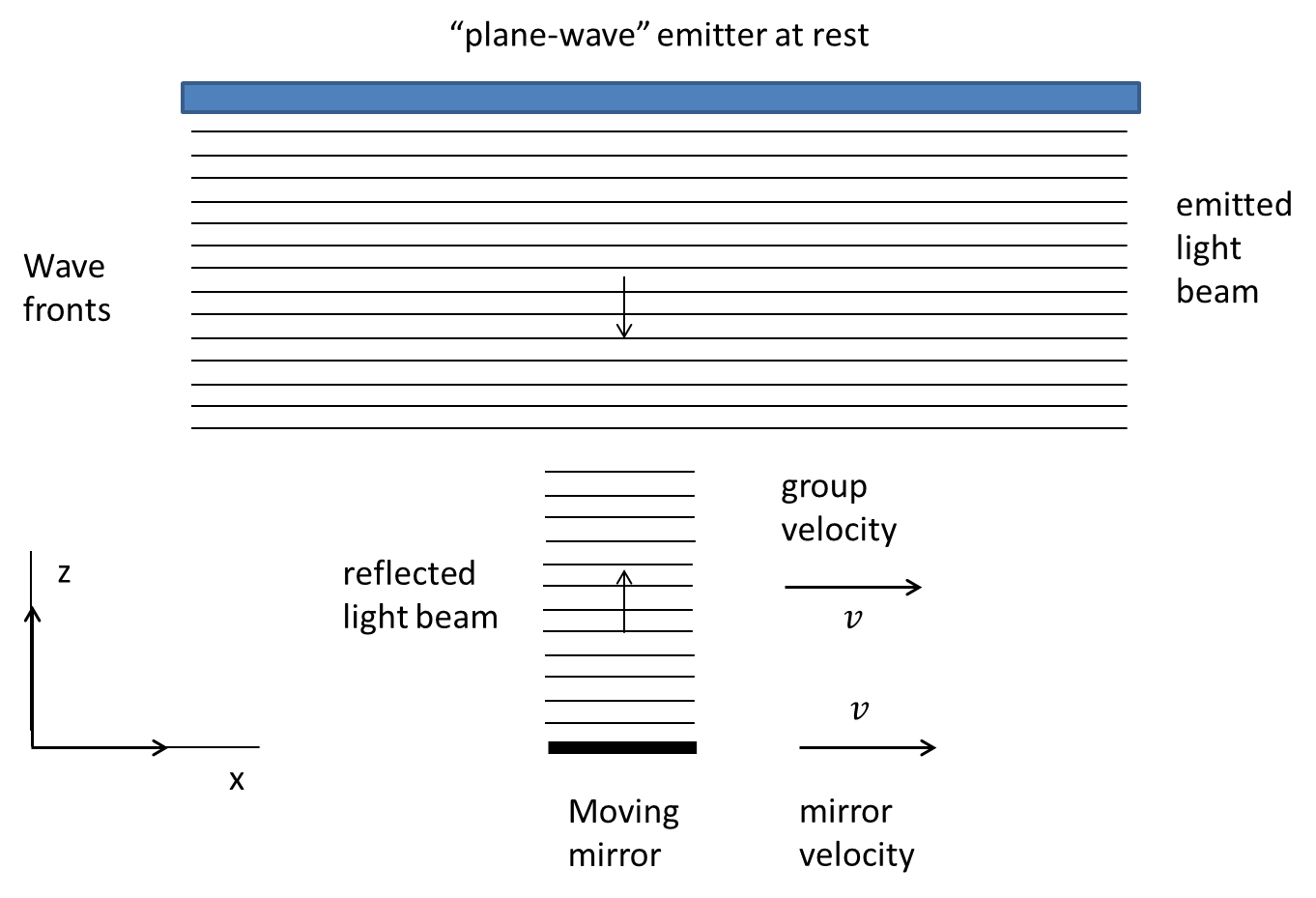}
	\caption{Tangentially moving mirror with a small aperture at normal incidence. 
		The effect of aberration is analyzed using absolute time coordinatization.}
	\label{B17}
\end{figure}

One of the key conclusions of the preceding discussion is a striking prediction in the theory of the aberration of light—specifically, the deviation of energy transport for light reflected from a mirror moving tangentially to its surface. When a plane wave (in absolute time coordinatization) falls such a mirror perpendicularly, the reflected beam exhibits a deviation in energy transport (see Fig. \ref{B17}). This phenomenon can be understood as a straightforward consequence of the Doppler effect\footnote{The Doppler effect serves merely as an intermediate mathematical description; when the full calculation is performed, only the geometric (aberration) effect remains.}.

A common misconception in the literature asserts that the direction of propagation in reflection from a tangentially moving mirror is not determined by the normal to the wavefront.\footnote{A widespread misunderstanding stems from assuming that the orientation of radiation wavefronts has objective significance. As Norton \cite{N} notes: "One might try to escape the problem by supposing that the direction of propagation is not always given by the normal to the wavefront. We might identify the direction of propagation with the direction of energy propagation, supposing the latter to transform differently from the wave normal under Galilean transformation. Whatever may be the merits of such proposals, they are unavailable to some trying to implement a principle of relativity. If the direction of propagation of a plane wave is normal to the wavefronts in one inertial frame, then that must be true in all inertial frames."}
This misunderstanding arises from a failure to distinguish between convention-dependent and convention-invariant aspects of the theory. The direction of energy transport is an objective, convention-invariant quantity. In contrast, the phase front orientation—corresponding to an observer’s plane of simultaneity—lacks objective meaning, as no experimental method can precisely determine this orientation due to the finite speed of light.

Since the phase front orientation has no physical reality within the angular range $v/c$, one might ask: why must we account for its exact orientation in electrodynamics calculations? The answer lies in the fact that, when analyzing radiation beam evolution within a single inertial frame, the phase front orientation appears unchanged. While this invariance lacks objective physical meaning, it serves as a useful tool in solving electrodynamics problems. A useful analogy can be drawn with gauge transformations in Maxwell’s electrodynamics.

\subsection{A Moving Mirror: Lorentz Transformations-Based Explanation}

Another clear way to explain the aberration of light from a moving mirror is by using a Lorentz boost, which accounts for the uniform translational motion of the mirror in the laboratory frame.
Maxwell’s equations are invariant under Lorentz transformations; however, the Lorentz boost introduces a space-time coordinates transformation of the form Eq. (\ref{GGT3}).
To minimize the mathematical complexity of the discussion, we will restrict our analysis to terms up to first order in $v/c$. This entails only a time shift of the form  $t \to t + xv/c^2$. This time shift corresponds to an effective rotation of the phase front of the incoming radiation by an angle of $v/c$. Within this framework, the mirror's reflection is analyzed using Maxwell’s equations.
 
For a plane wave of the form $\exp(i\vec{k}\cdot\vec{r} - i\omega t)$, the dispersion equation in Maxwell's electrodynamics simplifies to $k_z^2 + k_x^2 - \omega^2/c^2 = 0$. 

In the Lorentz-synchronized coordinates, the electric field of the incident wave transforms to: 

\[
\exp[i( - \omega_i z/c + v\omega_i x/c^2 - \omega_i t)]   .
\]

Using a Fourier decomposition, the problem reduces to that of wave reflection from a grating moving tangentially with velocity $v$. By invoking the equivalence between passive and active Lorentz transformations, we find that in the lab frame the diffracted wave components are given by:

\[
\omega = \omega_i + K_{\perp}v   ,    \qquad    k_z = \omega_i/c    ,    \qquad   k_x = v\omega_i/c^2 + K_{\perp}   .
\]

Substituting these expressions into the dispersion relation confirms that it remains satisfied, as expected.

As a result of the Doppler effect under Lorentz coordinatization, the reflection from a finite-sized moving mirror exhibits angular frequency dispersion. This implies that the reflected light beam has a group velocity along the $x$-direction given by: $d\omega/dk_x = v$, consistent with the prediction from Maxwell’s equations that the group velocity equals the phase velocity.

It is also insightful to analyze the electrodynamics of an emitter at rest in the case of Lorentz coordinatization. A rotation of the $x$-axis without changing the orientation of the $t$-axis is described by a Galilean transformation $x' = x -vt$, $t' = t$. Conversely, a rotation of the $t$-axis without changing the orientation of the $x$-axis is described by the transformation $x' = x$, $t' = t + xv/c^2$. This time transformation leads to the diagonal wave equation which describes mirror radiation.
To first-order, this transformation mathematically is similar to the Galilean transformation $x' = x +vt$, $t' = t$.

On the other hand, this time transformation introduces a new crossed term in the wave equation which describe emitter radiation, predicting a group velocity $d\omega/dk_x = - v$. However, since there is an angle $v/c$ between the wave vector $\vec{k}$
and the $z$-axis, and considering the initial condition, the radiated light beam from the stationary source propagates along the $x$-axis with a group velocity $d\omega/dk_x = 0$, as expected. This outcome is unsurprising—convention-independent physical results remain unchanged under the new variables.

\subsection{Discussion}

Physicists investigating the textbook approach to the aberration of light often find themselves puzzled by an apparent inconsistency. When analyzing the reflection of light from an infinite mirror moving normally to its surface, the standard textbook reasoning—based on two Lorentz reference frames—correctly predicts the variation in light frequency upon reflection. Relativistic kinematics successfully describes this phenomenon. However, a natural question arises: why does the same method yield incorrect results when the mirror moves tangentially to its surface?

The key to understanding this discrepancy lies in the nature of the motion. In the case of a normally moving mirror, the velocity is perpendicular to the plane of oscillating elementary sources (dipoles). As a result, the plane of simultaneity in absolute time coordinates retains the same orientation in Lorentz coordinates. For collinear motion, relativistic kinematics affects the system only at the second order.

Now, let us consider frequency measurement within the framework of special relativity. Suppose we use a Fabry-Perot interferometer or a grating spectrometer. In such cases, the measured frequency corresponds to the wavelength of the standing wave. This principle applies universally to all frequency measurements, which inherently obey the relationship between light frequency and interference patterns. Importantly, while the deviation of energy transport direction is a geometric effect, interference patterns themselves are independent of the chosen metric. In other words, in space-time geometry, phase remains a four-dimensional invariant. For collinear motion, both the Minkowski metric and the metric given by Eq. \ref{GGG11} yield identical predictions (see Chapter 14 for further details).

Another crucial aspect to consider is the size of the mirror. In the case of an infinite mirror moving normally to its surface, the motion of the mirror is a physically observable effect. Even for an infinite mirror, the problem remains time-dependent, and its velocity carries physical significance. This makes the infinite mirror a useful conceptual tool for analyzing the Doppler effect in such a scenario.

\subsection{Large Aperture Mirror}

\begin{figure}
	\centering
	\includegraphics[width=0.7\textwidth]{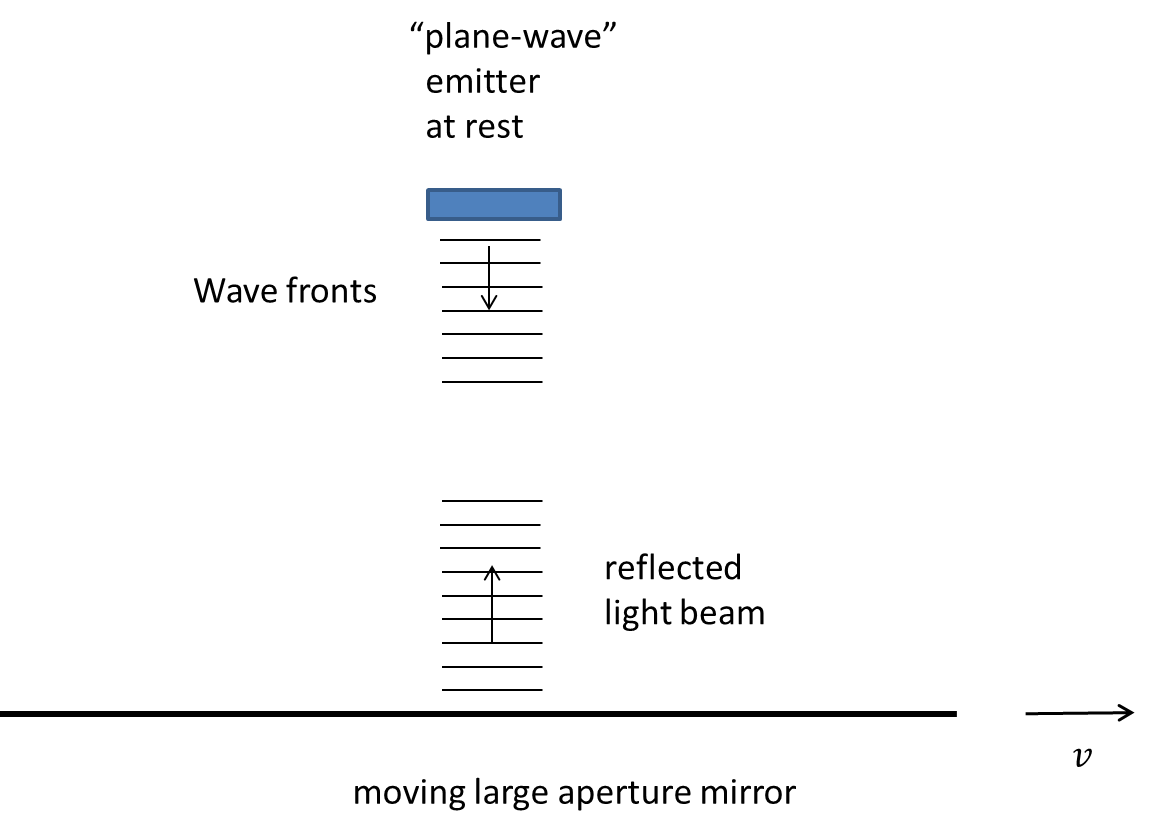}
	\caption{Tangentially moving mirror with a large aperture at normal incidence.
		When a light beam is incident normally on the mirror,  the reflected beam undergoes no deviation in energy transport.}
	\label{B117}
\end{figure}

In this chapter, we focus on a specific range of problem parameters where the effects of emitter edges are neglected. Although this treatment of aberration in light theory is an approximation, it holds significant practical importance.
We also consider the case where the transverse size of the emitter is much smaller than that of the moving mirror. It is important to note that this scenario does not occur in stellar aberration measurements. However, analyzing this case allows us to gain a clearer understanding of the underlying physical principles. Notably, it can be shown that in this situation, energy transport deviation is absent (see Fig. \ref{B117}). To address the moving mirror problem, we employ the Fourier transform method as a direct approach.

\subsection{Analysis of Transmission through a Hole in an Opaque Screen}

\begin{figure}
	\centering
	\includegraphics[width=0.7\textwidth]{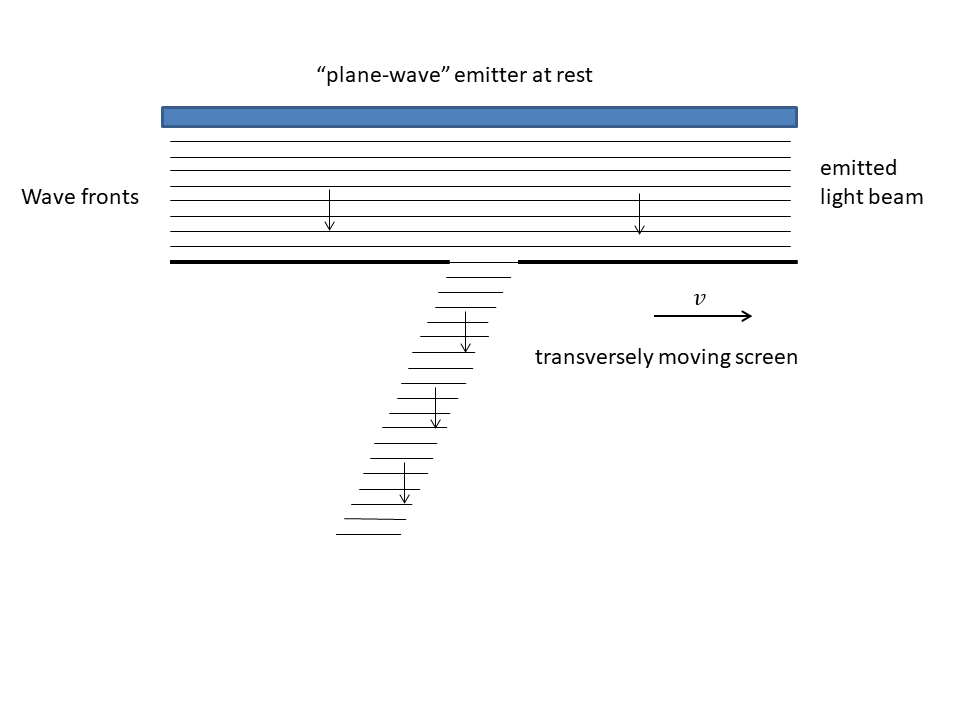}
	\caption{Aberration of light in an inertial frame of reference.  
		A tangentially moving screen with a hole allows the passage of light. According to textbooks, a monochromatic plane wave incident normally on the screen produces a transmitted oblique beam. }		
	\label{B3}
\end{figure}

Previously, we demonstrated that when a finite-aperture mirror moves tangentially to its surface and a plane wave of light is incident normally on the mirror, an aberration (deviation in energy transport) occurs in the reflected light. In this section, we consider a more practically significant problem: a screen moving with velocity $v$ along its surface in the lab frame. It is commonly assumed that light passing through a hole in such a moving opaque screen does not experience any deviation in energy transport (see Fig. \ref{B3}).

However, a common mistake in relativistic optics arises from aberration effects associated with a tangentially moving screen containing a hole. We analyze this system using a Fourier transform method similar to the one employed earlier. The screen with a hole acts as a diffraction grating, decomposing the incident plane wave into multiple diffracted plane-wave components. Each of these components corresponds to a Fourier mode of the transmitted light beam.

\begin{figure}
	\centering
	\includegraphics[width=0.4\textwidth]{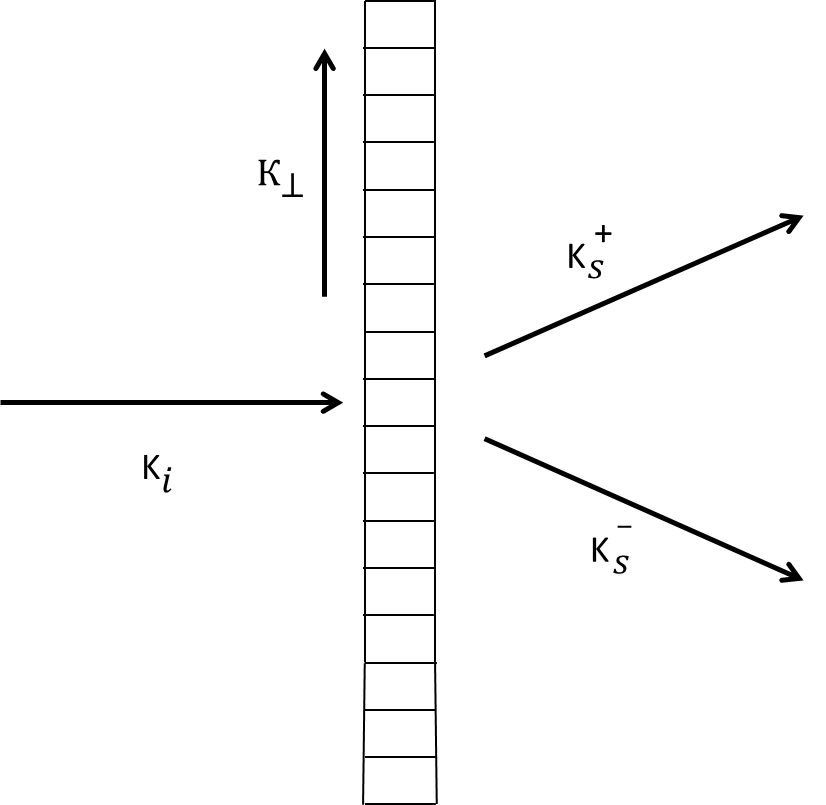}
	\caption{Bragg diffraction grating at normal incidence. The transmittance is sinusoidally modulated in space. }
	\label{G2}
\end{figure}

The gratings discussed thus far modulate the amplitude of an incident plane wave through a periodic reflection function. However, our analysis can be readily extended to gratings that modulate the amplitude of incident light using a periodic transmission function.
Consider a grating whose transmittance varies according to the function $T = g(K_{\perp})\sin{(\vec{K}_{\perp}\cdot\vec{r})}$, as illustrated in Fig. \ref{G2}. In this case, the transmittance exhibits a sinusoidal spatial modulation. Notably, all previously derived equations remain valid for the forward-scattered beams.

\begin{figure}
	\centering
	\includegraphics[width=0.7\textwidth]{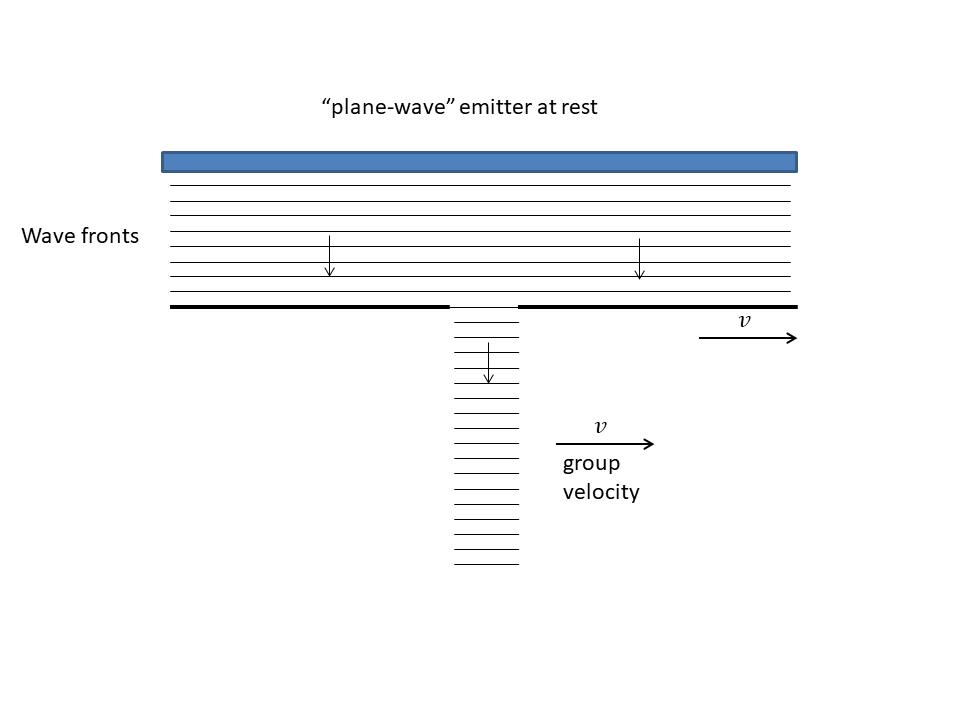}
	\caption{Aberration of light in an inertial frame of reference.
		A tangentially moving screen with a hole allows a normally incident monochromatic plane wave to pass through, generating a transmitted light beam. The explanation of the aberration effect relies on absolute time coordinatization.}
	\label{B1}
\end{figure}

Our approach reveals a remarkable prediction of the theory of aberration of light concerning the deviation of energy transport for light passing through a hole in a moving screen. Specifically, when a screen with a hole moves tangentially while a plane wave of light falls normally on it, the transmitted light exhibits a deviation in energy transport (see Fig. \ref{B1}).

\subsection{Spatiotemporal Transformation of the Transmitted Light Beam}

Suppose the transmitted light pulse propagates in the $x-z$ plane. Our focus is on the space-time intensity distribution within this plane. Spatiotemporal coupling naturally emerges in the transmitted radiation behind the screen due to the angular-frequency dispersion introduced during the transmission process.
In most cases, the emitted light beam can be accurately represented as the product of independent spatial and temporal factors. However, this assumption breaks down when the light must pass through an aperture in a moving opaque screen, as the transmission process disrupts the simple separation of space and time dependencies.

We begin by expressing the field of an emitted pulse as 

\[
E = b_i(x)\exp[ i\omega_i(z/c -t)]      . 
\]

The initial amplitude distribution, $b_i(x)$, in front of the moving screen is the optical replica of the emitter's aperture. The electric field of the transmitted pulse can be represented in the reciprocal domain as

\[
\bar{E}(\Delta k_x, \Delta\omega) = \bar{E}( K_{\perp},   K_{\perp}d\omega/dk_x)   ,
\]

which corresponds to the Fourier component of a beam with angular frequency dispersion,  where $d\omega/dk_x = v$. Taking the inverse Fourier transform, we obtain 

\[
E = b(x - v t)\exp[ i\omega_i(z/c -t)]   . 
\]

This represents the field immediately behind the moving screen. The beam profile, $b(x - vt)$, is the optical replica of the moving aperture.
Now, consider an  observer's screen positioned at rest at a distance $l$ behind the moving aperture. For simplicity, we assume that 
the Fresnel number is large, $N_F = D_h^2\omega/(cl) \gg 1$, allowing us to neglect diffraction effects. Here, $D_h$ is the characteristic aperture size.
From this, we conclude that the light spot on the observer's screen moves with the same velocity $v$ as the moving screen. Additionally, we observe that a particular quantity remains unchanged: the combination $x - vt$ is the same for both the spot and the aperture. What does this imply? A careful examination reveals that events occur simultaneously at two separate locations along the $z$-axis.
\footnote{Spatiotemporal coupling is typically discussed in the literature in the context of ultrashort laser pulse propagation through a grating monochromator. In this scenario, in addition to considering phase fronts, one must also account for planes of constant intensity—known as pulse fronts. In a grating monochromator, the different spectral components of the outgoing pulse propagate in different directions. The electric field of a pulse with angular dispersion can be expressed in the Fourier domain $[\Delta k_x,\Delta\omega]$ as $\bar{E}(\Delta k_x -p\Delta\omega, \Delta\omega)$.
Its inverse Fourier transform to the space-time domain $[x,t]$ results in  $E(x, t+px)$, representing a pulse with a pulse-front tilt. In our case of interest, we consider light diffracted by a tangentially moving set of gratings, where the Doppler effect induces frequency dispersion given by $d\omega/dk_x = v$. Consequently, the spatiotemporal coupling due to light transmission through a hole in a moving screen differs significantly from the conventional pulse-front tilt distortion.}

\subsection{Applicability of Ray Optics}

\begin{figure}
	\centering
	\includegraphics[width=0.6\textwidth]{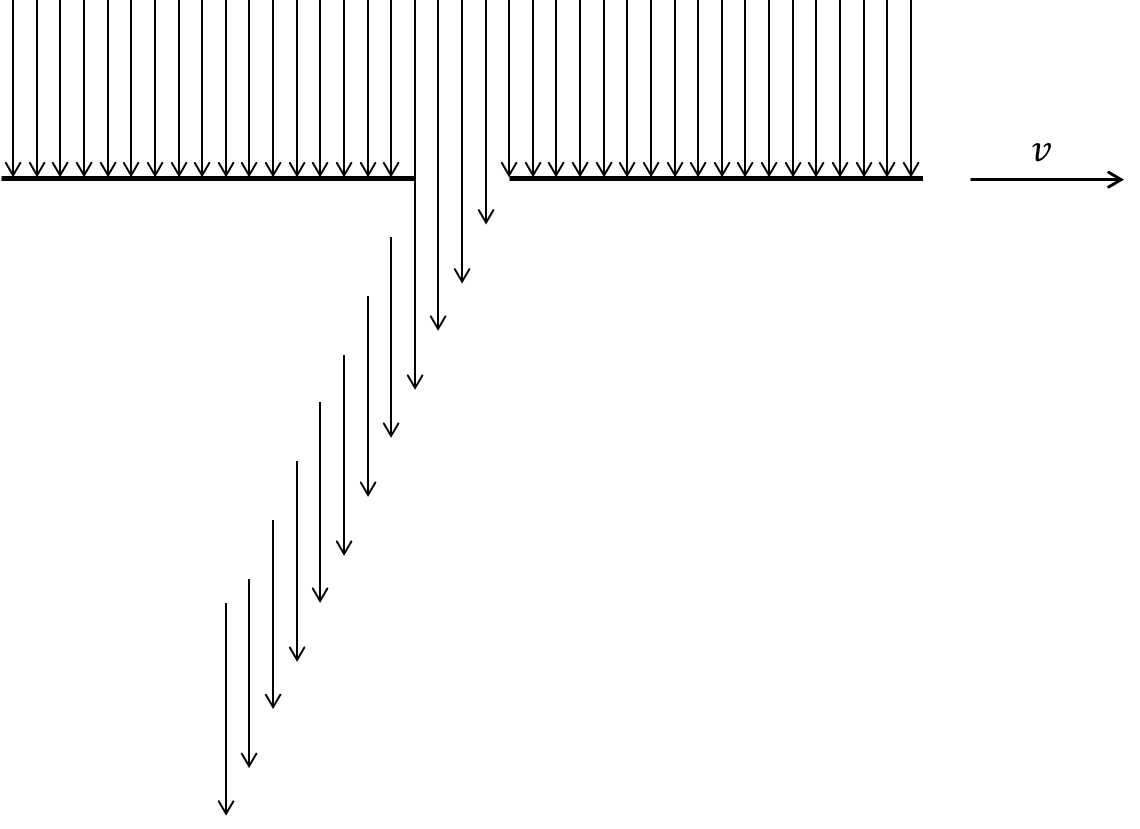}
	\caption{A tangentially moving screen with a hole in it. Light corpuscles fall normally on the screen and, according to the literature, produce an oblique light beam.}
	\label{B11}
\end{figure}

Today, it is commonly suggested that the phenomenon of light aberration can be explained using a corpuscular model of light. According to this model, light particles (corpuscles) fall a moving screen normally, generating an oblique light beam as shown in Fig. \ref{B11}. However, this argument remains incorrect and continues to persist. A satisfactory treatment of light aberration should be based on the electromagnetic wave theory.

Some experts argue that the applicability of the corpuscular model in the theory of light should be reinterpreted as the applicability of ray optics. \footnote{To quote Brillouin \cite{BRIL}: "Fig. 5 explains the situation assuming a simplified device consisting of a parallel plate moving with uniform velocity $v$ in the horizontal direction. Monochromatic light falls normally on the plate and generates an oblique ray." This oblique effect is demonstrated in Fig. \ref{B11}.} 
Let us examine what happens, according to ray optics, in our case of interest. When light rays fall normally on a moving screen, they generate an oblique ray beam, as illustrated in Fig. \ref{B11}.
In this scenario, treating light rays as tiny particles leads to a well-known effect.

Next, we discuss the region of applicability of ray optics. The use of ray optics in the theory of light aberration presents complexities. One might intuitively expect, based on textbook reasoning, that ray optics should apply to any spatially incoherent radiation. However, this assumption leads to incorrect results. Specifically, a completely incoherent source (such as an incandescent lamp or a star) consists of elementary point sources that are statistically independent and have varying offsets.

The radiation field produced by a completely incoherent source can be understood as a linear superposition of the fields of individual elementary point sources. Each elementary source generates an effective plane wave in front of an aperture. Consequently, the transmission process inevitably involves the diffraction of a plane wave through the aperture. Notably, any linear superposition of radiation fields from elementary sources retains fundamental single-source characteristics, such as deviations in the energy transport direction. This reasoning explains why ray optics is not applicable to the aberration theory of light originating from (spatially) completely incoherent sources.

To illustrate the applicability of ray optics in light aberration theory, we consider a specific class of spatially incoherent light beams. A method can be proposed to generate a ray beam from primary sources using an array of randomly phased lasers. Such a planar source produces rays that fall normally on the moving screen (within the laser Rayleigh range) and generate an oblique transmitted ray beam, as shown in Fig. \ref{B11}. Intuitively, a tangentially moving screen with a hole acts as a switcher for the lasers. A luminous spot moving at velocity $v$ can, in principle, be created more simply—so to speak, "manually."
By arranging laser-like sources along the $x$-axis and activating them sequentially from left to right with a given time lag, a luminous spot can be made to move at any desired velocity, even exceeding $c$. However, it is crucial to note that in this process, no information is transmitted along the $x$-axis, as each source radiates independently.

\subsection{Moving Large Aperture Emitter}

Let us now consider a "plane-wave" emitter that starts at rest in the lab inertial frame and is then accelerated to a velocity $v$ along the $x$-axis. An emitter with a finite aperture acts as an active medium, breaking the radiated beam into multiple diffracted plane-wave components. Each of these beams corresponds to a Fourier component of the emitted radiation. As discussed earlier in this chapter, the energy transport of coherent light from a tangentially moving emitter deviates from the expected path—a well-known phenomenon related to the aberration of light in an inertial frame of reference (Fig. \ref{B103}).

\begin{figure}
	\centering
	\includegraphics[width=0.7\textwidth]{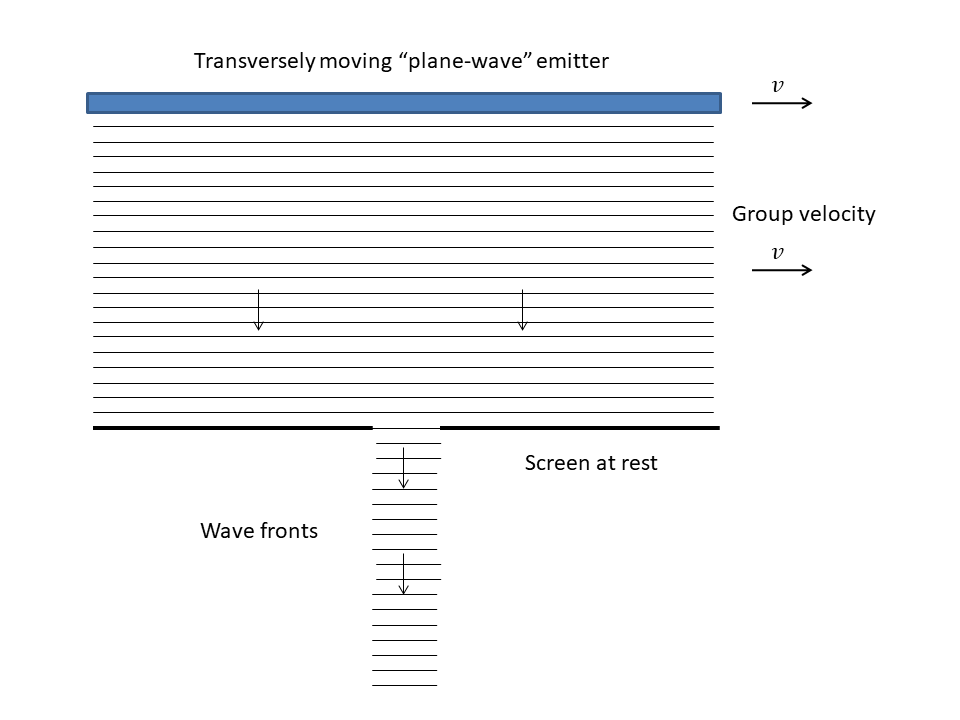}
	\caption{Aberration of light in an inertial frame of reference.
		A large-aperture plane-wave emitter moves tangentially to its surface, while
		the screen remains at rest. In this scenario, no aberration of light occur. }
	\label{B10}
\end{figure}

In our case of interest, the screen remains at rest in the lab frame, while the emitter moves tangentially to its surface at a constant velocity. For simplicity, we assume that the moving "plane-wave" emitter is significantly larger than the hole in the screen.
Now, consider an observer at rest relative to the screen, measuring the direction of energy transport (Fig. \ref{B10}).

If the screen is at rest while the light source is in motion, it is evident that the equations of electrodynamics must remain identical for all electromagnetic waves.
In our previous discussion, we established that a stationary screen must be described by Maxwell's electrodynamics. Consequently, the dispersion equation in the Lorentz coordinatization must be consistently applied to both incoming and scattered waves. In Maxwell’s electrodynamics, the dispersion equation is given by:
$k_x^2 + k^2_y + k^2_z - \omega^2 = 0$. A key characteristic of the discussed geometry is that, even after applying a Galilean transformation along the $x$-axis, 
the emitted light beam's wavefront remains perpendicular to the vertical direction $z$, as illustrated in Fig. \ref{B10}. 
A closer examination of the physics reveals that, in the inertial laboratory frame where the screen is stationary, the problem corresponds to a steady-state transmission. Since the Doppler effect is absent, the transmitted beam travels vertically, having lost its horizontal group velocity component. As a result, the transmission behaves as depicted in Fig. \ref{B10}.
It is important to emphasize the following point: when light passes through a small aperture, the resulting beam undergoes diffraction, perturbing its fields. This diffraction effect removes any information about the emitter’s motion, making the transmitted light independent of its original source movement.

\subsection{A Point Source in an Inertial Frame of Reference}

Above, we considered a single moving "plane wave" emitter in an inertial frame of reference. When analyzing the aberration of light in such a frame, two types of sources are particularly useful to consider:

(a) A "plane wave"-like emitter

(b) A point-like source (or, more generally, a spatially incoherent source)

A defining characteristic of a "plane wave" emitter is its ability to produce highly directional fields. Within the near zone, where $z \ll z_0  = D_e^2/ \lambdabar$, the wavefront remains nearly planar with minimal change. This allows for the analysis of light aberration from a single "plane wave" emitter without significant detector influence, provided the detector is sufficiently large. In contrast, point-like sources emit radiation isotropically, forming spherical wavefronts. As a result, the measuring instrument invariably influences the detected radiation.

The diffraction of a source field is typically categorized into Fresnel (near-zone) and Fraunhofer (far-zone) diffraction. In Fraunhofer diffraction, the wave phase is assumed to vary linearly across the detector aperture, which occurs, for example, when a plane wave impinges on the aperture at an angle to the optical axis. In contrast, Fresnel diffraction replaces this linear phase variation with a quadratic one.
In the far zone, both types of sources effectively produce a plane wave in front of the pupil detection system.

\begin{figure}
	\centering
	\includegraphics[width=0.8\textwidth]{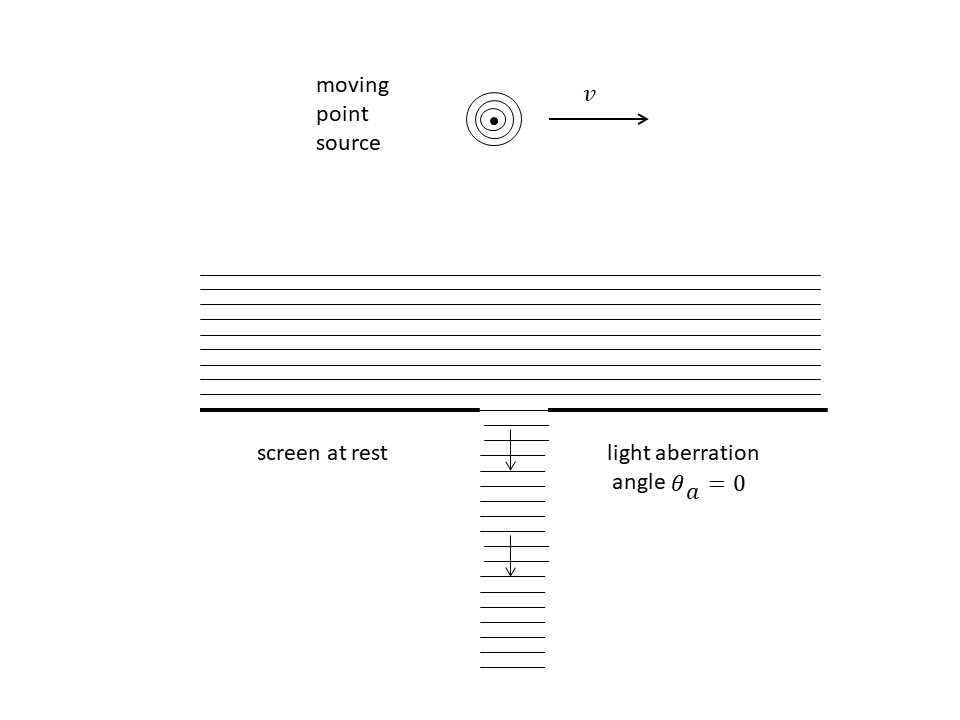}
	\caption{Aberration of light in an inertial frame of reference. In the Fraunhofer diffraction region, a point source produces in front of a hole aperture effectively a plane wave. In this case, the aberration of light phenomenon does not occur.}
	\label{B182}
\end{figure}

Let us now consider the case of a moving point source. Specifically, we examine a setup where the screen is at rest, while the point source moves tangentially. A key takeaway from the earlier discussion is that no aberration is observed for a large-aperture plane wave emitter moving tangentially in an inertial frame (see Fig.~\ref{B10}).

Remarkably, the same conclusion holds for a point source. This stems from the linearity of Maxwell's equations: the governing laws of electrodynamics are linear differential equations. As discussed previously, the most effective approach to analyzing a moving emitter is via the Fourier transform method. A point source can be decomposed into a superposition of spatial Fourier components—each corresponding to a radiated plane wave.

The current density associated with these components can be modeled as  $j_{pol} = j_0\sin{({\vec{K}}_{\perp}\cdot\vec{r})}$. Each elementary dipole source radiates a plane wave of the form $\exp(i\vec{k} \cdot \vec{r} - i\omega t)$, where the wavevector $\vec{k}$ is determined by the initial spatial modulation. Near normal incidence, we set $k_x = K_\perp$, where $K_\perp$ is the transverse spatial frequency corresponding to the modulated dipole density. For small angles, the incidence angle is approximately given by $\theta \simeq K_{\perp}/k_z \simeq K_{\perp}c/\omega$. 

Now, let us examing the behaviour of an incoming wave with a negligible phase gradient along $x$. In other words, a point source located in the far zone effectively generates a plane wave in front of the aperture.
The radiation transmitted through the aperture undergoes diffraction with a divergence angle on the order of $\lambdabar / D_h$, where $D_h$ is the characteristic size of the hole. In our discussion we assume that  the distance from the point source to the screen is sufficiently large and  the diffraction angle is significantly larger than the divergence of the incoming wave. This means we are taking the limit where $K_{\perp}/k_z << \lambdabar / D_h$. Here  
$K_{\perp}$ is  the spatial frequency corresponding to the elementary dipole sources.
Within the diffraction angle, individual elementary emitters cannot be distinguished. Consequently, the problem of a tangentially moving point source reduces to that of a tangentially moving, large-aperture plane wave emitter  (see Fig.~\ref{B182}).

This leads to a remarkable conclusion: the transmitted beam carries no information about the source's tangential motion. This principle lies at the heart of the binary star paradox discussed in Chapter~6.

\subsection{Experimental Test: Reflection from a Moving Grating}

Textbook treatments suggest there is strong theoretical evidence for the absence of an aberration effect when light reflects from a mirror moving transversely (parallel to its surface). Although this result was established long ago by Pauli \cite{PA} and Sommerfeld \cite{S}, a direct experimental verification remains absent.
This raises a natural question: why has it not been possible to experimentally detect the aberration effect in this configuration? Consider a mirror moving tangentially to its surface, assuming the mirror size is smaller than the transverse size of the incoming light beam. Notably, we analyze a setup where the aberration angle $\theta_a = v/c$ is significantly larger than the diffraction divergence $\lambda/D_m$ of the reflected radiation. Let us estimate the parameters for the Earth-based setup using tangentially moving mirror and an optical laser. For instance with a light wavelength of $\lambda = 0.3 ~ \mu\mathrm{m}$ and a mirror size $D_m$ of 10 cm, the mirror speed  
must satisfy $v \gg 1000 ~ \mathrm{m/s}$. Clearly, current experimental capabilities are insufficient to meet these demanding conditions. Fortunately, there exists a more practical alternative to this otherwise formidable
experimental challenge.

Previously, we analyzed the reflection of light from a mirror moving parallel
to its surface, relying solely on relativistic kinematics. After applying a
Galilean boost in the case of absolute time coordinatization, we employed
the metric given by Eq. \ref{GGG11} , which describes the electrodynamics of a moving
mirror as observed from an inertial frame.
Using a Fourier approach, we showed that the problem could be reduced to
the case of reflection from a tangentially moving grating. From the resulting
dispersion relation, we derived a Doppler shift in the diffraction maxima
due to the grating’s motion.

The consistency between our theoretical framework and experimental ob-
servations—such as the measured frequency shift by a tangentially moving
grating  \cite{ET}—provides further support for the validity of our approach.

In contrast, special relativity textbooks commonly conclude
that no Doppler effect occurs for light reflected from a tangentially moving
grating. This argument is typically made for a tangentially moving mirror
but applies equally to gratings. The issue stems from the widely accepted
assumption that the metric of a moving grating (or mirror) is the same as
the Minkowski metric of the stationary emitter. 
\footnote{The reasoning commonly presented in standard textbooks for a moving mirror can be directly extended to a moving grating. 
The argument that light reflected from a tangentially moving grating experiences no Doppler shift proceeds as follows. 
The reflection process is analyzed using two  Lorentz reference frames: the laboratory frame, in which the incident plane wave is defined, and the rest frame of the grating, which moves with velocity $v$. 
It is most convenient to determine the reflected wave vector in the grating's rest frame, where the reflection process is treated as stationary. 
The reflected wave in the laboratory frame is then obtained by applying the inverse Lorentz transformation. 
This procedure yields a reflected wave propagating away from the grating with the same frequency as the incident radiation. }

We now present a simple explanation of our central result.
The clearest understanding arises when we accept, as an experimental fact, the presence of a Doppler frequency shift in radiation diffracted from a moving grating.

Within our relativistically consistent framework—based on the equivalence of active and passive boosts within a single inertial frame—the diffracted radiation must satisfy the dispersion relation: $(1-v^2/c^2)k_x^2 + 2vk_x \omega/c^2 + k_z^2 -\omega^2/c^2 = 0$. The frequency shift $\Delta\omega = K_{\perp}v$ is exactly compensated by the cross term $2vk_x\omega/c^2$,  ensuring that the dispersion relation remains valid.

In contrast, conventional treatments in special relativity textbooks describe diffraction from a moving grating using Maxwell’s equations, leading to the dispersion relation:  $k_z^2 + k_x^2 - \omega^2/c^2 = 0$. This equation is isotropic and independent of velocity. As a result, it contradicts experimental observations: the Doppler frequency shift cannot be accounted for or compensated.

However, experimental evidence clearly demonstrates that the frequency of the reflected radiation depends on the tangential velocity of the grating. Doppler frequency shifts can be detected with high precision using advanced spectroscopic techniques, even at very low grating velocities. For instance, the authors of \cite{ET} used a Mach-Zehnder-type interferometer to measure these shifts.

This striking discrepancy suggests that the conclusions reached by Pauli \cite{PA}  and Sommerfeld \cite{S} are inconsistent with the principles of special relativity.

\subsection{Differences between the Light and Sound Aberration}

Having completed our discussion on the aberration of light in an initial inertial frame, we now turn to optical phenomena in an accelerated frame. However, before proceeding, we will first explore optical-acoustic analogies.
Sound is a periodic motion of air caused by vibrations of a source. The effect of sound aberration is best understood using a model of a single plane-wave emitter. Specifically, we consider a two-dimensional array of identical, coherent elementary sources uniformly distributed on a given $(x-y)$ plane $P$, all emitting waves simultaneously.
We analyze a scenario in which a finite-aperture mirror moves tangentially to its surface. For simplicity, we assume the mirror's size is small compared to the plane-wave emitter. An observer, at rest in the atmosphere frame, describes the emitted sound beam using the diagonal wave equation: 

\[
\nabla^2 f - \partial^2 f/\partial(v_st)^2  = 0     , 
\]

where $v_s$ is the speed of sound. 

The emitter radiates a plane wavefront in the vertical $z$-direction, and the moving mirror obeys the same wave equation as the stationary emitter.
The amplitude of the beam emitted by the oscillating sources can be understood as a superposition of spherical wavelets, as dictated by Huygens’ principle. The wavefront at any given instant is the envelope of these wavelets. Since the diagonal wave equation exhibits no intrinsic anisotropy, the energy transport of the reflected sound beam remains unchanged. A plane wave incident normally on the finite-aperture mirror generates a reflected oblique beam. Notably, in the atmosphere frame, the phase and group velocities remain equal.

Textbooks generally state that, to first order in $v/c$, there is no fundamental difference between the aberration of light and sound. The geometry of reflection from a moving mirror is identical (Fig. \ref{B33}). Now, consider a screen moving tangentially to its surface with velocity $v$ in the atmosphere frame. According to sound theory, energy transport for a sound beam transmitted through a hole in a moving screen remains undeviated. Consequently, it is commonly assumed that the energy transport for light transmitted through a hole in a moving opaque screen follows the same geometric behavior (Fig. \ref{B3}). In the next chapter, we will continue our discussion on the aberration of sound.

The conventional treatment of the aberration of light often leads physicists to draw misleading parallels between the aberration of sound and light in the first-order approximation. This is largely due to the fact that the wave equation governing a moving mirror (or screen) retains the same diagonal form as the wave equation of an emitter at rest. However, in sound theory, we operate within Newtonian space and time rather than the pseudo-Euclidean geometry of space-time. This distinction is crucial: the standard analysis of light aberration within a single inertial frame fails to account for the fundamental difference between the velocity of light and that of sound.

A rigorous relativistic treatment must be grounded in a space-time geometric approach. The Minkowski metric remains valid in an inertial frame where the observer is at rest, with time coordinates assigned through the slow clock transport method or Einstein synchronization, which is defined via light signals emitted by a dipole source at rest. In this Lorentz coordinate system, the Minkowski metric accurately describes the electrodynamics of a stationary light source from the perspective of an inertial observer.

Now, consider the case where a mirror in an inertial frame accelerates from rest to velocity $v$ along the $x$-axis. Synchronization is maintained using the same set of synchronized clocks, preserving the Minkowski metric for both the inertial observer and the stationary light source. In this scenario, the mirror undergoes a Galilean boost, represented by the transformation $x \to x - vt$, while keeping time unchanged in the observer’s Minkowski metric. This results in Eq. \ref{GGG11}, which describes the electrodynamics of the moving mirror from the inertial observer’s perspective.

In contrast, prior literature has often made the incorrect assumption that an inertial observer (i.e., an observer at rest in an inertial frame) and an accelerated mirror within the same frame share a common Minkowski metric—an assumption long treated as self-evident. 

Consequently, the widely accepted approach to special relativity in the first-order approximation effectively reduces to a classical theory akin to sound propagation.

\newpage

\section{Aberration of Light: Non-inertial Frame of Reference}

In this chapter, we revisit the problem of light transmission in non-inertial frames, focusing on the aberration of light. Using the Langevin metric in general relativity, we derive the aberration for a pulse of light traveling in an accelerated system. The acceleration is in
principle defined in terms of motion relative to the fixed stars. The implicit
”absolute” acceleration means acceleration relative to the fixed stars.

The aberration problem is fundamentally resolved by recognizing the inherent asymmetry between inertial and non-inertial observers. This asymmetry, akin to the well-known Sagnac effect, arises from the absolute nature of acceleration \cite{SA,SF,MAL1}. As Langevin noted in 1921, "any change of velocity, or any acceleration, has an absolute meaning" \cite{LAN}.     Mathematically, however, there is no distinction between calculations in general relativity and special relativity when space-time curvature is absent.

The usual arguments for the relativity of motion do not apply here, as inertial and non-inertial reference frames are not equivalent. The asymmetry paradox is typically resolved by recognizing acceleration as the defining factor. Although the duration of acceleration has a negligible effect on anisotropy within the accelerated frame, the acceleration itself fundamentally determines the problem. Interestingly, the lack of a dynamical explanation for this asymmetry in special relativity has puzzled some physicists. A useful perspective is to interpret the asymmetry as a consequence of the pseudo-gravitational effects experienced by an accelerated observer. From a pedagogical standpoint, treating the accelerated frame via the equivalence principle provides valuable insight, a topic we will explore further in Chapter 13.

\subsection{Absolute Time Coordinatization in Accelerated Systems}

We investigate the phenomenon of light aberration in accelerated systems within the framework of special relativity. In particular, we demonstrate that explaining optical effects in rotating frames does not require modifying special relativity or invoking general relativity. A rigorous application of special relativity is sufficient. However, to express electrodynamics in a non-inertial frame, one must take the additional step of defining the metric of that frame. Crucially, the metric tensor must remain a continuous function, leading naturally to the concept of smoothly tailoring the metric.

This issue in special relativity can be effectively addressed using an approach based on absolute time coordinatization. Toward the end of this section, we outline how special relativity may be mathematically extended to describe accelerating systems. In some situations, however, direct physical reasoning can yield rapid insights while remaining consistent with formal derivations.

Consider an inertial frame $S$, assumed to be at rest relative to the fixed stars. A second frame, $S_n$, along with  
an observer and measuring instruments, is accelerated from rest in $S$ to a velocity $v$ along the $x$-axis. In such 
accelerated systems, the notion of absolute simultaneity remains logically consistent with the smooth tailoring of the 
metric tensor. Synchronization in this context involves preserving the same uniformly synchronized clocks that were  
used when $S_n$ was initially at rest. It is well known that during acceleration relative to the fixed stars,
Einstein's synchronization procedure cannot be directly applied. As a result, by the time $S_n$ reaches uniform  motion, the metric in its reference frame will generally acquire a non-diagonal form.

We begin by treating the metric as the true measure of spacetime intervals for an accelerated observer in frame $S_n$,
with coordinates $(t_n,x_n,y_n,z_n)$.  To express this in the coordinates of an inertial observer in $S$, moving at 
velocity $-v$ relative to $S_n$, we apply the inverse Galilean transformation $x = x_n + vt$, keeping the time
coordinate unchanged: $t_n = t$. Substituting into the Minkowski metric, $ds^2 = c^2 dt^2 - d x^2 - dy^2 - dz^2$,
we obtain the transformed (Langevin) metric:

\begin{eqnarray}
&& ds^2 = c^2(1-v^2/c^2)dt_n^2 - 2vdx_ndt_n - dx_n^2 - dy_n^2 -dz_n^2 ~ .\label{GGG3}
\end{eqnarray}

It is crucial to note that Langevin metric, in fact, reflects measurements performed by an inertial observer using the Minkowski metric, subsequently re-expressed via the transformation $x \to x_n + vt$, $t \to t_n$. This  metric describes the electrodynamics of the accelerated light source with the viewpoint of the accelerated observer measurements as viewing this of the inertial observer. 

Inspecting Eq.(\ref{GGG3}) we can find the components of the metric tensor $g_{\mu\nu}$ in the coordinate system  of $S_n$. We obtain 

\[
g_{00} = 1-v^2/c^2    ,   \qquad  g_{01} = - v/c    ,  \qquad   g_{11} = - 1   .
\]

Note that the metric in Eq. (\ref{GGG3}) is not diagonal, since, $g_{01} \neq 0$, and this implies that time is not orthogonal to space. 
This result represents the Langevin metric, derived by matching the metric tensors of the accelerated frame and the initial inertial frame.

To interpret Eq.~(\ref{GGG3}) physically, recall that a new frame of reference can always be introduced via a passive coordinate transformation—a mere relabeling of coordinates used to describe physical events. For example, an observer in the inertial frame can introduce a comoving coordinate system to analyze radiation from a moving source. In this system, fields are expressed as functions of  $(t_n,x_n,y_n,z_n)$, which relate to the original coordinates $(t,x,y,z)$
through a passive Galilean transformation.

From the structure of the Langevin metric, it is evident that the electromagnetic field measured at position 
$x_n$ and time $t_n$ by a device at rest in the accelerated frame will match the field measured by an identical device at rest in the inertial frame at position $x = x_n +vt$ and time $t = t_n$. Substituting this coordinate relationship into the Minkowski metric of the inertial observer leads directly to Eq.~(\ref{GGG3}).

In describing physical phenomena in an accelerated frame, one must distinguish between coordinate quantities and physical quantities. The Langevin metric allows us to determine how these relate. For instance, the length of a physical rod $dx$ and the proper time interval $dt$ in the inertial frame coincide with the coordinate length and time in the accelerated frame: $dx_n = dx$, $dt_n = dt$. For a more detailed discussion on how to relate coordinate and physical quantities in non-inertial frames, refer to Chapter 13.

Let us now extend the mathematical framework of the special theory of relativity to include accelerated motion. 
To keep the mathematical complexity minimal, we assume that the reference frame $S_n$ starts at $t = 0$ and moves with a
constant acceleration $a > 0$ until $t = T$. Our goal is to determine an expression for velocity from the perspective of an inertial (non-accelerated) observer. 

In relativistic mechanics, uniformly accelerated motion satisfies the equation

\[
d(\gamma\vec{v})/dt = \vec{f}/m = \vec{a}  
\]

which integrates to $\gamma\vec{v} = \vec{a}t$. This leads to the velocity expression 

\[
dx/dt = at/\sqrt{1 + a^2t^2/c^2}    . 
\]

We emphasize that $dx(t)/dt$  represents the velocity of the accelerating frame as observed from the inertial reference frame.

Applying the inverse Galilean transformation, we have 

\[
dt = dt_n    ,   \qquad  dx = dx_n + at_ndt_n/\sqrt{1 + a^2t_n^2/c^2}     . 
\]

With this transformation, the metric in the uniformly accelerating observer’s frame takes the form:

\begin{eqnarray}
&& ds^2 = c^2dt_n^2(1 + a^2t_n^2/c^2)^{-1} - 2at_ndt_ndx_n(1 + a^2t_n^2/c^2)^{-1/2} -dx_n^2  ~ . \label{GGT17}
\end{eqnarray}

Since the metric tensor of space-time must be a continuous quantity, the coordinates in the accelerated and inertial stages must be matched accordingly.
At  $t = 0$, the metric naturally coincides with that of the $S$ frame, giving $g_{00} = 1$. During acceleration, we adopted an absolute time coordinatization. To ensure continuity of the metric tensor as the reference frame transitions from non-inertial to inertial motion, we employ the metric given in Eq. (\ref{GGT17}).
Thus, at $t = T$, the metric of the non-inertial frame must smoothly transition into the  accelerated frame metric. This can be achieved by relating the coordinates and time of the accelerated observer, $(x_n,t_n)$, to those of the inertial observer,  $(x,t)$, via the Galilean boost: $x_n = x - vt$, $t_n = t$, where 

\[
v = aT/\sqrt{1 + a^2T^2/c^2}    .
\]

For coordinate times $T <  t_n$, the reference frame of the accelerated system follows the metric given by Eq.(\ref{GGG3}). 
Importantly, $v$ represents the velocity of the frame $S_n$ as observed by an inertial observer during
the second (inertial) stage of $S_n$ motion.
It is evident that at $t = t_n = T$, the metric of the non-inertial frame continuously transitions into that of the (formerly accelerated) inertial frame.
For a broader discussion on the mathematical extension of special relativity to accelerated motion, we refer the reader to \cite{LOG}.

Textbooks describe inertial frames as moving at constant velocity relative to the fixed stars, with different inertial frames connected by Lorentz transformations. Consequently, one might expect the coordinates and time in frames $S$ and $S_n$ (during the inertial stage) to be related by a Lorentz transformation.
However, during the inertial segment of the trajectory, the metric in both frames remains diagonal in Lorentz coordinates. Thus, at $t_n = T$, the metric tensor of the accelerated frame, according to textbooks, must undergo an abrupt change from  

\[
g_{00} = (1 + a^2T^2/c^2)^{-1}    \to     g_{00} = 1     .
\]

\subsection{Asymmetry Between Inertial and Accelerated Frames}

The velocity of light emitted by a source at rest in the coordinate system $(t, x, y, z)$ of the inertial frame $S$ is $c$.  The Minkowski metric, given by Eq. (\ref{MM1}), implies a symmetry in the one-way speed of light in this frame. 
However, in the coordinate system $(t_n, x_n, y_n, z_n)$ 
of an accelerated frame $S_n$, the speed of light emitted by a
source at rest in $S_n$ differ from $c$
This is because $(t_n, x_n, y_n, z_n)$ is related to $(t, x, y, z)$ via a Galilean transformation.
This is readily verified if one recalls that the velocity of light in the reference system $S$ is equal to $c$. If $ds$ is the infinitesimal displacement along the world line of a ray of light, then  $ds^2 = 0$ and we obtain $c^2 = (dx/dt)^2$. In the accelerated reference system, since $x_n = x - vt$ and $t_n = t$, this expression takes the form $c^2 = (dx_n/dt_n + v)^2$, as seen by setting $ds^2 =0$ in Eq. (\ref{GGG3}). Consequently, in the accelerated coordinate system  $(ct_n,x_n)$, the velocity of light parallel to the x-axis, is $dx_n/dt_n = c - v$ in the positive direction, and $dx_n/dt_n = -c - v$ in the negative direction, as expected. 

We have identified a fundamental asymmetry between inertial and accelerated frames: Maxwell's equations do not hold from the perspective of an observer at rest with respect to an accelerated frame $S_n$. The metric given by Eq. (\ref{GGG3}), which corresponds to the accelerated frame $S_n$, predicts an asymmetry in the one-way speed of light along the direction of relative velocity.
Accelerations (with respect to the fixed stars) influence the propagation of light. In an accelerated system $S_n$, the velocity of light emitted by a source at rest must be modified by the acceleration—either added to or subtracted from—resulting in different speeds in opposite directions. In contrast, in an inertial frame $S$, the velocity of light emitted by a source at rest remains $c$.

One crucial aspect of describing an accelerated reference frame cannot be overstated: the metric applies to physical laws, not to physical facts. In other words, it is always possible to choose a set of variables in which the metric of the accelerated source remains diagonal.
We interpret the Langevin metric as indicating that the laws of electrodynamics, when expressed in an accelerated frame, take the form of anisotropic field equations. Here, "physical facts" refer specifically to the aberration of light emitted by a single "plane wave" source within the accelerated frame. To account for this effect, the electrodynamics equation must be integrated using the appropriate initial conditions for the radiation wavefront.

\begin{figure}
	\centering
	\includegraphics[width=0.8\textwidth]{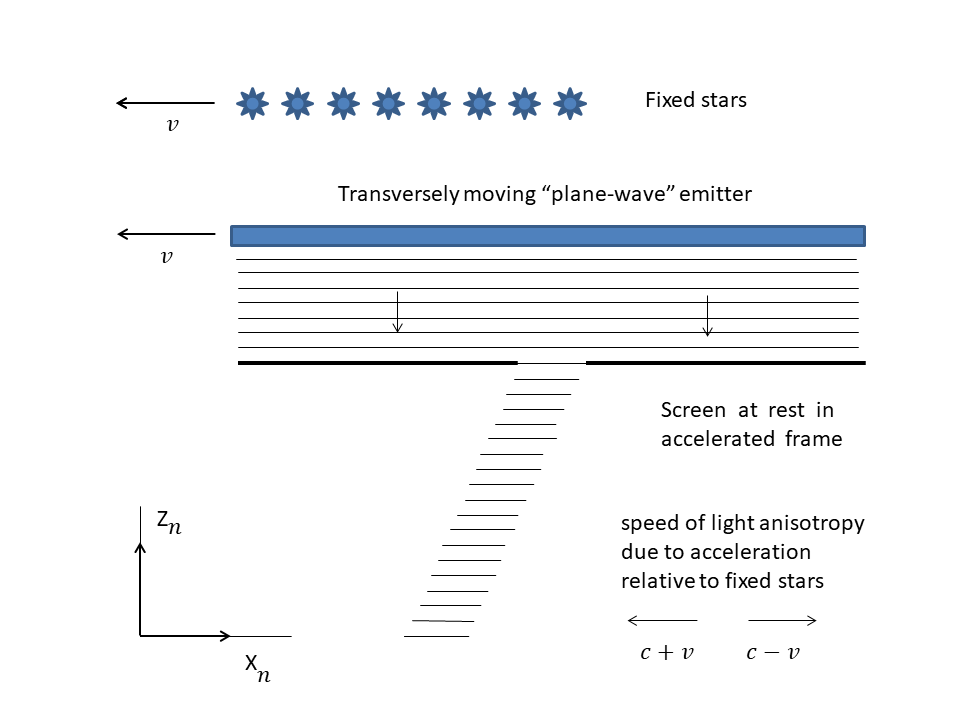}
	\caption{Aberration of light in an accelerated frame of reference $S_n$. The orientation of radiation wavefronts and the anisotropy of the speed of light are depicted using absolute time coordinatization ($t_n = t$) for the screen.}
	\label{B99}
\end{figure}

After the boost we can see that  acceleration does not affect the wavefront orientation.
In fact, the variables $(t,x,y,z)$ can be expressed in terms of the variables $(t_n,x_n,y_n,z_n)$ 
by means of Galilean transformation $x_n = x - vt$, $t_n = t$. This transformation ensures that, following acceleration, the emitted light beam's wavefront remains perpendicular to the vertical direction $z_n$, as illustrated in Fig. \ref{B99}. We will explore this topic further in Section 5.5.

\begin{figure}
	\centering
	\includegraphics[width=0.8\textwidth]{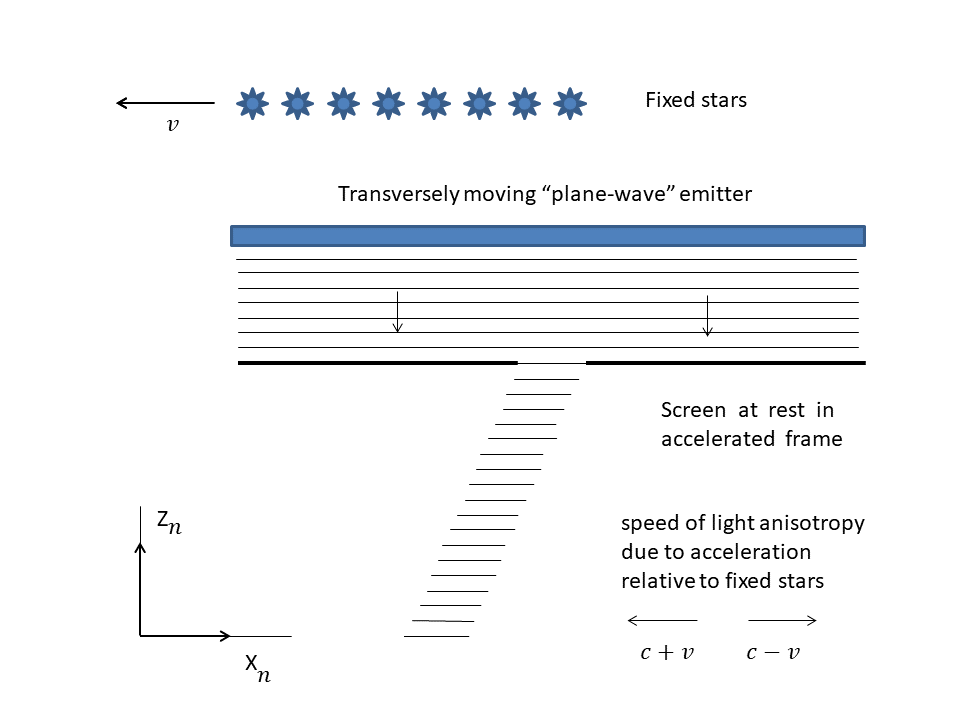}
	\caption{Aberration of light in an accelerated frame of reference $S_n$. The screen  at rest within the accelerated frame. The aberration of transmitted light is independent of the source speed.}
	\label{B999}
\end{figure}

It is useful to begin with an overview of some key results. 
Consider an observer in the accelerated frame $S_n$ performing an aberration measurement. To predict to the outcome, the
observer must use the non-diagonal metric given in Eq. (\ref{GGG3}). Due to the inherent asymmetry between inertial and accelerated frames, an intriguing prediction arises in the theory of light aberration. Specifically, if an opaque screen with a hole is initially at rest relative to the fixed stars and then set into motion, the apparent angular position of a "plane-wave" emitter observed through the aperture in the accelerated frame will shift by an angle $-v/c$. This effect is illustrated in Fig. \ref{B99}.
Notably, the source of this asymmetry is not the relative acceleration between the two frames but rather the difference in their individual acceleration histories with respect to the fixed stars.

In Chapter 4, we analyzed the emitter-screen problem in the reference frame $S$, which remains at rest (i.e., without any history of acceleration) relative to the fixed stars. We found that, in the case of a small aperture hole, no aberration arises from the emitter, regardless of its motion.
It is important to emphasize that the aberration of the light beam transmitted through the small aperture hole also 
remains independent of the emitter's motion in the accelerated frame $S_n$ (see Fig. \ref{B999}).
This is because the 
cross-term in the metric (Eq. \ref{GGG3}), which induces aberration in this particular case, depends solely on the velocity change of the accelerated frame relative to the initial inertial frame.

\subsection{Langevin Metric and Synchronization Using Light Signals}

Although the choice of a coordinate system has no intrinsic physical significance and serves primarily as a convenient means of describing events, it nevertheless reflects, to some extent, the assumptions underlying that description. In particular, different coordinatizations of an accelerated reference system arise from different conventions adopted for the synchronization of spatially separated clocks. The construction of a coordinate grid—namely, the assignment of four numbers $(x,y,z,t)$ to each spacetime event—is therefore an operational procedure. In this section, the assignment of coordinates is analyzed using light signals, a method especially well suited for discussing synchronization.

Synchronization by means of light signals was first proposed by Poincaré and later adopted by Einstein. The procedure may be summarized as follows. In general, nothing requires the velocity of light in one direction to be equal to that in the opposite direction; equality of the one-way speeds of light is a convention rather than an experimentally independent statement. 

Consider clocks located at two spatial points $A$ and $B$. Let synchronization be performed using light signals that propagate with velocity $c_1$ in the positive $x$-direction and velocity $c_2$ in the opposite direction. A light signal emitted from point $A$ at time $t_A$ reaches point $B$ at
\[
t_B = t_A + \frac{x_{AB}}{c_1}.
\]
After reflection at $B$, the signal returns to $A$ at
\[
t'_A = t_A + \frac{x_{AB}(c_1+c_2)}{c_1c_2}.
\]
Combining these relations gives
\[
t_B = t_A + \frac{c_2}{c_1+c_2}(t'_A-t_A).
\]
This leads to the synchronization convention introduced in \cite{RE},
\[
t_B = t_A + \epsilon_{AB}(t'_A - t_A),
\]
where
\[
\epsilon_{AB}=\frac{c_2}{c_1+c_2}, 
\qquad 
\epsilon_{BA}=\frac{c_1}{c_1+c_2}.
\]
For Einstein synchronization, $c_1=c_2=c$, and therefore $\epsilon_{AB}=1/2$. If $\epsilon_{AB}\neq 1/2$, the one-way speed of light from $A$ to $B$ differs from that from $B$ to $A$.

A pedagogically useful (though of limited applicability for precise scientific analysis) example of a nonstandard coordinatization is the so-called *radio* or *everyday* coordinatization \cite{LAK}. In this scheme one sets $\epsilon_{AB}=0$, meaning that distant clocks are synchronized by signals emitted from a single master clock without compensating for signal propagation time. Radio synchronization is therefore intrinsically asymmetric and leads to asymmetric one-way light velocities, formally corresponding to $c_1=\infty$,  $c_2 = c/2$.

We now determine the synchronization parameter $\epsilon$ for the accelerated frame discussed in the previous section. The coordinate velocity of a light signal in any direction is completely determined by the chosen coordinate system, or equivalently, by the spacetime metric. Hence the value of $\epsilon$ is fully specified once the metric is given.

In the present case spacetime is described by the Langevin metric, which governs electrodynamics for an accelerated light source as described in the accelerated frame. Within this metric, light propagates with coordinate velocities $c_1 = c-v$, $ c_2 = c+v $.

Substituting these expressions into the definition of $\epsilon$ yields

\[
\epsilon_{AB}=\frac{1}{2}\!\left(1+\frac{v}{c}\right), 
\qquad
\epsilon_{BA}=\frac{1}{2}\!\left(1-\frac{v}{c}\right).
\]

We now consider the round-trip velocity of light. Special relativity states that the two-way speed of light is invariant for inertial observers. Let us examine the corresponding quantity in the accelerated coordinate system. The total coordinate time between emission and reception of a light signal is
\[
\Delta t=\frac{x_{AB}}{c-v}+\frac{x_{AB}}{c+v}
      =\frac{2x_{AB}}{c\left(1-\frac{v^2}{c^2}\right)} .
\]
If one naively defines velocity using coordinate distance and coordinate time, one obtains a two-way light speed equal to $c(1-v^2/c^2)$, which appears to contradict the fundamental constant $c$. 

The apparent paradox arises because coordinate quantities do not necessarily correspond to physical measurements. A physical velocity can be defined only after specifying physical (proper) standards of length and time. In an accelerated system the coordinate velocity of light is entirely determined by the spacetime metric and therefore depends on the chosen synchronization convention.

It is important to emphasize that the Langevin metric describes measurements performed by an accelerated observer as interpreted by an inertial observer who retains the original set of synchronized clocks used when the reference system $S_n$ was at rest. Within this framework of absolute simultaneity, the inertial observer assigns to the accelerated observer one-way light speeds $c-v$ in the positive $x$-direction and $c+v$ in the negative direction.

The measured length of a moving object therefore depends not only on spacetime structure but also on the adopted synchronization convention. Analysis based on the Langevin metric shows that a coordinate distance $x_{AB}$ corresponds to the physical length
\[
x_{AB}/\sqrt{1 - v^2/c^2}.
\]
Since clocks remain at rest in the accelerated frame, their proper time evolves according to
\[
d\tau_n = dt_n\sqrt{1 - v^2/c^2}.
\]
When physical (proper) length and proper time are used, the round-trip velocity of light becomes $c$ as required. A more detailed discussion is given in Chapter~13.

\subsection{Electrodynamic Explanation of Light Aberration}

It is well known that electrodynamics fully complies with the principles of relativity.
For instance, we can derive the velocity of light in an accelerated frame without relying on the metric equation (\ref{GGG3}).
In an inertial frame, electromagnetic fields are expressed as functions of the independent variables $x, y, z$, and $t$. According to special relativity, Maxwell's equations remain valid in any Lorentz reference frame, and the electric field  $\vec{E}$ of an electromagnetic wave satisfies the wave equation:
$\Box^2\vec{E} =  \nabla^2\vec{E} - \partial^2\vec{E}/\partial(ct)^2  = 0$.
By applying a Galilean transformation, the variables $x,y,z,t$ can be expressed in terms of the coordinates $x_n, y_n, z_n, t_n$ in an accelerated frame. Consequently, the electromagnetic field of a wave emitted by a source at rest in the accelerated frame can be rewritten in terms of these transformed variables. As a result, the wave equation transforms into

\begin{eqnarray}
&& \Box^2\vec{E}_n = \left(1-\frac{v^2}{c^2}\right)\frac{\partial^2\vec{E}_n}{\partial x_n^2}  + 2\left(\frac{v}{c}\right)\frac{\partial^2\vec{E}_n}{\partial t_n\partial x_n}
+ \frac{\partial^2\vec{E}_n}{\partial y_n^2} + \frac{\partial^2\vec{E}_n}{\partial z_n^2}
- \frac{1}{c^2}\frac{\partial^2\vec{E}_n}{\partial t_n^2} = 0 ~ , \label{GGT4}
\end{eqnarray}

where coordinates and time are transformed according to a Galilean transformation. 
Acceleration affects the propagation of light. The coordinate velocity of light parallel to the $x_n$-axis is given by $dx_n/dt_n = c - v$ in the positive direction, and $dx_n/dt_n = -c - v$  in the negative direction.

We now consider the transmission through the hole in the opaque screen in the accelerated frame $S_n$,  Fig. \ref{B999}.   
We describe the aberration of light based on the observations made by an observer in the same accelerated frame as the screen. 
We have already discussed the explanation of the aberration based on electrodynamics.
Our earlier discussion is really about as far as anyone would normally need to go with the subject, but we are going to do it again. One reason is that one should know how to deal with what happens to waves in the accelerated frame from the viewpoint of electrodynamics equations. 
The screen containing a hole is a kind of diffraction grating that breaks up the radiated beam into a number of diffracted beams of plane waves. Each of these beams corresponds to one of the Fourier components into which a transmittance can be resolved.  Let us assume that the transmittance of the grating varies according to the law  $T = g(K_{\perp})\sin{({K}_{\perp}x)}$. 

Acceleration influences the field equations, transforming the wave equation into Eq. (\ref{GGT4}).
Consider a transmitted plane wave of the form $\exp(i\vec{k}_n\cdot\vec{r}_n - i\omega_n t_n)$. 
With a plane wave $\exp(i\vec{k}_n\cdot\vec{r}_n - i\omega_n t_n)$ with the  wavenumber vector $\vec{k}_n$ and the frequency $\omega_n$  equation Eq.(\ref{GGT4}) becomes: 

\begin{eqnarray}
(1-v^2/c^2)(k_n)_x^2 - 2v(k_n)_x \omega_n/c^2 + (k_n)_z^2 -\omega_n^2/c^2 = 0     ~ . \label{GGT45} 
\end{eqnarray}

The wavevector $\vec{k}_n$ is determined by the initial conditions before the acceleration. Transforming to the accelerated frame using Galilean transformation $x = x_n + vt$, $ t = t_n$ we find that the frequency shifts according to: $\omega_n = \omega_i - K_{\perp}v$, while the components of wavevector remain unchanged. Substituting these into dispersion relation confirms that it remains satisfied, as expected.

In our example, plane waves with different wavenumber vectors propagate away from the screen at different frequencies. Each transmitted wave satisfies the equation $\Delta\omega/\Delta k_x = - v$, regardless of the sign or magnitude of the transmission angle. This implies that a transmitted light beam with a finite transverse size moves along the $x_n$-direction with a group velocity of $d\omega/dk_x = - v$  (see Fig. \ref{B999}).

\subsection{A Simple Explanation of Light Aberration in Accelerated Frames}

Previously, we analyzed the aberration of light in an accelerated reference frame. Calculating this effect in a non-inertial frame is inherently complex. To tackle this challenge, we employed a metric tensor to derive the electrodynamics equations within the accelerated frame, using the Fourier transform method. A key consequence of the cross term in the wave equation is the emergence of a group velocity, which explains the aberration.

We now present a simpler and more intuitive explanation based on
electrodynamics. The aberration increment will be obtained by applying
Galilean transformation laws to the electromagnetic fields.

The flow of electromagnetic energy is described by the Poynting vector
\[
\vec{S}=\frac{c}{4\pi}\,\vec{E}\times\vec{B},
\]
which specifies both the direction and the rate of energy transport.

To analyze light aberration in an accelerated frame, we introduce a coordinate
system with a clearly defined reference direction. The reference directions are
chosen perpendicular to the direction of motion in both the inertial and the
accelerated frames. For simplicity, we assume that the aberrated light beam
remains confined to a single plane, so that its direction can be characterized
by one angle.

The discussion is simplified if we consider separately two polarization cases:
a beam whose electric field lies in the plane of incidence (the $xz$–plane) and
a beam whose electric field is perpendicular to this plane. We treat explicitly
the second case; the reasoning for the first is completely analogous.

We therefore assume that the electric field possesses only a $y$–component,
\[
\vec{E}=\vec{e}_y E .
\]
The magnetic field of the incoming plane wave is then
\[
\vec{B}=\vec{e}_z\times\vec{E},
\]
where $\vec{e}_z$ denotes the direction of the Poynting vector in the inertial
frame.

As shown earlier, consistency between the inertial and accelerated segments of
the motion requires matching the metric tensors in both descriptions. For our
present purpose it is sufficient to relate the coordinates of the accelerated
observer to those of the inertial observer by means of a Galilean boost.

The electrodynamics equations in the accelerated frame may therefore be obtained
by applying a Galilean transformation with velocity $-v$ to Maxwell’s equations.
The electromagnetic fields $\vec{E}_n$ and $\vec{B}_n$ measured in the
accelerated frame differ from the inertial fields $\vec{E}$ and $\vec{B}$.
To first order in $v/c$, the transformation laws read
\[
\vec{E}_n=\vec{E} +\frac{\vec{v}\times\vec{B}}{c},
\qquad
\vec{B}_n=\vec{B}-\frac{\vec{v}\times\vec{E}}{c}.
\]

Let the accelerated frame move with velocity
\[
\vec{v}=v\vec{e}_x .
\]
For polarization perpendicular to the $xz$–plane, the magnetic field in the
inertial frame has only an $x$–component. One immediately finds
\[
\vec{E}_n=\vec{E}=\vec{e}_y E ,
\]
and
\[
\vec{B}_n
=(\vec{e}_z\times\vec{e}_y)E
-(\vec{e}_x\times\vec{e}_y)\frac{vE}{c}
=-\vec{e}_x E-\vec{e}_z\frac{vE}{c}.
\]

The corresponding Poynting vector becomes
\[
\frac{c}{4\pi}\vec{E}_n\times\vec{B}_n
=
\left[
-(\vec{e}_y\times\vec{e}_x)
-(\vec{e}_y\times\vec{e}_z)\frac{v}{c}
\right]
\frac{cE^2}{4\pi}
=
\left[
\vec{e}_z-\vec{e}_x\frac{v}{c}
\right]
\frac{cE^2}{4\pi}.
\]

Thus the radiation propagates at an angle $-v/c$ with respect to the $z_n$–axis.
This angular deviation represents the aberration of light in the accelerated
frame.

We therefore arrive at an important conclusion: the direction of energy transport
is identical whether it is derived from the electrodynamics equations in the
accelerated frame or obtained from the Galilean transformation law for the
electromagnetic fields.  

In the first approach, the wave equation predicts that a finite transverse beam
acquires a group velocity
\[
\frac{d\omega}{dk_x}=-v .
\]
In the second approach, aberration appears as a change in the direction of the
Poynting vector. Both viewpoints describe the same physical effect. The group
velocity direction and the Poynting vector direction are therefore two
equivalent descriptions of one and the same phenomenon.

We now present an additional derivation of the Poynting vector direction in the
accelerated frame.

Acceleration modifies the field equations through a change in the time
derivative measured by an observer at rest in the accelerated frame. Consider
the coordinates $(t,x,y,z)$ of an inertial observer moving with velocity $-v$
relative to the accelerated observer. The Galilean transformation reads
\[
x_n=x-vt,\qquad
y_n=y,\qquad
z_n=z,\qquad
t_n=t .
\]

Taking partial derivatives gives
\[
\frac{\partial}{\partial t_n}
=\frac{\partial}{\partial t}
+v\frac{\partial}{\partial x},
\qquad
\frac{\partial}{\partial x_n}
=\frac{\partial}{\partial x}.
\]

The magnetic field in the accelerated frame satisfies the modified
electrodynamics equation
\[
\vec{\nabla}\times\vec{B}_n
=
\frac{1}{c}
\left(
\frac{\partial}{\partial t}
+v\frac{\partial}{\partial x}
\right)\vec{E}_n .
\]

Writing explicitly the relevant component yields
\[
-\frac{\partial B_z}{\partial x}
=
\frac{v}{c}\frac{\partial E_y}{\partial x}.
\]

Any experimental determination of aberration necessarily involves a light beam
of finite transverse size. Integrating this equation across the beam profile in
the $x$–direction gives
\[
\vec{B}_n=-\vec{e}_x E-\vec{e}_z\frac{vE}{c},
\]
which coincides exactly with the result obtained from the field transformation
argument.

This example illustrates an important physical point concerning the reality of
energy transport described by the Poynting vector. In any practical aberration
measurement the radiation beam has finite spatial extent, and the direction of
energy flow is uniquely defined and independent of synchronization conventions
or coordinate choices.

A frequent misconception arises when discussing ideal plane waves. For an
infinitely extended plane wave, the localization of energy transport cannot be
defined operationally, and therefore the precise direction of energy flow cannot
be measured experimentally. Only finite beams possess a physically meaningful
Poynting vector direction.

\subsection{Clock Resynchronization in Accelerated Systems}

It should be emphasized that there exists an alternative and equally
satisfactory explanation of light aberration in the accelerated reference
frame $S_n$. This interpretation is based on a re-synchronization of clocks
within the accelerated system.

When the system $S_n$ reaches a stage of motion with constant velocity, the
standard Einstein synchronization procedure can be applied. This method relies
on the assumption that light signals emitted by a source at rest propagate with
the same speed $c$ in all directions. Employing this synchronization scheme
allows us to introduce Lorentz coordinates associated with the screen.
Within this coordinate system, the transmission of radiation through the
aperture can be described directly by Maxwell’s equations. Consequently,
the spacetime interval in the accelerated reference frame $S_n$ acquires the
diagonal form given previously in Eq.~(\ref{MM1}) for the transmitted light
beam.

\begin{figure}
	\centering
	\includegraphics[width=0.8\textwidth]{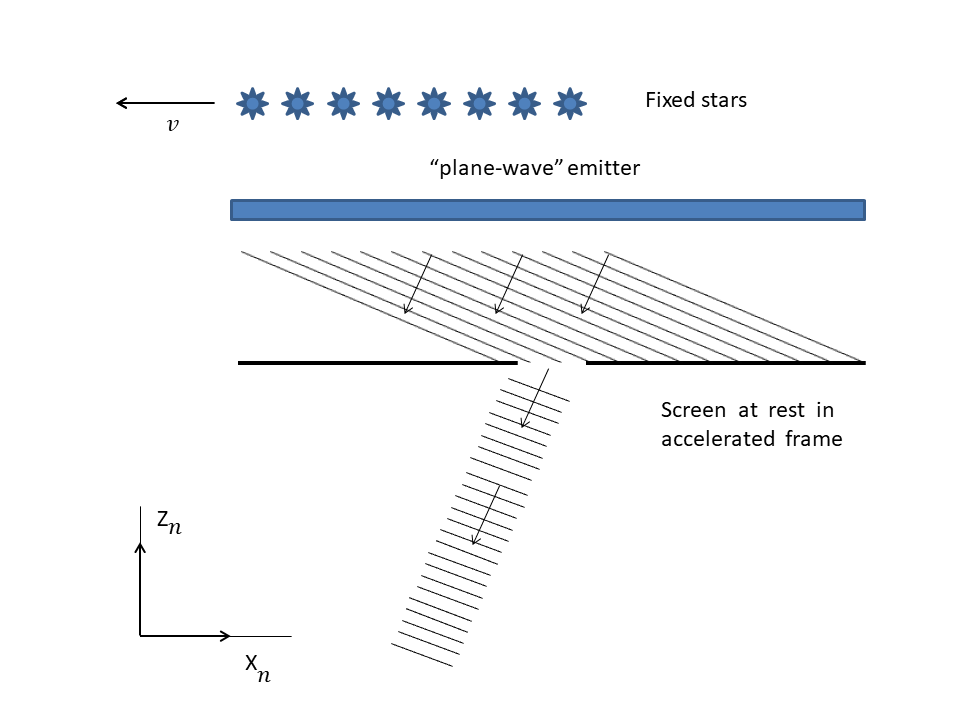}
	\caption{Aberration of light in the accelerated frame of reference $S_n$.
	In this frame the plane wavefront rotates after re-synchronization of the
	accelerated clocks. According to Maxwell’s electrodynamics, the transmitted
	light beam propagates at an angle $-v/c$, producing the aberration of light.}
	\label{B102}
\end{figure}

The Einstein-synchronized time coordinate $t'_n$ in the frame $S_n$ is obtained
by introducing a position-dependent time offset of the type discussed in
Eq.~(\ref{GGT3}), with the velocity $v$ replaced by $-v$. The synchronization
shift therefore reads
\[
t'_n = t_n - \frac{x_n v}{c^2},
\]
which represents the first-order approximation in $v/c$.

This time redefinition changes the hypersurfaces of simultaneity.
Geometrically, it corresponds to a rotation of the plane of simultaneity by an
angle $ - v/c$.

As a consequence, after clock re-synchronization the radiation wavefront
appears rotated in the accelerated frame, as illustrated in
Fig.~\ref{B102}.

The newly defined time coordinate restores the standard Maxwell form of the
electrodynamics equations, ensuring their direct applicability to light
propagation. The transmitted beam passing through the aperture therefore
propagates at an angle $\theta_a = - v/c$
which reproduces the aberration of light obtained previously by dynamical
methods.

Both approaches — the analysis based on field transformations and the one
based on clock re-synchronization — lead to identical physical predictions.
The difference between them lies only in the chosen synchronization convention.
A change of the four-dimensional coordinate system does not introduce new
physical effects; rather, it provides an alternative and often more transparent
description of the same physical phenomenon.

\subsection{Composition of Motions in Non-Inertial Frames}

Up to this point, our discussion has been restricted to the case of
nonrelativistic motion. We now extend the analysis to arbitrary velocities.

As emphasized earlier, the metric tensor must remain a continuous quantity
along the observer’s trajectory. The Langevin metric arises from matching the
metric tensors describing the inertial and accelerated segments of motion.
A smooth transition between these descriptions can be achieved by relating
the coordinates and time of the accelerated observer to those of the inertial
observer through the inverse Galilean boost. Under this transformation, the
Minkowski metric of the inertial frame transforms into the Langevin metric
appropriate for the accelerated frame.

Next, we consider the transformation of the direction of light propagation. 
The group velocity of light transforms similarly to that of a particle, following the Galileo velocity addition theorem.

Suppose that the accelerated frame moves with velocity $v$ along the $x$–axis.
According to the Galilean velocities addition, the aberration angle $\theta_a$ then satisfies
\[
\tan\theta_a=- v/c.
\]
For $v/c\ll1$, the aberration angle is small and may be approximated by
\[
\theta_a\approx - v/c.
\]

It is always possible to introduce coordinates in which the metric of the
accelerated system becomes diagonal. In such a Lorentz coordinatization,
light propagates with the invariant speed $c$, independent of the direction
of propagation and independent of the velocity of the emitting source.

Consider the transformation to Lorentz coordinates,
\[
t'_n
=
t_n\sqrt{1- v^2/c^2}
-
(v x_n/c^2)/\sqrt{1- v^2/c^2},
\]
\[
x'_n = x_n/\sqrt{1- v^2/c^2}.
\]

The velocity measured along the $x_n$–axis transforms according to
\[
dx'_n/dt'_n = (dx_n/dt_n)/[1- v^2/c^2  -  (v/c^2)dx_n/dt_n].
\]

Substituting the previously obtained group velocity
\[
dx_n/dt_n = - v,
\]
one finds
\[
dx'_n/dt'_n = - v.
\]
Thus, the group velocity of the light beam remains equal to $-v$
even in the Lorentz coordinatization.
However, when the propagation direction is expressed using the 
light velocity $c$, the aberration law takes the form
\[
\sin\theta_a= - v/c.
\]
This relation represents the relativistic correction to the classical
aberration formula when described in Lorentz coordinates.

Suppose now that an accelerated observer measures the aberration of light.
In describing physical phenomena in an accelerated frame, one must distinguish between coordinate quantities and physical quantities.
Because time intervals and spatial distances in Lorentz coordinates possess
direct operational meaning, the observer determines that the aberration angle
satisfies precisely the relation $\sin\theta_a= - v/c$.

\subsection{Discussion}

Let us now return to the topic of metric diagonalization and deepen our understanding of the relationship between the Lorentz coordinate systems $(x,y,z,t)$ and $(x'_n,y'_n,z'_n,t'_n)$.
We have already noted that the old coordinates  $(x_n,y_n,z_n,t_n)$ are found by matching the accelerated frame and inertial frame metrics; what does mismatching of the coordinates $(x,y,z,t)$ and $(x'_n,y'_n,z'_n,t'_n)$ mean, in terms of measurements made by the accelerated observer?
In both reference frames, the metric remains diagonal, and, according to standard textbooks, the coordinates should be related by the Lorentz transformation. At first glance, the problem appears entirely symmetrical, suggesting that both frames should be equivalent. However, as we demonstrated earlier, there is a fundamental distinction between an accelerated inertial frame and an inertial frame without a history of acceleration.

Where does the asymmetry originate?
The electrodynamics equation must be integrated with initial conditions. According to Maxwell's electrodynamics, coherent radiation is always emitted in the direction normal to the radiation wavefront. In the inertial frame, the 
wavefront of the emitted light beam is perpendicular to the vertical $z$-axis. However, when considering Lorentz 
coordinatization, the wavefront orientation is affected in the accelerated frame $S_n$.
According to relativistic kinematics, the extra phase chirp $d\phi/dx'_n =  k_x =  - v\omega_i/c^2$ is introduced as a consequence of this, the plane wavefront rotates after the acceleration.  
Then, the radiated light beam is propagated at the angle $-v/c$ with respect to the $z_n$-axis, yielding the phenomenon of the aberration of light in the accelerated frame $S_n$.

This phenomenon is purely kinematic and involves no forces. Many may find this counterintuitive.
We explore the resolution of this paradox in detail in Chapter 13, using a dynamical approach to explain the asymmetry 
between inertial and accelerated reference frames. Without proof, we state the key result: the apparent asymmetry arises from the presence of inertial (pseudo-gravitational) forces in the accelerated frame $S_n$. The principle of
equivalence allows non-inertial kinematic problems to be addressed using dynamical methods.

Wavefront rotation, which occurs when transitioning from an inertial to an accelerated frame (relative to the fixed stars), can be attributed to the pseudo-gravitational potential gradient during acceleration. A clock's rate depends on the local pseudo-gravitational potential, leading to an accumulated time difference between spatially separated clocks.

Suppose (for simplicity) that there is a constant acceleration $g$ along the $x_n$-axis during the time $T = v/g$.  
The pseudo-gravitational acceleration is simply equal to the gradient of scalar potential: $g = - \partial \phi/\partial x_n$.    
A clock at a higher gravitational potential (aligned with the acceleration direction) runs faster.  
The pseudo-gravitational potential difference between two points along the $x_n$-axis is given by

\[
\Delta \phi = \phi_1 - \phi_2 = (\partial \phi/\partial x_n) [x_n(1) - x_n(2)]   .  
\]

Once the fixed stars move with constant velocity, the potential gradient in $S_n$ becomes zero.
The accumulated time difference between two spatially separated clocks is

\[
t_n(1) - t_n(2) = - g[x_n(1) - x_n(2)] v/(c^2|g|) = v[x_n(1) - x_n(2)]/c^2    . 
\]

This time shift effectively rotates the plane of simultaneity by an angle $-v/c$.

\subsection{A Single Moving Emitter in an Accelerated Frame}

In Chapter 4, we examined the single-emitter problem within the initial inertial frame of reference, $S$. Our analysis revealed a deviation in the energy transport of light emitted by an accelerated source, which corresponds to a well-known phenomenon: the aberration of light from a tangentially moving emitter in an inertial frame of reference (see Fig. \ref{B103}). We demonstrated that this aberration can be analyzed using two distinct approaches: the covariant and the non-covariant formulations.

We now extend our analysis to the aberration of light emitted by a "plane wave" source moving within an accelerated reference frame. In this case, two different methods of coordinatization are possible:

(a) Consider a system $S_n$, initially at rest in the inertial frame $S$, which undergoes acceleration along the
$x$-axis until it reaches a velocity $v$. The simplest synchronization method, based on absolute time, involves 
maintaining the same set of uniformly synchronized clocks as when $S_n$ was at rest. However, acceleration relative to 
the fixed stars influences the propagation of light. As a result, by the time $S_n$ attains a constant velocity, the
velocity of light emitted from a source at rest in the accelerated frame $S_n$ will exhibit anisotropy along the
$x_n$-axis (see Eq. (\ref{GGG3})).

\begin{figure}
	\centering
	\includegraphics[width=0.8\textwidth]{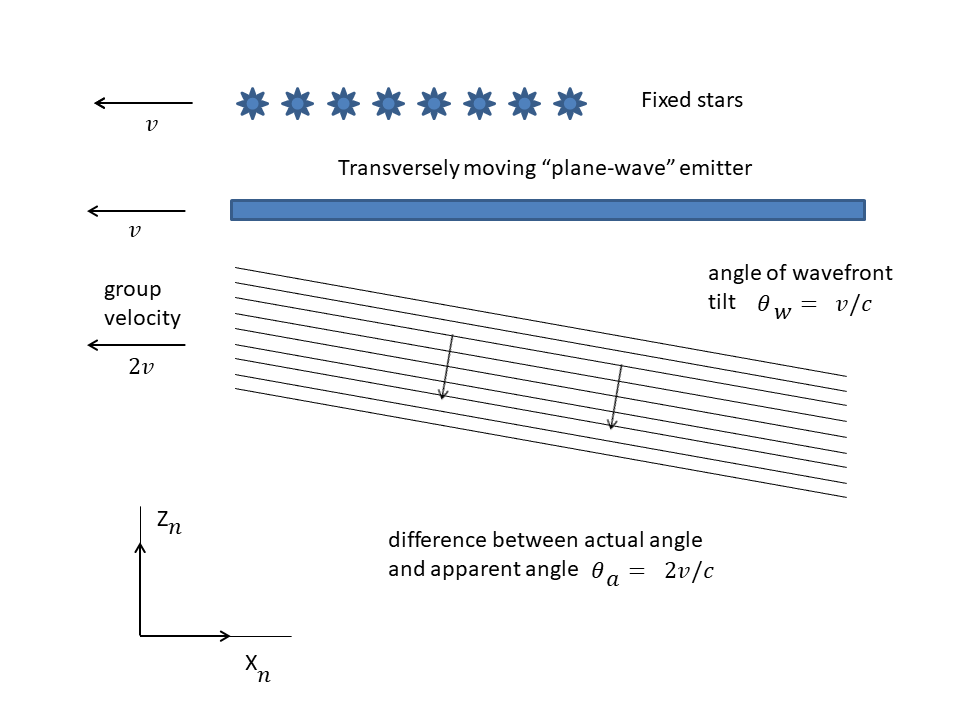}
	\caption{Aberration of light in an accelerated frame of reference $S_n$. The emitter remains at rest relative to the fixed stars. The wavefront orientation and group velocity are shown using Einstein synchronization.}
	\label{B105}
\end{figure}

(b) Once the system $S_n$ begins moving with a constant velocity,
Einstein's standard clock synchronization procedure 
can be applied. This synchronization method is defined using light signals emitted by a source at rest in the accelerated frame, under the assumption that light propagates isotropically with velocity $c$. By employing this Einstein synchronization procedure, we describe the light emitted from the source at rest (in the accelerated frame) using the standard Maxwell equations.

We now consider the case where the emitter is at rest in the initial inertial frame $S$ (i.e., at rest relative to the fixed stars), while the observer, who is at rest in the accelerated frame $S_n$, performs the measurement of energy transport direction.
It is important to emphasize that the aberration of light emitted by a single source also depends on the motion of the emitter within the accelerated frame $S_n$. Due to the fundamental asymmetry between inertial and accelerated frames, the theory of light aberration predicts a remarkable effect.
Specifically, if the emitter remains at rest relative to the fixed stars while the observer transitions from rest to 
motion with respect to the fixed stars, the apparent angular position of the "plane-wave" emitter, as seen in the 
accelerated frame, would jump by an angle of $ - 2v/c$. 
This aberration effect in the frame $S_n$ is illustrated in Fig. \ref{B105}.

Let us describe the radiation emitted by a source moving with velocity $-v$ in an accelerated frame, as observed by an 
accelerated observer. The acceleration relative to the accelerated frame influences the propagation of light. Assuming 
an absolute time coordinatization, we apply a Galilean boost by substituting $x_n \to x_n + vt_n$ while keeping the 
time coordinate unchanged. Applying this transformation to the Langevin metric yields

\begin{eqnarray}
&& ds^2 = c^2(1- 4v^2/c^2)dt_n^2 - 4vdx_ndt_n - dx_n^2 - dy_n^2 -dz_n^2 ~ .\label{GGG12}
\end{eqnarray}

This metric characterizes the electrodynamics of a stationary light source in an inertial frame from the perspective of measurements made by an accelerated observer.
From the formula, it is evident that the measurement of the electromagnetic field configuration in the moving 
system—expressed in terms of the coordinate $x_n$ and time $t_n$
of a measuring device at rest in the accelerated 
frame—yields the same result as that of the configuration at the point $x_n +vt_n$ and time $t_n = 0$.
A similar situation arises in the inertial frame. 

Let us now examine the fundamental asymmetry between inertial and accelerated frames in absolute time coordinatization. An inertial observer describes the electrodynamics of a stationary light source using the Minkowski metric. By substituting  $x = x_n + vt$ and $t_n  =  t$  into the Minkowski metric, we obtain the Langevin metric (Eq. \ref{GGG3}) in the comoving frame.

A natural question arises: what is the metric of the inertial observer from the perspective of the accelerated observer? To address this, we transform the coordinates $(t_n,x_n,y_n,z_n)$, which correspond to the accelerated observer $S_n$, using the inverse Galilean transformation. Specifically, we substitute $x_n =x-vt$ while keeping time unchanged ( $t_n = t$) into the Langevin metric (Eq. \ref{GGG3}). This transformation yields the Minkowski metric,
$ds^2 = c^2 dt^2 - d x^2 - dy^2 - dz^2$ 
which correctly describes the electrodynamics of a stationary light source in the inertial frame. Importantly, this transformation highlights how an inertial observer perceives the measurements of an accelerated observer. In particular, it illustrates the time dilation effect experienced by a physical clock at rest in the accelerated frame. Crucially, the slowdown of the accelerated clock is independent of the reference frame in which this effect is observed (see Chapter 13 for further details).

Another approach to solving the problem involves the Langevin metric diagonalization procedure. In this framework, the new coordinate system in the accelerated frame
$(x'_n,y'_n,z'_n,t'_n)$  is  interpreted such that an observer perceives a diagonal metric:

\[
ds^2 = c^2 (dt'_n)^2 - (d x_n')^2 - (dy_n')^2 - (dz_n')^2
\]

for an emitter at rest in the accelerated frame, and

\[
ds^2 = c^2(1- v^2/c^2)(dt'_n)^2 - 2vdx_n'dt_n' - (dx'_n)^2 - (dy'_n)^2 - (dz'_n)^2
\]

for a moving emitter. This reveals a symmetry (or reciprocity) in the metrics—and consequently in the electrodynamic equations—between the accelerated and inertial frames. This symmetry arises naturally from the pseudo-Euclidean geometry of space-time. The only difference between this metric and Eq. (\ref{GGG11}) is the sign of the velocity.
After diagonalization, it becomes evident that any apparent asymmetry stems from the initial conditions. Notably, the plane radiation wavefront of a moving emitter (regardless of its velocity) undergoes a rotation by an angle of $-v/c$ in the accelerated frame as a result of the metric diagonalization. This implies that the radiated light beam has a group velocity along $x$-direction given by $d\omega/dk_x = -2v$.

To predict the outcome of aberration measurements, one can alternatively analyze the metric of the moving emitter directly within the absolute-time coordinate system, given by Eq. (\ref{GGG12}). According to electrodynamics, the group velocity $-2v$ emerges as a direct consequence of the cross term $-4vdxdt$.

Now, let us consider the most general case in which 
the system $S_n$,
initially at rest in the inertial frame $S$, is 
accelerated to velocity $v$ along the $x$-axis.  
Simultaneously, an emitter in the accelerated frame $S_n$ is 
accelerated from rest to velocity $u$ along the $x_n$-axis. 
Suppose an observer in the accelerated frame $S_n$ performs
an aberration measurement.
To describe the radiation emitted by a source moving with velocity $u$ in the accelerated frame, we must account for the electrodynamics in terms of the accelerated observer. This can be done by employing the metric

\[
ds^2 = c^2[1- (v - u)^2/c^2]dt_n^2 - 2(v - u)dx_ndt_n - dx_n^2 - dy_n^2 - dz_n^2     . 
\]

A closer examination of the physics reveals that the aberration increment is related to the system parameters by the expression $\theta_a = - v/c + u/c$.  
Specifically, assume that light source moves with velocity $u = v$ relative to accelerated frame.   In this setup, diagonal metric applies and that the light beam undergoes no aberration.

How can this be? If the accelerated and inertial observers were to share a common three-dimensional space, presented above results for aberration increment would lead to a paradox — an apparent inconsistency in the theory. However, in Minkowski spacetime, no such paradox arises.

First, we should emphasize that we discuss  metric which describes the electrodynamics of the  light source with viewpoint of the accelerated observer measurements.
In special relativity simultaneity is relative. Moving reference systems therefore possess different simultaneity hypersurfaces.
Because of the relativity of simultaneity, inertial and accelerated observers do not share the same three-dimensional space in Minkowski spacetime.  
We will explore this topic further in Section 5.14.

\begin{figure}
	\centering
	\includegraphics[width=0.8\textwidth]{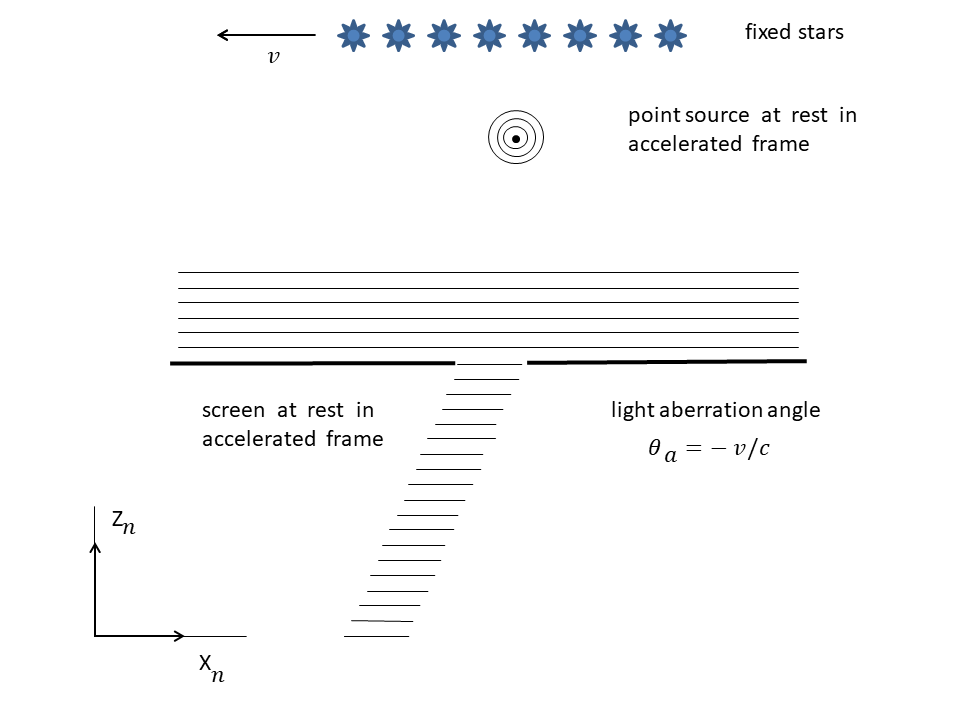}
	\caption{Aberration of light in an accelerated frame of reference. The point source is at rest within the accelerated frame, while the wavefront orientation is shown in absolute time coordinatization for the screen.}
	\label{B181}
\end{figure}

\begin{figure}
	\centering
	\includegraphics[width=0.8\textwidth]{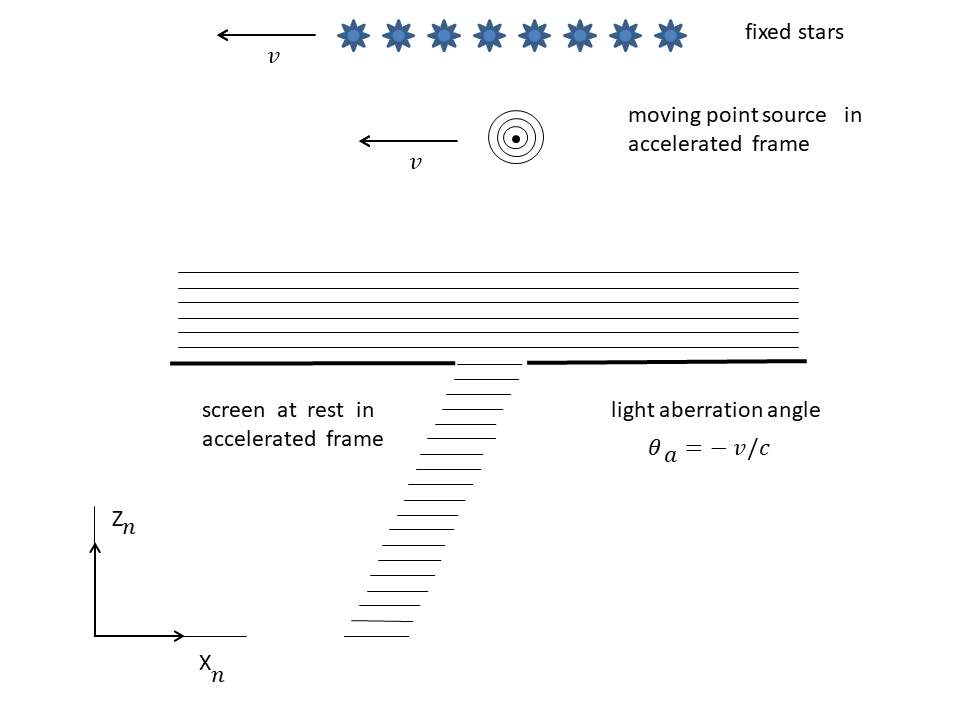}
	\caption{Aberration of light in an accelerated frame of reference. The point source remains at rest relative to the fixed stars.}
	\label{B181A}
\end{figure}

\subsection{A Point Source in an Accelerated Frame}

Now, let us return to observations of an accelerated observer.
A key characteristic of point-like sources is that radiation emitted at a given instant forms a spherical wave around the source, and any measuring instrument inevitably influences the observed radiation (see Fig. \ref{B181} - Fig. \ref{B181A}). The aberration of a point source is independent of the source's speed and is attributed solely to the observer’s motion relative to the fixed stars.
The cross term in metric Eq. (\ref{GGG3}) introduces anisotropy in the accelerated frame, altering the direction of transmitted radiation. As a result, a point source exhibits the same aberration pattern as fixed stars, meaning its apparent position shifts by an angular displacement identical to that of the fixed stars.

In conventional aberration theory, an intriguing puzzle arises concerning stellar aberration. Some binary star systems undergo velocity changes on timescales ranging from days to years. At certain times, the components of such systems can have significantly different velocities relative to Earth. However, it is well established that both components always exhibit the same aberration angle.
Rotating binary systems follow the same aberration pattern as all fixed stars, appearing to shift annually by the same universal aberration angle. This observation strongly supports our theoretical prediction for the aberration of light from a point source in a rotating frame. In the next chapter, we explore this experimental test in greater detail.

\subsection{Aberration of a Tilted Incoming  Plane Wave}

We now return to observations of an inertial observer.
A point source located in the far zone effectively generates a plane wave in front of the aperture.
Above, we considered only the case where the point source is positioned on the optical axis of the aperture (which is parallel to the 
$z$-axis), producing a normally incident plane wave.
The fundamental electrodynamic equations must hold for all electromagnetic waves.
In other words, the dispersion equation in the accelerated frame should be consistently applied to both the incoming and transmitted waves.
In Maxwell's electrodynamics, the dispersion relation simplifies to $k_x^2 + k_y^2 + k_z^2 - \omega^2 = 0$.
A notable feature of this case is that, even after a Galilean transformation along the $x$-axis, the dispersion relation in the accelerated frame retains the same diagonal form, $k_z^2 - \omega_i^2 = 0$, for a normally incident wave.

Now, let us examine the behavior of an incoming wave with a phase gradient along $x$. 
In other words, there is an angle $\alpha$ between the wavevector  $\vec{k}$ and the $z$-axis. 
For small angles, we can approximate $\alpha \simeq k_x/k_z$. Our goal is to express the parameters of the incoming wave in terms of the source offset  $x_o$ relative to the optical axis. The wave tilt angle can be written as $\alpha \simeq x_o/z_s$, where $z_s$ is the distance from the source to the screen. 

Next, we analyze the aberration of light emitted by a stationary point source in an accelerated frame. The wavevector of the incoming plane wave is determined by the initial conditions. Assuming that the emitted wavefront is initially tilted by an angle $\alpha$, we obtain 

\[
k_x =  \alpha k = \alpha\omega_i/c    ,       \quad     k_z =  \sqrt{\omega_i^2/c^2 - k_x^2}    . 
\]

Since acceleration modifies the field equations, we substitute the plane wave solution $\exp(i\vec{k}_n\cdot\vec{r}_n - i\omega_n t_n)$ into Eq. (\ref{GGT4}), yielding dispersion equation Eq. (\ref{GGT45}).  The wavevector $\vec{k}_n$ is determined by the initial conditions before the acceleration.  Using disprsion equation we find that the frequency shift according to: $\omega_n = \omega_i - k_x v$, while the components of wavevector remain unchanged. \footnote{ At first glance, the longitudinal Doppler effect in an accelerated frame reveals a fundamental assymetry between accelerated and inertial observers.  In the experiments, no effect was observed. 
The apparent paradox is resolved by recognizing that frequency measurements in all cases inherently follow the principles of interference.   
Consider a Fabry-Perot interferometer. In this setup, the frequency of light is effectively measured through the length of the standing wave it produces.   
The key point is that all methods used to measure interference - specifically, those involving standing wave  - are effectively a round-trip measurements. 
Interference experiments cannot detect accelerated frame velocity. This is because phase is a four-dimensional invariant - independent of the chosen inertial frame - due to the fundamental geometry of space-time. 
In contrast, the deviation in the direction of energy transport arises from a geometrical effect. In the Chapter 14 we will continue our discussion of relativistic measurements.}

A screen with a hole acts as a diffraction grating, splitting the incident plane wave into multiple diffracted 
components. The wavenumber of the transmitted plane wave along the 
$x_n$-direction is
$(k_n)_x = \alpha k + K_{\perp}$,
where  $K_{\perp}$ is the wavenumber associated with the sinusoidally modulated transmittance.
The dispersion equation then simplifies to 

\[
\omega_i\Delta \omega = - \omega_i K_{\perp}v + c^2 \alpha k K_{\perp}     . 
\]

This result indicates that the transmitted light beam propagates along 
$x_n$ with group velocity
$d\omega/dk_x = - v + c\alpha$.
Thus, the apparent angular position of the point source is related to the problem parameters by $\theta = -v/c + \alpha$ .
\footnote{A similar effect arises if the screen moves with velocity $v$ along the $x$-axis in an inertial frame. In Chapter 4, we considered the special case of a normally incident plane wave, but this approach generalizes to tilted wavefronts as well. The deviation in energy transport for light passing through a hole in the moving screen leads to an aberration increment given by $\theta = v/c + \alpha$.}

\subsection{Non-Reciprocity in the Aberration of Light Theory}

\begin{figure}
	\centering
	\includegraphics[width=0.8\textwidth]{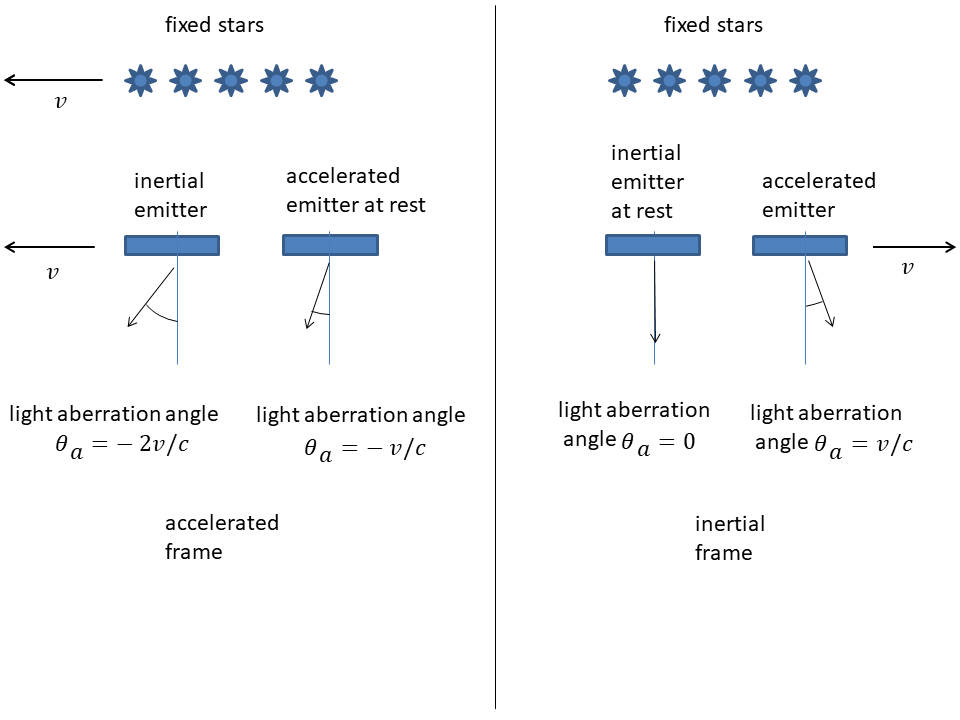}
	\caption{The aberration of light from stationary and moving sources.} 
	\label{B300}
\end{figure}

In this section, we continue our discussion on the aberration of light emitted by a single moving source. Now, let us consider a scenario with two identical emitters.
Suppose the first emitter remains at rest in an inertial frame, while the second emitter is initially at rest but then accelerates to a velocity $v$ along the $x$-axis. An observer, who is at rest in the inertial frame, measures the direction of energy transport.
Now we must be careful about the initial phasing
of these emitters. For example, we consider the case in which initially the
velocity component of the light beam along the x-axis is equal to zero.
The question then arises: How does the light beam from the moving emitter appear?
For the inertial observer, the angular displacement of the emitted radiation is given by $\theta_a = v/c$. This effect, illustrated in Fig. \ref{B300}, is a well-known example of the phenomenon of aberration of light.

\begin{figure}
	\centering
	\includegraphics[width=0.8\textwidth]{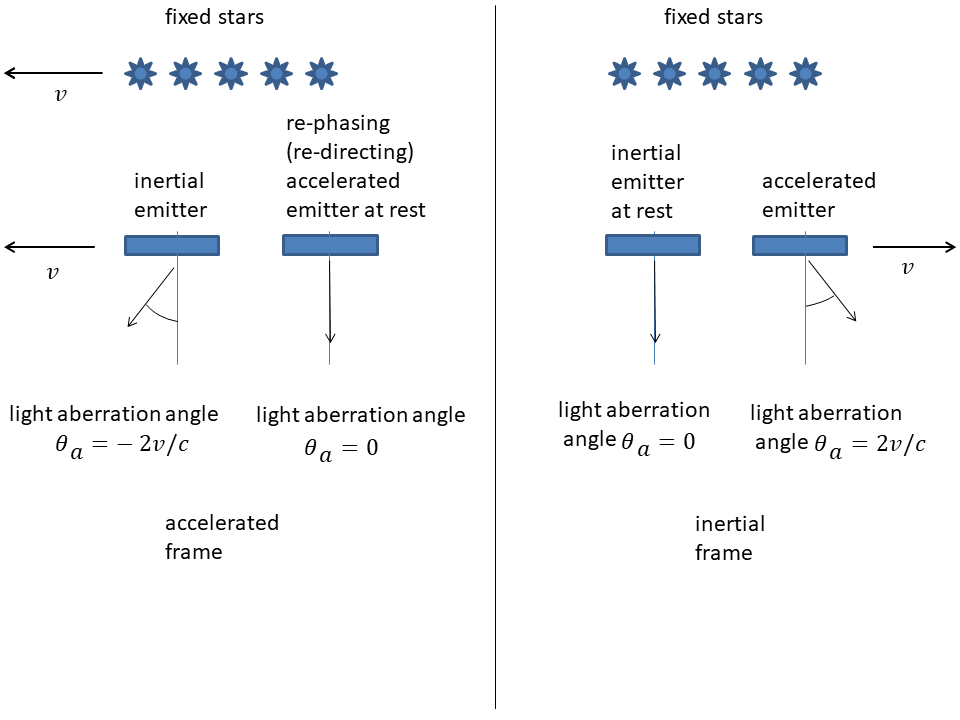}
	\caption{Aberration of light when an accelerated observer redirects an accelerated emitter. After the redirection, an inertial observer measures an angular displacement of $\theta_a = 2v/c$.}
	\label{B301}
\end{figure}

Let us analyze the scenario in which an accelerated observer redirects an accelerated emitter. We consider the specific 
case where, after redirection, the (group) velocity component of the light beam along the 
$x_n$-axis becomes zero.
Following this redirection, an inertial observer would measure an angular displacement of 
$\theta_a = 2v/c$.
Now, suppose the accelerated observer also conducts an aberration measurement. As shown in Fig. \ref{B301}, the aberration increment of the moving emitter is related to the system's parameters by the expression $\theta_a = - 2v/c$. .

\begin{figure}
	\centering
	\includegraphics[width=0.9\textwidth]{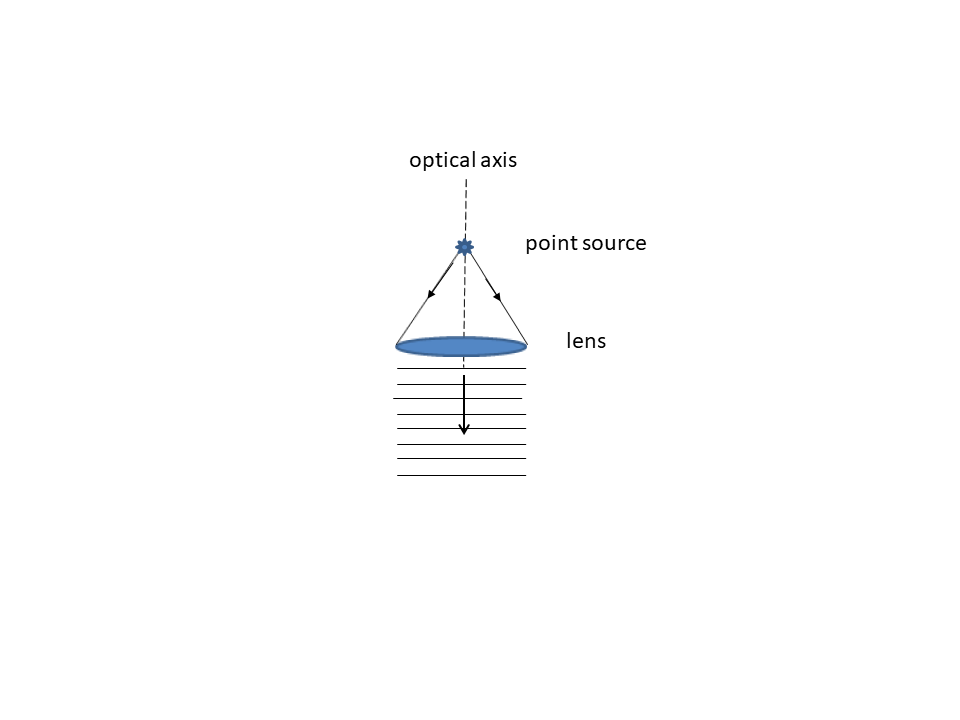}
	\caption{Diagram of a modified source of the "plane wave" emitter type. A point source of light is placed in the front focal plane of a lens.}
	\label{B409}
\end{figure}

\begin{figure}
	\centering
	\includegraphics[width=0.8\textwidth]{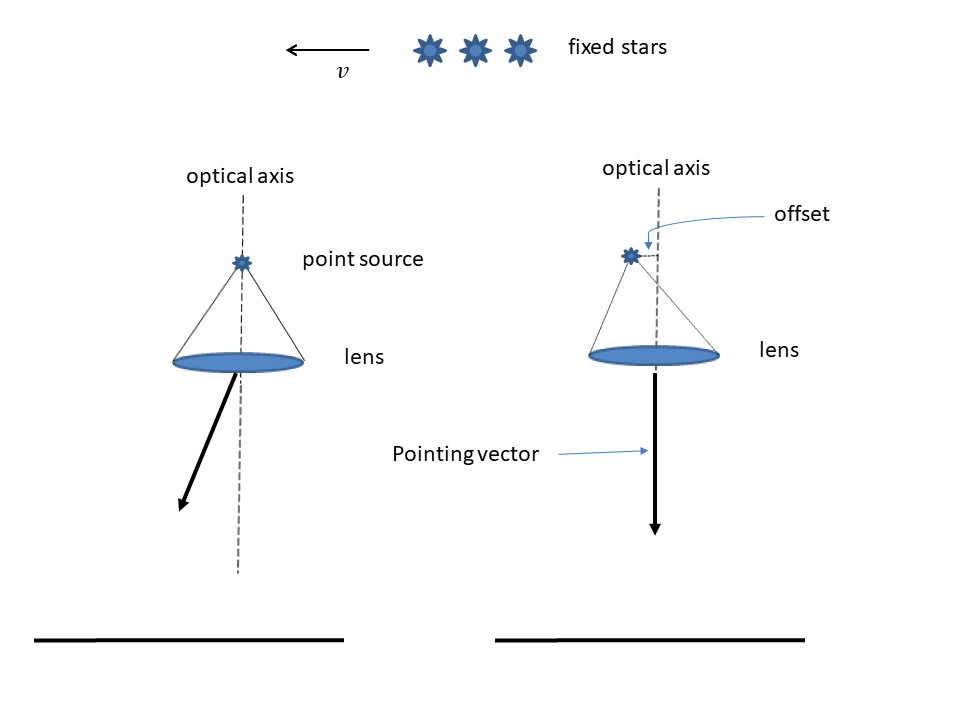}
	\caption{Illustration of the accelerated source redirection process. The source acceleration information is encoded in the point source offset relative to optical axis,          which is aligned with the $z_n$ axis.  }
	\label{BC30}
\end{figure}

\subsection{Redirection Procedure for an Accelerated Light Source}

To further explore the aberration of light in an accelerated frame, it is useful to consider a slightly modified source—a "plane wave" emitter. We adopt the model described in Section 4.2 because it is both relatively simple and sufficiently general to serve as a prototype for understanding aberration phenomena.
While no one has performed all the "thought experiments" exactly as described here, the outcomes can be confidently predicted based on the laws of special relativity, which are themselves grounded in experimental evidence.

Figure \ref{B409} illustrates the modified "plane wave" emitter setup. Though it appears more complex at first glance, it is practically feasible to construct. The setup consists of a point source of light positioned in the front focal plane of a lens. The source’s dimensions are assumed to be on the order of $\lambda$, where $\lambda$ is the wavelength of optical radiation.

Let us revisit the redirection procedure for an accelerated source.
A key question arises: where is the information about the source's acceleration recorded when redirecting light from an accelerated emitter?
To understand the redirection of a light source in an accelerated system, we must examine the source's internal mechanics and observe what occurs during the redirection process. Figure \ref{B409} illustrates a schematic of a "plane wave" emitter. In electrodynamics, redirection can be achieved by introducing an offset between the point source and the optical axis (which is aligned with the $z_n$- axis), as shown in Figure \ref{BC30}. This offset encodes the information about the source's acceleration.

\begin{figure}
	\centering
	\includegraphics[width=0.8\textwidth]{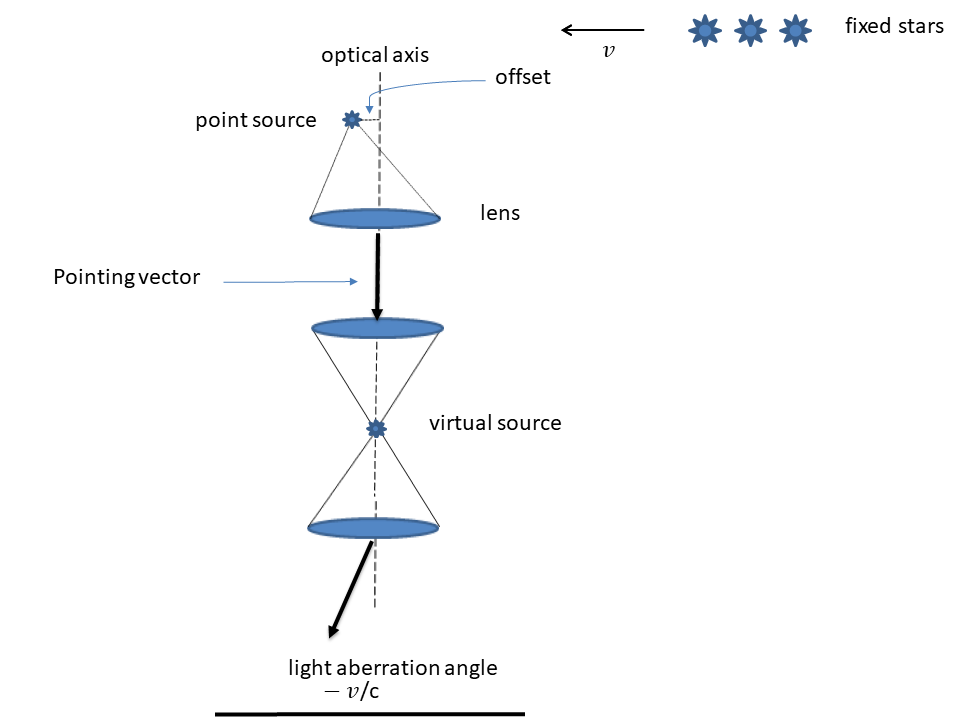}
	\caption{Geometry for propagation of a redirected light beam.  Redirection is achieved by introducing the offset between point souce and the optical axis, which is aligned with the $z_n$ axis.
	The configuration of first and second lenses follows the well-known two-lens imaging scheme.
	Virtual source (i.e image of point source) is positioned in the front focal plane of a third lens.}
	\label{P88}
\end{figure}

We want now to describe  the aberration of light in an accelerated reference frame in various circumstances. We should probably ask first about 
the propagation of a redirected light beam through  the succession of lenses (Fig. \ref{P88}).  Redirected light beam from plane-wave emitter is propagated along the $z_n$ axis and then is focusing by the  lens, making an image of the point source on the $z_n$-axis. Then we treat this image as the virtual point source for the  next (last)  lens. In our case of interest the virtual source is placed a focal distance in front of the  last lens. Due to the inherent asymmetry between inertial and accelerated frames  a transmitted light beam moves along the $x_n$ direction with a group velocity $- v$.

How can we demonstrate that the direction of the energy transport is deviated at the exist of this sequence of lenses? 
Acceleration modifies the field equations through a change in the time
derivative measured by an observer at rest in the accelerated frame. Using a Galilean transformation, the inertial coordinates can be related to the non-inertial ones as follows: $x_n = x - vt, ~ y_n = y, ~ z_n = z, ~ t_n = t $. This allows all fields to be re-expressed in terms of the variables $x_n, y_n, z_n, t_n$. Taking partial derivatives accordingly, we find:
$\partial/{\partial t_n} = \partial/{\partial t} + v\partial/{\partial x}$, $\partial/\partial x_n = \partial/\partial x$.  Thus the magnetic field in the accelerated frame satisfies the modified electrodynamics equation $c\vec{\nabla}\times\vec{B}_n = (\partial/\partial t + v\partial/\partial x)\vec{E}_n$.

A point source located in the front focal plane of the first lens effectively genrates a plane wavefront that illuminates the input of second lens. Our goal is to express the parameters of the second lens input in terms of the source offset $x_0$ relative to the optical axis. The wave tilt angle can be written as $\theta \simeq x_0/f$, where $f$ is the focal length.
Evidently, when the  wavefront is tilted by an angle $\theta = v/c$, we obtain   the usual Maxwell's equation $c\vec{\nabla}\times\vec{B}_n = \partial \vec{E}_n/\partial t $  behind the first lens. In this case, redirected light beam is propagated along the optical axis.

Conider next the last lens output. 
Any experimental determination of aberration necessarily involves a light beam
of finite transverse size. In our case of interest, this is last lens aperture.  
Consider next the case of  a light beam whose electric field is perpendicular to the incident plane.
Integrating  modified electrodynamics equation across the beam profile in the $x$–direction gives $\vec{B}_n=-\vec{e}_x E-\vec{e}_z vE/c$. The flow of electromagnetic energy is described by the Pointing vector $\vec{S} = c\vec{E_n}\times\vec{B_n}/(4\pi)$. Thus the output radiation propagates at an angle $\theta_a = - v/c$ with respect to the $z_n$-axis (see Section 5.5 for more details).

\subsection{Two Metrics}

In special relativity, an important theorem states that applying a Galilean boost yields the metric of an accelerated emitter in the initial inertial frame, given by Eq. \ref{GGG11}. In this framework, a non-accelerated light source is described using the Minkowski (diagonal) metric.
To transform the coordinates $(t, x, y, z)$ of an inertial observer, we use the equivalence of active and passive pictures. By substituting $x \to x - vt, t \to t$ into the Minkowski metric, we obtain Eq. \ref{GGG11}, which describes the electrodynamics of a moving emitter from the perspective of an inertial observer.

On the other hand, the Langevin metric describes the electrodynamics of an accelerated emitter from the viewpoint of an accelerated observer measurements. A key principle is that the metric tensor must remain continuous. This problem in special relativity can be effectively addressed using an absolute time coordinatization, which relies on Galilean boosts.
To transform the inertial observer's coordinates to the accelerated frame, we apply the inverse Galilean transformation: substituting $x \to x_n + vt, t \to t_n$  into the Minkowski metric yields the Langevin metric, Eq. (\ref{GGG3}). Notably, the Langevin metric Eq.\ref{GGG3} of the accelerated emitter in the accelerated frame is the same as metric Eq.\ref{GGG11} of the accelerated emitter in the inertial frame. The difference is only that the sign of velocity $v$. 
These two metrics are two sides of the same coin.

Notably, this theorem is absent from conventional formulations of special relativity. Here, for the first time, we 
present the metric given in Eq. \ref{GGG11}. It is surprising that this result has not previously been reported, 
especially considering that the Langevin metric—derived by matching metric tensors across frames using Galilean 
boosts—is already widely used in applications involving optical phenomena in rotating frames.

In the problem under discussion, a puzzle remains. We introduce a comoving coordinate system in the initial inertial frame and analyze radiation from the moving source in terms of the new (comoving) coordinate labels.
After performing a passive transformation, the source coordinates transform as  $x' = x -vt, t' = t$. According to the equivalence of active and passive transformations within a single inertial frame, Maxwell's equations remain valid in the comoving coordinate system. Consequently, the electric field $\vec{E}'$ of an electromagnetic wave satisfies the equation  $\Box'^2\vec{E}' =  \nabla'^2\vec{E}' - \partial^2\vec{E}'/\partial(ct')^2  = 0$. This implies that the Minkowski metric, $ds^2 = c^2 (dt')^2 - (d x')^2 - (dy')^2 - (dz')^2$, governs the electrodynamics of the accelerated light source in the comoving coordinate system. 

However, a conceptual tension emerges: Why does the Minkowski metric apply in the comoving description (within an inertial frame), while the Langevin metric governs the physics in the truly accelerated frame? This apparent inconsistency requires careful examination.

The key lies in recognizing that inertial and accelerated observers do not share the same three-dimensional space in Minkowski spacetime. Although this might seem paradoxical if both observers were assumed to share a common spatial geometry, special relativity resolves the issue: observers following different worldlines inhabit different instantaneous 3-spaces due to the relativity of simultaneity.

To analyze this, we begin by examining the relationship between passive and active transformations. A passive Galilean boost within a single inertial frame, given by $x' = x -vt, t' = t$, simply re-expresses the results of measurements made by the inertial observer using a new set of coordinates. 

In the case of active transformations within the same inertial frame, the motion of fixed stars relative to the observer and their instruments remains unchanged. This establishes the equivalence between active and passive Galilean boosts in inertial frames.

In contrast, the Langevin metric describes measurements made by an accelerated observer. Under an inverse Galilean boost, the motion of fixed stars relative to the accelerated observer’s space-time grid $(t_n, x_n, y_n, z_n)$ is no longer preserved. This highlights the fundamental difference in how motion is perceived in non-inertial frames.

When observers adopt absolute-time coordinatization, relativistic kinematics becomes remarkably simple. In this approach, all asymmetries between the initial inertial and accelerated frames are absorbed into their respective frame metrics.

Alternatively, consider observers using Lorentz coordinatization. Here, the frame metrics become diagonal, and the complexities of special relativity manifest as counterintuitive kinematic effects. Notably, in this coordinatization, the inertial and accelerated observers experience different three-dimensional spatial geometries.

On the other hand, the standard textbook treatment of Lorentz transformations implicitly assumes that the 
$x_ny_nz_n$-axes of the accelerated observer remain parallel to the $xyz$-axes of the inertial observer.
In other words, it presumes that both observers share a common three-dimensional space. This assumption, often taken for granted, suggests that the simultaneous acceleration of a rigid reference frame has direct physical significance. However, this is a misconception.

For example, conventional derivation of the aberration of light effect overlooks the conventionality of distant simultaneity. There exists an inherent ambiguity in the spatial position along the $z_n$-axis (in the $x$-direction)
due to uncertainties in the timing of acceleration. This uncertainty can be quantified as $z_nv/c$, revealing that the associated events are space-like separated.

Consequently, the orientation of the $z_n$-axis lacks an exact objective meaning, due to the relativity of simultaneity. The idea that the direction of the $z_n$-axis relative to the inertial $z$-axis is physically well-defined is, at best, an approximation. This relationship can be visualized as a vertical $z_n$ axis with an angular uncertainty (or "blurring") given by: $\Delta\theta = v/c$. This inclination angle has no precisely defined physical meaning, since—given the finite speed of light—there is no experimental method to determine such a spatial displacement.

Let us illustrate these ideas with the example of a plane-wave emitter moving along the $x$-axis  with velocity $v$
in an initial inertial frame. Assume that clocks are synchronized according to the Einstein synchronization procedure, using light signals from a stationary dipole source. The electrodynamics of the moving source in this inertial frame is described by the metric in Eq. \ref{GGG11}. As previously shown, the radiated beam exhibits aberration: due to the motion of the source, the direction of energy propagation is altered.

However, when we introduce a comoving coordinate system via a passive Galilean boost, this aberration disappears. In this comoving system, the light beam propagates along the  $z$-axis. Crucially, because the boost is performed within a single inertial frame, the comoving coordinate axes $(x', y', z')$ remain parallel to the original inertial axes
$(x, y, z)$. That is, both systems inhabit the same three-dimensional space.

Let us now examine the relationship between the coordinate system of the inertial frame and that of the comoving frame. To detect aberration in the inertial frame, one must first establish a coordinate system, which requires a physical, operational procedure. In this sense, the spatial grid  $(x, y, z)$ of the inertial frame represents a physical reality. By contrast, the comoving coordinate grid  $(x', y', z')$ introduced through the passive Galilean boost is a purely mathematical construct—there is no need to consider how it would be physically realized.

It is important to emphasize that a passive Galilean boost within a single inertial frame merely provides an alternative parametrization of the same observations made by an inertial observer using real, physical reference axes.

Now consider the perspective of an accelerated observer. As before, to detect any aberration effect within the accelerated frame, a coordinate system must be established. This requires the observer to define a practical, operational method for assigning spatial coordinates  $(x_n, y_n z_n)$.

The accelerated observer finds that the angular displacement of the beam is negative, given by $\theta_a =-  v/c$.
This result leads to a key insight: from the viewpoint of the inertial observer, the $z_n$-axis of the accelerated observer is not parallel to the inertial $z$-axis. In other words, the inertial and accelerated observers inhabit different three-spaces within Minkowski spacetime.

This subtle but important distinction reflects the breakdown of simultaneity in relativistic physics and anticipates deeper discussions that will follow in Chapter 7.

\subsection{Ether Theory and the Aberration of Light Effect}

The framework of classical physics is rooted in the concept of absolute space, often linked to the hypothetical "ether"—a substance once thought to provide an absolute frame of rest. In this section, we will show that the initial inertial frame discussed earlier is not physically equivalent to absolute space or the ether. Instead, it represents an abstract idealization: the concept of absolute space-time.

To deepen our analysis, we revisit the classical ether theory in its original form, which posits the existence of a stationary ether that determines the speed of light. Since the ether is presumed immobile, any frame moving relative to it would experience anisotropy in light propagation. A non-inertial frame, accelerating with respect to the  ether, would thus detect an "ether wind" when measuring the speed of light.

\begin{figure}
	\centering
	\includegraphics[width=0.8\textwidth]{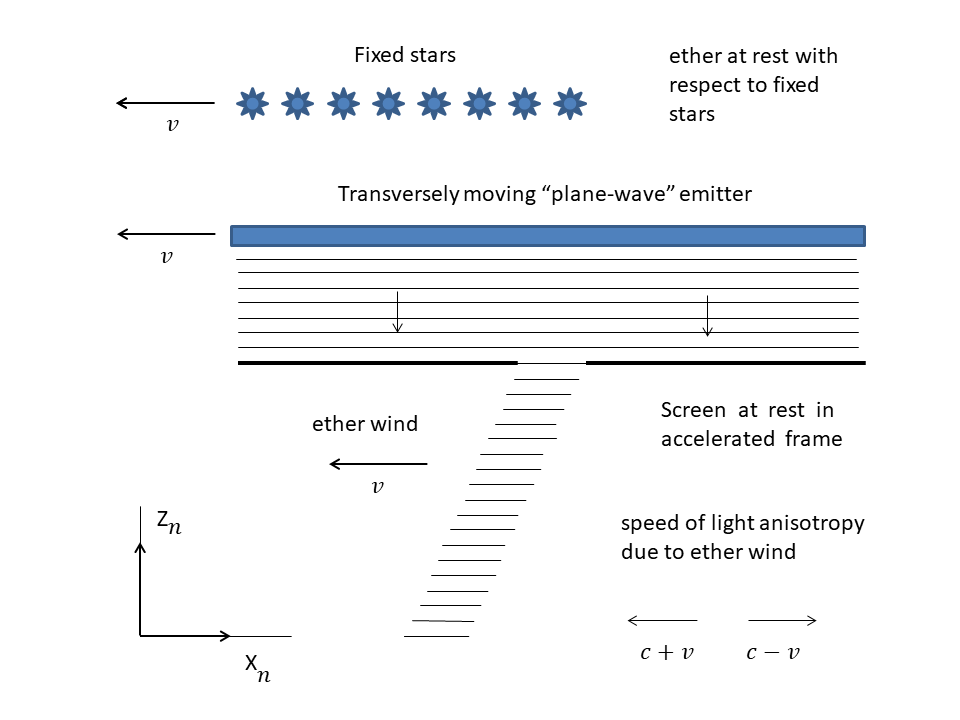}
	\caption{Aberration of light in an accelerated frame. According to the ether theory, the ether wind induces anisotropy in a frame accelerated relative to the ether,  which causes a change in the direction of radiation.}
	\label{B101}
\end{figure}

We begin by examining the influence of the ether wind on light speed. According to the ether hypothesis, the velocity of light observed from an accelerated frame would be $c + u$ for a beam propagating in the same direction as the 
accelerated system and $-c + u$ for a beam propagating in the opposite direction, where $u = -v$ represents the ether velocity in the accelerated reference frame. The aberration effect arises as a consequence of the ether wind's influence on the speed of light (Fig. \ref{B101}).

\subsubsection{Lorentz's Theory of Corresponding States}

In 1895, Hendrik Lorentz derived a first-order version of the so-called "theorem of corresponding states" to provide a general explanation for the null results of first-order optical interference experiments. This theorem establishes a relationship between pairs of electromagnetic field configurations: one in a system at rest in the ether and the other in uniform motion through it. Lorentz demonstrated that if a particular field configuration is allowed by the laws of electrodynamics, its corresponding configuration in the moving system must also be valid.
The theorem works as follows. The mathematical description of the configuration in the moving system in terms of "local time" (instead of "real time" ) is the same as the description of its corresponding state in the system at rest in terms of real time.  
Lorentz showed that two field configurations related to one another in many of their observable properties, in particular, they give rise to identical interference patterns. Thus, it is not possible to detect the orbital velocity using interference (e.g. Michelson-Morley) experiments \footnote{For discussion of the relationship between the generally accepted way of looking at special relativity and Lorentz's ether theory we suggest reading the paper \cite{JAN}.}.

We now examine the application of Lorentz’s pre-relativistic theory to the aberration of light in a non-inertial frame of reference. Previously, we analyzed the aberration of light in accelerated systems within the framework of special relativity, considering only first-order effects. Since light is inherently a relativistic phenomenon, its behavior in non-inertial frames requires careful treatment.

In this chapter, we have already discussed the asymmetry between inertial and accelerated frames. A key observation was that the metric tensor must remain continuous. In special relativity, this issue is best addressed through an absolute time coordinatization approach. Maxwell’s equations hold in the initial inertial frame, where the optical system (e.g., a light source and lens, as shown in Fig. \ref{B409}) is at rest, and time coordinates are assigned via slow clock transport. To ensure continuity, the metric of a non-inertial frame associated with an accelerating optical system (relative to the initial inertial frame) must transition smoothly into the Langevin metric. This requires relating the accelerated observer's coordinates $(x_n,y_n,z_n,t_n)$ to those of the inertial observer $(x,y,z,t)$ using the inverse Galilean transformation: $x_n = x - vt, y_n = y, z_n = z$, $t_n = t$. Under this transformation, the wave equation takes the form of Eq. \ref{GGT4}, where the coordinates and time adhere to Galilean transformation rules. The Langevin metric, along with the initial conditions, then governs optical effects in the accelerated frame.

In the preceding discussion, we analyzed the aberration of light in an accelerated frame using the Fourier transform method. The fundamental asymmetry arises from the electrodynamics equation (Eq. \ref{GGT4}). One consequence of the cross-term in the wave equation is the emergence of group velocity, which contributes to the aberration. However, an alternative approach provides an equally satisfactory explanation. This involves clock resynchronization, wherein Einstein’s synchronization method is used to define a Lorentz coordinate system for the optical setup at rest in the accelerated frame. In this framework, Maxwell’s equations describe the field configuration, and the resynchronized 
time $t'_n$ is given by $t'_n = t_n - x_n v/c^2$ in the first-order approximation. After resynchronization, the asymmetry shifts to the initial condition: the plane wavefront of the emitter undergoes a rotation by an angle $-v/c$ in the accelerated frame.

At first glance, Lorentz's treatment of an optical setup in an inertial frame moving through the ether appears similar to our approach to such a setup in an accelerated frame (relative to the fixed stars). We now analyze this similarity.
One key observation is that our initial inertial frame—one without a history of acceleration relative to the fixed stars—can be interpreted as what Lorentz considered a frame at rest in the ether. In Lorentz’s theory, and more broadly in pre-special relativity physics, the relationship between an observer at rest (using coordinates $(x,y,z,t)$)) and an observer bound to a moving body (using $(x_n,y_n,z_n,t)$ ) was governed by the Galilean transformation:  $x_n = x - vt, y_n = y, z_n = z, t_n = t$. In Lorentz’s framework, a Galilean boost is empirically equivalent to a smooth adjustment of the metric tensor. Under this transformation, the metric of the initial inertial frame transitions into the Langevin metric of the accelerated frame. Of course, Lorentz did not conceptualize his theory in this way, as Minkowski spacetime was not yet available to him.

Due to this Galilean boost, the homogeneous wave equation for the field in the moving frame (Eq. \ref{GGT4}) lacks a standard form. The primary complication arises from the presence of a cross-term, which makes solving the equation more challenging. To circumvent this difficulty, Lorentz observed that simplification is always possible. In solving the wave equation, he introduced the variable $t'_n = t_n - x_n v/c^2$, which he called "local time." In Lorentz's theory, this time shift was a mathematical device used to transform the electrodynamics equations into Maxwell's form.

At first sight, Lorentz’s theorem of corresponding states for first-order phenomena seems sufficient to provide an exact treatment of light aberration in an accelerated frame. One might naively expect that Lorentz's theory would predict, in full generality, that observations of light aberration from an earth-based incoherent source should reveal Earth's orbital motion (Fig. \ref{BC30}). However, this leads to an important question:
Why did Lorentz conclude, based on his theorem of corresponding states, that no first-order effects of Earth’s orbital motion could be detected?

\subsubsection{Analogy with the Sound Theory}

A comparison between Lorentz ether theory and the theory of sound may be insightful in this context.
In Section 4.15, we analyzed the problem of a sound emitter in the atmosphere frame, specifically considering a screen moving along its surface with velocity $v$. According to the theory of sound, energy transport remains unaffected when sound passes through a hole in the moving screen. When a plane wave falls the screen normally, it generates a transmitted oblique sound beam, as illustrated in Fig. \ref{B3}. In this scenario, the group velocity matches the phase velocity. The diagonal wave equation, $\nabla^2 f - \partial^2 f/\partial(v_st)^2  = 0$, is always valid in the atmosphere frame. However, this equation is not applicable from the perspective of an observer at rest in an accelerated frame. Accelerations relative to the atmosphere frame influence the propagation of sound, leading to a transformation of the wave equation into

\begin{eqnarray}
&&\left(1-\frac{v^2}{v_s^2}\right)\frac{\partial^2f}{\partial x_n^2}  + 2\left(\frac{v}{v_s}\right)\frac{\partial^2f}{\partial t\partial x_n}
+ \frac{\partial^2f}{\partial y_n^2} + \frac{\partial^2f}{\partial z_n^2}
- \frac{1}{c^2}\frac{\partial^2f}{\partial t^2} = 0 ~ , \label{GGT11}
\end{eqnarray}

where coordinates and time are transformed according to the Galilean transformation $x_n = x - vt$, $ t_n = t$. 
To describe sound propagation, the wave equation must be integrated with the initial conditions of the sound wavefront. Notably, acceleration does not influence the wavefront’s orientation, meaning the emitted sound beam remains perpendicular to the $z_n$ axis. Since the air remains stationary in the atmospheric frame, any reference frame moving relative to it experiences anisotropy in sound propagation. As a result, an observer in such a moving frame perceives an "air wind" when measuring the speed of sound. According to the wave equation (Eq. \ref{GGT11}), the sound velocity observed from an accelerated reference frame is $v_s -v$ for a beam propagating in the same direction as the accelerated system and  $- v_s  - v$ for a beam propagating in the opposite direction. Here, $-v$ represents the wind velocity in the accelerated frame. This effect, known as aberration, arises due to the influence of the perceived air wind on the speed of sound (Fig. \ref{B101}).

Let us now analyze the aberration of sound radiated by a single "plane-wave" emitter in the atmospheric frame.
The wavefront radiated by the plane-wave emitter at a given instant time is constructed as an envelope of all the spherical wavelets from the point sources on the emitter that have radiated until this instant. 
It is not difficult to see that when the emitter is moving at a constant velocity $v$ along the $x$ axis, the wavefront propagates at a speed $v_s$ at an angle $\theta = v/v_s$ from the vertical. 
This phenomenon arises due to the motion of the emitter, which causes constructive interference to occur when the projection of the phase velocity onto the $x$-axis matches the velocity of the emitter. According to classical kinematics, this results in an additional phase chirp given by $d\phi/dx = k_x = vk/v_s = v\omega_s/v_s^2$.
Consequently, the plane wavefront undergoes rotation as the emitter accelerates. As a result, the radiated sound beam propagates at an angle  $v/v_s$, manifesting as the aberration of sound in the atmospheric frame—analogous to the aberration of light.

Now, consider the perspective of an accelerated observer. The wave equation must be integrated with the initial   
condition set by the sound wavefront. A Galilean boost does not affect the orientation of the wavefront, meaning that 
the wavefront of the emitted sound beam remains tilted at an angle $v/v_s$ relative to the $z_n$-axis. The wave vector of the 
radiated plane wave is determined by this initial condition, leading to $k_x = vk/v_s = v\omega_s/v_s^2$.
Furthermore, the non-diagonal wave equation (Eq. \ref{GGT11}) implies that the radiated sound beam moves along the 
$x_n$-axis with a group velocity given by
$d\omega/dk_x = - v + v_s^2k_x/\omega_s = 0$. 
For further details, refer to Section 5.10.

Thus, we derive the same addition theorem for sound beam velocities as for particle velocities in classical physics. According to Lorentz, the wave equation for a moving light (or sound) emitter in the ether (or atmospheric) frame is identical to that of an emitter at rest.
Lorentz concluded that Earth's orbital motion could not be determined using aberration measurements from Earth-based sources. Consequently, ether and sound theories were merely variations of the same classical framework. However, this analysis overlooks a fundamental distinction: the velocity of light differs inherently from the velocity of sound.

When Lorentz formulated his pre-relativistic theory he assumed a space-time background with a well-defined geometrical structure: Newtonian space-time.
A close look at the physics of this subject shows that there is a real difference between the inertial frame without an accelerational history relative to the fixed stars and the Lorentz classical ether at rest. 
The main difference between the initial inertial frame and the absolute space of Newton is that special relativity taught us to think in terms of a unified space-time model.
In special relativity, the absolute space and the absolute time of Newton are fused into absolute space-time.

\newpage

\section{Stellar Aberration}

\subsection{The Corpuscular Model of Light and Stellar Aberration}

It is commonly understood that the aberration of light can be interpreted through the corpuscular model of light, drawing an analogy to how a moving observer perceives the oblique fall of raindrops. This classical kinematic approach to the calculation of stellar aberration has been employed in astronomy for nearly three centuries \footnote{The phenomenon of the annual apparent motion of celestial objects, known as stellar aberration, was first discovered by Bradley in 1727, who also explained it using the corpuscular model of light \cite{BR}.}.

However, in the 20th century, it was discovered that Newton's dynamical laws for light were incorrect, necessitating the introduction of electromagnetic wave theory to rectify these errors.
As is well-known in textbooks, the physical basis of stellar aberration arises from the finite speed of light and the change in its direction when observed from a different reference frame. This can be explained as a consequence of the velocity addition formula applied to a light beam when the observer shifts between reference frames.

For an observer on Earth, the velocity relative to the solar reference frame is approximately $v = 30 ~ \mathrm{km/s}$, corresponding to Earth's motion around the Sun. To describe the effect of stellar aberration, it is sufficient to 
consider only the first-order approximation, $v/c = 10^{-4}$.
The conventional treatment of stellar aberration remains deeply connected to classical (Newtonian) kinematics, relying on the erroneous assumption that observers on Earth and the Sun share a common 3-space.

\subsection{The Wave Theory of Stellar Aberration}

Most authors treat light propagating through the telescope barrel as a stream of photons rather than as a wave. \footnote{A common misconception is that the rays of light coming from a star simply fall onto the telescope tube without interacting with its sides. As French \cite{FR} states: "Regarding light as composed of a rain of photons, we can easily calculate the change in the apparent direction of a distant object such as a star." Similar views are also found in recent textbooks. For instance, Rafelski \cite{RAFELSKI} writes: "We consider a light ray originating from a distant star... The following discussion addresses the observation of well-focused light rays, not (spherical) plane wave light... This view of the experimental situation is accurate because the light emitted by a star consists of an incoherent flux of photons produced in independent atomic processes." This conclusion, however, is incorrect. We emphasize that the design of a telescope's optical system is based on classical diffraction theory, which stems directly from classical electromagnetic theory. The resolution of a telescope is fundamentally limited by the diffraction of light waves, and the telescope always affects the measured radiation due to the unavoidable diffraction of starlight by the telescope aperture. In a well-corrected optical system with a circular pupil, the size of the Airy disk (i.e., the size of the image of a point source) is inversely proportional to the pupil’s diameter.}

By questioning the validity of this standard reasoning, we argue that a proper treatment of stellar aberration should rely on coherent wave optics. It is easy to show that light from a distant star is approximately coherent over a circular area whose diameter, in practical terms, is far larger than the diameter of the telescope. Thus, the telescope samples only a tiny portion of the star's coherent light, meaning the waveforms can be approximated as (flat) plane waves. \footnote{While stars are considered incoherent sources, the mutual intensity function produced by an incoherent source is fully described by the Van Cittert-Zernike theorem \cite{G}. Any star can be considered as very distant from the Sun, and telescopes typically operate in the far zone of such sources. For instance, consider Sirius, one of the closest stars to Earth. The coherent area of light from Sirius has a diameter of about 6 meters, a correlation observed by Brown and Twiss in 1956 \cite{BT}.}

\begin{figure}
	\centering
	\includegraphics[width=0.6\textwidth]{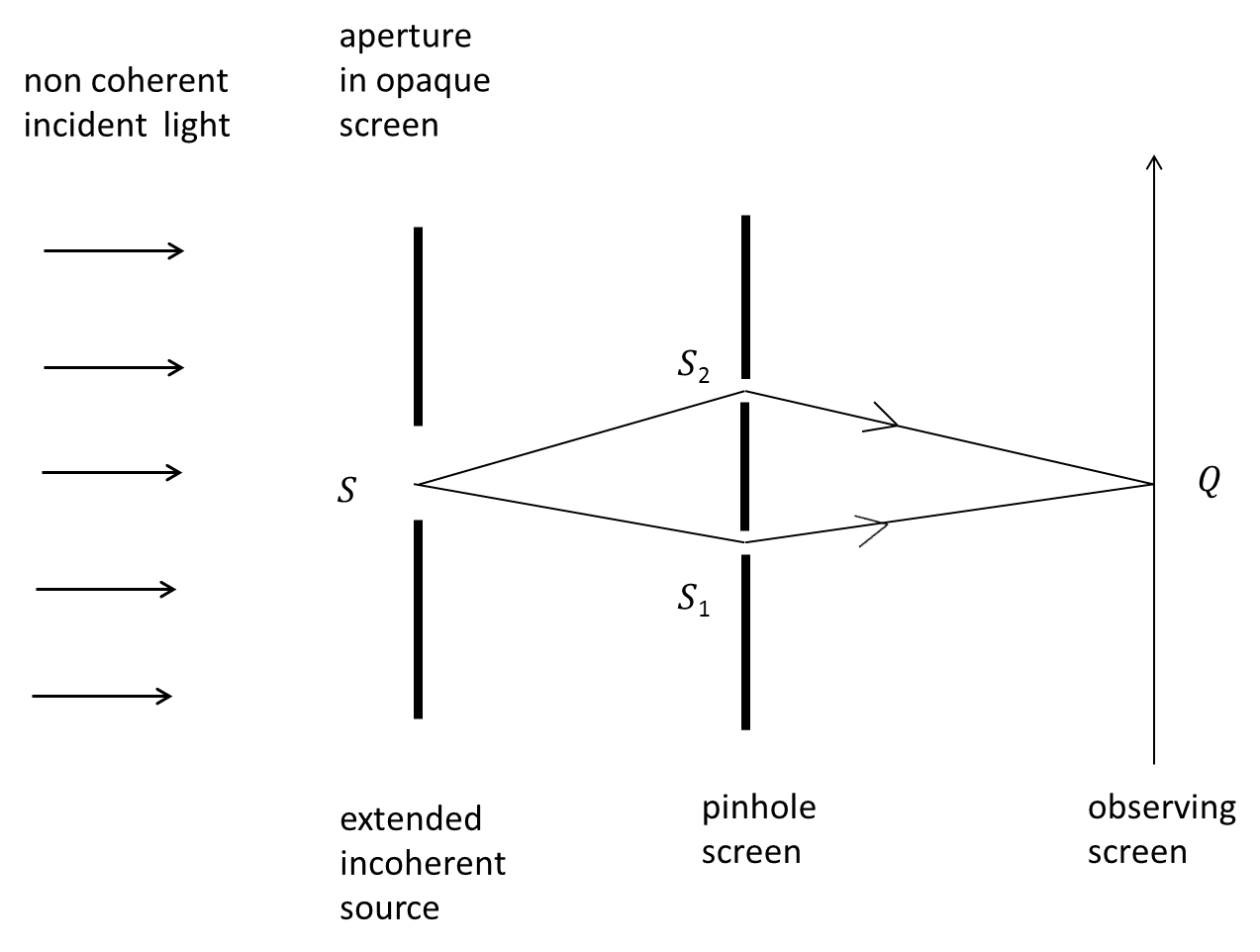}
	\caption{Young's interferometer.}
	\label{B14a}
\end{figure}

Let's verify whether this assertion holds true. The spatial coherence of a light beam generally refers to the coherence between two points in the field illuminated by the light source. To understand spatial coherence more clearly, we can refer to Young's two-pinhole experiment (see Fig. \ref{B14a}). In its simplest form, the degree of coherence between two points is described by the contrast of the interference fringes obtained when these points are treated as secondary 
sources. Consider a source $S$ illuminating two pinholes, $S1$ and $S2$,
as shown in Fig. \ref{B14a}. If the source is 
perfectly incoherent, no interference fringes can be observed by placing two pinholes in the plane of the source. However, it has been demonstrated that when the two pinholes are positioned sufficiently far from the incoherent source, interference fringes with good contrast can be observed. It is often stated that the spatial coherence of light beams increases with distance "simply through propagation." It would be valuable to find an elementary explanation that allows us to visualize the physical process behind this phenomenon.

\begin{figure}
	\centering
	\includegraphics[width=0.8\textwidth]{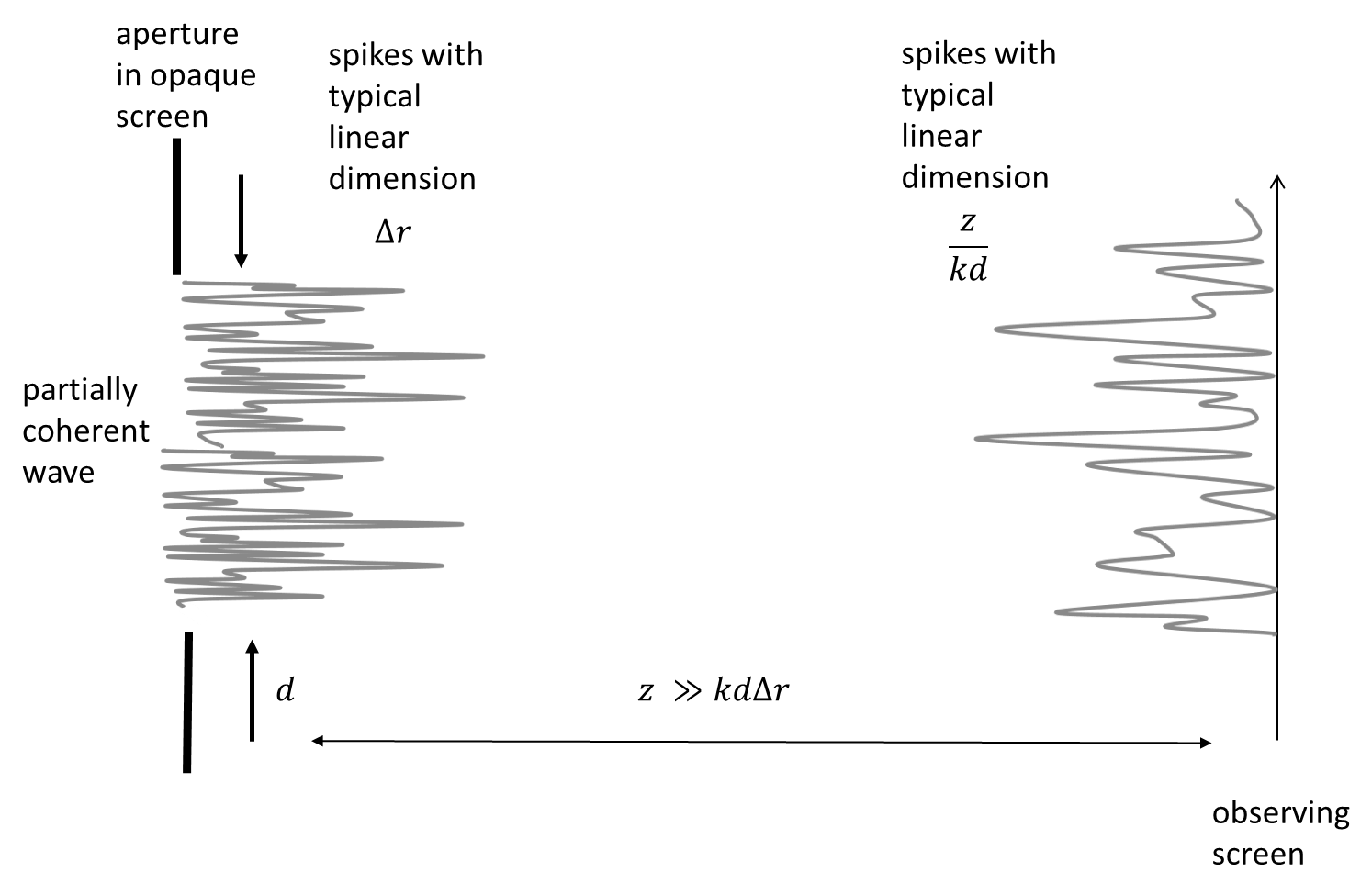}
	\caption{Geometry for propagation of spatial coherence.}
	\label{B141}
\end{figure}

Consider a quasi-monochromatic wave incident on an aperture in an opaque screen, as depicted in Fig. \ref{B14a}. In general, this wave may exhibit partial coherence. As the wave propagates through space, its detailed structure evolves, and similarly, the spatial coherence structure also changes. In this context, the transverse coherence function is said to "propagate." Given the spatial coherence at the aperture, our goal is to determine the spatial coherence on the observation screen, which is located at a distance $z$ beyond the aperture.
Stellar radiation is inherently stochastic, and for any starlight beam, there exists a characteristic linear dimension, $\Delta r$, which defines the scale of spatially random fluctuations. Fig. \ref{B141} illustrates the spiky pattern that appears on the aperture in an opaque screen. When  $\Delta r \ll d$, where $d$ is the aperture size, the radiation beyond the aperture remains partially coherent, as shown in Fig. \ref{B141}. In this case,  $\Delta r$ corresponds to the typical linear dimension of the spikes.

\begin{figure}
	\centering
	\includegraphics[width=0.7\textwidth]{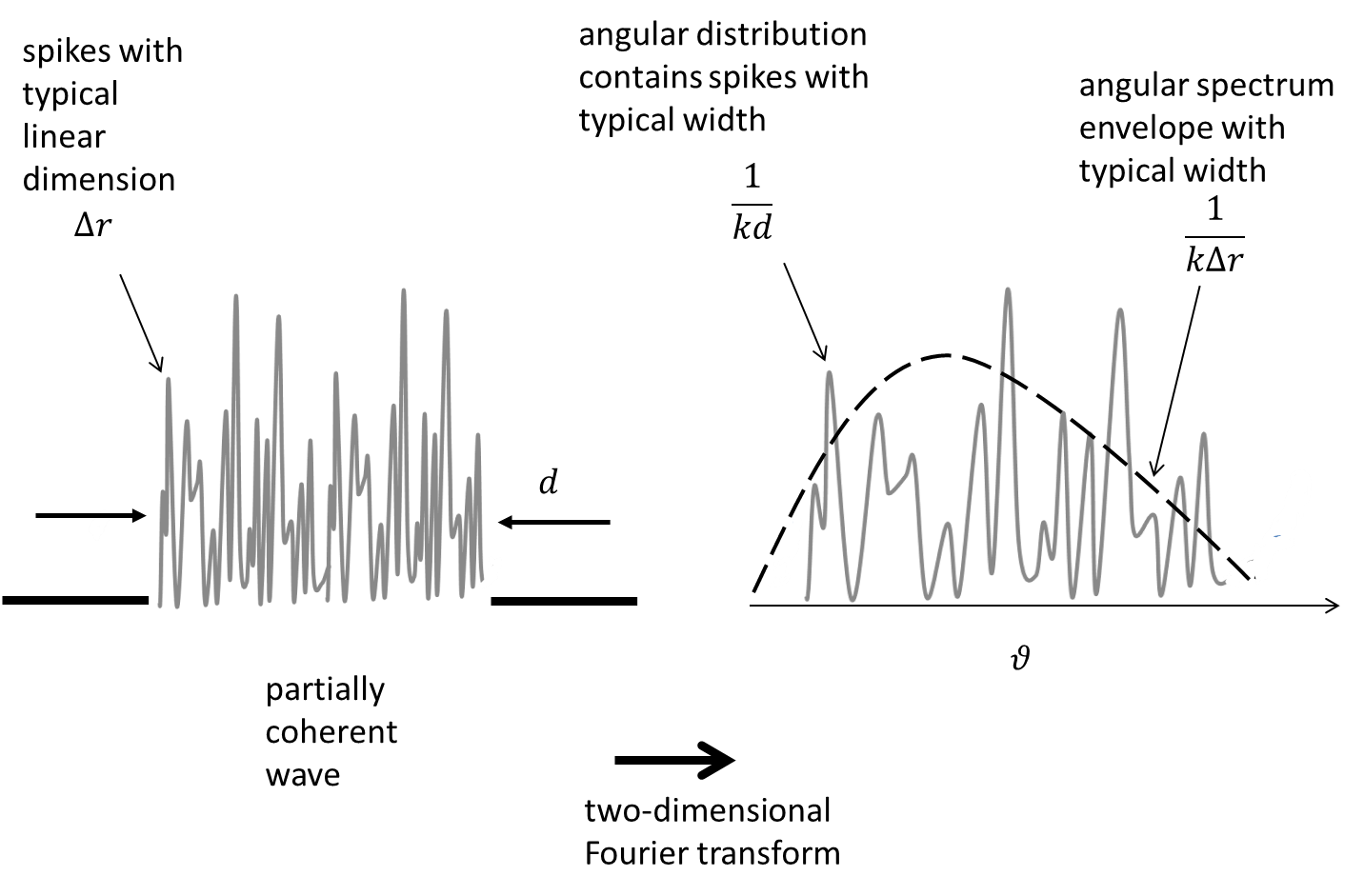}
	\caption{Reciprocal width relation of Fourier transform pair. Free space basically acts as a Fourier transformation. This means that the field in the far zone is, phase factor and proportionality factor aside, the spatial Fourier transform of the field at the source plane.}
	\label{B142}
\end{figure}

First, we aim to calculate the instantaneous intensity distribution observed across a parallel plane located a distance 
$z$ beyond the aperture. The observed intensity distribution can be determined by taking a two-dimensional Fourier transform of the field. The radiation field across the aperture can be expressed as a superposition of plane waves, all 
with the same wavenumber $ k = \omega/c$. The ratio $k_{\perp}/k$ represents the sine of the angle $\theta$ between the $z$-axis and the propagation direction of the plane wave. In the paraxial approximation, we have $k_{\perp}/k = \sin \theta \sim \theta$. If the radiation beyond the aperture is partially coherent, a spiky angular spectrum is expected. The characteristics of these spikes can be easily described using Fourier-transform notation. The typical width of the angular spectrum envelope is expected to be on the order of $(k\Delta r )^{-1}$, where $\Delta r$  is the coherence length. Additionally, the angular spectrum of a source with transverse size $d$ will exhibit spikes with a typical width of about $(kd)^{-1}$, which follows from the reciprocal width relations of Fourier transform pairs (see Fig. \ref{B142}).
It is the linear dimension $d$ of the source that determines the coherent area of the observed wave, which is given by
$z/(kd)$. Therefore, if the screen is placed sufficiently far from the incoherent source, such that $z \gg \Delta r d /\lambdabar$, a coherent area of large linear dimension can be observed.

A star is an incoherent source, characterized by spatial fluctuations on the scale of $\Delta r  \sim \lambdabar$.
Such a source radiates in all directions. Additionally, the angular spectrum of the starlight contains spikes with a typical width of $(kd)^{-1}$, where $d$ represents the star's diameter.
In astronomical observations using a telescope, each spike observed on the Earth's surface has a finite thickness, approximately equal to the spatial coherence length of the starlight. Therefore, the coherent area of the observed starlight on the Earth's surface can be estimated to be of the order of  $z/(kd)$, where $z$ is the distance from the star to the Earth. The star chosen by Bradley for this observation was Draconis, one of the nearest stars. For this case, the diameter of the coherent area on the Earth's surface is roughly estimated to be around 100 meters.

\subsection{Astronomical Observations}


It is widely accepted that the theory of relativity aligns with Bradley's discovery of stellar aberration, interpreting it as a consequence of an observer's motion relative to a light source. 

In his paper on relativity, Einstein derived the aberration formula by considering the velocity $v$ as the relative velocity of the star-Earth system. This interpretation has been adopted by numerous textbook authors.
As Moeller \cite{M} states: "This phenomenon, known as aberration, was observed by Bradley, who noticed that stars appear to undergo an annual collective motion in the sky. This apparent motion arises because the observed direction of a light ray from a star depends on the Earth's velocity \textit{relative} to the star."

\begin{figure}
	\centering
	\includegraphics[width=0.7\textwidth]{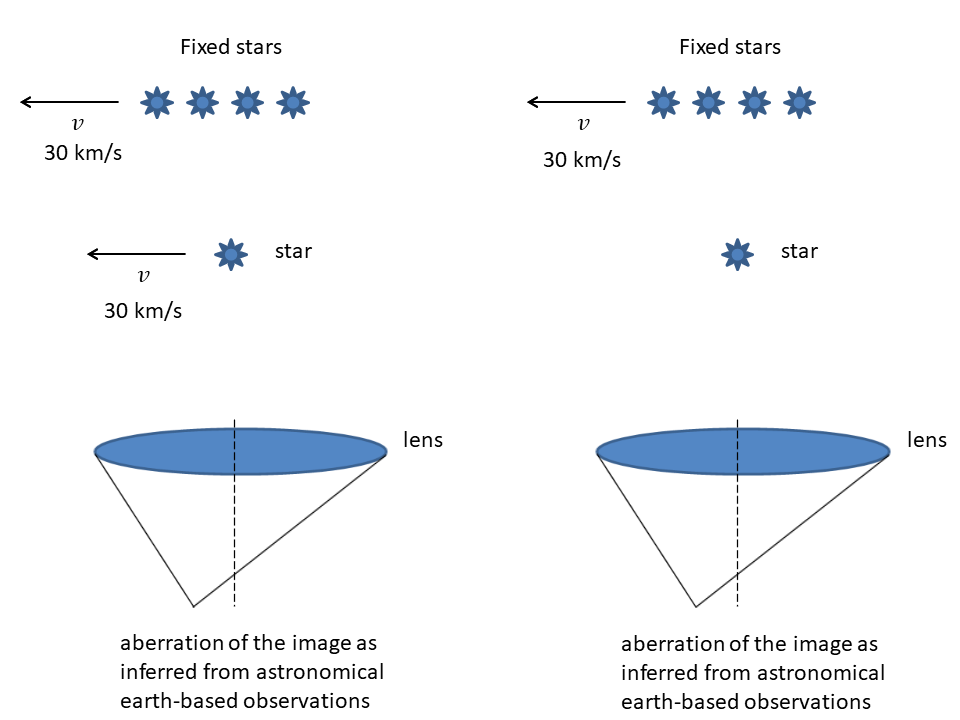}
	\caption{Aberration shift inferred from astronomical observations. The plane wavefronts of starlight entering the telescope are focused by the lens to form a diffraction spot in the focal plane. Two scenarios are considered, each with different velocities of an Earth-based telescope and a star. In the first scenario (left), the star remains stationary relative to the fixed stars. In the second scenario (right), the star moves at the same velocity as the Earth.}
	\label{A1}
\end{figure}

\begin{figure}
	\centering
	\includegraphics[width=0.7\textwidth]{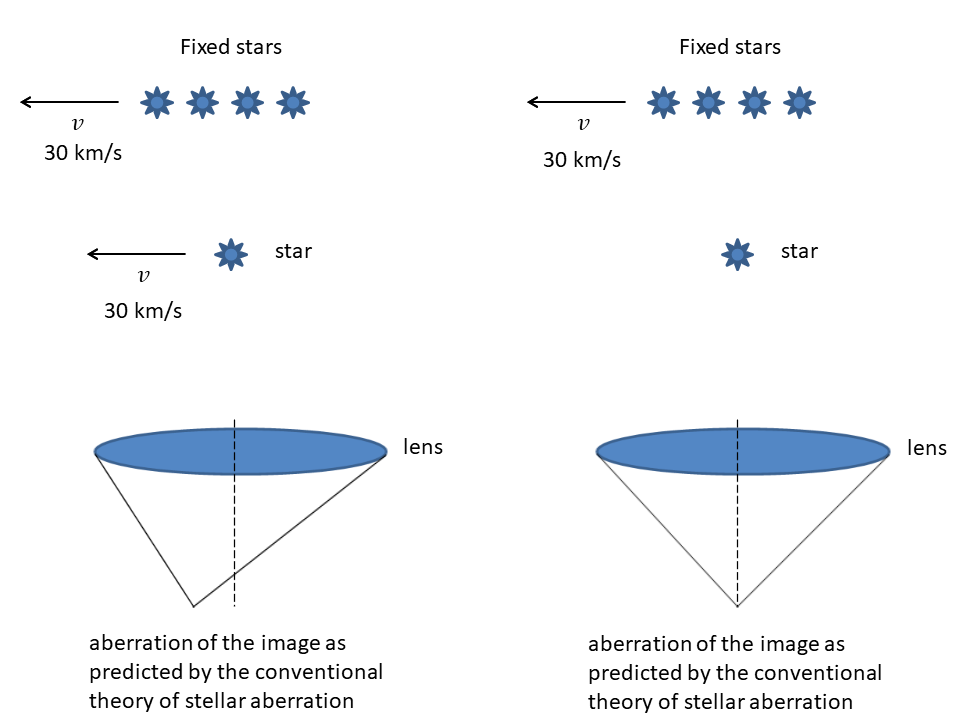}
	\caption{Predicted aberration shift according to the conventional theory of stellar aberration.} 
	\label{A2}
\end{figure}

The asymmetry between the cases where either the star or the Earth-based telescope is in motion becomes evident when considering the separation of binary stars.
Spectroscopic binaries exhibit velocities exceeding that of Earth's orbit around the Sun. These stars revolve around their common center of mass within just a few days—a period during which Earth's motion remains practically constant. If the changing velocities of the binary components were comparable to Earth's orbital velocity, their separation should be easily detectable. However, this is not observed (Fig. \ref{A1}-Fig. \ref{A2}).

In 1950, Ives \cite{I} was the first to highlight a significant challenge that binary star observations pose to the theory of relativity. He argued that the idea of aberration being purely a function of relative motion is contradicted by the existence of spectroscopic binaries with velocities comparable to Earth's orbital velocity. Yet, these binaries exhibit no aberration effects different from other stars.
For instance, the spectroscopic binary Mizar A has well-established orbital parameters. If aberration were solely due to relative velocity, its angular separation should be observable at 1'10". However, empirical measurements show a value below 0.01", which is clearly inconsistent with the perspective presented in standard textbooks \cite{PS, SE}. 

For planets, the observed direction involves two separate steps: 

Step 1: Light-time correction (dominant for planets). This already includes the effect of the planet’s motion. 

Step 2: Aberration (same as for stars). This step depends only on Earth's velocity, not on the planet's velocity.

So what do observations actually show?

High-precision astrometry (e.g., ephemerides, meridian observations, modern CCD astrometry) confirms:
After light-time correction: The aberrational displacement of planets is the same as for stars - magnitude  20.5 arcseconds. It depends only on Earth’s velocity.

Key fact (confirmed by astronomical observations)

1. Binary stars → large relative velocities, no aberrational effect in their separation.

2. Planets → large relative velocities with respect to Earth, yet aberration remains 20. 5 arcseconds.

3. Stars → arbitrary velocities → same aberration constant. 

This directly contradicts a naive “relative velocity” interpretation.

\subsection{Stellar Aberration in the Context of Special Relativity}

In Chapter 4, we critically reexamined the textbook claim that wavefronts and raindrops experience the same aberration. Using the theory of relativity, we demonstrated that when a transversely moving mirror reflects a normally incident plane wave of light, the energy transport deviates. This effect arises because, in the absolute time coordinatization, the wave equation depends on the velocity vector.
As a result, the energy transport velocity differs from the phase velocity.
According to Babinet's principle, this remarkable prediction should also apply to light transmitted through a hole in a moving opaque screen or, consequently, through the moving open end of a telescope barrel.

The binary star paradox is resolved by recognizing that when light passes through the end of a telescope barrel, diffraction perturbs its fields. As a result, the light beam no longer carries information about the star’s motion relative to the fixed stars. If the telescope were at rest relative to the fixed stars and the observed star began moving, its apparent position in the telescope would not suddenly shift by any angle.

However, a different challenge arises when explaining Earth-based observations—specifically, the apparent shift in the positions of fixed stars as the Earth’s telescope changes its motion relative to them. It is crucial to emphasize that aberration is caused by changes in the telescope's velocity, not the star's.

The resolution of this issue lies in the fundamental asymmetry between Earth-based and Sun-based observers—namely, the acceleration of the Earth-based observer relative to the fixed stars. Some experts, such as Selleri, have explicitly recognized this, noting that aberration can be explained in terms of variations in the Earth's absolute velocity due to its orbital motion, while the star–Earth relative velocity is irrelevant. Thus, planetary acceleration plays a key role in the phenomenon \cite{SE}.
We derive stellar aberration using the Langevin metric. Although stellar aberration exhibits asymmetry, it does not contradict special relativity. This is because the heliocentric (Sun-based) reference frame is inertial, whereas the geocentric (Earth-based) reference frame is non-inertial.

\subsection{Heliocentric and Earth-Based Frames}

In the discussion above, we demonstrated that when a small hole is present in an opaque screen at rest relative to the fixed stars, and a point source moves tangentially to the screen, there is no aberration—no deviation in the transport of energy—for light passing through the hole. The transmission appears as illustrated in Fig. \ref{B182}.
The absence of any effect from the motion of the light source relative to the fixed stars in this setup suggests a fundamental issue for stellar aberration theory in the heliocentric frame of reference.

Now, consider an observer at rest with respect to the telescope who measures the direction of energy transport. A careful analysis reveals that in the heliocentric frame, where the telescope is at rest, we are dealing with a steady-state transmission problem. How does the transmitted light beam behave in this case?
The beam appears to travel along the telescope’s axis because it has lost its horizontal group velocity component. This situation corresponds to a telescope aligned perpendicularly to the phase front of incoming starlight—pointed directly at the star. If the star’s motion is parallel to the phase front (i.e., perpendicular to the telescope axis), then starlight entering the telescope’s aperture will propagate through its full length without deviation.

A key conclusion from this discussion is that stellar aberration is absent in this configuration. Specifically, binary star components remain unresolved, meaning their velocities do not affect the aberration phenomenon.

\begin{figure}
	\centering
	\includegraphics[width=0.45\textwidth]{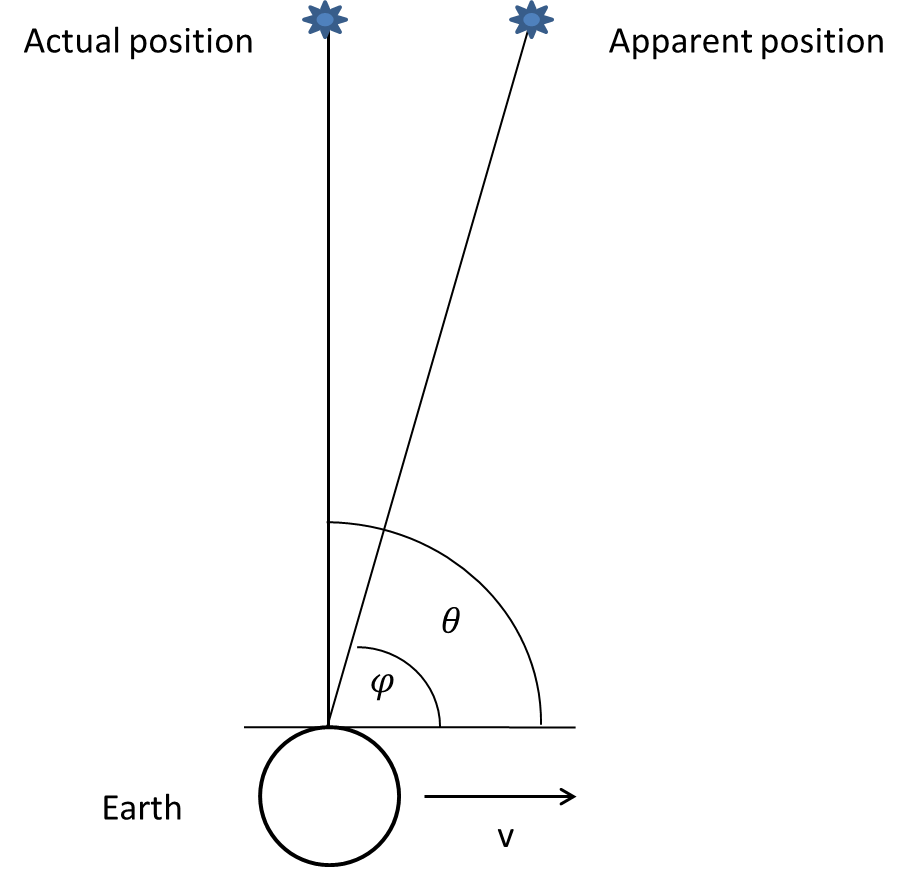}
	\caption{The direction of a star as seen from the earth. }
	\label{B12}
\end{figure}

The analogy between the obliquity of raindrops and stellar aberration is incorrect. A complete description of all Earth-based experimental observations of stellar aberration is possible only through the theory of relativity and wave optics. Our theory predicts stellar aberration effects that are in full agreement with Bradley's results (Fig. \ref{B12}).

Due to the asymmetry between inertial and rotating frames, the theory of light aberration makes a striking prediction: if a telescope is at rest relative to the Earth while the Earth rotates relative to the fixed stars, the observed direction of a star from Earth differs from that seen by a hypothetical observer at rest with respect to the Sun. Specifically, the apparent angle is smaller than the actual angle. The difference between the actual and apparent 
angles, $\theta_a$,
is given by the relation: 
$\theta_a = v/c$, 
where $v$ is the Earth's orbital velocity around the 
Sun. This effect arises from the anisotropy introduced by the cross-term in the metric equation (Eq. \ref{GGG3}), which influences the direction of radiation (aberration) in the rotating frame. The transmission through the telescope aperture in this frame is illustrated in Fig. \ref{B181} - Fig. \ref{B181A}.

In the geocentric frame of reference, stellar aberration is independent of the star's speed and originates solely from the observer's motion relative to the Sun. According to our interpretation, there are multiple types of aberration, with stellar aberration in the Earth-based frame being just one of them. We have demonstrated that all Earth-based experiments can be explained by two factors: the effect of the measuring instrument (i.e., the physical influence of the telescope on the measurement) and the acceleration of the Earth-based observer relative to the fixed stars.


The aberration of light is a geometric phenomenon. To detect its effects within an Earth-based frame, a well-defined coordinate system with a reference direction is necessary.

A conventional approach to the aberration of light relies on an implicit assumption of a reference direction, though this assumption often goes unrecognized. Traditionally, the physical interpretation of aberration is based on measurements using rods that remain at rest in the observer’s frame. This local frame of reference is so intuitively accepted that it is rarely questioned or explicitly discussed in aberration theory.

However, a more fundamental spatial description exists beyond the reliance on measuring rods. In an Earth-based frame, the natural reference axis is defined by the gravitational field vector. The plumb line, pointing toward the Earth's center, establishes the nadir, forming the most fundamental local coordinate system. For instance, James Bradley used a vertically mounted telescope to study stellar aberration. He selected the star Draconis because it transited almost exactly at the zenith. The traditional plumb line provided a sufficiently accurate zenith reference for observing stellar aberration.

\subsection{Limits of Applicability}

The applicability of our stellar aberration theory is broader than one might expect. Previously, for simplicity, we assumed that a star image is represented by a point spread function in the image plane of a telescope—in other words, that the input signal is effectively a plane wave.

Now, suppose a telescope can resolve details in the star image. A complete understanding of the object-image relationship requires accounting for diffraction effects. Diffraction convolves the ideal image with the Fraunhofer diffraction pattern of the telescope pupil.

A star is a spatially incoherent source. In statistical optics, stellar emission is described classically at the level of Maxwell's equations, while the emitting medium—an ensemble of atoms—is treated using quantum mechanics. A star consists of elementary point sources, each statistically independent, with characteristic dimensions on the order of the radiation wavelength, $\lambda$. Within an elementary source volume $(\lambda^3)$, a vast number of atoms contribute to emission. Semi-classical theory models these atoms as coherent, interacting radiating dipoles. The induced macroscopic dipole moment in an elementary source produces classical electromagnetic radiation. In statistical optics, electromagnetic fields are treated classically until they interact with the atoms of a photosensitive material, where their interaction is quantized. This avoids the need to quantize the electromagnetic field itself—only the field-matter interaction is quantized. The photodetector converts the continuous cycle-averaged classical intensity into a sequence of discrete photocounts.

Because a star is composed of statistically independent elementary point sources at different spatial offsets, each elementary source effectively produces a plane wave at the telescope pupil. An offset in an elementary source tilts the far-zone field. The total radiation field from a completely incoherent source is thus a linear superposition of the fields from these individual point sources. Since the image of each elementary source is a point spread function, the telescope inevitably influences the measurement of any completely incoherent source.

Importantly, any linear superposition of radiation fields from elementary point sources preserves single-source characteristics, such as independence from source motion. This reasoning confirms that our stellar aberration theory remains valid for imaging any completely incoherent source.

\newpage

\section{The Concept of Ordinary Space in Special Relativity}

\subsection{Inertial frame view of Observations of a Non-Inertial Observer}

In this chapter, we introduce a novel approach to the theory of light aberration in non-inertial reference frames, offering a fresh perspective on this complex subject.

We begin by emphasizing a key principle: the laws of physics in an initial inertial frame must fully account for all physical phenomena, including those observed by non-inertial observers.

This principle was already applied in Chapter 5, where we analyzed phenomena in non-inertial (e.g., rotating) reference frames using the Langevin metric. This metric captures the electrodynamics of an accelerated light source from the viewpoint of measurements made by an accelerated observer—but interpreted from the perspective of an inertial observer.

Our findings there can be summarized as follows:

1. Inertial Frame Setup

We assume that the inertial frame has no history of acceleration and that spacetime, as perceived by an observer at rest, is described by the standard diagonal Minkowski metric:  $ds^2 = c^2 dt^2 - d x^2 - dy^2 - dz^2$.
Applying a Galilean transformation $x \to x-vt$, $ t \to t$ we obtain the metric of an accelerated emitter in the inertial frame, as presented in Eq. (\ref{GGG11}). In this coordinate system $(t,x,y,z)$, the devices of the inertial observer remain governed by the standard diagonal metric.

2. Non-Inertial Frame Transformation

To describe an accelerated observer with coordinates $(t_n,x_n,y_n,z_n)$,  we apply the inverse Galilean transformation:
$x \to x_n + vt$, with $t \to t_n$ unchanged. Substituting this into the Minkowski metric yields the Langevin metric, as shown in Eq. (\ref{GGG3}).

However, we argue that physical phenomena in accelerated frames can also be effectively analyzed within an inertial frame using the framework of standard Einsteinian special relativity and its well-established kinematic effects.

For instance, consider the Sagnac effect. From the perspective of an inertial laboratory frame, its interpretation becomes straightforward. The phase difference between counter-propagating waves arises from the relativistic velocity addition formula. That is, as viewed from the lab frame, the Sagnac effect emerges purely as a kinematic consequence of special relativity. This analysis assumes the speed of light in the lab frame is $c$, which—according to special relativity—is independent of the source’s motion when using Lorentz coordinates.

In contrast, analyzing the Sagnac effect within a rotating frame is more intricate. In such cases, authors typically employ a spacetime approach using a metric tensor—namely, the Langevin metric—within a four-dimensional Minkowski framework, to compute the propagation time differences between counter-propagating waves.

This leads us to a natural question:
Can we analyze the aberration of light in the same way?
Indeed, we propose applying the Langevin metric to describe light aberration in non-inertial frames as well.

That said, this approach requires a clock resynchronization procedure in the inertial frame. To maintain Lorentz coordinates for a moving light source, one must diagonalize the metric given in Eq. (\ref{GGG11}), as detailed in Section 3.5. The Lorentz coordinates can then be expressed in terms of the original coordinates $(t,x,y,z)$, using the transformation given in Eq. (\ref{GGT3}).

\subsection{Relativity of Simultaneity}

Let us first apply the classical kinematic method to compute the aberration of light effect in a non-inertial frame, as observed from an inertial frame. Textbook authors derive the same velocity addition theorem for light beams as for particles in classical physics. This result is particularly striking, as the study of light aberration is deeply rooted in classical kinematics. When terms of order $v^2/c^2$ are neglected, the Galilean vectorial velocity addition law is used. However, applying classical kinematics to aberration calculations leads to a significant error. According to our approach, a remarkable prediction emerges regarding light aberration in an accelerated frame (see Fig. \ref{B300}). To correctly predict the aberration measurement, the accelerated observer must use the Langevin metric. The cross-term in this metric induces anisotropy in the accelerated frame, which alters the radiation direction.
In contrast, relying solely on velocity addition (whether Galilean or Einsteinian) suggests that acceleration does not break the symmetry between the accelerated and initial inertial frames.

The error in the last argument arises from the incorrect assumption that an inertial observer and an accelerated observer share a common 3-space. According to special relativity, an inertial observer and an accelerated observer (relative to the fixed stars) have distinct 3-spaces. The classical kinematics method used to compute the aberration of light effect is flawed because it treats the speed of light as a classical velocity, disregarding relativistic effects. In the classical approach, no fundamental distinction is made between the aberration of light and the aberration of raindrops.

It is particularly interesting to observe that geometric effects in our ordinary space are closely linked to the relativity of simultaneity.
In Fig. \ref{B102}, the transmitted light beam propagates at an angle of  $ - v/c$, resulting in the phenomenon of light aberration. This naturally raises the question: relative to what does a light beam propagate within an accelerated frame when it undergoes an angular displacement of $ - v/c$?
Consider an observer in the accelerated frame measuring the direction of the light beam. If the plane wavefront of the transmitted beam is focused by a lens, it will form a diffraction spot in the focal plane along the optical axis. Determining the direction of the optical axis with respect to the frame's axes is equivalent to measuring the angular displacement.
To detect the aberration of light within the accelerated frame, it is essential to establish a coordinate system with a well-defined reference direction. This necessitates a detailed examination of the method by which coordinates are assigned—a process that inherently involves a physical procedure. Ultimately, this method must provide a consistent framework for defining coordinates in both inertial and accelerated reference frames.

In ordinary space, an accelerated frame moves relative to an inertial frame along the line of motion, and conversely, the inertial frame moves relative to the accelerated frame along the line of motion. The angle between the observer's coordinate system axis and the line of motion is a fundamental geometric parameter in ordinary space.
By using the line of motion as the reference axis, the accelerated observer can define a second reference axis. To achieve this, we must provide a practical and operational method for assigning an axis perpendicular to the motion line axis. One possible approach is to use a light beam as a reference direction.
We define the second reference direction as follows: Suppose the aberration direction within the accelerated frame is determined relative to the fixed direction of a light beam emitted from a "plane-wave" source at rest in the accelerated frame. In other words, the coordinate system is established using the electromagnetic axis and the line of motion. For simplicity, we assume that the motion of the aberrated light beam lies in the same plane, with its angular position described by a single angle.

When the accelerated system transitions to uniform motion at a constant velocity, Einstein's standard clock synchronization procedure can be applied.
This synchronization method relies on light signals emitted by a source at rest, under the assumption that light propagates isotropically with constant speed $c$. In this framework, the accelerated observer analyzes the light beam from the stationary emitter using the standard Maxwell equations. According to classical electrodynamics, light is emitted perpendicular to the wavefront of the radiation.
We consider a specific configuration in which the  $x_n$-axis lies parallel to the wavefront of the reference light beam.
The aberration of the light beam transmitted through an aperture can be characterized by the angle between the directions of the reference and transmitted beams, which quantifies the aberration increment.

\begin{figure}
	\centering
	\includegraphics[width=0.8\textwidth]{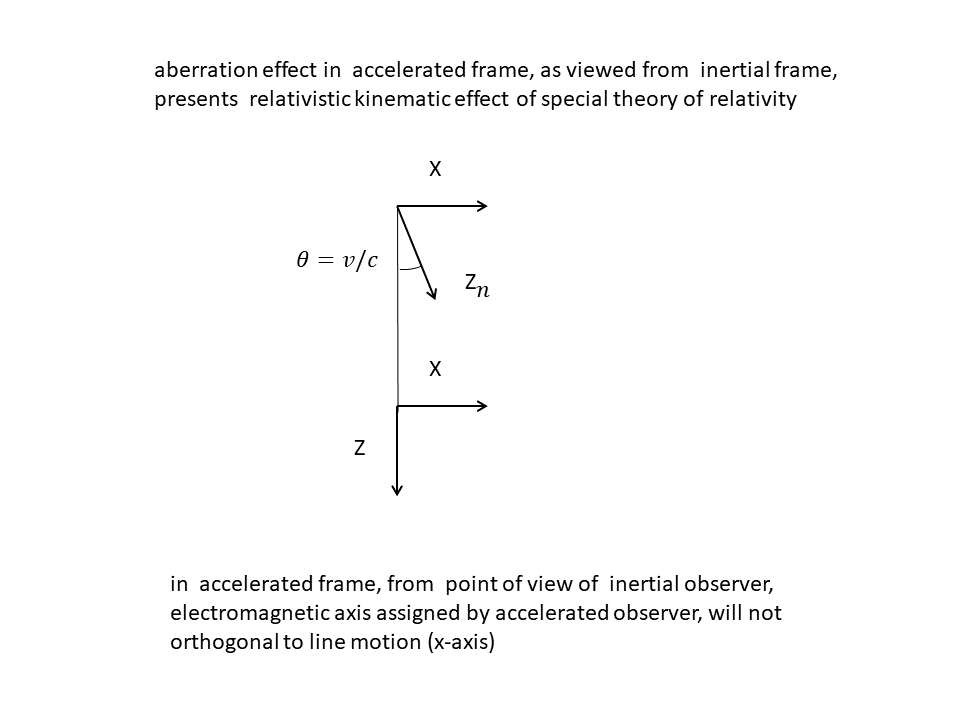}
	\caption{Inertial frame view of observations of the accelerated observer.}
	\label{B203}
\end{figure}

Let us examine the relationship between reference directions in the inertial frame and those in the accelerated frame. In the context of our study on the aberration of light, this approach relies on the use of local reference light sources.
To define reference directions perpendicular to the motion in both frames, we employ two local light sources: one remains at rest in the accelerated frame, while the other is stationary in the inertial frame. In essence, the reference electromagnetic axis in each frame is established by an individual light beam.
However, reference axes can also be defined using more practical methods than light beams. For example, in an Earth-based frame, the reference axis may be aligned with the gravitational field vector, such as the standard direction of a plumb line.
The theory of relativity is founded on the equivalence of all local physical frames of reference within a single inertial frame. Consequently, when the aberration angle arises, it manifests identically across all local reference systems.

Now, let us return to the observations of an accelerated observer as viewed from an inertial frame. 
In Chapter 3, we discussed how absolute time coordinatization can be transformed into Lorentz coordinatization. Within the inertial frame, this transformation can be understood as a change of the time variable according to  $t \to t + xv/c^2$. The combined effect of the Galilean transformation and this variable change results in the Lorentz transformation when using absolute time coordinatization in the inertial frame. Consequently, in this framework, the speed of light remains independent of the motion of the source.

Simultaneity of events is relative for both inertial and accelerated frames. Events occurring at different locations along the direction of relative motion cannot be simultaneous in both frames when using Lorentz coordinatization—an idea analogous to Einstein’s train-embankment thought experiment.
On one hand, the wave equation remains invariant under Lorentz transformations. On the other hand, employing Lorentz coordinatization introduces a time transformation,  $t \to t + xv/c^2$, which results in a rotation of the radiation phase front by an angle $v/c$.

From the preceding discussion, it follows that in an accelerated system, as viewed by an inertial observer, the electromagnetic axis assigned by the accelerated observer will not be parallel to the $z$-axis of the inertial frame (see Fig. \ref{B203}). We conclude that, in the accelerated frame, the transmitted light beam propagates along the $z_n$-axis of the accelerated frame with an angular displacement of $-v/c$.

We determine the reference directions in both the inertial and accelerated frames using two local light sources, leading to intriguing consequences.
First, both the accelerated and inertial observers can directly measure the angular displacement of a transmitted light beam relative to the reference light beam in the inertial frame. Consider an observer in the inertial frame measuring the direction of the transmitted light relative to a reference light beam from an emitter at rest in that frame. The inertial observer finds that the angular displacement is positive and equal to $\theta_a = v/c$.

Now, we examine the angular displacement of the transmitted light relative to the reference beam within the accelerated frame. On one hand, the angular displacement of the transmitted light with respect to the reference beam from an emitter at rest in the accelerated frame is $-v/c$. On the other hand, the angular displacement of the inertial reference light beam is $-2v/c$ (see Fig. \ref{B105}). This implies that the accelerated observer can directly measure the angular displacement of the transmitted light with respect to the inertial frame’s reference light beam.
Specifically, the relative angular displacement is
$\theta_a(\mathrm{transmitted ~ beam}) - \theta_a(\mathrm{inertial~ reference~ beam})  = - v/c + 2v/c = v/c$.
This result is consistent with the measurements of the inertial observer, as expected.
\footnote{So far, we have been discussing the electromagnetic axis. Within a single inertial frame, an observer can directly measure the aberration of transmitted light with respect to an arbitrary reference axis, since all local reference systems are equivalent. However, if we wish to determine the reference direction of an accelerated frame relative to that of the initial inertial frame, we must take into account the physical realization of these reference axes. The accelerated observer can directly measure the angular displacement of the transmitted light relative to the inertial frame’s reference axis only by using the inertial frame’s reference light beam. If, instead, the reference directions in both frames are defined using two relativistic electron sources, an important distinction arises. Unlike a reference light beam, a reference relativistic electron beam does not permit an exact measurement of the transmitted light’s angular displacement within our experimental setup.}

\subsection{The Non-Existence of Instantaneous Three-Dimensional Space}

We derived the results for a non-inertial observer's observations using Lorentz transformations.
Let us follow the standard textbook assumption that the linear motion of observers remains the same in both the accelerated and inertial frames. At first glance, if the reference axes are orthogonal to the direction of motion, they should remain parallel to each other. However, due to the angular displacement $v/c$, this leads to an apparent paradox.

\begin{figure}
	\centering
	\includegraphics[width=0.8\textwidth]{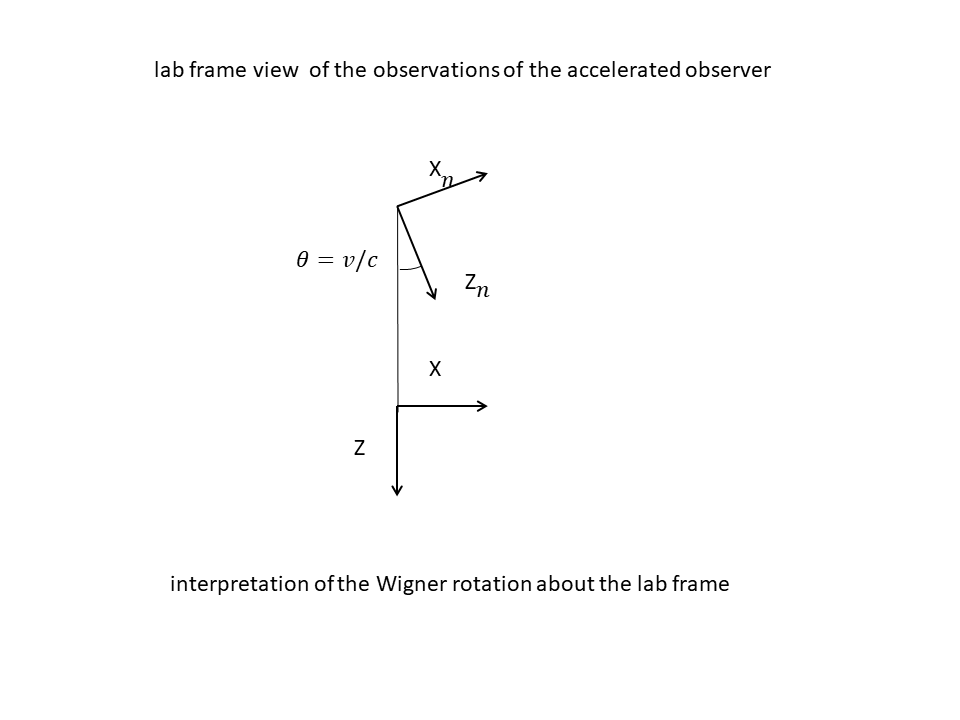}
	\caption{The inertial observer sees the rotation of the moving axes with respect to the lab frame axes, which is closely associated with the relativity of simultaneity.}
	\label{B303}
\end{figure}

The key to resolving this paradox, as noted above, lies in recognizing that the theory of special relativity eliminates the concept of absolute simultaneity. In the case of the relativity of simultaneity, space and time become intertwined—an observer’s spatial measurements inherently include a subtle influence of time as perceived by another observer.

A closer examination reveals that the commonly accepted assumption of a shared $(x)$ line motion among observers is based on the mistaken belief that they share a common 3-space. This is a misconception.
The correct approach is to use the light beam as the reference direction. We have already established that accelerated observer can directly measure the angular displacement between the inertial $(z)$ and accelerated $(z_n)$ electromagnetic reference directions, and this measurement aligns with the inertial observer's perspective. In other words, the displacement between electromagnetic axes is a physical reality.

From the inertial observer’s viewpoint, the moving frame's axes rotate relative to the inertial frame axes, as illustrated in Fig. \ref{B303}. This rotation is a relativistic kinematic effect.
One can directly verify that the accelerated frame's axes are rotated by an angle  $\theta_w = v/c$ (to first order in $v/c$) relative to the inertial frame axes (see the next chapter for further details).

\subsection{Acceleration of a Rigid Body in Special Relativity}

\begin{figure}
	\centering
	\includegraphics[width=0.85\textwidth]{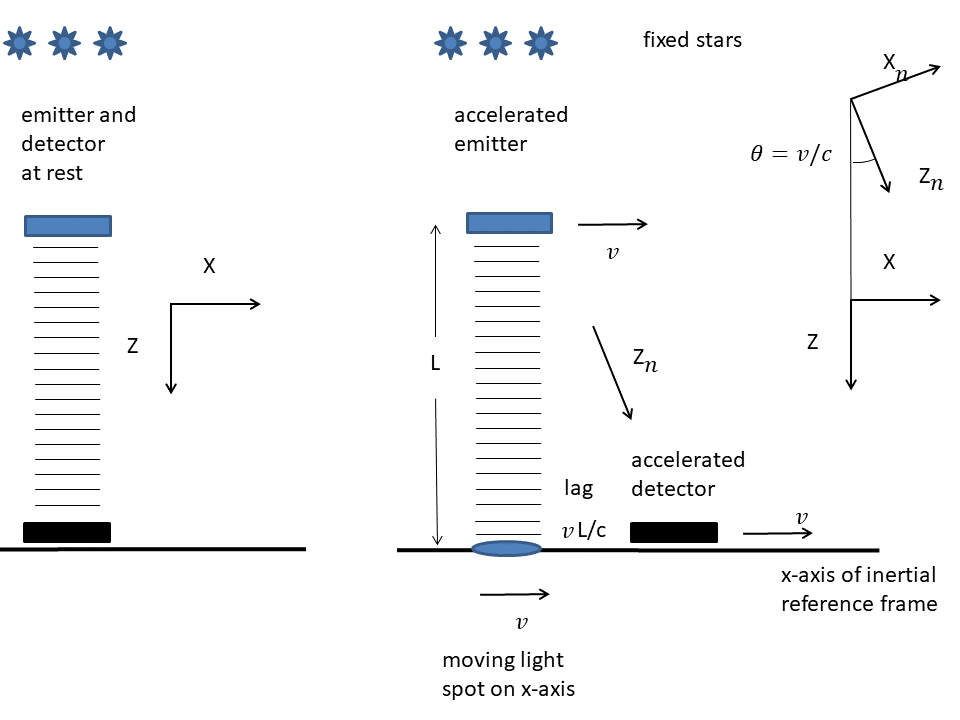}
	\caption{Aberration of light in an inertial frame of reference. The orientation of the wavefront lacks an exact objective meaning. For simplicity, only the wavefront orientation in the absolute time coordinatization of the moving source is depicted here.}
	\label{B334}
\end{figure}

Now, let us explore further consequences of the relativity of simultaneity. The concept of rigid motion, as understood in Newtonian mechanics, cannot be directly carried over to the special theory of relativity.
First, let us study the relativistic kinematics of rotation of the moving frame axes with respect to the inertial frame axes. Consider two emitter-detector setups in an initial inertial frame. Suppose one setup is accelerated from rest to velocity $v$ along the $x$-axis, as shown in Fig. \ref{B334}. The explanation of aberration relies on absolute time coordinatization \footnote{An alternative explanation involves a clock re-synchronization procedure in the initial inertial frame. Under Einstein’s synchronization, the time for the moving source is obtained by introducing an offset factor $xv/c^2$ in the first-order approximation. This time shift results in a rotation of the radiation wavefront by an angle $v/c$ for both sources. In this synchronization scheme, the radiation of the moving source is described using Maxwell’s equations.}.
A key distinction between relativistic and Newtonian kinematics is that, in the inertial frame, the accelerated emitter-detector setup undergoes a shearing motion. While counterintuitive, this effect is not paradoxical—it arises naturally from the relativity of simultaneity and the geometric structure of spacetime in special relativity.

\begin{figure}
	\centering
	\includegraphics[width=0.85\textwidth]{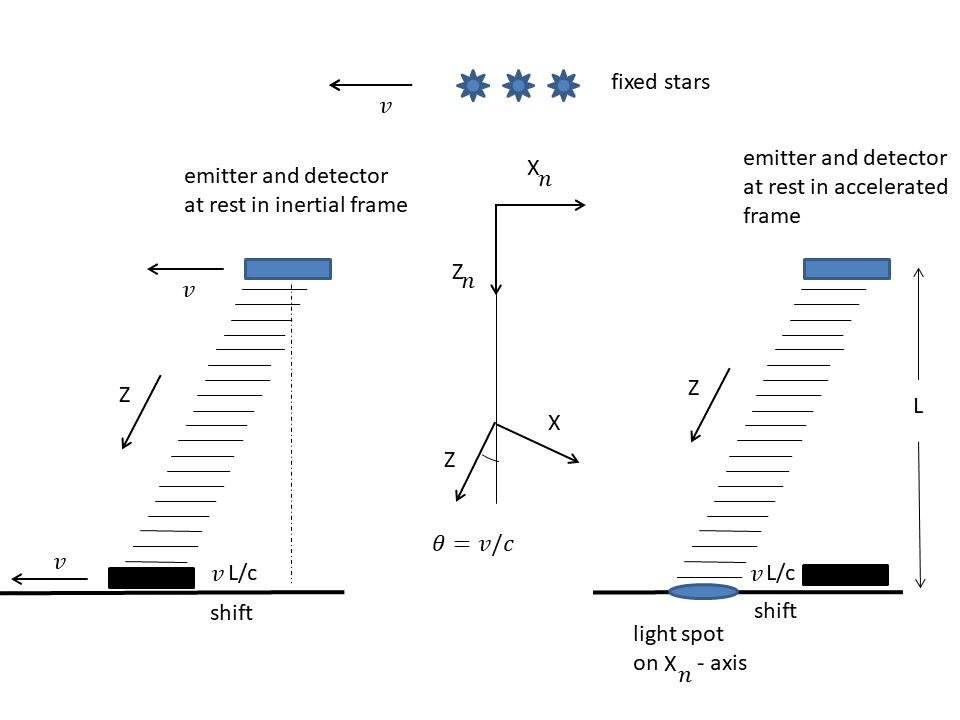}
	\caption{Aberration of light in an accelerated frame of reference. In an accelerated system, the accelerated detector does not receive light radiated by the accelerated emitter.}
	\label{B335}
\end{figure}

We have demonstrated that the aberration of light in an accelerated frame, as observed from an inertial frame, can be derived from the Lorentz transformations. In other words, the aberration effect in the accelerated frame, viewed from the initial inertial frame, is a kinematic consequence of special relativity (Fig. \ref{B335}).
It is shown that the relativity of simultaneity is responsible for aberrations to first order in $v/c$. Furthermore, we identified a fundamental asymmetry between inertial and accelerated frames: acceleration (relative to the inertial frame) influences the propagation of light in the accelerated frame. Specifically, in an accelerated system, an accelerated detector does not receive light emitted by an accelerated source (Fig. \ref{B335}). In contrast, a detector at rest in the inertial frame continues to receive light (Fig. \ref{B334}, left).

\begin{figure}
	\centering
	\includegraphics[width=0.85\textwidth]{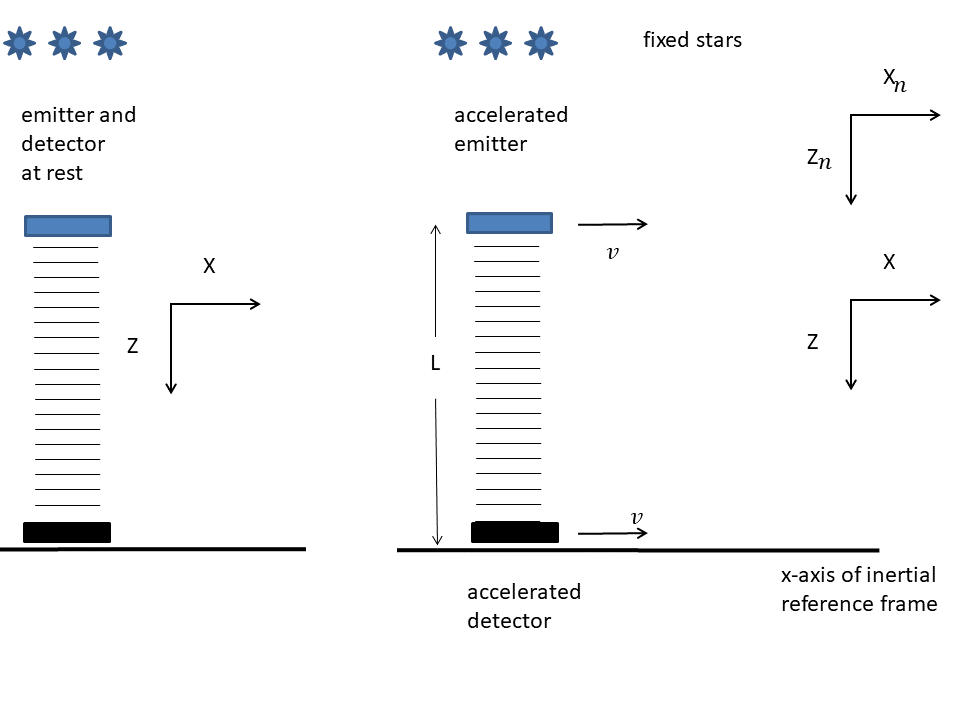}
	\caption{Aberration of light in an inertial frame of reference, as predicted by conventional theory.}
	\label{B336}
\end{figure}

\begin{figure}
	\centering
	\includegraphics[width=0.85\textwidth]{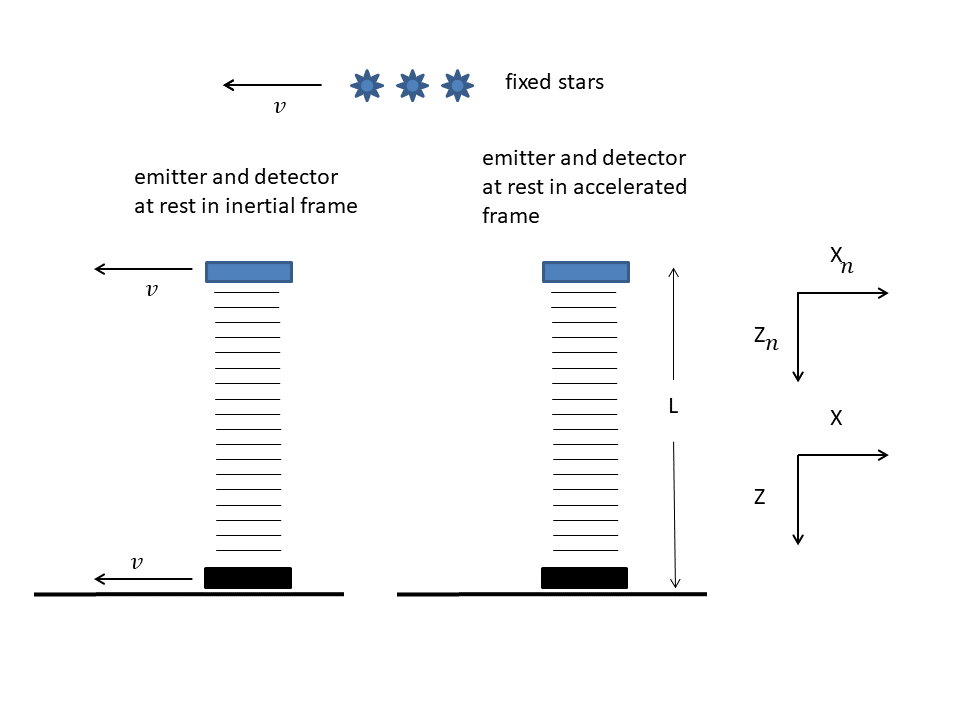}
	\caption{Aberration of light in an accelerated frame of reference, as predicted by conventional theory.  Acceleration does not disrupt the motional symmetry between the accelerated  and inertial reference frames.}
	\label{B337}
\end{figure}

Common textbook presentations of special relativity typically follow the (3+1) approach, which relies on the implicit assumption that distant clocks are synchronized according to the absolute simultaneity convention. According to these textbooks, when a reference frame at rest is set into motion, all points in the three-dimensional reference grid are assumed to move simultaneously. 
The standard derivation of the aberration of light effect is based on the hidden assumption that the 
$(x_x,y_n,z_n)$ axes of the accelerated frame remain parallel to the $(x,y,z)$ axes of the inertial frame. In other words, it is incorrectly presumed that both accelerated and inertial observers share a common three-dimensional space (see Fig. \ref{B336} - Fig. \ref{B337}). According to conventional theory, acceleration does not disrupt the motional symmetry between the accelerated and inertial reference frames. However, our findings (Fig. \ref{B334} - Fig. \ref{B335}) contradict the textbook predictions.
A critical flaw in the widely accepted derivation of the aberration of light effect is the omission of Wigner rotation, which, in our analysis, is fundamentally linked to the relativity of simultaneity. In the next chapter, we will explore the theory of Wigner rotation in detail.

\subsection{Discussion}

Let us return to the observations of an inertial observer.
According to the special theory of relativity, an orthogonal emitter-detector setup undergoes a shearing motion during acceleration, as illustrated in Fig. \ref{B334}. This effect is purely kinematic and does not involve any forces.

How can this be? The answer is not straightforward, but one way to approach the issue is as follows:
Our ordinary intuition often fails when dealing with two space-like separated events. The commonly accepted assumption of rigidity in an accelerated frame is based on the belief that the simultaneous acceleration of the reference space grid has a direct physical meaning. However, this is a misconception. When considering two distant events, we must account for the conventionality of distant simultaneity over a time interval of $l/c$, where $l$ is their spatial separation. The position of the accelerated emitter along the $x$-axis relative to the accelerated detector lacks an exact objective meaning. Due to the finite speed of light, no experimental method can precisely determine this position. Instead, there is an inherent uncertainty (blurring) in the relative position along the $x$-direction, quantified as $lv/c$, which arises from uncertainties in the timing of acceleration.

A relevant example involves a light clock—a rod with mirrors at each end. Textbooks typically describe how, from the perspective of an inertial observer, the light inside the orthogonal light clock follows a zigzag path. However, this explanation implicitly assumes that the $(x_n, y_n, z_n)$ coordinate axes of the moving observer remain parallel to the $(x, y, z)$ axes of the inertial observer. This assumption is incorrect. To properly analyze the behavior of light clocks, we must consider both the phase and initialization of the system. The term "phase" refers to the zigzag pattern of the light path, which cannot be directly measured in the lab frame. Moreover, the relative position of the accelerated bottom mirror with respect to the accelerated top mirror also lacks an exact objective meaning. Here, too, an uncertainty of order $lv/c$ arises, where $l$ is the rod length.
In the Chapter 14, we will continue our discussion of orthogonal light clock operation.

\newpage

\section{Kinematics of Wigner Rotation}

The focus of this chapter is the Wigner rotation. \footnote{It is well known that the composition of non-collinear Lorentz boosts does not result in a simple boost but rather in a Lorentz transformation that includes both a boost and a spatial rotation. This rotation is known as the Wigner rotation \cite{WI1, WI2}. It is sometimes referred to as the Thomas rotation (see, e.g., \cite{M, JACK}).}
We previously introduced the Wigner rotation in the last chapter, where our discussion primarily centered on the aberration of light. However, the Wigner rotation is also closely related to the aberration of particles.
As a relativistic kinematic effect, the Wigner rotation describes how the coordinate axes of a reference frame, moving along a curvilinear trajectory, rotate with respect to the axes of an inertial Lorentz frame.
The primary objective of this chapter is to derive an expression for this effect. Throughout, we have aimed to minimize mathematical complexity while maintaining clarity.

\subsection{Composition of Lorentz Boosts}

Let us now consider a relativistic particle accelerating in an initial inertial frame and analyze its evolution within Lorentz coordinate systems. Since the particle’s permanent rest frame is not inertial, we introduce an infinite sequence of comoving frames to address this difficulty. At each instant, the rest frame is a Lorentz frame centered on the particle, moving alongside it. As the particle’s velocity changes at an infinitesimally later instant, a new Lorentz frame, centered on the particle and moving at the updated velocity, is adopted.

Let us denote the three inertial frames by $K, K', K"$. 
The initial inertial frame is $K$, while $K'$ is the rest frame of the particle, which has a velocity $\vec{v} = \vec{v}(\tau)$ relative to $K$, and $K"$ represents the particle's rest frame at the next instant of proper time $\tau + d\tau$, moving with an infinitesimal velocity $d\vec{v'}$ relative to $K'$. 
All three inertial frames are assumed to be Lorentz frames. 
To clarify velocity measurements in different inertial frames: an observer in $K$ measures the velocity of frame $K'$ as $\vec{v}$, while an observer in $K'$ measures the velocity of frame $K"$ as $d\vec{v'}$.

Now, let us examine the transformation of velocity in the theory of relativity. For a rectilinear motion along  the $z$ axis, the velocity transformation follows the equation:
$v_z(\tau + d\tau) = (dv'_z+v_z)/(1 + v_zdv'_z/c^2)$. 
Just as with Galilean transformations, collinear Lorentz boosts commute—the order in which collinear boosts are applied does not affect the final result. This means that the composition of two Lorentz boosts in the same direction yields another Lorentz boost in that direction, with the total velocity given by the formula above.

However, when Lorentz boosts are applied in non-collinear directions, they no longer commute. While the composition of two Galilean boosts or two spatial rotations yields another transformation of the same kind (i.e., a boost or a rotation), the combination of two non-parallel Lorentz boosts results in a transformation that is not purely a boost—it includes a rotation known as the Wigner rotation.

To illustrate this, suppose frame $K'$ moves with velocity $v_z$ along the $z$-axis relative to $K$ and
in $K'$, the particle experiences an infinitesimal acceleration in the $x'$-direction,  perpendicular to the trajectory of the initial inertial frame $K$. That is, frame $K"$ moves relative to $K'$ with a small velocity $dv'_x$ along the $x'$-axis.

From the perspective of an inertial observer, time dilation occurs in the Lorentz frame moving with velocity
$\vec{v}_z$ relative to the initial inertial frame, given by $dt/\gamma = d\tau$, where $\gamma = 1/\sqrt{1 - v_z^2/c^2}$.
Due to this time dilation, the velocity increment in the initial inertial frame $dv_x$ corresponds to a velocity $dv'_x = \gamma dv_x$  in $K'$.

\subsection{Wigner Rotation}

Numerous incorrect expressions for the Wigner rotation appear in the literature, underscoring the need for a careful and transparent treatment of this topic. Rather than deriving all transformation matrices for four-vector components, we adopt a geometric perspective that offers clear and intuitive insight into the phenomenon.

The mathematics involved is notably straightforward, relying only on basic algebra. The key challenge lies in bridging the conceptual gap that arises from the fact that different Lorentz frames cannot be described within a shared three-dimensional space.

\begin{figure}
	\centering
	\includegraphics[width=0.9\textwidth]{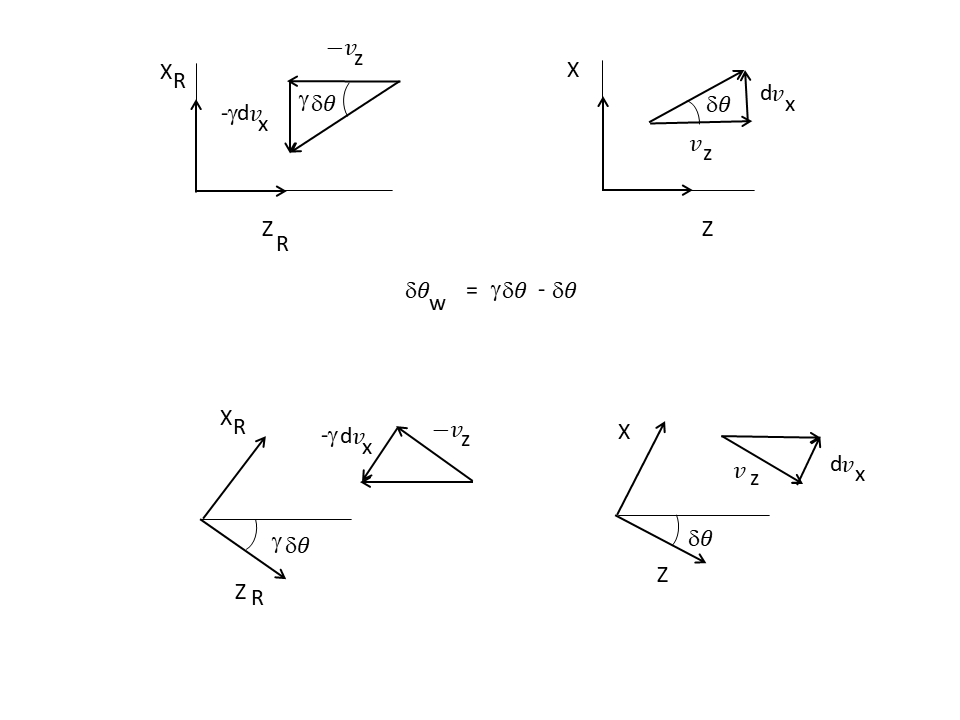}
	\caption{Illustration of the Wigner rotation interpreted with respect to the axes of the proper frame.}	
	\label{B78}
\end{figure}

Consider a succession of inertial frames $K \to K' \to K"$. From the perspective of the initial inertial frame, an observer in the frame $K"$ sees frame $K$ moving with velocity 
$-v_z$ and velocity increment $ - \gamma d\vec{v}_x$. The corresponding rotation of the velocity direction in the $K"$ frame is given by $d\theta_{K"} = \gamma dv_x/v_z$ (Fig. \ref{B78}). In the $K$ frame, the velocity rotation angle is $dv_x/v_z = d\theta$, with both rotations occurring in the same direction. 
If the proper and lab observers were to share a common three-dimensional space, this discrepancy would lead to a paradox — an apparent inconsistency in the theory. However, in Minkowski spacetime, no such paradox arises: the lab’s 3-space is rotated relative to the proper 3-space by an angle 

\[
d\Phi_{K"} = \gamma dv_x/v_z - dv_x/v_z = (\gamma - 1)\d\theta
\]

in the same direction as the velocity vector's rotation in the proper frame.

This angular difference  

\[
d\Phi_{K"}  =  (\gamma - 1)\d\theta  
\]

is the Wigner rotation — the rotation of the original frame’s axes as seen from the proper frame $K"$. In vector form, this can be expressed as 

\[
d\vec{\Phi}_{K"} = (\gamma - 1)\vec{v}\times d\vec{v}/v^2   . 
\]

Several key insights can be drawn regarding the theory of Wigner rotation. Notably, the expression for the Wigner rotation angle in the proper frame, $d\vec{\Phi}_{K"}$ can be reformulated as 

\[
d\vec{\Phi}_{K"} = (1-1/\gamma)\vec{v}_{K"}\times d\vec{v}_{K"}/v_{K"}^2   , 
\]

where $\vec{v}_{K"} =  - \vec{v}$ is the velocity of the original frame relative to the proper frame, and $d\vec{v}_{K"} = - \gamma  d\vec{v}$ is its differential change. This alternative form emphasizes the expression of the Wigner rotation in terms of quantities measured in the proper frame.

Transforming this expression back to the lab frame $K$ yields

\[
d\vec{\Phi}_{K} = (1-1/\gamma)\vec{v}\times d\vec{v}/v^2
\]

representing the Wigner rotation angle of the proper frame’s axes as observed in the lab frame. According to the principles of special relativity, this quantity must remain form-invariant under Lorentz transformations. \footnote{The correct expression for the Thomas (Wigner) rotation was first obtained by V. Ritus \cite{R2}. 
In deriving expressions for the Thomas (Wigner) rotation, the majority of authors (see e.g. \cite{JACK}) was supposedly guided by the incorrect expression for Thomas (Wigner) rotation from  Moeller's monograph \cite{M}. The expression obtained by Moeller is given by  
$\vec{\delta \Phi} =  (1 - \gamma) \vec{v}\times d\vec{v}/v^2  =  (1 - \gamma) \vec{\delta \theta}$ 
(and subsequently expression for Thomas precession   $\Omega_\mathrm{T} =  (1 - \gamma)\omega_0 $). It should be noted that, in his monograph, Moeller stated several times that this expression is valid in the lab Lorentz frame. Clearly, this expression and correct result differ both in sign and magnitude. An analysis of why Moeller obtained an incorrect expression for the Thomas (Wigner) rotation in the lab frame is the focus of Ritus paper \cite{Rit}. As analyzed in Ritus's work \cite{Rit}, Moeller's error is not computational but rather conceptual.
The correct expression, derived decades earlier by several researchers, remained largely unnoticed amid the proliferation of inaccurate results, as discussed in the review \cite{MAL2}.} We note that owing to the relativistic effect of time dilation in the reference frame that moves to the lab frame, the Wigner rotation angle in the proper frame is always $\gamma$ time higher than in the lab frame. \footnote{In 1986, M. Stranberg obtained an expression for the Thomas (Wigner) rotation correct both in the lab inertial frame and the comoving  reference frame  \cite{STRAN}. His paper is one of the few that explicitly states that the angle of Thomas (Wigner) rotation in the comoving frame is $\gamma$ times higher than in the lab frame \cite{MAL2}.}

We derived the exact relation  

\[
d\vec{\Phi} = (1-1/\gamma)\vec{v}\times d\vec{v}/v^2
\]

using only rudimentary knowledge of special relativity.  
In standard textbooks on relativity, the spatial rotation that arises from the composition of two non-parallel Lorentz boosts is typically introduced through an algebraic approach. This reliance on algebra, without sufficient geometrical interpretation, is one reason why many authors arrive at an incorrect expression for the Wigner rotation. These treatments often describe the rotation of a moving object without addressing its geometrical significance, leading to serious difficulties in both the interpretation of the computations and the physical meaning of the results.

\subsection{Measuring Wigner Rotation}

The Wigner rotation is a relativistic kinematic effect that arises from the geometry of pseudo-Euclidean space-time. It manifests as a rotation of the spatial coordinate axes of a particle’s proper frame when the particle follows a curvilinear trajectory in the Lorentz lab frame. In this context, we derive the expression for the infinitesimal rotation resulting from the acceleration of a relativistic particle:

\begin{eqnarray}
&& d\vec{\Phi} = (1-1/\gamma)\vec{v}\times d\vec{v}/v^2 =	\left(1 - \frac{1}{\gamma} \right) \vec{\delta \theta}   ~ .\label{WG1}
\end{eqnarray}

Here $d\vec{v}$ represents the infinitesimal change in velocity due to acceleration, 
$d\Phi$ is the Wigner rotation angle describing the rotation of the particle’s proper frame relative to the Lorentz lab frame, and $d\theta$ is the infinitesimal orbital angle. 
A natural question arises: how can this rotation be observed or measured? Since a moving coordinate system changes its position over time, determining its orientation becomes nontrivial. To explore the physical meaning and possible experimental interpretation of this rotation, we illustrate the problem of defining the orientation of a moving frame using a simple example. The Wigner rotation, when interpreted in the context of the original inertial frame, is intimately related to length contraction. \footnote{The length of a moving object depends not only on the structure of space-time but also on our choice of synchronization convention. As with the orientation of radiation phase fronts, such quantities do not possess an exact objective meaning. Under the standard clock synchronization, they are typically regarded as "relativistic kinematic effects." The Wigner rotation similarly arises as a coordinate effect, lacking an absolute, observer-independent significance.}

\begin{figure}
	\centering
	\includegraphics[width=0.9\textwidth]{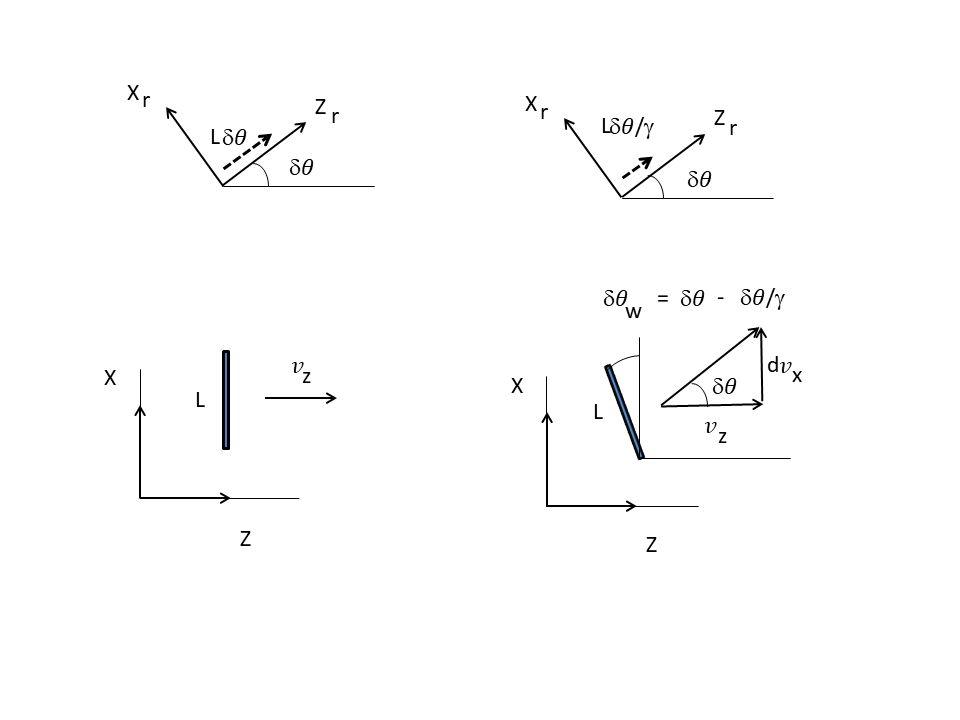}
	\caption{Illustration of the Wigner rotation interpreted with respect to the axes of the lab frame.}
	\label{B77}
\end{figure}

Suppose that, after performing a Lorentz boost along the $z$-axis with velocity $v_z$,
an observer in the lab frame rotates the coordinate system by an angle $\vec{\delta \theta} = \vec{v_z}\times d\vec{v_x}/v_z^2$,  such that the new $x_r$-axis is orthogonal to the vector $\vec{v_z} + d\vec{v_x}$, 
and the $z_r$-axis is aligned with it.
Now, consider a rod oriented along the $x$-axis in the comoving frame. The motion takes place in the $(x,z)$-plane, and 
the rod is initially perpendicular to the velocity $\vec{v_z}$. After rotating the lab frame axes, the projection of the 
rod onto the $z_r$-axis becomes simply
$l\delta\theta$, where $l$ is the length of the rod in the comoving frame, and also in the lab frame following the first boost along the $z$-axis. After a second infinitesimal Lorentz boost with velocity $d\vec{v_x}$,
this projection undergoes Lorentz contraction and becomes $l\delta\theta/\gamma$ (Fig. \ref{B77}). Assume now that the lab observer fixes the orientation of the comoving frame axes. In the ultrarelativistic limit $\gamma \to \infty$, the projection of the rod onto the $z_r$-axis vanishes, and thus  the comoving frame axes become parallel to the rotated lab frame axes $(x_r,z_r)$.
However, for an arbitrary (finite) velocity, the comoving frame axes are not aligned with the rotated lab frame axes. Due to the contracted projection, the angular deviation is $-\vec{\delta \theta}/\gamma$.
Consequently, one can directly verify that the comoving frame axes are rotated relative to the original lab frame axes  $(x,z)$ by an angle $\delta\theta - \delta\theta/\gamma$, which is precisely the Wigner rotation angle, as given by Eq.(\ref{WG1}). This example illustrates how the Wigner rotation arises naturally from the composition of Lorentz transformations.

\subsection{Discussion}

Numerous incorrect expressions for the Wigner (Thomas) rotation can be found in the literature. Here, we examine the underlying reasons behind these errors.
In 1959, Bargmann, Michel, and Telegdi (BMT) formulated a consistent relativistic theory for spin dynamics as observed in the laboratory frame, which was later confirmed experimentally \cite{BMT}.
A common misconception is that the BMT equation inherently includes the standard (but incorrect) expression for the Wigner rotation in the Lorentz lab frame. Many physicists, having learned about Wigner (Thomas) rotation from well-known textbooks, argue:
"The highly precise measurements of the electron’s magnetic-moment anomaly, conducted on relativistic electrons, rely on the BMT equation—where the Wigner rotation is an integral part—thus serving as an experimental confirmation of the standard expression for the Wigner rotation."
This belief, however, is both widespread and misleading. It is important to clarify that the results in the Bargmann-Michel-Telegdi paper were derived using a semiclassical approximation of the Dirac equation. The Wigner rotation was not explicitly considered, as the Dirac equation allows for a complete description of spin dynamics without separating it into Larmor and Wigner components.

It should be noted that the application of Wigner (Thomas) rotation theory in relativistic spin dynamics is complex. Bargmann, Michel, and Telegdi began their analysis in the particle's rest frame, deriving the equation of motion for angular momentum. They then generalized it to the Lorentz lab frame before transforming it back to the rest frame. This raises an interesting question: Why is it convenient to transform this equation to the rest frame at that instant?
In experimental practice, working with the proper spin is preferred due to the clear physical interpretation of the 
three-dimensional spin pseudo-vector $\vec{s}$. Unlike the three-dimensional momentum $\vec{p}$, which has well-defined 
components in every reference frame, angular momentum is specific to a single frame and does not transform. 
Thus, any meaningful statement about spin implicitly refers to the particle’s instantaneous rest frame. For example, when we say that a particle’s spin in the lab frame makes an angle $\phi$ with its velocity, we actually mean that in the particle’s rest frame, the spin vector forms the same angle with the direction of the lab frame’s motion.

Physical phenomena in non-inertial frames can still be analyzed within the framework of standard special relativity by applying well-established relativistic kinematic effects. In the context of spin dynamics, this eliminates the need for measurements using non-inertial devices.

The conventional approach to explaining spin dynamics in the lab frame is quite unusual. As previously noted, the laws of physics in the lab frame account for all physical phenomena, including observations made by non-inertial observers. In particular, spin orientation measurements with respect to the lab frame axes can be determined within the proper frame. 
The BMT equation uses a spin quantity defined in the proper frame but observed with respect to the lab frame axes.
That means that we know the orientation of the proper spin with respect to the lab coordinate system which is moving with velocity $-\vec{v}$ and acceleration $-\gamma d\vec{v}/d\tau$ in the proper frame.

Thus, the spin rotation measurements by a polarimeter in the lab frame is interpreted from the viewpoint of the proper observer as viewing of the lab observer.
This inherent complexity has led textbook authors to derive incorrect expressions for Wigner rotation—an issue that will be discussed further in Chapter 19.

Additionally, it is crucial to examine why errors in Wigner rotation theory went undetected for so long. Wigner (or Thomas) rotation is typically presented within the context of spin kinematics as a peculiar effect of special relativity. This limited perspective likely explains why, until recently, researchers failed to recognize the significant discrepancies between different works.

The Wigner rotation is a fundamental relativistic phenomenon, on par with time dilation and length contraction. Special relativity demonstrates that the aberration effect in the rotating Earth-based frame, as observed from the inertial Sun-based frame (in Lorentz coordinatization), is a kinematic consequence of the theory. In this context, the shift in aberration images within the Earth-based frame, when viewed from the Sun-based frame, is governed by the Wigner rotation.
According to Wigner rotation theory, observers following different trajectories experience different 3-spaces. However, previous literature has incorrectly assumed that an Earth-based observer and an inertial (e.g., Sun-based) observer share a common 3-space—an assumption that has long been taken for granted.

\newpage

\section{Aberration of Relativistic Particles}

We now examine the phenomenon of aberration for relativistic particles.

Consider a particle source initially at rest in an inertial frame, emitting particles that travel with a constant velocity  $V$ along the $z$-axis. Suppose both the observer (along with their instruments) and the particle source are uniformly accelerated from rest to a velocity $v$ along the $x$-axis, under the condition that  $v \ll V$. Our goal is to understand how the aberration of the particle beam is perceived by the observer in this accelerated frame.

To approach this question, we begin by summarizing the key result.

The central question is: How should the aberration of a particle beam be described when the source is at rest in the accelerated frame? To predict the outcome of such a measurement, the observer must refer to the Langevin metric, given in Eq. (\ref{GGG3}). Within this framework, the aberration angle $\theta_a$ is related to the system parameters by the expression:  

\[
\theta_a  = - (1 - 1/\gamma)v/V  , 
\]

where  $\gamma = 1/\sqrt{1 - V^2/c^2}$. is the Lorentz factor of the particles. This relativistic expression smoothly reduces to the classical result  $\theta_a  = 0$ in the limit $V^2/c^2 \to 0$.

It is crucial to note that a correct description of aberration in the accelerated frame requires incorporating the full metric tensor, even to first order. This is because the cross term in the Langevin metric—representing the leading-order deviation from the Minkowski metric—plays an essential role in capturing the non-inertial kinematics of relativistic particle motion.

In the ultrarelativistic limit  $V/c \to 1$, the aberration increment approaches the simple and intuitive form: $\theta_a = - v/c$.

\subsection{Explanation of Electron Aberration Based on Electrodynamics}

The aberration of particle trajectories can be effectively addressed using dynamical arguments. Electromagnetic forces, which determine the behavior of an emitted electron beam, are influenced by the acceleration of the source in such a way that they cause a deviation in the direction of electron transport. To analyze these effects, we can derive the electromagnetic fields in an accelerated frame through a Galilean transformation.

For readers already familiar with special relativity from standard textbooks, it may be advantageous to approach the aberration of particles—both theoretical and experimental—using a microscopic perspective. From this standpoint, one can calculate relativistic quantities directly from the fundamental theories of matter, without explicitly invoking relativistic kinematics. The aberration of particles associated with the transformation from an inertial to an accelerated frame can be understood as the result of a force acting on the system. Importantly, this issue is unrelated to the concept of the ether and is fully contained within the framework of special relativity.
Let us consider a specific case in which an electron gun, serving as a source of relativistic electrons, is initially at 
rest in an inertial frame and then accelerated to a velocity $v$ along the $x$-axis. The electromagnetic forces 
governing the properties of the emitted electron beam are modified by this acceleration, resulting in a change in the 
beam's transport direction. The underlying mechanism behind the change in electron momentum 
$\vec{p}$, when both the source and the observer (along with their measuring instruments) are at rest in the accelerated frame, can be fully explained within the framework of electrodynamics.

We can derive the electrodynamics equations in an accelerated reference frame by applying a Galilean transformation. When analyzing Maxwell's equations under such a transformation, a key question arises: What are the transformation laws for the electromagnetic fields $\vec{E}$ and $\vec{B}$?

The electric and magnetic fields $\vec{E}_n$ and $\vec{B}_n$ observed in the accelerated frame—where the particle source is at rest—differ from the fields $\vec{E}$ and $\vec{B}$ observed in the original inertial frame, prior to the Galilean boost. Our goal is to determine the relationship between these sets of fields.

To proceed, we transform the coordinates $(t, x, y, z)$ of the inertial observer $S$, moving with velocity $-v$ relative to the accelerated observer $S_n$, using the Galilean transformation:  $x_n =x-vt$, $t_n = t$. Substituting these into the Minkowski metric, $ds^2 = c^2 dt^2 - d x^2 - dy^2 - dz^2$, we obtain the so-called Langevin metric in the accelerated frame Eq. (\ref{GGG3}):  $ds^2 = c^2(1-v^2/c^2)dt_n^2 - 2vdx_ndt_n - dx_n^2 - dy_n^2 -dz_n^2$. This transformed metric allows us to study electrodynamics equations in a non-inertial frame using the Galilean transformation with velocity $-v$. In the specific case where $\vec{B} = 0$, the transformed fields are:  

\[
\vec{E}_n = \vec{E},   \qquad   \vec{B}_n = - \vec{v}\times\vec{E}/c . 
\]
Here, $\vec{v}$ is the velocity of the source in the original inertial frame. This result captures the leading-order relativistic correction to the magnetic field observed in the accelerated frame.

\begin{figure}
	\centering
	\includegraphics[width=0.7\textwidth]{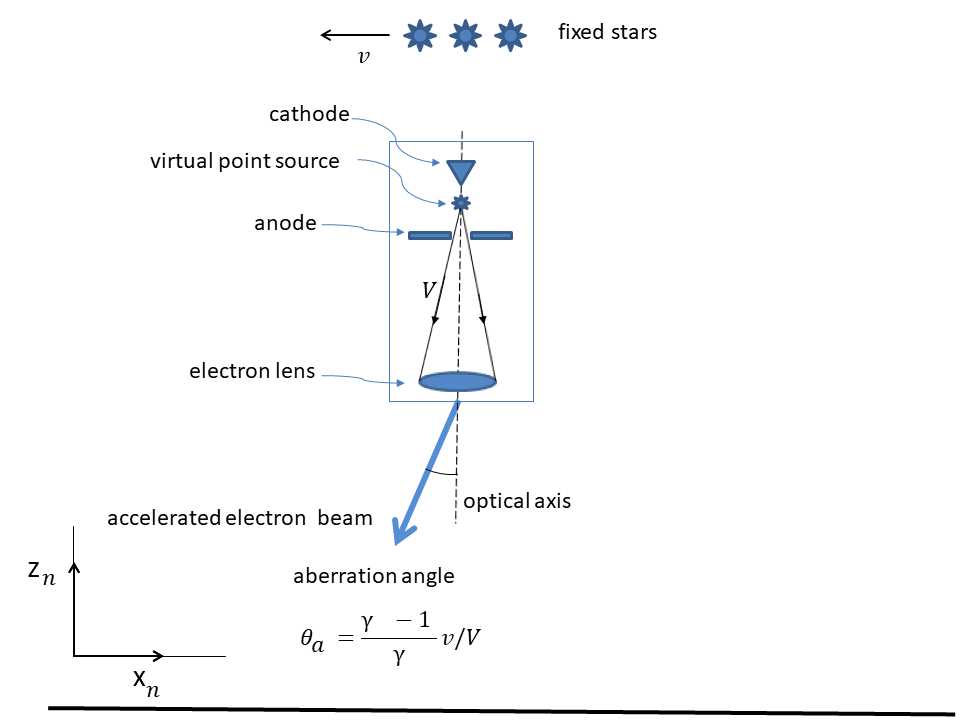}
	\caption{Electron aberration observed when both the electron source and the observer are at rest in an accelerated frame. }
	\label{B558}
\end{figure}

The accelerated observer experimentally determines the Lorentz force $\vec{F}_n$ acting on a charge $e$ moving with velocity $\vec{V}$ in the region of the electron gun, where electric and magnetic fields $\vec{E}_n$ and $\vec{B}_n$ are present. The force is given by: 

\[
\vec{F}_n =  e\vec{E} - e\vec{V}\times\vec{v}\times\vec{E}/c^2 . 
\]

In the special case where $\vec{v} \cdot \vec{V} = 0$ and $\vec{E} \cdot \vec{V} = EV$, the magnetic term simplifies to: 

\[
 - e\vec{V}\times\vec{v}\times\vec{E}/c^2 = -\vec{v}(e\vec{V}\cdot\vec{E}/c^2) . 
\]

Using the relativistically correct equation of motion, $\vec{F}_n = d\vec{p}/dt$, we expect that after a short time interval $\d t$, the emitted electron will acquire a transverse momentum in the accelerated frame given by: 

\[
\d\vec{p}_{\perp} = -\vec{v}(e\vec{V}\cdot\vec{E}/c^2)\d t. 
\]

The differential change in kinetic energy of the accelerated electrons is:
$\d T = \vec{F}\cdot\vec{V}\d t = \vec{F}\cdot \d\vec{s}$. Upon integration, this yields:

\[
T = e\int \vec{E}\cdot \d\vec{s} = mc^2/\sqrt{1 - V^2/c^2} - mc^2. 
\]

Given our condition $\vec{v} \cdot \vec{V} = 0$, the aberration angle increment becomes:  $\theta_a = p_{\perp}/p$, where the total momentum of the accelerated electron is: $p = mV/\sqrt{1 - V^2/c^2}$. Substituting, we find:  $\theta_a = - (1 - 1/\gamma)v/V$. Expressed in vector form, the aberration increment is:

\[
\vec{\theta}_a = - (1 - 1/\gamma)\vec{V}\times\vec{v}/V^2. 
\]

The aberration of particles resulting from the transformation between an inertial and an accelerated coordinate system can be interpreted as a manifestation of the Lorentz force. The electromagnetic forces that govern the behavior of an emitted electron beam are influenced by acceleration relative to the fixed stars, leading to a deviation in the beam’s transport direction, as illustrated in Fig. \ref{B558}.

\subsection{Magnetic Field Measurements in an Accelerated Electron Source}

We now turn to an apparent paradox.
Special relativity introduces concepts that may seem counterintuitive at first. Yet it remains a rigorously mathematical theory, and it leads to a surprising conclusion: when we consider an accelerated electron source, the electromagnetic fields in the accelerated frame are not static—the problem becomes time-dependent. This is a direct implication of special relativity.

We derived the electrodynamics equations in an accelerated frame by applying the inverse Galilean transformation (with velocity $-v$) of Maxwell's equations. Under a Galilean transformation, and assuming $\vec{B} = 0$ in the original frame,
the transformed fields become $\vec{E}_n = \vec{E}$, $\vec{B}_n = - \vec{v}\times\vec{E}/c$. The electric field $\vec{E}_n$ and the induced magnetic field  $\vec{B}_n$ are both static, meaning they do not vary with time—the system is entirely time-independent. Many physicists familiar with standard electrodynamics might argue, based on textbook knowledge, that a static magnetic field must arise from electric currents, which in turn require moving charges. Since no charges are in motion in this case, the presence of a magnetic field appears paradoxical.
This represents a fundamentally new kind of situation, distinct from traditional electrodynamics in an initial inertial frame.

How can that be?
Even when the electron source and the observer are at rest in the accelerated frame, the time derivative $\partial/\partial t_n$ is generally nonzero. This can be readily demonstrated by transforming from the coordinates $(t, x, y, z)$ of an inertial observer $S$—moving with velocity $-v$ relative to the non-inertial observer $S_n$—to the coordinates $(t_n, x_n, y_n, z_n)$ of $S_n$. Using a Galilean transformation, the inertial coordinates can be related to the non-inertial ones as follows: $x_n = x - vt, ~ y_n = y, ~ z_n = z, ~ t_n = t $. This allows all fields to be re-expressed in terms of the variables $x_n, y_n, z_n, t_n$. Taking partial derivatives accordingly, we find:

\[
\partial/{\partial t_n} = \partial/{\partial t} + v\partial/{\partial x}   ,    \qquad   \partial/{\partial x_n} = \partial/{\partial x}   . 
\]

Therefore, if the fields are time-independent in the inertial frame (i.e., $\partial/\partial t = 0$), it follows that: $\partial/\partial t_n = v\partial/\partial x_n$. Thus, $\partial/\partial t_n$ is not necessarily zero, even though the source and observer are at rest in the accelerated frame.

Specifically, if both the electron source and the observer are at rest in the accelerated frame, one might intuitively expect that everything should remain unchanged over time—that is, the electric field should not vary. However, special relativity tells us otherwise. The time derivative of the electric field, $\partial \vec{E} / \partial t_n$, is nonzero because of the spatial variation of the field, $\partial \vec{E} / \partial x_n \neq 0$, and the relation $\partial \vec{E}/\partial t_n = v\partial \vec{E}/\partial x_n$. This seemingly paradoxical result is simply a manifestation of the four-dimensional formalism of relativity. It carries no physical contradiction or special consequence—just a shift in perspective due to the geometry of spacetime.

To clarify this, it may help to draw an analogy with a familiar concept in classical electromagnetism: the use of gauge transformations. For example, when we work in the Coulomb gauge, the scalar potential appears to propagate instantaneously. However, this is merely a mathematical artifact of the gauge choice. In a complete and self-consistent calculation, all unphysical instantaneous effects cancel out, preserving causality and agreement with physical observation.

Now let us return to the clock resynchronization procedure and try to get a better understanding what happens if we employ synchronization scheme allows us to introduce Lorentz coordinates associated with the accelerated setup, Fig. \ref{B558}.

A natural starting point is to assume that the time derivative $\partial/\partial t_n$ originates from a Galilean boost. To generalize, we now consider the case of Lorentzian coordinatization.
When the reference frame $S_n$ begins moving at a constant velocity, Einstein’s standard procedure for clock synchronization can be applied. In this context, the time coordinate $t^{(L)}_n$ in the frame $S_n$—as defined by Einstein synchronization—is obtained by introducing an offset term $x_n v / c^2$ and applying the first-order approximation: $t^{(L)}_n = t_n - x_nv/c^2$. This transformation allows us to reinterpret the new time coordinate in the accelerated frame in such a way that Maxwell’s equations remain valid, particularly in describing the aberration of electrons. Taking partial derivatives with respect to the transformed coordinates yields:

\[
\partial/{\partial t^{(L)}_n} = v\partial/{\partial x_n}   ,  \qquad   \partial/{\partial x^{(L)}_n} = \partial/{\partial x_n} + (v/c)\partial/{\partial t_n} . 
\]

Hence, to first order in $v/c$, we find: $\partial/\partial t^{(L)}_n = v\partial/\partial x^{(L)}_n$ neglecting terms of order $v^2/c^2$. It is important to note that changing the (four-dimensional) coordinate system does not give rise to new physical phenomena. The underlying physics remains invariant under such transformations.

When the situation is described as we have done here, the paradox seems to disappear entirely; it becomes quite natural that, in general, space and time can no longer be treated as separate entities.
The argument for a paradox typically goes like this: If we could measure the magnetic field $\vec{B}_n$ using an ordinary magnetometer, then a paradox would indeed arise. However, such a measurement is not possible. As discussed earlier, the magnetic field $\vec{B}_n$ induces a change in the transverse momentum of test particles, described by

\[
\theta_a = p_{\perp}/p = (1 - 1/\gamma)v/V. 
\]

For non-relativistic particle velocities, applying a binomial expansion 
yields  

\[
\vec{\theta}_a = - [V^2/(2c^2)] \vec{V}\times \vec{v}/V^2    , 
\]

indicating that the momentum perturbation is a 
relativistic effect. According to special relativity, as $V^2/c^2 \to 0$, the change in particle momentum vanishes. 
Thus, a classical magnetometer using test particles would detect no trace of the magnetic field. Moreover, it is not 
just that this particular classical magnetometer cannot detect $\vec{B}_n$;
if the theory of relativity holds, any classical magnetometer—regardless of its operating principle—would similarly fail to reveal any evidence of the field's effect.

To understand magnetic field measurements in an accelerated electron source, we begin by examining how a classical magnetometer behaves when observed from an accelerated frame. Since a full treatment is complex, we consider a simplified, conventional type of magnetometer. One practical approach to observing nuclear magnetism is through the phenomenon of nuclear magnetic resonance (NMR), where a proton resonance apparatus can serve effectively as a proton resonance magnetometer. We present a general analysis of this classical type of magnetometer, independent of its experimental precision. In the initial inertial  frame, a current loop possessing a magnetic moment $\vec{m}$ acquires an electric dipole moment when it moves with velocity $\vec{v}$. This dipole moment is given by:

\[
\vec{d} = \vec{v}\times\vec{m}/c. 
\]

Since the magnetic properties of materials arise from atomic-scale current loops, a magnetized body undergoing acceleration relative to the inertial frame should exhibit an induced electric polarization in that frame. As a result, the potential energy of a moving particle with intrinsic magnetic moment $\vec{m}$, in the presence of a magnetic field $\vec{B}$ and electric field $\vec{E}$, is given by:  

\[
U_p = - \vec{m}\cdot\vec{B} - \vec{d}\cdot\vec{E}   .  
\]

To analyze this in an accelerated frame, we apply the Galilean transformation (with velocity $-\vec{v}$) to Maxwell’s equations. Under this transformation, assuming an initial zero electric dipole moment ($\vec{d} = 0$), the magnetic and electric dipole moments in the accelerated frame become: 

\[
\vec{m}_n = \vec{m},          \qquad           \vec{d}_n  = - \vec{v}\times\vec{m}/c.  
\]

From vector algebra, we use the identity 

\[
(\vec{a} \times \vec{b}) \cdot \vec{c} = \vec{a} \cdot (\vec{b} \times \vec{c})
\]

to evaluate terms in the potential energy expression: 

\[
(-\vec{v}\times\vec{E})\cdot\vec{m} = -\vec{v}\cdot(\vec{E}\times\vec{m}),  \qquad (-\vec{v}\times\vec{m})\cdot\vec{E} = -\vec{v}\cdot(\vec{m}\times\vec{E}). 
\]
Thus, the additional contribution to the potential energy evaluates to zero, as expected.
Consequently, a conventional magnetometer cannot detect the magnetic field $\vec{B}_n$ in the accelerated frame.

\subsection{Aberration of Particles from a Moving Source}

\begin{figure}
	\centering
	\includegraphics[width=0.8\textwidth]{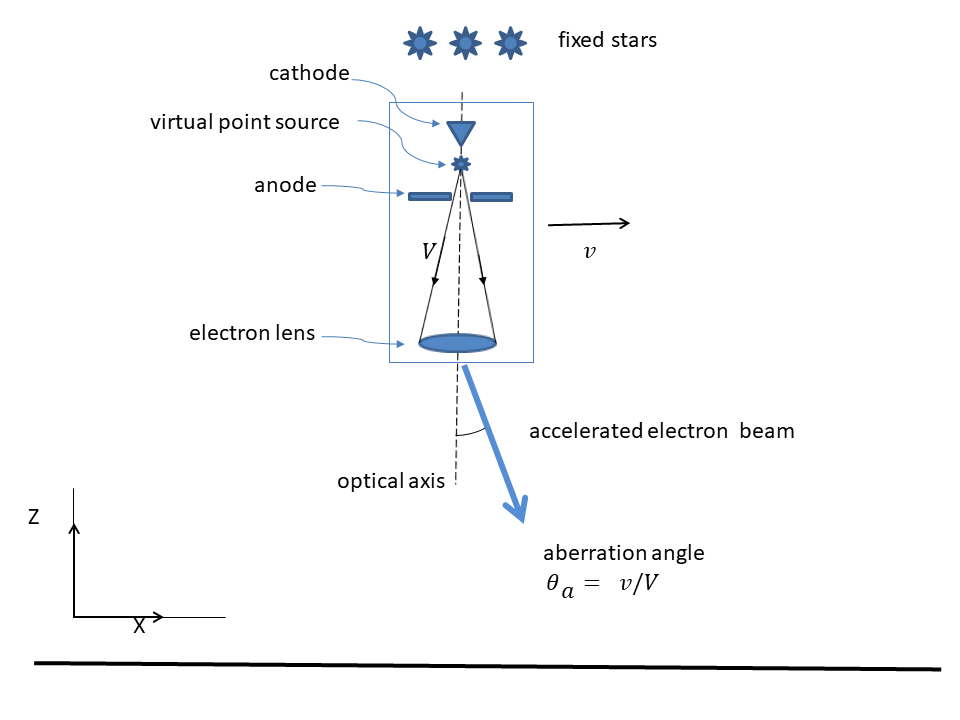}
	\caption{Electron aberration from a moving source in an inertial frame. The electron gun moves perpendicular to the optical axis.}
	\label{B557}
\end{figure}

Let us examine what occurs when the electron gun—serving as the source of a relativistic electron beam—is accelerated from rest to a velocity $v$ along the $x$-axis in the initial inertial frame (see Fig. \ref{B557}). In other words, we consider an active (physical) Lorentz boost.
In an active boost, we are observing the evolution of the same physical system over time, analyzed from the perspective of a single reference frame. The simplest method of clock synchronization in this context is to retain the same set of uniformly synchronized clocks used when the particle source was at rest. In other words, we continue to use the synchronization method based on clock transport or Einstein synchronization, which relies on light signals emitted by a stationary dipole source.
This approach is typically the most convenient, as it allows for a smooth and consistent definition of the metric tensor. It also preserves simultaneity and follows the convention of absolute time—or absolute simultaneity.

We can determine the electric field $\vec{E}$ and magnetic field $\vec{B}$ observed in the inertial frame where the source particle is moving with velocity $\vec{v}$ by applying the Galilean transformation. When analyzing the equations of electrodynamics under such a transformation, a natural question arises: What are the transformation laws for the electromagnetic fields?
The equivalence between the active and passive viewpoints stems from the fact that moving a physical system in one direction is equivalent to moving the coordinate system in the opposite direction by the same amount. Consequently, the fields in the inertial frame before an active Galilean boost are identical to the fields $\vec{E}'$ and $\vec{B}'$ in the comoving coordinate system.
Under a Galilean transformation, and assuming $\vec{B}' = 0$ in the comoving coordinate system, the transformed fields in the lab frame are given by:

\[
\vec{E} = \vec{E'},    \qquad      \vec{B} =  \vec{v}\times\vec{E'}/c. 
\]

This relation captures only first-order effects in $v/c$, where $\vec{v}$ is the velocity of the source relative to the initial inertial frame.

An inertial observer experimentally measures the Lorentz force $\vec{F}$ acting on a charge $e$ moving with velocity $\vec{V}$ in the region of an electron gun, where both electric and magnetic fields, $\vec{E}$ and $\vec{B}$, are present. The force is given by: $\vec{F} =  e\vec{E'} + e\vec{V}\times\vec{v}\times\vec{E'}/c^2$, where $\vec{E}'$ is the electric field in the coordinate system moving with velocity $\vec{v}$ relative to the inertial frame. In the special case where $\vec{v} \cdot \vec{V} = 0$ and $\vec{E}' \cdot \vec{V} = E'V$, the magnetic component of the force simplifies to:
$  e\vec{V}\times\vec{v}\times\vec{E'}/c^2 = \vec{v}(e\vec{V}\cdot\vec{E'}/c^2)$. The differential change in the kinetic energy of the accelerated electrons is: $dT = \vec{F}\cdot\vec{V}dt = \vec{F}\cdot d\vec{s}$. Integrating this expression yields the total kinetic energy: $T = e\int \vec{E'}\cdot d\vec{s} = mc^2/\sqrt{1 - V^2/c^2} - mc^2$. The total transverse momentum $p_{\perp}$ of the accelerated electron beam, to first order in $v/c$, is given by: 

\[
p_{\perp} = p_0 + \Delta p_{\perp} = mv + vT/c^2 = mv/\sqrt{1 - V^2/c^2}  , 
\]

where $p_0 = mv$ is the initial transverse momentum of the emitted electron. Since we have chosen $\vec{v} \cdot \vec{V} = 0$, the aberration angle $\theta_a$ becomes: $\theta_a = p_{\perp}/p$, where $p = mV/\sqrt{1 - V^2/c^2}$ is the momentum of the accelerated electron. Substituting, we obtain: $\theta_a = v/V$. This can be written in vector form as: 

\[
\vec{\theta}_a = \vec{V}\times\vec{v}/V^2 . 
\]

It is important to note that this result holds for an arbitrary Lorentz factor $\gamma = 1/\sqrt{1 - V^2/c^2}$.

We analyzed the effect of particle aberration by considering terms only up to first order in $v/c$.
A basic explanation of this effect is well-known: in the inertial frame, particle aberration can be readily understood through the Galilean transformation of velocities between reference frames.
It is important to emphasize that the dynamical reasoning presented here clarifies the physical meaning of particle aberration in the inertial frame. The  kinematics effects are only an interpretation of the behavior of the electromagnetic fields.  

Let us now return to the magnetic field measurements in the accelerated electron source, as shown in Fig. \ref{B557}. In the initial inertial frame, the electron source is in motion and its evolution is analyzed from the perspective of an inertial reference system. A key feature of this situation is that the velocity of the source carries real physical significance.
Indeed, the problem is explicitly time-dependent.
For a moving electron source, special relativity tells us that a magnetic field arises according to $\vec{B} = \vec{v} \times \vec{E} / c$, leading to a change in the transverse momentum of test particles. This change can be quantified as $\theta_a = p_{\perp} / p = v / V$, a result that remains valid even in the non-relativistic limit as $V^2 / c^2 \to 0$.
This brings us to an interesting question: what does an inertial observer measure when using a conventional magnetometer?
Given that charges in the source are moving, the resulting current produces a magnetic field described by $\vec{B} = \vec{v} \times \vec{E} / c$. This field is indeed detectable with a standard magnetometer. To understand this more concretely, consider the operation of an NMR magnetometer. When at rest in the lab frame, it measures the additional potential energy experienced by a proton with magnetic moment $\vec{m}$ in the field $\vec{B}$, given by $-\vec{m} \cdot \vec{B} \neq 0$. We come to the conclusion that the magnetic field generated by a moving source in an inertial frame is a real field  in the sense we have defined it.

\subsection{Inertial Frame View of Observations by a Non-Inertial Observer}

Earlier, we demonstrated that the aberration of light observed in an accelerated frame, when viewed from the initial inertial frame, is a well-known kinematic effect described by the special theory of relativity.
Can we not interpret the aberration of particles in an accelerated frame in a similar fashion?
As we shall see, the effect of particle aberration in accelerated frames can indeed be understood within an initial inertial frame using the framework of standard special relativity—specifically, by employing the theory of Wigner rotation.

The Wigner rotation is a relativistic kinematic effect in which the coordinate axes of a particle's proper reference frame appear to rotate relative to an initial (Lorentz) inertial frame. This rotation arises due to successive Lorentz boosts in different directions. The Wigner rotation angle in the inertial frame can be expressed in vector form as 

\[
\vec{\theta_w} = (1-1/\gamma)\vec{V}\times \vec{v}/V^2   ,  
\]

where $\vec{V}$ is the vector of particle velocity in the initial Lorentz frame before acceleration, $\vec{v}$ is the vector of small velocity change ($v \ll V$) due to acceleration, $\gamma$ is the Lorentz factor associated with $V$, and
$\theta_w$  is the Wigner rotation angle of the 
proper frame axes as observed from the (Lorentz) inertial frame. 

\footnote{This expression for the Wigner rotation is derived in Chapter 8 for the limiting case of an infinitesimal rotation angle, $\delta\theta = v/V$, corresponding to a small change in the particle's velocity vector. The derivation is based on the composition of Lorentz boosts in the      scenario: an electron is accelerated in the initial inertial frame from rest to velocity $\vec{V}$, then from $\vec{V}$ to $\vec{V} + \vec{v}$.  In the context of an accelerated electron source, the electron is first accelerated to velocity $\vec{v}$, then from $\vec{v}$ to $\vec{V} + \vec{v}$. A careful analysis shows that both orderings yield the same Wigner rotation expression. For example, one can use the reasoning presented in Section 8.3}   We now examine how this formula applies in the specific case where both the observer and the particle source are accelerated to a velocity $v$ in the inertial frame.

To find the rotation magnitude in the stated problem, we introduce a composition of Lorentz boosts. Let $S$ be the initial inertial  frame of reference, $S_n$ an accelerated frame with velocity $\vec{v} = \vec{e}_x v$ relative to the initial inertial frame, and $S'$ is a frame which moves relative to the $S_n$ with velocity $\vec{V}$ along the $z_n$-axis. Two sequential boosts from the inertial frame $S$ to $S_n$ and then to $S'$ are equivalent to the boost from $S$ to $S'$ and the subsequent rotation. The frame $S'$ is rotated through the angle $\theta_w$ with respect to the frame $S$.  Now we must be careful about the direction of rotation.
There is a good mnemonic rule. The rule says that the direction of the Wigner rotation in the Lorentz lab frame is the same as the direction of the velocity rotation in the initial inertia frame.
We can easily understand that the coordinate axes of the frame $(x', y', z')$ will be parallel to the coordinate axes of the accelerated frame $(x_n, y_n, z_n)$. As viewed from the initial inertial frame, the coordinate axes of the accelerated frame $(x_n, y_n, z_n)$ are also rotated through the angle $\theta_w$ with respect to the coordinate axes of the initial inertial frame $(x, y, z)$. Our results show that we cannot remain with the framework of parallel axes $z_n, z$ when considering the inertial frame view of particle beam observations of the accelerated (up to the velocity $v$) observer in perpendicular geometry.

Consider an electron gun–detector setup initially at rest in the initial inertial frame of reference. Suppose this system is then uniformly accelerated along the $x$-axis until it attains a velocity $v$. Now, from the perspective of an observer who remains at rest in the original inertial frame, the direction of the electron beam's propagation appears to shift. This observer measures an angular displacement of the beam given by $\theta_a = v/V$, where $V$ is the velocity of the particles emitted by the gun. This effect, known as particle aberration, is a well-established phenomenon.
A key distinction between relativistic and Newtonian kinematics emerges when we examine the behavior of the emitter–detector system under acceleration. From the initial inertial frame’s viewpoint, the accelerated setup undergoes a shearing motion: the detector no longer remains aligned with the electron beam after acceleration. For instance, if the gun–detector assembly was originally aligned along the $z_n$-axis, the detector fails to intercept the beam once the acceleration ends. In the accelerated frame, this misalignment is reflected in the angular deviation, and we obtain an expression for the aberration increment: 

\[
\theta_a = - \theta_w =  - (1 - 1/\gamma)v/V. 
\]

In the special case of light this expression simplifies to: $\theta_a = - v/c$.  \footnote{It's straightforward to verify that the infinitesimal Wigner rotation angle $\theta_w = (1 - 1/\gamma)v/V$ is always less than $v/c$ regardless of the particle velocity $V$. Specifically, $\theta_w = (v/c)(1 - \sqrt{1 - \beta^2})/\beta$, where $\beta = V/c$  and function $(1 - \sqrt{1 - \beta^2})/\beta$ is strictly less than 1 for $0 <\beta <1$. As discussed in Chapter 5, there is an inherent uncertainty in the relative position along the $z_n$-axis in the $x$-direction, amounting to  $z_nv/c$, due to the ambiguity in the timing of acceleration. This uncertainty arises because the relevant events are space-like separated. Consequently, the orientation of the $z_n$ axis relative to the inertial frame’s $z$-axis carries an angular uncertainty of $v/c$. The Wigner inclination angle $\theta_w$ always lies within this uncertainty bound.}

The formula  $\vec{\theta}_a = - (1 - 1/\gamma)\vec{V}\times\vec{v}/V^2$ 
matches the result previously derived for observations made by an accelerated observer within the framework of electrodynamics.
According to special relativity, the aberration of particle trajectories observed in an accelerated frame—when viewed from an initial inertial frame (using Lorentz coordinates)—is a purely kinematic effect. In such cases, the apparent shift in particle directions (i.e., aberration) within the accelerated frame is governed by the Wigner rotation.
The theory of Wigner rotation reveals that observers following different worldlines possess distinct notions of 3-space. This insight challenges a widespread but incorrect assumption in earlier literature: that a non-inertial observer (e.g., Earth-based) and an inertial observer (e.g., Sun-based) share a common 3-space.
It is important to emphasize that the dynamical arguments presented above clarify the physical meaning of the Wigner rotation in a rotating frame. The relativistic kinematic effects observed in the rotating frame—when interpreted from the perspective of an inertial observer—are simply manifestations of how electromagnetic fields transform under relativistic motion.

\newpage

\section{The Aberration of Light from a Laser Source}

\subsection{Theory of a Laser with an Optical Resonator}

Suppose an Earth-based observer attempts to measure the aberration of light using a laser as the radiation source. It is important to emphasize that the Earth's orbital motion cannot be detected in such an experiment. To understand this point, we consider a simplified model of a laser whose optical resonator consists of plane mirrors.

Laser oscillation occurs only when the gain acquired by the electromagnetic wave during a round trip through the active medium exceeds the total losses caused by reflection and diffraction. Diffraction losses therefore play a decisive role: they determine the oscillation threshold and simultaneously shape the spatial energy distribution inside the resonator during operation.

Our goal is to analyze how diffraction influences the electromagnetic field in a Fabry--Pérot interferometer operating in free space. The obtained results are equally applicable to gas lasers provided that the interferometer is immersed in the active medium.

The electric field $\vec{E}$ in a passive resonator satisfies the wave equation
\[
\nabla^{2}\vec{E}-\frac{\partial^{2}\vec{E}}{\partial(ct)^{2}}=0 .
\]
For simplicity, we assume circular polarization. This assumption does not restrict the generality of the analysis because the system possesses polarization degeneracy.

The electric field in a plane resonator can be represented as a superposition of longitudinal modes,
\[
E_x+iE_y=\sum_m \widetilde{E}_m(x,y,t)
e^{-i\omega_m t}\sin(K_m z+\delta),
\]
where
\[
K_m=\frac{m\pi}{l}-\frac{i}{nl}, \qquad
\omega_m=\frac{m\pi c}{l}, \qquad
\delta=\frac{i}{n}.
\]
Here $n$ is the complex refractive index of the mirror material ($|n|\gg1$), $l$ is the mirror separation, and $m\gg1$ is an integer longitudinal mode number. This representation satisfies the Leontovich impedance boundary condition at the mirror surfaces $z=0$ and $z=l$.\footnote{Approximate boundary conditions of this type were introduced by Leontovich for media with large refractive index modulus; see~\cite{Le}.}

We assume that the field varies only slightly during one resonator round trip,
\[
\left|\frac{\partial\widetilde{E}_m}{\partial t}\right|\frac{l}{c}
\ll |\widetilde{E}_m|,
\]
which implies that the oscillation frequency is close to the natural longitudinal resonance frequencies $\omega_m$.
We therefore seek a solution of the form
\[
\widetilde{E}_m=\Phi(x,y)\exp(\Lambda t).
\]
Substitution into the wave equation yields
\[
\nabla_\perp^{2}\Phi
+2i\omega_m\!\left(\frac{\Lambda}{c^{2}}+\frac{2}{nl}\right)\Phi=0 .
\]

In a plane resonator, diffraction at the mirror edges can be taken into account by treating the space between the mirrors as a waveguide terminated by an open end and applying diffraction theory to this opening. When the resonator length contains nearly an integer number of half-wavelengths, the effect of radiation leakage may be incorporated through an equivalent boundary condition imposed on the transverse amplitude:
\[
\Phi+(1+i)\beta_0
\sqrt{\frac{cl}{4\omega_m}}
\frac{\partial\Phi}{\partial\eta}=0,
\]
where $\eta$ denotes the outward normal to the imaginary side surface of the resonator and $\beta_0=0.824$.\footnote{These are impedance-type boundary conditions of resonance form~\cite{Va}. The constant $\beta_0$ is related to the Riemann zeta function through $\beta_0=-\zeta(1/2)/\sqrt{\pi}$; see~\cite{Va}.}

Since the boundary conditions are independent of polarization, the resonator exhibits polarization degeneracy. The problem of excitation of an open plane resonator can therefore be reduced to the classical problem of a closed resonator with equivalent boundary conditions.

Consider now a resonator formed by two circular plane mirrors of radius $R$ separated by a distance $l$. The system possesses axial symmetry about the $z$ axis. We seek solutions in polar coordinates in the form
\[
\Phi(\vec r)=\Phi_\nu(r)\cos(\nu\phi), \qquad
\Phi_\nu(r)\sin(\nu\phi),
\]
where $\nu$ is an integer azimuthal index. The radial eigenfunctions satisfy
\[
r^{2}\frac{d^{2}\Phi_{\nu j}}{dr^{2}}
+r\frac{d\Phi_{\nu j}}{dr}
+(k_{\nu j}^{2}r^{2}-\nu^{2})\Phi_{\nu j}=0,
\]
together with the impedance boundary conditions. The resulting eigenfunctions form a complete orthogonal set.

In the first approximation for the small parameter
\[
M=\frac{1}{\sqrt{N}}, \qquad
N=\frac{\omega_m R^{2}}{cl}\gg1,
\]
where $N$ is the Fresnel number, the eigenfunctions take the form
\[
\Phi_\nu(r)=
\frac{J_\nu(k_{\nu j}r)}{J_{\nu+1}(\mu_{\nu j})},
\]
with $\mu_{\nu j}$ denoting the $j$th zero of the Bessel function $J_\nu$,
\[
k_{\nu j}=\frac{\mu_{\nu j}(1-\Delta)}{R},
\qquad
\Delta=\frac{(1+i)\beta_0 M}{2}.
\]

Each mode with $\nu>0$ is fourfold degenerate. Two independent angular functions,
$\cos(\nu\phi)$ and $\sin(\nu\phi)$, exist for a given azimuthal number, and each admits two independent polarization states because the boundary conditions do not depend on polarization.

The rigorous three-dimensional theory of a plane Fabry--Pérot resonator shows that the radiation field decomposes into a discrete set of modes, each characterized by a specific decay rate (decrement) and a definite transverse field distribution on the mirror surface. For a mode with transverse wave number $k_{\nu j}$, the eigenvalue is
\[
\Lambda=
-\frac{2c}{nl}
-i\,\frac{\mu_{\nu j}^{2}(1-2\Delta)}{2R^{2}} .
\]
Because the decay rate increases exponentially with $\mu_{\nu j}^{2}$, higher-order modes experience stronger diffraction losses, and only low-order modes survive in a passive open resonator.

The normal modes supported by circular plane mirrors are denoted $TEM_{\nu j}$ modes, where $\nu$ specifies the angular order (sinusoidal azimuthal variation) and $j$ the radial order. The fundamental mode $TEM_{00}$ therefore represents the dominant operating mode of such resonators.

\subsection{Absent of Light Aberration from an Earth-Based Laser}

We have shown that acceleration influences the field equations. Within the resonator, the electric field of an electromagnetic wave satisfies the wave equation, which transforms into Eq. (\ref{GGT4}) under a Galilean transformation of space and time coordinates.
At first glance, measuring the diffraction losses of the dominant mode may seem equivalent to determining the angular displacement. However, it is important to recognize that the difference between the Langevin and Minkowski metrics does not affect the parameters of a laser incorporating an optical resonator.

In our earlier discussion of the Langevin metric, we demonstrated that applying a Galilean transformation to the wave equation yields a more complex, anisotropic form—namely, Eq. (\ref{GGT4}). This raises a fundamental question: how can we solve this form of the electrodynamics equation, particularly in the presence of boundary conditions? The key lies in recognizing that the underlying electrodynamics theory for optical resonators remains Lorentz-covariant. To address the anisotropic terms and simplify the field equations, we employ a mathematical transformation. By introducing a suitable change of variables, these anisotropic terms can be eliminated from the transformed equations, effectively restoring the Minkowski metric in the accelerated frame.

It is important to stress that this transformation is purely a mathematical convenience, not a reflection of any underlying physical reality. The variables used carry no intrinsic physical meaning; they serve only to streamline the problem-solving process. Crucially, the physical content of the theory remains unchanged—results expressed in the transformed coordinates are entirely equivalent to those in the original frame. Changing the four-dimensional coordinate system introduces no new physical effects.

A comprehensive analysis of experiments performed in an accelerated frame confirms that optical phenomena within a resonator are unaffected by acceleration relative to the fixed stars. This invariance arises because all radiation measurements in such systems are based on standing-wave phenomena—specifically, round-trip or two-beam interference methods.

\begin{figure}
	\centering
	\includegraphics[width=0.7\textwidth]{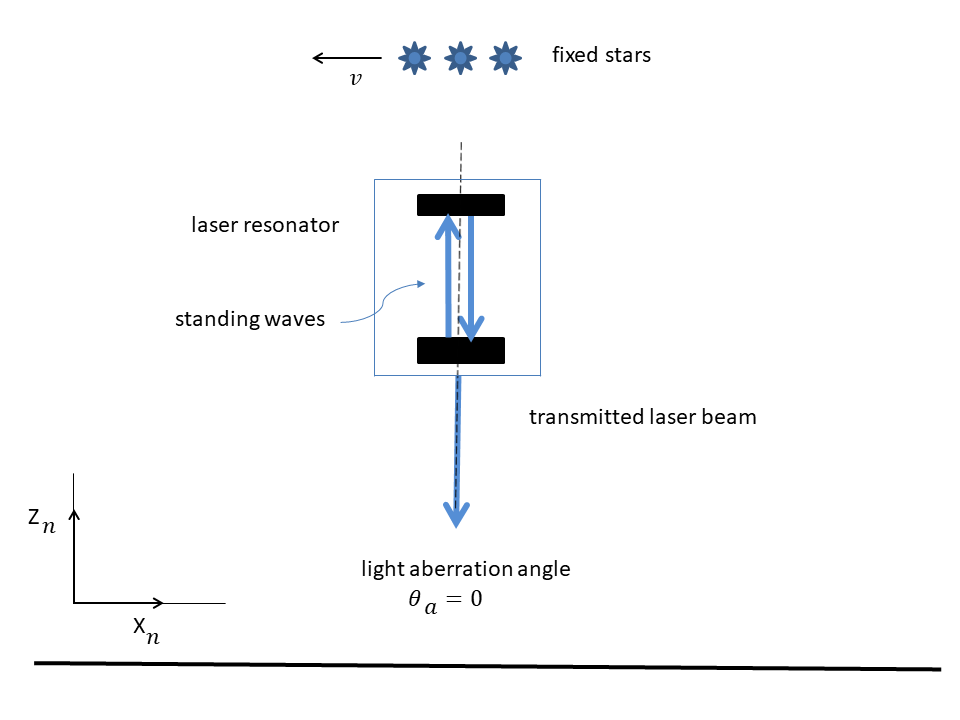}
	\caption{Light aberration in an accelerated frame of reference, assuming the laser operates in a steady-state regime. }
	\label{B554}
\end{figure}

Let us now consider the aberration of light emitted from a laser resonator at rest in an accelerated frame. Upon analyzing the data, the accelerated observer detects no shift in the direction of energy propagation, as illustrated in Fig. \ref{B554}. This absence of aberration indicates that the cross-term in the Langevin metric, which would normally cause such an effect, is effectively averaged out over the complete round-trip trajectory of the radiation within the resonator.

\begin{figure}
	\centering
	\includegraphics[width=0.7\textwidth]{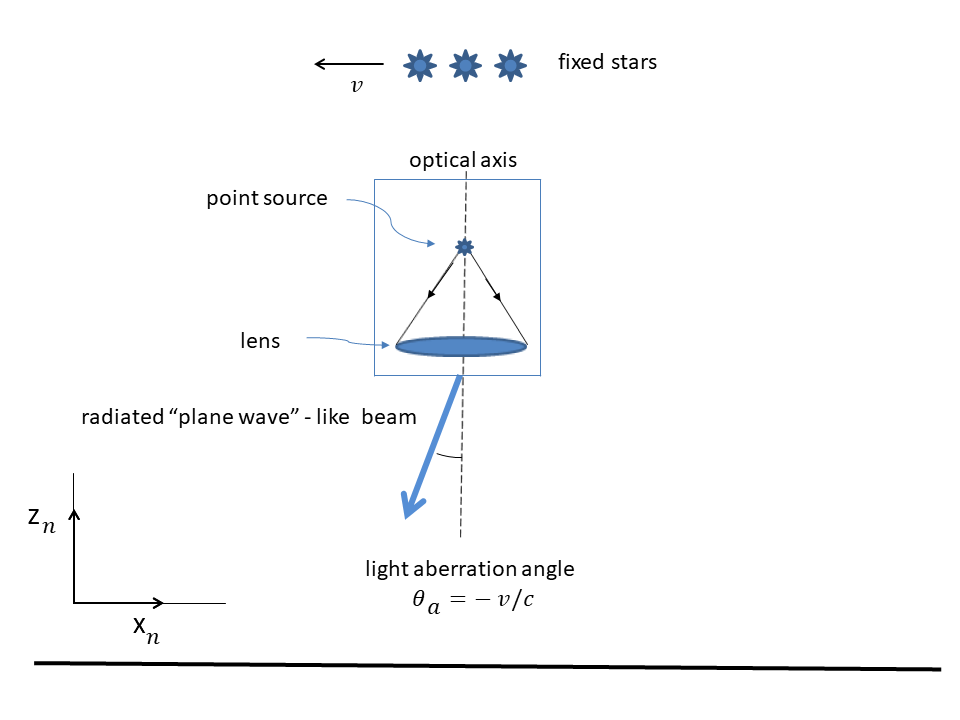}
	\caption{Light aberration from a plane-wave emitter in an accelerated frame of reference.}
	\label{B556}
\end{figure}

The distinction between a "plane wave" emitter and a laser source leads to a striking prediction in the theory of light aberration. Specifically, if a "plane wave" emitter initially at rest with respect to the fixed stars begins to move, the apparent angular position of the emitter—as observed from the accelerated frame—undergoes a sudden shift by an angle of $-v/c$ (see Fig. \ref{B556}). Understanding this phenomenon requires the framework of special relativity; classical theory is insufficient.

But how can such a difference between a "plane wave" emitter and a laser source arise?
To answer this, we must consider: What happens to an accelerated optical resonator? In previous discussions of optical resonators, we applied electrodynamic boundary conditions in 3-space, which define the direction of energy propagation along the optical axis—perpendicular to the mirror surfaces. According to special relativity, acceleration does not alter the orientation of the mirrors relative to the axes of the accelerated frame (i.e., the 3-space of the observer).
Thus, an observer co-moving with an accelerating laser source will find that the direction of energy transport (along the optical axis) remains unchanged relative to the observer’s own frame, regardless of the motion with respect to the fixed stars (see Fig. \ref{B554}).

Special relativity tells us that both accelerated and inertial observers share the same absolute space-time but differ in their respective metrics, and therefore in their associated 3-spaces. In the case of an optical resonator, the direction of energy propagation is dictated by boundary conditions within 3-space. In contrast, the emission from a source—such as a "plane wave" emitter—is influenced by initial conditions. For an accelerated observer, these conditions involve a nontrivial mixture of space and time coordinates, a consequence of the pseudo-gravitational field experienced during acceleration.
The solution lies in using the absolute time coordinate. After acceleration, the wavefront of the light emitted by the incoherent source is perpendicular to the direction $z_n$. Applying this initial condition, together with the Langevin metric in an accelerating reference frame, allows us to fully explain the observed aberration of light from a "plane wave" emitter (see Fig. \ref{B556}).

\begin{figure}
	\centering
	\includegraphics[width=0.7\textwidth]{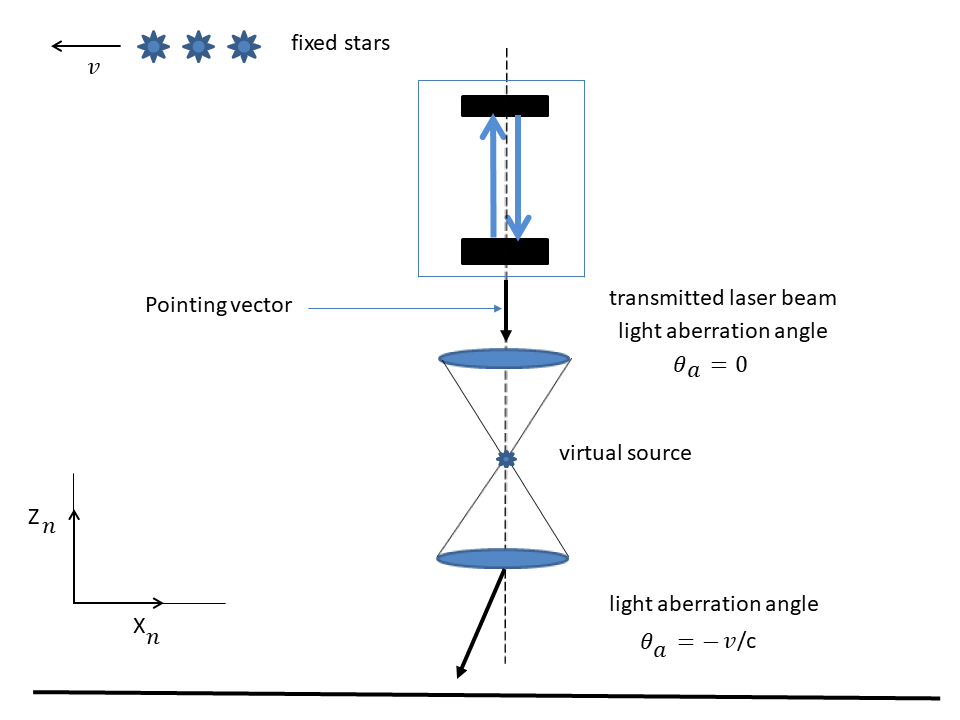}
	\caption{Geometry for propagation of a laser beam in an accelerated frame of reference. Virtual source (i.e. focus of laser beam) is positioned in the front focal plane of second lens.}
	\label{C554}
\end{figure}

\subsection{Focusing of a Laser Beam in an Accelerated Frame of Reference} 

In this section we analyze the propagation of a laser beam through a system of lenses separated by a finite distance, see Fig.~\ref{C554}. As an instructive example, we consider a lens of focal length $f$ that focuses a laser beam with a plane wavefront, and we compare the result with the prediction obtained in the initial inertial frame.

A laser output beam with a large beam diameter and an effectively infinite radius of curvature provides a good approximation to a plane wave. It is important to emphasize that, in an accelerated frame, a redirected plane-wave emitter (Fig.~\ref{BC30}) is equivalent to a laser source (Fig.~\ref{B554}).

Let us now examine the consequences of this equivalence under different circumstances. Consider a system consisting of two identical lenses. An interesting situation arises when the distance between the lenses is equal to $2f$. What happens in this case?

We begin with a parallel laser beam and determine the position of the focus produced by the first lens. The parallel beam from the laser source is focused by the first lens on the $z_n$-axis at a distance $f$ (Fig.~\ref{C554}). This focused beam can then be treated as originating from a virtual point source for the second lens.

Acceleration modifies the field equations. In particular for polarization perpendicular to the $x_nz_n$-plane, in the accelerated frame the magnetic field satisfies the modified electrodynamics equation  $c\vec{\nabla}\times\vec{B}_n = (\partial/\partial t + v\partial/\partial x)\vec{E}_n$.
Any experimental determination of aberration necessarily involves a light beam of finite transverse size. In the situation considered here, this limitation is determined by the aperture of the second lens. As a result, the refocused laser beam emerging from the second lens propagates at an angle $-v/c$ with respect to the $z_n$-axis.

\subsection{A Moving Laser Source in an Inertial Frame of Reference}

\begin{figure}
	\centering
	\includegraphics[width=0.7\textwidth]{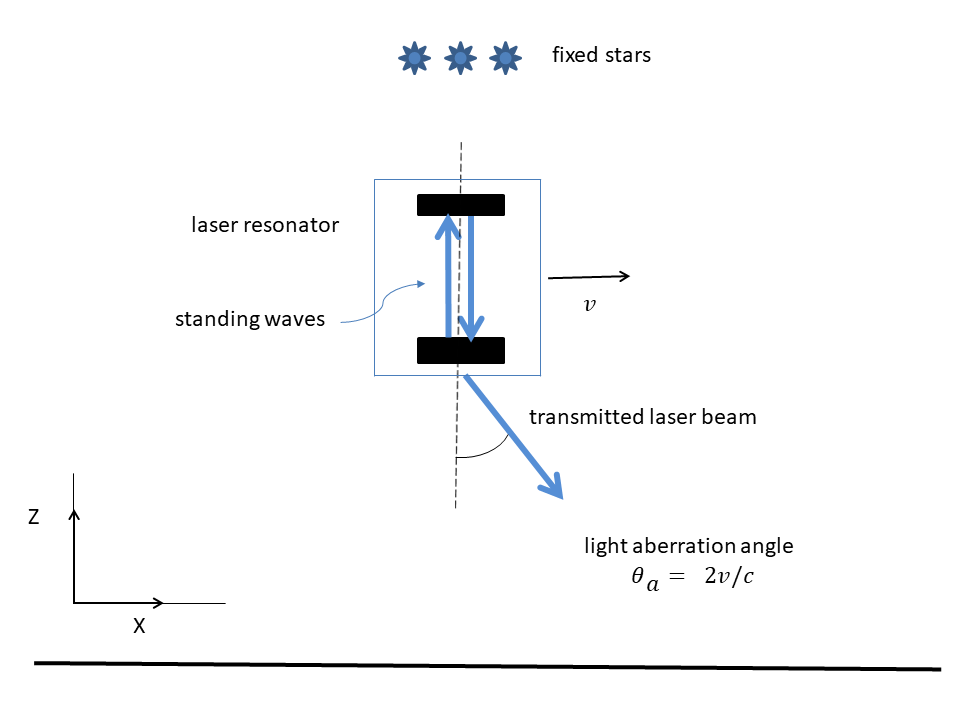}
	\caption{Light aberration caused by a moving laser source in the initial inertial frame of reference. The laser resonator moves parallel to the mirrors, and it is assumed that the accelerated laser operates in a steady-state regime}
	\label{B552}
\end{figure}

Let us consider the radiation emitted by an accelerated laser source in the inertial frame. The light aberration effect can be described within the framework of standard special relativity, utilizing the relativity of simultaneity. When a laser moves transversely (i.e., parallel to the mirrors), there is a deviation in the energy transport of the light emitted from the optical resonator. In the first-order approximation in  $v/c$ the inertial observer would observe an angular displacement of $\theta_a = 2v/c$, Fig. \ref{B552}.

\begin{figure}
	\centering
	\includegraphics[width=0.7\textwidth]{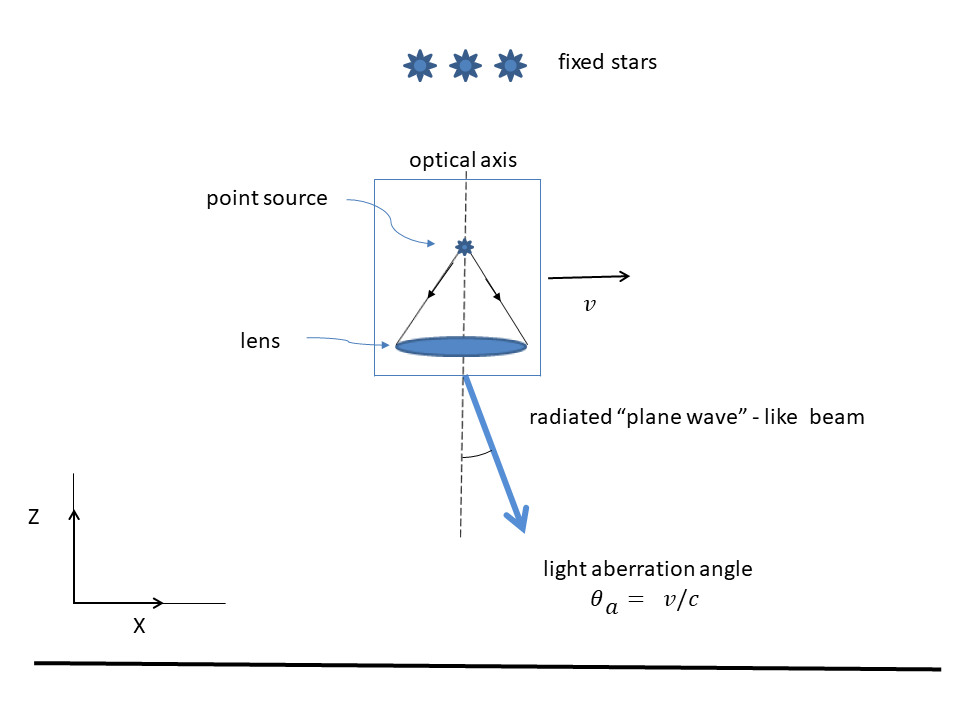}
	\caption{Light aberration from a moving "plane wave" emitter in the initial inertial frame of reference.}
	\label{B553}
\end{figure}

When discussing light aberration, it is important to distinguish between the effect from a "plane wave" emitter and that from the laser source. Consider a "plane wave" emitter that is accelerated from rest to velocity $v$ in a direction perpendicular to its optical axis. \footnote{It is assumed that after acceleration, the laser moves at constant velocity and operates in a steady-state regime.} 
Suppose an observer, at rest with respect to the inertial frame of reference, measures the direction of light transport. How does the light beam from the moving "plane wave" emitter appear to this observer? The inertial observer would find an angular displacement of $\theta_a = v/c$, as illustrated in Fig. \ref{B553}.

The first point to make regarding measurements in an inertial frame is that there is an intuitively plausible way to understand the aberration of light from a moving "plane wave" emitter. An elementary explanation of this effect is well-known and can be understood in terms of the transformation of velocities between different reference frames. The aberration of light in the initial inertial frame can also be readily explained within the framework of the corpuscular theory of light. This is plausible when considering that a light signal represents a certain amount of electromagnetic energy. Like mass, energy is conserved, so a light signal shares several characteristics with material particles. Consequently, we should expect the group velocities of light signals to follow the same addition theorem as particle velocities. A more detailed treatment based on the wave theory of light confirms this expectation.

In the case of a moving laser source, intuition suggests that the aberration increment would be given by $\theta_a = v/c$. However, special relativity introduces an additional factor of 2. The standard derivation of the aberration of light effect does not account for the mixing of positions and time in an accelerated frame. It incorrectly assumes that accelerated and inertial observers share a common 3-space. In contrast, from the perspective of the initial inertial frame, the coordinate axes of the accelerated frame are rotated by an angle $v/c$ relative to those of the inertial frame.

\subsection{A Moving Laser Source in an Accelerated Frame of Reference}

Let us describe the radiation emitted by a laser source moving in an accelerated frame, as observed by an accelerated observer. First, we analyze the scenario in which a laser source is at rest in the accelerated frame. As discussed previously, the accelerated observer detects no shift in the direction of energy propagation, as illustrated in Fig.\ref{B554}.

It is instructive to further illustrate the difference between a plane-wave emitter and a laser source within the framework of special relativity. Suppose an accelerated observer redirects an emitter that is initially at rest in the accelerated frame. Figure. \ref{B409} shows a schematic representation of a "plane-wave" emitter. In electrodynamics, such redirection can be achieved by introducing an offset between the point source and the optical axis (aligned with the $z_n$-axis), as shown in Fig. \ref{BC30}. This offset encodes information about the source acceleration. 

We consider the specific case in which, after redirection, the (group) velocity component of the light beam along the $x_n$-axis becomes zero. In other words, both the laser source and the emitter, when at rest in the accelerated frame, radiate along the $z_n$-axis. In this situation, an inertial observer would find no difference in the deviation of energy transport: the aberration increment is $\theta_a = 2v/c$ for both sources.

\begin{figure}
	\centering
	\includegraphics[width=0.7\textwidth]{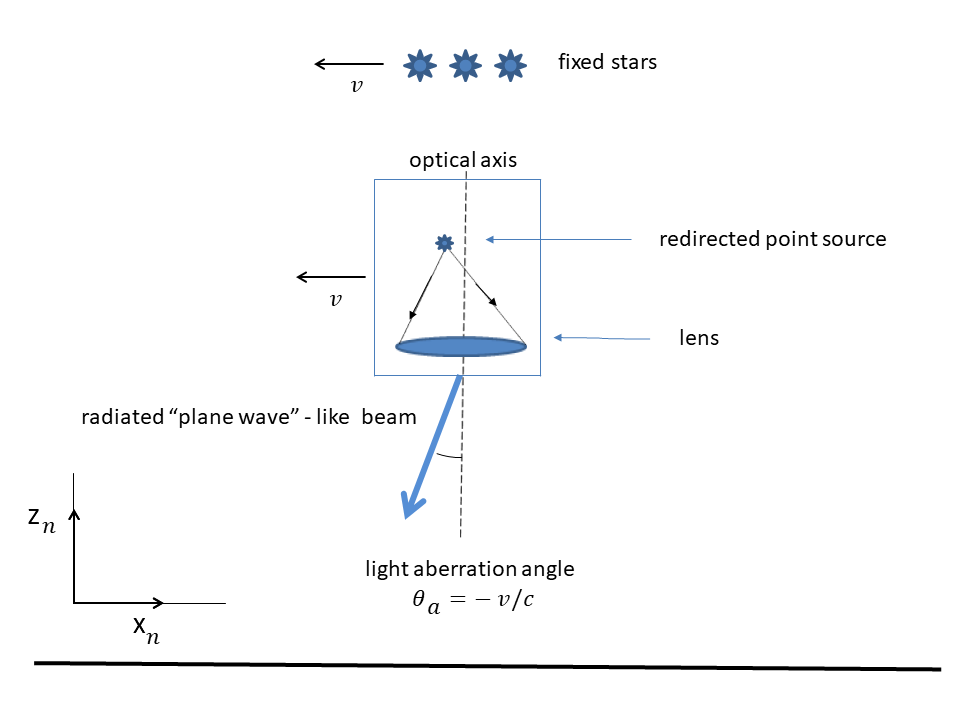}
	\caption{Aberration of light in an accelerated frame of reference. The redirected plane-wave emitter, initially at rest in an accelerated frame, accelerating to velocity $-v$ along the  $x_n$-axis. }
	\label{B1556}
\end{figure}

\begin{figure}
	\centering
	\includegraphics[width=0.7\textwidth]{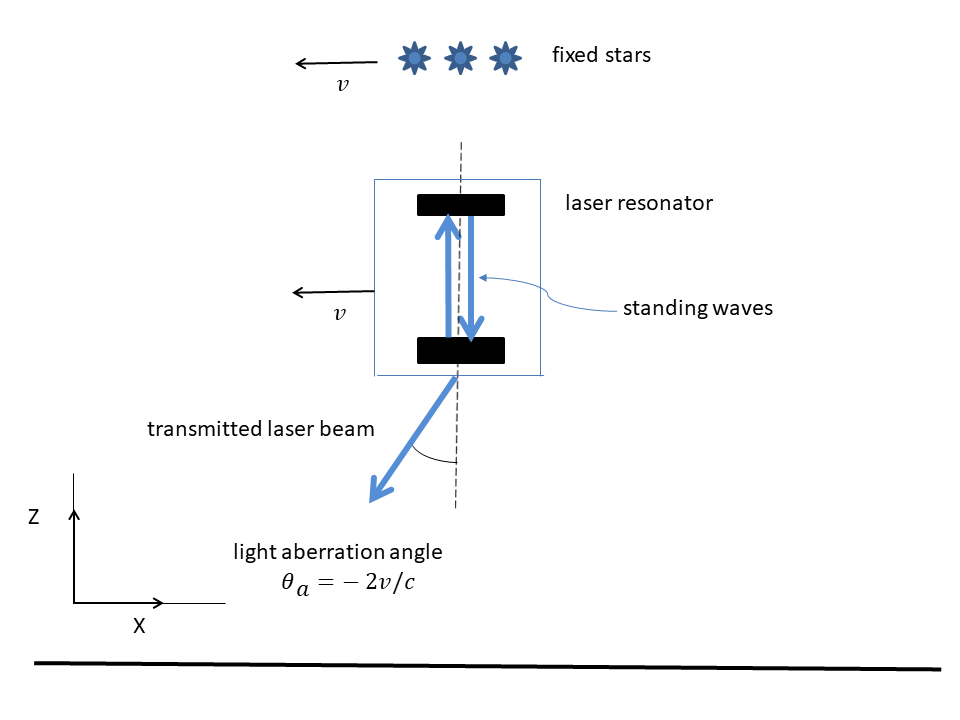}
	\caption{Aberration of light in an accelerated frame of reference. The laser source, initially at rest in an accelerated frame, accelerating to velocity $-v$ along the  $x_n$-axis.  It is assumed that the accelerated laser operates in a steady-state regime.}
	\label{B1552}
\end{figure}

Let us now examine what occurs when the redirected plane-wave emitter is accelerated from rest to velocity $-v$ along the $x_n$-axis in the accelerated frame. Suppose an observer at rest with respect to the accelerated frame measures the direction of light propagation. How does the light beam from the moving redirected plane-wave emitter appear to this observer? The accelerated observer finds an angular displacement of $\theta_a = -v/c$, as illustrated in Fig.~\ref{B1556}.

From the inertial observer’s viewpoint, the emitter is at rest in the inertial frame. The radiated light beam propagates at an angle $v/c$ relative to the inertial-frame axes as a result of the redirection procedure, i.e., due to the offset between the point source and the optical axis of the lens in the emitter (Fig.~\ref{BC30}).

Now suppose the laser source is accelerated from rest in the accelerated frame to velocity $-v$ along the $x_n$-axis. The accelerated observer finds that the angular displacement of the laser beam is $\theta_a = -2v/c$, as illustrated in Fig.~\ref{B1552}. Consequently, in the inertial frame the laser source is at rest, and the inertial observer detects no shift in the direction of energy propagation.

To interpret these results physically, recall that in an optical resonator the direction of energy propagation is determined by boundary conditions in three-dimensional space. According to special relativity, acceleration does not alter the orientation of the resonator mirrors relative to the axes of the laser’s proper frame (i.e., the three-space of the laser). Thus, an observer co-moving with an accelerating laser source finds that the direction of energy transport (along the optical axis) remains unchanged relative to the observer’s own frame, regardless of the motion with respect to distant stars.

\newpage

\section{ Earth-Based Setups to Detect the Aberration Phenomena}

\subsection{The Potential of  Earth-Based Electron Microscopes}

Above, we discussed the new relativistic particle kinematics in rotating frames of reference. Since we live on a rotating Earth, the distinction between relativistic kinematics in rotating and non-rotating frames holds both theoretical and practical importance. We propose an aberration-of-particle experiment using an Earth-based particle source. The results of such experiments can be interpreted within the framework of special relativity. However, obtaining an accurate solution in the Earth-based frame requires the use of the metric tensor, even for first-order approximations.

We propose an experiment using  commercial 200-kV scanning transmission electron microscope (STEM) as a particle source. 
One could use a vertically oriented optical setup.  
We suppose that the electron beam is imaged by a lens to a spot that lies in the image plane. 
Measurement of the spot shift  with respect to the optical axis (which is parallel to the zenith-nadir axis) is equivalent to the determination of the angular displacement. 
Due to the high stability, the aberration increment could be observed. \footnote{The dedicated STEM with high stability was made  \cite{STEM}. It was developed based on a 200-kV commercial instrument. The electron source was installed in the anti-seismic room. Temperature fluctuations in the room were within 0.2 $^{o}C$. The magnetic field in the room was less than $0.2 mG$ near the source column.   
	To prevent thermal drift, a shield box was placed over the high-voltage generator. The column was covered with rubber sheets to reduce the temperature fluctuations. 
	The image drift measured during a period of 61 minutes was $ \simeq 6 ~\mathrm{nm}$. The drift speed and direction during this period were approximately constant. The average drift speed was 0.12 nm/min. It is substantially smaller than that conventional commercial microscope, in which the drift speed is about 2 nm/min.}

Consider the inertial  Sun-based reference system. In this system, there is the Earth which rotates around the Sun with orbital velocity $v$. Developing into powers of $v/c$ we can classify effects for velocities $v \ll c$ as of the first order, of the second order, or higher orders. 
Clearly, in the case of the rotating (around the Sun) Earth-based frame, we consider the small expansion parameter $v/c \simeq 10^{-4}$ neglecting terms of order of $v^2/c^2$.   

Suppose that an Earth-based observer performs an aberration of particle measurement.
The aberration angle varies with a one-year temporal period. 
An approximate formula to express the aberration angle can be found by using the Langevin metric in the Earth-based frame or by using the Wigner rotation theory in the Sun-based frame.
In the aberration of particles, the aberration angle is the apparent angular deviation of the position of the particle source relative to the real location of the source. 
The reference axis in the Earth-based frame can be formed by a plumb line, which is the most fundamental local Earth-based coordinate system.

When measured  with an Earth-based particle source, the aberration angle 
is $\vec{\theta}_a = - \vec{\theta}_w = - (1 - 1/\gamma)\vec{V}\times \vec{v}/V^2$, where $\vec{v}$ is the orbital velocity of the Earth relative to the Sun, $\vec{V}$ is the velocity of the accelerated electron, and $\gamma = 1/\sqrt{1 - V^2/c^2}$ is the particle relativistic factor. 
This means that the angle of Wigner rotation depends on $V^2/c^2$. 
It should be noted that the presented formula can be considered only as a first-order approximation in $v/c$. In the theory of (infinitesimal) Wigner rotation, we must also consider the relativistic parameter $V/c$. 
If a particle motion velocity is non-relativistic, the binomial expansion yields  $\vec{\theta}_a = - [V^2/(2c^2)] \vec{V}\times \vec{v}/V^2$. 

Let us calculate the parameters for the STEM setup. The Wigner rotation angle parameter (constant of aberration) defines the scale of the aberration increment and is equal to $\theta_w = (1 - 1/1/\gamma)v/V \simeq 40$ $\mu\mathrm{rad}$. Here $\gamma \simeq 1.39$, $V/c = 0.7$, $v/c \simeq 100$ $ \mu\mathrm{rad}$.  Earth moves around the sun, with a consequent change of the direction of $\vec{v}$. Therefore the angle of aberration changes. The total change of the order of $40 \mu\mathrm{rad}$. It is oscillatory with a period of one year.

\begin{figure}
	\centering
	\includegraphics[width=0.9\textwidth]{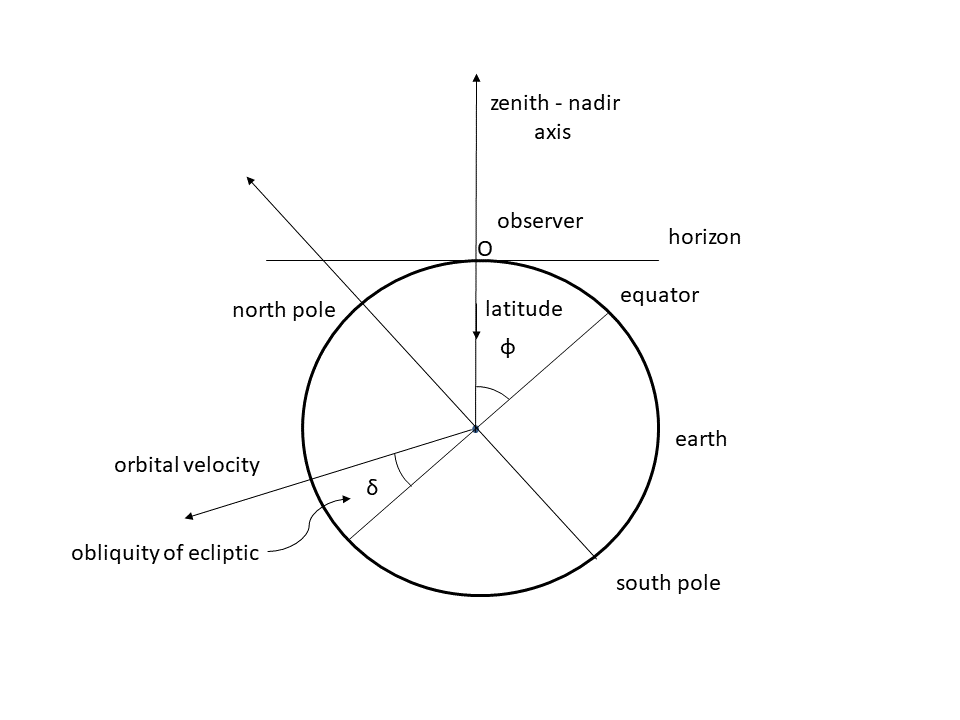}
	\caption{Definitions related observer's position on the earth. }
	\label{B408}
\end{figure}

A specific aspect of our case study needs further investigation. The proposed method of measuring the angle of aberration involves the use of Earth-based sources and has a big advantage. The rotation of the Earth on its axis should produce a corresponding shift of the image. Obviously, it is important that observation should be recorded in the shortest possible time.
In principle, records could be taken over a period of one day. The aberration shift depends only upon the value of $v_{\perp}$, the component of the orbital velocity perpendicular to the earth's rotation axis  \footnote{
	The constant of diurnal aberration defines the scale of the aberration increment and is equal to $\theta_d = (1 - 1/1/\gamma)v_e/V \simeq 0.6$ $ \mu\mathrm{rad}$. Here  $\gamma \simeq 1.39$, $V/c = 0.7$, $v_e/c \simeq 1,55$ $\mu\mathrm{rad}$ where $v_e = 0.45$ km/s is the linear velocity of the Earth rotation at equator.}. In practice, the measurement procedure is complicated by a number of factors, Fig. \ref{B408}. 
A correction has to be made for the observer's position on the Earth's surface and the obliquity of the ecliptic  $\delta = 23^o$. The image appears to move in an ellipse.
The shape of the aberration ellipse depends on the observer's latitude $\phi$. For an observer on the equator, the ellipse degenerates into a line segment, and for an observer at the pole of the Earth, the ellipse is a circle. 
At the value  $l = 2$ mm of the focal length, the major axis of the aberration ellipse
of the order of 80 ($\cos \delta)$ nm. \footnote{Electron-optical system of the STEM contains an electron gun and several magnetic lenses, stacked vertically to form a lens column. The illumination system of the instrument comprises the electron gun, together with two condenser lenses that focus the electrons onto the spacemen.   
Electrons entering the lens column appear to come from a virtual source.  The first condenser lens (C1) is a strong magnetic lens, with a focal length $f_1$ of typically 2 mm. Using a virtual electron source as its object, C1 produces a real image. Because the lens is located $D_1 = 15$ cm below the object (virtual source), its object distance is 15 cm  $ \gg f_1$ and the image distance $\simeq f_1$. The second condenser lens (C2) is a weak magnetic lens ($f_2 \simeq 10$ cm) that provides little magnification but allows the diameter of illumination at the specimen to vary continuously over a wide range. The distance between lens centers is typically 10 cm. Spacemen is located $D_3 = 25$ cm below the C2.} The value of the minor axis depends on the observer's position and is given by 80 ($\cos \delta \sin \phi ))$ nm.  When the latitude is $\phi = 35^o$, 
the value of the scale of image shift per hour depends on the local time and varies from a value of about 10 nm/60 min 
to about  20 nm/60 min through the day. \footnote{Experimental results (see Fig. 44 p.167 in \cite{STEM}) shows that the image shift is quite close to the theoretical prediction $ \simeq 10 ~\mathrm{nm}$. It is important to note that we used the typical  focal length (2 mm) in our estimations. The details of the optical setup may influence the image shift.}

\subsection{A Thought Experiment with a  Point-Like Source}

\begin{figure}
	\centering
	\includegraphics[width=0.9\textwidth]{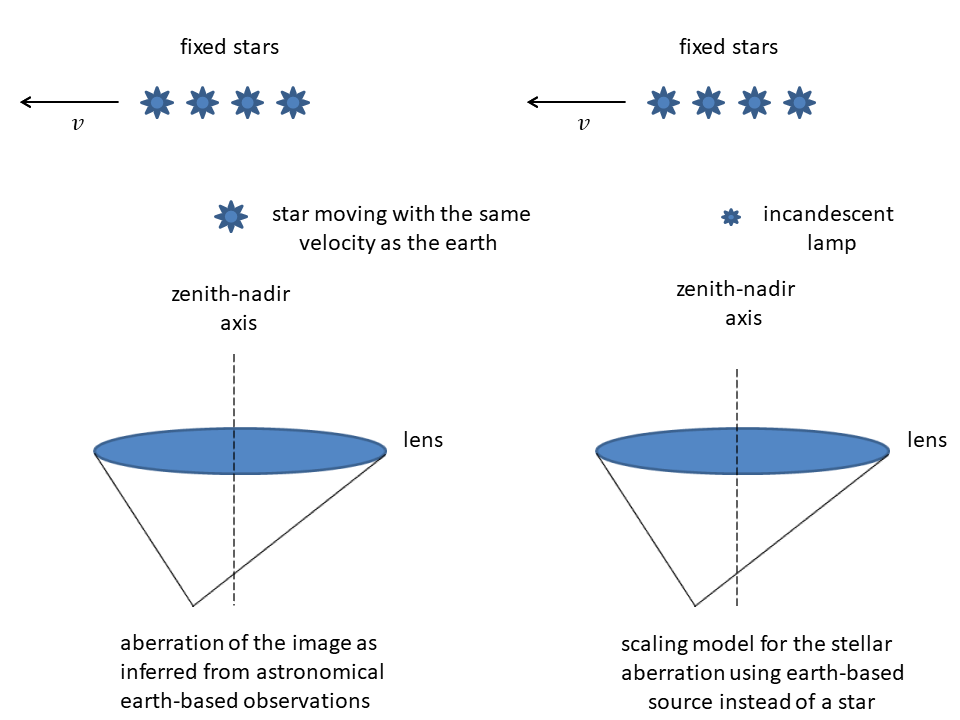}
	\caption{Scaling model for stellar aberration using an Earth-based setup with an incandescent lamp. Based on astronomical observations, we know what to expect even without invoking special relativity.}
	\label{B405}
\end{figure}

We now turn to a simple scaling model for stellar aberration, illustrated in Fig. \ref{B405}.
Consider a scale transformation involving both the linear dimensions of the source and the distance between the source and the observer. This approach allows us to derive a condition for optical similarity between the aberration of light from a distant star and that from an Earth-based source.
Such a transformation can assist in the preliminary design of an Earth-based experimental setup, where an incandescent lamp is used in place of a star.

Scaling model for stellar aberration using an Earth-based setup with an incandescent lamp. Based on astronomical observations, we know what to expect even without invoking special relativity. Remarkably, by measuring the aberration of light alone—without referencing anything external to Earth—it is possible to determine the Earth's orbital speed around the Sun.

A star ( or an incandescent lamp) is a system of elementary statistically independent point sources.
Point-source field diffraction can be divided into categories - the Fresnel (near-zone) diffraction and Fraunhofer  ( or telescopic) diffraction. 
If the Fraunhofer (or equivalently, far zone) approximation  $z > D^2/\lambda$ is satisfied, then the quadratic phase factor is approximately unity over the detection aperture. Here $D$ is the aperture width, $\lambda$ is the wavelength of the radiation, and $z$ is the distance between the source and observer.  In the far zone, a point source produces in front of pupil detection effectively a plane wave. 
The plane wavefronts of starlight entering the telescope are imaged by the lens to a diffraction spot that lies in the focal plane. 
It is worth noting that we consider the divergence of the transmitted radiation, $\lambda/D$, that is relatively small compared to the aberration angle, $\theta_a = v/c$.  

An incandescent lamp is a completely incoherent source. The character of the mutual  intensity function produced by an incoherent source is fully described by the
Van Cittert-Zernike theorem \cite{G}.
In our case of interest, imaging system is situated in the far zone $D^2/(\lambda z) < 1$. It is the source linear dimension $d$ (source diameter) that determines the coherent area of the observed radiation $(z\lambda/d)^2$. 
The condition for neglecting the transverse size of the source, $d$, formulated as the requirement that the  
change of the correlation function along the aperture be less than unity, that is $d < D$. 
\footnote{
	In all cases of practical interest, telescopes are situated in the far zone of the distant stars. If the far zone approximation  $z > d$ is satisfied, then the  Van Cittert-Zernike theorem takes its simplest form. For example, the coherence area of the light emitted by the circular incoherent source  is $A_c = 4\lambda^2z^2/(\pi d^2)$. 
	The minimal coherent area of light observed on the earth's surface from the nearest star has a diameter of about $\lambda z/d \simeq$ 10 m.  This diameter is larger than a typical telescope aperture $D$.
	In this situation, a star can be considered as a point source and a star image is a point spread function in the image plane of a telescope. Our scaling model is based on the far zone approximation. 
	We have only to change the distance and the source diameter. Assuming small parameter conditions $ Dd/(z\lambda) < 1$,  $D^2/(z\lambda) < 1$,  the scaling approach yields the following requirements: $D^2/(z\lambda) < 1$,  $d < D$.}

Let us estimate the parameters for an Earth-based setup using an incandescent lamp.
We aim to describe a series of thought experiments—hypothetical, yet physically plausible.
At optical frequencies, the conditions required for the Fraunhofer approximation can be quite stringent. For instance, with a wavelength of $\lambda = 0.3$ $\mu$m  and an aperture width of $D = 3$ cm, the observation distance must satisfy $ z > 3000$ m. 
Obviously, the required conditions are met in a number of important problems. 
No one has ever done such an experiment, but we know what would happen from the astronomical observations.

\subsection{Practical Setup Using a Point-Like Optical Source}

It should be noted that Fraunhofer diffraction patterns can be observed at distances much closer than implied by equation  $z > D^2/\lambda$ provided the aperture is illuminated by a spherical wave converging towards the observer, or if a positive lens is properly situated between the observer and the aperture.

\begin{figure}
	\centering
	\includegraphics[width=0.9\textwidth]{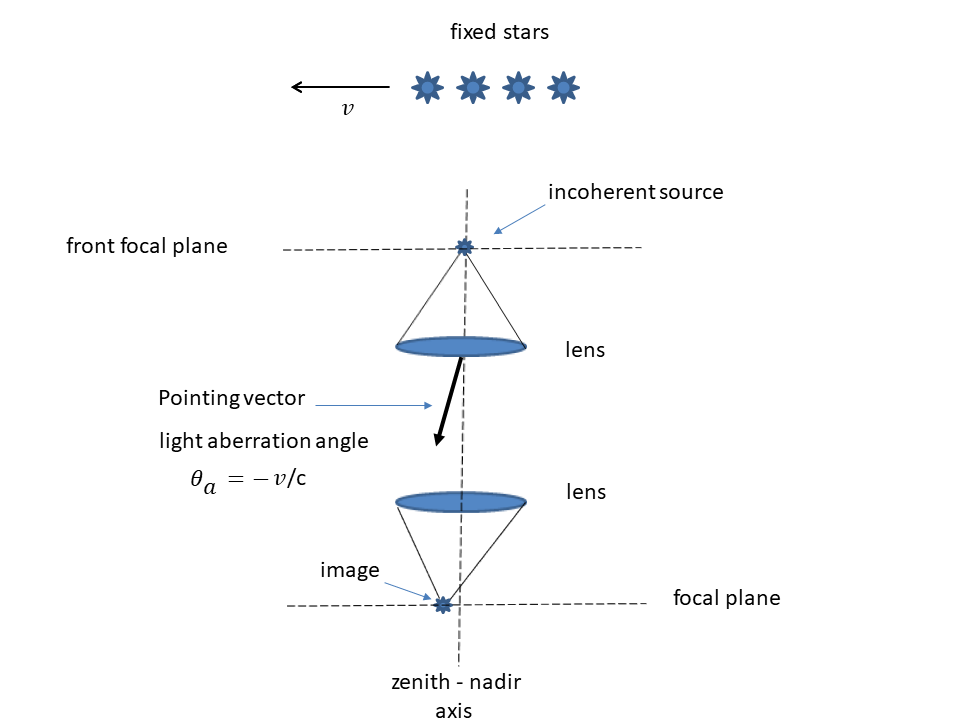}
	\caption{Earth-based optical setup for detecting the aberration of light. The setup employs a two-lens system to form an image using an incoherent light source at optical frequencies.}
	\label{B404}
\end{figure}

The operational principle of the Earth-based optical setup designed to detect light aberration is schematically illustrated in Fig.~\ref{B404}. The configuration follows the well-known two-lens imaging scheme, which allows magnification to be adjusted by changing the focal length of the second lens. For simplicity, however, we assume in the following discussion that both lenses have identical focal lengths, corresponding to a 1:1 imaging configuration. Two-lens systems are widely used in optical imaging because they provide greater flexibility for image manipulation than single-lens arrangements.

Within this two-lens configuration, we now address the relatively straightforward task of characterizing the radiation emitted by a point source in the image plane, assuming that the lenses shown in Fig.~\ref{B404} are identical.

To measure the aberration of light, the linear dimension $d$ of the source must be made extremely small. We assume that the characteristic size of the “point” source is on the order of the optical wavelength $\lambda$. Within an elementary source volume of order $\lambda^3$ there exists an enormous number of atoms. In optics, radiation from such a “point” source is described classically by Maxwell’s equations, while the emitting medium—viewed as an ensemble of atoms—is treated quantum mechanically. Owing to modern lithography techniques, it is relatively straightforward to fabricate an unresolved point source at optical wavelengths, enabling the generation of sufficiently bright radiation.

Consider a diffracting aperture that is circular, and let the radius of the aperture be $a$. The physical properties of the lens can be combined in a single number $f$ called the focal length. We assume that the image is produced by a diffraction-limited optical system (i.e. a system that is free from aberrations).  
Once the wavelength is fixed, the resolution only depends on the numerical aperture $NA = a/f \ll 1$ of the system. 
The response of the system at point $(x_i,y_i)$ of the image plane to a $\delta$ function input at coordinates  
$(x_o,y_o)$ of the object plane is called the point spread function of the system. In our case of interest,
one obtains the following amplitude point spread function: $A(x_i,y_i) = J_1(\alpha)/\alpha$,
which is the diffraction pattern of a circular aperture, where $\alpha = 2\pi a r_i/(f\lambda)$, 
$x_0 = 0, y_0 = 0$, $r_i = (x_i^2 + y_i^2)^{1/2}$. The first zero of the Airy's pattern is at $r_i = 0.6 \lambda/NA$.

Let us estimate the parameters for the optical setup shown in Fig.\ref{B404}. Suppose $a = 1.5$ cm and $\lambda = 0.3$ $ \mu$m. We will use $NA = 0.1$. The resolution analysis is reduced to the theoretical framework of standard imaging theory. One can take advantage of well-known resolution criteria like Rayleigh criteria i.e. $0.6 \lambda/NA = 2$ $\mu$m.
Here we analyze a vertically oriented optical setup. \footnote{An alternative configuration is a horizontally  oriented optical setup. Measurement of the spot shift with respect to the optical axis, which is parallel to the meridian direction, is equivalent to the determination of an angular displacement. 
The image appears to move in an ellipse. In this orientation, for an observer on the equator, the ellipse is a circle, and for an observer at the pole of the earth, the ellipse degenerates into a line segment.  }
The constant of aberration defines the scale of the aberration increment, and is equal to $v/c = 100$ $ \mu$rad.  
Measurement of the spot shift with respect to the optical axis (which is parallel to the nadir direction) is equivalent to the determination of the angular displacement. 
As already pointed out, the proposed method of measuring the angle of aberration involves the use of Earth-based sources and has a big advantage. The rotation of the earth on its axis should produce a corresponding shift of the image. In principle, records could be taken over a period of one day. The image appears to move in an ellipse.
The major axis of the  ellipse is about 14 $\mu$m. The value of the minor axis depends on the observer's position and is given by $14 ( \sin \phi)$ $\mu$m.  \footnote{The small image size—approximately 2 $\mu$m in the numerical case considered—places significant limitations on the use of a CCD as a detector. To address this, one promising alternative is photon detection using a microchannel plate photomultiplier (MCP-PMT). A feasible approach involves using a commercially available precision 2D translation stage with a resolution of 0.5 $\mu$m to control the position of a submicron aperture in an opaque screen. For instance, the aperture diameter could be set to 1  $\mu$m or smaller. The screen is mounted horizontally on the translation stage, which is positioned in the image plane. The MCP-PMT can then detect photons transmitted through the aperture. This setup offers the advantage of compactness, with the overall height of the optical configuration being approximately one meter.}

\subsection{O the Possibility of Creating a Point Source in an Imaging System}

One possible method for creating a ``point'' source at optical wavelengths in a two-lens imaging scheme is to use a focused laser beam, as illustrated in Fig.~\ref{C554}. In this case, we implicitly assume that the diffraction angle of the virtual source exceeds the numerical aperture $NA$ of the two-lens imaging system, i.e.,
 $\lambdabar/w_0 > NA$, where  $w_0$ is the waist spot size of the focused Gaussian laser beam. 
 
 A further development of this approach is to employ a masked laser source. In this configuration, the point source is formed by a screen with a pinhole illuminated by a non-focusing laser beam. The screen is placed in the object plane of the two-lens imaging system. The diffraction angle of the radiation transmitted through the pinhole must exceed the numerical aperture of the system, i.e.,  $\lambda/D_h > NA$, where  $D_h$ is the diameter of the pinhole. 

In this kind of experiments it is often important to know the diffraction of electromagnetic waves by a small hole in an plane conducting screen.  The diffraction of electromagnetic radiation by a hole small compared with the wavelength is treated theoretically in \cite{BETH}.  The solution holds for a circular hole in a perfectly conducting plane screen.  It is practical interest to discuss the case of normal incident plane wave.   The total cross section is

\[
A = 64 k^4 R^6/(27\pi)  .
\]

Here $k = 1/\lambdabar$, and $R$ is the hole radius. The result for the diffracted field is entirely different that of the Kirchhoff theory because presented in \cite{BETH} solution satisfies the boundary condition while Kirchhoff's does not. This result is exact when (conducted) screen thickness is small compared with wavelength. In practice this condition may not be satisfied and one has to to use the results of paper \cite{BETH} carefully.

\subsection{A Finite-Size Incoherent Source in an Imaging System} 

We derived our results under the assumption that the dimensions of the source are on the order of the wavelength, $\lambda$. It should be emphasized, however, that the resolution of the proposed optical techniques is not fundamentally limited by the size of the incoherent source.

The radiation field produced by an incoherent source can be represented as a linear superposition of the fields emitted by many elementary point sources. The image of each elementary point source is described by a point spread function (PSF). For an optical system illuminated with incoherent light, the relationship between the object distribution and the resulting image distribution is given by a convolution integral. Within this framework, the optical system is completely characterized by its impulse response, namely the image of a point source. The rounding or softening of the source’s edges is a characteristic effect of the convolution ("smoothing") process.

It should be noted that any linear superposition of radiation fields from elementary point sources preserves the characteristic properties of the individual point-source images, such as the oscillatory image shift occurring with a period of one day. This observation explains why the proposed technique can also be applied to finite-size incoherent sources: the aberration angle can still be reconstructed from measurements of the image shift. In particular, objects with sharp edges are useful for detecting aberration-induced image displacements.

An alternative realization of a “point” source is an incoherent source with physically sharp edges. In this case, the achievable resolution is determined by the edge width of the source. For the example considered here, the edge width should be on the order of  $1 \mu$m.  One possible implementation is to use a masked spatially incoherent UV light source. A mask with a small aperture is placed at a distance $l$ from the spatially incoherent UV source. In this configuration, it is important to account for the angular filtering of the radiation emitted by the elementary point sources. The best resolution of the incoherent imaging system is obtained when the filtering angle $d/l$ exceeds the numerical aperture $NA$ of the two-lens imaging system, i.e., $d/l > NA$, where $d$ is the diameter of the mask aperture.

\newpage

\section{The Static Magnetic Field in Various Situations}

In discussing the induced electromagnetic fields in an accelerated frame, we have so far considered only the case of a particle source producing a static electric field. 
We now turn to the complementary situation and show why no net induced electromagnetic field arises in the case of an accelerated magnet.

Although this scenario does not directly occur in particle aberration measurements, its analysis provides valuable insight into the underlying physical principles.

\subsection{A Moving Magnet: Induced Electric Dipoles}

Let us consider a magnet - serving as the source of a static magnetic field - that is accelerated from rest to a velocity $\vec{v}$ in an inertial frame. 
Since there are no magnetic charges, any static magnetic field produced by steady currents can be regarded as arising from a distribution of magnetic dipoles. A current loop with magnetic moment $\vec{m}$ acquires, when moving with velocity $\vec{v}$, an induced electric dipole moment $ \vec{d}$. 

The Lorentz transformations give us a simple answer for what we see if a magnet is accelerated in an inertial frame.
From the Minkowski spacetime viewpoint, a three-vector  $\vec{m}$  is understood as a spacelike part of antisymmetric tensor (bi-vector) $m_{\mu\nu}  = 
(\vec{m},\vec{d})$. In Lorentz coordinates, there exists a kinematical constraint 

\[
m_{\mu\nu}u_{\nu} = 0 ,
\]  

which expresses the fact that electrical dipole $\vec{d}$ is vanishes in the magnet rest frame. Here  $u_{\nu} = dx_{\nu}/d\tau $ is the four-velocity of magnet. 
The coordinate-independent proper time $\tau$ is a parameter describing the evolution of the physical system under the relativistic laws of motion.

Now we have to translate back to three-vector notation.
We should probably ask first about the  3D motion of a test particle in a magnet field.
The electric field associated with a time-varying magnetic field is governed by Maxwell’s equation
\[
\vec{\nabla}\times \vec{E} =
-\frac{1}{c}\frac{\partial \vec{B}}{\partial t}.
\]

For a rigidly moving magnetic field profile, integration along the direction of motion yields
\[
\vec{E} = - \frac{1}{c}\,\vec{v}\times \vec{B}.
\]


To first order in $v/c$, the dipole moments transform as components of the antisymmetric tensor $m_{\mu\nu}$, giving
\[
\vec{d} = \frac{1}{c}\,\vec{v}\times \vec{m}.
\]

Thus, a moving magnet develops an induced electric polarization. The electric field associated with these induced dipoles is
\[
\vec{E} = \frac{1}{c}\,\vec{v}\times \vec{B}.
\]

We therefore identify two contributions to the electric field in a moving magnet:
\begin{itemize}
\item the field arising from $\partial \vec{B}/\partial t$,
\item the field produced by the induced electric dipoles.
\end{itemize}

These two contributions exactly cancel. Hence, the total electric field of a uniformly moving magnet is zero.

Now let us  see how we can write the four-vector of Lorentz force in three vector notations.
Based on the equivalence between active and passive Lorentz boosts within a single inertial frame we have the simple result.
The expression for  ``magnetic force''   $\vec{F} = e( \vec{v}^{(p)} - \vec{v} )  \times\vec{B}/c $ gives us answer for what we see if we move any system of fixed magnetic dipoles.  Here $\vec{v}^{(p)}$ is the velocity of the test particle in the inertial frame.

We now ask: how should the electromagnetic field be described when the magnet is at rest in an accelerated frame?

Starting from Maxwell’s equations in an inertial frame and applying an inverse Galilean transformation with velocity $-v$, we obtain, assuming $\vec{E}=0$ in the original frame,
\[
\vec{B}_n = \vec{B}, \qquad \vec{E}_n = \frac{1}{c}\,\vec{v}\times \vec{B}.
\]

The magnetic field becomes time-dependent in the accelerated frame due to spatial variation,
\[
\frac{\partial \vec{B}}{\partial t_n} = v\,\frac{\partial \vec{B}}{\partial x_n}.
\]

At the same time, the dipole moments transform as
\[
\vec{m}_n = \vec{m}, \qquad \vec{d}_n = -\frac{1}{c}\,\vec{v}\times \vec{m}.
\]

The electric field associated with these dipoles is therefore
\[
\vec{E} = -\frac{1}{c}\,\vec{v}\times \vec{B},
\]
which exactly cancels the field arising from $\partial \vec{B}/\partial t_n$. 

According to Galilean transformation, the force is equal to  $\vec{F} = e( \vec{v}^{(p)} - \vec{v}^{(m)} )  \times\vec{B}/c $  even in the accelerated frame. 
Thus, relativistic dynamics and electrodynamics give a simple answer: even in the accelerated frame, the electric field is zero, while the magnetic force depends only on the relative velocity.

\subsection{A Moving Magnetic Dipole: Lorentz-Transformation-Based Explanation}

We now examine more closely the physical origin of the induced electric dipole moment.

Formally, the transformation $\vec{d} = \vec{v}\times \vec{m}/c$ follows from the Lorentz transformation of the antisymmetric tensor $m_{\mu\nu} = (\vec{m}, \vec{d})$. However, it is instructive to understand this effect in more physical terms.

The transformation gives us a simple unswer for what we see if we move any sistem of fixed charges. For example, 
the reciprocal relation,
\[
\vec{m} = -\frac{1}{c}\,\vec{v}\times \vec{d},
\]
has a simple interpretation: a moving electric dipole produces a current density $\vec{j} = \rho \vec{v}$,  where $\rho$ is the charge density distribution of electric dipole.

How can we explain the induced electric dipole in terms of real moving charges? 
To understand the origin of the induced electric dipole, we examine the transformation of charge and current densities. To first order in $v/c$, Lorentz transformations give
\[
\vec{j} = \vec{j}' + \rho'\vec{v}, \qquad
\rho = \rho' + \frac{\vec{j}'\cdot \vec{v}}{c^2}.
\]

The second relation shows that a moving current distribution acquires an additional charge density of purely relativistic origin. This term is responsible for the induced electric dipole moment.

\medskip

Let us use the expession for current density $\rho$ to find the  electric dipole of a moving magnetic dipole.
We take the rectangular loop with the current $I$  and the area $A = ab$. We choose our coordinates as follow. There are no current in $z$-direction and there are currents in $x$-direction on the two sides of length $a$. This loop is moving along $x$-axis with velocity $v$.

The dipole moment, in this case, is the total charge on one rod times the separation between them:
$d = j' A_w ab v/c^2 = vIab/c^2  = vm/c $,
where $A_w$ is the area of a cross section of the wire and $m = Iab/c$ is the magnetic dipole of the rectangular current loop. In vector form, this can be expressed as
$\vec{d} = \vec{v}\times\vec{m}/c$, in agreement with the tensor transformation.

\subsection{A Moving Magnetic Dipole: Galilean Transformations-Based Explanation}

We would like now to describe an apparent paradox. Suppose we change clock synchronization and introduce the absolute time coordinatization. 
In the absolute time coordinatization, solving the dynamics problem in the lab frame does not involve Lorentz transformations. This means that, within the lab frame, particle motion appears exactly as predicted by classical mechanics, which assumes an absolute time framework. 

The Galilean transformation law is

\[
\vec{x} = \vec{x}' + \vec{v}t'  , \qquad    t = t' .  
\]

We now ask about the charge density produced by the moving current density. From kinematical viewpoint the answer is follows

\[
\vec{j}  = \vec{j}'  +  \rho'\vec{v}  , \qquad   \rho = \rho'  .
\]

Based on Galilean kinematics we would expect that there is no electric dipole related with moving magnetic dipole.

This result is especially surprising, because we are based on the use of the relativistic dynamics even in the absolute time coordinatization. Where does electric dipole come from?  Interesting that the approximate (to the first order in $v/c$) relativistic formula $\rho = \rho' + \vec{j}'\cdot\vec{v}/c$ includes the second order relativistic term. Actually the term  $\vec{j}'\cdot\vec{v}/c$ proportional to product of velocity $\vec{v}'$ of moving charges in magnetic dipole and traslation velocity $\vec{v}$.

At first glance, the mechanism of moving magnetic dipole described may not seem to involve second order relativistic effects. However, it is implicitly present in the assumption that mass of a moving charges corresponds to its relativistic mass in the corrected second Newton's law.  We can now use our knowledge that described in the Lorentz coordinatization second order (i.e.  $\propto c^{-2}$) effect  linearly depends on translation velocity $v$ to find the same effect in absolute time coordinatization.

We will consider a simplified magnetic dipole - small current loop. 
In normal conductor, the electric currents come from the motion of negative electrons -called the conduction electrons. 
We let the charge density of conduction electrons be $\rho'$and their velocity  be $\vec{v}'$ in magnetic dipole at rest in the inertial frame.   
Now we turn our attention to what happens in inertial frame, in which magnetic dipole moves with velocity $\vec{v}$.

The mass of the moving electrons changes with the velocity $\vec{v}$

\[
M      = m_e/\sqrt{1 - |\vec{v} + \vec{v}' + \Delta\vec{v}'|^2/c^2} . 
\]

Where $\Delta\vec{v}'$ is the velocity change of conducting electrons.

Now we can ask about the electron mass change. For obvious reason, the velocity change of conducting electron into the current loop $\Delta\vec{v}'$ is the second order ( $\propto  c^{-2}$) term. We can, therefore, write to  the second order  term which linearly depends on  $\vec{v}$. That is $\Delta M  =  m_e \vec{v}\cdot\vec{v}'/c^2$  .

In the non-covariant (3+1) approach, the dynamics in the lab frame are formulated without reference to Lorentz transformations.
The four-dimensional equations of motion are decomposed into three spatial and one temporal component, effectively eliminating any mixing between space and time in the dynamical equation. 

It is important to emphasize that Einstein’s velocity addition is fully compatible with conventional  non-covariant  relativistic dynamics within the inertial frame, which itself relies on the absolute time convention. Crucially, Einstein’s velocity addition inherently incorporates relativistic time dilation. Similarly, conventional non-covariant approach  also accounts for relativistic time dilation, albeit implicitly, through the use of relativistic mass for moving particles A more detailed discussion is given in Chapter 17.

Applying   Einstein’s  (three-vector)) longitudinal velocity addition, we find that the velocity change of conducting electrons is

\[
\Delta\vec{v}' =  - \vec{v}'(\vec{v}\cdot\vec{v}')/c^2 .
\]

If we call $\vec{v}'+\Delta \vec{v}'$ the velocity  and $\rho' +\Delta\rho'$ the charge density  of electrons into the moving curret loop  
we have equality

\[
\rho'\Delta\vec{v}'  = - \Delta\rho\vec{v}' ,
\]

because current is the same.  So we have  $\rho'(\vec{v}\cdot\vec{v}')/c^2 = \Delta\rho$. The total charge density is then

\[
\rho = \rho' + \vec{j}'\cdot\vec{v}/c^2  .
\]

We conclude that the charge density in the moving magnetic dipole  in the absolute time coordinatization is just equal to the charge density in the Lorentz coordinatization as  it must be.

\subsection{Faraday's ``Flux Rule''}

We now turn to Faraday's flux rule, which states that the electromotive force (emf) in a circuit is equal to the rate of change of magnetic flux through it, regardless of whether the change arises from motion of the circuit, variation of the field, or both.

In standard textbook treatments, two distinct mechanisms are invoked:
\begin{itemize}
\item the magnetic force term $e\vec{v}\times \vec{B}/c$ for moving circuits,
\item the induced electric field from $\partial \vec{B}/\partial t$ for time-varying fields.
\end{itemize}

As emphasized by Feynman, this dual explanation is conceptually unsatisfactory. \footnote{As Feynman  noted \cite{FEY}:
``We know of no other place in physics where such a simple and accurate general principle requires for its real understanding an analysis in term of two different phenomena. }

\medskip

The analysis developed here provides a unified relativistic interpretation. A moving magnet develops induced electric dipoles whose fields exactly compensate the field arising from $\partial \vec{B}/\partial t$. It is clear that, according to special relativity, when the magnet is moved, we have only  magnetic force  on the test charged particle. 
As a result, the electromagnetic interaction depends only on the relative motion between the magnet and the charges.

Generally, the Lorentz force law, together with measurements of the acceleration components of test particles, serves to define the components of the electric and magnetic fields. Once these field components are determined from the acceleration of test particles, they can be used to predict the accelerations of other particles.

For example, suppose we want to determine the fields in our inertial frame when we are at rest. What do we observe? In general, the force per unit charge is given by
 $\vec{F}/e = \vec{E} + \vec{v}\times\vec{B}/c $. From the acceleration of a test particle, we infer the magnetic field $\vec{B}$ and, in additional, an electric field   $\vec{E} = - \vec{v} \times\vec{B}/c $. Within this approach, the ``flux rule'' must therefore be understood as the combination of two completely independent effects.

The relativistic approach provides another way to determine the fields of a moving magnet while remaining entirely within a single inertial frame. Relativistic dynamics and electrodynamics give a simple answer: the electric field is always zero, while the magnetic force depends only on the relative velocity.
Thus, relativistic transformations express the same physics in a different language.

These two approaches are equivalent in all their physical predictions. However, they are not psychologically identical when we attempt to pass to the accelerated frame.
It is crucial to note that Langevin metric describes the electrodynamics of the accelerated magnet with the viewpoint of the accelerated observer measurements as viewing this of the inertial observer. 

The conventional approach leads to a peculiar conclusion: when we pass to the accelerated frame, the electric field is no longer zero, and in addition there is a magnetic force on the test particle at rest in the accelerated frame. These two contributions exactly cancel each other. How can this be?

Intuition would seem to suggest that everything is at rest, so the magnetic force term should vanish, and the electric field should also be zero.

However, there is another way of interpreting the situation. Relativistic theory indicates that this intuition is in fact correct: both contributions are absent altogether, exactly as one would naturally expect.

\newpage

\section{Special Relativity and Reciprocal Symmetry}

\subsection{Pseudo-Gravitational Fields and the Langevin Metric}

This chapter examines the dynamical content underlying special-relativistic
effects.  Although relativity is often presented primarily as a theory of
kinematics, many relativistic phenomena acquire their clearest interpretation
when viewed from a dynamical perspective.  In this sense, the metric structure
of space--time reflects how physical fields and material systems are perceived
by different classes of observers.

A frequently discussed issue is the absence, within special relativity itself,
of an explicit dynamical interpretation of the Langevin metric.
The apparent asymmetry between inertial and accelerated frames may be understood
as arising from a pseudo-gravitational field experienced by an accelerated
observer.  The purpose of this chapter is to demonstrate that non-inertial
reference systems in flat (pseudo-Euclidean) space--time can be consistently
described using concepts closely related to those employed in general relativity,
without invoking genuine gravitational curvature.

Let $S$ denote an inertial reference frame at rest with respect to the fixed
stars, with Lorentz coordinates $(ct,x,y,z)$.  Consider a second system $S_n$
initially at rest relative to $S$ that undergoes uniform acceleration along
the $x$-axis until reaching velocity $v$.  The inverse Galilean transformation
\[
x = x_n + vt, \qquad t = t_n ,
\]
defines coordinates comoving with the accelerating system.
After the acceleration phase, $S_n$ becomes an inertial frame moving with
velocity $v$ relative to $S$, while $S$ moves with velocity $-v$ relative to
$S_n$.

At first sight the two inertial frames appear completely symmetric, in
accordance with the principle of relativity.  Each observer measures the
Minkowski interval
\[
ds^2=c^2dt^2-dx^2-dy^2-dz^2
\]
in their own coordinates.  The question therefore arises:  
where does the observable asymmetry between inertial and accelerated motion
originate?

The resolution follows from the equivalence principle.
An accelerated observer experiences a pseudo-gravitational field that acts on
all bodies proportionally to their mass.
This field is not a real gravitational interaction but an inertial effect
associated with acceleration.
A uniformly accelerating frame may therefore be treated as an inertial frame
immersed in a uniform pseudo-gravitational potential.

To simplify the discussion we retain terms only up to order $(v/c)^2$.
Assume that, during an interval
\[
T=\frac{v}{|g|},
\]
the system $S$ experiences constant acceleration
$g=-|g|$ along the negative $x_n$ direction.
From the viewpoint of $S_n$, the inertial frame $S$ undergoes free fall in a
pseudo-gravitational field whose acceleration satisfies
\[
g=-\frac{\partial\phi}{\partial x_n}.
\]
At the end of the acceleration stage one finds
\[
x_n=\frac{gT^2}{2}+x ,
\]
and the corresponding pseudo-gravitational potential difference becomes
\[
\phi=-\frac{v^2}{2},
\]
where the potential is chosen to vanish at $x_n=x$.
Once uniform motion is reached, the pseudo-gravitational field disappears.

Although no real gravitational field is present, the equivalence principle
allows us to employ gravitational reasoning.
In a gravitational potential $\phi$, clocks run at different rates.
If light of frequency $\omega_1$ emitted at potential $\phi_1$ reaches a
point with potential $\phi_2$, its locally measured frequency becomes
\[
\omega_2=\omega_1\!\left[1+\frac{\phi_1-\phi_2}{c^2}\right].
\]
Because frequency is inversely proportional to proper time,
the relation between proper time and coordinate time in the accelerated frame
is
\[
dt^{(p)} = dt_n\!\left(1+\frac{\phi}{c^2}\right).
\]
Clocks situated deeper in the potential therefore run more slowly.

Consider now a freely falling reference frame initially located at potential
$\phi_1$ and descending toward a region with lower potential $\phi_2$.
During free fall the frame retains the time rate associated with $\phi_1$.
To an observer at $\phi_2$, clocks in the freely falling frame appear to run
faster, and radiation emitted within that frame is observed to be blue shifted.

An analogous effect occurs in uniformly accelerated systems.
Using $\phi=-v^2/2$, one obtains
\[
dt^{(p)} = dt_n\!\left(1-\frac{v^2}{2c^2}\right),
\]
where $dt_n$ is the coordinate time of the accelerated frame but coincides
with proper time in the inertial frame $S$.
Thus an accelerated observer interprets inertial clocks as running faster,
a result that explains the asymmetry of aging purely through pseudo-gravitational
time dilation.

The relativity of simultaneity also acquires a dynamical interpretation.
Two clocks separated along the direction of acceleration experience different
effective potentials,
\[
\Delta\phi
=\frac{\partial\phi}{\partial x_n}
\,[x_n(1)-x_n(2)] ,
\]
leading to the accumulated time offset
\[
t^{(p)}(1)-t^{(p)}(2)
=\frac{v[x_n(1)-x_n(2)]}{c^2}.
\]
Hence simultaneity differences arise from integrated pseudo-gravitational
effects during acceleration.

The Lorentz deformation of lengths can be derived in the same manner.
Integrating the time-offset relation for an infinitesimal velocity change
$dv$ gives
\[
dx^{(p)} = dx_n\!\left(1+\frac{v^2}{2c^2}\right),
\]
showing that spatial geometry in the accelerated frame differs from that in
the inertial frame even in the absence of physical forces.
Relativistic time dilation, length deformation, and simultaneity shifts
therefore emerge as mutually connected consequences of a generalized
equivalence principle.

Combining these results, the physical space--time variables
$(x^{(p)},t^{(p)})$ relate to the initial inertial frame coordinates
$(x_n,t_n)$ as
\[
cdt^{(p)} = dt_n\!\left(1-\frac{v^2}{2c^2}\right)
-\frac{vdx_n}{c^2}\!\left(1+\frac{v^2}{2c^2}\right),
\qquad
dx^{(p)} = dx_n\!\left(1+\frac{v^2}{2c^2}\right).
\]
Extending the result to arbitrary velocity yields
\[
dt^{(p)}= dt_n \sqrt{1-v^2/c^2} - \frac{vdx_n/c^2}{\sqrt{1-v^2/c^2}},
\qquad
dx^{(p)}=
\frac{dx_n}{\sqrt{1-v^2/c^2}} .
\]

We have above derived an expression of physical time  $dt^{(p)}$ and distance  $dx^{(p)}$ in terms of physical time   $dt_n$ and distance $dx_n$ of the initial inertial frame.

The measurable differentials $dt^{(p)}$ and $dx^{(p)}$ define the
physical space--time interval,
\[
ds^2=(cdt_n^{(p)})^2-(dx_n^{(p)})^2
=(1-v^2/c^2)c^2dt_n^2-2v\,dx_n dt_n-dx_n^2 ,
\]
which is precisely the Langevin metric. For light signal the interval is zero, and the physical velocity of light in the accelerated frame is $dx^{(p)}/dt^{(p)} = c$. 


The crucial point is that the metric determines the *form of physical laws*,
not the physical phenomena themselves.
Observable effects—such as light aberration in an accelerated frame—follow
from solving the dynamical field equations together with appropriate initial
conditions.
The Langevin metric provides the geometric framework within which those
dynamical processes unfold.

\subsection{The Equivalence Principle and Gravitational Physics}

We should add a further remark concerning the equivalence principle.\footnote{For a general discussion of the equivalence principle, we recommend \cite{Log2}.}
It is commonly believed that the equivalence principle lies at the foundation of general relativity. However, this perspective is not entirely accurate. Gravity, as understood through Einstein’s insight, is not merely a universal force acting identically on all bodies—it is fundamentally tied to the geometry of space-time itself.

To appreciate this idea, one must consider the role of the Riemann curvature tensor, which, in a precise mathematical sense, represents the gravitational field. The vanishing of this tensor is the necessary and sufficient condition for the absence of gravity. This fact alone shows that the traditional formulation of the equivalence principle has only limited applicability.

While the equivalence principle holds for mechanical processes, it breaks down when applied to electrodynamical phenomena. This distinction implies that internal measurements can, in principle, distinguish between an inertial frame and a frame in free fall within a homogeneous gravitational field.

For instance, consider a free electric charge in such a gravitational field. Relative to an inertial frame, it experiences uniform acceleration and consequently emits radiation—regardless of whether the gravitational field is weak or strong. In contrast, in a uniformly accelerated frame in flat space-time (i.e., in the absence of gravity), a freely falling charge does not accelerate relative to the inertial frame and, according to classical electrodynamics, does not radiate. Thus, in this context, the equivalence principle is explicitly violated.

Real experiments performed on Earth combine pseudo-gravitational effects
arising from acceleration with genuine gravitational fields.
In weak fields, where $|\phi|/c^2\ll1$, general relativity reduces to
Newtonian gravity.
The space--time metric then takes the form
\[
ds^2=(1+2\phi/c^2)c^2dt^2
-(1-2\phi/c^2)(dx^2+dy^2+dz^2),
\]
with $\phi=-GM/r$.

For rotating systems such as the Earth orbiting the Sun, inertial effects are
described by the Langevin metric,
\[
ds^2=(1-v^2/c^2)d(ct)^2
-2(\vec v\!\cdot d\vec r)dt
-(d\vec r)^2 .
\]
Unlike the static gravitational metric, the accelerated-frame metric contains
non-diagonal terms reflecting inertial motion.
This difference explains why electromagnetic processes can distinguish between
true gravitation and acceleration.

For an Earth-based observer both effects coexist.
To the adopted order of approximation the invariant interval becomes
\[
ds^2=(1+2\phi/c^2-v^2/c^2)d(ct)^2
-2(\vec v\!\cdot d\vec r)dt
-(1-2\phi/c^2)(d\vec r)^2 .
\]
This expression represents the superposition of the weak solar gravitational
field and the Langevin metric associated with Earth's motion.
Because both contributions are small, the Langevin metric alone provides an
excellent approximation for many terrestrial relativistic experiments.

\subsection{Time Dilation}

Special relativity is often described as a reciprocal theory, since all inertial
reference frames are physically equivalent.  Nevertheless, the phenomenon of
time dilation provides an instructive example showing that reciprocity can be
subtle once acceleration enters the discussion.

Consider two reference frames.  Frame $S$ is inertial and taken to be at rest
relative to the fixed stars.  A second system, $S_n$, initially coincides with
$S$ but undergoes acceleration along the $x$-axis until it reaches velocity $v$.
As discussed previously, the transition from an inertial to an accelerated
description may be introduced through the Langevin transformation
\[
t_n=t, \qquad x_n=x-vt ,
\]
which has the form of a Galilean boost.
Under this transformation the Minkowski metric continuously transforms into
the Langevin metric of the accelerated frame $S_n$
(see Chapter~5).

Although coordinate clocks are fixed in the accelerated coordinates,
they measure the same coordinate time as clocks in the inertial frame.
However, the appearance of the mixed term $dt_n dx_n$ in the Langevin metric
implies that clocks at different spatial positions in $S_n$ are not
Einstein-synchronized by light signals.

Suppose two identical clocks coincide at $t=0$ in frame $S$.
The first clock remains at rest in $S$, while the second begins moving with
constant velocity $v$ along the $x$-axis.
We now describe the situation from the viewpoint of the accelerated frame
$S_n$, in which the second clock is permanently at rest.

In frame $S_n$ the first clock moves with velocity
\[
\frac{dx_n}{dt_n}=-v .
\]
Using the Langevin metric
\[
ds^2=c^2(1-v^2/c^2)dt_n^2-2v\,dx_n dt_n-dx_n^2 ,
\]
the proper time of the moving clock becomes
\[
d\tau=\frac{ds}{c}=dt_n .
\]
Thus the proper time of the inertial clock coincides with the coordinate time
of the accelerated frame.

The second clock, which is at rest in $S_n$, satisfies $dx_n=0$, giving
\[
d\tau_n=\sqrt{1-v^2/c^2}\,dt_n .
\]
Hence the clock at rest in the accelerated frame runs slower.
Importantly, this slowdown is independent of the frame used for the analysis;
time dilation is therefore not a purely reciprocal effect once acceleration
is involved.

Consider now two frames, $S_n^{(1)}$ and $S_n^{(2)}$, both accelerated relative
to the inertial frame $S$ until they reach equal speeds $v$ in opposite
directions.
From the viewpoint of $S$ the configuration is perfectly symmetric.
Consequently, clocks at rest in $S_n^{(1)}$ and $S_n^{(2)}$ must experience the
same time dilation relative to clocks in $S$.

Let one clock be at rest in $S_n^{(1)}$ and another in $S_n^{(2)}$.
From the perspective of $S_n^{(1)}$ the second clock moves with velocity
$dx_n/dt_n=-2v$.
Substituting this motion into the Langevin metric yields
\[
d\tau_2=\sqrt{1-v^2/c^2}\,dt_n .
\]
Thus
\[
d\tau_1=d\tau_2 ,
\]
even though the two frames possess nonzero relative velocity.
This example demonstrates an important point:
\emph{relative velocity alone does not necessarily produce time dilation}.
The decisive factor is the history of acceleration.

The result admits a natural interpretation.
Each accelerated frame contains a pseudo-gravitational potential arising from its
non-inertial motion.
Clocks that share identical acceleration histories experience identical
pseudo-gravitational potentials and therefore run at the same rate, even when
they subsequently occupy different inertial frames.

Finally, consider a hierarchical acceleration process.
Frame $S_n$ first accelerates relative to $S$ to velocity $v$.
Two additional frames, $S_n^{(1)}$ and $S_n^{(2)}$, then accelerate relative
to $S_n$ with equal speeds $v$ in opposite directions.

From the viewpoint of $S_n$ the situation is asymmetric.
Frame $S_n^{(1)}$ moves with velocity $-v$, giving
\[
d\tau_n^{(1)}=dt_n .
\]
Frame $S_n^{(2)}$, however, moves with velocity $v$ relative to $S_n$.
Using the Langevin metric one finds
\[
d\tau_n^{(2)}=\sqrt{1-4v^2/c^2}\,dt_n .
\]

How does this appear in the original inertial frame $S$?
There, frame $S_n^{(2)}$ moves with velocity $2v$.
Applying the Minkowski metric immediately gives
\[
d\tau_n^{(2)}=\sqrt{1-4v^2/c^2}\,dt ,
\]
in agreement with the previous result.

The simplicity of these relations reflects the use of Galilean boosts and
Galilean velocity addition within the accelerated description.
Despite this simplicity, the examples illustrate a fundamental lesson:
time dilation is governed not merely by instantaneous relative velocity,
but by the dynamical structure of acceleration and the associated
pseudo-gravitational fields.

\subsection{The Light-Clock and Observations by a Non-Inertial Observer}

To understand how a clock slows down in an accelerated system, we examine its internal workings using a simplified example—a light clock. This device consists of a meter stick with mirrors at both ends; a ``tick'' corresponds to a light pulse bouncing between the mirrors.  

\begin{figure}[h!]
	\centering
	\includegraphics[width=0.85\textwidth]{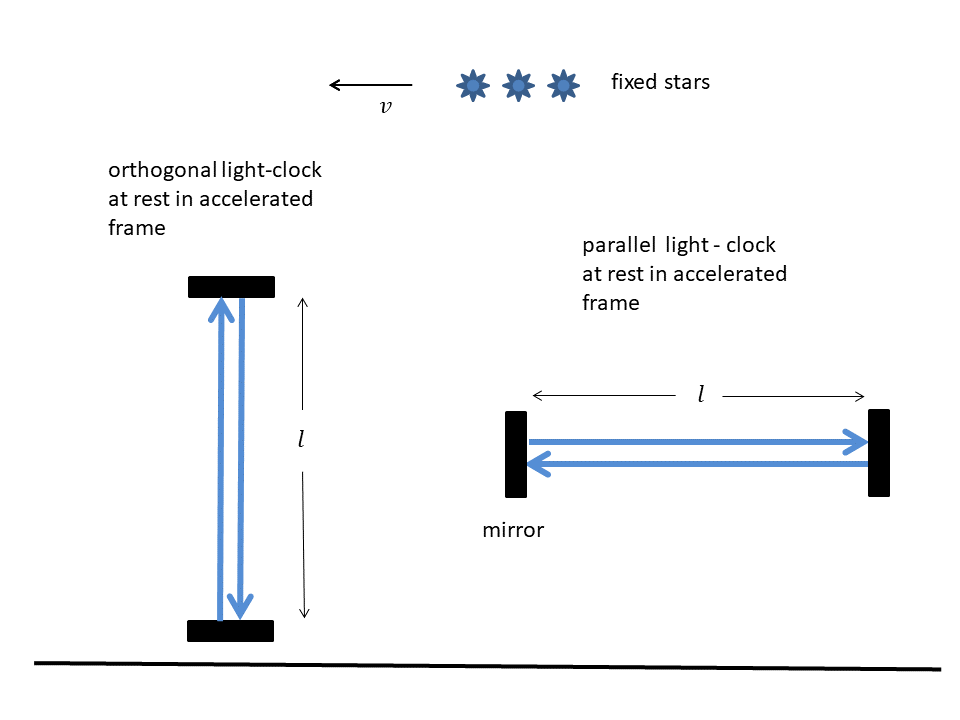}
	\caption{Parallel and orthogonal light-clock setups.}
	\label{B222}
\end{figure}

Consider two configurations of the light clock: one with its length parallel to the direction of motion, and the other oriented orthogonally. According to the principle of relativity—pseudo-Euclidean spacetime geometry—all clocks, regardless of construction, measure time consistently in all inertial frames. We will demonstrate that, irrespective of the light clock’s orientation relative to its motion, the elapsed time measured remains consistent. Two examples of these configurations are illustrated in Fig.~\ref{B222}.

We now analyze the behavior of a physical light clock under acceleration, from the perspective of an observer co-moving with the accelerated frame. First, consider the light beam traveling between two mirrors aligned parallel to the direction of motion relative to the fixed stars.  

Within the Langevin metric, the speed of light emitted by an accelerated source is no longer simply $c$. For an infinitesimal displacement $ds$ along the worldline of a light beam, setting $ds^2 = 0$ in Eq.~(\ref{GGG3}) gives
\[
c^2 = \left(\frac{dx_n}{dt_n} + v\right)^2,
\]
implying that, in the accelerated coordinate system $(ct_n,x_n)$, the light velocity along the $x_n$-axis is  $dx_n/dt_n = c - v$ in the positive direction, and $dx_n/dt_n = -c - v$  in the negative direction. 

Let $l_n$ denote the rod length in this frame. The time for light to traverse from the left mirror to the right is 
\[
t_1 = \frac{l_n}{c-v},
\]
and for the return path 
\[
t_2 = \frac{l_n}{c+v}.
\] 
The total interval is
\[
\Delta t_n = t_1 + t_2 = \frac{2l_n}{c(1 - v^2/c^2)}.
\]

To determine the length of physical rods, we use rods identical to those in the inertial frame but now at rest in the accelerated frame. Spatial distances are interpreted via $\sqrt{-ds^2}$, representing the length of an infinitesimal rod whose endpoints are simultaneous in the rod’s rest frame. From Eq.~(\ref{GGG3}), for $dt_n = 0$, we have $dx_n = dx$, indicating that a rod at rest in the inertial frame maintains its length in the accelerated frame. However, simultaneity differs for rods at rest in the accelerated frame, so their lengths generally differ from their inertial-frame counterparts.

The accelerated frame’s spatial geometry is described by the differential spatial line element $dl_n$, which provides a coordinate-independent measure of distance. By contrast, setting $dt_n = 0$ in Eq.~(\ref{GGG3}) yields $dl^2 = dx_n^2$, which differs from $dl_n$.  

In coordinates $(t_n,x_n,y_n,z_n)$, the spacetime interval is
\[
ds^2 = c^2 g_{00} dt^2 + 2c g_{01} dx dt + g_{11} dx^2,
\]
which can be decomposed as
\[
ds^2 = c^2 \Big(\sqrt{g_{00}}\, dt + \frac{g_{01}}{c \sqrt{g_{00}}} dx \Big)^2 + \Big(g_{11} - \frac{g_{01}^2}{g_{00}}\Big) dx^2.
\]  
Here, $dT^2 = \big(\sqrt{g_{00}}\, dt + g_{01} dx/(c \sqrt{g_{00}})\big)^2$ is time-like and represents physical time in the accelerated frame, while $dX^2 = ( - g_{11} + g_{01}^2/g_{00}) dx^2$ is space-like, corresponding to physical length. These quantities are physically meaningful and invariant under coordinate transformations within the accelerated frame.

From Eq.~(\ref{GGG3}), the three-dimensional spatial line element satisfies
\[
dl_n^2 = \frac{dx_n^2}{1 - v^2/c^2},
\]
showing that the physical length $dl_n$ in the accelerated frame is longer than the corresponding coordinate distance $dx_n$. This asymmetry reflects the effect of acceleration relative to the fixed stars, indicating that space in an accelerated frame is inherently non-Euclidean.  

The Langevin metric describes measurements made by an accelerated observer as interpreted by an inertial observer employing absolute simultaneity—maintaining the same synchronized clocks used when the light clock was at rest. Within this framework, the inertial observer sees the accelerated observer measuring the light velocity along the $x$-axis as $c - v$ (positive) and $-c - v$ (negative), and the meter stick with mirrors appears contracted along the motion direction.  

Hence, the total time for the parallel light clock in the accelerated frame is
\[
\Delta t_n = \frac{2l}{c\sqrt{1 - v^2/c^2}}.
\]
Since $\Delta t = 2l/c$ in the inertial frame, this demonstrates that clocks at rest in the accelerated frame tick more slowly than those in the original inertial frame.

Next, consider a light clock accelerated to velocity $v$ perpendicular to the light beam. In this orthogonal orientation, length contraction along the motion direction does not occur. For a light pulse, $ds^2 = 0$ implies
\[
\left(\frac{dz_n}{dt_n}\right)^2 = c^2(1 - v^2/c^2),
\]
yielding a total light travel time
\[
\Delta t_n = \frac{2l}{c \sqrt{1 - v^2/c^2}}.
\]  

Since $t_n = t$ and $\Delta t = 2l/c$ represent the time indicated by a physical clock at rest in the inertial frame, this result shows that clocks at rest in the accelerated frame run slower compared to those in the original inertial frame.

\subsection{Metrics in Accelerated Frames}

In this section, we continue our exploration of non-reciprocity in special relativity. We begin by examining the implications of assuming that the inertial frame has no history of acceleration and that the space-time metric, as perceived by an observer at rest, is diagonal.

Consider two emitters within this inertial frame: the first remains stationary, while the second moves with velocity
$v$. A Galilean boost transforms the metric associated with the moving emitter, resulting in the form given in Eq. (\ref{GGG11}). In the coordinates of the inertial frame, the non-accelerated emitter is described by the standard diagonal Minkowski metric: $ds^2 = c^2dt^2 - dx^2- dy^2 - dz^2$. 

Now, let us shift our attention to the accelerated frame in which the second emitter is at rest, and the first appears to move with velocity $-v$. This raises a fundamental question: Does the principle of reciprocity still apply in this context?

Conventional theory holds that acceleration does not intrinsically violate the symmetry of motion between inertial and accelerated frames. Therefore, an observer in the accelerated frame should also describe the metric of the stationary (second) emitter as diagonal. By reciprocity, the moving (first) emitter should then be described in this frame by a metric analogous to Eq. (\ref{GGG11}), differing only by the sign of $v$.

However, this leads to a deeper inquiry into the principle of reciprocity itself.
The principle of reciprocity states that all inertial frames are equivalent. However, this relativity principle does not hold and the Lorentz covariance of all physical equations  does not dictate the reciprocal symmetry of nature.

A geometric analysis of space-time in the accelerated frame reveals this asymmetry. Specifically, within the accelerated frame, the Langevin metric—given in Eq. (\ref{GGG3})—emerges as the correct description of the second emitter’s space-time. This metric governs electrodynamics in the accelerated frame and departs from the simple diagonal form of inertial frames.

This departure has significant consequences: Maxwell’s equations, valid in inertial frames, no longer retain their form for an observer at rest in the accelerated frame. This breakdown signals a deeper structural difference between inertial and accelerated reference frames. Notably, because we adopt an absolute time coordinatization, the initial conditions—such as the orientation of radiation wavefronts—remain identical in both frames.

We now arrive at a paradox. The equivalence of all inertial frames is often supported by the claim that, after an appropriate coordinate transformation (e.g., diagonalization of the metric tensor), both the inertial and accelerated frames yield diagonal metrics. This seems to imply symmetry: both frames appear to be related by a Lorentz transformation and should be physically equivalent.
\footnote{Many physicists argue that the universality of Maxwell's equations across inertial frames rules out the existence of any privileged frame. As Dieks notes \cite{Di}, “As we already mentioned, it is a basic principle of the special theory of relativity that the line element $ds$ supplies all information about the physics of the situation, as described in the given coordinates.”}

However, a key asymmetry becomes evident when we consider the physical history of the frames. While the inertial frame remains unaccelerated, the accelerated frame retains a memory of its dynamical past. This memory is encoded in the metric structure. After diagonalization, the metric tensor in the accelerated frame undergoes a discontinuous shift—from a form such as $g_{00} = (1 - v^2/c^2) < 1, g_{01} = - v/c$ to the standard inertial values: $g_{00} = 1, g_{01} = 0$. This discontinuity highlights the fact that the Lorentz transformation does not preserve the smoothness of the metric across such transitions. Hence, the apparent symmetry between frames breaks down when physical considerations—specifically, the history of acceleration—are taken into account.

This leads us to a subtle but important conclusion: although the mathematical form of the metric may appear similar in both frames, the physical context—particularly the presence or absence of past acceleration—introduces observable differences. The accelerated frame thus retains a “memory” of its non-inertial history, breaking the naive equivalence suggested by purely kinematic symmetry arguments.

It is important to note that any space-time symmetry of a theory is a dynamical symmetry of the theory. Equivalence of  active and passive boosts  within an inertial frame is a manifestation of the principle.
In Chapter 3, we analyzed the electrodynamics of a moving emitter using the metric given in Eq. (\ref{GGG11}). Within a single accelerated frame, during the inertial phase after acceleration, the symmetry principle—namely, the equivalence of active and passive Galilean boosts—is upheld. For further details, refer to Section  5.9.

\newpage

\section{Relativistic Measurements}

\subsection{The Concept of Time in Special Relativity}

The theory of relativity reveals that our intuitive understanding of time does not align with its true nature. To clarify the concept of time, we adopt the principle that "each physical quantity is defined by the method used to measure it." Let us explore how this principle applies.
Consider the case of length. We have a standard unit of length, and we can measure the length of any object accordingly. However, this measurement refers to objective physical length only when the object is at rest. For moving objects—such as a rod in motion—length becomes a convention-dependent quantity. It relies on the chosen method of synchronization and lacks an exact, objective meaning.
With this in mind, we can now turn to the problem of defining time within the framework of special relativity.

Here is a simple example that illustrates the concept clearly. Suppose we know the law governing muon decay in its rest frame. When we apply a Lorentz transformation to this decay law, we find that, in the laboratory frame, the characteristic lifetime of the particle appears to increase from $\tau_0$ to $\gamma \tau_0$. This result can be interpreted as follows: after traveling a distance of $\gamma v \tau_0$, the number of muons observed in the lab frame is reduced to half of the original population.
In this context, the time measured corresponds to a "length". This leads to an intriguing question: could this interpretation apply to all time measurements within the framework of special relativity? We argue that the answer is yes.

Our next example concerns frequency measurements.
Consider a Fabry-Perot interferometer. In this setup, the frequency of light is effectively measured through the "length" of the standing wave it produces. Frequency measurements in such cases inherently follow the principles of interference.
Another example that illustrates the connection between frequency and interference is the grating spectrometer. Here, the frequency is inferred from the position of the light spot along the dispersion direction.
Although their designs differ, there is no fundamental physical distinction between a diffraction grating and an interferometer—both rely on the interference of incident and reflected light waves to perform frequency measurements.

\subsection{Comparison of Clock Rates between Two Inertial Frames}

Let us now explore how to compare the clock rates between the initial inertial frame $S$ and the accelerated frame $S_n$.
It is generally understood that when comparing clock rates between two frames $S$ and $S_n$ in relative motion, 
one cannot simply compare the readings of a single clock in each frame. This is because clocks from different frames coincide at the same spatial point only once. Therefore, in at least one of the frames, multiple clocks—assumed to be synchronized—must be used for meaningful comparison.

We have already pointed out that, within the framework of special relativity, all time measurements ultimately reflect what is essentially a measurement of "length."
When it comes to measuring the length of objects at rest, we are dealing with an objective physical quantity—one that is independent of any particular convention. Since our empirical access is limited to such length measurements, it is difficult to accept the standard textbook claim that one can only compare the reading of a single clock in one frame with the readings of multiple synchronized clocks in another.
Let us now apply our operational interpretation of time measurements to the example of a light clock, in order to see how this approach functions in practice.

\begin{figure}
	\centering
	\includegraphics[width=0.85\textwidth]{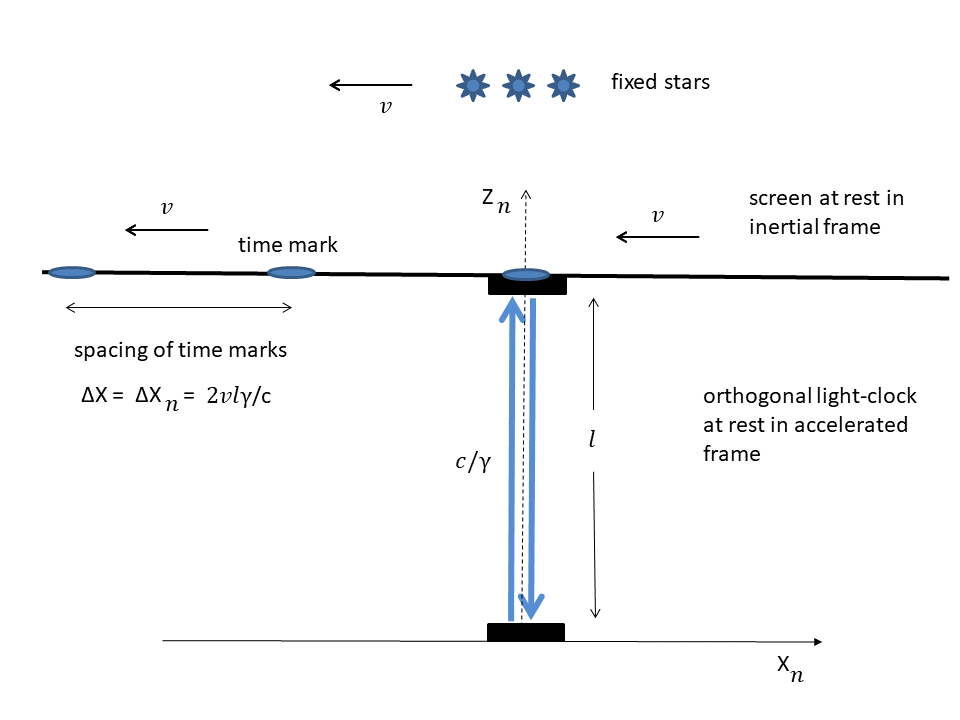}
	\caption{The  "light clock" thought experiment. Time marks can be imprinted on the moving screen by clock. Observations made by an observer in the same accelerated frame as the light clock. The anisotropy of speed of light presented in the absolute time coordinatization. The distance $\Delta x_n$ through which the screen in the lab frame falls in the time (i.e. spacing of the time marks) is $\Delta x_n = \Delta x = 2vl\gamma/c$.}
	\label{B878}
\end{figure}

Let us consider a specific thought experiment. We examine a situation in which time marks can be imprinted on a moving 
object (a screen) by a clock,  as sketched in Fig. \ref{B878}. To begin, we describe an orthogonal light clock from the perspective of an observer in the 
same accelerated frame as the clock. The screen is at rest in the initial inertial frame. Suppose the top mirror of the 
light clock is semitransparent, allowing time marks to be recorded on the (moving) screen, thereby measuring the time 
intervals between clock "clicks." In the case of the Langevin metric the speed of light emitted 
transversely by the accelerated source is given by $c\sqrt{1 - v^2/c^2}$. Accordingly, the time interval between the 
emission and return of the light signal is $\Delta t_n = 2l/[c\sqrt{1 - v^2/c^2}]$. During this interval, the object 
(i.e., the screen in the initial inertial frame) falls a distance $\Delta x_n  = \Delta x = 2lv/[c\sqrt{1 - v^2/c^2}]$,
which corresponds to the spacing between consecutive time marks.

Taking into account the Langevin metric (Eq.~\ref{GGG3}), we find that at $dt_n = dt = 0$, it follows that  $dx_n = dx$, i.e. the length of the physical rod in the inertial frame coincidence with the coordinate distance in the accelerated frame. Since $t_n = t$ and $\Delta t = 2l/c$  represent the time indicated by a physical clock at rest in the initial inertial frame, it follows that clocks at rest in the accelerated frame run slower compared to those in the inertial frame. In fact, the rate of the accelerated clock as measured by the inertial observer corresponds to the spacing between time marks on the screen in the initial inertial frame.

\begin{figure}
	\centering
	\includegraphics[width=0.75\textwidth]{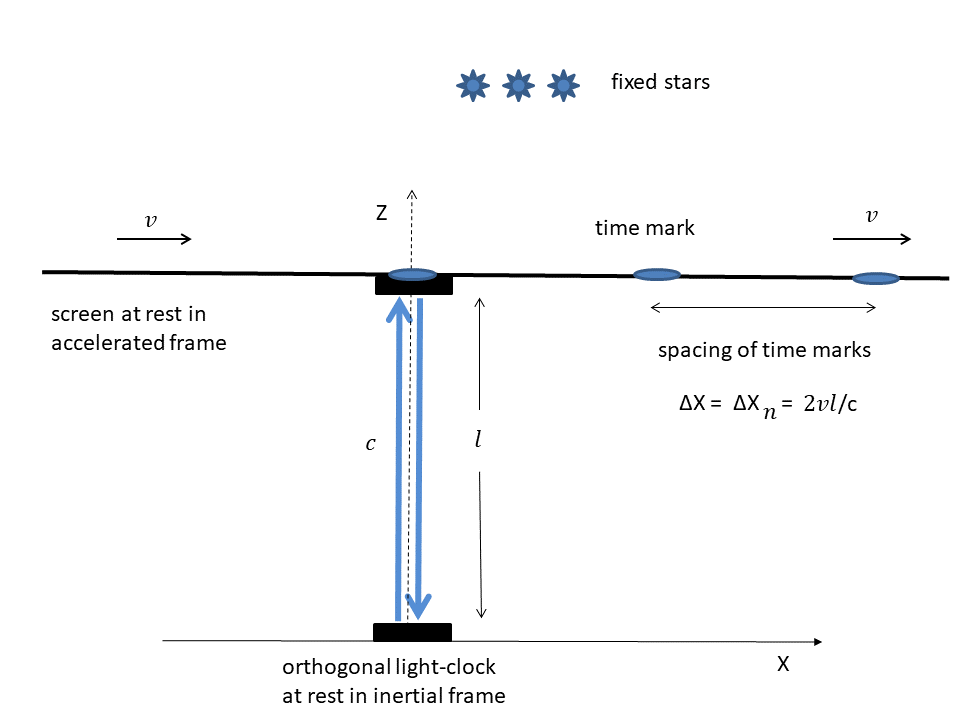}
	\caption{The  "light clock" thought experiment. Time marks can be imprinted on the moving screen by clock. Observations made by an observer in the same inertial frame as the light clock. The distance $\Delta x_n$ through which the screen in the accelerated frame falls in the time (i.e. spacing of the time marks) is $\Delta x_n = \Delta x = 2vl/c$.}
	\label{B879}
\end{figure}

Second, we describe the orthogonal light clock from the perspective of an observer who is initially at rest in the same inertial frame as the clock, Fig. \ref{B879}.  According to this observer, the time interval between the emission and reception of the light signal is given by $\Delta t = 2l/c$. During this interval, the screen in the accelerated frame falls a distance
$\Delta x  = \Delta x_n = 2lv/c$, which corresponds to the spacing between the time marks.

An analysis using standard measuring rods in an accelerated frame to examine its geometrical properties reveals that the coordinate distance $dx_n$ corresponds to a physical length of $dx_n/\sqrt{1 - v^2/c^2}$. This relationship directly reflects the asymmetry in the length of the meter stick, as derived from the Langevin metric.
This result indicates that the physical spacing between time marks, as measured in the accelerated frame, experiences compression in the direction of motion relative to the fixed stars.  Since $\Delta x_n = \Delta x$, and $\Delta x = 2vl/c$ represents the physical time rate shown by a clock at rest in the inertial frame, it follows that the accelerated observer perceives a shortening of the mark spacing, given by $\Delta l_n = 2lv\sqrt{1 - v^2/c^2}]/c$. Here $\Delta l_n$ 
denotes the physical spacing of time marks in the accelerated frame.
This thought experiment clearly demonstrates the absence of reciprocity in time dilation.

\subsection{Orthogonal Light Clock Initialization}

The slowing of a clock in a moving system is a highly nontrivial phenomenon.
Consider a light clock that is initially at rest in an inertial frame and is then accelerated to a velocity $v$ transverse to the direction of the optical pulse. An observer at rest in the inertial frame measures the direction of light propagation.

In the previous section, we demonstrated that the apparent time between successive clicks is longer for an accelerated clock. We now continue the discussion of light-clock operation in frames $S$ and $S_n$, focusing on the \emph{phase} of the light clock. Here, the term ``phase'' refers to the zigzag shape of the path followed by the light beam. As already discussed in Figs.~\ref{B878}–\ref{B879}, the rate of a moving clock can be measured by recording the time marks it imprints on a laboratory screen. We found that the rate of the moving clock—namely, the time interval between emission and reception of the light pulse as measured by a laboratory observer—is represented by the spacing between these time marks on the screen. The question we now address is whether an analogous statement holds for phase measurements. We shall show that it does.

To illustrate the idea, suppose the inertial observer uses two screens and employs the up-and-down imprinting method, that is, a configuration in which a moving clock imprints time marks on both screens. As discussed in Chapter~7, the orientation of the coordinate axes of an accelerated frame relative to those of a Lorentz inertial frame is governed by the Wigner rotation.

\begin{figure}
	\centering
	\includegraphics[width=0.85\textwidth]{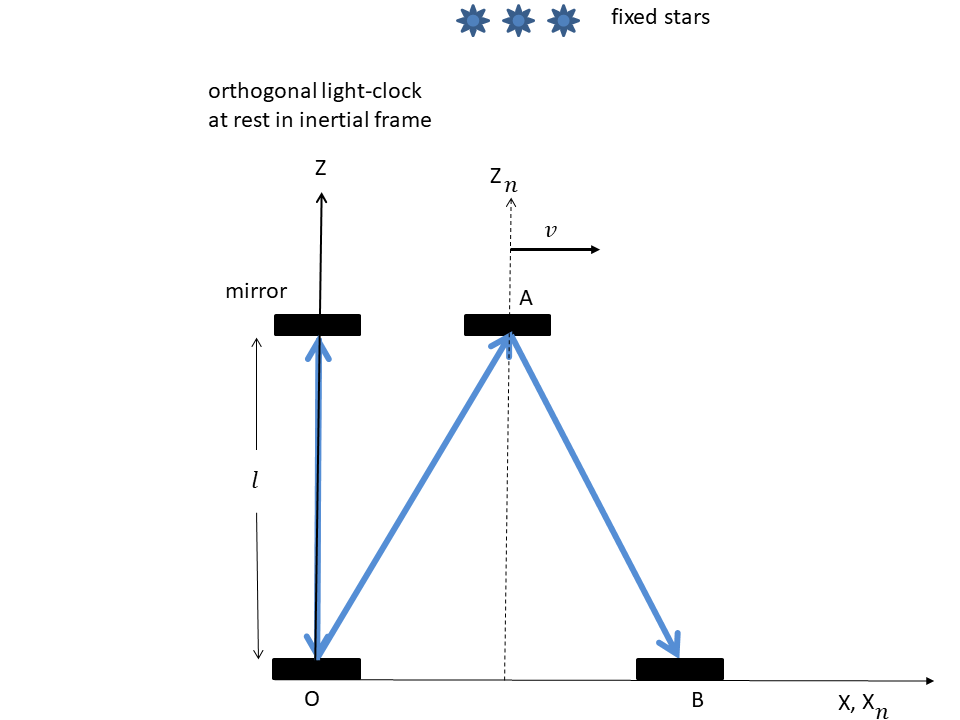}
	\caption{The ``light clock'' thought experiment. Calculation of the time interval between emission and reception of a light signal in the reference frame $S$, relative to which the light clock moves. Textbook treatments assume that the axes $(x_n,y_n,z_n)$ of the moving frame remain parallel to the axes $(x,y,z)$ of the laboratory frame.}
	\label{B877}
\end{figure}

We begin by examining the standard textbook reasoning. According to the usual account, when an inertial observer watches an orthogonal light clock pass by, the light traveling between the mirrors follows a zigzag path, as illustrated in Fig.~\ref{B877}. This path is typically represented as an isosceles triangle.

However, there is a fundamental problem with this commonly accepted picture. The standard explanation relies on the implicit assumption that the axes $(x_n,y_n,z_n)$ of the moving observer remain parallel to the axes $(x,y,z)$ of the laboratory observer. In other words, it is assumed that both observers share a common three-dimensional space. This assumption is incorrect. Observers following different worldlines in Minkowski spacetime necessarily possess different three-spaces.

Relativity theory teaches us that the relationships between spatial positions and temporal intervals are not aligned with our intuitive expectations. When dealing with distant events, we must confront the conventionality of distant simultaneity over time intervals of order $l/c$, where $l$ is the spatial separation between the events.

The textbook assumption of rigidity for an accelerated reference frame is based on the belief that a simultaneous acceleration of the entire spatial grid $(x_n,y_n,z_n)$ has direct physical meaning. In fact, the position of the first mirror relative to the second along the $x$-direction in an accelerated frame has no precise objective meaning. Owing to the finite speed of light, there exists no operational procedure by which this position could be unambiguously determined. As a result, there is an intrinsic uncertainty (or blurring) in the relative position along the $x$-direction of order $lv/c$, where $l$ is the rod length. This uncertainty arises from the indeterminacy in the moments at which different parts of the system undergo acceleration.

\begin{figure}
	\centering
	\includegraphics[width=0.75\textwidth]{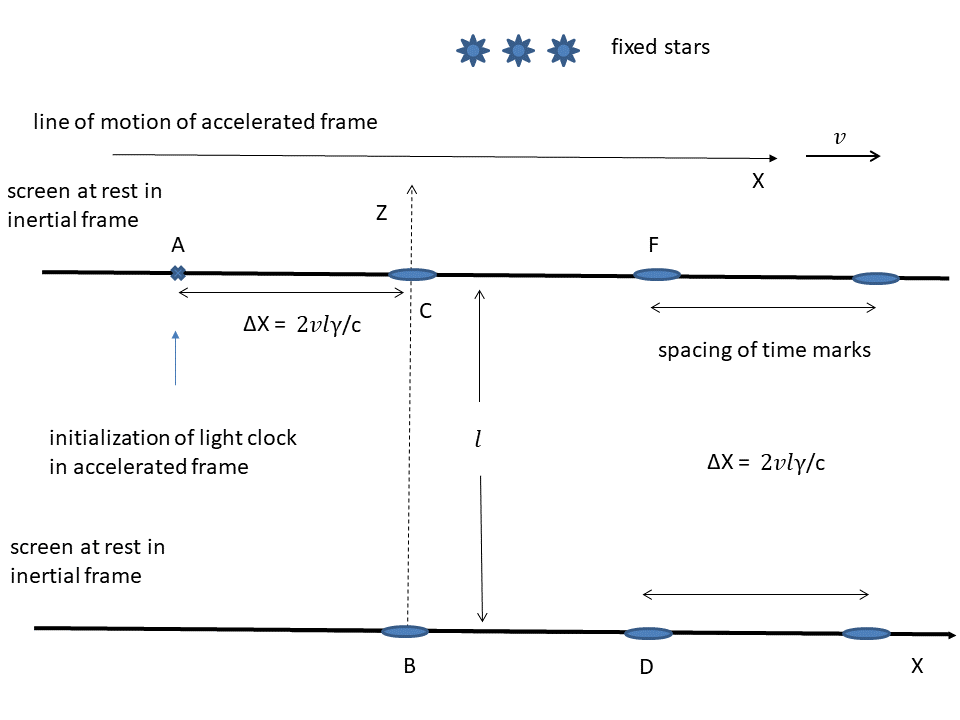}
	\caption{The ``light clock'' thought experiment. Phase of the orthogonal light clock in the accelerated frame, as observed from the inertial laboratory frame. }
	\label{B880}
\end{figure}

\begin{figure}
	\centering
	\includegraphics[width=0.75\textwidth]{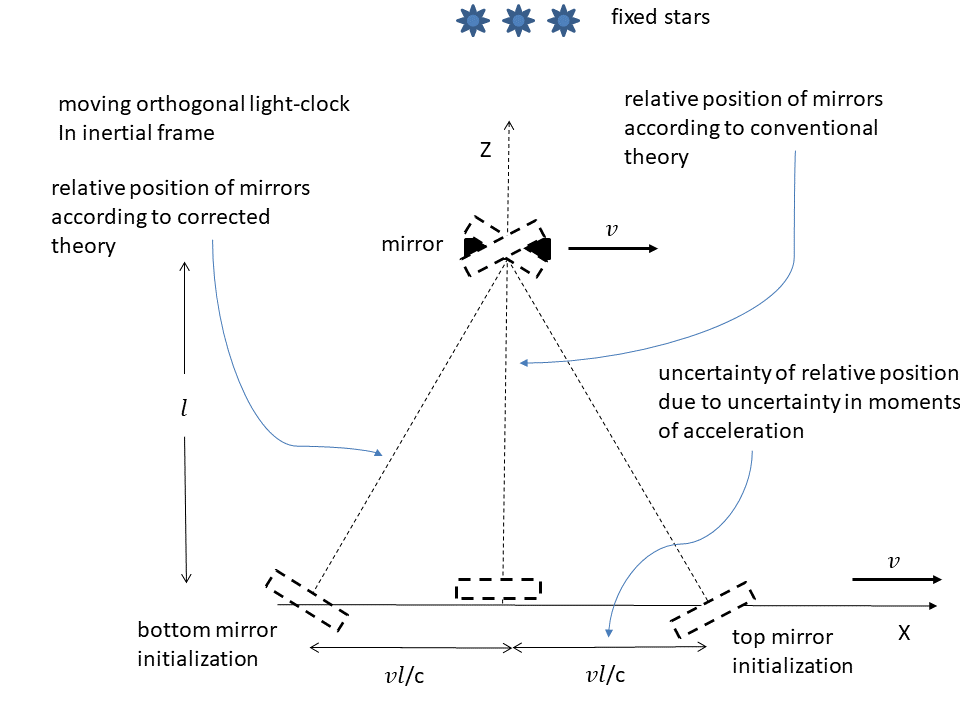}
	\caption{The ``light clock'' thought experiment. Observation made in the inertial laboratory frame. The relative position of the accelerated mirrors along the $x$-axis has no exact objective meaning. The orientation of the accelerated frame axes relative to the laboratory axes is governed by the Wigner rotation. For simplicity, we describe this effect only to first order in $v/c$.}
	\label{B882}
\end{figure}

\begin{figure}
	\centering
	\includegraphics[width=0.75\textwidth]{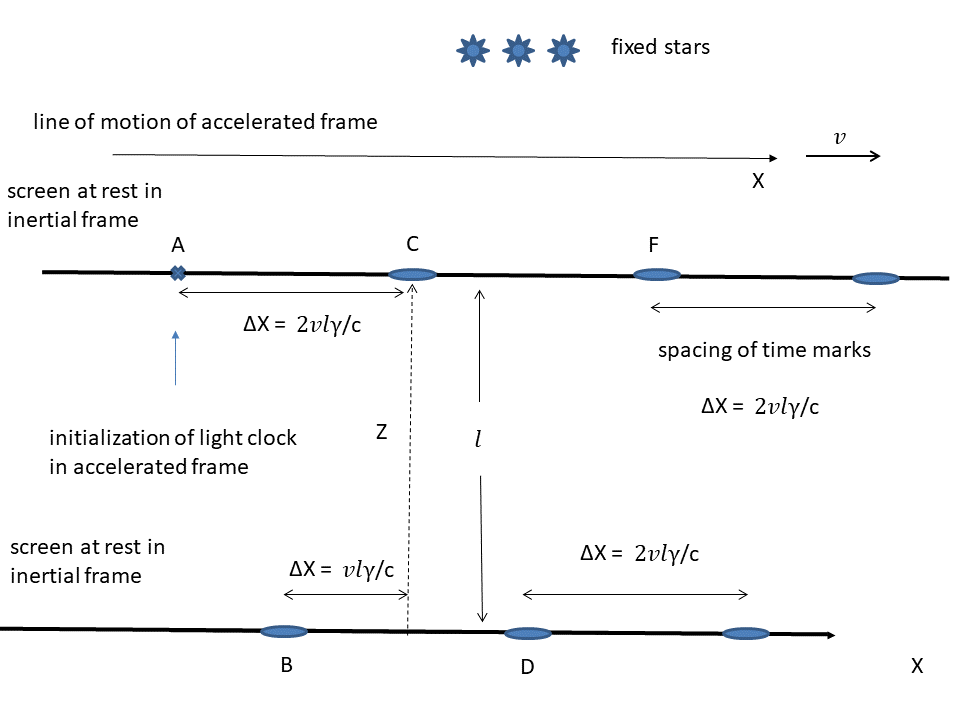}
	\caption{The ``light clock'' thought experiment. Phase of the orthogonal light clock as described in textbooks, based on the incorrect assumption that accelerated and inertial observers share a common three-space.}
	\label{B881}
\end{figure}

According to conventional treatments, when a reference frame initially at rest is set into motion, all points of its three-dimensional grid begin to move simultaneously. Consequently, the rigidity of the orthogonal light clock assumed in textbook discussions cannot be regarded as a genuine kinematical property of relativistic motion. Our result, illustrated in Fig.~\ref{B880}, stands in clear contradiction to the textbook prediction shown in Fig.~\ref{B881}. The standard description of orthogonal light-clock operation fails to account for the Wigner rotation, which in this context is closely tied to the relativity of simultaneity (see Fig.~\ref{B882}).

We now briefly describe a concrete mechanism for initializing an orthogonal light clock in an accelerated frame. To understand how light propagates between the mirrors, it is necessary to analyze the internal dynamics of the clock as viewed from the inertial frame. For this purpose, we consider a particularly simple initialization procedure that is sufficient in principle.

Although the Fabry–Perot interferometer discussed in Chapter~10 may at first appear quite different from the light clock used in special relativity, the two systems are in fact closely related. Consider an optical cavity formed by two plane mirrors. Each mirror is assumed to be semitransparent so that the time marks produced by the circulating pulse can be recorded on two external screens. The light clock is initialized by a short laser pulse emitted from an external laser that is at rest in the initial inertial frame and coupled into the cavity through either the upper or the lower semitransparent mirror. We assume that the transverse size of the mirrors is much smaller than the transverse extent of the laser pulse.

After applying a  Galilean boost in the case of absolute time coordinatization, we employed
the metric given by Eq. \ref{GGG11}, which describes the electrodynamics of a moving
light clock as observed from the initial inertial frame.  The absolute time synchronization method involves maintaining the same set of uniformly synchronized clocks used when the light clock was at rest—essentially preserving Einstein synchronization, which is defined using light signals from a stationary dipole source.

In the absolute-time coordinatization, Maxwell’s equations can be applied directly to the laser radiation. In this description, there is an angle $\alpha = v/c$ between the wavevector $\vec{k}$ and the $z$-axis. The question then arises: how can one demonstrate that such a laser beam will successfully initiate the operation of the moving light clock?

A clear explanation of the initialization mechanism follows from the results presented in Section~4.5. When a plane wave (in the absolute-time coordinatization) is incident normally on such a mirror, the reflected and transmitted beams exhibit a deviation in the direction of energy transport (see Fig.~\ref{B17}). This effect can be understood as a straightforward consequence of the Doppler effect. In particular, the transmitted light beam propagating along the $x$-axis has a group velocity given by $d\omega/dk_x = v$. 
Let us now consider an incoming laser wave with a phase gradient along the  $x$-axis.  If the laser beam is initially tilted by the angle $v/c$, one finds that the transmitted light beam acquires a group velocity component along the $x$-direction given by  $d\omega/dk_x = 2v$.

Suppose the initialization pulse strikes the upper mirror at point $A$ in the inertial frame (Fig.~\ref{B880}). Special relativity predicts that the light travels to the lower mirror, reaching it at point $B$, and is then reflected back to the upper mirror, which has meanwhile moved to position $C$. This sequence $(A \to B \to C)$ constitutes the first tick of the clock. Subsequent ticks $(C \to D \to F \to \dots)$ arise from further reflections between the mirrors.

This behavior is deeply counterintuitive, reflecting the fact that different observers possess different three-spaces embedded in four-dimensional spacetime. Spatial measurements performed by one observer generally involve mixtures of space and time when described by another. In particular, the relative position of the accelerated mirrors along the $x$-direction lacks a well-defined operational meaning. The resulting uncertainty in their relative position is again of order $lv/c$ (Fig.~\ref{B882}).

To continue the analysis, we consider the trajectory of a light ray inside the moving clock following top-mirror initialization. The path $(A \to B \to C \to D \to F \to \dots)$ traced by the light beam is shown in Fig.~\ref{B880}. The time marks define a right triangle $ABC$. Within special relativity, time measurement reduces to length measurement: the time between successive clicks corresponds to the length of side $AC$, while $\angle ABC$ is the aberration angle.

As discussed in Section~5.12 (see Fig. \ref{B301}), for a single redirected moving emitter, the aberration angle at first order in $v/c$ is $2v/c$. 
The next question is: what is the  $\angle ABC $ for arbitrary velocity?
In special relativity, we are unrestricted in our choice of a coordinate system. 
There are many ways of clocks synchronization.  \footnote{We note that the synchchronization changes have no intrinsic significance; their meaning is pure conventional. The convention-independent results, such as position of the down mirror at the location on $B$ (i.e. the possition of first mark at the location on $B$ after top mirror initialization at the location on $A$) remain unchanged in the new variables. }  One way, involving very little calculation, would be  the absolute time synchronization convention.


Extending our analysis to arbitrary velocities, and employing absolute-time synchronization, we find that the aberration angle becomes $\theta_a =   2\arctan v/c$. 
The advantage of the absolute-time coordinatization is that inertial-frame measurements are described by the standard Minkowski metric,
 $ds^2 = c^2dt^2 - dx^2- dy^2 - dz^2$.  Since spatial lengths at rest in the inertial frame retain direct physical meaning, the laboratory observer finds the length of side $AC$ to be  $L_{AC} = l\tan 2\arctan v/c =  2lv/[c\sqrt{1 - v^2/c^2}]$.  Thus, the time between clicks of the moving clock is increased by precisely this factor, in agreement with the time-dilation result derived previously.

Finally, we reverse the perspective. 
It is necessary to analyze the internal dynamics of the clock at rest in the initial inertial frame as viewed from the accelerated frame.
In this frame, the zigzag light path again forms a right triangle $A'B'C'$, with the time between clicks equal to the length of side $A'C'$. 
The angle $\angle A'B'C'$ corresponds to  the light clock moving with velocity $-v$  in the accelerated frame. 

However, the angle $\angle A'B'C'$ differs from $\angle ABC$.  We recall that   $\angle ABC$  includes both the aberration of the moving source and the Wigner rotation. These two angles are equal and $\angle ABC$ is given by the formula $\angle ABC = 2 \arctan v/c$.  

In contrast, the metric Eq. (\ref{GGG12}) characterizes  the electrodynamics of a stationary light source in an inertial frame from the perspective of measurements made by an accelerated observer. From this formula, it is evident that the radiated light pulse has a group velocity along $x_n$-direction given by $-2v/c$ and the angle of aberration $\angle A'B'C' = \arctan 2v/c$.  Consequently, the coordinate distance $\Delta x_n = L_{A'C'} = 2lv/c$.

Since physical spacing between time marks, as measured in the accelerated frame, experiences compression in the direction of motion relative to the fixed stars, it follows that accelerated observer perceives a shortening of mark spacing, given by   $\Delta l_n  =  L_{A'C'}/\gamma =
2lv\sqrt{1- v^2/c^2}/c$ in agreement with the result previously obtained using the Langevin metric.

Readers who find it difficult to form a mental picture of the moving light clock should not be discouraged. Special relativity places severe demands on intuition. Most apparent paradoxes associated with light clocks arise from space–time mixing. Once this is properly accounted for within pseudo-Euclidean spacetime geometry, the paradoxes disappear, and the theory reveals itself to be logically consistent and fully satisfactory.

\subsection{Light Aberration from a Moving Source }

Let us consider an emitter located in the 
 $(x,y)$  plane of an inertial frame and accelerated from rest to a velocity $v$ along the $x$-axis.
Suppose an observer at rest in this inertial frame measures the direction of energy transport. The question then arises: how can this angular deviation be measured experimentally? A light beam radiated by a source moving with group velocity  $v$ shifts its position along the  $x$-axis with time. The central issue is whether this displacement can be interpreted experimentally as an aberration effect.

To address this, we consider a simplified emitter–detector setup in the initial inertial frame that captures the essential physical principles. Consider the configuration shown in Fig.~\ref{B103}. Our interpretation of aberration relies on absolute-time coordinatization. In Chapter 4 we analyzed only the nonrelativistic case; here we extend the analysis to arbitrary velocities.

Assume the emitter radiates a pulse of nearly monochromatic radiation with duration  $T_p$.
A detector at rest is positioned at a distance $L$ along $z$-axis
from the moving emitter. We assume that the aberration shift is large compared with the emitter size  $D_e$, $D_e \ll vL/c$. 
To resolve the aberration shift, we must also require $cT_p \ll L$. Within the small diffraction angle approximation,
 $\lambda/D_e \ll v/c$, we obtain a second small parameter, $D_e/L \ll v/c$. 

It is useful to discuss the relation between these two small parameters. Their combination, $N_F = D_e^2/(\lambda L)$, is the Fresnel number. In the regime of interest here, there is no restriction on $N_F$. At first glance, one might think that the aberration shift could be determined with arbitrary precision by increasing  $L$. In practice, however, the measurement accuracy is limited by the finite angular aperture of the emitter,  $\lambda/D_e$. Thus, the propagation direction of the light pulse cannot be determined more accurately than this angle.

At this stage, one may ask why these technical details are important. The reason is that textbook discussions of light aberration usually rely on a different experimental configuration. It is widely believed—incorrectly—that an observer at rest in an inertial frame and a moving source share a common Minkowski metric. For example, a point source moving transversely with velocity 
$v$ is often assumed to produce, in the far zone, a plane wave whose phase front is tilted by an angle $v/c$ (to first order).
According to this viewpoint, the angular position of a moving point source observed through a telescope aperture in the inertial frame shifts by $\theta_a = v/c$. If this plane wave is focused by a lens, it forms a diffraction spot in the focal plane displaced relative to the optical axis. Measuring this displacement is then interpreted as measuring aberration.

However, experimental evidence clearly shows that aberration occurs when the telescope changes its velocity relative to the fixed stars, not when the source changes its transverse velocity. As discussed in Chapter 4, light transmitted through a telescope aperture carries no information about the tangential motion of the source.

Let us now return to our specific experimental setup. Extending the analysis to arbitrary velocities and using absolute-time synchronization, we consider the transformation of the direction of light propagation. The group velocity of light transforms in the same way as a particle velocity, following Galilean velocity addition. The aberration angle  $\theta_a$ therefore satisfies
 $\tan\theta_a = v/c$.

We choose our coordinates as follow. 
Suppose the emitter radiates a pulse at point $A$ with coordinates $x = 0$, $z = L$ and the pulse is detected at point $B$
 on the $x$-axis. In this configuration, measurement of the aberration angle reduces to a length measurement. The quantity
$L\tan \theta_a$ equals the length of segment  $OB$, while $\angle OAB$ equals the aberration angle, where  $O$ is the origin.

In absolute-time coordinatization, inertial-frame measurements are described by the standard Minkowski metric. The inertial observer therefore finds the physical length  $L_{OB} = L\tan \theta_a =  Lv/c$.

Our analysis instead indicates that the correct interpretation of aberration measurements requires a consistent spacetime geometric treatment. The Minkowski metric remains valid for an inertial observer using Einstein synchronization with light emitted by a source at rest. In these Lorentz coordinates, electrodynamics is described by the standard Maxwell equations.

If an emitter accelerates from rest to velocity $v$ along the $x$-axis, the inertial observer maintains synchronization using the same clock network, preserving the Minkowski metric.  In this scenario, the emitter undergoes a Galilean boost $x \to x - vt$, yielding Eq.~\ref{GGG11}, which describes the electrodynamics of the moving emitter from the inertial observer’s viewpoint. Previous literature often assumes—incorrectly—that an inertial observer and an accelerated emitter share the same Minkowski metric.

It is also instructive to analyze aberration using Lorentz coordinatization. In appropriately chosen variables, the metric associated with the accelerated source can be diagonalized. Using the transformation 
$t_L = t \sqrt{1 - v^2/c^2} + (vx/c^2)/\sqrt{1 - v^2/c^2}$, $x_L = x/\sqrt{1 - v^2/c^2}$, 
the velocity transformation becomes 

\[
dx_L/dt_L = (dx/dt)/[1 - v^2/c^2 + vdx/dt)]   . 
\]

Substituting
$dx/dt = v$ gives  $dx_L/dt_L = v$.  In Lorentz coordinates, light propagates with velocity $c$
independent of direction and source velocity.
Thus the group velocity component along $x$ remains $v$, while expressing this in terms of the isotropic light speed gives $\sin \theta_a = v/c$.

Now, suppose an inertial observer measures the aberration of light. 
How do physical notions of length and time measured at rest in the inertial frame manifest in the coordinate system $(t_L,x_L,y_L,z_L)$?  
The relation  $dl^2 = ( - g_{11} + g_{01}g_{01}/g_{00})dx_L^2$
establishes the connection between the physical length 
and the four-dimensional spacetime metric.
In the new variables, inertial observer  has the spacetime interval given by Eq. (\ref{STI}).

Using these relations, we find that coordinate distance  $dx_L$ corresponds to physical length  $dl = dx_L\sqrt{1 - v^2/c^2}$, as expected.

We therefore conclude that the measured aberration angle satisfies $\tan\theta_a = v/c$.  This agreement underscores the physical objectivity of the result: the length measured at rest (physical length) is a measurable, convention-independent quantity and must remain invariant under changes of coordinatization.

In contrast, conventional aberration theory predicts 

\[
\sin\theta_a = v/c   ,   \qquad  \tan\theta_a = v/(c\sqrt{1 - v^2/c^2})   ,
\]

\[
L_{OB} = L\tan \theta_a =  Lv/(c\sqrt{1 - v^2/c^2})   .  
\]

Now , we examine the light aberration from a moving source within the accelerated frame. Consider the case where an accelerated observer redirects radiation from an accelerated emitter such that the group velocity component along the $x_n$-axis becomes zero; that is, the beam propagates along the $z_n$-axis (see Section 5.12).

After acceleration, once the system $S_n$
moves with constant velocity, Einstein synchronization can be applied using light signals emitted by a source at rest in the accelerated frame, assuming isotropic light propagation with velocity $c$. In this framework, the new coordinate system in the accelerated frame is interpreted such that an observer perceives a diagonal metric:

\[
ds^2 = c^2 (dt'_n)^2 - (d x_n')^2 - (dy_n')^2 - (dz_n')^2 
\]

for an emitter at rest in the accelerated frame,  and

\[
ds^2 = c^2(1- v^2/c^2)(dt'_n)^2 + 2vdx_n'dt_n' - (dx'_n)^2 - (dy'_n)^2 - (dz'_n)^2
\]

for an emitter accelerated to velocity $v$ along $x_n'$-axis. In this scenario, the emitter undergoes a Galilean boost $x_n' \to x_n' - vt_n'$, yielding Eq.~\ref{GGG11}, which describes the electrodynamics of the moving emitter from the accelerated observer’s viewpoint.  This reveals a symmetry (or reciprocity) in the metrics—and consequently in the electrodynamic equations—between the accelerated and inertial frames. After diagonalization, it becomes evident that any apparent asymmetry stems from the initial conditions (see Section  5.9 for more details).

Since spatial lengths at rest in the accelerated frame retain direct physical meaning in Lorentz coordinatization, the accelerated observer measures   

\[
L_{O_n B_n} = L\tan \theta_a =  Lv/c   . 
\]

Thus, no difference should exist between aberration measurements for a moving emitter in the initial inertial frame and for a moving redirected emitter in the accelerated frame.

\subsection{Relarivistic Measurements of Distances}

The measurement of the length of a moving rod requires careful reconsideration. We begin by examining the reasoning commonly presented in standard textbooks. There are two distinct approaches to the problem of length measurements for a moving rod.

The first is the invariant approach, which explains length transformations by applying Lorentz transformations to spacelike four-vectors and determining how the measured length depends on the velocity of the rod relative to a given inertial frame. It is important to emphasize that Lorentz length contraction, contrary to the commonly held view, is not itself a consequence of the transformation of a spacelike four-vector.

In this section, we discuss both approaches, beginning with the invariant one.

The invariant treatment of a moving rod is based on the invariance of the four-dimensional interval $ds$. Textbook discussions often analyze length measurements using the Minkowski expression for the interval.

Suppose that the interval between two events is spacelike, so that $ds^2<0$. Then there exists an inertial frame $S'$ in which the two events are simultaneous. If the events occur at points lying on the $x'$-axis, the spacetime interval reduces to

\[
ds^2 =-dx'^2,
\]

so that in frame $S'$ the interval corresponds to a purely spatial distance. In the initial inertial frame $S$, described by Lorentz coordinates, the interval is

\[
ds^2=c^2dt^2-dx^2.
\]

Consequently,

\[
dx^2=c^2dt^2+dx'^2.
\]

It follows immediately that the length $|dx'|$ measured in frame $S'$ is smaller than the corresponding coordinate distance $|dx|$ in frame $S$. Using the inverse Lorentz transformation,

\[
dt=\frac{dt'+vdx'/c^2}{\sqrt{1-v^2/c^2}},
\]

and noting that $dt'=0$, we obtain

\[
|dx|=\frac{|dx'|}{\sqrt{1-v^2/c^2}}.
\]

This result demonstrates that the transformation of spatial distances follows directly from the geometric structure of spacetime. In the invariant approach, measurements performed in different inertial frames refer to the same spacetime interval, that is, to the same four-dimensional geometric quantity.

A fundamental difficulty arises when this argument is interpreted as a description of physical length measurements. In standard textbook treatments, the metric associated with a moving rod is implicitly assumed to be the same Minkowski metric used by an observer at rest in the initial inertial frame.

How should one account for the relative velocity between the observer and the rod?

One possibility is to assign a Lorentz coordinates to describe the observer measurements. However, for the moving rod, this Lorentz  coordinates would act as an absolute time coordinatization, and its motion would be described by a Galilean transformation in this framework.  Suppose that inertial observer measures the length of moving rod. Taking to account the Minkowski metric of the inertial observer, we find that the length of physical rod at rest in the inertial frame possess direct operational meaning. Accordingly, the measured   length of the moving rod  is equal to the proper length $|dx'|$.

It is always possible to introduce a suitable change of variables in the initial inertial frame and thereby construct Lorentz coordinates adapted to the moving rod. In these coordinates one finds

\[
|dx|=\frac{|dx'|}{\sqrt{1-v^2/c^2}},
\]

which corresponds to a coordinate lengthening rather than a contraction.

When describing physical phenomena in an inertial frame, it is essential to distinguish between coordinate quantities and physical quantities. After the change of variables, the Minkowski metric no longer directly describes the observer's measurements. Instead, the observer's spacetime interval is given by Eq.~(\ref{STI}). From this relation one finds that the coordinate length $|dx|$ corresponds to the physical length

\[
|dx|\sqrt{1-v^2/c^2}.
\]

As a result, the observer again obtains the physical rod length $|dx'|$. Thus, if length is regarded as a well-defined physical quantity associated with spacetime geometry, no distinction should arise between the measured length of a moving rod and that of a rod at rest.

The situation changes when one adopts Einstein's operational definition of the length of a moving object. According to this definition, the length is the spatial separation between the endpoints of the rod measured simultaneously in the observer's frame.

For a rod moving with velocity $v$, the Lorentz transformation gives

\[
dx'=\frac{dx-vdt}{\sqrt{1-v^2/c^2}}.
\]

To determine the rod's length according to Einstein's definition in frame $S$, one must consider two events that are simultaneous in $S$. Therefore,

\[
dt=0.
\]

Substituting this condition into the preceding expression yields

\[
|dx|=|dx'|\sqrt{1-v^2/c^2}.
\]

This is the familiar Lorentz contraction. It arises directly from the relativity of simultaneity and from Einstein's operational procedure for measuring the length of a moving rod.

It is widely believed that experiments directly demonstrate Lorentz length contraction. This claim is frequently repeated in textbooks but is, in fact, misleading. The essential point is that Lorentz contraction is inherently coordinate-dependent and therefore does not represent an invariant geometric property of spacetime.

Nevertheless, if the measurement is carried out according to Einstein's prescription—that is, by recording the positions of both ends of the rod simultaneously in the observer's frame—the observed result is unquestionably the Lorentz-contracted length.

Comparison with the invariant approach shows that the rod's proper length and the length defined by Einstein's simultaneity procedure are distinct spacetime quantities. They are not related by a Lorentz transformation acting on a single spacetime object. Consequently, Lorentz contraction is not a direct consequence of the fundamental postulate that inertial frames are related by Lorentz isometries.

As early as 1967, Gamba \cite{GA} argued that the concept of Lorentz contraction plays no essential role in the formulation of special relativity. Nevertheless, it continues to be presented in most textbooks as one of the central features of the theory.

\subsection{Interference Phenomena}

We now turn our attention to a specific aspect of the phenomenon of interference. To explore this, let’s examine the processes of reflection and refraction within the framework of special relativity.
At this point, we are prepared to analyze what happens when light emitted by a moving source passes through a stationary medium, such as glass. The key task is to calculate the interference between the incident and reflected radiation.
At first glance, one might assume that calculating both refraction and reflection in a single inertial frame would require accounting for differences in the metrics, and consequently the electrodynamics equations, for a moving emitter and a stationary glass. This suggests that the possibility of calculating interference effects is not immediately apparent.

\begin{figure}
	\centering
	\includegraphics[width=0.85\textwidth]{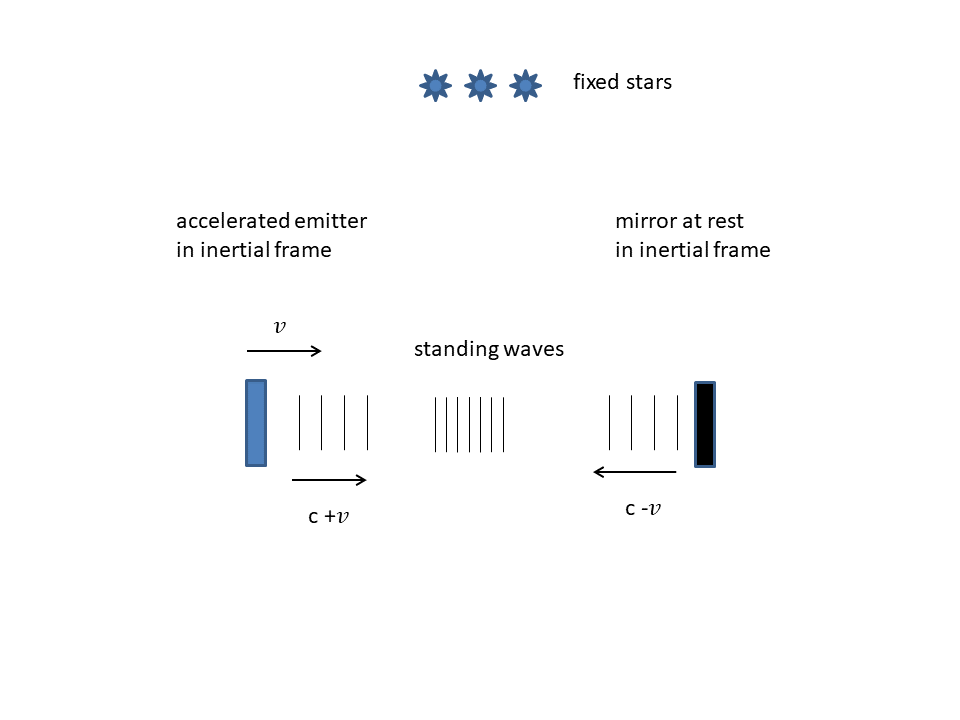}
	\caption{The light emitted by a moving source reflects off a stationary mirror in a collinear arrangement. Interference occurs between the incident radiation and the radiation resulting from the material.}
	\label{B555}
\end{figure}

If the mirror is stationary and the emitter is in motion, it is evident that the equations governing electrodynamics must be the same for all electromagnetic waves. In other words, a consistent metric must be applied to both the incident and scattered waves. However, this raises a paradox: how can the electrodynamics equations be identical for both the incident and reflected waves, given that they are governed by two different metrics?

In a previous discussion, we addressed the issue of two distinct metrics in the context of light aberration. In Chapter 4, we focused on the first-order approximation and highlighted that the apparent paradox is resolved. One aspect of the aberration setup can be easily understood based on concepts we have already explored. It is advantageous to begin with the simpler case of a normally incident plane wave, and only later in Section 5.10 do we introduce the theory that includes tilted incoming plane waves. This approach is beneficial because the unique geometry of light aberration ensures that, even after applying a Galilean transformation, the electrodynamics equations in an absolute time coordinate system will retain Maxwell's form for a normally incident plane wave. Now, let us examine the situation in a collinear geometry, as illustrated in Figure \ref{B555}.

We begin by summarizing the central idea. The apparent paradox is resolved by recognizing that both the Minkowski metric and the metric given by Eq. (\ref{GGG11}) yield identical predictions. The key point is that all methods used to measure interference—specifically, those involving standing waves—are effectively round-trip measurements.
In contrast, the deviation in the direction of energy transport arises from a geometric effect. Since empirical measurements grant access only to the round-trip (two-way) average speed of light, the one-way speed of light remains a matter of convention without intrinsic physical meaning. The two-way speed of light, however, is directly measurable and therefore possesses physical significance.

We now turn to an important aspect of interference phenomena. It is commonly believed that, in a setup involving a moving light source and a stationary glass medium, the incident wave and the wave scattered by the dipoles of the glass cannot interfere—as required by the electron theory of dispersion—because they travel at different velocities in an absolute time framework. This notion, however, is incorrect.
Physically, an incident wave of a given frequency will excite electrons in the glass to oscillate at that same frequency, regardless of the wave’s velocity. These oscillating electrons then re-emit radiation at the same frequency. Thus, the incident and scattered waves at any given location share the same frequency and can interfere. The difference in their velocities manifests as a relative phase shift that varies spatially. This phase variation influences both the velocity and amplitude envelope of the resulting wave formed by the superposition of the incident and scattered components.

The following simple analysis supports these ideas. We consider the case where the metric is non-diagonal, as given in Eq. \ref{GGG11}. and is  identical for both the incident and reflected waves.
Let the incoming wave be represented by $\exp[i(\omega t - kx)]$, where the phase velocity is $\omega/k = c + v$. Similarly, let the scattered (outgoing) wave be represented by $\exp[i(\omega t + k'x + \phi)]$, having the same amplitude and frequency, but a different velocity $\omega/k' = c - v$, and a phase offset $\phi$.
The superposition of these two waves is given by: 

\[
\exp i(\omega t - kx)  +  \exp i(\omega t + k'x + \phi) = 
\]

\[
2\cos[(k+k')x/2 +\phi/2]\exp i[\omega t - (k-k')x/2 +\phi/2]
\]

Here, the cosine term acts as an amplitude envelope that is stationary in space. Its periodicity is inversely proportional to the difference in the wave numbers $k$ and $k'$ of the two components. This expression can be further simplified by substituting the wave numbers in terms of $\omega$, $c$, and $v$, leading to:  

\[
2\cos[\omega x/[c(1 - v^2/c^2)] +\phi/2]\exp i[\omega t - xv\omega/[c^2(1 - v^2/c^2)]  +\phi/2]   .
\]

Consider a source at rest emitting waves with a natural frequency $\omega_0$. In the laboratory frame, after applying a Galilean transformation, the velocity of the incoming wave becomes $c+v$. Given this, the frequency observed in the lab frame is: 

\[
\omega = \omega_0/[1 - v/(c + v)] = \omega_0(1 + v/c)   .
\]

This frequency shift corresponds to the well-known Doppler effect. Incorporating this into the expression for the superposition of two waves, we obtain:

\[
2\cos[\omega_0 x/[c(1 - v/c)]  +\phi/2]
\]

\[
\times \exp i[\omega_0(1+v/c) t - xv\omega_0/[c^2(1 - v/c)]  +\phi/2]   .
\]

Suppose an observer in the laboratory conducts a standing wave measurement. It is important to analyze which aspects of the measured data depend on the chosen synchronization convention and which remain invariant. We emphasize that time oscillation lacks intrinsic physical meaning—its interpretation arises solely from a chosen convention. In particular, consider the time shift $xv\omega_0(1 + v/c)/c^2$ appearing in the phase term 

\[
\exp i[\omega_0(1+v/c) t - xv\omega_0/[c^2(1 - v/c)]  +\phi/2]      . 
\]

This shift is closely tied to the issue of synchronizing distant clocks. Furthermore, from a physical standpoint, the absolute scale of time—or equivalently, frequency—cannot be directly identified.

Suppose the laboratory observer measures the wavelength of a standing wave. The relation
$dl_i^2 = (- g_{11} + g_{01}g_{01}/g_{00})dx^2$ describes the connection between the spatial line element and the metric. Using Eq.~(\ref{GGG11}), we find $dl_i^2 =  dx^2/(1 - v^2/c^2)$, which defines the spatial geometry in the context of the  non-diagonal metric. When analyzing the measured data, the laboratory observer finds that the wave number is given by

\[
\sqrt{1 - v^2/c^2} \omega_0/[c(1 - v/c)] = \omega_0\sqrt{1 + v/c}/[c\sqrt{1 - v/c}]   .
\]

Notably, this is the same factor one obtains under the assumption of a diagonal metric. In other words, the laboratory observer will measure the same two-way speed of light regardless of the choice of metric representation.

The reason a laboratory observer measures the same wavelength lies in the underlying principle of interference. All interference phenomena are governed by a fundamental law known as the relativistic invariance of phase. This principle states that a specific quantity—the phase—remains unchanged under transformations between reference frames. The invariance of phase is a powerful tool for analysis, as it allows us to draw meaningful conclusions (as will be illustrated in the next section) without requiring detailed knowledge of the full set of transformation formulas.
However, a puzzle remains in the scenario under discussion.

Let us consider a dipole source that accelerates in the laboratory inertial frame to a velocity $v$ along the $x$-axis. After acceleration, assuming synchronization in the lab frame remains unchanged, we encounter a complex situation involving the dynamics and electrodynamics of moving charges.

In conventional (non-covariant) particle tracking, the particle’s motion can be modeled as a sequence of infinitesimal Galilean transformations, with a corrected form of Newton's second law applied at each step. Interestingly, this approach does not invoke Lorentz transformations. Within the inertial lab frame, the particle’s motion appears exactly as classical mechanics would predict—following absolute time, without invoking the relativity of simultaneity. That is, time and position remain distinct and do not mix in this description.

In contrast, Maxwell’s equations are valid in the inertial frame only when Lorentz coordinates are used. This raises a fundamental question: How can classical mechanics (with Galilean kinematics) be used alongside electrodynamics governed by Einstein’s kinematics? This apparent inconsistency calls for explanation.

The resolution lies in the nature of the coupling between particles and fields in collinear geometries. In such configurations—where both the particle motion and the emitted radiation are aligned along the same axis—the standard Maxwell equations and the corrected Newtonian dynamics are sufficient to explain interference phenomena within a single inertial frame.

In this context, the dynamical evolution of the particle can be alternatively interpreted as a sequence of infinitesimal Lorentz transformations, rather than Galilean ones.
But how is this possible? As discussed in Section 3.6, the key insight lies in the fact that collinear Lorentz boosts form a commutative group—just as Galilean boosts do. This commutative property plays a critical role in the analysis.

In the case of collinear geometry, where the particle moves along the same line as the emitted radiation, the velocity vector is perpendicular to the radiation's wavefront (or the plane of simultaneity). In this configuration, the motion of the source influences only higher-order kinematic terms—those proportional to powers of $v/c$.

At first glance, conventional particle tracking in collinear geometry appears to be entirely Newtonian. However, relativistic effects do manifest, albeit subtly, through the implicit assumption that the mass of the moving particle corresponds to its relativistic mass.

\newpage

\section{ A Preferred Inertial Frame and Second Order Experiments }

\subsection{Second-Order Optical Experiments and Special Relativity}

An illustrative example of time dilation is observed in the resonance absorption of gamma rays in Mössbauer rotating disk experiments. When measuring the resonance absorption for the same relative velocity between the source and the observer, a blue shift is detected when the source is positioned at the center of the disk, while a red shift appears when the source is located at the rim.
This result reveals an asymmetry in the transverse Doppler effect.\footnote{The first experimental confirmation of time dilation via resonance absorption of gamma rays was reported by Champeney, Isaak, and Khan in 1965 \cite{CHAM}.} Notably, when both the source and observer are situated at the rim, no frequency shift is observed.
In such a configuration, a pulse emitted by one accelerated observer on the rotating disk, which appears redshifted at the center due to time dilation, is received by a diametrically opposite observer on the rim without any frequency change. This occurs despite both observers moving at the same speed $v$ in opposite directions in the laboratory frame. Their mutual time dilation effects effectively cancel the frequency shift in this scenario.

At first glance, the transverse Doppler effect in an accelerated frame reveals a fundamental asymmetry between 
accelerated and inertial observers. This leads to the question: is it experimentally possible to determine the state of 
motion of an inertial frame $S$ from the perspective of an accelerated frame
$S_n$,  using the second-order Doppler 
redshift and blueshift of moving sources? Specifically, can this effect be used to determine the motion of the Earth relative to a Sun-centered inertial frame?

Let us examine the resonance absorption of gamma rays in a Mössbauer rotating disk experiment. Consider a setup where the absorber is located at the center of a disk, which remains at rest relative to the Earth-based frame, while the emitter is positioned on the edge of the disk. The emitter thus moves in a circular path around the absorber with a constant tangential velocity $\vec{w}$. To keep the mathematical complexity manageable, we restrict our analysis to terms up to second order in $w/c$ and $v/c$, where $v$ is the orbital velocity of the Earth.
The frequency of the emitter changes with the velocity $\vec{w}$ in a manner as it suggests itself from the transverse Doppler effect. Indeed, the inner frequency of the emitter depends on its velocity $\vec{w}$, so that (using Langevin metric in the Earth-based frame) 

\[
\omega (t_n) = \omega_0\sqrt{1 - |\vec{v} + \vec{w}|^2} . 
\]

Expanding this expression and neglecting higher-order terms, we obtain 

\[
\omega (t_n) = \omega_0[1 - v^2/(2c^2) - w^2/(2c^2) - \vec{v}\cdot\vec{w}(t_n)/c^2] . 
\]

Since the physical clock is at rest in the earth-based frame, the reading of its proper time is $d\tau = (1 -v^2/2) dt_n$. We therefore expect that radiation emitted with a constant intrinsic frequency $\omega_0$ will arrive at the absorber with a time-dependent frequency, given by 

\[
\omega (\tau) = \omega_0 [1 - w^2/(2c^2) - \vec{v}\cdot\vec{w}(\tau)/c^2]  .
\]

Because the emitter moves in a circular path, the scalar product 
$\vec{v} \cdot \vec{w}(\tau)$ 
varies periodically, leading to periodic fluctuations in the observed frequency at the disk center. Consequently, the resonance absorption measured at the absorber will also vary periodically with the emitter's position along the disk's circumference.

In the actual experiment, no effect on absorption was observed when the disk was set into rotation. This negative result can be explained by considering that only a portion of the total frequency fluctuation is initially taken into account. A more refined analysis reveals the necessity of including the frequency shift caused by the so-called radial Doppler effect. We analyze a configuration based on perpendicular geometry, where, at first glance, the radial Doppler effect would appear to be absent. However, as discussed in previous chapters, the aberration of light in the Earth-based reference frame implies that the radial component of the emitter's velocity is given by $\vec{w} \cdot \vec{v} / c$. As a result, the emitted radiation reaching the center of the disk exhibits a periodically varying frequency shift

\[
\delta \omega =\omega_0 \vec{v}\cdot\vec{w}(\tau)/c^2   ,
\]

which exactly matches the periodic variation in the internal resonance frequency. Thus, the two effects — the intrinsic frequency variation and the radial Doppler shift — cancel each other out, leading to no net observable effect.

The second-order Doppler redshift observed in moving atoms is a key prediction of special relativity.
The first experimental verification of the transverse Doppler effect was conducted in 1938 by Ives and Stilwell  \cite{IS}. In their experiment, ionized hydrogen molecules were accelerated in a cathode tube to energies of up to 28 keV. The frequencies of light emitted in both the forward (parallel) and backward (antiparallel) directions relative to the motion of the particles were measured using a stationary spectrograph.
Let us now analyze this experiment. We discuss a setup based on a parallel geometry.
Radiation emitted by the moving atoms, exhibiting Doppler-shifted spectral lines, was observed alongside radiation from stationary atoms present in the same working volume, which produced an unshifted reference line.
By including this reference, Ives and Stilwell effectively circumvented the challenging task of directly measuring the asymmetry between the red- and blue-shifted lines. Instead, they compared both to the stationary reference line, simplifying the analysis.

At first glance, the Earth's orbital motion relative to the Sun can be examined through this type of experiment.\footnote{In the Ives and Stilwell experiment, the effect arising from the Earth's orbital motion relative to the Sun was within the limits of experimental accuracy. This limitation is due to the high velocity of the emitter ($w \gg v$). In this discussion, we present a general analysis of such experiments, independent of specific experimental limitations.} Specifically, if the internal frequency of an atom depends on its velocity $\vec{w}$, then (using the Langevin metric in the Earth-based frame) we have  $\omega (t_n) = \omega_0\sqrt{1 - |\vec{v} + \vec{w}|^2}$  ,
which implies that the frequency of radiation recorded by a spectrograph should, in principle, depend on the Earth's orbital velocity $\vec{v}$.
In rotating disk experiments, the observable effect due to the transverse Doppler shift is counteracted by another relativistic effect. For the case of perpendicular geometry, we have previously explained the experimental results by accounting for both the aberration of light and the radial (i.e., classical) Doppler effect. In contrast, in the collinear geometry, the aberration of light is absent, and only the radial Doppler effect needs to be considered. By analyzing the geometry of this configuration, we find that the observed frequency (i.e., the frequency of the radiation reaching the detector) is given by 

\[
\omega  = \omega_0[1 - v^2/(2c^2) - w^2/(2c^2) + \vec{v}\cdot\vec{w}(t_n)/c^2]/[ 1 - w/(c + \vec{w}\cdot\vec{v}/w)]   ,
\]

where, according to the Langevin metric, the coordinate velocity of light in the Earth-based frame is 

\[
c^* = c + \vec{w} \cdot \vec{v} / w   .
\]

Thus, just as in the case of the rotating disk experiment, we find that in measurements of the second-order Doppler redshift for moving atoms, the two effects caused by the Earth's orbital motion effectively cancel each other. As a result, no net observable effect remains.

Such cancellations are typically understood to arise from deep underlying principles.
Given the series of negative results—like the one obtained by Michelson and Morley—it seems reasonable to assume that the consistent failure of various optical experiments to detect the Earth’s orbital velocity $v$ is no coincidence.
This has led to the widely accepted view that a fundamental law of nature prevents the determination of $v$ through laboratory-based experiments.
However, in the present case, no such profound implication appears to be at play.
This outcome reflects a specific limitation of second-order optical experiments.
As discussed in earlier chapters, we already know that it is possible to detect the Earth’s orbital velocity—without referencing anything external—by using the phenomenon of light aberration.

\subsection{Features of Second Order Optical Experiments}

The principle of relativity has been experimentally confirmed, though only to a limited extent. In second-order experiments, the effects caused by Earth's orbital motion cancel each other out, leaving no observable impact. This suggests that second-order experiments inherently prevent us from determining Earth's velocity relative to the Sun through laboratory measurements. Let us examine this claim more closely.

The key effects under consideration involve the interference of two light beams. A light source is split into two beams using a suitable apparatus, and their overlap produces an interference pattern. Since both beams originate from the same source, second-order interference experiments cannot detect Earth's orbital velocity. This is because phase is a four-dimensional invariant—independent of the chosen inertial frame and coordinate system—due to the fundamental geometry of space-time.

This discussion shows that special relativity correctly predicts the observed null fringe shift in Earth-based frames. One of the most well-known interference experiments that demonstrates this is the Michelson-Morley experiment.
Next, we turn to the second-order Doppler shift in moving atoms. The frequencies of emitted light were measured using a grating spectrograph. Notably, there is no fundamental distinction between a diffraction grating and an interferometer in this context.

An intriguing example of this effect is the measurement of resonance absorption of gamma rays in a Mössbauer rotating disk experiment. To illustrate this effect, consider a source positioned at a great distance from a thin absorber plate. We aim to determine the resulting field at a distant point on the opposite side of the absorber.
According to electrodynamics, the total electric field is the sum of contributions from the external source and from the fields generated by charges within the absorber. The absorber consists of atoms with nuclei, and when the external field acts on these nuclei, it induces motion in their charges. Moving charges generate additional fields, acting as new radiators. Consequently, the two plane waves that interfere propagate in the same direction, leading to interference between the incident and coherently forward-scattered waves.
Our model of the nuclear oscillator includes a damping force. With damping taken into account, the refractive index becomes complex, resulting in destructive interference. Thus, this too is an interference phenomenon.

\subsection{Wigner Rotation and the Trouton–Noble Experiment}

Next, we consider additional second-order experiments not directly related to light propagation but interpretable via general dynamical principles. A particularly significant example is the Trouton–Noble experiment \cite{TN}, which aimed to detect the rotational motion of a charged parallel-plate capacitor suspended at rest in the Earth's frame. The goal was to measure the Earth's velocity relative to the Sun (i.e., relative to the fixed stars). The fundamental idea behind the experiment can be described as follows.

Consider two opposite point charges $+e$ and $-e$ in the Earth-based frame, with the radius vector pointing from $-e$ to $+e$ denoted by $\vec{r}_n$. If the charges are at rest, the force acting on $+e$ is
\[
\vec{F}_n = e \vec{E} = -\frac{e^2 \vec{r}_n}{r_n^3}.
\]
Since this force acts along $\vec{r}_n$, the resulting torque vanishes: $\vec{M}_n = \vec{r}_n \times \vec{F}_n = 0$.

If the charges move with velocity $\vec{v}$ relative to the Sun-based frame, the positive charge experiences both the electric field  from $-e$ and the magnetic feld
\[
\vec{B} = \frac{e (\vec{v} \times \vec{r})}{c r^3}.
\]
The magnetic force becomes
\[
\vec{F} = \frac{e^2}{c^2 r^3} \left[ \vec{v} \times (\vec{v} \times \vec{r}) \right] 
\]
Using the vector identity $\vec{v} \times (\vec{v} \times \vec{r}) = \vec{v} (\vec{v}\cdot\vec{r}) - v^2 \vec{r}$, the torque on the pair is
\[
\vec{M} = \frac{e^2 (\vec{v}\cdot \vec{r})}{c^2 r^3} (\vec{r} \times \vec{v}).
\]
Denoting the angle between $\vec{v}$ and $\vec{r}$ by $\theta$, the magnitude of the torque is
\[
M = \frac{e^2 v^2}{c^2 r} \sin\theta \cos\theta.
\]

In the actual experiment, a charged condenser was suspended from an elastic string and oriented such that $\theta = \pi/4$—the line perpendicular to the capacitor plates formed an angle of $\pi/4$ with the assumed direction of the Earth's orbital velocity, $\vec{v}$. The experiment yielded a null result: no torque proportional to $v^2/c^2$ was observed.

The absence of torque in the Earth-based frame is fully consistent with special relativity. Electrodynamic equations in the Earth frame can be derived via a Galilean transformation (velocity $-v$) applied to Maxwell's equations. Assuming $\vec{B} = 0$ in the Sun-based frame, the transformed fields are
\[
\vec{E}_n = \vec{E}, \quad \vec{B}_n = - \frac{\vec{v} \times \vec{E}}{c}.
\]
Consequently, the induced magnetic field in the Earth-based frame does not exert any torque on the capacitor, which is at rest in that frame.

The apparent paradox arises because a magnetic field produces a torque in the Sun-based frame, suggesting different mechanical equations for force and torque in different inertial frames.\footnote{Jackson \cite{JJ} discusses this apparent paradox for a single charged particle. A similar situation occurs in standard treatments of the Trouton–Noble experiment \cite{JJ}, arising from the hidden assumption that all inertial frames share a common three-dimensional space. A fully relativistic explanation requires accounting for Wigner rotation.} The root of the paradox is the flawed assumption that the Earth-based and Sun-based observers share a common 3-space. In special relativity, observers on different worldlines possess different three-dimensional hypersurfaces of simultaneity. Standard treatments implicitly assume that the axes $(x_n, y_n, z_n)$ of the moving Earth-based observer are parallel to the axes $(x, y, z)$ of the Sun-based observer. A consistent explanation emerges only when the equations of motion are expressed covariantly, properly transforming coordinates between frames.

The key to resolving the paradox is recognizing that spatial measurements by the Earth-based observer mix space and time coordinates as viewed from the Sun frame. The angle between the velocity vector $\vec{v}$ and the radius vector $\vec{r}_n$ in the Earth frame does not equal the angle $\theta$ in the Sun frame. This apparent rotation is linked to length contraction. For a rod along the $x$-axis in the Earth frame, a Lorentz boost along $x$ with velocity $v$ contracts its length in the Sun frame to $l_n/\gamma$, where $l_n$ is the proper length and $\gamma$ the Lorentz factor. The relation between the angles is
\[
\gamma \tan \theta_n = \tan \theta,
\]
and for $v^2/c^2 \ll 1$,
\[
\Delta \theta = \theta_n - \theta = - \frac{v^2}{2c^2} \sin \theta \cos \theta.
\]
Hence, the torque in the Earth frame is reduced to
\[
M_n = \frac{e^2 v^2}{2 c^2 r} \sin \theta \cos \theta.
\]

An additional correction arises from the Wigner rotation of the Earth-based frame. The perpendicular component of velocity, $v_\perp = v \sin \theta$, induces a rotation of the frame axes as seen from the Sun frame. The Wigner rotation rate is
\[
\frac{d \vec{\Phi}}{dt} = \frac{v^2}{2 c^2} \frac{\vec{v}}{v^2} \times \frac{d\vec{v}}{dt}.
\]
This rotation aligns with the rotation of the velocity vector in the Sun frame. Therefore, the rate of change of the angle $\theta$ is
\[
\frac{d\theta}{dt} = \frac{v^2}{2 c^2} \frac{e^2 \sin \theta}{m v r^2},
\]
with $m$ the particle mass. The corresponding correction to the torque is
\[
m r \frac{d v_\perp}{dt} = m v \cos \theta \frac{d \theta}{dt} = \frac{e^2 v^2}{2 c^2 r} \sin \theta \cos \theta.
\]
Combining all effects, the net torque on the capacitor in the Earth-based frame vanishes, resolving the Trouton–Noble paradox.

\subsection{Time Dilation in Moving Atomic Clock}

The next topic we need to address is the time dilation effect.
We have already discussed this subject in Chapter 13. Here's a brief summary of our findings:

1. We established that the slowing down of a physical clock that is accelerated with respect to the fixed stars does not depend on the choice of the reference frame in which the effect is measured. In other words, the time dilation effect is absolute, not relative.

2. Due to the asymmetry between the initial inertial and accelerated frames, special relativity makes a striking prediction. Suppose frame $S$ is at rest relative to the fixed stars, and frame $S_n$ is accelerated with respect to
$S$ until it reaches a speed $v$. Now consider two clocks in frame $S_n$ both symmetrically accelerated to speed $v$,
but moving in opposite directions. Surprisingly, the time dilation experienced by these two clocks in $S_n$ is not the same. This discrepancy arises because the clocks have different acceleration histories relative to the fixed stars.

A natural question arises: is it possible to experimentally determine the state of motion of the inertial frame $S$
from the perspective of the accelerated frame $S_n$, using the time dilation effect observed in a moving (atomic) reference clock? At first glance, it may seem that the Earth's orbital motion could be detected through time dilation experiments. While this is theoretically possible, a more detailed analysis is required to properly address the question.

The proper time interval can be associated with phenomena such as particle lifetimes, atomic transition periods, and nuclear half-lives. The key point, however, is that this interval is entirely determined by the motion of a clock within its initial inertial frame.

In 1971, Hafele and Keating conducted a landmark experiment in which four atomic clocks were synchronized with a laboratory reference clock and then flown around the Earth aboard commercial aircraft—two traveling eastward and two westward. \footnote{This experiment provided the first experimental confirmation of time dilation as predicted by special relativity, using atomic clocks in circular motion. An important follow-up came from the Global Positioning System (GPS), which also involves relativistic effects. GPS satellites, numbering around 80, orbit the Earth at velocities of approximately 4 km/s. The system demonstrates that satellite clocks tick more slowly due to their relative motion, with a measured discrepancy of about 7.1 microseconds per day. Interestingly, while there is a noticeable time difference between the clocks on Earth and those in orbit, no asymmetries are observed among the satellites themselves. This suggests that relative velocity alone does not account for time dilation. Furthermore, the experiment revealed no effect of Earth’s orbital velocity on clock rates, a result attributed to the fact that both the satellites and Earth-based clocks share a common orbital path around the Sun, rendering orbital effects unobservable in this context \cite{HK, GPS}.} Upon return, the airborne clocks were compared with the laboratory reference clock. The results showed that the clocks carried on the planes had run slower than those left stationary on Earth. The observed discrepancies matched the predictions of special relativity, confirming the role of relative acceleration in time dilation. Notably, the experiment found no measurable influence of Earth's orbital motion on clock rates, reinforcing the idea that time dilation effects are tied to local motion relative to the chosen inertial frame.

The null result of the experiment described above can be understood by considering that the setup involves a circular path geometry. A more careful analysis reveals that it is essential to account for the fact that the moving clocks are ultimately compared to a reference clock located at the same point and at rest relative to the Earth. The moving clock travels around the reference clock along a circular path with velocity $w$. We will analyze the effect by expanding terms up to second order in  $w/c$ and $v/c$, where $v$ is the Earth's orbital velocity. The inner rate of the moving clock depends on velocity $w$, so that $d\tau = dt_n[1 - v^2/(2c^2) - w^2/(2c^2) + \vec{v}\cdot\vec{w}(t_n)/c^2]$. Since the reference clock is at rest in the Earth-based frame, the reading of its proper time is  $d\tau = dt_n(1 - v^2/2)$.
Analyzing the geometry of the situation, we see that for motion along a closed circular path, the cross term averages out over a complete cycle. Specifically, the integral of the mixed term vanishes:

\[
 \int \vec{v}\cdot\vec{w}(t_n) dt_n = \oint\vec{v}\cdot d\vec{s} = 0   .
\] 
 
Therefore, no observable effect from the Earth’s orbital velocity remains, which explains the null result.

\newpage

\section{The Principle of Relativity and Modern Cosmology}

In the preceding discussion, we explored the fundamental concepts necessary for understanding the phenomenon of the aberration of light. At first glance, the source-observer asymmetry associated with stellar aberration appears to contradict the principle of relativity. This principle, first introduced by Poincaré—who coined the term—states that an observer in uniform motion relative to the fixed stars cannot, by any internal measurement, detect this motion, provided they do not observe external celestial objects.

Now, let us examine the stellar aberration experiment more closely. One might argue that in the case of stellar aberration, the key distinction is that the accelerated (Earth-based) observer does, in fact, observe external celestial objects—the stars. However, the motion of the stars relative to Earth is never accompanied by any aberration. The aberration shift, as inferred from astronomical observations, remains even when a star moves with the same velocity as Earth. Thus, when considering Earth-based observations of stellar aberration, we cannot rely solely on the notion of "looking outside to the stars."

In Chapter 11, we introduced a simple scaling model for stellar aberration and derived a condition for optical similarity between the aberration of light from a distant star and that from an incoherent Earth-based source. This analysis revealed that even without observing external celestial objects, it is possible to determine Earth's orbital velocity around the Sun through aberration of light measurements.

There is no fundamental conflict between the structure of special relativity and the aberration of light phenomena. As discussed in previous chapters, special relativity does not require the principle of the irrelevancy of velocity formulated earlier. The principle of special relativity applies to physical laws, not to physical facts. It asserts that the same laws must hold in all inertial frames, which we interpret to mean that these laws should be expressed by equations that retain the same form in all inertial frames.

In special relativity, the transformations preserving this form-invariance between inertial frames are the Lorentz transformations. \footnote{In the special theory of relativity the form-invariant transformations between
	inertial frames are Lorentz transformations \cite{Ein}. For a general discussion about the
	Principle of Relativity we suggest reading the paper \cite{MU}. } According to the special principle of relativity, all inertial frames are equivalent concerning physical laws, but not necessarily concerning physical facts.

The existence of absolute acceleration in nature implies the existence of absolute velocity.

A closer examination of the problem of time dilation in an accelerated frame reveals the fundamental asymmetry between an accelerated frame and an initial inertial frame (i.e., one without a history of acceleration). The proper time of any particle moving arbitrarily within the initial inertial frame always runs slower than the physical time of that frame. This is a fundamental characteristic of pseudo-Euclidean space-time.

Uniform motion is not relative in this framework, allowing us to determine a particle's absolute velocity within the initial inertial frame. The formalism of classical physics relies on the structure of absolute space, which, in turn, must be associated with some underlying "substance"—the ether—which serves as an absolute rest frame.

Could space-time itself be the new ether? It shares similar properties, notably providing an "absolute rest reference frame." The key distinction between the initial inertial frame and Newtonian absolute space lies in the shift introduced by special relativity: we now conceive of a unified space-time model, without an inherent separation between space and time.

While the initial inertial frame is not physically identical to absolute space, it represents the concept of absolute space-time.

The principle of relativity has a long history in physics. In Newtonian mechanics, physical laws remain invariant under Galilean transformations. However, from a mathematical perspective, the Lorentz transformation is fundamentally different from the Galilean transformation. Notably, Lorentz boosts alone do not form a group, whereas the set of Galilean boosts does. This reflects the essence of Galilean relativity: all inertial frames are equivalent, meaning that no one frame is preferred over another. The underlying reason for this symmetry is that the equations of motion in Newtonian mechanics do not explicitly depend on velocity, ensuring that a system's internal dynamics remain unchanged under a Galilean boost.

Historically, Galileo's relativity principle was mistakenly assumed to apply universally across all of physics. However, the principle of relativity does not hold across the entire domain of Lorentz-covariant physical laws, and Lorentz covariance itself is not a fundamental symmetry of nature. Since the space-time metric tensor is a continuous quantity, an accelerated inertial frame is connected to its initial state by a Galilean transformation rather than a Lorentz transformation. This distinction is particularly evident in electrodynamics: Galilean transformations do not preserve the form-invariance of Maxwell's equations under a change of inertial frames. The reason is that electrodynamic equations explicitly depend on velocity, leading to the breakdown of Galilean relativity in this context.

According to special relativity, a fundamental difference exists between an accelerated inertial frame (relative to the fixed stars) and an inertial frame without an acceleration history. This distinction is closely linked to inertial forces and forms the core idea of the equivalence principle. Understanding the relationship between inertia and gravity remains an open question in fundamental physics.

Now we want to discuss a serious difficulty associated with the ideas of modern cosmology.
The difficulty we speak of is associated with the concept of source-observer asymmetry when applied to the motion relative to the cosmic microwave background (CMB).
One may say that perhaps there is no use worrying about these difficulties since there are so many things about the universe that we still don't understand. 
We live in a universe governed by the unknown. Dark energy and dark matter together make up nearly 96 percent of the universe’s total energy-mass, yet their true nature remains a mystery. However, it is essential to recognize that the foundational assumptions of modern cosmology are only approximations. As our understanding deepens, these assumptions will inevitably evolve.

We will not delve into the principles of modern cosmology at this time but will assume their validity and focus on discussing some of their implications.

According to the cosmological principle, the universe should appear isotropic — lacking any preferred direction — to a comoving observer with no peculiar motion relative to the cosmic fluid of the expanding universe. However, if such an observer has a peculiar motion, it can introduce a dipole anisotropy in the observed properties of certain objects. This anisotropy can, in turn, be used to infer the observer's peculiar velocity.

A well-known example of this phenomenon is the anisotropy observed in the cosmic microwave background (CMB) radiation. The CMB dipole is almost universally attributed to kinematic effects, specifically the Doppler shift caused by the relative motion between the Earth (the observer) and the reference frame in which the CMB appears nearly isotropic. By subtracting this dipole component, the CMB rest frame is established as the fundamental reference frame of the universe.

The observed dipole indicates that the solar system is moving at 370 km/s relative to the observed universe in the direction of galactic longitude $l = 264^o$ and latitude $b = 48^O$. This is quite far from the galactic rotation direction of 250 km/s towards $l = 90^o$ and $b = 0$.  
The motion relative to the cosmic microwave background results from the sum of many components of velocity due to the gravitational attraction of various mass concentrations.
The existence of clusters and super clusters of galaxies and our motion is a natural consequence of the large-scale organization of matter. The peculiar velocity consists of the five vector contributions: the motion of the earth around the sun (30 km/s), the hypothetical circular motion around our galaxy (250 km/s), the motion of our galaxy in the local group, and the motion of the local group       
with respect to the cosmic microwave background. The origin of the velocity of the local group is still uncertain and has been under discussion over the past two decades. This peculiar velocity is believed to be generated by the spatial inhomogeneities of mass (mainly dark matter) distribution in nearly large-scale structures. The velocity of the local group with respect to the cosmic microwave background is estimated to be 500 km/s. 

Now let us see how aberration of light varies with the speed relative to the rest frame of the universe.
Consider first the textbook explanation. 
It is very easy to understand how the aberration of light effect comes about from the point of view of the conventional aberration of light theory.
The existence of an aberration of the line of sight by motion can be recognized by considering an observer in a car driving  through a rainstorm: raindrops falling vertically appear to be oblique. 
An important application of the Galilean velocity transformation law is provided by stellar aberration, the change in the apparent direction of a star caused by the Earth's motion around the Sun.
The apparent positions of all fixed stars are thus always a little displaced in the direction of the Earth's motion at that moment, and hence describe a small elliptical figure during the annual revolution of the Earth around the Sun.  
What about the motion relative to the CMB rest frame?
According to textbooks, if the Earth's motion were uniform, the aberration effect would be undetectable since the "true" direction of the star is unknown. 
Indeed, who can say where a given star should be?  
On the other hand, the true direction of an Earth-based source is known. But, according to conventional theory, the aberration of light phenomenon does not exist in an aberration of light experiments using an Earth-based light source. 

We are now in a position to understand the nature of our difficulty with the motion relative to the CMB.
It is believed that the dipole anisotropy is produced by the Doppler effect due to acceleration (during the billions of years) of the solar system with respect to the rest frame of the universe. This acceleration is believed to be generated by the spatial inhomogeneities of (Dark matter) mass distribution in nearly large-scale structures.
A correct solution of the aberration problem in the Earth-based frame requires the use of metric tensor even in first-order experiments since the crossed term in the Langevin metric plays a fundamental role in the non-inertial kinematics of a light ( or relativistic particle ) beam produced by the Earth-based source. According to the relativistic theory of aberration presented in this book, the proper rotation of the Earth on its axis should produce a corresponding shift of the image. The aberration shift will depend also upon the value of $v_{\perp}^{CMB}$, the component of the solar system velocity (relative to the CMB) perpendicular to the Earth's rotation axis. 
The rotation of the Earth produces aberration in an amount larger enough to be taken into account in precise observation work using an electron microscope as an Earth-based particle source. 
Experimental results show (see Chapter 11) that the image shift is quite close to the theoretical prediction for the (30 km/s) orbital velocity and clearly indicate that the signal associated with motion (370 km/s) relative to the CMB  does not exist.
The simplest explanation is that the CMB dipole might be of  non-kinematical origin.

More recent observations by astronomers also cast doubt on the CMB dipole, being the ultimate representative of the solar peculiar velocity.
In recent years observations have emerged hinting at an anisotropic universe.  
The discovery of the preferred direction in the universe was serendipitous. 
Until very recently 
the velocity of the solar system in the rest frame of the universe is inferred from CMB temperature dipole anisotropy. Obviously, an independent measurement of this velocity is needed to fully establish the kinematical origin of the dipole. 
Another such quantity that could be employed to look for departures from isotropy is the angular distribution of distant radio sources in the sky. This could provide an independent check on the interpretation of CMB dipole.
The radio data clearly indicate that significantly larger dipoles exist in the rest frame of the radio galaxies. 
While the velocity of the solar system inferred from the CMB temperature dipole anisotropy is 370 km/s, the radio dipole measurements find the speed of motion to be around 1000 km/s (i.e. to be around three times larger than that of CMB). From the Hubble diagram of quasars, the motion of the solar system is derived, which is out to be the largest value ever found, $8\cdot10^3$ km/s. \footnote{Recent observations of the cosmic dipole anisotropies is revolutionized our understanding of the universe. 
	The various dipoles, including CMB dipole, all pointing along the same direction, suggest a preferred direction in the universe, raising thereby uncomfortable questions about the cosmological principle, the basis of the standard model in modern cosmology \cite{CMB1, CMB2, CMB3}.}

On the other hand, 
a common direction for all these dipoles, determined from completely independent surveys by different groups employing different techniques, indicates that these dipoles are not resulting from some systematics in the observations or the data analysis, but could instead suggest an inherent anisotropy. This is totally unexpected in a standard model of the universe. We have a preferred direction, aligned with the CMB dipole, in the universe. That is, going to the CMB rest frame, we see an anisotropic background. 
There is a difficulty with such some sort of an "axis" of the universe which, in turn, would be against the cosmological principle.
Three independent dipole vectors pointing along the same particular direction could imply an anisotropic universe, violating the cosmological principle, a cornerstone of modern cosmology.

Let us be conservative and say that there are two kinds of the Earth velocity. The total velocity with respect to the initial (i.e. privilege ) inertial frame could be the sum of the orbital velocity and the velocity of the Sun. The velocity of the Sun consists the two contributions: the circular motion around our galaxy and the motion of our galaxy. In Earth-based experiments where we measure the Earth's velocity with respect to the privileged frame by seeing the aberration of light (or particles), we are measuring the recorded (in the accelerational history) velocity. So the velocity with respect to the privilege frame consists of two contributions: a recorded motion plus an unrecorded motion. We know that there is definitely a recorded (orbital) motion.  
It is therefore impossible to get all the Earth's velocity to be recorded in the accelerational history in the way we hoped. Clearly something else has to be added.

Let's take a closer look at why the Earth moves relative to the initial inertial frame. For the Earth to be in motion, a force must have acted upon it at some point. To deepen our understanding, it is useful to examine the origins of these forces in greater detail.
We must say immediately that orbital motion is the result of non-gravitational forces. It is well known that a full 98 percent of all the angular momentum in the solar system is concentrated in the planets, yet a staggering 99.8 percent of all the mass in our solar system  is in our sun. Perhaps the first scientifically respectable theory of the origin of the solar system was given by Hoyle (1960) \cite{HOYLE}. 
He invoked the action of the magnetic field to transfer angular momentum from the central body, the Sun, to the ejected matter which eventually formed the planets. In contrast, the unrecorded motion around our galactic and the motion of our galactic is the result of the action of gravity.

We began by discussing the gravitational interaction between the Earth and the Sun. Our confidence in the theory of gravity is so strong that we use it to describe the forces acting between entire galaxies. However, we may be making too great an extrapolation, applying our limited understanding of gravity to cosmic scales. 
Perhaps the entire difficulty is that a modification of gravity may be responsible for unrecorded velocities.
Given the many unresolved mysteries surrounding dark matter, dark energy, and cosmic anisotropy, a deeper exploration of fundamental physics is essential to uncovering the true nature of the universe.

\newpage

\section{Relativistic Particle Dynamics}

\subsection{Manifestly Covariant Formulation}

The equations of dynamics can be formulated as tensor equations within Minkowski space-time. Upon selecting a coordinate system, these equations may be expressed in terms of components rather than abstract geometric objects. Leveraging the geometric structure of Minkowski space-time, one can identify the class of inertial frames and, for any such frame, adopt a Lorentz frame with orthonormal basis vectors.
In any Lorentz coordinate system, the law of motion takes the form:

\begin{eqnarray}
&& m\frac{d^2 x_{\mu}}{d\tau^2} = e F^{\mu\nu}\frac{dx_{\nu}}{d\tau}~ ,\label{DDE}
\end{eqnarray}
where here the particle's mass and charge are denoted by $m$ and $e$ respectively.
The electromagnetic field is described by a second-rank, antisymmetric tensor with components $F^{\mu\nu}$. The coordinate-independent proper time $\tau$ is a parameter describing the evolution of the physical system under the relativistic laws of motion, Eq. (\ref{DDE}).

The covariant equation of motion for a relativistic charged particle, subject to the four-force
$K_{\mu} = e F^{\mu\nu}dx_{\nu}/d\tau$, as given in Eq.~(\ref{DDE}), represents a relativistic extension of Newton's 
second law. While the classical form $md\vec{v}/dt = \vec{f}$  remain valid in the particle's instantaneous comoving
Lorentz frame, the relativistic formulation embeds these three spatial equations into the four-dimensional Minkowski spacetime. \footnote{The notion of embedding relies on the fact that Newton’s second law can always be applied in any Lorentz frame in which the particle is instantaneously at rest \cite{CT}. That is, if a comoving Lorentz frame is defined at a given instant, the particle’s evolution can be accurately predicted in this frame over an infinitesimal time interval. Geometrically, Newton’s law holds rigorously on a hyperplane orthogonal to the particle’s world line.
As the particle moves, the hyperplane—and its normal vector	$u_{\mu}$—tilts accordingly along the world line. To formalize this embedding, one introduces a projection operator $\hat{P}_\perp$ that continuously projects vectors in Minkowski space onto hyperplanes orthogonal to the world line. This operator is given by $(\hat{P}_\perp)_{\mu\nu} = \eta_{\mu\nu} - u_{\mu}u_{\nu}/u^2$ \cite{CT}. The presence of $\hat{P}_\perp$ highlights that only three of the four components of the covariant equation are independent. }   

The immediate generalization of $md\vec{v}/dt = \vec{f}$ to an arbitrary Lorentz frame is given by Eq.~(\ref{DDE}), as can be verified by reducing the equation to the particle's rest frame. In Lorentz coordinates, the four-velocity 
$u^{\mu} = dx^{\mu}/d\tau$ satisfies the kinematic constraint $u^{\mu}u_{\mu} = c^2$. Due to this condition, the four-dimensional equation of motion, Eq.~(\ref{DDE}), contains only three independent components. Using the explicit form of the Lorentz force, one can show that Eq.~(\ref{DDE}) automatically preserves the constraint $u^{\mu}u_{\mu} = c^2$, as required. To demonstrate this, we take the scalar product of both sides of the equation of motion with $u_{\mu}$. Exploiting the antisymmetry of the electromagnetic field tensor, $F^{\mu\nu} = - F^{\nu\mu}$, we find
$u_{\mu}d u^{\mu}/d\tau = eF^{\mu\nu}u_{\mu}u_{\nu} = 0$. This result implies that the quantity  $Y = (u^2 - c^2)$
is conserved, since $dY/d\tau = 0$.

\subsection{Conventional Particle Tracking}

Having expressed the equation of motion in 4-vector form, Eq.~(\ref{DDE}), and determined the components of the 4-force, we have, on the one hand, ensured compliance with the principle of relativity and, on the other hand, obtained a complete set of four equations describing particle motion. This represents a covariant, relativistic generalization of Newton’s three-dimensional equation of motion, formulated with the particle's proper time as the evolution parameter.

We now aim to describe the motion of a particle in the Lorentz laboratory frame, using the lab time $t$ as the evolution parameter. To begin, we consider the standard treatment found in textbooks. Let us determine the spatial components (i.e., the first three components) of the four-force. For this purpose, we examine the spatial part of the dynamical equation, Eq.~(\ref{DDE}): 

\[
\vec{Q} = (dt/d\tau) d(m\gamma\vec{v})/dt = \gamma d(m\gamma\vec{v})/dt    . 
\]

The prefactor $\gamma$ appears due to the change in the evolution parameter from the proper time $\tau$—natural for covariant formulations—to the lab frame time $t$, which is necessary for introducing the conventional three-force vector
$\vec{f}$. In this context, the spatial part of the four-force is related to the three-force by $\vec{Q}=\gamma\vec{f}$. Explicitly, the relativistic form of the three-force is given by

\begin{eqnarray}
&& \frac{d}{dt}\left(\frac{m\vec{v}}{\sqrt{1-v^2/c^2}}\right) = e\left(\vec{E} + \frac{\vec{v}}{c}\times \vec{B}\right)~ .\label{DDDE1}
\end{eqnarray}

The time component of the four-force equation is 

\begin{eqnarray}
&& \frac{d}{dt}\left(\frac{m c^2}{\sqrt{1-v^2/c^2}}\right) = e\vec{E}\cdot\vec{v} ~ .\label{DFE1}
\end{eqnarray}

The evolution of the particle is governed by these four equations, together with the invariant constraint

\begin{eqnarray}
&& \mathcal{E}^2/c^2 - |\vec{p}|^2 = mc^2  ~ .\label{DCFE1}
\end{eqnarray}

According to the non-covariant (3+1) approach we seek the initial value solution to these equations.
Using the explicit expression for Lorentz force we find that the three  equations Eq.(\ref{DDDE1}) automatically imply the constraint  Eq.(\ref{DCFE1}),
once this is satisfied initially at $t = 0$.
In this (3+1) formalism, the four-dimensional equations of motion are decomposed into three spatial and one temporal component, effectively eliminating any mixing between space and time in the dynamical equation Eq.(\ref{DDE}). 
This approach to relativistic particle dynamics relies on the use of three independent equations of motion  Eq.(\ref{DDDE1}) for three independent coordinates and velocities, "independent" meaning that  equation Eq.(\ref{DFE1}) (and constraint Eq.(\ref{DCFE1})) are automatically satisfied.

One might expect that the particle’s trajectory in the lab frame, denoted as $\vec{x}(t)$ and derived from the previous reasoning, should correspond to the covariant trajectory $\vec{x}_{cov}(t)$. However, identifying them in this way leads to paradoxical results. Specifically, the trajectory $\vec{x}(t)$ fails to account for relativistic kinematic effects. In the non-covariant (3+1) approach, the dynamics in the lab frame are formulated without reference to Lorentz transformations. As a result, the motion of a particle in a constant magnetic field, when described from the lab frame, appears identical to its Newtonian counterpart—relativistic effects are absent in this formulation. In conventional particle tracking, the lab-frame trajectory $\vec{x}(t)$ can be interpreted as arising from a sequence of Galilean boosts, which follow the motion of the particle as it accelerates under the influence of the magnetic field. The standard Galilean velocity addition rule is applied to determine the boosts at each instant, tracing the particle’s progression along its curved path.

The old kinematics is especially surprising, because we are based on the use of the covariant approach. Where does it come from?
The previous commonly accepted derivation of the equations for the particle motion in the three-dimensional space from the covariant equation Eq.(\ref{DDE}) includes one delicate point. 
In Eq.(\ref{DDDE1}) and Eq.(\ref{DFE1}) the restriction  $\vec{p}  = m\vec{v}/\sqrt{1 - v^2/c^2}$ has already been imposed. One might well wonder why because in the accepted covariant approach, the solution of the dynamics problem for the momentum in the lab frame makes no reference to the three-dimensional velocity. Equation  Eq.(\ref{DDE}) tells us that the force is the rate of change of the momentum $\vec{p}$, but does  not tell us how momentum varies with speed. 
The four-velocity cannot be decomposed into $u = (c\gamma, \vec{v}\gamma)$ when we deal with a particle accelerating along a curved trajectory in the Lorentz lab frame.  

Actually, the decomposition  $u = (c\gamma, \vec{v}\gamma)$ comes from the relation 

\[
u_\mu = dx_\mu/d\tau = \gamma dx_\mu/dt = (c\gamma, \vec{v}\gamma)    . 
\]

In other words, the presentation of the time component as the relation 
$d\tau = dt/\gamma$ between the $\tau$ and lab time $t$ is based on the hidden assumption that the type of clock synchronization, which provides the time coordinate $t$ in the lab frame, is based on the use of the absolute time convention. 

It is important to stress at this point that the situation when only one clock in the comoving frame is involved in dynamics cannot be realized. 
The Newton law can be written down in the proper frame only when a space-time coordinate system $(x', y', z', \tau)$ has been specified. The type of clock synchronization which provides the time coordinate in the Newton equation 
has never been discussed in textbooks. It is clear that without an answer to the question about the method of distant clock synchronization  used, not only the concept of acceleration but also the dynamics law has no physical meaning. 
A proper frame approach to relativistic particle dynamics is forcefully based on a definite (Einstein) coordinatization assumption.  After this, the dynamics theory states that the equation of motion in the proper frame is  $md^2\vec{x'}/d\tau^2 = \vec{f}'$. It should be stressed that in the case of velocity increment  $\Delta \vec{v'} = \Delta\vec{x'}/\Delta\tau$, we also deal with the distant events. It can be said with some abuse of language that a 3-velocity vector is always a "spatially extended" object.  Let us return to our consideration of the relation $d\tau = dt/\gamma$ between the $\tau$ and lab time $t$.
The calculation carried out in the case of a spatially extended object shows that the temporal coincidence of two distant events has absolute character: $\Delta \tau = 0$ implies $\Delta t = 0$.

It should be noted that usually the notation "$\tau$" arose from the proper time. Now let us recall the standard concept of the object's proper time. Let a point-like object move uniformly and rectilinearly relative to the lab frame $K$. The proper frame $K'$ can be fixed to the moving object. The object is at rest in this frame, so that events happening with this object are registered by one clock. This clock counts the proper time at the point where the object is located.  
As we already know, the proper time on moving  object flows slower than the time $t$. This phenomenon was called time dilation. The proper time can also be introduced for a point particle moving with acceleration.
This standard interpretation of the proper time $\tau$ has nothing to do with dynamics and one may wonder where this contradiction existing between the name and the content of time in dynamic law comes from. 

To avoid being overly abstract for too long, we have introduced a concrete example.
Consider the case where the velocity increment is not aligned with the direction of uniform motion. For instance, imagine a particle moving "upward" with a velocity increment of $\Delta y'/\Delta\tau$ in the frame $S_n$, while the frame $S_n$ itself is moving "horizontally" with velocity $v$ in the lab frame.
According to standard textbook treatments, one would apply the relation $d\tau = dt/\gamma$, leading to the familiar result: $\Delta v_y = \Delta v'_y/\gamma$. However, this approach can lead to paradoxical outcomes. Specifically, it implies a trajectory that does not appropriately mix spatial and temporal components.
One of the key conclusions drawn from the discussion in Chapter 7 is that special relativity does not allow for an absolute notion of simultaneity. Consequently, there is no well-defined concept of an instantaneous three-dimensional space. The standard expression for the transformation of a velocity increment, $\Delta v_y = \Delta v'_y/\gamma$, rests on a hidden assumption—that the $y'$-axis and $y$-axis remain parallel. In reality, the Lorentz transformation induces a Wigner rotation, which causes these axes to become oblique.
This insight immediately reveals a fundamental distinction between non-covariant and covariant descriptions of point particle trajectories. (We will explore velocity transformations in more detail in Section 17.6.)

\subsection{Incorrect Expansion of the Relation  $d\tau = dt/\gamma$  for Arbitrary Motion}

Textbook authors are fundamentally mistaken in their widespread belief about the general validity of the standard momentum–velocity relation. According to the theory of relativity, the equation $\vec{p}_{cov}  = m\vec{v}_{cov}/\sqrt{1 - v^2_{cov}/c^2}$ does not hold for particles moving along curved trajectories in the Lorentz lab frame.
Many experts who have studied relativity through conventional textbooks may find this claim counterintuitive or even unsettling at first. 

But how can such an unconventional momentum–velocity relation arise? We know that the components of the four-momentum
$p_\mu = (E/c,\vec{p})$ transform between Lorentz frames in the same way as the components of the four-position vector
$x = (x_0,\vec{x})$. However, when translating from four-vector notation back to the familiar three-dimensional velocity vector $\vec{v}$, defined via  $\vec{v} = d\vec{x}/dx_0$, subtleties emerge. Surprises are to be expected in this transition, especially when dealing with non-rectilinear motion. 

It is well established that for rectilinear accelerated motion, the standard momentum–velocity relation does hold. In such cases, the combination of the usual relation and the covariant velocity transformation law (Einstein’s velocity addition) is consistent with the covariant momentum transformation. Both the non-covariant and covariant approaches yield the same trajectory in the Lorentz lab frame.

We can see why by examining the transformation of the 3-velocity in the theory of relativity. For a rectilinear motion, this transformation is performed as
$v = (v'+V)/(1 + v'V/c^2)$.  
The relativistic factor $1/\sqrt{1 - v^2/c^2}$ is given by:

\[
1/\sqrt{1 - v^2/c^2} = (1 + v'V/c^2)/(\sqrt{1 - v'^2/c^2}\sqrt{1 - V^2/c^2})    . 
\]

The new momentum is then simply $mv$ times the above expression. But we want to express the new momentum in terms of the primed momentum and energy, and we note that 

\[
p = (p' + \mathcal{E}'V/c^2)/\sqrt{1 - V^2/c^2}    . 
\]

Thus, for a rectilinear motion, the combination of  Einstein's addition law for parallel velocities and the usual momentum-velocity relation is consistent with the covariant momentum transformation.

It is well known that collinear Lorentz boosts, like their Galilean counterparts, form a commutative group. This implies that the result of applying successive Lorentz boosts in the same direction is independent of the order in which they are applied. However, Lorentz boosts in different directions do not commute and do not form a group. In contrast, Galilean boosts always form a group, regardless of direction.

A key relativistic effect arises when composing non-collinear Lorentz boosts: the result is not simply another boost but a combination of a boost and a spatial rotation, known as the Wigner rotation. This phenomenon has no analogue in non-covariant physics. One of its consequences is the emergence of a non-trivial relationship between momentum and velocity: 

\[
\vec{p}_{cov}  \neq m\vec{v}_{cov}/\sqrt{1 - v^2_{cov}/c^2}    , 
\]

which also lacks a non-covariant counterpart. Here  $\vec{p}_{cov} $ is  the three-vector part of four-vector $p_{\mu}$. Relativity further reveals that this unusual momentum-velocity relationship is tied to motion along curved trajectories. In such cases, the concept of a single, shared "ordinary space" breaks down, leading to a fundamental distinction between covariant and non-covariant particle trajectories.

\subsection{Standard Integration of the 4D Covariant Equation of Motion}

Attempts to solve the dynamical equation, Eq.~(\ref{DDE}), in a manifestly covariant form can be found in the literature. However, the resulting trajectories often do not account for relativistic kinematic effects. Consequently, they cannot be identified with $\vec{x}_{cov}(t)$, even though they may initially appear to follow a covariant prescription.

We begin by analyzing the textbook treatment of covariant equation integration. Consider, for example, the motion of a charged particle in a given electromagnetic field. Of particular practical interest is the simplest case: a uniform electromagnetic field, where the field tensor $F^{\mu\nu}$ is constant throughout the space-time region of interest. Specifically, we focus on the motion of a particle in a constant, homogeneous magnetic field, represented by the tensor components
\[
F^{\mu\nu} = B\,(e^{\mu}_2 e^{\nu}_3 - e^{\nu}_2 e^{\mu}_3),
\]
where $e^{\mu}_2$ and $e^{\mu}_3$ are orthonormal spacelike basis vectors, $e_2^2 = e_3^2 = -1$, $e_2 \cdot e_3 = 0$. In the lab frame, where $e^{\mu}_0$ is taken as the time axis and $e^{\mu}_2$, $e^{\mu}_3$ are spatial vectors, the field is purely magnetic, with magnitude $B$ parallel to the $e_1$ axis. 

Let the initial four-velocity be 
\[
u^{\mu}(0) = \gamma c\, e^{\mu}_0 + \gamma v\, e^{\mu}_2,
\]
where $v$ is the particle's initial velocity relative to the lab along $e_2$ at $\tau = 0$, and $\gamma = 1/\sqrt{1-v^2/c^2}$. The components of the equation of motion are then
\[
\frac{du^{(0)}}{d\tau} = 0, \quad 
\frac{du^{(1)}}{d\tau} = 0, \quad 
\frac{du^{(2)}}{d\tau} = - \frac{eB}{mc} u^{(3)}, \quad 
\frac{du^{(3)}}{d\tau} = \frac{eB}{mc} u^{(2)}.
\]
We seek the initial-value solution to these equations as done in standard references.\footnote{Textbook authors, such as in \cite{RAFELSKI, CT, GF}, conclude from the covariant equations that a charged particle in a constant magnetic field moves in a uniform circular trajectory. This analysis does not account for relativistic kinematics, applying instead the Galilean law of velocity addition. The non-relativistic kinematics are obtained via $d\tau = dt/\gamma$, yielding the familiar non-covariant trajectory.}  

A distinctive aspect of relativistic dynamics is the presence of constraints. The evolution of the particle obeys $md u^{\mu}/d\tau = eF^{\mu\nu}u_{\nu}$, but also to the constraint $u^2 = c^2$.

However, this condition can be enforced only initially (at $\tau = 0$). If $Y(\tau) = u^2 - c^2$ vanishes initially, $Y(0) = 0$, then the Lorentz-force equation guarantees that $Y(\tau) = 0$ for all $\tau$.  

Integrating with respect to proper time, we obtain
\[
u^{\mu}(\tau) = \gamma c\, e^{\mu}_0 + \gamma v\,[\,e^{\mu}_2 \cos(\omega\tau) + e^{\mu}_3 \sin(\omega\tau)\,], 
\quad \omega = \frac{eB}{mc}.
\]
Here, $\gamma$ remains constant, indicating that the particle's energy is conserved. Integrating once more yields the trajectory
\[
X^{\mu}(\tau) = X^{\mu}(0) + \gamma c \tau\, e^{\mu}_0 + R\, [\, e^{\mu}_2 \sin(\omega\tau) - e^{\mu}_3 \cos(\omega\tau) \,], 
\quad R = \frac{\gamma v}{\omega}.
\]
This gives the time-dependent particle position $[0, X^{(2)}(t), X^{(3)}(t)]$, since $\tau = t/\gamma$. The motion is uniform circular, consistent with classical expectations, but derived within a fully covariant framework.  

One might expect that this trajectory should correspond to $\vec{x}_{cov}(t)$. However, $[0,X^{(2)}(t), X^{(3)}(t)]$ does not incorporate relativistic kinematics. The standard integration of the four-dimensional covariant equation, Eq.~(\ref{DDE}), yields a trajectory mathematically identical to that from Newtonian dynamics: relativistic effects are absent, and the Galilean law of velocity addition is effectively applied.  

The issue lies in a subtle conceptual point rather than a computational error. The first integration with respect to proper time $\tau$ yields the four-momentum, which is convention-invariant and physically meaningful. The concept of three-dimensional velocity only arises during the second integration. In textbooks, the initial condition
\[
u^{\mu}(0) = \gamma c\, e^{\mu}_0 + \gamma v\, e^{\mu}_2
\]
represents four-momentum components, not velocities. When the second integration is performed using $d\tau = dt/\gamma$, one recovers the conventional, non-covariant trajectory.  

This mirrors the (3+1) non-covariant particle tracking results. From the four-component equation, we see that only three equations are independent. The time component is treated simply as the parametrization
\[
d\tau = \frac{dt}{\gamma},
\]
yielding the corrected Newton equation, Eq.~(\ref{DDDE1}), in terms of lab time $t$. This approach relies on three spatial coordinates and velocities, unconstrained, and implicitly assumes an absolute-time convention for clock synchronization in the lab frame.

\subsection{Convention-Invariant Particle Tracking}

So far, we have described the motion of a particle in three-dimensional space using the vector-valued function $\vec{x}(t)$. This function defines a prescribed curve (or path) along which the particle moves. The motion along this path is parameterized by $l(t)$, where $l$ represents a particular parameter—specifically, the arc length in the case of interest.

It is important to distinguish between the concepts of path and trajectory. A trajectory provides a complete description of a particle’s motion by specifying its position as a function of time. In contrast, a path is a purely geometric construct, independent of time. Paths can take various forms, such as straight lines, circular arcs, or helical curves.

If we connect the origin of a Cartesian coordinate system to a point on the path, the resulting vector is called the position vector, denoted by $\vec{x}(l)$. The derivative of this vector with respect to arc length, $d\vec{x}(l)/dl$, yields the tangent vector to the curve at that point. The direction of this tangent vector is determined by the orientation of the curve as traced by increasing arc length.

As established in Chapter 2, the path $\vec{x}(l)$ has an exact, objective meaning—it is invariant under changes in convention. Similarly, the components of the four-momentum vector $mu = (\mathcal{E}/c, \vec{p})$ are also convention-invariant. In contrast, the trajectory $\vec{x}(t)$—as measured in a laboratory frame—is convention-dependent, reflecting the conventionality inherent in the definition of velocity. As such, it lacks the same level of objective meaning as the path $\vec{x}(l)$.

We now describe how to determine the position vector $\vec{x}(l)_{cov}$ in the convention-invariant particle tracking approach. Consider a particle moving in a uniform magnetic field, with no electric field present. Using Eq.~(\ref{DDE}), we obtain:

\begin{eqnarray}
&& \frac{d\vec{p}}{d\tau} =  \frac{e}{mc} ~ \vec{p}\times \vec{B}, ~ ~\frac{d\mathcal{E}}{d\tau} = 0 ~ ~ .
\end{eqnarray}

Since $d\mathcal{E}/d\tau = 0$ and the constraint Eq. (\ref{DCFE1})  holds, it follows that $dp/d\tau = 0$, where 

\[
p = |\vec{p}| = m|d\vec{x}_{cov}|/d\tau       .
\]

The unit vector $\vec{p}/p$ satisfies: 

\[
\vec{p}/p =  d\vec{x}_{cov}/|d\vec{x}_{cov}| = d\vec{x}_{cov}/dl   ,
\]

where $|d\vec{x}_{cov}| = dl$ is the differential path length.

From this we derive:

\begin{eqnarray}
&& \frac{d^2\vec{x}_{cov}}{dl^2} =  \frac{d\vec{x}_{cov}}{dl}\times \left (\frac{e\vec{B}}{pc}\right)~ .\label{CPP}
\end{eqnarray}

These equations are identical to those obtained using the non-covariant particle tracking approach. Consequently, $\vec{x}(l)_{cov} = \vec{x}(l)$, as expected. Both formulations describe the same physical reality.
Since the curvature radius of the trajectory in a magnetic field—and therefore the three-momentum—has an objective (i.e., convention-invariant) meaning, both approaches yield the same physical results.

\subsection{Einstein's Velocities Addition Vs. Covariant Addition}

In conventional particle tracking, the trajectory of a particle, denoted by $\vec{x}(t)$, is often described from the laboratory frame as the result of successive Galilean boosts. These boosts account for the motion of a particle accelerating in a constant magnetic field. The classical Galilean rule for the addition of velocities is used to determine the appropriate boost at each instant, effectively tracking the particle’s motion along its curved trajectory.

However, we cannot rely on Newtonian kinematics for mechanics while simultaneously applying Einsteinian kinematics to electrodynamics. A consistent and accurate description requires solving the dynamical equations in their covariant form. Only this approach ensures the correct coupling between Maxwell's equations and the particle trajectories observed in the laboratory frame. In this covariant framework, the trajectory  $\vec{x}_{cov}(t)$ is understood, from the lab frame, as the result of successive Lorentz transformations. This introduces relativistic kinematic effects that are absent in the classical treatment. 

Let us apply our algorithm for reconstructing $\vec{x}_{cov}(t)$ to a few examples to observe its performance in practice. To begin, we revisit the standard textbook derivation of velocity transformations between reference frames.
Assume a frame $K'$ moves relative to the $K$ system with velocity $V$ along the $x$ axis. Let $\vec{v} = d\vec{x}/dt, $ be the vector of the particle velocity in the $K$ system and $\vec{v'} = d\vec{x'}/dt'$ the velocity vector of the same particle in the $K'$ system. From Lorentz transformation we have  

\[
dx = \gamma(dx' + Vdt')   ,   \qquad   dy = dy'      ,     \qquad  dz = dz'     ,   \qquad   dt = \gamma(dt' + Vdx'/c^2)     , 
\]

where $\gamma = 1/\sqrt{1 - V^2/c^2}$. Dividing the spatial components by the time component yields the relativistic velocity transformation formulas: 

\[
v_x = (v_x' + V)/(1 + Vv'_x/c^2)    ,      \qquad      v_y = v'_y/[\gamma(1 + Vv'_x/c^2)]      ,   
\]

\[
 v_z = v'_z/[\gamma(1 + Vv'_x/c^2)]     .
\]

These expressions describe the relativistic law of composition of velocities and form the basis of velocity transformations in special relativity, as presented in standard texts.

We begin with the simple case of relativistically combining perpendicular velocities. While there is no length contraction in directions perpendicular to motion, time dilation still affects the observed velocities.
As a result, in the frame $K$, the particle's velocity components are
$v_x = V$, $v_y = v_y'/\gamma$. It is important to clarify what is meant by "perpendicular" in this context. Two velocity vectors can be meaningfully said to be perpendicular only if they are defined within the same reference frame. Therefore, when textbook authors assert that  $\vec{e}_{y'}v_y'$ and   $\vec{e}_x V$ are perpendicular, what they actually rely on—often implicitly—is the assumption that the spatial axes $(x',y',z')$ and $(x,y,z)$ are parallel. In other words, they assume that the observers in frames  $K$ and $K'$ share a common 3-space. This is a subtle but important misconception.

This is a good point to make some general remarks about Einstein’s velocity addition.
It should be noted that presented above the commonly accepted derivation of velocity composition does not follow from the composition of Lorentz boosts. In particular, second boost from $K'$ to particle proper frame does not Lorentz boost. Einstein's velocity addition is based on the relation $dt' = dt/\gamma$, which implicitly assumes an absolute time convention for second boost. However, it  does not account for the interplay between transverse positions $y'$ and time $t'$, which is inherent in Lorentz boost matrices.

In contrast, every Lorentz boost represents a transformation between reference frames within a Lorentzian framework (i.e., under Lorentz coordinatization including also transverse direction). While Lorentz boosts intertwine space and time coordinates, Einstein’s velocity addition relies on the assumption of a shared three-dimensional space across all inertial frames. The consequence of this difference is the Wigner rotation accompanying the composition of non-parallel 3D velocities.

One of the consequences of the non-commutativity of non-collinear Lorentz boosts is the absence of a shared, common notion of ordinary space. Suppose that in the Lorentz frame  $K'$, a traveler observes a particle moving along the  $y'$ axis. In other words, the particle has a velocity component  $v'_y$ in this frame, while the frame $K'$ itself is moving with velocity $V$ along the $x$ axis relative to initial Lorentz frame $K$. The proper frame  $K''$ is rotated by an angle  $\theta_w$ relative to frame $K$. The coordinate axes of the proper frame remain parallel to those of frame $K'$. From the perspective of frame $K$, the coordinate axes of frame $K'$ are also rotated by the angle  $\theta_w$ with respect to those of frame $K$.

Let us consider the simple case where the boost velocity $V$  approaches the speed of light (i.e., the ultrarelativistic limit), while the transverse velocity  $v'_y$ remains small. This means we are taking the limit where

\[
\gamma \gg 1 \gg v'_y/c    ~ , 
\]

treating  $v'_y/c$  as a small parameter and keeping                                                                                                                                                                                                                                                                            
terms up to second order. In this regime, the axes of the  $K'$ frame appear rotated with respect to those of
$K$  by an angle  $\theta_w = v'_y/(\gamma c)$. Notably, this rotation is in the same direction as the rotation of the particle’s velocity vector in $K$.  
We now consider the physical consequences of this rotation. First, note that the transverse velocity component  $v'_y$ acquires a projection onto the $x$-axis when viewed from frame  $K$.
As a result, the component of the particle’s velocity along the  $x$-axis in frame  $K$ is slightly reduced from $V$.

The boost from frame $K'$ to $K''$, as observed from $K$, can be understood as a sequence of infinitesimal Lorentz transformations, each involving a small velocity increment $\Delta v'_y/\gamma$. To provide a complete analysis, we must also account for the Wigner rotation which occurs in the same direction as the rotation of the velocity vector of $K''$ as seen from $K$. The coordinate axes of the frame $K''$ remain parallel to those of frame $K'$. From the perspective of frame $K$, the coordinate axes of frame $K'$ are also rotated by the same angle with respect to those of frame $K$.

The infinitesimal velocity increment of $K''$ as observed from  $K$ can be decomposed into two physically meaningful components:
1) a transverse component of $\Delta v_{\perp} \simeq \Delta v'_y/\gamma$, perpendicular to the original velocity $\vec{V}$, and
2) a tangential component of $\theta_w\Delta v'_y/\gamma\simeq \theta_w\Delta v_{\perp}$ which is aligned with the direction of $\vec{V}$, and arises due to the Wigner rotation.
Here $\theta_w$ denotes the Wigner rotation angle. In the ultra-relativistic limit we approximate it as $\theta_w  \simeq v_{\perp}/c$. Notably, analyzing the transverse component is considerably more straightforward than handling the tangential one. In fact, to second order in $v'_y/c$, no relativistic corrections affect the transverse velocity component.

We now proceed to compute the net effect of accumulating these infinitesimal increments. This process is equivalent to integrating over the transverse velocity: $\int \theta_w (v_{\perp})dv_{\perp}$.  Evaluating this from $v_{\perp} = 0$ to $v_{\perp} = v'_y/\gamma$ gives the total tangential correction due to the Wigner rotation. This results in a tangential velocity component of  $V - (v'_y)^2/(2\gamma^2c)$.
Thus, to second order in  $v'_y/c$,  the total speed of frame $K''$ with respect to $K$ remains $V$.
This is intuitively reasonable, since  we are using Lorentz coordinatization and in our relativistic asymptotic  $V$ approaching the speed of light.

This result stands in contrast with the textbook prediction, which states that the total speed of the particle in frame 
$K$ increases from $V$ to $V + (v'_y)^2/(2\gamma^2c)$.  Our result contradicts this claim. The standard derivation of velocity addition neglects the mixing of transverse spatial coordinate $y'$ with time $t'$ in Lorentz transformations. This omission leads to an incorrect conclusion about the magnitude of the resulting velocity in initial inertial frame $K$.

It is important to highlight another effective approach to covariant particle tracking. Here, we revisit the motion of an accelerated relativistic particle from the perspective of an inertial lab observer—without changing reference frames. Specifically, we consider a particle undergoing acceleration within the lab frame, representing an example of an active boost. In this scenario, we track the system's evolution entirely within a single (lab) reference frame.

The simplest synchronization method maintains the same set of uniformly synchronized clocks used before the acceleration, without any adjustments. This approach is based on the absolute time (or absolute simultaneity) convention, which preserves simultaneity across space. Under this convention, the standard Galilean velocity addition rule can be applied instant by instant to track the particle’s motion as it accelerates.

As discussed in Chapter 3, absolute time coordinatization can be transformed into Lorentz coordinatization. By combining Galilean transformations with appropriate changes of variables, we can derive the Lorentz transformation within the framework of absolute time. To further illustrate this idea, we now examine how distant lab clocks resynchronize during the particle’s acceleration. This analysis reveals a direct connection between the transformation of particle velocity (in Lorentz coordinates) and time dilation.

As previously emphasized, the Lorentz coordinate system is a conceptual construct. The notion of synchronized clocks in this system is purely hypothetical, used primarily for the application of Maxwell’s equations in radiation calculations.

Suppose that, before acceleration, we choose a Lorentz coordinate system in the initial inertial frame $K$. Immediately after the acceleration, particle velocity changes by an infinitesimal amount $d\vec{v}$ along the $y$-axis.
If clock synchronization is fixed, this corresponds to adopting the absolute time convention. To maintain Lorentz coordinates in the $K$ frame—as discussed previously—we must perform a clock resynchronization by introducing an infinitesimal time shift. The simplest case arises when the $v_y = v'_y/\gamma$ is very small, allowing us to work up to second-order terms in $v_y^2/c^2$. This restriction greatly simplifies the calculations. We observe that relativistic corrections to the boost velocity $v_y$ only appear at order  $v_y^3/c^3$ and can be neglected. This enables us to defer clock resynchronization until after the total second boost.

To maintain a Lorentz coordinate system in the lab frame after the acceleration, a resynchronization of clocks is required. In this context, operations involving the rule clock structure in the lab frame can be interpreted as a change of variables governed by the transformation in Eq. \ref{GGT3}:

\[
y_L = \gamma_y y     ,   \qquad     t_L = (t/\gamma_y + \gamma_y yv_y/c^2)     , 
\]

where 

\[
\gamma_y = 1 + v_y^2/(2c^2) 
\]

and the particle's displacement is given by $y = v_yt$.

This transformation effectively rescales the time variable—adjusting the synchronization of all clocks from
$t$ to $\gamma_y t$.  
As a result, we see that the particle's speed after acceleration is unaffected by its transverse motion.
Although no second-order relativistic correction appears in the transverse ($y$-axis) velocity component, the longitudinal component is modified. Specifically, the longitudinal velocity  $V$  changes to $V/\gamma_y$ with $v_z = V - v_y^2/(2c)$. 

Thus, after clock resynchronization, the total electron speed in the lab frame remains $V$—a result that aligns perfectly with the outcome derived earlier using a sequence of Lorentz transformations in changing frames. All pieces fit together seamlessly. We will explore the topic of relativistic velocity composition in greater detail in the next chapter

This analysis is carried out in the limit of a small Wigner rotation angle, $\theta_w = v'_y/(\gamma c) \ll 1/\gamma$. Even within this simplified regime, we can clearly demonstrate the fundamental difference between covariant velocity addition and Einstein's velocity addition.

To properly interpret the physical meaning of covariant velocity addition as presented here, it is essential to recognize that Einstein’s velocity addition is fully compatible with conventional particle tracking methods, which are based on the absolute time convention. However, the only rigorous way to maintain consistency between Maxwell’s equations and particle trajectories is to formulate and solve the equations of motion in a fully covariant manner.

\newpage

\section{Mathematical Analysis of Relativistic Velocity Composition}

In this chapter, we shall explore the subject of relativistic velocity addition more mathematically. Textbook authors often assume that the principle of relativity requires Einstein’s velocity addition law to obey group composition. However, the non-associativity of non-collinear velocity addition is not widely recognized. Many physicists, having studied relativity from standard textbooks, might find this claim surprising, since Einstein’s velocity addition is commonly assumed to be associative. Yet, a straightforward algebraic exercise reveals its non-associative nature.

The root cause of this issue lies in the presence of the Wigner rotation. In fact, the breakdown of commutativity and associativity in Einstein’s velocity addition law naturally manifests as Wigner rotation, offering deeper insight into the structure of relativistic transformations.

\subsection{Parametrization of Lorentz Transformations}

Einstein formally defined the special theory of relativity through two explicit postulates. Minkowski, in 1908, recognized that these postulates could be reformulated as a single geometric axiom: spacetime possesses a pseudo-Euclidean geometry. As a consequence of this spacetime geometry, Lorentz coordinate systems are related via Lorentz transformations.

Consider a four-vector, $x = (x_0 , \vec{x})$, with
squared-length $x^2 = \eta_{ij} x^i x^j = (x_0)^2 - |\vec{x}|^2$  where $\eta_{ij} = \mathrm{diag}[1,-1,-1,-1]$ is the Minkowski space-time metric. Since Lorentz transformations preserve the space-time interval, they satisfy  
$x' = \Lambda x$ such that $\eta_{ij} = \Lambda^m_i\Lambda^n_j \eta_{mn} $. 
This condition defines the most general $4 \times 4$ Lorentz transformation matrix $\Lambda$. The set of all such matrices forms the symmetry group of the Minkowski metric under matrix multiplication, denoted by $O(1,3)$. In particular, $O(1,3)$ is a Lie group known as the Lorentz group.

It is widely accepted that the relationship between inertial frames is fully determined by their relative velocities and orientations. Consequently, a general Lorentz transformation, denoted $\Lambda(\vec{v}, \vec{\theta})$, is parameterized by a velocity vector $\vec{v}$ and a rotation vector $\vec{\theta}$, representing a combined boost and spatial rotation. Here, $\vec{v}$ is a vector in three-dimensional Euclidean space, while $\vec{\theta}$ belongs to the Lie algebra of the rotation group $O(3)$, which consists of all possible spatial rotations in three dimensions. This parameterization of the Lorentz group is standard in the literature.

Now, we turn to a significant conceptual and structural challenge associated with this parametrization of the Lorentz group. A comparison with the Galilean group provides valuable insight.

It is well known that the Galilean transformation $G(\vec{v}, \vec{\theta})$, which satisfies the group composition law, is isomorphic to the semi-direct product group $\mathbb{R}^3 \rtimes O(3)$, where $\mathbb{R}^3$ represents the group of Galilean boosts and $O(3)$ denotes the rotation group. In this structure, $O(3)$ acts as a normal subgroup, and the composition of transformations remains straightforward.

In contrast, the composition of Lorentz transformations is significantly more intricate. The essential distinction between the Galilean and Lorentz groups lies in their group-theoretic structure: the Lorentz group is a ``simple'' Lie group, meaning it does not admit a nontrivial decomposition into a direct or semi-direct product of boosts and rotations. In particular, $O(3)$ is not a normal subgroup of the Lorentz group $O(1,3)$. As a result, the combined action of Lorentz boosts and spatial rotations does not obey the same algebraic structure as in the Galilean case.

This leads to an important conclusion: standard textbook analyses of inertial frames often overlook the fundamental geometric difference between Newtonian space and time, and the pseudo-Euclidean geometry of Minkowski spacetime.

In classical mechanics, the components of the velocity three-vector transform under Galilean transformations in the same way as the components of the position vector $\vec{x}$. This correspondence arises because, in Newtonian physics, all inertial observers share a common, absolute three-dimensional space. Time is also absolute, so the spatial position and velocity vectors can be consistently defined across all inertial frames.

However, in special relativity, this intuition fails. When transitioning from the four-dimensional spacetime formalism back to the three-dimensional velocity vector $\vec{v}$, defined as $\vec{v} = d\vec{x}/dx^0$, subtle but profound issues emerge. The Lorentz group $O(1,3)$ acts on four-dimensional spacetime, and all of its subgroups also act on four-vectors, not directly on spatial vectors. Therefore, there is no concept of a universal or common instantaneous three-dimensional spatial frame among all inertial observers within the framework of the Lorentz group.

This absence of a shared simultaneity structure underlies many of the non-intuitive features of relativistic kinematics, including the non-associativity and non-commutativity of velocity addition.

\subsection{The Preferred (1 + 3) Decomposition of Spacetime}

Let us return to the Lorentz transformations and clarify the distinction between reference systems and coordinate systems. A reference system must be material, whereas a coordinate system has no intrinsic connection to physical mass. In particular, one may change coordinates without altering the underlying material reference system.

An inertial system must therefore possess nonzero mass; that is, it must be associated only with timelike vectors. Special relativity postulates that any two material (timelike) reference systems are related by a Lorentz isometry. Such an isometry acts on all vectors in spacetime, including lightlike and spacelike ones, even though only timelike vectors can represent reference systems.

The formal axiom of special relativity—that any two reference systems are related by a Lorentz isometry—does not involve the concept of relative velocity. To extract a notion of relative velocity from a Lorentz isometry, one must factor the Lorentz group $O(1,3)$ by a rotation subgroup $O(3)$.

What does this imply? A careful examination shows that each choice of a rotation subgroup corresponds to a choice of a privileged reference system. Consequently, within the Lorentz isometry itself, the relative velocity between two reference systems is not uniquely defined. Uniqueness of relative velocity requires the selection of one reference system as preferred.

Thus, there exists a unique relative velocity if and only if a preferred reference system (absolute space) is chosen. Only the preferred observer—that is, a preferred $(1+3)$ decomposition of spacetime—determines a unique relative velocity for each pair of massive bodies.

Let us now examine further consequences of adopting a preferred $(1+3)$ decomposition of spacetime. A standard representation of Lorentz transformations employs $4\times4$ matrices acting on spacetime four-vectors. Any Lorentz transformation can, in principle, be decomposed into a rotation and a boost. However, the explicit form of this decomposition depends on the chosen reference frame, which in the matrix formalism corresponds to selecting a distinguished timelike basis vector, typically denoted by $e_0$.

For example, in his widely used textbook, Barut defines a Lorentz boost (see Eq.~(1.28) and the accompanying discussion in \cite{BA2}) as

$$  L_{\vec{v}} = (L_{-\vec{v}})^{-1} = \left( \begin {array}{cccc}
\gamma& -\beta\gamma & 0 & 0 \\
-\beta\gamma& \gamma& 0& 0 \\
0 & 0 & 1& 0\\
0 & 0& 0& 1\end{array}\right) .$$

where $\beta = v/c$ and $\gamma = 1/\sqrt{1-\beta^2}$. This boost corresponds to motion along the $x$-axis and is defined relative to a specific choice of basis.

The Lorentz transformations between inertial frames cannot be derived without the reciprocal inverse axiom 
 $L_{\vec{v}} = (L_{-\vec{v}})^{-1}$ .

The relativistic reciprocity principle states that if an inertial frame $K'$ moves with velocity $\vec{v}$ relative to another inertial frame $K$, then the velocity of $K$ relative to $K'$ is $-\vec{v}$. This principle treats the inverse of a velocity as its negation, $\vec{v}^{-1}=-\vec{v}$, in close analogy with Galilean kinematics. It is usually presented in textbooks as self-evident.

Moreover, a Lorentz boost is not fully determined by the velocity parameter $\vec{v}$ alone. Velocity is intrinsically relational: it is the velocity of one massive body relative to another. However, in Einstein’s formulation of special relativity, this relational character is obscured by the postulate that every pair of reference systems is connected by a Lorentz isometry. The very concept of velocity presupposes a distinction between space and time—that is, a $(1+3)$ decomposition of spacetime—which requires the selection of a specific timelike direction. This direction identifies the observer through the choice of a basis vector $e_0$ representing time in that observer’s rest frame.

We now describe the geometric meaning of the basis vector $e_0$ from the viewpoint of Minkowski spacetime. Any material body in Minkowski spacetime is associated with a four-velocity vector, referred to here as the Minkowski (absolute) four-velocity. This is a future-directed timelike vector $e_0$, normalized to the speed of light.

As made explicit by Barut and other standard references \cite{BA2}, the Lorentz boost is expressed as a $4\times4$ matrix relative to a chosen orthonormal basis. In the standard formulation, the preferred inertial frame is associated with the basis vector $e_0=(1,0,0,0)$. Rotations act within the three-dimensional spatial subspace orthogonal to $e_0$, and any boost involving a velocity vector $\vec{v}$ orthogonal to $e_0$ is implicitly defined with respect to this preferred frame.\footnote{From the Minkowski spacetime viewpoint, a three-vector $\vec{v}$ is understood as a spacelike four-vector $v_\mu=(0,\vec{v})$.} Thus, although Lorentz transformations are covariant under changes of inertial observers, their explicit matrix representations—and even the definition of velocity itself—depend on the choice of a preferred observer.

Successive Lorentz boosts are composed by matrix multiplication, provided that both boosts are expressed in the same basis. If $L_{\vec{v}_1}$ is the boost from frame $K_0$ to frame $K_1$, and $L_{\vec{v}_2}$ is the boost from frame $K_1$ to frame $K_2$, their composition is $L_{\vec{v}_2}L_{\vec{v}_1}$. Here, $\vec{v}_1$ is the velocity of $K_1$ relative to the preferred frame $K_0$, and $\vec{v}_2$ is the velocity of $K_2$ relative to $K_1$, as seen from $K_0$.

It is important to stress that the second velocity $\vec{v}_2$ is not a genuine velocity measured by the body $K_1$, which itself moves relative to the laboratory frame $K_0$ with velocity $\vec{v}_1$. Rather, $\vec{v}_2$ is the velocity of $K_2$ relative to $K_1$, as reconstructed by the preferred observer $K_0$.

Finally, $(1+3)$ decompositions of spacetime based on different choices of the timelike vector $e_0$ are mutually incomparable, since they correspond to distinct three-dimensional spatial hypersurfaces for which no canonical identification exists. In the standard Lorentz-group-based velocity addition, all velocities are implicitly assumed to lie within the same preferred spacelike “space,” namely the space defined by the preferred observer.

\subsection{Observer-Independence and Lorentz Covariance} 

It is crucial to distinguish between observer-independence and Lorentz covariance. Although Lorentz covariance refers to the invariance of physical laws under Lorentz transformations, it does not imply that all inertial frames are physically equivalent in a pseudo-Euclidean spacetime. The geometry of Minkowski space allows for Lorentz covariance without guaranteeing the equivalence of all inertial observers. This subtle distinction opens the door to the possibility of a physically preferred inertial frame, thereby challenging the traditional interpretation of the principle of relativity.

A deeper tension emerges between the principle of relativity—which asserts the equivalence of all inertial frames—and the geometric structure of Minkowski spacetime. 
In fact, only one of these principles can hold unconditionally.
\footnote{For a more detailed analysis, see “How Do You Add Relative Velocities?” by Oziewicz in \cite{Z, Z1}. Oziewicz argues that within pseudo-Euclidean spacetime, the equivalence of all reference frames is not possible.}

As discussed in earlier chapters, further mathematical considerations support the view that pseudo-Euclidean geometry does not entail the complete equivalence of all inertial frames. One key insight comes from analyzing the continuity of the metric tensor under coordinate transformations. The Langevin metric, for instance, arises from a continuous deformation of the Minkowski metric when transitioning between an inertial frame and an accelerated frame. This smooth transition can be modeled via a Galilean boost relating the coordinate systems of the inertial and accelerated observers.

However, restoring the diagonal Minkowski form in the accelerated frame often requires an abrupt coordinate transformation, resulting in a discontinuity in the metric tensor. This discontinuity signals a breakdown in the symmetry between inertial frames: if a Lorentz transformation fails to preserve the smoothness of the metric, then the equivalence of the frames involved is compromised.

This raises a natural and important question: How can observers identify a preferred inertial frame?

Within the structure of the Lorentz group, relative velocity between frames enters exclusively through the boost transformations. However, a Lorentz boost is always defined relative to a specific reference frame—typically one associated with a distinguished timelike vector. Special relativity, in its standard formulation, does not specify such a frame; yet physical observations suggest the existence of a unique inertial frame—one not subjected to prior accelerations, such as relative to the fixed stars.

There is, however, no intrinsic contradiction between the mathematical framework of special relativity and the physical asymmetry introduced by selecting a preferred frame. The theory is built on the requirement that physical laws remain invariant under Lorentz transformations. But special relativity is not a self-contained theory in the sense of fully determining physical phenomena from dynamical laws alone—initial conditions must also be specified. The principle of relativity asserts that the form of physical laws is independent of velocity, but not necessarily that initial conditions must be velocity-independent.

While Lorentz coordinatization appears to establish symmetry among inertial frames, the asymmetry emerges from the history of acceleration. Once a frame has undergone acceleration, information about that acceleration is not encoded in the transformation laws themselves but in the initial conditions from which they operate.

Consider, for instance, the relativistic aberration of particles emitted from a source at rest in an accelerated frame. As discussed in Chapter 9, special relativity predicts that—alongside a static electric field—a magnetic field also arises in the accelerated frame, leading to deviations in the observed trajectory of particles. This deviation is a measurable consequence of the frame’s prior acceleration.

The mathematical framework of special relativity remains entirely self-consistent when formulated relative to an initial inertial frame—one that has not undergone prior acceleration with respect to the distant stars. Such a frame functions as a *privileged reference system*. As demonstrated in Chapter 7, when the laws of physics are formulated relative to this frame, they accurately account for all observed phenomena, including those reported by accelerated observers.

In particular, phenomena like the aberration of light in an accelerated frame can be fully derived using standard tools of special relativity—specifically, Lorentz transformations and Wigner rotations—when analyzed from the standpoint of the initial inertial frame.

\subsection{Discussion}

Let us return to the privileged $(1+3)$ decomposition of spacetime.
The relativistic velocity addition law ($\oplus$) can be written as

\begin{eqnarray}
&&   \vec{v} \oplus \vec{u}
= \frac{\vec{u}+\vec{v}}{1+\vec{v}\!\cdot\!\vec{u}}
+\frac{\vec{u}\!\cdot(\vec{v}\times\vec{u})}
{(1+\sqrt{1-u^2})(1+\vec{v}\!\cdot\!\vec{u})}  ~ .\label{EAV}
\end{eqnarray}

(see, for example, Møller \cite{M}).  
Our goal is to clarify the interpretation of the quantities appearing in this expression.

At first sight, relativistic velocity addition seems to be formulated entirely in terms of ordinary three–dimensional vector algebra involving scalar and vector products.  
However, within relativity this interpretation is far from obvious.  
Because spacetime possesses a pseudo-Euclidean geometry, one must ask:  
where are these scalar and vector products defined — in spacetime itself or in some associated three–dimensional space?

The evolution of a reference system, understood as a material body, is represented by a worldline in spacetime, while any inertial reference system is mathematically characterized by a timelike unit four-velocity vector.  
If three–velocities describe mutual motion of observers, they are spacelike vectors belonging to the *private three-space* of each observer.

Consider three material bodies.  
An observer on body $A$ measures the relative velocity $\vec{u}$ of body $B$, while an observer on body $B$ measures the relative velocity $\vec{v}$ of body $C$.  
These measurements imply that the normalized four-velocity $e_A$ of body $A$ is orthogonal to $\vec{u}$, and the four-velocity $e_B$ of body $B$ is orthogonal to $\vec{v}$.  
Consequently, observer $A$ cannot directly measure the velocity $\vec{v}$.  
Indeed,
\[
\vec{u}\cdot e_A = 0, \qquad
\vec{v}\cdot e_B = 0,
\]
and the observers possess different four-velocities.

In special relativity simultaneity is relative.  
Moving reference systems therefore possess different simultaneity hypersurfaces, so the velocities $\vec{u}$ and $\vec{v}$ are tangent to different instantaneous three-spaces, which are generally not parallel.  

This observation raises an important question concerning the meaning of the velocity parameter appearing in the Lorentz transformation $L_{\vec v}$.  
It is usually assumed that a Lorentz transformation relating two inertial observers is an isometry of Minkowski spacetime parametrized by a reciprocal relative velocity.  
In particular, one adopts the property $L_{\vec v}=(L_{-\vec v})^{-1}$.


However, the reciprocity condition $\vec v^{-1}=-\vec v$ is not a direct consequence of pseudo-Euclidean spacetime geometry; rather, it is introduced as an additional physical assumption.  
When simultaneity is observer-dependent, moving material bodies possess different private three-spaces, and the velocity $\vec v$ and its inverse $\vec v^{-1}$ belong to different instantaneous spaces.

A careful  examination shows that the difficulty arises because the six-dimensional Lorentz group admits no unique factorization onto a three-dimensional space of Einstein velocities.

According to Oziewicz \cite{Z, Z1}, a two-body system possesses a relative velocity only with respect to a preferred third reference system — an initial inertial frame providing a privileged $(1+3)$ splitting of spacetime.  
The three-dimensional space of Einsteinian velocities must therefore be identified with the spatial section of this privileged inertial system.

Let us examine the physical meaning of these statements through examples.
We consider two reference systems: an initial inertial (Sun-based) frame and an accelerated (Earth-based) frame.  
Because of the relativity of simultaneity, inertial and accelerated observers do not share the same three-dimensional space in Minkowski spacetime.

It is important to emphasize that the Langevin metric represents measurements performed by an inertial observer within Minkowski spacetime and subsequently rewritten using the transformation
\[
x \rightarrow x_n + vt, \qquad
t \rightarrow t_n .
\]

This  metric describes the electrodynamics of the accelerated light (or electron) source with the viewpoint of the accelerated observer measurements  \textit{as viewing this of the inertial observer }. 

Within the special relativity, relative velocities between frames enters exclusively through the preferred (1 + 3) decomposition of spacetime.
In this context,  the reciprocity condition $\vec v^{-1}=-\vec v$ is a direct consequence of the inverse Galilean boost.
An inertial observer moving at velocity $dx_n/dt_n = -v$ relative to accelerated observer \textit{as viewing this of the inertial observer }. 

It is always possible to introduce coordinates in which the metric of the
accelerated system becomes diagonal. In such a Lorentz coordinatization,
light propagates with the invariant speed $c$, independent of the direction
of propagation and independent of the velocity of the emitting source.

Consider the transformation to Lorentz coordinates,
\[
t'_n
=
t_n\sqrt{1- v^2/c^2}
-
(v x_n/c^2)/\sqrt{1- v^2/c^2},
\]
\[
x'_n = x_n/\sqrt{1- v^2/c^2}.
\]

The velocity measured along the $x_n$–axis transforms according to
\[
dx'_n/dt'_n = (dx_n/dt_n)/[1- v^2/c^2  -  (v/c^2)dx_n/dt_n].
\]

Substituting the previously obtained velocity
\[
dx_n/dt_n = - v,
\]
one finds
\[
dx'_n/dt'_n = - v.
\]

All velocities   $dx/dt = v$, $dx_n/dt_n = -v$ and $dx_n'/dt_n' = -v$  are implicitly assumed to lie within the same preferred spacelike “space,” namely the space defined by the  observer  in initial inertial frame.

The relation between privileged and non-privileged frames can be clarified by comparison with Newtonian mechanics.  
In Newtonian theory, acceleration relative to the fixed stars is absolute.  
There is no relativity of acceleration, and absolute rotation or acceleration can be detected by purely internal experiments.

For example, the dynamics of a body observed in the Earth-based frame reveals fictitious forces arising from rotation.  
Even without observing distant stars, Earth's rotation can be detected through measurements such as the Coriolis effect.  
Thus, in Newtonian mechanics the system of fixed stars plays the role of a privileged frame.

At the same time, uniform motion relative to the fixed stars remains relative: absolute velocity cannot be determined.  
All inertial frames are equivalent because Newtonian equations of motion do not explicitly depend on velocity; internal dynamics remain invariant under Galilean boosts.

The situation changes fundamentally in electrodynamics.  
Electromagnetic fields are primary physical entities, and Maxwell’s equations are fully compatible with special relativity.  
Because these equations explicitly depend on velocity of light, Galilean relativity breaks down.  
The velocity of light transforms according to the same kinematical rules as particle velocities.

Within special relativity there exists a crucial distinction between an inertial system that has undergone acceleration and one that has not.  
This distinction is connected with acceleration relative to the initial inertial frame (or fixed stars).  
Although the duration of the acceleration stage may be negligible, the accumulated effect of acceleration — namely the final velocity relative to the initial inertial system — determines observable anisotropies in the accelerated frame.

\subsection{Einstein Velocity Addition and Conventional Particle Tracking}

Composition of relativistic velocities, also  named as "Einstein's addition of velocities" is somewhat confusing due to
fact that this addition is nonassociative. This is contrast with associativity of the Lorentz group.
Most texts on special relativity present the relativistic velocity addition formula only for parallel velocities. In this simplified case, Einstein's velocity addition $(\oplus)$ is both commutative and associative, making it a commutative group operation analogous to the Galilean velocities addition $(+)$. However, in the general case, Einstein's velocity addition $\oplus$ does not form a group.

For composite velocities, $\vec{u}\oplus\vec{v}$, reciprocity implies 

\[
(\vec{u}\oplus\vec{v})^{-1} = - (\vec{u}\oplus\vec{v}) = (-\vec{v})\oplus(-\vec{u}) = - (\vec{v}\oplus\vec{u})   . 
\]

However, since Einstein's velocity addition $\oplus$ is generally non-commutative, this leads to a contradiction, known as the reciprocity paradox. This paradox is closely related to the Mocanu paradox, as discussed by Mocanu \cite{MOCA} and was latter resolved by Ungar \cite{UNG} who demonstrated the crucial role of the Wigner rotation in the transformation. \footnote{It is sometimes referred to as the Thomas rotation}
Furthermore, Ungar \cite{UNG} also discovered that Einstein's velocity addition is non-associative meaning that 

\[
(\vec{w}\oplus(\vec{v}\oplus\vec{u})) \neq ((\vec{w}\oplus\vec{v})\oplus\vec{u})    . 
\]

Being non-associative, the Einstein's addition $\oplus$ is not a group operation. It is generally accepted by the physics community that non-associativity is paradoxical.

Let us illustrate non-commutativity of Einstein's velocity addition using only simple case of relativistically combining perpendicular velocities.
Consider a coordinate system $K_1$ moving with respect to $K_0$ with a velocity $\vec{v}_{01}$ and a frame $K_2$ moving with respect to $K_1$ with a velocity $\vec{v}_{12}$. 
The question then arises: what is the velocity transformation from $K_0$ to $K_2$?

Now, consider the case, where $\vec{v}_{01}$ and $\vec{v}_{12}$ are perpendicular. The "perpendicular" means that velocity $\vec{v}_{12}$ is perpendicular to line motion of frame $K_0$ in $K_1$. The velocity $\vec{v}_{02}$ of frame 
 $K_2$ as observed from   $K_0$ is given by 
 
\[ 
 \vec{v}_{02} = \vec{v}_{01} + \vec{v}_{12}\sqrt{1 - v_{01}^2/c^2}      .
\]

According to reciprocity principle, accepted in all textbooks, an observer in $K_2$ sees frame $K_1$ moving with velocity $-\vec{v}_{12}$, while an observer in $K_1$ sees frame $K_0$ moving with velocity  $-\vec{v}_{01}$. Similarly, applying Einstein's velocity addition, we find that the velocity $\vec{v}_{20}$ of frame 
$K_0$ as observed from $K_2$ is: 

\[
\vec{v}_{20} = - \vec{v}_{12} - \vec{v}_{01}\sqrt{1 - v_{12}^2/c^2}    .
\]

According to textbooks, $\vec{v}_{20}$ should be reciprocal of $\vec{v}_{02}$, but  they point in different directions. 
This reciprocity discrepancy leads to the Mocanu contradiction and highlights the non-associativity of Einstein's velocity addition.

It is important to stress that the commonly accepted derivation of velocity composition does not follow from the composition of Lorentz boosts. Specifically, if $L_{\vec{v}_{01}}$ is the boost from frame $K_0$ to frame $K_1$, and $L_{\vec{v}_{12}}$ is the boost from $K_1$ to $K_2$, then their  Lorentz boost formula composition for coordinates is given by: $L_{\vec{v}_{12}}L_{\vec{v}_{01}}$.  However, Einstein's velocity addition is based on the Lorentz boost matrix formula  for $L_{\vec{v}_{01}}$ only and implicitly assumes an absolute time convention for the second active boost  of $K_2$ with velocity $\vec{v}_{12}$ within frame $K_1$ (see Section 17.6 for more details). 
Einstein's velocity addition does not account for the interplay between  positions along  $\vec{v}_{12}$ direction and time $t_1$, which is inherent in Lorentz coordinatization. 

It is generally  assumed that an observer at rest in $K_1$  and the moving frame $K_2$ share the same Minkowski metric, thereby neglecting the need for any clock re-synchronization procedure. However, this standard approach is flawed, as it fails to account for the principle of equivalence between active and passive boosts within a single inertial frame (see Chapter 3 for more details). The standard expression for the transformation of velocities 
rests on a hidden assumption that 
$(x_1, y_1, z_1)$ axes of the moving $K_1$ frame remain parallel to the $(x_0, y_0, z_0)$ axes of the initial inertial frame $K_0$.

 We emphasize that the addition of non-collinear velocities in Lorentz-coordinatized spacetime is governed by the Wigner rotation. It is crucial to recognize that Einstein’s velocity addition does not directly follow from the full structure of Lorentz transformations   $L_{\vec{v}_{12}}L_{\vec{v}_{01}}$. Textbook derivations of transverse (perpendicular) velocity addition often invoke the relation  $dt_0 = dt_1/\gamma_{01}$, yet fail to account for the Wigner rotation, which introduces a coupling between transverse position and time.  All  Wigner rotations are implicitly assumed to lie within the same  theree-dimensional space  defined by the observer in the preferred  frame  $K_0$.

Despite this, Einstein’s velocity addition is widely used in relativistic dynamics raising a reasonable question: How can this be? It is important to emphasize that Einstein’s velocity addition is fully compatible with conventional particle tracking within the initial inertial frame, which itself relies on the absolute time convention. Crucially, Einstein’s velocity addition inherently incorporates relativistic time dilation. Similarly, conventional particle tracking also accounts for relativistic time dilation, albeit implicitly, through the use of relativistic mass for moving particles.

Non-covariant approach presents no fundamental challenges in mechanics. The preference for a non-covariant formulation within the framework of dynamics arises from its simplicity and practicality. 
What is essential is the consistent application of the chosen convention across both dynamics and electrodynamics.
In an absolute-time coordinatization, the equations of electrodynamics are not isotropic. This poses a challenge, as electrodynamics involves partial differential equations, whereas dynamics deals only with ordinary differential equations. Only by adopting a kinematic framework based on Lorentz coordinatization can we ensure the correct coupling between the isotropic Maxwell equations and particle trajectories in the laboratory frame.

\subsection{Lorentz-Covariant Particle Tracking}

In covariant particle tracking, a particle’s trajectory is described from the perspective of an initial inertial frame as a result of successive Lorentz boosts. Let us now apply this approach to a concrete example and assess its practical implications.

Suppose an observer in the initial inertial frame $K_0$ establishes a coordinate system using Einstein’s synchronization procedure, which relies on light signals emitted by a dipole source at rest and assumes that light propagates isotropically at speed $c$. This procedure enables the construction of a Lorentz coordinate system in $K_0$.

Now, consider two additional inertial frames, $K_1$ and $K_2$, both moving relative to $K_0$. The simplest synchronization method is to retain the same set of synchronized clocks across all frames. This preserves simultaneity and leads to an absolute-time coordinatization based on $K_0$.

To describe physical phenomena across these frames, the preferred observer uses a metric tensor to derive the (electro)dynamical equations.
According to the equivalence between passive and active Lorentz boosts within a single inertial frame, the metric in Eq.~\ref{GGG11} characterizes the measurements of the preferred observer.

Let us now examine velocity composition within the absolute-time coordinatization framework. Assume that frame $K_1$ moves with velocity $\vec{v}_{01}$ relative to $K_0$, and frame $K_2$ moves with velocity $\vec{v}_{12}$ relative to $K_1$. Importantly, both $\vec{v}_{01}$ and $\vec{v}_{12}$ are defined from the perspective of the observer in $K_0$. 
\footnote{The velocity $v_{12}$ used here denotes the relative velocity as measured by a preferred observer, and should not be confused with the $v_{12}$ in the preceding section, which refers to the relative velocity measured in frame $K_1$ when observing this observer from frame $K_0$.} In this setup, velocity transformation is particularly straightforward: under absolute-time coordinatization, Galilean velocity addition applies. Thus, the velocity of $K_2$ as observed from $K_0$ is simply $\vec{v}_{02} = \vec{v}_{12} + \vec{v}_{01}$.

This observation highlights a key advantage: when the preferred observer adopts the absolute-time coordinatization, relativistic kinematics becomes extremely simple. In this formulation, all the complexity of relativistic particle tracking is absorbed into the frame metrics.

Consider now two electrons, both moving with the same velocity $\vec{v}_{02}$ relative to the preferred frame. The first electron is accelerated directly from rest to $\vec{v}_{02}$ within the preferred frame. The second reaches the same final velocity through two steps: first accelerated to $\vec{v}_{01}$, then from $\vec{v}_{01}$ to $\vec{v}_{02}$. Despite having the same final velocity, the two electrons have different acceleration histories. 
As a result, the electrodynamics of these two electrons will differ due to differences in their respective metrics. The contrast between these seemingly equivalent cases is both subtle and intriguing.

This paradoxical asymmetry between the two setups is most naturally understood from the perspective of electrodynamics. Our everyday intuition often breaks down when confronted with the counterintuitive nature of special relativity. The dynamics governed by electromagnetic fields are, in this context, encoded in the language of pseudo-Euclidean geometry, which can obscure their physical implications.

Let us consider the case where the velocity $\vec{v}_{01}$ approaches the speed of light, while the velocity $\vec{v}_{12}$ is perpendicular to $\vec{v}_{01}$ and remains small. More specifically, we are interested in the limit where $\gamma_{01} \gg 1 \gg v_{12}/c \gg 1/\gamma_{01}$.

In this setup, the first electron is accelerated in parallel geometry, reaching a final velocity $\vec{v}_{02}$. The second electron, in contrast, is finally accelerated in a direction perpendicular to the ultrarelativistic velocity $\vec{v}_{01}$. It is well known that transverse acceleration in the ultrarelativistic regime results in radiation that is enhanced by a factor of approximately $\gamma_{01}^2$ compared to the case of parallel acceleration.

Thus, in the ultrarelativistic limit, the first electron (undergoing parallel acceleration) emits negligible radiation. In contrast, the second electron emits significant synchrotron radiation due to its transverse acceleration. From the standpoint of electrodynamics, these two configurations are fundamentally different. In particular, the radiation emitted by the second electron carries information about its acceleration history.

Now let us consider a scenario where the deflection angle $v_{12}/c$ of the second electron is smaller than $1/\gamma_{01}$. In this regime, synchrotron radiation is effectively absent. This leads to a compelling question: in the absence of radiation, where—within the framework of absolute time coordinatization—is the information about the second electron’s acceleration physically located?

According to conventional particle tracking, both electrons end up with the same final velocity after acceleration. However, the second electron underwent a qualitatively different acceleration process, raising subtle questions about how this information is encoded in the system when no radiation is emitted.

One might be tempted to assert: “Since an electron is a structureless particle, this situation seems paradoxical.” However, this is not accurate. 
For a rapidly moving electron, the transverse components of the electric and magnetic fields are nearly equal in magnitude and orthogonal to one another—effectively indistinguishable from those of a real radiation beam.   When the electron experiences transverse acceleration, its electromagnetic field is perturbed in a way that reflects this acceleration.
\footnote{Certain assumptions are necessary when applying relativistic kinematics in practice. If the electron undergoes a sudden acceleration, there is a minimum distance—known as the formation length—required for the development of the (virtual laser-like) self-field of an ultrarelativistic electron. Only when the distance traveled after the first acceleration exceeds this formation length can a second acceleration (or 'boost') be considered kinematically independent. If this condition is not met, the two accelerations effectively combine into a single boost from rest to a final velocity $v_{02}$ and the electron’s acceleration history will appear nearly identical from the perspective of relativistic electrodynamics.}

Under the framework of classical (Galilean) kinematics, the orientation of the virtual radiation phase front remains unchanged. However, since Maxwell's equations are not invariant under Galilean transformations—as discussed throughout this book—the adoption of such kinematics necessitates the use of anisotropic field equations. Consequently, although the phase front remains planar, the direction of propagation is no longer perpendicular to it. In other words, the (virtual) radiation beam’s direction of motion diverges from the normal to its phase front.

Within the absolute time coordinatization, electrodynamics predicts that the virtual radiation beam of the second electron propagates along the direction of $\vec{v}_{02}$, with a phase front tilt $v_{12}/c$. This tilt is the crucial clue to resolving the paradox. The key insight is this: The information about the electron's acceleration is not lost—it is embedded in the perturbation of the electron’s self-fields. Mathematically, this information is encoded in the difference between metrics of these two electrons 

This rises a reasonable question: why does there is no virtual phase front perturbation in the situation with first electron acceleration? The key distinction is in collinear geometry - the first electron accelerates perpendicular to the plane of the virtual radiation wavefront (i.e., the plane of simultaneity). In the case of the boost with velocity $v_{12}$ the second electron undergoes acceleration along the virtual radiation wavefront.  

Absolute time coordinatization in the preferred frame can be transformed into Lorentz coordinatization. 
In the Chapter 3 we consider a general method for metric diagonalization in the case when particle accelerated from rest to velocity $v$ in the preferred frame. The change variables according to Eq. \ref{GGT3} completed by Galilean transformation is mathematically equivalent to the Lorentz transformation with the same velocity. As a result, covariant particle tracking in the case of the first electron is straightforward. Similar there is no problem with the first boost of the second electron. To maintain a Lorentz coordinate system in the preferred frame after the second boost with velocity $v_{12}$, additional changes in rule-clock structure is required as described in Section 17.6. Thus, after addition variable change, the total speed of the second electron decries in the Lorentz coordinatization from $v_{02}$ to $v_{01}$. This result stands in contrast with the speed of the first electron which remains $v_{02}$. 
In covariant particle tracking, information about difference in acceleration history between two setups is recorded in the covariant velocities difference.

An equally valid interpretation exists within the non-covariant framework, which tracks particle motion using absolute time and applies Galilean transformations to field equations.

The velocity of an electron and the speed
of light are convention-dependent. In absolute-time coordinatization, the speed of
light differs from the electrodynamics constant $c$ because Maxwell’s equations are
not invariant under Galilean transformations. However, the dimensionless ratio
of electron velocity to light velocity remains convention-independent, meaning
it is unaffected by distant clock synchronization or variations in clock rhythms.
In electrodynamics, parameters related to synchrotron radiation, which have direct physical significance, depend on this ratio.

According to non-covariant particle tracking, the second electron's velocity is equal $v_{02}$. However, Maxwell’s equations are not invariant under Galilean transformations, leading to a change in the speed of light. Specifically, the velocity of light increases from $c$, before the second boost with velocity $v_{12}$, to $[c +  v_{12}^2/(2c)]$ after the second boost.  

This variation arises because, under the absolute time convention, clocks are not resynchronized after the boost. As a result, the speed of light differs from the electrodynamical constant $c$ after the second boost. Nevertheless, the ratio of the electron velocity to the speed of light remains invariant across different synchronization conventions—that is, it does not depend on the method of distant clock synchronization or the rate of the clocks. 

Our calculations demonstrate that both covariant and non-covariant treatments—when the correct coupling between fields and particles is used—yield the same prediction for the relativistic electrodynamics phenomena. This prediction is convention-invariant and depends solely on the dimensionless parameter

\[
v_{coord}/c_{coord} = [v_{01} + v_{12}^2/(2c)]/[c +  v_{12}^2/(2c)] = v_{01}/c      , 
\]

where $v_{coord}$ is the coordinate velocity of the second electron, and $c_{coord}$ is the coordinate velocity of light.  Further details are provided in the final two chapters.

\subsection{Acceleration-History Dependence of Relativistic Transformations}

It is useful to illustrate the non-associative character of transitions between inertial frames within the preferred $(1+3)$ decomposition of spacetime by means of a relatively simple example.

Consider two scenarios. In the first scenario, an electron is accelerated in the preferred frame from rest to velocity $\vec{V}$, then from $\vec{V}$ to $\vec{V}+\vec{v}$, and finally back to velocity $\vec{V}$. Covariant particle tracking employs Lorentz coordinates and Lorentz transformations. Let $S$ denote the initial inertial frame and $S'$ the inertial frame comoving with the electron moving at velocity $\vec{V}$ relative to $S$. A Lorentz boost transforms a spacetime event four-vector $X$ according to

\[
X' = L(\vec{V})X.
\]

Let us consider the ultrarelativistic limit in which $V$ approaches the speed of light, while $\vec{v}$ remains small and perpendicular to $\vec{V}$. The complete sequence of transformations between inertial frames,

\[
S \rightarrow S' \rightarrow S'' \rightarrow S''',
\]

may be written as

\[
X''' = L(-\vec{v})L(\vec{v})L(\vec{V})X.
\]

In the second scenario, the electron is first accelerated to velocity $\vec{v}$, then decelerated back to rest, and finally accelerated from rest to velocity $\vec{V}$. The corresponding overall transformation is

\[
X''' = L(\vec{V})L(-\vec{v})L(\vec{v})X.
\]

At first sight, the velocities appearing in these compositions of Lorentz boosts seem to be ordinary three-dimensional vectors. Within relativity, however, this interpretation is not entirely straightforward, since one must specify the spatial hypersurface on which these vectors are defined. A careful mathematical analysis shows that $(1+3)$ decompositions of spacetime based on distinct three-dimensional spatial hypersurfaces are  mutually incomparable.

As already emphasized, a two-body system possesses relative velocities only with respect to a preferred third reference system—namely, the initial inertial frame that provides the privileged $(1+3)$ decomposition of spacetime. Consequently, the velocity vectors $\vec{V}$, $\vec{v}'$, and $-\vec{v}'$ must all be regarded as elements of the three-dimensional space associated with this privileged inertial frame.

From the standpoint of electrodynamics, the two scenarios are fundamentally different. This conclusion is physically intuitive because the electron experiences different acceleration histories in the two cases. In the first scenario, the electron undergoes substantial transverse acceleration and consequently emits significant synchrotron radiation. In contrast, in the second scenario the radiation emitted is negligible.

At first sight, this result merely illustrates the non-commutativity of Lorentz boosts:

\[
L(\vec{V})L(-\vec{v})L(\vec{v})
\neq
L(-\vec{v})L(\vec{v})L(\vec{V}).
\]

What, then, can be said about non-associativity?

One might be tempted to consider

\[
L(-\vec{v})
\left[
L(\vec{v})L(\vec{V})
\right]
\neq
\left[
L(-\vec{v})L(\vec{v})
\right]
L(\vec{V}).
\]

The point we wish to emphasize is the distinction between Lorentz covariance and observer equivalence. In the conventional formulation of special relativity, all inertial frames are regarded as physically equivalent, and the composition of Lorentz transformations is described entirely by the associative structure of the Lorentz group.

The situation changes if the pseudo-Euclidean geometry of spacetime is taken as the fundamental axiom while physical processes are analyzed relative to a preferred inertial frame. Lorentz covariance requires that the laws of physics retain the same form under Lorentz transformations, but it does not by itself imply reciprocal symmetry of physical histories. An electron retains information about its acceleration history relative to the preferred frame, and this is entirely consistent with relativistic electrodynamics.

From this perspective, the expression

\[
\left[
L(-\vec{v})L(\vec{v})
\right]
L(\vec{V})
\]

cannot be interpreted as an independent physical process.  Any change of velocity, or any acceleration, has an absolute meaning. The implicit ”absolute” acceleration means acceleration relative to the fixed stars.
The intermediate stages represented by the sequence of inertial frames $S$, $S'$, $S''$, and $S'''$ depend on the actual acceleration history of the electron. Consequently, when discussing the boost $L(\vec{V})$ following the sequence $L(-\vec{v}')L(\vec{v}')$, one is in fact referring to the complete physical process represented by

\[
L(\vec{V})L(-\vec{v})L(\vec{v}).
\]

The essential point is that the behavior of the electron cannot be described independently of its acceleration history. This conclusion emerges from the requirement of continuity of the metric tensor describing phenomena in an accelerated frame. Such continuity is closely connected with a fundamental principle: all physical phenomena observed in accelerated frames must ultimately be describable within the initial inertial frame. Consequently, a complete physical description must retain information about the acceleration history that connects the successive inertial states.


We should perhaps make a few further remarks on the question of how an electron ``remembers'' its previous acceleration. 
We have already seen that an ultrarelativistic electron cannot be regarded as a structureless point particle in the conventional sense. As its velocity approaches the speed of light, its electromagnetic field becomes increasingly concentrated into a narrow forward-directed beam. This virtual radiation field possesses a macroscopic transverse extent, and its spatial distribution is described by the Ginzburg--Frank formula \cite{GIF} (see Appendix A4). When the electron undergoes transverse acceleration, these self-fields are perturbed in a manner that reflects the acceleration process. The information about the electron's acceleration is therefore not lost; instead, it remains encoded in the perturbation of its own electromagnetic field.

Now, let us return to observations of an accelerated observer (as viewing this of the preferred observer).
Let us consider a specific case in which an electron is initially at rest in the initial inertial frame and then accelerated to a velocity $v$ along the $x$-axis. We derived the electrodynamics equations in an accelerated frame by applying the inverse Galilean transformation (with velocity $-v$) of Maxwell's equations. 
The electric and magnetic fields $\vec{E}_n$ and $\vec{B}_n$ observed in the accelerated frame—where the particle source is at rest—differ from the Coulomb's field observed in the initial inertial frame, prior to the Galilean boost. 
Under a Galilean transformation, to first order in $v/c$, the transformed fields become $\vec{E}_n = \vec{E}$, $\vec{B}_n = - \vec{v}\times\vec{E}/c$. The electric field $\vec{E}_n$ and the induced magnetic field  $\vec{B}_n$ are both static, meaning they do not vary with time—the system is entirely time-independent.

This seemingly paradoxical result carries no physical contradiction. The special relativity tells us that the time derivative of the electric field, $\partial \vec{E} / \partial t_n$, is nonzero because of the spatial variation of the field, $\partial \vec{E} / \partial x_n \neq 0$, and the relation $\partial \vec{E}/\partial t_n = v\partial \vec{E}/\partial x_n$. 

When the electron experiences acceleration in the initial inertial frame, its Coulomb's field in the proper frame is perturbed in a way that reflects this acceleration. For rapidly moving electron, the transverse components of the electric and magnetic fields in the accelerated frame are nearly equal in magnitude.

\subsection{Algebraic Composition of Lorentz Transformations} 

However, a mathematical puzzle remains. Many mathematicians reject the non-associative law of Einstein velocity addition. Their argument is straightforward: since the Lorentz group $O(1,3)$ is an associative group under matrix multiplication, the corresponding Lobachevskian factor space $O(1,3)/O(3)$ should also possess an associative composition law, even though $O(3)$ is not a normal subgroup. 

This argument may be expressed more concretely as follows. Lorentz boosts are represented by $4\times4$ matrices, and matrix multiplication is strictly associative. Therefore, no mathematical non-associativity appears to exist at the level of Lorentz transformations themselves. 

On the other hand, we have argued that the physical composition of successive Lorentz boosts is non-associative because of the Wigner rotation. Yet the standard decomposition of two non-collinear boosts, 

\[ 
L(\vec{v})L(\vec{u})= L(\vec{w})R(\vec{u},\vec{v}), 
\]

is itself merely a factorization of an associative matrix product. How, then, can the physical non-associativity discussed above be reconciled with the associative algebra of matrices? 

There is, in fact, no contradiction. The apparent paradox disappears once Lorentz transformations are interpreted within the preferred $(1+3)$ decomposition of spacetime. 

At first sight, the parameters of a Lorentz boost matrix appear to be ordinary three-vector velocities. Indeed, the very notion of velocity presupposes a distinction between space and time, that is, a $(1+3)$ decomposition of spacetime. However, these velocity parameters are not vectors belonging to a fixed Euclidean three-space. Rather, they are the spatial components of space-like four-vectors defined with respect to the three-space associated with the observer's inertial frame. The essential point is that the velocity vector appearing in the boost matrix $L(\vec{v})$ is not a fixed vector defined in the preferred three-space. Instead, it is defined with respect to the coordinate axes of the immediately preceding inertial frame. The orientation of these axes within the preferred three-space depends on the entire sequence of preceding Lorentz transformations. Whenever that sequence contains a Wigner rotation, the axes of the  inertial frame—and therefore the velocity vector specifying the boost—are rotated with respect to the preferred frame. 

Consequently, the parameters of the  boost matrix implicitly depend on the previous transformation history. Thus, the information carried by the Wigner rotation is encoded in the parameters of the  Lorentz boost itself. Throughout this discussion, Wigner rotations are understood as rotations within the unique three-dimensional space associated with the preferred inertial frame. 

Although matrix multiplication remains strictly associative, the composition of successive Lorentz boosts is history-dependent because the parameters entering the  boost are determined by the preceding sequence of transformations. Once this dependence is properly taken into account, the composition of Lorentz boosts becomes mathematically non-associative. A careful examination therefore shows that the apparent mathematical difficulty originates in the geometry of Minkowski spacetime itself. Because successive non-collinear boosts generate Wigner rotations in the preferred three-space, the boost parameters cannot be regarded as independent variables. Instead, they depend on the complete history of the preceding transformations.

This interpretation differs from several attempts to restore associativity by introducing alternative mathematical formalisms, most notably those of Oziewicz \cite{Z} and Ungar \cite{UNG}. In these approaches, the non-associativity of Einstein velocity addition is regarded as a mathematical feature requiring reformulation.\footnote{It is widely assumed that only relative motion has physical significance. Consequently, many physicists, having learned special relativity from standard textbooks, argue that non-associative velocity addition is paradoxical because, for a system of four or more bodies, adding three non-parallel velocities may yield two different relative velocities between the same pair of bodies. This objection, however, is based on the implicit assumption that all inertial frames are physically equivalent regardless of their acceleration histories. From the viewpoint developed in this book, this assumption is incorrect. The principle of special relativity does not imply the equivalence of inertial frames that have experienced different acceleration histories.}
From the present viewpoint, however, the non-associativity is not a mathematical inconsistency but a physical manifestation of the fact that an electron retains information about its acceleration history relative to the preferred inertial frame. Relativistic kinematics cannot be separated from relativistic electrodynamics. The dependence of the final state on the acceleration history is therefore not a defect of the theory but a direct consequence of the behavior of electromagnetic fields.

To clarify the origin of the apparent non-associativity of successive Lorentz boosts associated with the Wigner rotation, it is useful to compare two physically distinct ways of combining boosts in a specific experimental setup. For this purpose, we consider a somewhat idealized experiment involving particle aberration.

We consider the behaviour of electrons in the experimental setup shown in Figs. \ref{B558}--\ref{B557}.

Let us consider two scenarios.

In the first scenario, an electron gun is accelerated in the preferred frame $S$ from rest to velocity $v$ along the $x$-direction. Subsequently, in the accelerated frame $S_n$, electrons are accelerated from rest to velocity $V$ along the $z_n$-axis. We consider the particular case in which, after this acceleration, the velocity component of the electron beam along the $x_n$-axis vanishes.

An observer at rest in the initial inertial frame $S$ measures an angular displacement
\[
\theta_a=\frac{v}{V}+\left(1-\frac{1}{\gamma}\right)\frac{v}{V}.
\]

As before, we retain only terms up to first order in $v/c$ in the analysis of particle aberration.

For a moving electron gun, a simple intuitive argument would suggest that the aberration angle should be
\[
\theta_a=\frac{v}{V}.
\]
Special relativity, however, introduces an additional contribution,
\[
\left(1-\frac{1}{\gamma}\right)\frac{v}{V}.
\]
This additional term is a manifestation of the Wigner rotation. A conventional treatment of particle aberration does not explicitly account for the mixing of spatial coordinates and time in the accelerated frame. In effect, it assumes that the accelerated and initial inertial observers share a common three-dimensional space.

From the viewpoint of the initial inertial frame $S$, however, the spatial axes of the accelerated frame are not parallel to those of $S$. In particular, the $z_n$-axis is rotated relative to the $z$-axis by the Wigner angle
\[
\theta_w=\left(1-\frac{1}{\gamma}\right)\frac{v}{V}.
\]
Thus, the observed angular displacement consists of two contributions: the ordinary particle-aberration term $v/V$ and the additional contribution associated with the Wigner rotation.

Finally, the reference system $S_n$, together with the electron gun and the electron beam, is decelerated back to rest with respect to the preferred frame $S$. The corresponding overall transformation of the electron beam is
\[
X'''=L(-v\vec e_x)L(V\vec e_{z_n})L(v\vec e_x)X.
\]

The question then arises: how does the electron beam emitted by the gun appear after the complete cycle? For an inertial observer in $S$, the final angular displacement of the beam is
\[
\theta_a=\left(1-\frac{1}{\gamma}\right)\frac{v}{V}.
\]
This residual angular displacement is a direct manifestation of the Wigner rotation.

In the second scenario, the electron gun is again first accelerated to velocity $v$ along the $x$-axis. However, in the accelerated frame $S_n$, the electrons are now accelerated from rest to velocity $V$ along the direction of the $z$-axis of the initial inertial frame $S$. We consider the case in which, before the acceleration of the electron beam, its velocity component along the $x$-axis is zero.

An observer at rest in $S$ then finds the angular displacement
\[
\theta_a=\frac{v}{V}.
\]

This effect, illustrated in Fig. \ref{B557}, is the familiar phenomenon of particle aberration.

Finally, the reference system $S_n$, together with the electron gun and the electron beam, is decelerated back to rest with respect to the preferred frame $S$. The corresponding overall transformation is
\[
X'''=L(-v\vec e_x)L(V\vec e_z)L(v\vec e_x)X.
\]

For the inertial observer in $S$, the electron beam emitted by the gun after the complete cycle propagates along the $z$-axis. In other words, after deceleration, the $x$-component of the beam velocity again vanishes, just as it did before the initial acceleration.

From the mathematical standpoint, the two scenarios are fundamentally different. In the second scenario, all velocity vectors appearing in the boost matrices are fixed vectors defined with respect to the same preferred three-space. The composition
\[
L(-v\vec e_x)L(V\vec e_z)L(v\vec e_x)
\]
therefore involves boosts whose velocity parameters are defined independently of the preceding transformations.

The first scenario is different. Here, the velocity vector appearing in the boost
\[
L(V\vec e_{z_n})
\]
is defined with respect to the coordinate axes of the intermediate inertial frame $S_n$. When this transformation is viewed from the preferred frame $S$, one cannot simply identify the axes $z_n$ and $z$. The orientation of the $z_n$-axis within the preferred three-space depends on the preceding Lorentz transformation $L(v\vec e_x)$.

As viewed from the initial inertial frame $S$, the $z_n$-axis of the accelerated frame is rotated through the Wigner angle $\theta_w$ with respect to the $z$-axis of $S$. Consequently, the information about this rotation is already encoded in the velocity parameter of the boost $L(V\vec e_{z_n})$. Replacing $\vec e_{z_n}$ by the fixed vector $\vec e_z$ therefore changes the physical meaning of the transformation. In this sense, the composition
\[
L(-v\vec e_x)L(V\vec e_{z_n})L(v\vec e_x)
\]
cannot be represented as an ordinary composition of boosts whose velocity parameters are all defined in one fixed three-space. The apparent non-associativity arises from this dependence of the intermediate boost parameter on the preceding Lorentz transformation, and the Wigner rotation provides the geometrical manifestation of that dependence.

The particle aberration observed in the initial inertial frame can also be understood within the framework of electrodynamics.

The electromagnetic forces governing the motion of the emitted electron beam are affected by the acceleration of the source relative to the fixed stars, resulting in a change in the beam-transport direction. As discussed earlier, the magnetic field
\[
\vec B_n=-\frac{\vec v\times\vec E}{c}
\]
induces a change in the transverse momentum of the accelerated electrons. In the first scenario, the accelerated observer redirects the electron beam emitted by the accelerated source. We consider the particular case in which, after redirection, the velocity component of the beam along the $x_n$-axis vanishes.\footnote{When the particle-aberration effect is evaluated only to first order in $v/c$, there is no distinction between the $x$ and $x_n$ directions.}

Experimentally, this redirection can be produced by introducing a weak magnetic dipole (corrector) at the exit of the electron gun to compensate for the induced transverse momentum. After this correction, an observer at rest in the initial inertial frame $S$ observes an angular displacement
\[
\theta_a=\frac{v}{V}+\left(1-\frac{1}{\gamma}\right)\frac{v}{V}.
\]

Thus, the additional contribution to the particle-aberration angle is not merely a consequence of a formal manipulation of Lorentz transformations. It reflects the geometrical fact that the spatial axes of the intermediate frame are rotated relative to the initial inertial frame. The Wigner rotation therefore enters directly into the experimentally observable direction of the electron beam.

\subsection{Polar Decomposition and Non-Associativity of Lorentz Boosts}

Let us focus on the composition of two non-collinear Lorentz boosts and, in particular, on the polar decomposition of their product. The polar decomposition is commonly used to decompose an element of the Lorentz group into a boost and a spatial rotation. This decomposition depends on a choice of the privileged inertial system, which, in the matrix formulation, is specified by choosing a timelike normalized four-velocity vector $e_0$ of the basis. The rotation appearing in the decomposition acts within the corresponding spatial section of this privileged inertial system.

It is important to emphasize that the polar decomposition does not provide a factorization of the Lorentz group $O(1,3)$ into two independent groups, one consisting of boosts and the other of rotations. In particular, the set of pure boosts is not a subgroup of the Lorentz group. Consequently, the product of two boosts is, in general, not a boost but a Lorentz transformation containing a rotational component.

In the textbooks,  the relativistic velocities eddition in the polar decomposition is given by the Einstein's velocity addition. 
For instance, Ferraro  \cite{FER} states: ``For simplcity we shall consider the composition of two boosts $B(\vec{\beta}_1)$ and $B(\vec{\beta}_2)$ along mutually perpendicular directions  $\vec{\beta}_1 = \beta_1\vec{e}_x$,   $\vec{\beta}_2 = \beta_2\vec{e}_y$. ... the matrix resulting from the composition of two boosts is not symmetric, therefore it cannot be a boost. However it is easy to prove that result can be written as the succesive application of a unique boost $B_c$ and spatial rotation $R$:  

\[
 B(\vec{\beta}_2) B(\vec{\beta}_1) = RB_c(\vec{\beta}_c ).
\]

... Therefore, velocity $\vec{\beta}_c$ of boost $B_c$ is relativistic composition of velocities $\vec{\beta}_1$ and  $\vec{\beta}_2$. In summary, the composition of two mutually perpendicular Lorentz boosts is equal to a unique boost whose velocity is the relativistic composition 

\[
\vec{\beta}_c = \beta_1\vec{e}_x +  \sqrt{1 - \beta^2_1}\beta_2\vec{e}_y  ~ ,
\]

following by a spatial rotation in the plane defined by the velocities to be composed, ... ``

Similar assertions appear in more recent mathematical literature. As illustration, consider the following statement by Guilini from a 2026 paper \cite{GU}: ``The task is to polar decompose the product of boosts
$B(\vec{\beta}1)$ and $B(\vec{\beta}2)$:
\[
B(\vec{\beta}2)B(\vec{\beta}1) = B(\vec{\beta})R(\vec{D})  .
\]
The boost parameter $\vec{\beta}$ and the rotation matrix $\vec{D}$ on the right-hand side are uniquely
determined by $\vec{\beta}1$ and $\vec{\beta}2$. ... Hence the boost contained in the polar
decomposition of the product of boosts with parameters $\vec{\beta}1$ and $\vec{\beta}2 $ is just the boost
with Einstein added parameters $\vec{\beta} = \vec{\beta}1 \oplus \vec{\beta}2$.  ...  The rotation matrix $\vec{D}$ resulting from polar decomposition of the product of two boosts  $B(\vec{\beta}1)$ and $B(\vec{\beta}2)$ is called  Thomas rotation and denoted by $T( \vec{\beta}1, \vec{\beta}2)$.''

As analyzed above in this chapter, error in standart polar decomposition  is not (matrix agebra) computational but rather conceptual.  This standard velocities addition overlooks a critical point. We have already argued that the mathematical composition of succesive Lorentz boosts is non-associative because of the Wigner rotation.

It is interesting to observe that error in polar decomposition theorem  is closely linked to the non-associativity of  Lorentz boosts composition - even when only two velocities are involved. The following example illustrates this connection in a particularly transparent way.

As in Ferraro \cite{FER}, we begin with the simple case of the relativistic composition of perpendicular velocities. Consider the product
\[
L(\vec{e}_{y_2}\beta_2)
L(\vec{e}_{x_1}\beta_1).
\]

Let us consider the asymptotic regime in which the first boost velocity $\beta_1$ approaches the speed of light while the second velocity $\beta_2$ remains small. Thus,
\[
\gamma_1\gg1,
\qquad
\beta_2\ll1,
\]
and we retain terms through second order in $\beta_2$.

According to the standard Einstein velocity-addition formula, the speed of the particle in the initial inertial frame increases from $\beta_1$ to
\[
\beta_1+\frac{\beta_2^2}{2\gamma_1^2}.
\]

Now consider the second boost written as two successive collinear boosts,
\[
L(\vec{e}_{y_2}\beta_2/2)
L(\vec{e}_{y_2}\beta_2/2).
\]
In general, two successive collinear Lorentz boosts are not exactly equivalent to a single boost with velocity $\vec{e}_{y_2}\beta_2$. However, in the present asymptotic regime the Einstein velocity-addition law for parallel velocities approaches the Galilean composition law to the required accuracy. One might therefore be tempted to write
\[
L(\vec{e}_{y_2}\beta_2/2)
L(\vec{e}_{y_2}\beta_2/2)
=
L(\vec{e}_{y_2}\beta_2),
\]
and hence
\[
L(\vec{e}_{y_2}\beta_2)
L(\vec{e}_{x_1}\beta_1)
=
L(\vec{e}_{y_2}\beta_2/2)
L(\vec{e}_{y_2}\beta_2/2)
L(\vec{e}_{x_1}\beta_1).
\]

We have now three sucsecive Lorentz boosts. What, then,  can  be said about non-associativity? 
The error in standard polar decomposition is now clear. Unlike Galilean boost composition, Lorentz boosts composition   
is non-associative

\[
\big[
L(\vec{e}_{y_2}\beta_2/2)
L(\vec{e}_{y_2}\beta_2/2)
\big]
L(\vec{e}_{x_1}\beta_1)
\neq
L(\vec{e}_{y_2}\beta_2/2)
\big[
L(\vec{e}_{y_2}\beta_2/2)
L(\vec{e}_{x_1}\beta_1)
\big].
\]

The boost from frame $K_1$ to $K_2$, as observed from initial inertial frame $K_0$, can be understood as sequence of two Lorentz transformations, each involving a small velocity increment $\beta_2/(2\gamma_1)$.  
The first task is to polar decompose the product of boosts $L(\vec{e}_{y_2}\beta_2/2)L(\vec{e}_{x_1}\beta_1)$.
The composition of two perpendicular Lorentz boosts is equal to a boost whose velocity is the relativistic composition
\[
\vec{\beta}_c = \vec{e}_{x_1} \beta_1 +  \vec{e}_{y_1}\beta_2/(2\gamma_1)   ~ ,
\]
following by a spatial rotation $R(\vec{e}_z\theta_w)$, where $\theta_w = \beta_2/(2\gamma_1)$ is the angle of Wigner rotation.  According to the standard velocity-addition formula, the speed in the initial inertial frame therefore increases from $\beta_1$ to
\[
\beta_1+\frac{\beta_2^2}{8\gamma_1^2}.
\]

The crucial point, however, is that the orientation of the $y_2$ axis in the initial frame $K_0$ is determined by the preceding Lorentz transformation
\[
L(\vec{e}_{y_2}\beta_2/2)
L(\vec{e}_{x_1}\beta_1).
\]
As viewed from $K_0$, the $y_2$ axis has been rotated by the Wigner angle $\theta_w$ with respect to the original $y_1$ axis. Consequently, the velocity associated with the second boost is not simply another vector in the original $y_1$ direction. Its direction is determined by the rotated spatial frame.

Taking this Wigner rotation into account, the second boost produces a different increment in the speed measured in $K_0$. To the order considered here, the total speed after the two successive transverse increments becomes
\[
\beta_1+\frac{\beta_2^2}{4\gamma_1^2}.
\]

Thus, splitting the nominal second boost
\[
L(\vec{e}_{y_2}\beta_2)
\]
into two successive boosts
\[
L(\vec{e}_{y_2}\beta_2/2)
L(\vec{e}_{y_2}\beta_2/2)
\]
does not reproduce the velocity increment obtained by treating the second boost as a single transformation. The difference arises precisely because the intermediate Wigner rotation changes the orientation of the spatial axes with respect to the initial inertial frame.

This observation becomes even more striking if the second boost is represented as a sequence of infinitesimal Lorentz transformations. If one successively accumulates infinitesimal transverse velocity increments while consistently accounting for the Wigner rotation generated at each step, the transverse increments do not simply add as ordinary vectors in the initial spatial frame. In the limiting procedure, the rotational contribution compensates the apparent second-order increase predicted by a naive application of the Einstein velocity-addition formula.
If we will  proceed to compute the net effect of accomulating these infinitesemal increments  we come to the conclussion that the total speed of particle with respect to initial inertial frame $K_0$ remains $\beta_1$.

This example illustrates the central conceptual point of the present section. The Einstein velocity-addition formula is obtained by assigning a velocity parameter to the boost factor in a polar decomposition. But the boost factor cannot be regarded independently of the rotational factor when successive non-collinear transformations are considered. The Wigner rotation changes the spatial frame in which subsequent velocity increments are defined. Consequently, the boost parameter obtained after a composition depends on the history of the preceding transformations.

The standard approach, however, treats the Wigner rotation as a secondary effect that appears only after the velocity composition has been performed. The analysis above suggests the opposite viewpoint: the Wigner rotation is an intrinsic part of the composition process itself. It determines the orientation of the spatial axes relative to which subsequent velocity increments are defined and therefore cannot consistently be separated from the velocity composition.

This distinction is particularly important when one attempts to construct relativistic kinematics by repeatedly composing finite or infinitesimal velocity increments. There have been numerous attempts in the literature to derive analytical results in relativistic kinematics starting from the standard Einstein velocity-addition formula, Eq.~\ref{EAV}. From the viewpoint developed here, such derivations are incomplete unless the accompanying Wigner rotation and its effect on the orientation of the successive spatial frames are incorporated from the outset.

\subsection{Einstein's Velocity Addition and  the Wigner Rotation}

For completeness, it is worth noting earlier attempts to address the limitations of Einstein velocity composition. In a series of mathematically rigorous works, Ungar proposed a refined framework to resolve the apparent paradox of non-associativity in relativistic velocity addition (see \cite{UNG1, UNG2} and references therein).

In this refined framework, the composition of velocities is not solely represented by $\vec{u} \oplus \vec{v}$, but rather by the pair 
$[\vec{u} \oplus \vec{v}, \mathrm{R}(\vec{u}, \vec{v})]$ , 
where $\mathrm{R}(\vec{u}, \vec{v})$ denotes the rotation arising from the non-commutative nature of Lorentz boosts. 

A common textbook derivation of the Wigner rotation begins by decomposing the product of two Lorentz boosts, $L(\vec{v})$ and $L(\vec{u})$, as follows:  

\[
L(\vec{v})L(\vec{u}) = L(\vec{w})R(\vec{u},\vec{v})     , 
\]

where $R(\vec{u},\vec{v})$ is identified as the Wigner rotation, and the resultant boost $L(\vec{w})$ corresponds to the Einstein velocity addition $\vec{w} = \vec{u} \oplus \vec{v}$.

Notably, the binary operation ($\oplus$) in Ungar approach remains the standard Einstein velocity addition; what changes is the recognition that a rotation must accompany it to fully capture the transformation.

However, this standard approach is mathematically incomplete, as it treats the rotation $R$ as a secondary effect rather than an intrinsic component of the velocity composition itself. As shown above, the Wigner rotation is not a mere additive correction—it is fundamentally embedded in the structure of relativistic velocity addition.

Consider, for example, the composition of two perpendicular velocities $\vec{\beta}_1$ and $\vec{\beta}_2$ as discussed earlier. Suppose $\gamma_1 \gg 1 \gg \beta_2$. Treating $\beta_2$ as a small parameter and retaining terms up to second order, we observe that the intermediate frame $K_1$ appears to rotate with respect to $K_0$. This effect is a manifestation of the Wigner  rotation.

Ungar’s approach—based on the Einstein velocity addition within an improved composition law—suggests that the total speed  of $K_2$ in the Lorentz  frame $K_0$ increases from $\beta_1$ to  $\beta_1 +  \beta_2^2/(2\gamma_1^2)$. 

However, this standard velocities addition overlooks a critical point. Above we found that in Lorentz coordinatization, the total velocity vector of frame $K_2$ in $K_0$, after a boost with velocity $\vec{\beta}_2$ in $K_1$, undergoes a simple rotation by an angle $\beta_2/(\gamma c)$ without any change in its magnitude, $\beta_{02} = \beta_1$.

This leads us to a natural question: Why did this error in relativistic kinematics remain undetected in experiments so long?
It is important to distinguish between dynamics and electrodynamics. 
From the four component equation Eq. \ref{DDE}, we see that  in dynamics only path is important.  The path $\vec{x}(l)$ has an exact, objective meaning—it is invariant under changes in convention. Similarly, the three -dimensional momentum vector $\vec{p}$ is also convention-invariant. In contrast, the trajectory $\vec{x}(t)$, as measured in a laboratory frame, is convention-dependent, reflecting the conventionality inherent in the definition of three dimensional velocity vector $d\vec{x}/dt$.  
While dynamical theory includes the concept of a particle’s trajectory, this quantity is not directly measured but instead serves as a tool in the analysis of electrodynamics problems.  

Now, let us address the important problem of velocity addition in relativity. It is evident that the velocity of light transforms just as a particle’s velocity does. We cannot use one set of kinematics for a particle’s velocity and another for the velocity of light. That is, Maxwell’s equations can be consistently applied in the lab frame only if particle trajectories are described as the result of Lorentz-covariant particle tracking.
Consequently, special relativity dictates that if Maxwell’s equations are to hold in the lab frame, particle trajectories must incorporate correct relativistic velocities addition.

Experimental results of synchrotron radiation experiments support our correction to the conventional theory of relativistic velocity addition; further details are provided in the final two chapters.

\newpage

\section{Relativistic Spin Dynamics}

\subsection{Kinematics of Spinning Particles}

We have shown that covariant particle tracking in the preferred frame can be achieved without switching reference frames. Initially, we employed conventional particle tracking—i.e., absolute time coordinatization—within the preferred frame. Then, by applying Galilean transformations alongside appropriate redefinitions of space and time variables, we derived the covariant particle trajectory, $\vec{x}(t)_{cov}$. From the standpoint of electrodynamics involving relativistically moving charges, this approach ensures consistency between Maxwell's equations and particle trajectories within the preferred frame.

There is, however, one important case where tracking requires a change of reference frame: the description of a particle’s angular momentum (spin) relative to the preferred frame. In the non-relativistic regime, spin is represented as a three-dimensional pseudovector $\vec{s}$. For a relativistically moving particle, by contrast, angular momentum must be described using the antisymmetric four-tensor $S_{\mu\nu}$.
In experimental practice, however, the spin vector $\vec{s}$ remains the preferred object due to its intuitive physical interpretation as a three-dimensional pseudovector. As a result, any meaningful statement about $\vec{s}$ is understood to refer implicitly to the particle’s instantaneous rest frame—that is, its proper frame.

As previously noted in Chapter 8, spin orientation measurements with respect to the preferred frame axes can be determined within the proper frame. In other words, spin rotation is interpreted from the viewpoint of the proper observer as the apparent rotation of the preferred frame axes.

From a purely kinematic standpoint, spin rotation within the proper frame is absent. However, the preferred frame undergoes relativistic motion relative to the proper frame, which gives rise to the Wigner (Thomas) rotation of the preferred frame axes  in the proper frame as seen from the preferred frame. This rotation is given by  $d\vec{\phi}_W = (\gamma - 1)d\theta$, 
where $d\theta$ is the orbital deflection angle of the particle in the preferred frame (see Chaper 8 for more details). The direction of the preferred frame’s rotation in the proper frame, aligns with the direction of velocity rotation in the proper frames. Accordingly, the spin  rotates with respect to the preferred frame axes.

\subsection{Magnetic Dipole  at Rest in an Electromagnetic Field}

The equations governing spin dynamics can be formulated as tensor equations in Minkowski space-time. Here, we focus on the case of a particle with a magnetic moment $\vec{\mu}$  in a microscopically homogeneous electromagnetic field.
Notably, the torque influences only the spin, while the force affects only the momentum. Consequently, the overall motion of the system in any reference frame is determined solely by its charge, independent of its magnetic dipole moment—a topic addressed in the Chapter 17. Our current focus is on the dynamics of spin motion.

Let us first consider  the spin precession for a nonrelativistic charged particle. 
The proportionality of magnetic moment $\vec{\mu}$ and angular momentum $\vec{s}$ has been confirmed in many "gyromagnetic" experiments on many different systems. The constant of proportionality is one of the parameters charactering a particular system. It is normally specified by giving the gyromagnetic ratio or $g$ factor, defined by $\vec{\mu} =  ge\vec{s}/(2mc)$. This formula says that the magnetic moment is parallel to the angular momentum and can have any magnitude. For an electron $g$ is very nearly 2.

Suppose  that a particle is at rest in an external magnetic field $\vec{B}_R$. The equation of motion for the angular momentum in its rest frame is  $d\vec{s}/d\tau = \vec{\mu}\times \vec{B}_R =
eg\vec{s}\times\vec{B}_R/(2mc) = \vec{\omega}_s\times \vec{s}$. In other words, the spin precesses around the direction of the magnetic field with the frequency $\omega_s = -eg\vec{B}_R/(2mc)$.
In the same nonrelativistic limit the velocity processes around the direction of $\vec{B}_R$ with the frequency $\omega_p = -(e/mc)\vec{B}_R$: $d\vec{v}/d\tau = (e/mc)\vec{v}\times\vec{B}_R$. Thus, for $g = 2$ spin and velocity precess with the same frequency, so that the angle between them is conserved.

\subsection{Derivation of the Covariant (BMT) Equation of Spin Motion}

We aim to demonstrate that while studying relativistic particle dynamics, we simultaneously explored a wide range of related topics. As an example, consider the motion of spin in a given field.

To derive the equation governing spin motion, we start from the well-established spin dynamics in the particle's rest frame and apply known relativistic transformation laws. Since spin is inherently defined in the rest frame, constructing covariant expressions requires introducing a four-dimensional quantity associated with spin that transforms appropriately under Lorentz transformations.

There are two general ways to generalize the spin vector $\vec{s}$—and, by extension, the Larmor equation $d\vec{s}/d\tau = \vec{\mu} \times \vec{B}_R$—to arbitrary frames: one approach uses an axial four-vector, while the other employs an antisymmetric rank-2 tensor.

The most commonly used representation is the four-(pseudo)vector $S^\alpha$, often referred to as the four-spin vector. It is defined such that, in the rest frame, its spatial components coincide with the components of $\vec{s}$, while its time-like component vanishes. When normalized by its invariant magnitude, $S^\alpha$ becomes the polarization four-vector. Being space-like, its spatial components remain non-zero in all inertial frames.
\footnote{An arguably more physically natural representation is the antisymmetric four-tensor. The motion of a spinning point particle with mass $m$ and charge $e$ in Minkowski spacetime is then described by its polarization tensor $D_{\mu\nu}$—an antisymmetric four-tensor that incorporates both intrinsic magnetic and electric dipole moments. It is defined as $D_{\mu\nu} = geS_{\mu\nu}/(2mc)$, where $S_{\mu\nu}$ is the intrinsic angular momentum tensor.}

Let the spin of the particle be represented in the rest frame by $\vec{s}$.
The four-vector $S^{\alpha}$ is by definition required to be purely spatial at time $\tau$ in an instantaneous Lorentz rest frame $R(\tau)$ of the particle and to coincide at this time with the spin $\vec{s}(\tau)$ of the particle; that is $S_R^{\alpha}(\tau) = (0,\vec{S}_R(\tau)) = (0,\vec{s}(\tau))$. At a later instant $\tau + \Delta\tau$ in an instantaneous inertial rest frame $R(\tau +\Delta\tau)$, we have similarly
$S_R^{\alpha}(\tau+\Delta\tau) = (0,\vec{S}_R(\tau+\Delta\tau)) = (0,\vec{s}(\tau+\Delta\tau))$.

The BMT equation \cite{BMT} is the manifestly covariant equation of motion for a four-vector spin $S^{\alpha}$ in an electromagnetic field $F^{\alpha \beta}$:

\begin{eqnarray}
&& \frac{dS^{\alpha}}{d\tau} = \frac{ge}{2mc} \left[F^{\alpha\beta}S_{\beta} + \frac{1}{c^2}u^{\alpha}\left(S_{\lambda}F^{\lambda\mu}u_{\mu}\right) \right] - \frac{1}{c^2}u^{\alpha}\left(S_{\lambda}\frac{du^{\lambda}}{d\tau}\right) ~ ,\label{DDS}
\end{eqnarray}

where $u_\mu = dx_\mu/d\tau$ is the four-dimensional particle velocity vector.
With Eq.(\ref{DDE}), one has 

\begin{eqnarray}
&& \frac{dS^{\alpha}}{d\tau} = \frac{e}{mc} \left[\frac{g}{2}F^{\alpha\beta}S_{\beta} +  \frac{g-2}{2c^2}u^{\alpha}\left(S_{\lambda}F^{\lambda\mu}u_{\mu}\right) \right]  ~ ,\label{DDSA}
\end{eqnarray}

The BMT equation is valid in any Lorentz frame and, together with the covariant force law, consistently describes the motion of a charged particle with spin and magnetic moment. The covariant equation of spin motion for a relativistic particle subjected to the four-force $Q^{\mu} = e F^{\mu\nu}u_\nu$ (as expressed in the Lorentz lab frame, Eq.~(\ref{DDS})) serves as a relativistic generalization of the classical spin dynamics equation in the particle’s rest frame. This generalization embeds the three Larmor spin precession equations into the four-dimensional structure of Minkowski space-time.

In Lorentz coordinates, there exists a kinematic constraint $S^{\mu}u_{\mu} = 0$, which expresses the orthogonality between the four-spin and the four-velocity vectors. Due to this condition, the four-dimensional dynamical equation, Eq.~(\ref{DDS}), effectively contains only three independent equations of motion. By substituting the explicit form of the Lorentz force, one finds that Eq.~(\ref{DDS}) inherently satisfies the constraint $S^{\mu}u_{\mu} = 0$, as required. To demonstrate this, note that in any Lorentz frame, the time component of the spin satisfies $S^0 = \vec{S}\cdot\vec{v}$. Although $S^0$  vanishes in the particle's rest frame, its derivative with respect to proper time,
 $dS^0/d\tau$, does not necessarily vanish. In fact, the condition $d(S^{\mu}u_{\mu})/d\tau = 0$  leads directly to
$dS^0/d\tau = \vec{S}\cdot d\vec{v}/d\tau$. 
The immediate generalization of  $d\vec{S}/d\tau =  eg\vec{S}\times\vec{B}_R/(2mc)$  and 
$dS^0/d\tau = \vec{S}\cdot d\vec{v}/d\tau$
to arbitrary Lorentz frames is Eq.(\ref{DDS}) as can be checked by reducing to the rest frame.
A methodological parallel can thus be drawn between this formulation and the relativistic generalization of Newton's second law.

To fully grasp the significance of embedding the spin dynamics law within Minkowski space-time, it is important to recall that the spin dynamics equation, as presented earlier in the Lorentz comoving frame, is characterized as a phenomenological law. Its formulation does not provide a microscopic interpretation of a particle’s magnetic moment. In other words, the spin dynamics law is generally accepted as phenomenological, with the magnetic moment introduced in an ad hoc manner. The coordinate system in which the classical equations of motion for a particle’s angular momentum are valid is referred to as the Lorentz rest frame. The relativistic generalization of the three-dimensional equation $d\vec{s}/d\tau =  eg\vec{s}\times\vec{B}_R/(2mc)$ to an arbitrary Lorentz frame enables accurate predictions of spin behavior across different reference frames.

\subsection{Changing Spin Variables}

When Bargmann, Michel, and Telegdi first formulated the correct laws of spin dynamics, they derived a manifestly covariant equation in Minkowski space-time, Eq. (\ref{DDSA}), describing the motion of the four-spin $S^{\mu}$. Their approach closely resembled the four-tensor equations already established in relativistic particle dynamics.

To apply Eq. (\ref{DDSA}) to concrete physical scenarios, it is necessary to express the standard three-dimensional spin vector $\vec{s}~$ in terms of the four-vector $S$.  The relationship between  $\vec{s}$ and $S$ follows directly from the Lorentz transformation:
$\vec{s} = \vec{S} + S^{0}\vec{p}c/(\mathcal{E} +  mc^2 )$. 
With the help of this relation, one can work out the equation of motion for $\vec{s}~$. Let us restrict our treatment of spinning particle dynamics to purely transverse magnetic fields. 
This means that the  magnetic field vector $\vec{B}$ is oriented normally to the particle line motion. 
In this practically important case one has, after somewhat lengthy calculations \cite{BMT}:

\begin{eqnarray}
&& \frac{d\vec{s}}{d\tau} = \vec{\Omega}\times\vec{s} = - \frac{e}{2mc} \left[\left(g-2 + \frac{2}{\gamma}\right)\gamma\vec{B} \right]\times\vec{s}   ~ ,\label{DDSS1}
\end{eqnarray}

What must be recognized is that in the accepted covariant approach (indeed, Eq.(\ref{DDSA}) is manifestly covariant), the solution of the dynamics problem for the spin in the lab frame makes no reference to the three-dimensional velocity. In fact, the Eq.(\ref{DDSS1}) includes  relativistic factor $\gamma$  and vector $\vec{v}/c$, which are actually notations:  $\gamma = \mathcal{E}/(mc^2)$,   $\vec{v}/c = \vec{p}c/\mathcal{E}$. 
All quantities $\mathcal{E}, \vec{p}, \vec{B}$ are defined in the lab frame and possess exact, objective meaning—that is, they are independent of any choice of convention. The evolution parameter $\tau$ is likewise measured in the lab frame and has a similarly objective interpretation. For example, it is straightforward to show that $d\tau = m dl/|\vec{p}|$, where  $dl$ denotes the differential path length.

We now have the equation in a form that is convenient for solving. Consider a charged, spinning particle moving through a bending magnet of length $dl$ in the lab frame. The orbital deflection angle of the particle is given by $d\theta = - eBdl/(|\vec{p}|c)$. Additionally, the proper time interval is  $d\tau = mdl/|\vec{p}|$. From Eq.~(\ref{DDSS1}), we can express the spin rotation angle relative to the lab frame axes as  $\Omega d\tau$, which becomes: $ \Omega d\tau = [(g/2 - 1)\gamma d\theta  + d\theta]$.

The spin vector $\vec{s}$
directly represents the spin as perceived in a comoving system. In the lab frame, if we say that a particle's spin makes an angle $\phi$ with its velocity, this means that in the particle's rest frame, the spin makes the same angle with the direction of motion of the lab frame.
This leads us to conclude that the conventional approach used to describe spin dynamics in the lab frame is rather unconventional. Specifically, the measurement of spin rotation in the lab frame is interpreted, from the proper observer's perspective, as viewing this of the lab observer.

\subsection{An Alternative Approach to BMT Theory}

Above we described the BMT equation, Eq.(\ref{DDSS1}), in the standard manner. It uses a spin quantity defined in the proper frame but observed with respect to the lab frame axes.
That means that we know the orientation of the proper spin with respect to the lab coordinate system which is moving with velocity $-\vec{v}$ and acceleration $-\gamma d\vec{v}/d\tau$ in the proper frame.
Let's look at what the equation Eq.(\ref{DDSS1}) says in a little more detail. It will be more convenient if we rewrite this equation as

\begin{eqnarray}
&& d\vec{s} =  \vec{\Omega} d\tau \times\vec{s} =  -eg\gamma \vec{B}d\tau/(2mc)\times\vec{s}
+  e(\gamma - 1) \vec{B}d\tau/(mc)\times\vec{s} ~ .\label{DDSS11}
\end{eqnarray}

Now let's see how we can write the right-hand side of Eq.(\ref{DDSS11}). 
The first term is that we would expect the spin rotation due to a torque with respect to the proper frame axes $d\vec{\phi}_L = -eg\gamma \vec{B}d\tau/(2mc) = (g/2)\gamma d\vec{\theta}$.
Here $d\vec{\theta} = -eBdl/(|\vec{p}|c)$ is the angle of the velocity rotation  in the lab frame. 
It has also been made evident by our analysis in Chapter 8 that the angle of rotation $d\vec{\phi}_W =  -e(\gamma - 1) \vec{B}d\tau/(mc) = (\gamma - 1)d\vec{\theta}$ 
corresponds to the Wigner rotation of the lab frame axes with respect to the proper frame axes. 
With these definitions, we have

\[
d\vec{s} =  \vec{\Omega} d\tau \times\vec{s} = d\vec{\phi}_L \times\vec{s} -  d\vec{\phi}_W \times\vec{s}      .
\]

We now present a new approach to the BMT theory, offering an alternative method for addressing this complex problem. It is important to note that $d\vec{\phi}_L$ and $d\vec{\phi}_W$ represent rotations with respect to the axes of the proper frame. However, our primary goal is to determine the spin motion relative to the laboratory frame axes. In doing so, we must pay close attention to the signs of the rotations, as they play a critical role in the analysis.

A helpful mnemonic can be used to remember the signs of various rotational effects. It consists of three parts:

1. The direction of velocity rotation in the proper frame is the same as the direction of velocity rotation in the lab frame.

2. The direction of lab frame rotation as observed in the proper frame is the same as the direction of velocity rotation in the proper frame.

3. When  $g > 0$ (which is the case for an electron, where $g$ is positive and approximately equal to 2), the direction of spin rotation due to a torque in the proper frame matches the direction of velocity rotation in the proper frame.

We now turn to the question of determining the proper spin rotation relative to the lab frame axes.
This is straightforward: the relevant rotation angle is given by the difference $d\vec{\phi}_L - d\vec{\phi}_W $.
With this, we begin to grasp the fundamental framework of spin dynamics.
It becomes clear why the Wigner rotation of the lab frame axes, as seen from the proper frame, must be accounted for when analyzing spin dynamics relative to the lab frame.

Why is the new derivation of the BMT equation so simple?
The key reason is that the decomposition of the particle's spin motion relative to the lab frame axes into dynamic and kinematic components can only be properly realized in the particle's proper frame. In this frame, there is no need to consider a relativistic "generalization" of the phenomenological classical equation of motion for angular momentum.
This approach allows the spin dynamics problem to be separated into two distinct parts: the trivial (Larmor) dynamic problem and the kinematic problem of the Wigner rotation of the lab frame axes in the proper frame.

\subsection{Spin Tracking}

By expressing the spin motion equation in four-vector form, Eq. (\ref{DDSA}), and determining the components of the four-force, we have not only ensured compliance with the principle of relativity but also obtained the four-component formulation of the spin motion equation. This represents a covariant relativistic generalization of the conventional three-dimensional equation for the motion of a magnetic moment, where the particle's proper time serves as the evolution parameter. Next, we aim to describe the spin motion relative to the Lorentz lab frame, using the lab time  $t$ as the evolution parameter.

When going from the proper time $\tau$ to the lab time $t$, the frequency of spin precession with respect to the lab frame can be obtained  using the well-known formula $d\tau = dt/\gamma$. We then find

\begin{eqnarray}
&& \frac{d\vec{s}}{dt} = \vec{\omega}_s\times\vec{s} = - \frac{e}{2mc} \left[\left(g-2 + \frac{2}{\gamma}\right)\vec{B} \right]\times\vec{s}   ~ .\label{DDSS3}
\end{eqnarray}

The frequency of spin precession can be written in the form 
$\omega_s = \omega_0[1 + \gamma (g/2 - 1)]$,
where $\omega_0$ is the particle revolution frequency.

The old kinematics comes from the relation $d\tau = dt/\gamma$. 
The presentation of the time component simply as the relation $d\tau = dt/\gamma$ between proper time and coordinate time is based on the hidden assumption that the type of clock synchronization that  provides the time coordinate $t$ in the lab frame is based on the use of the absolute time convention.

In the Chapter 17, we saw that the particle path $\vec{x}(l)$ has an exact objective meaning i.e. it is convention-invariant. The spin orientation $\vec{s}$  at each point of the  particle path $\vec{x}(l)$  has also exact objective meaning. In contrast to this, and consistently with the conventionality of the three-velocity, the function $\vec{s}(t)$ describing the spinning particle  in the lab frame has no exact objective meaning.

We now describe how to determine the spin orientation along the path $\vec{s}(l)$ in convention-invariant spin tracking. Starting with the covariant equation (Eq. \ref{DDS}), we obtain Eq. (Eq. \ref{DDSS1}). By using the relation $d\tau = mdl/|\vec{p}|$, we arrive at the convention-invariant equation of spin motion:

\begin{eqnarray}
&& \frac{d\vec{s}}{dl}  = - \frac{e\mathcal{E}}{m|\vec{p}|c^3} \left[\left(\frac{g}{2}-1 + \frac{mc^2}{\mathcal{E}}\right)\vec{B} \right]\times\vec{s} 
= \left[\left(\frac{g}{2} - 1\right)\frac{\mathcal{E}}{mc^2} + 1\right]\frac{d\vec{\theta}}{dl}\times\vec{s} ~ ,\label{DDSS4}
\end{eqnarray}

where $l$ is the path length, used as the evolution parameter. These three equations precisely correspond to the equations for components of the proper spin vector derived from the non-covariant spin tracking equation (Eq. \ref{DDSS3}).

It is important to note that there are two distinct approaches—covariant and non-covariant—that yield the same spin orientation $\vec{s}(l)$ along the path. Both approaches describe the same physical reality correctly, meaning that the orientation of the proper spin $\vec{s}$ at any point along the particle's path in the magnetic field is objective and convention-invariant.

Let us now consider an example that illustrates why it is not necessary to determine experimentally the time-dependent function $(\vec{s}(t))$ of a spinning particle. A particularly instructive example is provided by accelerator physics.

Suppose that we want to calibrate the beam energy in an electron storage ring by measuring the spin-precession frequency of a polarized beam. The precession frequency can be determined, for example, by the method of beam-resonance depolarization using an oscillating electromagnetic field.\footnote{A method for measuring the particle energy in an electron--positron storage ring by means of resonance depolarization using a radio-frequency longitudinal magnetic field is described in \cite{DE}. There are many possible forms of depolarizers; we mention here one particularly simple implementation, in which the radio-frequency longitudinal magnetic field is produced by a current-carrying loop surrounding a ceramic section of the vacuum chamber.}

Suppose that the observer in the laboratory frame performs the beam-energy measurement. We should then ask which parts of the measured data depend on the choice of synchronization convention and which do not. Physically meaningful results must, of course, be invariant under a change of convention.

At first sight, this appears to be a typical time-dependent measurement involving the function $(\vec{s}(t))$. However, the numerical value assigned to the spin-precession frequency $(\omega_s)$ is not, by itself, a convention-invariant quantity. Its value depends on the choice of time coordinate: in particular, an arbitrary rescaling of the time coordinate changes the numerical value of any frequency defined with respect to that time. Thus, the function $(\vec{s}(t))$, regarded as a function of a particular coordinate time, is not itself an invariant observable.

Our physical problem, however, is not to determine an abstract function $(\vec{s}(t))$, but to determine the energy of the electron beam. For this purpose, the revolution frequency $(\omega_0)$ must also be measured using the same space--time grid. The physically relevant quantity is then the dimensionless ratio
\[
\frac{\omega_s}{\omega_0}.
\]
This ratio is convention-invariant: it is unaffected by the synchronization of distant clocks and is also unchanged by a rescaling of the time coordinate. Consequently, measurement of the dimensionless spin-precession frequency $(\omega_s/\omega_0)$ allows us to determine the convention-invariant beam energy $(\mathcal{E})$. The factor $(g/2-1)$, which characterizes the anomalous magnetic moment of the electron, can be calculated within quantum electrodynamics.

Let us return to the experimental situation of beam-energy calibration in a storage ring. The spin vector $(\vec{s}(t))$ makes an angle $(\phi)$ with the particle velocity. From Eq.~\ref{DDSS4}, we have expressed this angle in terms of the orbital angle $(\theta)$, so that
\[
\phi=\phi(\theta).
\]
It is therefore natural to use the orbital angle $(\theta)$, rather than the coordinate time (t), as the evolution parameter.
Suppose that the
depolarizer is placed at an azimuth $\theta_0$. A spin rotates by $(\Delta\phi)$ per revolution

\[
\Delta\phi = \phi(\theta_0 +2\pi) - \phi(\theta_0) .
\]

The point is
that depolarizer measurements are made to determine the observable $\Delta\phi$.

Let us see how equation Eq. \ref{DDSS4} gives the observable $\Delta\phi$. It can be written
in integral form
\[
\Delta\phi = \int \d\theta [(g/2 -1)\mathcal{E}/((mc^2)] = 2\pi(g/2 - 1)\mathcal{E}/(mc^2) .
\] 

These results reveal an important feature of the measurement. The convention-invariant quantity $(\Delta\phi)$ is, in fact, a geometric parameter: it is the accumulated rotation of the spin relative to the orbital motion over one complete revolution. It does not require us to assign an absolute meaning to the coordinate-time dependence of $(\vec{s}(t))$.

Thus, in experiments involving spin dynamics in a storage ring, there is no need to determine experimentally the function $(\vec{s}(t))$ itself. What is measured is a convention-invariant geometric quantity---the spin rotation accumulated during one revolution. The beam energy can then be inferred directly from this invariant observable.

\newpage

\section{Relativity and Electrodynamics}

The differential form of Maxwell's equations describing electromagnetic phenomena in the Lorentz lab frame is given by Eq.(\ref{CD11}). Now let us use these equations to discuss the phenomena called radiation.
To evaluate radiation fields arising from external sources in Eq. (\ref{CD11}), we need to know the velocity $\vec{v}$ and the position $\vec{x}$ as a function of the lab frame time $t$. As discussed above, it is generally accepted that one should solve the Maxwell's equations in the lab frame with current and charge density created by particles moving along non-covariant trajectory like $\vec{x}(t)$.
The trajectory  $\vec{x}(t)$, which follows from the solution of the corrected Newton's second law Eq. (\ref{N1}) under the absolute time convention, does not include, however, relativistic kinematics effects.

We now turn to the motion of a particle in a given magnetic field. Maxwell's equations, in their standard form, are valid only within Lorentz reference frames. According to the proper coupling between fields and particles, Eq.~\ref{CD} implies the following expressions for the charge and current densities:

\begin{eqnarray}
&&\rho(\vec{x}, t) =  e\delta(\vec{x} - \vec{x}_{cov}(t)) ~,\cr &&
\vec{j}(\vec{x},t) = e\vec{v}_{cov}(t)\delta(\vec{x} - \vec{x}_{cov}(t))~, \label{CD1}
\end{eqnarray}

where $\vec{v}_{cov} = d\vec{x}_{cov}/dt$. Here, the covariant trajectory of the particle, as observed in the laboratory frame, results from a sequence of Lorentz transformations.

In this book, we emphasize the significant role of the Wigner rotation as a regulator of the velocity addition law. As previously discussed, the four-velocity of an accelerating particle moving along a curved path in the Lorentz lab frame cannot, in general, be decomposed as $u = (c\gamma, \vec{v}\gamma)$. \footnote{Consider, for instance, a typical textbook treatment \cite{SCHE}, which discusses the projection of an arbitrary world line onto the Lorentz lab frame basis:
"The particle, which is assumed to curry the charge $e$, creates the current density   
$j(x) = ec\int d\tau  u(y) \delta^4(y - x(\tau))$. [...] Furthermore, in any frame of reference $K$, one recovers the expected expressions for the charge and current densities by integrating over $\tau$ by means of relation $d\tau = dt'/\gamma$ between proper time and coordinate time and using the formula $\delta(y_0 - x_0(\tau)) = \delta(ct - ct') = \delta(t-t')/c$,  $j_0(t,y) = ce\delta^{(3)}(y-x(t)) \equiv c\rho(t,y)$, $j^i(t,y) = ev^i(t)\delta^{(3)}(y-x(t))$, $i = 1,2,3$".} The presentation of the time component simply as the relation $d\tau = dt'/\gamma$ between proper time and coordinate time is based on the hidden assumption that the type of clock synchronization, which provides the time coordinate $t$ in the lab frame, is based on the use of the absolute time convention.

\subsection{The Dipole Approximation}

The distinction between covariant and non-covariant particle trajectories was never fully understood. As a result, physicists failed to recognize the contribution of relativistic kinematic effects to radiation. This oversight naturally raises an important question: why did this error in radiation theory remain undetected for so long?

To address this question, we will analyze the subject more mathematically.
For an arbitrary velocity parameter $v/c$, performing covariant calculations of the radiation process is highly challenging. However, in certain cases, significant simplifications arise. One such case is the non-relativistic radiation regime.
The non-relativistic asymptotic limit offers a fundamental simplification for covariant calculations. This is because the non-relativistic assumption justifies the dipole approximation, which is of great practical importance. When considering only the dipole component of radiation, all details of the electron trajectory are effectively ignored. Consequently, dipole radiation is completely insensitive to the distinction between covariant and non-covariant particle trajectories.
\footnote{Similarly, the ultrarelativistic asymptotic limit also simplifies covariant calculations. This is due to the paraxial approximation, which naturally follows from the ultrarelativistic assumption. More details on this effect will be discussed in the next chapter.}

We want now to solve electrodynamics equations mathematically in a general way and consider the radiation associated with the succeeding terms in (multi-pole) expansion of the field in powers of the ratio $v/c$. 
Radiation theory is naturally developed in the space-frequency domain, as one is usually interested in radiation properties at a given position in space and at a certain frequency.\footnote{In this book we define the relation between temporal and frequency domain via the following definition of Fourier transform pair:
	$\bar{f}(\omega) = \int_{-\infty}^{\infty} dt~ f(t) \exp(i \omega t ) \leftrightarrow
	f(t) = \frac{1}{2\pi}\int_{-\infty}^{\infty} d\omega \bar{f}(\omega) \exp(-i \omega t)$.}

Suppose we are interested in the radiation generated by an electron and observed far away from it. In this case, it is possible to find a relatively simple expression for the electric field. We indicate the electron velocity in units of $c$ with $\vec{\beta}$, the electron trajectory in three dimensions with $\vec{r}(t)$ and the observation position with $\vec{r}_0$. Finally, we introduce the unit vector

\begin{equation}
\vec{n} =
\frac{\vec{r}_0-\vec{r}(t)}{|\vec{r}_0-\vec{r}(t)|}
\label{enne}
\end{equation}
%

pointing from the retarded position of the electron to the observer. In the far zone, by definition, the unit vector $\vec{n}$ is nearly constant in time. If the position of the observer is far away enough from the charge, one can make the expansion $\left| \vec{r}_0-\vec{r}(t) \right|= r_0 - \vec{n} \cdot \vec{r}(t)$.

We then obtain the following approximate expression for the the radiation field  in the space-frequency domain:

\begin{eqnarray}
\vec{\bar{E}}(\vec{r}_0,\omega) &=& {i\omega e\over{c
		r_0}}\exp\left[-\frac{i \omega}{c}\vec{n}\cdot\vec{r}_0\right]
\int_{-\infty}^{\infty}
dt~{\vec{n}\times\left[\vec{n}\times{\vec{\beta}(t)}\right]}\exp
\left[i\omega\left(t+\frac{\vec{n}\cdot
	\vec{r}(t)}{c}\right)\right] \cr &&  \label{revwied}
\end{eqnarray}
where $\omega$ is the frequency, $(-e)$ is the negative electron charge and we make use of Gaussian units. 
A different constant of proportionality in Eq.(\ref{revwied}) and the well-known textbooks is to be ascribed to the use of different units and the definition of the Fourier transform.

Let us now examine in greater detail how the dipole radiation term arises.
In the integrands of the expression for the radiation field amplitude, Eq. (\ref{revwied}), the time argument $\vec{r}(t)\cdot(\vec{n}/c)$ can be neglected if the trajectory of the charge changes insignificantly during this interval. Determining the conditions under which this approximation holds is straightforward.
Let $a$ represent the characteristic size of the system. Then, the time delay can be estimated as $\vec{r}(t)\cdot(\vec{n}/c) \sim a/c$. To ensure that the charge distribution remains essentially unchanged over this duration, it is necessary that $a \ll \lambda$, where $\lambda$ is the wavelength of the emitted radiation.
This condition can also be expressed in an alternative form: $v \ll c$, where $v$ denotes the characteristic velocity of the charges in the system.

We consider the radiation corresponding to the zeroth-order term in the expansion of  Eq. (\ref{revwied}) in powers of
$\vec{r}(t)\cdot(\vec{n}/c)$. In this approximation, all details of the electron trajectory  $\vec{r}(t)$ are neglected. This is the essence of the dipole approximation, where the spatial scale of the electron's motion is assumed to be much smaller than the wavelength of the emitted radiation. Under this condition,  Eq. (\ref{revwied}) yields fields that closely resemble those predicted by instantaneous, non-retarded theories. Therefore, it is appropriate to use a non-covariant approach when analyzing dipole radiation.

However, this is only the first and most practically significant term. The remaining terms indicate that there are higher-order corrections to the dipole radiation approximation. Calculating these corrections requires detailed knowledge of the electron's trajectory. In particular, to determine the correction to dipole radiation, we must use the covariant form of the trajectory rather than relying on a non-covariant approach. Nevertheless, corrections to multipole radiation are generally expected to be small. For instance, the covariant correction to quadrupole radiation is typically regarded as a contribution of higher order than the quadrupole term itself.

\subsection{ An Illustrative Example}

It is beneficial to illustrate errors in standard coupling fields and particles in accelerator and plasma physics using a relatively simple example, where the core physical concepts remain clear and unobscured by unnecessary mathematical complexities. This example is primarily intended for readers with limited knowledge of accelerator and synchrotron radiation physics. Fortunately, the error in standard coupling fields and particle interactions can be explained in a straightforward manner.

\begin{figure}
	\centering
	\includegraphics[width=0.9\textwidth]{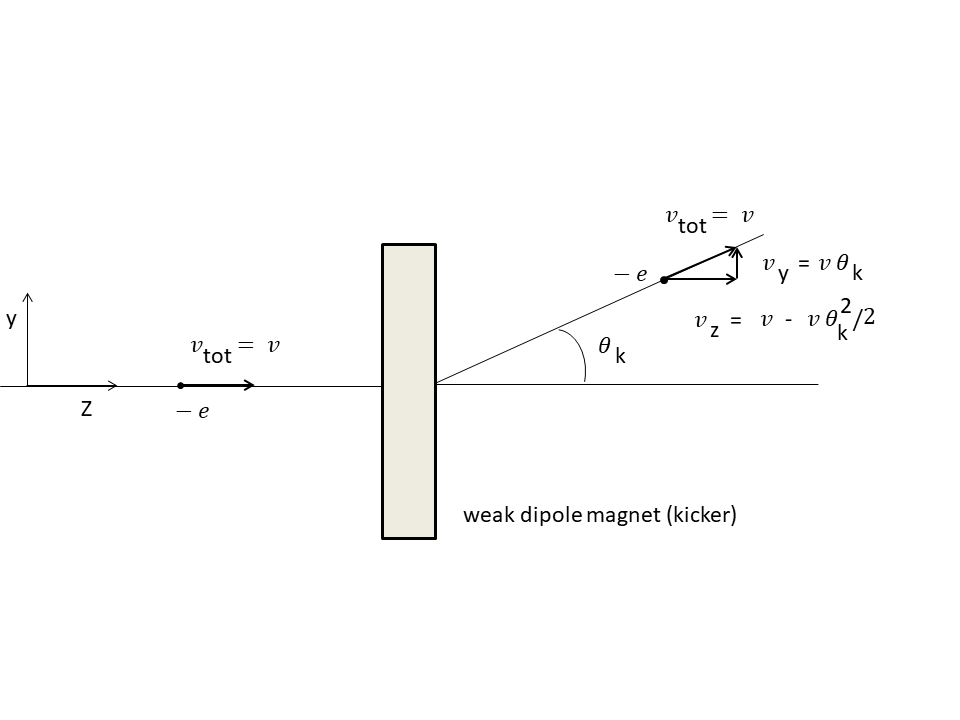}
	\caption{Illustration of the difference between covariant and non-covariant trajectories. The setup shows the motion of a relativistic electron accelerated by a kicker field, assuming $\gamma \gg 1 \gg \gamma v_y/v$ for simplicity. In the non-covariant particle tracking framework, the magnetic field $B\vec{e}_x$ affects only the direction of the electron's motion without altering its speed.}
	\label{B81}
\end{figure}

An electron kicker setup serves as a practical example for illustrating the distinction between covariant and non-covariant trajectories. Consider a simple case where an ultrarelativistic electron, moving with velocity  $v$ along the $z$-axis in the inertial (lab) frame, is subjected to a weak dipole magnetic field directed along the  $x$-axis. 
For simplicity, we assume that the kick angle is small compared to $1/\gamma$, where $\gamma = 1/\sqrt{1 - v^2/c^2}$ is the Lorentz factor.  This corresponds to the limit $\gamma \gg 1 \gg \gamma v_y/v$. We begin with non-covariant particle tracking calculations. The electron's trajectory, derived from the corrected form of Newton’s second law under the absolute time convention, does not account for relativistic effects. Consequently, as in classical Newtonian kinematics, the Galilean vector addition of velocities is employed. Non-covariant particle dynamics predicts that the electron’s direction changes as a result of the kick, while its speed remains constant (see Fig. \ref{B81}). According to this model, the magnetic field $B\vec{e}_x$ affects only the direction of motion and not the magnitude of the velocity.
After the kick, the components of the beam’s velocity are $(0,v_y,v_z)$, where $v_z = \sqrt{v^2 - v_y^2}$. In the ultrarelativistic limit, where $\gamma \gg 1$, we can use a second-order approximation to obtain 

\[
v_z = v [1- v_y^2/(2v^2)] = v [1- v_y^2/(2c^2)]    . 
\]

In contrast, covariant particle tracking—which relies on Lorentz coordinates—yields different results for the electron's velocity. To explore this, consider a sequence of passive Lorentz transformations used to track the motion of a relativistic electron accelerated by a kicker field. Let $S$ be the lab frame and $S'$ a frame comoving with the electron at velocity $\vec{v}$ relative to $S$. Upstream of the kicker, the electron is at rest in the $S'$ frame. This implies that $S'$  is related to $S$ by a Lorentz boost  $L(\vec{v})$, with $\vec{v}$ aligned along the $z$ axis. The boost transforms a four-vector event  $X$ in space-time according to $X' = L(\vec{v})X$. Now, let us analyze the particle’s dynamics from the perspective of the $S'$ frame. In this frame, the electron remains at rest, while the kicker moves toward it with velocity $-\vec{v}$. Due to its motion, the kicker’s magnetic field generates an electric field perpendicular to it. As the kicker interacts with the stationary electron in $S'$, the particle experiences a combination of perpendicular electric and magnetic fields. 

We consider a small expansion parameter,  $\gamma v_y/c \ll 1$, and retain terms up to second order—specifically, terms of order $(\gamma v_y/c)^2$—while neglecting higher-order contributions such as $(\gamma v_y/c)^3$. This corresponds to employing the second-order kick angle approximation. It is straightforward to show that the acceleration in crossed fields leads to a particle velocity $v'_y =\gamma v_y$ along the $y$-axis and $v'_z = - v(\gamma v_y/c)^2/2$ along the $z$-axis. Within this second-order approximation, relativistic corrections to velocity composition—typically arising at third order—do not appear.

We begin by analyzing the textbook treatment of the composition of motions. Let $S"$ be a reference frame fixed with respect to a particle located downstream of the kicker. As is well known, non-collinear Lorentz boosts do not commute. However, in our second-order approximation, we can neglect the difference between $\gamma v_y/c$ and $\gamma_z v_y/c$, 
where $\gamma_z =1/\sqrt{1-v_z^2/c^2}$. Here $v_z = v(1 - \theta_k^2/2)$ and $\theta_k = v_y/v = v_y/c$ in the ultrarelativistic regime of interest.  Therefore, we can approximate the transformation from coordinates $X'$ in frame $S'$ to coordinates $X"$ in $S"$ using a sequence of two commuting non-collinear Lorentz boosts:

\[
X"=L(\vec{e'}_y v'_y) L(\vec{e'}_z v'_z)X' =  L(\vec{e'}_z v'_z)L(\vec{e'}_y v'_y)X'    , 
\]

where $\vec{e'}_y$ and $\vec{e'}_z$ are unit vectors  along the $x'$ and $z'$ axis respectively.  It is important to note that, as seen by an observer in $S'$, the axes of frame $S"$ remain parallel to those of $S'$. The full transformation from the lab frame $S$ to the downstream particle frame $S"$  can thus be written as: 

\[
X" = L(\vec{e'}_y v'_y)L(\vec{e'}_z v'_z)L(\vec{e}_z v)X. 
\]

In the special case where the velocities are collinear, the composition of boosts simplifies: 

\[
L(\vec{e'}_z v'_z)L(\vec{e}_z v) = L(\vec{e}_z v_z)    , 
\]

leading to the overall transformation:  

\[
X" = L(\vec{e'}_y v'_y)L(\vec{e}_z v_z)X      . 
\]

Textbooks typically emphasize that, within the Lorentz framework, a magnetic field can only alter the direction of an electron’s motion, but not its speed.

\begin{figure}
	\centering
	\includegraphics[width=0.9\textwidth]{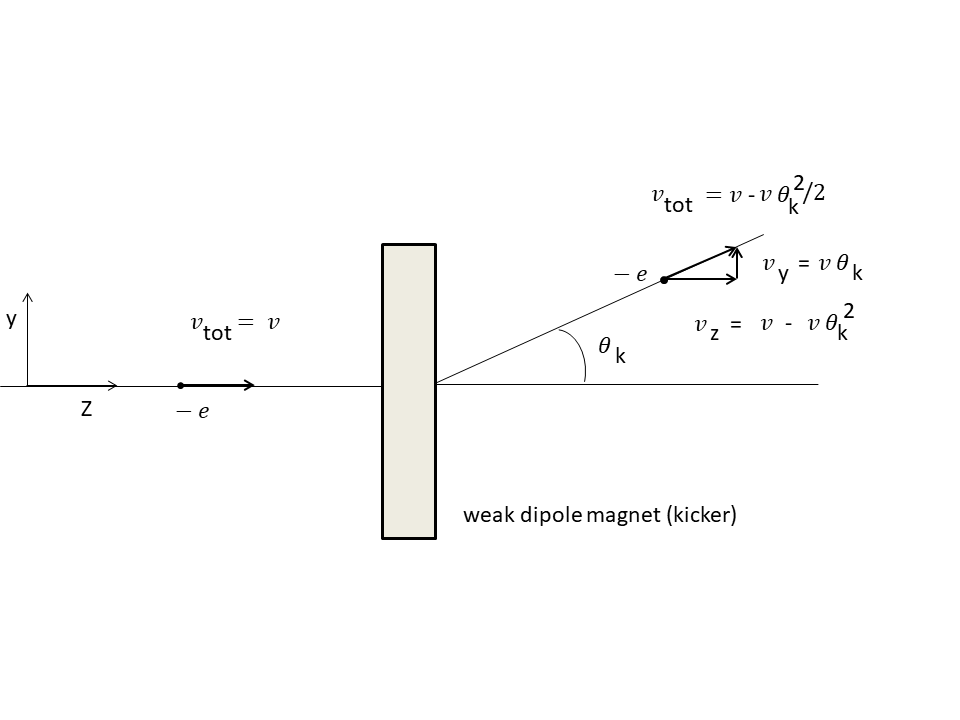}
	\caption{Trajectory of a relativistic electron accelerated by a kicker field.
		To describe the kicker’s influence on the electron’s motion using a Lorentz transformation, it is necessary to resynchronize the clock by introducing a time shift and applying a time scale adjustment.}
	\label{B82}
\end{figure}

According to standard textbooks, in relativistic kinematics, the composition of perpendicular velocities is analyzed using Einstein’s velocity addition theorem. We previously discussed the law of velocity composition in Section 17.6, where it was shown that, within the Lorentz coordinatization, velocity addition is governed by the Wigner rotation.
In the ultrarelativistic approximation, a simple result emerges:  $v_{tot} = v_z = v [1- v_y^2/(2c^2)]$. This implies that a Lorentz transformation induces a rotation of the particle’s velocity  $v_z$ by an angle approximately equal to
$v_y/c$  (Fig. \ref{B82}). This result highlights that the apparent contradiction is rooted in the presence of the Wigner rotation.

Now let us revisit the kinematics of a relativistic electron accelerated by the kicker field, analyzing it from the perspective of an inertial lab observer without changing reference frames.
Specifically, we consider an electron in the lab frame undergoing acceleration due to the kicker field. 
The simplest synchronization method involves maintaining the same set of uniformly synchronized clocks as in the pre-kick state, without modification. This approach, based on the absolute time (or absolute simultaneity) convention, preserves simultaneity. The standard Galilean velocity addition rule is then applied to track the electron’s motion instant by instant as it moves through a constant magnetic field.

By combining Galilean transformations with variable changes, we ultimately derive the Lorentz transformation within the framework of absolute time coordinatization in the lab frame. This analysis establishes a direct link between the electron’s speed reduction after the kick (in Lorentz coordinates) and time dilation.

Suppose that, upstream of the kicker, we choose a Lorentz coordinate system in the lab frame. Immediately after the electron enters the magnetic field, its velocity changes by an infinitesimal amount $d\vec{v}$ along the $y$-axis.
At this initial stage, Eq.~(\ref{DDDE1}) allows us to express the differential  $d\vec{v}$ in terms of the differential $dt$ within the Lorentz coordinate system defined upstream of the kicker. If clock synchronization is fixed, this corresponds to adopting the absolute time convention. To maintain Lorentz coordinates in the lab frame—as discussed previously—we must perform a clock resynchronization by introducing an infinitesimal time shift. The simplest case arises when the kick angle $\theta_k$ is very small, allowing us to work up to second-order terms in $(\theta_k \gamma)^2$. This restriction greatly simplifies the calculations for two reasons. First, relativistic corrections due to the composition of non-collinear velocity increments only appear at order  $(\gamma \theta_k)^3$ and can therefore be neglected. Second, time dilation effects also enter only at higher orders.
Thus, Eq.~(\ref{DDDE1}) enables us to express the small velocity change  $\Delta\vec{v}$ due to the kick in the initial Lorentz coordinate system and defer clock resynchronization until after the kicker. Consequently, the post-kick motion can be described as a composition of two Lorentz boosts in the perpendicular  $y$ and $z$ directions. The first boost imparts a velocity  $v\theta_k\vec{e}_y$ along the $y$-axis,  while the second applies an additional velocity  $- (v\theta_k^2/2)\vec{e}_z$ along the $z$ axis. The second-order approximation ensures that the two boosts commute.

To maintain a Lorentz coordinate system in the lab frame after the electron receives a transverse kick, a resynchronization of clocks is required. In this context, operations involving the rule clock structure in the lab frame can be interpreted as a change of variables governed by the transformation in Eq. (\ref{GGT3}):

\[
y_L = \gamma_y y   ,    \qquad   t_L = (t/\gamma_y + \gamma_y yv_y/c^2)    , 
\]

where $\gamma_y = 1 + v_y^2/(2c^2)$ and the electron's displacement is given by $y = v_yt$.

This transformation effectively rescales the time variable—adjusting the synchronization of all clocks from
$t$ to $\gamma_y t$, with $\gamma_y \simeq 1 + \theta_k^2/2$.  As a result, it becomes evident that the electron's downstream speed is no longer independent of its transverse motion in the magnetic field (see Fig. \ref{B82}). Although no second-order relativistic correction appears in the transverse ($y$-axis) velocity component, the longitudinal component is modified. Specifically, the longitudinal velocity  $v_z$  changes to $v_z/\gamma_y$ with $v_z = v(1 - \theta_k^2/2)$, leading to  $v_z/\gamma_y = v(1 - \theta_k^2)$. As a result, the total electron speed in the lab frame, after clock resynchronization downstream of the kicker, decreases from $v$ to $v(1 - \theta_k^2/2)$.

\subsection{Electron Motion Accelerated by a Kicker in a Bending Magnet}

\begin{figure}
	\centering
	\includegraphics[width=0.95\textwidth]{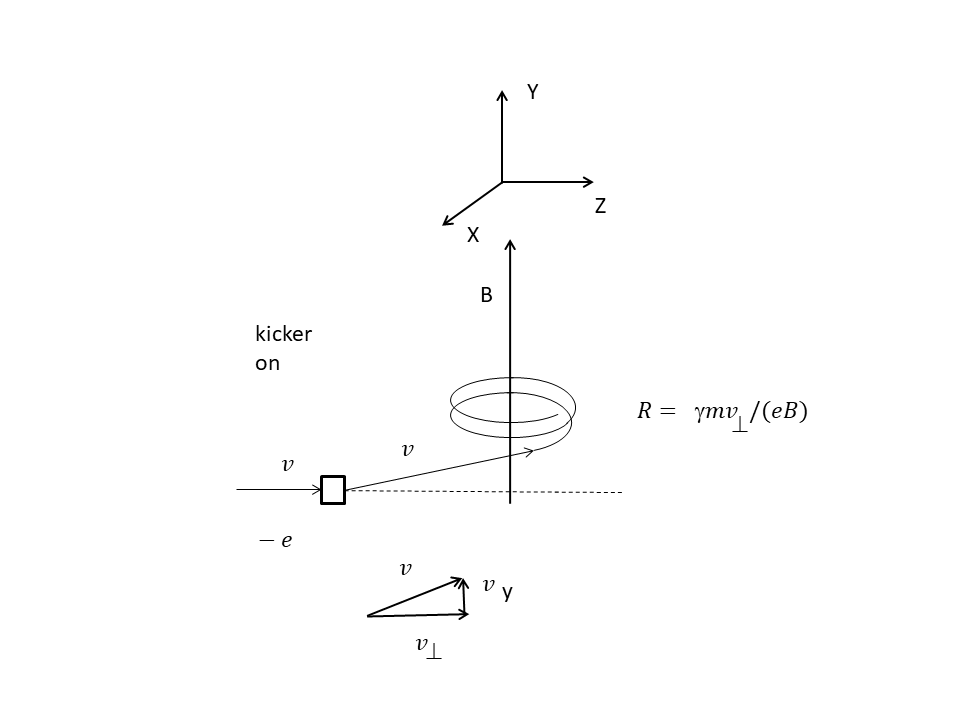}
	\caption{ Geometry of radiation production in a bending magnet. In non-covariant particle tracking, the kicker’s magnetic field $B\vec{e}_x$ can change the direction of the electron’s motion but does not affect its speed.} 
	\label{B96}
\end{figure}

In our relativistic, yet non-covariant, analysis of electron motion in a magnetic field, the electron maintains the same velocity—and therefore the same relativistic factor $\gamma$—both upstream and downstream of the kicker. Suppose the electron then enters a bending magnet, i.e., a region with a uniform magnetic field directed along the $y$-axis (see Fig. \ref{B96}). The resulting motion within the bending magnet closely resembles that predicted by non-relativistic dynamics, with the only difference being the presence of the relativistic factor $\gamma$ in the expression for the cyclotron frequency: $\omega_c = eB/(m\gamma)$. The curvature radius $R$ of the trajectory follows from the relation  $v_{\perp}/R  = \omega_c$, where $v_{\perp} = v(1 - \theta_k^2/2)$ is the component of the velocity perpendicular to the magnetic field $\vec{B} = B\vec{e}_y$. According to non-covariant particle tracking, the correction to the radius $R$ after the kick is of order $\theta_k^2$.

At first glance, one might expect that in covariant particle tracking, the total speed of an electron in the lab frame—after passing through the kicker—would decrease from  $v$ to $v(1 - \theta_k^2/2)$. This, in turn, would suggest a corresponding decrease in the magnitude of the three-momentum $|\vec{p}|$ from $m\gamma v$ to $m\gamma v(1 - \gamma^2\theta_k^2/2)$, based on our approximation. However, such a change in momentum would imply a correction to the bending radius $R$ on the order of $\gamma^2\theta_k^2$, leading to a stark contradiction with the radius calculated using noncovariant tracking. Since the curvature radius in the bending magnet has an objective, convention-invariant meaning, this apparent discrepancy poses a paradox. The resolution lies in recognizing that, in Lorentz coordinates, the momentum three-vector $\vec{p}$ transforms as the spatial part of the four-momentum $p_{\mu}$ under Lorentz boosts. To analyze the motion of the relativistic electron affected by the kicker, we consider a composition of Lorentz boosts that follow the particle's trajectory. Under such a transformation, the longitudinal component of the momentum remains effectively unchanged—accurate to order $\theta_k^2$.

Let us verify the validity of this assertion. The four-momentum is given by  $p_{\mu} = [\mathcal{E}/c, \vec{p}]$. Consider the Lorentz frame $S'$, which is at rest with respect to the electron upstream of the kicker. In the special case where the electron is initially at rest in this frame, its four-momentum is $p_{\mu}' = [mc , \vec{0}]$. We now focus on the dynamics in frame  $S'$.  Acceleration due to the crossed electric and magnetic fields of the kicker induces a velocity component $v_y' =  \gamma v_y$ along the $y$-axis, and a second-order longitudinal component $v_z' = - v(\gamma v_y)^2/2$ along the $z$-axis. After the interaction with the kicker, the four-momentum in  $S'$ becomes:

\[
p_{\mu}' = [mc + m v_y'^2/(2c), 0,  mv_y',mv_z']     , 
\]

where the expression is evaluated to second order in $(\gamma v_y/c)^2$, consistent with our approximation scheme.
We observe that the transverse acceleration results in an increase in the time-like component of the four-momentum—that is, the energy of the electron. Specifically, the energy increases from  

\[
mc^2            \to     \quad  mc^2 + m (\gamma v_y)^2/2     . 
\]

Recall that frame $S'$ is related to the lab frame $S$  via a Lorentz boost. Under a boost to a frame moving with velocity
$\vec{v} = - v\vec{e}_z$, the longitudinal component of the momentum (which is normal to the magnetic field of the bending magnet) transforms as: 

\[
p_z = \gamma(p_z' + vp_0'/c) = \gamma mv         . 
\]

This shows that, within our approximation, the momentum component along the $z$-axis remains unchanged. Furthermore, applying the Lorentz transformation to the time component yields: 

\[
p_0 = \gamma(p'_0 + vp'_z) = \gamma mc   , 
\]

consistent with our earlier result.

It is intriguing to examine the implications of having two distinct approaches—covariant and noncovariant—that yield the same particle three-momentum. The key insight is that both frameworks accurately describe the same physical reality. The curvature radius of the trajectory in a magnetic field, and consequently the three-momentum, possesses an objective meaning, making it invariant under different conventions.

\subsection{Redshift of the Synchrotron Radiation Critical Frequency}

Next, we explore the intriguing problem of synchrotron radiation emission in a bending magnet, both with and without the influence of a preceding kick. We will focus on the physical interpretation without delving into computational details (see the next chapter for a more thorough analysis).
Consider the setup illustrated in Fig. \ref{B96}. An ultrarelativistic electron is initially moving along the $z$-axis in the laboratory frame. Before entering a uniform magnetic field directed along the $y$-axis (i.e., a bending magnet), the electron receives a small transverse kick from a weak dipole field oriented along the $x$-axis.
An electron undergoing acceleration along a curved trajectory emits electromagnetic radiation. At relativistic speeds, this radiation is concentrated into a narrow cone, tangent to the electron’s path. Furthermore, the radiation amplitude is significantly enhanced in this direction—a phenomenon known as Doppler boosting. Synchrotron radiation arises specifically when a relativistic electron is accelerated by a bending magnet. In the ultrarelativistic limit, the key features of the emitted spectrum can be intuitively understood to depend primarily on the small difference between the speed of the electron and the speed of light.

\begin{figure}
	\centering
	\includegraphics[width=0.9\textwidth]{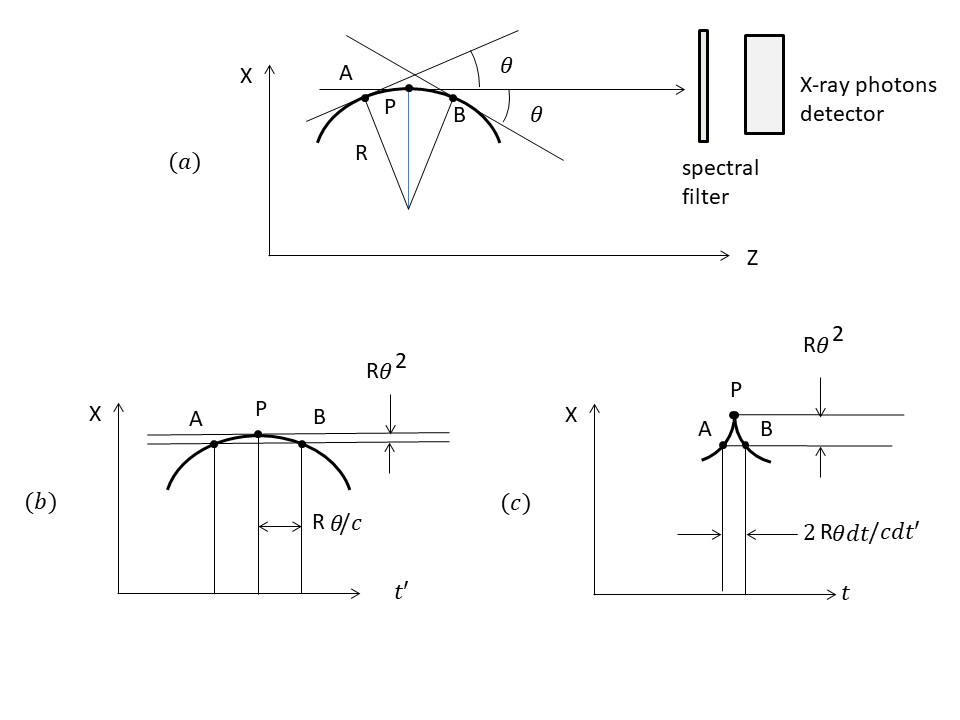}
	\caption{Geometry for synchrotron radiation from a bending magnet.}
	\label{B84}
\end{figure}

We now turn to the radiation emitted by an ultrarelativistic electron traversing a bending magnet. Consider the case in which the electron is moving toward the observer (Fig. \ref{B84}). The electromagnetic source follows a trajectory
$\vec{x}(t')$ over time, where $t'$ denotes the emission time of the radiation. However, because electromagnetic signals propagate at a finite speed (the speed of light), a signal emitted at time $t'$ from position  $\vec{x}(t')$ reaches the observer at a later time $t$. Consequently, the observer perceives the motion of the source as a function of $t$, not $t'$. It is important to note that the prime notation here indicates retarded time and should not be confused with primes denoting quantities in a Lorentz-transformed frame, as used in subsequent sections.

Let the coordinates of the electron be denoted by  $(x,y,z)$, where the $z$-axis is aligned with the direction of observation. We continue to assume that the detector is located far from the radiation source. At a given moment $t'$, the electron's position components are $x(t')$, $y(t')$, and $z(t')$. The distance to the observer is approximately

\[
R(t') = R_0 - z(t')      . 
\]

If we denote the observation time by $t$, then $t'$ is not equal to $t$; rather, it is delayed by the time it takes light to travel from the electron to the observer.
Disregarding a constant delay—equivalent to shifting the time origin—we find the relation 

\[
ct = ct' - z(t')      .
\]

Our goal is to express $x$ as a function of the observation time $t$, rather than the emission time $t'$.
Assuming  $c - v \ll c$, we can use the approximation: 

\[
dt/dt' = (c - v\cos \theta)/c  \simeq (1 - v/c +\theta^2/2) \simeq (1/2)(1/\gamma^2 + \theta^2)   , 
\]

where $\theta$ is the observation angle (Fig. \ref{B84}).

\begin{figure}
	\centering
	\includegraphics[width=0.9\textwidth]{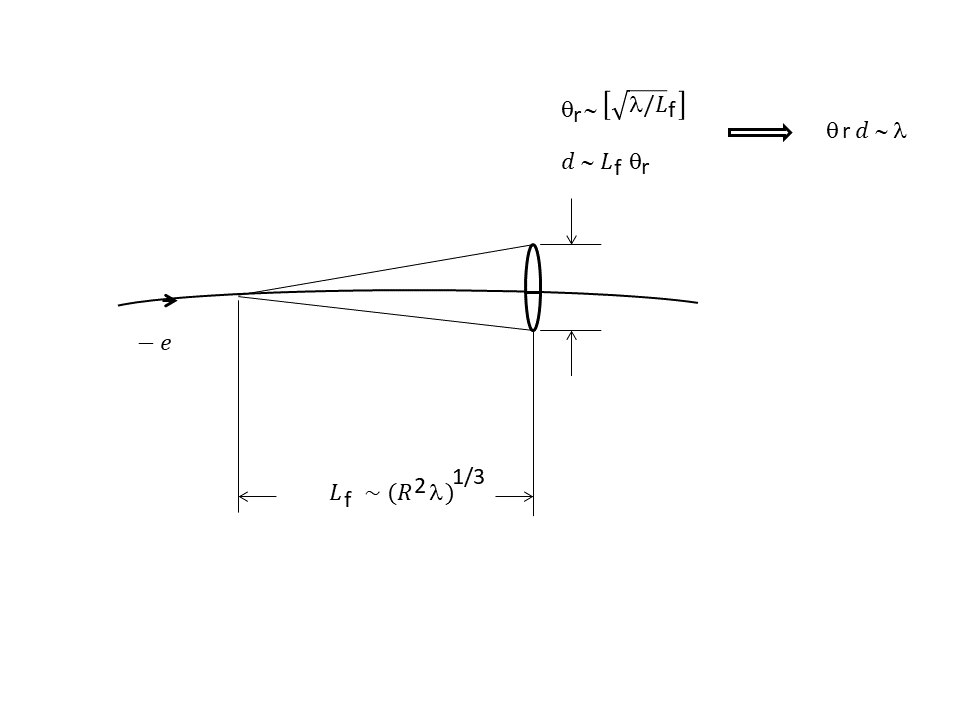}
	\caption{Formation length for the bend. }
	\label{B98}
\end{figure}

This relation implies that the observer perceives a time-compressed motion of the electron as it travels from point $A$ to point $B$, corresponding to an apparent spatial interval of $2R\theta dt/dt'$. 
Let us assume (this assumption will be justified in a moment) $\theta^2 > 1/\gamma^2$. In this case, one has 
$2R\theta dt/dt' \simeq R\theta^3$. 
One can distinguish between radiation emitted at point $A$ and radiation emitted at point $B$ only when compressed distance $R\theta^3 \gg \lambdabar$, i.e. for $\theta \gg (\lambdabar/R)^{1/3}$, where $\lambda$ is the filtered radiation wavelength. This means that, as concerns the radiative process, we cannot distinguish between point $A$ and point $B$ on the bend such that $R\theta < (R^2\lambdabar)^{1/3}$. It does not make sense at all to talk about the position where electromagnetic signals are emitted within $L_f =  (R^2\lambdabar)^{1/3}$ (here we assume that the bend is longer than $L_f$). This characteristic length is called the formation length for the bend.    
Physically, the formation length can be interpreted as the effective longitudinal size of the single-electron radiation source in the space-frequency domain. Importantly, radiation from a single electron is always diffraction-limited. The spatial coherence condition is expressed through the space-angle product: $\theta_r d \simeq \lambdabar$, where $d$ being the transverse size and $\theta_r$ the divergence of the source. Since $d \simeq L_f\theta_r$ it follows that the divergence angle $\theta_r$ is strictly related to $L_f$ and $\lambdabar$: $\theta_r \simeq \sqrt{\lambdabar/L_f}$. One may check that using $L_f = (R^2\lambdabar)^{1/3}$, one obtains $\theta_r \simeq  
(\lambdabar/R)^{1/3}$ as it must be. In particular, at $\theta_r \simeq 1/\gamma$ one obtains  the characteristic wavelength $\lambdabar_{cr} \simeq R/\gamma^3$, which is the familiar result for radiation from a bending magnet (see Fig. \ref{B98}).

It is evident from the preceding discussion that, according to conventional synchrotron radiation theory, introducing a kick to the electron's motion results merely in a rigid rotation of the angular distribution of the emitted radiation, aligning it with the new direction of motion. This conclusion is intuitive: the electron retains its speed after the kick and, due to the Doppler effect, emits radiation along the new trajectory.

However, a more rigorous treatment—based on the proper coupling between fields and particles—reveals a noteworthy prediction of synchrotron radiation theory for the setup described above. Specifically, there is a red shift in the critical frequency of the synchrotron radiation observed in the kicked direction. To demonstrate this effect, we turn to a covariant formulation that explicitly incorporates Lorentz transformations.
When the kick is applied, covariant particle tracking predicts a nonzero red shift in the critical frequency. This shift arises because, in Lorentz coordinates, the electron’s velocity decreases from $v$ to $v - v\theta_k^2/2$, while the speed of light remains unchanged at the electrodynamic constant $c$. The resulting red shift in the critical frequency  $\omega_{cr} \simeq R/(c\gamma^3)$ can be approximated by the relation 

\[
\Delta\omega_{cr}/\omega_{cr} \simeq - \gamma^2 v_y^2/c^2 = - \gamma^2\theta_k^2      . 
\]

Here, we observe a second-order correction in $\theta_k^2$ that is, however, multiplied by a large factor $\gamma^2$.

To validate the predictions of our synchrotron radiation theory, we propose an experimental test using third-generation synchrotron radiation sources. Synchrotron radiation from bending magnets spans a broad frequency range; however, employing narrow-bandwidth sources offers a more promising approach for studying redshifts in the radiation spectrum. Narrower bandwidths enhance the sensitivity of output intensity to redshift effects and reduce the demands on beam kicker strength and photon beamline aperture.
Undulators, which generate quasi-monochromatic synchrotron radiation, are particularly well-suited for this purpose. By forcing the electron beam into a periodic undulating path, they induce interference effects that significantly narrow the emitted bandwidth. Since undulators typically consist of many periods, the resulting bandwidth scales inversely with the number of periods. Consequently, insertion devices at third-generation facilities provide an effective means to boost sensitivity to redshift signals, even at small kick angles, $\theta_k < 1/\gamma$.
In essence, we argue that, despite decades of experimental work, synchrotron radiation theory remains insufficiently confirmed, and a more precise test is both timely and necessary.

\newpage

\section{Synchrotron Radiation}

\subsection{Introductory Remarks}

Accelerator physics has traditionally been framed within the context of classical (Newtonian) kinematics, which is fundamentally incompatible with Maxwell’s equations. Here, we aim to revisit this perspective by applying modern insights from dynamics and electrodynamics to explore a central question in greater detail:
Why did the error in synchrotron radiation theory remain undetected for so long?

The electromagnetic radiation phenomena under consideration are inherently complex. In a general setup, covariant calculations of the radiation process can be highly intricate. However, certain configurations allow for significant simplifications. One such case is the synchrotron radiation setup. Much like the non-relativistic limit, the ultrarelativistic limit introduces essential simplifications into covariant calculations. This is because the ultrarelativistic regime justifies the paraxial approximation: the formation length of the radiation is much longer than its wavelength, and as a result, radiation is emitted within a narrow cone of angles on the order $1/\gamma$ 
or smaller. Consequently, the small-angle approximation becomes valid.
In this regime, the transverse velocity is small compared to the speed of light, allowing us to use a second-order relativistic approximation for the transverse motion. Rather than using a small total velocity parameter $(v/c)$ as in the non-relativistic case, we now deal with a small transverse velocity parameter  $(v_{\perp}/c)$. The next step is to analyze the longitudinal motion using the same approach. Notably, in synchrotron radiation, the longitudinal dynamics are particularly straightforward. When expanding transformations up to second order in  $(v_{\perp}/c)^2$, relativistic corrections to the longitudinal motion are absent within this approximation.

In the covariant framework, relativistic kinematic effects associated with synchrotron radiation appear as successive corrections at different orders of approximation: 

$\cdot$ First order  $(v_{\perp}/c)$: Effect such as the relativity of simultaneity arises.

$\cdot$ Second order $(v_{\perp}/c)^2$: Effects such as time dilation and the relativistic correction to the velocity addition law emerge. Relativistic correction in the law of composition of velocities, which already appears in the second order, results directly from the time dilation.

The first-order kinematic term $(v_{\perp}/c)$ plays a significant role primarily in the description of coherent radiation from a modulated electron beam. In a storage ring, however, the longitudinal positions of electrons within a bunch are essentially uncorrelated. As a result, the radiation emitted by different electrons is also uncorrelated. In this case, the total radiated power is simply the sum of the individual contributions—intensities are added rather than electric fields. For incoherent synchrotron radiation, where one considers the motion of a single ultrarelativistic electron in a constant magnetic field, relativistic effects influence only the second-order kinematic terms, specifically $(v_{\perp}/c)^2$.

\subsection{Paraxial Approximation of the Radiation Field}

The general method to derive the frequency spectrum is to transform the electric field from the time domain to the frequency domain by the use of the Fourier transform. 
First, let us rewrite Eq. (\ref{revwied}) as follows

\begin{eqnarray}
\vec{\bar{E}}(\vec{r}_0,\omega) &=& -{i\omega e\over{c
		r_0}}\exp\left[- \frac{ i \omega}{c}\vec{n}\cdot\vec{r}_0\right]
\int_{-\infty}^{\infty}
dt~ [\vec{\beta}(t) - \vec{n}] \exp
\left[i\omega\left(t+\frac{\vec{n}\cdot
	\vec{r}(t)}{c}\right)\right] \cr &&  \label{rrevwied}
\end{eqnarray}

Eq. (\ref{revwied}) and Eq. (\ref{rrevwied}) are
equivalent but include different integrands. This is no mistake,
as different integrands can lead to the same integral (see Appendix A1).

We call $z_0$ the observation distance along the optical axis of the system, while $(x_0,y_0)$ fixes the transverse position of the observer.  
Using the complex notation, in this and in the following sections we assume that the temporal dependence of fields with a certain frequency is of the form:

\begin{eqnarray}
\vec{E} \sim \vec{\bar{E}}(z_0,x_0,y_0,\omega) \exp(-i \omega t)~.
\label{eoft}
\end{eqnarray}
With this choice for the temporal dependence we can describe a plane wave traveling along the positive $z$-axis with

\begin{eqnarray}
\vec{E} = \vec{E}_a \exp\left(\frac{i\omega}{c}z_0 -i \omega t\right)~.
\label{eoftrav}
\end{eqnarray}
In the following, we will always assume that the ultra-relativistic approximation is satisfied, which is the case for SR setups. As a consequence, the paraxial approximation applies too. The paraxial approximation implies a slowly varying envelope of the field with respect to the wavelength. It is therefore convenient to introduce the slowly varying envelope of the transverse field components as

\begin{equation}
\vec{\widetilde{E}}(z_0,x_0,y_0,\omega) = \vec{\bar{E}}(z_0,x_0,y_0,\omega) \exp{\left(-i\omega z_0/c\right)}~. \label{vtilde}
\end{equation}

We will now replace all vectors by their components to obtain directional dependency of the synchrotron radiation. 
The emission angle $\theta = \sqrt{\theta_x^2 + \theta_y^2}$ is taking with respect to the $z$-axis. 
Here $\theta_x = x_0/z_0$ is the observation angle  projected onto the $x-z$ plane, $\theta_y = y_0/z_0$ is the observation angle projected onto the $y-z$ plane.
The components of the unit vector $\vec{n}$ can be approximated by $n_z = 1 - \theta_x^2/2 - \theta_y^2/2$, 
$n_x = \theta_x$, 
$n_y = \theta_y$, 
so $\vec{n}\cdot(\vec{r}_0 - \vec{r}) = (z_0  - z)( 1 -\theta_x^2/2 - \theta_y^2/2) - x\theta_x
- y\theta_y$.
We consider the motion in a static magnetic field.
According to conventional particle tracking the magnitude of the velocity is constant and is equal $v = ds/dt = \mathrm{const.}$, where $s(z)$ is the longitudinal coordinate along the path.  
The transverse components of the envelope of the field in Eq. (\ref{rrevwied}) in the far zone and paraxial approximation finally becomes

\begin{eqnarray}
\vec{\widetilde{{E}}}(z_0, \vec{r}_0,\omega) &=& -{i
	\omega e\over{c^2}z_0} \int_{-\infty}^{\infty} dz' {\exp{\left[i
		\Phi_T\right]}}  \left[\left({v_x(z')\over{c}}
-\theta_x\right){\vec{e_x}}
+\left({v_y(z')\over{c}}-\theta_y\right){\vec{e_y}}\right] \label{generalfin}
\end{eqnarray}
where the total phase $\Phi_T$ is

\begin{eqnarray}
&&\Phi_T = \omega \left[{s(z')\over{v}}-{z'\over{c}}\right] \cr &&
+ \frac{\omega}{2c}\left[z_0 (\theta_x^2+\theta_y^2) - 2 \theta_x x(z') - 2 \theta_y y(z') + z'(\theta_x^2+\theta_y^2)\right]~
. \label{totph}
\end{eqnarray}
Here $v_x(z')$ and $v_y(z')$ are the horizontal and the vertical components of the transverse velocity of the electron, $x(z')$ and $y(z')$ specify the transverse position of the electron as a function of the longitudinal position, $\vec{e}_x$ and $\vec{e}_y$ are unit vectors along the transverse coordinate axis.

\subsection{Undulator Radiation}

Traditionally, courses on synchrotron radiation theory begin by following the historical development of the field, starting with radiation from bending magnets. In contrast, we will begin this chapter by exploring a more advanced topic: the theory of synchrotron radiation from undulator setups, where covariant calculations of the radiation process become notably straightforward.

To generate specific characteristics of synchrotron radiation, special insertion devices known as undulators are often employed. In such setups, the resonance approximation—which is always applicable—significantly simplifies the theoretical treatment. This approximation complements, rather than replaces, the paraxial one, and leverages another large parameter: the number of undulator periods $N_w \gg 1$. 

\begin{figure}
	\centering
	\includegraphics[width=0.9\textwidth]{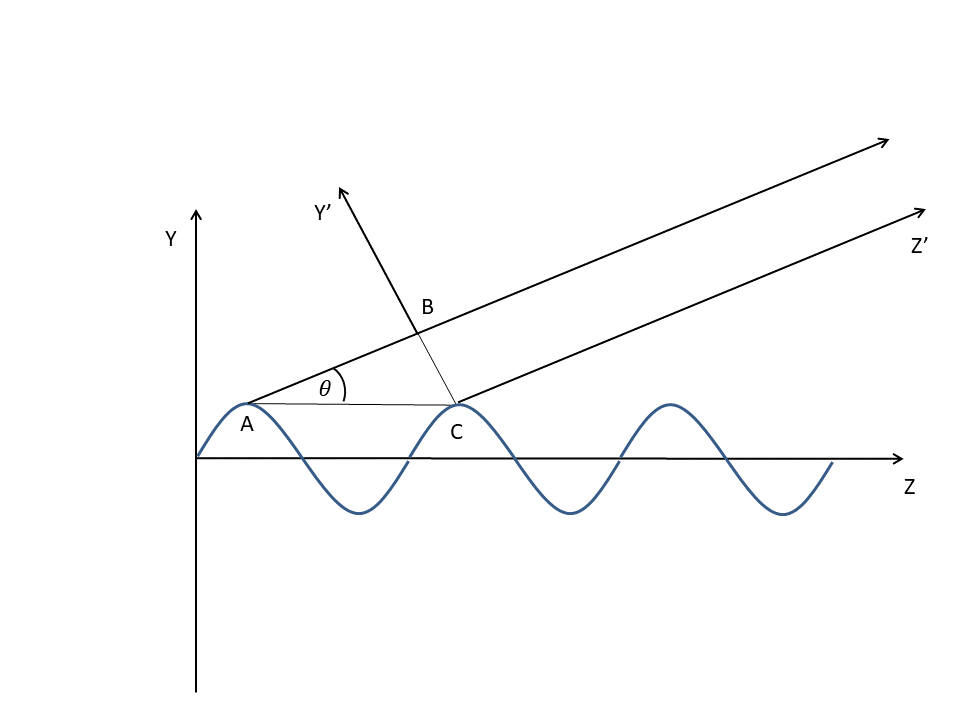}
	\caption{ Constructive interference of radiation from the successive poles }
	\label{BC94}
\end{figure}

\begin{figure}
	\centering
	\includegraphics[width=1.\textwidth]{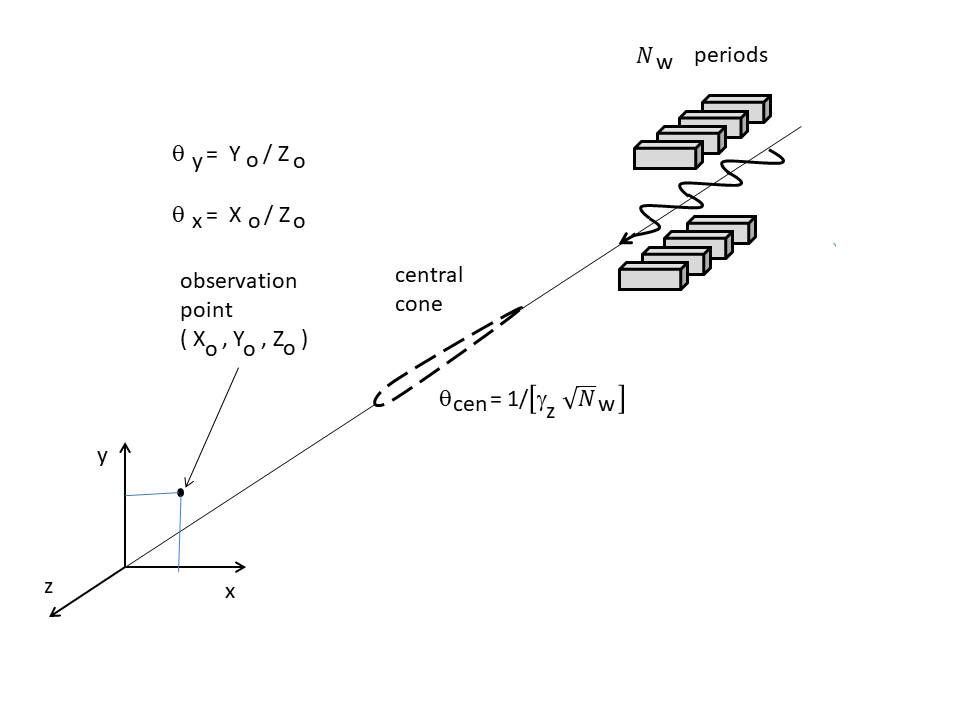}
	\caption{ Geometry for radiation production from an undulator }
	\label{B94}
\end{figure}

The frequency of the radiation emitted by a particle traversing an undulator can be derived by analyzing the interference between radiation produced at successive undulator periods (see Fig. \ref{BC94}). This frequency is subject to Doppler shifting, and the shortest wavelength is observed along the  particle's velocity vector. Under the resonance approximation, radiation at this shortest wavelength is emitted within an angular cone much narrower than $1/\gamma$
(see Fig. \ref{B94}). This sharply defines the relevant observation angles. Outside the diffraction angle, the intensity falls to zero with an accuracy $1/N_w$.

This leads us to an important question: Why did the error in insertion device theory remain undetected for so long? We address this in detail below, focusing specifically on radiation within the central cone (see Fig. \ref{B94}). In this approximation, the electron trajectory is effectively neglected. Regardless of the strength of the undulator (i.e., the undulator parameter), the amplitude of the electron's oscillation remains much smaller than the diffraction-limited size of the radiation at the undulator exit—again, a consequence of $N_w \gg 1$. As a result, undulator radiation theory, in this context, produces fields closely resembling those predicted by dipole-like instantaneous emission models. For practical purposes, especially when describing radiation into the central cone, the conventional (non-covariant) approach has therefore been considered adequate.

However, there is a specific scenario in which the conventional theory breaks down. The covariant formulation predicts a non-zero red shift in the resonance frequency when the electron motion experiences perturbations—specifically, transverse kicks with respect to the longitudinal axis. Experimental observations support this correction for spontaneous undulator emission, validating the covariant approach in such cases.

\subsubsection{Conventional Theory}

Equation (\ref{generalfin}) provides a general framework for characterizing the far-field radiation from an electron following an arbitrary trajectory.
In this section, we present a straightforward derivation of the frequency-domain representation of the radiated field produced by an electron traversing a planar undulator. The magnetic field along the undulator axis is given by

\begin{eqnarray}
\vec{B}(z) = \vec{e}_y B_w\cos(k_wz) ~,\label{mf}
\end{eqnarray}

Here  $k_w=2\pi/\lambda_w$, and $\lambda_w$ is the undulator period.
The Lorentz force is used to derive the equation of motion of the electron in the presence of a magnetic field. Integration of this equation gives

\begin{eqnarray}
v_x(z) = -{c \theta_s} \sin(k_w z) = -\frac{c \theta_s}{2
	i}\left[\exp(ik_w z)-\exp(-i k_w z) \right]~. \label{vxpl_ap1}
\end{eqnarray}

Here $\theta_s=K/\gamma$, where $K$ is the deflection
parameter defined as

\begin{eqnarray}
K = \frac{e\lambda_w B_w}{2 \pi m c^2}~,\label{Kpar_ap1}
\end{eqnarray}
$m$ being the electron mass at rest and $B_w$ being the maximal
magnetic field of the undulator on the axis.

In this case, the electron path is given by

\begin{eqnarray}
x(z) = r_w\cos(k_w z)~,\label{oa}
\end{eqnarray}

where $r_w = \theta_s/k_w$ is the oscillation amplitude. 

We write the undulator length as $L = N_w \lambda_w$, where
$N_w$ is the number of undulator periods. With the help of Eq.
(\ref{generalfin}) we obtain an expression, valid in the far zone:

\begin{eqnarray}
&& {\vec{\widetilde{E}}}= {i \omega e\over{c^2 z_0}}
\int_{-L/2}^{L/2} dz' {\exp\left[i
	\Phi_T\right]\exp\left[i\frac{\omega \theta^2 z_0}{2c}\right]}
\left[{K\over{\gamma}} \sin\left(k_w z'\right)\vec{e}_x
+\vec{\theta}\right]~. \cr && \label{undurad_ap1}
\end{eqnarray}
Here

\begin{eqnarray}
&& \Phi_T = \left({\omega \over{2 c \bar{\gamma}_z^2}}+ {\omega
	\theta^2 \over{2  c }}\right) z' -
{K\theta_x\over{\gamma}}{\omega\over{k_w c}}\cos(k_w z') -
{K^2\over{8\gamma^2}} {\omega\over{k_w c}} \sin(2 k_w z')
~,\cr &&\label{phitundu_ap1}
\end{eqnarray}
where the average longitudinal Lorentz factor $\bar{\gamma}_z$ is
defined as

\begin{equation}
\bar{\gamma}_z = \frac{\gamma}{\sqrt{1+K^2/2}}~. \label{bargz_ap1}
\end{equation}
The choice of the integration limits in Eq. (\ref{undurad_ap1})
implies that the reference system has its origin in the center of
the undulator.

Typically, calculating the intensity distribution using Eq. (\ref{undurad_ap1}) alone is insufficient, as it neglects additional contributions—both interfering and non-interfering—from other segments of the electron's trajectory. To obtain the total field in a given setup, one must account for the full electron path and include these extra terms alongside Eq. (\ref{undurad_ap1}). However, there are specific scenarios where the contribution from Eq. (\ref{undurad_ap1}) dominates over the others. In such cases, Eq. (\ref{undurad_ap1}) can be considered to have an independent physical significance.

One such situation arises when the resonance approximation is applicable. This approximation does not replace the paraxial approximation, which is based on the condition $\gamma^2 \gg 1$, but is instead used in conjunction with it. It exploits another typically large parameter—the number of undulator periods, $N_w \gg 1$. Under these conditions, the integral over $dz'$ in Eq. (\ref{undurad_ap1}) simplifies significantly. This simplification occurs regardless of the frequency of interest, due to the long integration range compared to the characteristic scale of the undulator period.

A well-known expression for the angular distribution of the first harmonic field in the far zone (see Appendix A2 for a 
detailed derivation) can be derived from Eq. (\ref{undurad_ap1}). This expression is axisymmetric and can therefore be 
represented as a function of a single observation angle,  $\theta$, where $\theta^2 = \theta_x^2 + \theta_y^2$. The resulting distribution for the slowly varying envelope of the electric field is given by:

\begin{eqnarray}
&&{\vec{\widetilde{E}}}= - \frac{K \omega e }{2
	c^2 z_0 \gamma} A_{JJ} \exp\left[i\frac{\omega \theta^2 z_0}{2c}\right]
\int_{-L/2}^{L/2} dz' \exp\left[i \left(C + {\omega \theta^2
	\over{2  c }}\right) z' \right] \vec{e}_x~,\cr &&\label{undurad5finale}
\end{eqnarray}

Here $\omega = \omega_r + \Delta \omega$, $C =  k_w \Delta\omega/\omega_r$ and

\begin{eqnarray}
\omega_r = 2 k_w c \bar{\gamma}_z^2~, \label{res_ap1}
\end{eqnarray}
is the fundamental resonance frequency. Finally $A_{JJ}$ is defined as

\begin{eqnarray}
A_{JJ} = J_0\left(\frac{ K^2}{4 + 2K^2}\right) -
J_1\left(\frac{K^2}{4 + 2K^2}\right) ~,\label{AJJ}
\end{eqnarray}

$J_n$ being the $n$-th order Bessel function of the first kind. The integration over longitudinal coordinate can be carried out leading to the well-known final result:

\begin{eqnarray}
&&{\vec{\widetilde{E}}}(z_0, \vec{\theta}) = -\frac{K \omega e
	L  }{2 c^2 z_0 \gamma} A_{JJ}\exp\left[i\frac{\omega \theta^2
	z_0}{2c}\right] \mathrm{sinc}\left[\frac{L}{2}\left(C+\frac{\omega
	\theta^2}{2c} \right)\right] \vec{e}_x~,\cr && \label{generalfin4}
\end{eqnarray}
where $\mathrm{sinc}(\cdot) \equiv \sin(\cdot)/(\cdot)$.
Therefore, the field is horizontally polarized and azimuthal
symmetric. Eq. (\ref{generalfin4}) describes a field with a spherical wavefront centered in the middle of the undulator.

\subsubsection{Why Did the Error in Insertion Device Theory Remain Undetected so Long?}

We have seen that, in full generality, the expression for the undulator field in the far zone under the ultrarelativistic (i.e., paraxial) approximation can be written as Eq. (\ref{undurad2_ap1}).
Within the resonance approximation ($N_w \gg 1$), and for frequencies near the first harmonic, this expression simplifies to the well-known form given by Eq. (\ref{generalfin4}), where the field is horizontally polarized and exhibits azimuthal symmetry. The angular divergence of the radiation is significantly smaller than the characteristic angle $1/\bar{\gamma}_z$.

This narrow divergence arises from the resonance nature of undulator radiation, mathematically expressed through the sinc-like term $\sin(\cdot)/(\cdot)$. To characterize the angular width of the radiation peak around the forward direction ($\theta = 0$), we introduce a small angular displacement $\Delta\theta$. By identifying the first zero of the $\sin(\cdot)/(\cdot)$ function at $C = 0$, we can define the natural angular width of the first harmonic radiation, denoted $\theta_c$.
The resulting cone of radiation with aperture $\theta_c$ is commonly referred to as the central cone. It can be shown that the angular width satisfies $\theta_c^2 = 1/(2N_w\bar{\gamma}_z^2) \ll 1/\bar{\gamma}_z^2$.

We now aim to determine the characteristic transverse size of the radiation field at the exit of the undulator. Radiation emitted by the magnetic poles interferes coherently along the undulator axis, with constructive interference occurring within an angle of approximately  $\sqrt{c/(\omega L_w)}$. This corresponds to a transverse interference size at the undulator exit on the order of $\sqrt{cL_w/\omega}$. Meanwhile, the amplitude of the electron's transverse oscillation is given by  $r_w = c\theta_s/k_w = cK/(\gamma k_w)$. Comparing this with the interference size, we find

\[ 
 r_w^2/(cL_w/\omega)  = K^2\omega/(L_wk_w^2\gamma^2) = K^2/[\pi N_w(1 + K^2/2)] \ll 1  , 
\]

 where we used the relation 
$\gamma^2 = (1 + K^2/2)\bar{\gamma}_z^2$. This inequality holds regardless of the value of  $K$, since $N_w \gg 1$.
Therefore, the electron's oscillation amplitude is always much smaller than the diffraction-limited size of the radiation field at the undulator exit.

We analyze the radiation associated with the first-order term in the expansion of Eq. (\ref{undurad3_ap1}) in powers of the small parameter $ [K \theta_x \omega/(\gamma k_w c)]$ (see Eq. (\ref{drop})). However, this approximation neglects the term $K\theta_x\omega\cos(k_w z')/(\gamma k_w c)$, and as a result, all information about the transverse electron trajectory in the phase factor of Eq. (\ref{phitundu_ap1}) is lost. Under this approximation, the scale of the electron's orbit is much smaller than the radiation's diffraction size, and Eq. (\ref{generalfin4}) yields fields that closely match those predicted by dipole radiation theory. Therefore, we consider the non-covariant approach sufficient for describing transverse electron motion in this context.

Several important points can be made regarding the result discussed above. As previously explained, by considering only the radiation emitted within the central cone, we overlook critical information about the electron's transverse motion. To present a complete analysis, we must also account for the effects of acceleration along the $z$-direction—that is, along the undulator axis. We assume that the transverse velocity  $v_{\perp}(z)$  is small compared to the speed of light $c$, and we introduce $v_{\perp}/c$ as a small expansion parameter. In this framework, we neglect terms of order 
$(v_{\perp}/c)^3$, corresponding to a second-order relativistic approximation for the transverse dynamics.

It is worth noting that, under the ultrarelativistic approximation, analyzing the longitudinal motion is considerably simpler than analyzing the transverse motion. In a constant magnetic field, the electron acquires a transverse velocity 
$v_{\perp}$, and its longitudinal velocity is reduced by  $\Delta v_z = -v(v_{\perp}/c)^2/2$, where $v$ is the total velocity.

When we apply the relevant transformations consistent with a second-order approximation in  $(v_{\perp}/c)^2$, no relativistic correction to the longitudinal motion arises. Consequently, within this level of approximation, the distinction between covariant and non-covariant constrained electron trajectories has no impact on the undulator radiation observed in the central cone.

\subsubsection{Influence of the Kick According to Conventional Theory}

Equation (\ref{generalfin4}) can be extended to describe a particle with an initial offset $\vec{l}$ and deflection angle $\vec{\eta}$ relative to the longitudinal axis, under the assumption that the magnetic field in the undulator is independent of the particle’s transverse position (see Appendix A3). Although this generalization can be derived directly from Eq. (\ref{generalfin2}), it is often more efficient to apply certain intuitive geometrical arguments that are consistent with a rigorous mathematical treatment.

Consider first the effect of a transverse offset $\vec{l}$ with respect to the longitudinal axis $z$. Since the particle experiences the same magnetic field, the far-field radiation pattern is simply shifted by $\vec{l}$. As a result, Eq. (\ref{generalfin4}) can be generalized by replacing the transverse observation coordinate $\vec{r}_0$ with $\vec{r}_0 - \vec{l}$. Equivalently, the observation angle $\vec{\theta} = \vec{r}_0 / z_0$ should be substituted with $\vec{\theta} - \vec{l}/z_0$, leading to:

\begin{eqnarray}
\widetilde{{E}}\left(z_0,  \vec{l}, \vec{\theta}\right)&=&
-\frac{K \omega e L} {2 c^2 z_0 \gamma}
A_{JJ}\exp\left[i\frac{\omega z_0}{2 c}
\left|\vec{\theta}-\frac{\vec{l}}{z_0}\right|^2\right]
\mathrm{sinc}\left[\frac{\omega L
	\left|\vec{\theta}-\left({\vec{l}}/{z_0}\right)\right|^2}{4
	c}\right] ~.\label{undurad4bisgg0}
\end{eqnarray}

Let us now examine the effect of a deflection angle $\vec{\eta}$. Assuming the magnetic field experienced by the electron remains independent of its transverse position, the electron’s trajectory continues to follow a sinusoidal path. However, the effective undulator period is modified and becomes $\lambda_w/\cos(\eta) \simeq (1 + \eta^2/2)\lambda_w$, due to the projection of the undulator axis onto the new trajectory. This leads to a relative red shift in the resonant wavelength given by $\Delta \lambda / \lambda \sim \eta^2/2$.

In typical scenarios of interest, the deflection angle can be approximated as $\eta \sim 1/\gamma$. Consequently, the relative red shift scales as $\Delta \lambda / \lambda \sim 1/\gamma^2$. This should be compared with the intrinsic relative bandwidth of the resonance, which is approximately $\Delta \lambda / \lambda \sim 1/N_w$, where $N_w$ denotes the number of undulator periods. For instance, if $\gamma > 10^3$, the red shift induced by the deflection angle becomes negligible in all practical situations.

Therefore, introducing a deflection angle effectively corresponds to a rigid rotation of the entire system. When performing this rotation, it is important to note that the phase factor in Eq. (\ref{undurad4bisgg0})—which represents a spherical wavefront emanating from $z = 0$—remains invariant under rotation. In contrast, the argument of the $\mathrm{sinc}(\cdot)$ function in Eq. (\ref{undurad4bisgg0}) is altered by the rotation, as the point $(z_0, 0, 0)$ is mapped to $(z_0, -\eta_x z_0, -\eta_y z_0)$. As a result, after the rotation, Eq. (\ref{undurad4bisgg0}) transforms accordingly:

\begin{eqnarray}
&&\widetilde{{E}}\left(z_0, \vec{\eta}, \vec{l},
\vec{\theta}\right)= -\frac{K \omega e L A_{JJ}} {2 c^2 z_0 \gamma}
\exp\left[i\frac{\omega z_0}{2 c}
\left|\vec{\theta}-\frac{\vec{l}}{z_0}\right|^2\right]
\mathrm{sinc}\left[\frac{\omega L
	\left|\vec{\theta}-\left({\vec{l}}/{z_0}\right)-\vec{\eta}\right|^2}{4
	c}\right] \cr &&\label{undurad4bisgg00}
\end{eqnarray}

Finally, in the far-zone regime, we can take the limit $l/z_0 \ll 1$, which justifies neglecting the term ${\vec{l}}/{z_0}$ in the argument of the $\mathrm{sinc}(\cdot)$ function, as well as the quadratic phase term  $\omega l^2/(2 c z_0)$.   As a result, Eq. (\ref{undurad4bisgg00}) simplifies further, yielding a generalized form of Eq. (\ref{generalfin4}) in its final expression:

\begin{eqnarray}
\widetilde{{E}}\left(z_0, \vec{\eta}, \vec{l},
\vec{\theta}\right)&=& -\frac{K \omega e L A_{JJ}} {2 c^2 z_0 \gamma}
\exp\left[i\frac{\omega}{c}\left( \frac{z_0 \theta^2}{2}-
\vec{\theta}\cdot\vec{l} \right)\right]
\mathrm{sinc}\left[\frac{\omega L
	\left|\vec{\theta}-\vec{\eta}\right|^2}{4 c}\right] ~.
\label{undurad4bisgg}
\end{eqnarray}

It is evident from the above that, within the framework of conventional synchrotron radiation theory, considering the radiation from a single electron at a detuning $C$ from resonance, the effect of a kick is merely a rigid rotation of the angular distribution in the direction of the new electron trajectory. This is intuitively reasonable, since after the kick, the electron maintains its velocity and emits radiation in the new, kicked direction due to the Doppler effect. Following this rotation, Eq. (\ref{generalfin4}) transforms into Eq. (\ref{undurad4bisgg}).

\subsubsection{Influence of the Kick According to Correct Coupling of Fields and Particles}

According to the correct coupling between fields and particles, undulator radiation theory makes a striking prediction regarding radiation from a single electron, both with and without a transverse kick. Specifically, when a kick is introduced, the theory predicts a red shift in the resonance wavelength of the undulator radiation along the direction of the electron's velocity.

To demonstrate this effect, we consider a covariant treatment that explicitly employs Lorentz transformations. Within this framework, introducing a kick leads to a non-zero red shift in the resonance frequency. This shift arises because, in Lorentz coordinates, the electron’s velocity is reduced—from $v$ to  $v - v\theta_k^2/2$ -while the speed of light remains constant at its electrodynamical value, $c$.

Consequently, the expression given in Eq. (\ref{totph3}) requires correction: it should use the modified electron velocity $v - v\theta_k^2/2$ rather than the unperturbed velocity $v$. The resulting shift in the total phase $\Phi_T$, appearing under the integral in Eq. (\ref{generalfin2}), is given by $\Delta \Phi_T = \omega\theta_k^2 z'/(2c)$,
where we have used the ultrarelativistic approximation.

Now, consider the case where the electron moves along a straight, constrained trajectory parallel to the undulator axis, without any kick. The corresponding radiation field in the far zone is described by Eq. (\ref{generalfin4}). Referring back to Eq. (\ref{undurad4bisgg}), we find that the conventional undulator radiation theory provides the following expression for the radiation field after the kick:

\begin{eqnarray}
&&{\vec{\widetilde{E}}} = -\frac{K \omega e
	L  }{2 c^2 z_0 \gamma} A_{JJ}\exp\left[i\frac{\omega \theta^2
	z_0}{2c}\right] \mathrm{sinc}\left[\frac{L}{2}\left(C + \frac{\omega
	\left| \vec{\theta}  -\vec{\theta}_k\right|^2}{2c} \right)\right] \vec{e}_x~.\cr && \label{generalfins}
\end{eqnarray}

The covariant equations indicate that, when the kick is applied, the resulting radiation field is described by the following formula:

\begin{eqnarray}
&&{\vec{\widetilde{E}}} = -\frac{K \omega e
	L  }{2 c^2 z_0 \gamma} A_{JJ}\exp\left[i\frac{\omega \theta^2
	z_0}{2c}\right] \mathrm{sinc}\left[\frac{L}{2}\left(C+ \frac{\omega\theta_k^2}{2c} + \frac{\omega
	\left| \vec{\theta}  -\vec{\theta}_k\right|^2}{2c} \right)\right] \vec{e}_x~,\cr && \label{generalfin6}
\end{eqnarray}

This formula is nearly, but not exactly, of the same form as Eq. (\ref{generalfins}). The key difference lies in the additional term $\omega\theta_k^2/(2c)$ appearing in the argument of the $\mathrm{sinc}$ function. Special attention should be paid to the shift in resonance frequency between the undulator radiation setups with and without a kick.

Recalling the definition of the detuning parameter, $C = k_w\Delta\omega/\omega_r$, we can express the redshift in resonance frequency as $\Delta\omega/ \omega_r =  - \omega_r\theta_k^2/(2k_w c)$. Alternatively, this redshift can also be written in terms of the relativistic factor as $\Delta\omega/ \omega_r =  - \gamma^2\theta_k^2/(1+K^2/2)$. 
This reveals a second-order correction in $\theta_k^2$, which, however, is amplified by the large factor 
$\bar{\gamma}_z^2$. 

We are now prepared to more generally examine how the field expression is modified by the introduction of a kick. Suppose that, in the absence of a kick, the electron moves along a trajectory that makes an angle $\vec{\eta}$ with respect to the undulator axis. The corresponding field is given by Eq. (\ref{undurad4bisgg}).
Let $\vec{\theta}_k$ denote the kick angle of the electron relative to its initial trajectory. According to the conventional approach, the expression for the field after the kick becomes:

\begin{eqnarray}
&&{\vec{\widetilde{E}}} = -\frac{K \omega e
	L  }{2 c^2 z_0 \gamma} A_{JJ}\exp\left[i\frac{\omega \theta^2
	z_0}{2c}\right] \mathrm{sinc}\left[\frac{L}{2}\left(C + \frac{\omega
	\left| \vec{\theta}  -\vec{\eta} -\vec{\theta}_k\right|^2}{2c} \right)\right] \vec{e}_x~.\cr && \label{generalfinss}
\end{eqnarray}

In contrast, the covariant approach gives

\begin{eqnarray}
&&{\vec{\widetilde{E}}} = -\frac{K \omega e
	L  }{2 c^2 z_0 \gamma} A_{JJ}\exp\left[i\frac{\omega \theta^2
	z_0}{2c}\right] \mathrm{sinc}\left[\frac{L}{2}\left(C+ \frac{\omega\theta_k^2}{2c} + \frac{\omega
	\left| \vec{\theta}  - \vec{\eta} - \vec{\theta}_k\right|^2}{2c} \right)\right] \vec{e}_x~,\cr && \label{generalfinnn}
\end{eqnarray}

This brings us to an intriguing situation. According to conventional theory, the resonance wavelength depends solely on the observation angle relative to the electron’s velocity direction.
However, Eq. (\ref{generalfinss}) shows that for any kick angle $\vec{\theta}_k$ and for any angle $\vec{\eta}$ between the undulator axis and the initial electron velocity, the radiation emitted along the velocity direction experiences no redshift. This highlights a critical distinction between the conventional and covariant formulations.

In contrast, the result from the covariant approach—Eq. (\ref{generalfinnn})—explicitly depends on the magnitude of the kick angle $\theta_k$. In this case, radiation along the velocity direction exhibits a redshift only when the kick angle is nonzero.

We are thus led to an important conclusion:
when the electron is accelerated in the lab frame upstream of the undulator, the covariant trajectory retains this information.

\subsubsection{Experimental Test of SR Theory in 3rd Generation Light Source}

One way to highlight the incompatibility between the standard approach to relativistic electrodynamics, which typically involves Maxwell's equations, and the particle trajectories derived from non-covariant particle tracking is through a direct laboratory test of synchrotron radiation theory.

Let us now explore the potential of synchrotron radiation sources to validate the predictions of a revised synchrotron radiation theory. In modern synchrotron radiation sources, the electron beam emittance is sufficiently small that one can disregard the effects of finite beam size and angular divergence, particularly in the soft X-ray wavelength range. This makes it feasible to model the synchrotron radiation source using the approximation of a filament electron beam, thereby enabling the use of analytical models for single electron synchrotron radiation fields.

\begin{figure}
	\centering
	\includegraphics[width=0.9\textwidth]{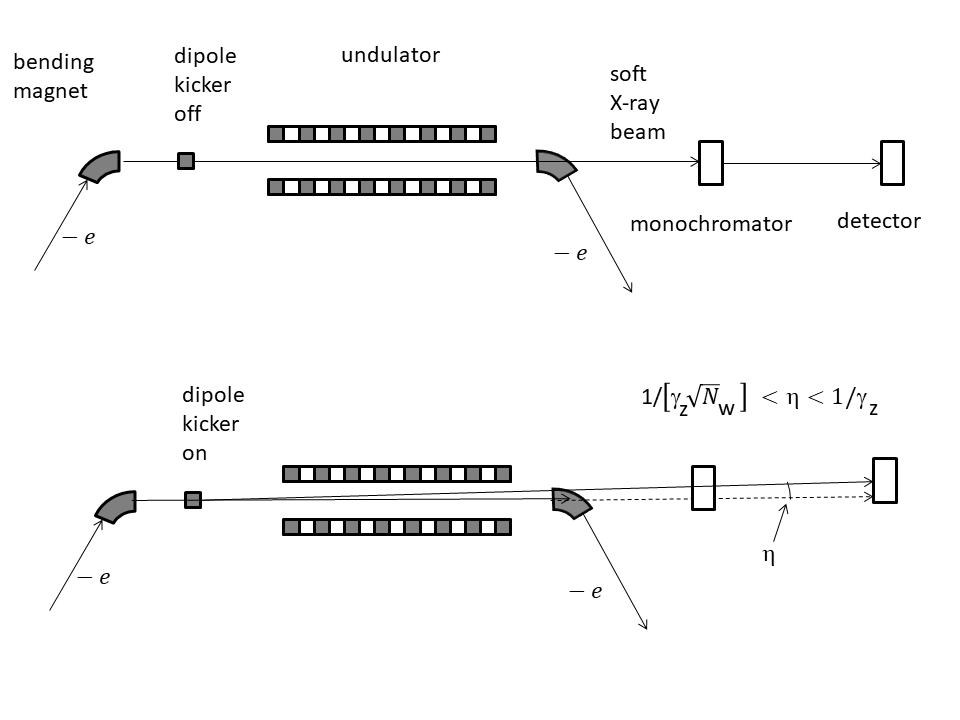}
	\caption{ Illustration of the proposed experimental setup to test synchrotron radiation theory using a third-generation light source. Top: Electron beam without a kick. Bottom: Electron beam kicked by an angle $\eta$. In both configurations, the X-ray pulse is filtered through a monochromator, and the total energy is recorded by a detector as a function of undulator detuning.}
	\label{B86}
\end{figure}

The basic setup for the test experiment is illustrated in Fig. \ref{B86}. The soft X-ray undulator beamline should be tuned to a minimum photon energy, typically corresponding to the "water window" wavelength range. The radiation pulse passes through a monochromator filter $F$, and its energy is then measured by the detector. Precise monochromatization of the undulator radiation is not required in this case; a monochromator line width of $\Delta\omega/\omega \simeq 10^{-3}$ is sufficient.

To conduct the proposed test experiment, it is essential to control the beam kick, for example, using a corrector magnet. Without the kick, the maximum pulse energy detected will align with the monochromator line tuned to resonance. However, when the kick is applied, conventional synchrotron radiation theory predicts no redshift in the resonance wavelength. In contrast, one of the immediate consequences of the corrected theory is the occurrence of a non-zero redshift.

The proposed experimental procedure is relatively simple, as it involves relative measurements in the electron beam's velocity direction, both with and without the transverse kick. This measurement is crucial because conventional theory predicts the absence of redshift, a result that, to our knowledge, has never been observed at synchrotron radiation facilities. However, an XFEL-based experiment confirms our correction for spontaneous undulator emission—further details can be found in the next chapter.

\subsection{Synchrotron Radiation from Bending Magnets}

Consider a relativistic electron moving along a circular orbit. In the standard treatment, the observer is located in a vertical plane tangent to the electron’s circular trajectory at the origin, positioned at an angle $\theta$ above the orbital plane. In this configuration, the $z$-axis is not fixed but varies with the observer’s position. Notably, the electron’s motion exhibits cylindrical symmetry about the vertical axis through the center of the orbit. Owing to this symmetry, it is sufficient—when calculating the spectral and angular distribution of emitted photons—to consider an observer in this specific geometric arrangement, without resorting to a more general configuration. It is important to emphasize the distinction between the general geometric setup used here and the more restrictive arrangement often assumed in standard synchrotron radiation treatments of bending magnet radiation.

To proceed, we begin with a broad overview of the main results. The goal is to solve the electrodynamics problem using Maxwell’s equations in their conventional form. Since the particle undergoes relativistic acceleration, its dynamics must be analyzed within the framework of special relativity. However, defining Lorentz coordinates for the laboratory frame becomes nontrivial in the presence of acceleration. The only consistent method to introduce Lorentz coordinates in this context is to employ individual coordinate systems—so-called "ruler-clock" structures—at each point along the electron’s trajectory.

Several notable effects emerge from the cylindrical symmetry and the use of the paraxial approximation. In our analysis, we demonstrate that the derivation of bending magnet radiation does not require covariant particle tracking. Nonetheless, there is one key instance in which conventional theory fails: the covariant approach predicts a nonzero redshift of the critical frequency when perturbations influence the electron’s motion along the magnetic field—that is, when the trajectory deviates from the nominal orbit.

\subsubsection{Conventional Theory}

\begin{figure}
	\centering
	\includegraphics[width=0.9\textwidth]{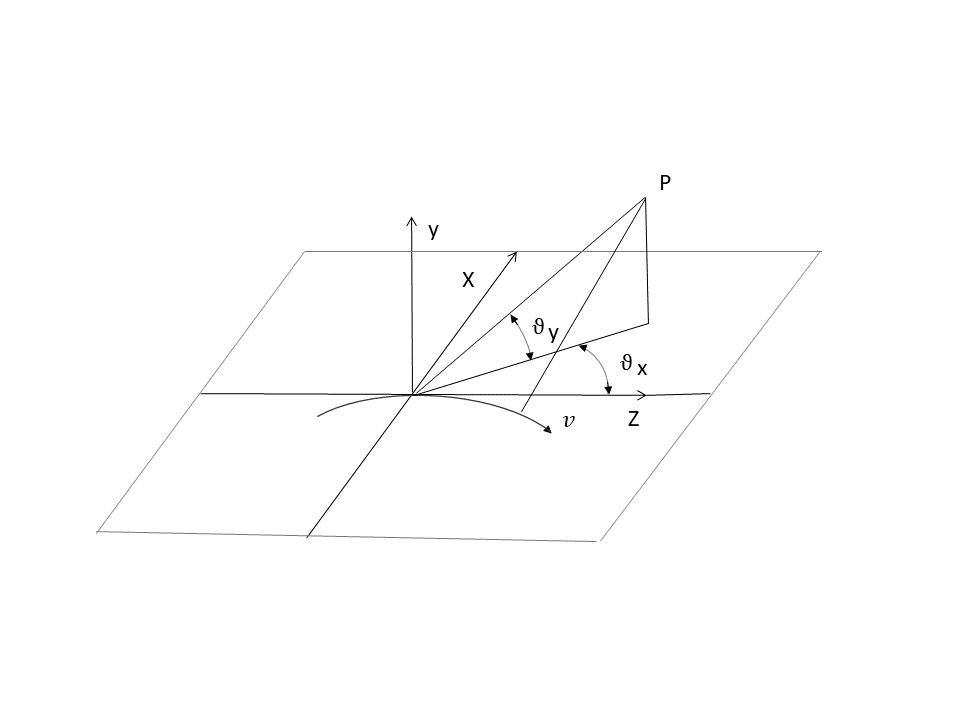}
	\caption{ Geometry for the analysis of bending magnet radiation.}
	\label{BM1}
\end{figure}

Consider a single relativistic electron moving along a circular orbit and an observer as sketched  in Fig. \ref{BM1}.
In conventional treatments, the horizontal observation angle $\theta_x$ is typically assumed to be zero.
This simplification arises because most textbooks focus on calculating the intensity radiated by a single electron in the far field—specifically, the square modulus of the field amplitude—without addressing more complex scenarios such as source imaging.

Equation (\ref{generalfin}) can be used to calculate the far-zone radiation field emitted by a relativistic electron moving along a circular arc. Assuming a geometry with a fixed $z$-coordinate, the transverse position of the electron can be expressed as a function of the curvilinear abscissa $s$ as follows:

\begin{equation}
\vec{r}(s) = -R\left(1-\cos(s/R)\right) \vec{e_x}
\label{trmot}
\end{equation}
and

\begin{equation}
z(s) = R \sin(s/R) \label{zmot}
\end{equation}
where $R$ is the bending radius.

Since the integral in Eq. (\ref{generalfin}) is performed along $z$ we should invert $z(s)$ in Eq. (\ref{zmot}) and find the explicit dependence $s(z)$:

\begin{equation}
s(z) = R \arcsin(z/R) \simeq z + {z^3\over{6R^2}} \label{sz}
\end{equation}
so that

\begin{equation}
\vec{r}(z) = - {z^2\over{2 R}} \vec{e_x}~,\label{rpdis}
\end{equation}
where the expansion in Eq. (\ref{sz}) and Eq. (\ref{rpdis}) is justified, once again, in the framework of the paraxial approximation.

With Eq. (\ref{generalfin}) we obtain the radiation field amplitude in the far zone:

\begin{eqnarray}
\vec{\widetilde{E}}= {i \omega e\over{c^2 z_0}}
\int_{-\infty}^{\infty} dz' {e^{i  \Phi_T}} \left({
	z'+R\theta_x\over{R}}\vec{e_x}
+\theta_y\vec{e_y}\right)~\label{srtwo}
\end{eqnarray}
where

\begin{eqnarray}
&&\Phi_T = \omega \left[
\left({\theta_x^2+\theta_y^2\over{2c}}z_0\right)
+\left({1\over{2\gamma^2c}} +
{\theta_x^2+\theta_y^2\over{2c}}\right)z' \right.  \cr && \left. +
\left({\theta_x\over{2Rc}}\right)z'^2 +
\left(1\over{6R^2c}\right)z'^3\right]~.\label{phh2}
\end{eqnarray}

One can easily reorganize the terms in Eq. (\ref{phh2}) to obtain

\begin{eqnarray}
&& \Phi_T = \omega\left[
\left({\theta_x^2+\theta_y^2\over{2c}}z_0\right)-{R\theta_x\over{2c}}\left({1\over{\gamma^2}}
+{\theta_x^2\over{3}} +\theta_y^2\right) \right. \cr && \left.
+\left({{1\over{\gamma^2}}+\theta_y^2}\right){\left(z'+R\theta_x\right)\over{2c}}
+ {\left(z'+R\theta_x\right)^3\over{6 R^2 c
}}\right]~.\label{phh2b}
\end{eqnarray}
With redefinition of $z'$ as $z' + R \theta_x$ under integral we obtain the final result:

\begin{eqnarray}
&& \vec{\widetilde{E}}= {i \omega e\over{c^2 z_0}} e^{i\Phi_s}
e^{i\Phi_0} \int_{-\infty}^{\infty} dz'
\left({z'\over{R}}\vec{e_x}+\theta_y\vec{e_y}\right) \cr &&
\times
\exp\left\{{i\omega\left[{z'\over{2\gamma^2c}}\left(1+\gamma^2\theta_y^2\right)
	+{z'^3\over{6R^2c}}\right]}\right\}~,\label{srtwob}
\end{eqnarray}
where
\begin{equation}
\Phi_s ={\omega z_0\over{2c}}\left(\theta_x^2+\theta_y^2
\right)\label{phis}
\end{equation}
and

\begin{equation}
\Phi_0 = -{\omega R \theta_x\over{2c}}\left( {1\over{\gamma^2}}
+{\theta_x^2\over{3}} +\theta_y^2 \right)~.\label{phio}
\end{equation}
In standard treatments of bending magnet radiation, the phase term $\exp(i\Phi_0)$ is absent. 
The horizontal observation angle $\theta_x$ is always equal to zero.

\subsubsection{Why Did the Error in Synchrotron Radiation  Remain Undetected so Long?}

Our case of interest involves an ultrarelativistic electron undergoing circular acceleration.
As previously noted, conventional (non-covariant) particle tracking describes the dynamical evolution in the laboratory frame using the absolute time convention. In this framework, simultaneity is treated as absolute, requiring only a single set of synchronized clocks in the lab frame to describe the electron's accelerated motion. However, adopting the absolute time convention leads to significantly more complex field equations, which vary with the particle's velocity—that is, they differ at each point along the electron’s trajectory.
This complexity is precisely why the covariant approach is preferred when treating both dynamics and electrodynamics.

We begin by considering an electron moving along a circular trajectory that lies in the  $(x,z)$-plane and is tangent to the $z$-axis. Due to the cylindrical symmetry of the problem, it is not necessary to consider a general observation point to calculate the spectral and angular distributions of the emitted photons. Instead, we assume the observer is located in the vertical plane tangent to the trajectory at the origin. Within the ultrarelativistic (paraxial) approximation, we perform transformations accurate up to order  $v^2_x/c^2$, which significantly simplifies the calculations. This second-order approximation allows us to neglect higher-order terms while retaining essential relativistic effects.

In the lab frame, manipulations involving the rule-clock structure can be interpreted as a change of variables governed by the transformation in Eq. (\ref{GGT3}):  $x_L = \gamma_x x $, $t_L = (t/\gamma_x +   \gamma_x xv_x/c^2)$, where
$\gamma_x = 1 + v_x^2/(2c^2)$ reflects the second-order approximation.

This combination of a Galilean transformation and a change of variables effectively results in a transverse Lorentz transformation. Since this modified Galilean approach is mathematically equivalent to a Lorentz transformation, applying these variable changes naturally leads to the correct form of Maxwell's equations.

To retain Lorentz coordinates in the laboratory frame, as previously discussed, it suffices to apply a time transformation given by

\[
\Delta t = t_L - t =  - [v_x^2/(2c^2)]t + xv_x/c^2    . 
\]

At this order of expansion, the relativistic correction to the particle's offset $x$ does not appear; such corrections arise only at order $v_x^3/c^3$. Therefore, for our case of interest, $x_L = x$.

Although this time transformation was initially derived for a specific scenario, the result generalizes to any transverse velocity direction. In general, we can write: 

\[
\Delta t = t_L - t =  - [|\vec{v}_{\perp}|^2/(2c^2)]z'/c + \vec{r}_{\perp}\cdot \vec{v}_{\perp}/c^2     . 
\]

To complete our analysis, we now consider the relativistic correction to longitudinal motion. We emphasize again that when evaluating the transformations up to second order in  $(v_{\perp}/c)^2$, no correction to the longitudinal motion appears at this level of approximation.

In summary, we have demonstrated a covariant approach applicable to arbitrary trajectories—providing a general framework for analyzing the system directly in the space-frequency domain under the paraxial approximation.

Let us now explore how to apply the covariant method to a specific case. We will use our understanding of the relativistically correct approach for calculating synchrotron radiation emission to determine the photon angular-spectral density distributions from a bending magnet.

In the ultrarelativistic limit, the electron undergoes uniform transverse acceleration, given by  $a = v^2/R \simeq c^2/R$. Under this approximation, the electron's velocity and displacement in the transverse direction can be expressed as: $v_x = at = az'/v = az'/c$, $x = at^2/2 = az'^2/(2c^2)$. With these expressions, we now have all the quantities needed to evaluate the relativistic time shift:  

\[
\Delta t =  t_L - t = - a^2z'^3/(2c^5) + a^2z'^3/(2c^5) = 0     . 
\]

There is no time difference! This means that covariant particle tracking is not required to derive the radiation from a bending magnet.

Such a precise cancellation often hints at a deeper underlying principle. However, in this case, it appears to be a mere coincidence—there is no evident profound significance.

This cancellation is not surprising when one analyzes the general expression for the radiation field from a bending magnet in the far zone, given by Eq.~(\ref{srtwob}).

In our previous discussion of undulator radiation, we learned that the relativistic correction appears only when the transverse electron trajectory is included in the total phase $\Phi_T$ under the integral Eq.(\ref{generalfin}). Referring back to Eq.(\ref{totph}) for the phase factor $\Phi_T$, we see that the term which depends on the transverse position of the electron can be written as $\exp i(\omega/c)[\theta_x x(z') +  \theta_y y(z')]$. We conclude that the observation angle $\theta_x$  in the total phase factor under the integral must be related to the contribution of the transverse electron trajectory. 

Now, examining Eq.~(\ref{srtwob}), we observe that the phase factor includes only the $\theta_y$ component of the observation angle. This implies that the transverse constrained motion of the electron,  $x(z')$ in the bending magnet does not contribute to synchrotron radiation.
Therefore, it is justified to use a non-covariant approach when analyzing the constrained electron motion along the nominal orbit in the $(x,z)$-plane.

We emphasize that the cancellation of the relativistic time shift and the independence of the Fraunhofer propagator (more precisely, the paraxial approximation of the Green’s function for the inhomogeneous Helmholtz equation in the space-frequency domain) with respect to the observation angle $\theta_x$ in the far zone are two aspects of the same phenomenon. Both arise from the cylindrical symmetry inherent in the motion of an electron along a circular arc.

Due to this symmetry, calculating the spectral and angular photon distributions in the far field does not require placing the observer at a general position. Instead, it suffices to consider an observer located in the vertical plane tangent to the circular trajectory at the origin. In this configuration, the observation angle $\theta_x = 0$, while $\theta_y$ corresponds to elevation above the orbital plane. In other words, within this specific geometry, the $z$-axis is effectively defined by the observer's position, Fig. \ref{BM1}.

\begin{figure}
	\centering
	\includegraphics[width=0.9\textwidth]{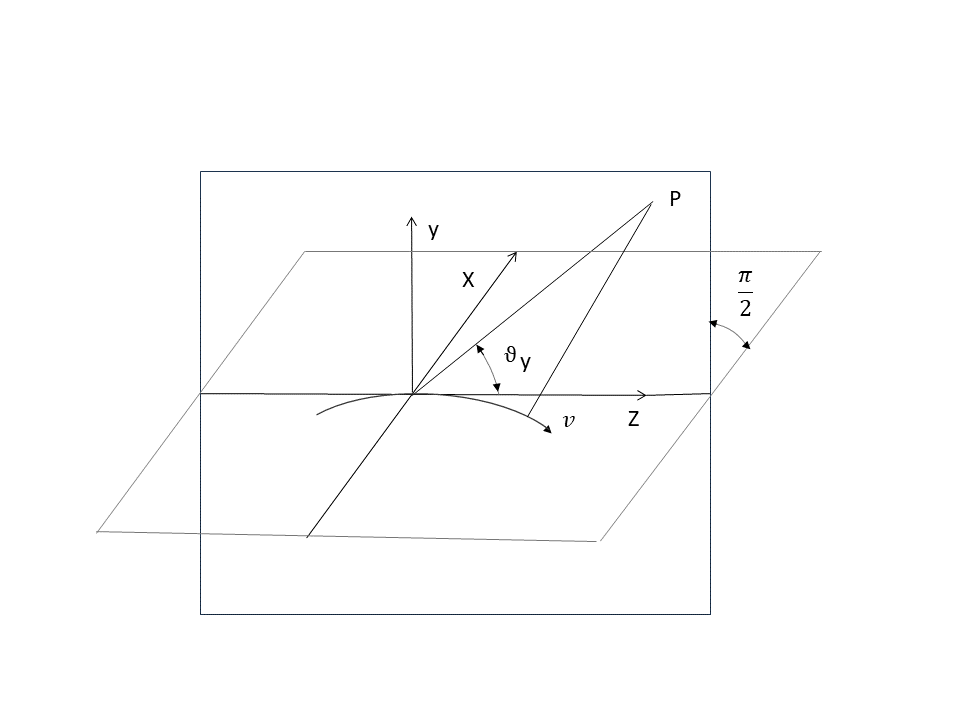}
	\caption{ Standard geometry for the analysis of synchrotron radiation from a bending magnet. }
	\label{BM2}
\end{figure}

However, this approach offers limited insight into the near-field radiation, where the Fresnel propagator must be used. In the near zone, the radiation field is sensitive to the detailed, constrained motion of the electron. Far-field arguments, though, are useful in demonstrating that synchrotron radiation from bending magnets is unaffected by differences between non-covariant and covariant electron trajectories. Te near-zone field can be calculated from the knowledge of the far-zone field. This is made possible through the use of the space-frequency domain and by exploiting the paraxial approximation \cite{GE}. The cancellation of the relativistic time shift ensures consistent results in both the far and near zones, as expected \footnote{It is important to emphasize that the inverse field problem—based solely on far-field data—cannot be solved without invoking the paraxial approximation.}.

\subsubsection{Influence of the Kick According to Conventional Theory}

Up to this point, we have examined the case of an electron moving along a circular trajectory confined to the $(x,z)$-plane and tangent to the $z$-axis. The resulting phase difference in the emitted fields is influenced by both the observer’s position and the specifics of the electron’s motion. We now shift focus to the radiation emitted by a single electron traversing a bending magnet, allowing for arbitrary angular deflection and spatial offset relative to the nominal orbit.

An approximate representation of the electron’s path—given by Eqs. (\ref{shiftilt}) and (\ref{curvabs}) in Appendix A3—can be used to characterize the emitted field for a general trajectory. By applying Eqs. (\ref{sz}) and (\ref{rpdis}), we derive an approximate expression for $s(z)$:

\begin{equation}
s(z) = z+ {z^3\over{6 R^2}}+{z^2 \eta_x\over{2 R}} +{z
	\eta_x^2\over{2}}+{z \eta_y^2\over{2}} \label{szangle}
\end{equation}
so that

\begin{equation}
\vec{v}_{\perp}(z) =   \left(- {v z\over{R}}+ v \eta_x  \right)
\vec{e_x} + \left(v \eta_y  \right)
\vec{e_y}~\label{vapprangle}
\end{equation}
and

\begin{equation}
\vec{r}(z) = \left(- {z^2\over{2R}}+ \eta_x z + l_x\right)
\vec{e_x} + \left(\eta_y z +l_y \right)
\vec{e_y}~.\label{trmot2}
\end{equation}

It is clear that the offsets $l_x$ and $l_y$ are consistently subtracted from $x_0$ and $y_0$, respectively. This implies that a shift in the particle trajectory on the vertical plane is equivalent to an opposite shift of the observer. Taking this into account, we define the modified angles $\bar{\theta}_x = \theta_x -l_x/z_0$ and $\bar{\theta}_y = \theta_y - l_y/z_0$ in order to obtain

\begin{eqnarray}
\vec{\widetilde{E}}= {i \omega e\over{c^2 z_0}}
\int_{-\infty}^{\infty} dz' {e^{i \Phi_T}} \left({
	z'+R(\bar{\theta}_x-\eta_x)\over{R}}\vec{e_x}
+{(\bar{\theta}_y-\eta_y)}\vec{e_y}\right)~\label{srtwoang}
\end{eqnarray}
and

\begin{eqnarray}
&& \Phi_T =
\omega \left({\bar{\theta}_x^2+\bar{\theta}_y^2\over{2c}}z_0 \right)
+{\omega\over{2c}}\left({1\over{\gamma^2}} +
\left(\bar{\theta}_x-\eta_x\right)^2  +
\left(\bar{\theta}_y-\eta_y\right)^2\right)z' \cr && +
\left({\omega(\bar{\theta}_x-\eta_x)\over{2Rc}}\right)z'^2 +
\left(\omega\over{6R^2c}\right)z'^3~.\label{phh2ang}
\end{eqnarray}
One can easily reorganize the terms in Eq. (\ref{phh2ang}) to
obtain

\begin{eqnarray}
&&\Phi_T =
\omega\left({\bar{\theta}_x^2+\bar{\theta}_y^2\over{2c}}z_0\right)-
{\omega R(\bar{\theta}_x-\eta_x)\over{2c}} \cr &&\times
\left({1\over{\gamma^2}} +(\bar{\theta}_y-\eta_y)^2
+{(\bar{\theta}_x-\eta_x)^2\over{3}}\right) \cr &&
+\left({{1\over{\gamma^2}}+(\bar{\theta}_y-\eta_y)^2}\right)
{\omega\left(z'+R(\bar{\theta}_x-\eta_x)\right)\over{2c}} \cr  &&+
{\omega \left(z'+R (\bar{\theta}_x-\eta_x)\right)^3\over{6 R^2 c
}}~.\label{phh2angfin}
\end{eqnarray}
Redefinition of $z'$ as $z'+R(\bar{\theta}_x-\eta_x)$ gives the
result

\begin{eqnarray}
&&\vec{\widetilde{E}}= {i \omega e\over{c^2 z_0}} e^{i \Phi_s} e^{i
	\Phi_0} \int_{-\infty}^{\infty} dz'
\left({z'\over{R}}\vec{e_x}+(\bar{\theta}_y-\eta_y)\vec{e_y}\right)
\cr && \times
\exp\left\{{i\omega\left[{z'\over{2\gamma^2c}}\left(1+\gamma^2
	(\bar{\theta}_y-\eta_y)^2\right)
	+{z'^3\over{6R^2c}}\right]}\right\}~,\label{srtwoang2}
\end{eqnarray}
where
\begin{equation}
\Phi_s = {\omega z_0
	\over{2c}}\left(\bar{\theta}_x^2+\bar{\theta}_y^2
\right)\label{phisang}
\end{equation}
and

\begin{equation}
\Phi_0 = - {\omega R(\bar{\theta}_x-\eta_x)\over{2c}}
\left({1\over{\gamma^2}} +(\bar{\theta}_y-\eta_y)^2
+{(\bar{\theta}_x-\eta_x)^2\over{3}}\right)~.\label{phioang}
\end{equation}
In the far zone we can neglect terms in $l_x/z_0$ and $l_y/z_0$, which leads to

\begin{eqnarray}
&&\vec{\widetilde{E}}= {i \omega e\over{c^2 z_0}} e^{i \Phi_s} e^{i
	\Phi_0} \int_{-\infty}^{\infty} dz'
\left({z'\over{R}}\vec{e_x}+\left(\theta_y-\eta_y
\right)\vec{e_y}\right) \cr && \times
\exp\left\{{i\omega\left[{z'\over{2\gamma^2c}}\left(1+\gamma^2
	\left(\theta_y-\eta_y\right)^2\right)
	+{z'^3\over{6R^2c}}\right]}\right\}~,\label{srtwoang2bis}
\end{eqnarray}
where
\begin{equation}
\Phi_s = {\omega z_0 \over{2c}}\left(\theta_x^2+\theta_y^2
\right)\label{phisangbis}
\end{equation}
and

\begin{eqnarray}
\Phi_o \simeq - {\omega R({\theta_x}-\eta_x)\over{2c}}
\left({1\over{\gamma^2}} +(\theta_y-\eta_y)^2 
+{(\theta_x-\eta_x)^2\over{3}}\right)-{\omega\over{c}}(l_x
\theta_x+l_y\theta_y) ~.\label{phioangbis}
\end{eqnarray}

It is evident from the discussion above that the field distribution in the far zone depends solely on the observation angle relative to the direction of the electron’s velocity.

According to the conventional (though incorrect) approach to coupling fields and particles, radiation theory predicts the behavior of bending magnet radiation from a single electron, both with and without an applied kick. Specifically, when a kick is introduced, the angular distribution in the far zone is predicted to undergo a rigid rotation.

\subsubsection{Influence of the Kick According to Correct Coupling of Fields and Particles}
	
Let us now discuss the covariant treatment, which explicitly employs Lorentz transformations. Consider the radiation from a bending magnet, emitted by a single electron that receives a transverse kick with respect to the nominal orbit in the $(x,z)$-plane. Due to the kick, there is an additional translation along the $y$-axis with a constant velocity $v_y = v\theta_k$. The corresponding offset of the electron can be written as: $y = \theta_k z'$. We now incorporate the velocity and offset into the relativistic time shift: 

\[
\Delta t = t_L - t = -\theta_k^2z'/(2c)
+ \theta_k^2z'/c = \theta_k^2z'/(2c)   . 
\]

Thus, the resulting shift in the total phase accumulated along the path is given by: $\Delta \Phi_T = \omega\theta_k^2 z'/(2c)$. This result is consistent with our redshift calculation in the undulator case when a kick is introduced, as expected.

We would like to make a historical note: the distinction between covariant and non-covariant particle trajectories was not fully understood in the past. As a result, accelerator physicists did not recognize that relativistic kinematic effects contribute to synchrotron radiation.

This raises an important question: how can storage rings function effectively under such assumptions?

The key point is that in this context, the dynamics of the electron beam are primarily governed by synchrotron radiation emitted from bending magnets. Due to the cylindrical symmetry of the system, both covariant and non-covariant descriptions of electron motion along a circular arc produce similar synchrotron radiation characteristics—with one important exception.

The covariant framework predicts a small redshift in the critical frequency of the emitted radiation when the electron motion experiences perturbations in the vertical direction. However, since synchrotron radiation from bending magnets spans a broad frequency range, the overall output intensity is largely insensitive to this redshift.

\subsection{Problem-Solving: Multiple Trajectory Kicks}

We now consider the case of $n$ arbitrarily spaced kickers, each introducing a distinct rotation angle. Our goal is to explore how covariant particle tracking can be applied in this scenario and to gain insight into the resulting dynamics—particularly when an undulator is placed downstream of the kicker configuration.

Formally, calculating the radiation from the undulator requires accounting for all trajectory perturbations experienced by the electron since its generation. This may seem surprising at first, but it aligns with the general expression for the radiation field from a single electron, as given by Eq.~(\ref{revwied}).

It is important to note that, in principle, the entire history of the electron - from $t' = -\infty$ to $t' = \infty$
-must be considered, since the integral in Eq.~(\ref{revwied}) extends over this range. However, the physical interpretation of this statement depends on the specific scenario. In practice, the integration should be restricted to the time intervals during which the electron contributes significantly to the radiation field.

Ultimately, it is electrodynamics itself that determines which portions of the particle’s trajectory are relevant for calculating the undulator radiation and which may be neglected. The most crucial general statement in this context is that the trajectory must be computed using covariant methods—provided one wishes to employ the standard formulation of Maxwell’s equations.

Let us consider the ultrarelativistic assumption $1/\gamma^2 \ll 1$, which is commonly valid in typical synchrotron radiation setups. In general, introducing a small parameter into a physical theory leads to significant simplifications. Specifically, the ultrarelativistic approximation corresponds to a paraxial regime, under which Eq. (\ref{revwied}) simplifies to Eq. (\ref{generalfin})

Let us now consider a scenario in which the deflection angle of the first bending magnet upstream of the undulator is much larger than $1/\gamma$. In other words, we examine a standard synchrotron radiation setup, where an electron enters the system through a bending magnet, traverses a straight section, passes through an undulator, continues along another straight section, and exits via a second bending magnet.

Although the integration in Eq.(\ref{generalfin}) is performed from $-\infty$ to $\infty$, the only (edge) part of the trajectory into the bending magnets contributing to the integral is of the order of the radiation formation length $L_f$. 

Mathematically, this is reflected in the behavior of  $\Phi_T(z')$ in Eq.(\ref{generalfin}), which exhibits increasingly rapid oscillations as $z'$ exceeds the formation length. At the critical wavelength, the formation length is approximately $R/\gamma$, where $R$ is the bending radius.  This corresponds to an angular interval along the orbit of roughly $\Delta\theta \simeq 1/\gamma$. 

Typically, the critical wavelength of the radiation from a bending magnet in a synchrotron radiation source is about 0.1 nm. The formation length in this case is only a few millimeters.

Note that for ultrarelativistic systems, the formation length is generally much longer than the radiation wavelength. This seemingly counterintuitive result arises because, in ultrarelativistic systems, it is not possible to localize radiation sources within a macroscopic portion of the trajectory.

The formation length can be interpreted as the longitudinal size of a single electron source. It is meaningless to specify the exact location where electromagnetic signals are emitted within the formation length. As a result, in the context of the radiative process within a bending magnet, we cannot distinguish between radiation emitted at point
$A$ and radiation emitted at point $B$ if the distance between these two points is shorter than the formation length 
$L_f$. 

Now, consider the case of a straight section of length $L$ placed between the bending magnet and the undulator. The same reasoning applied to the bending magnet can be used to define a region of the trajectory in which it is indistinguishable to the observer whether radiation originates from different points. Just as in the bending magnet case, the observer perceives a time-compressed motion of the source. For straight-line motion, the apparent time corresponds to an apparent distance $\lambdabar \gamma^2$.  At the critical wavelength, the bending magnet formation length $L_f \simeq R/\gamma$ is of the same order as the straight section formation length $\lambdabar \gamma^2$.

Intuitively, bending magnets can be seen as "switchers" for the ultrarelativistic electron trajectory. While we focus on bending magnets here, other configurations could also serve as switchers, provided they share a common characteristic: the switching process must depend exponentially on the distance from the starting point. In this context, a characteristic length $d_s$ can be associated with any switcher.

Consider, for example, a plasma accelerator, where an electron is accelerated by high-gradient fields. In this case, the accelerator itself acts as a switcher for the relativistic electron trajectory, since acceleration in the GeV range occurs over a distance of only a few millimeters. In the (soft) X-ray range, the acceleration distance $d_a$ is shorter than the formation length $\lambdabar \gamma^2$ for the following straight section. In this particular case length $d_a$ plays the role of the characteristic length of the switcher $d_s$, which switches on the ultrarelativistic electron trajectory.

\begin{figure}
	\centering
	\includegraphics[width=1.\textwidth]{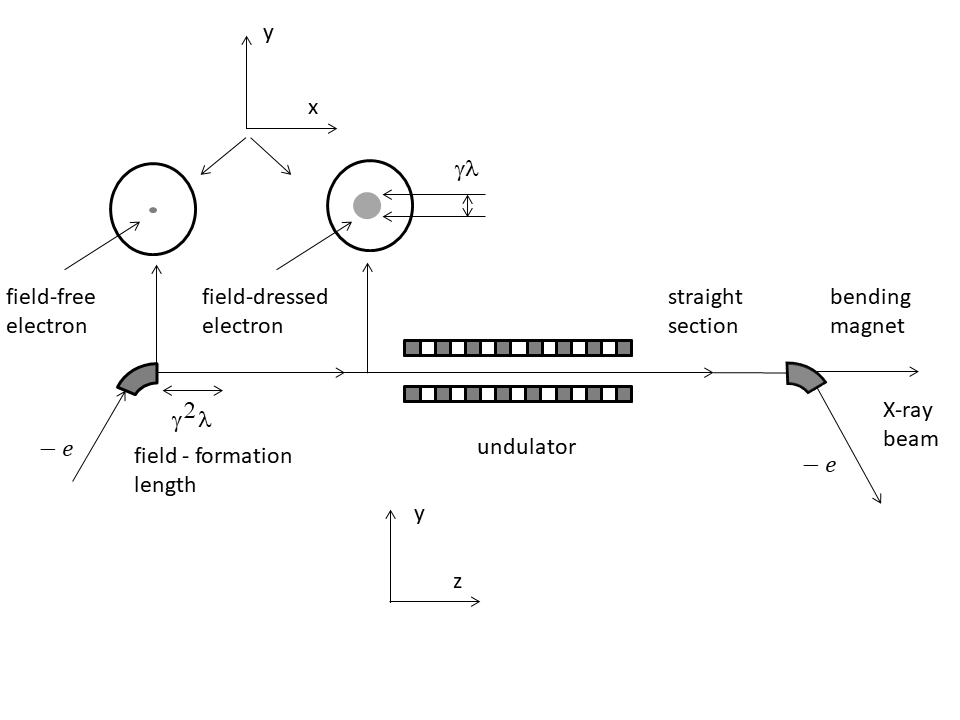}
	\caption{ Standard undulator radiation setup. As the electron traverses a bending magnet, it emits synchrotron radiation, effectively shedding the Frank-Ginzburg fields associated with its relativistic motion. At the exit of the bending magnet, the electron is effectively “naked,” having lost its field dressing. A new field configuration forms within a characteristic *formation length* at the beginning of the straight section downstream. For third-generation synchrotron radiation sources, the transverse size of this field-dressed ultrarelativistic electron is typically on the order of a few microns..}
	\label{B90}
\end{figure}

Let us now return to our examination of the standard synchrotron radiation setup and analyze the radiation process in an insertion device, such as an undulator. The "creation" of the relativistic electron occurs within a distance on the order of $\lambdabar \gamma^2$ from the start of the straight section upstream of the undulator. It is assumed that the length of the straight section, $L$, is much longer than the formation length, $\lambdabar \gamma^2$, which is always the case in the X-ray range. When the switching distance $d_s \lesssim \lambdabar \gamma^2 \ll L$, the specifics of the switcher become irrelevant for describing the radiation from the undulator installed within the straight section (see Fig. \ref{B90}).

Downstream of the switcher, the electron moves uniformly. The fields associated with an electron traveling at a constant velocity exhibit interesting behavior as the speed of the charge approaches that of light. Specifically, in the space-frequency domain, the fields of a relativistic electron become equivalent to those of a beam of electromagnetic radiation. For a rapidly moving electron, the transverse electric and magnetic fields are nearly equal in magnitude and mutually perpendicular. These fields are indistinguishable from the radiation fields of a beam. This virtual radiation beam has a macroscopic transverse size on the order of $\lambdabar \gamma$ (see Appendix A4).

At the exit of the switcher, we have a "naked" (or "field-free") electron, meaning an electron that is not accompanied by any virtual radiation fields. Within a distance on the order of $\lambdabar \gamma^2$ downstream of the switcher, the electron undergoes a process of field formation, becoming a "field-dressed" electron—one that is now accompanied by the fields generated by its fast motion.

The electron's trajectory can be divided into two fundamentally distinct segments: before and after the switcher. When the electron is accelerated upstream of the switcher in the lab frame, information about this acceleration is embedded in the first segment of its covariant trajectory. However, this pre-acceleration history—along with the associated electromagnetic fields of the ultrarelativistic electron—is effectively "washed out" during the switching process. As a result, the electron enters the straight section in a "naked" state, devoid of its previous field configuration.

We begin by describing the field formation process along the straight section downstream of the switcher, using a covariant framework. The first step is to synchronize distant clocks within the lab frame, employing the standard Einstein synchronization procedure. Within this Lorentzian lab frame, we assume the electron follows a rectilinear trajectory at constant velocity $v$; this serves as our initial condition. From here, the evolution of the electromagnetic field can be analyzed using Maxwell's equations.

When one analyzes the process of "field-dressed" electron formation from the viewpoint of the noncovariant approach, one assumes the same initial conditions (rectilinear trajectory with velocity $v$) for the electron motion. Then one solves the electrodynamics problem of field formation by using the usual Maxwell's equations. We already mentioned that the type of clock synchronization that results in the time coordinate $t$ in an electron trajectory $\vec{x}(t)$ is never discussed in accelerator physics.
However, we know that the usual Maxwell's equations are only valid in the Lorentz frame. The noncovariant approach is obviously based on a definite synchronization assumption, but this is actually a hidden assumption. In our case of interest, within the lab frame, the Lorentz coordinates are then automatically enforced. 
So one should not be surprised to find that in this simple case of rectilinear motion, there is no difference between covariant and noncovariant calculations of the initial conditions at the undulator entrance. 

Due to the nature of undulator radiation, the radiation field within the central cone can be accurately calculated using the instantaneous (dipole-like) approximation. This justifies the use of a noncovariant approach to describe the constrained electron motion through the undulator.

In conclusion, for a standard synchrotron radiation setup, it does not matter whether one adopts the covariant approach with Einstein synchronization or the noncovariant approach based on absolute time. Both yield the same results for the radiation field within the central cone.

Let us now examine the behavior of a weak dipole magnet (commonly referred to as a kicker) installed in the straight section upstream of the undulator. This kicker is characterized by a small kick angle such that $(\gamma\theta_k)^2 \ll 1$. What should we expect in terms of undulator radiation under these conditions?

At first glance, the setup appears similar to the switcher configuration: the electron trajectory is again divided into two segments—before and after the kicker. However, there is a crucial difference. In this case, electrodynamics dictates that both segments of the trajectory must be considered when calculating the resulting undulator radiation.

As the electron passes through the kicker, there is no significant synchrotron radiation. More precisely, any emitted radiation is indistinguishable from the electron’s self-fields. As a result, the virtual radiation fields are not washed out as they are in the switcher scenario. The electron emerges from the kicker still "field-dressed," but with a perturbed electromagnetic field that now carries information about the acceleration experienced relative to an inertial frame.

According to conventional theory—rooted in Newtonian kinematics—the Galilean vector law of velocity addition is applied. Within this non-covariant framework, the electron’s direction changes after the kick, but its speed remains constant. In contrast, covariant particle tracking—based on Lorentz coordinates—leads to a different prediction: the electron’s speed is slightly reduced, from $v$ to $v(1 - \theta_k^2/2)$. This discrepancy arises because the covariant approach properly accounts for the relativistic addition of non-parallel velocities.

In standard electrodynamics, the usual algorithm involves solving Maxwell’s equations using particle trajectories derived from non-covariant tracking. According to this method, the undulator radiation emitted along the post-kick velocity direction exhibits no redshift in resonance frequency, regardless of the kick angle $\theta_k$.

However, when the coupling between fields and particles is treated correctly within the covariant framework, a notable prediction emerges. Specifically, synchrotron radiation theory indicates a redshift in the resonance frequency of the undulator radiation in the kicked direction. This redshift is given by: 

\[
\Delta\omega_{r}/\omega_{r} = - \gamma^2\theta_k^2/(1+K^2/2)    . 
\]

This result underscores the importance of using covariant dynamics in accurately modeling radiation phenomena in relativistic beamlines.

\subsection{Helical Trajectories and Synchrotron Radiation }

The observation of a redshift in bending magnet radiation inherently suggests a similar issue within the conventional theory of cyclotron radiation. In the ultrarelativistic regime, well-established analytical expressions describe the spectral and angular distribution of radiation emitted by an electron moving in a uniform magnetic field, where the motion has a non-relativistic component parallel to the field and an ultrarelativistic component perpendicular to it.

According to the conventional framework—just as in the case of bending magnet radiation—the angular and spectral distribution of the emitted radiation depends on the total velocity of the particle, a consequence of the Doppler effect. In contrast, the covariant approach predicts a redshift of the critical frequency that arises specifically when there are perturbations in the electron’s motion along the direction of the magnetic field.

It is important to note that cyclotron-synchrotron radiation is a fundamental emission process in both plasma physics and astrophysics. Therefore, the corrections we propose have significant implications extending well beyond synchrotron radiation facilities.

\subsubsection{Existing Theory}

\begin{figure}
	\centering
	\includegraphics[width=1.\textwidth]{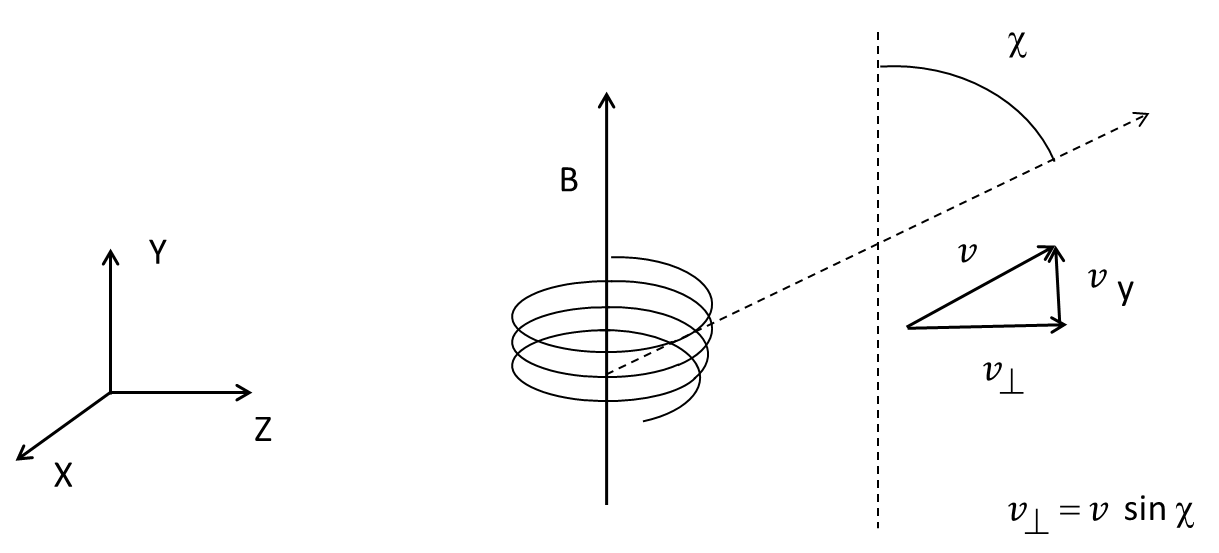}
	\caption{ Geometry for radiation production from helical motion. }
	\label{B92}
\end{figure}

Let us now examine the relativistic cyclotron radiation in greater detail. Here, we present only the final results and discuss their connection to the conventional synchrotron radiation theory associated with bending magnets. In the case of uniform translational motion with non-relativistic velocity along the direction of the magnetic field (see Fig. \ref{B92}), a widely accepted expression in astrophysics describes the angular and spectral distributions of radiation emitted by an ultra-relativistic electron following a helical trajectory\footnote{The angular and spectral distributions of radiation from an ultra-relativistic electron on a helical orbit were derived in \cite{WW, EP}. These results are now standard examples in the literature (see, e.g., \cite{GI}) and are not elaborated upon here.}

\begin{eqnarray}
{\vec{\widetilde{E}}}(\chi, \alpha)  &&  \sim
\Bigg\{ \vec{e}_x
\left[(\xi^2 + \psi^2) K_{2/3} \left(\frac{\omega}{2\omega_c}
\left(1+\frac{\psi^2}{\xi^2}\right)^{3/2}\right)\right]\cr   && - i \vec{e}_y \left[(\xi^2 +
\psi^2)^{1/2} \psi K_{1/3}\left(\frac{\omega}{2\omega_c}
\left(1+\frac{\psi^2}{\xi^2}\right)^{3/2}\right)\right] \Bigg\}~,  \label{EF}
\end{eqnarray}

where $K_{1/3}$ and $K_{2/3}$ are the modified Bessel functions, $\xi = 1/\gamma$, $\psi = \chi - \alpha$, ($\chi$
is the angle between $\vec{v}$ and $\vec{B}$ and $\alpha$ that between $\vec{n}$ and $\vec{B}$);
the angle $\psi$ is the angular distance between the direction of the electron velocity $\vec{v}$ and the direction of observation $\vec{n}$.
Here the $\omega_c$ is defined by $3eB\gamma^2/(2mc)$.

We have already discussed the radiation emitted by an ultrarelativistic electron following a helical trajectory in the previous section. Equation (\ref{srtwoang2bis}) presents the result we derived earlier for synchrotron radiation from a single electron undergoing an angular deflection relative to the nominal orbit. At first glance, Eq. (\ref{EF}) appears different from Eq. (\ref{srtwoang2bis}). However, the two become equivalent upon introducing the small deflection angle $\eta_y = \pi/2 - \chi$ and the observation angle $\theta_y = \pi/2 - \alpha$, under the assumption that the observer lies in the vertical plane tangent to the trajectory (i.e., $\theta_x = \eta_x = 0$). The integrals in Eq. (\ref{srtwoang2bis}) can then be expressed in terms of modified Bessel functions:

\begin{eqnarray}
&&\int_{0}^{\infty}x\sin[(3/2)\alpha(x+x^3/3)]dx = (1/\sqrt{3})K_{2/3}(\alpha)
~ ,\cr &&
\int_{0}^{\infty}\cos[(3/2)\alpha(x+x^3/3)]dx = (1/\sqrt{3})K_{1/3}(\alpha)
 ~ . 
\end{eqnarray}

Then, making the necessary variable changes, the formula reduces to Eq.( \ref{EF}).

\subsubsection{Methodology of Solving Problems Involving Boosts}

The derivation leading to Eq. (\ref{EF}) is rather involved, so it is useful to present an independent and more 
intuitive argument. A particularly simple way to analyze radiation from ultrarelativistic helical motion leverages special relativity and requires minimal calculation.

In the case of uniform translational motion, radiation analysis can often be simplified by identifying a reference frame in which the problem is already solved—such as the frame where the particle undergoes circular motion—and then transforming the result back to the original frame.

The reference system $S'$ in which the electron moves in circular motion can be transformed to a reference system $S$ in which the electron proceeds following a helical trajectory. Eq. ( \ref{EF}) holds, indeed, in the frame $S$ for a particle whose velocity is 

\[
(v_x,v_y,v_z)  = (v_0\sin \chi\sin\phi,v_0\cos\chi, v_0\sin \chi\cos\phi )   . 
\]

The Lorentz transformation, which leads to the value $v_y = v_0\cos\chi$ for the $y$-component of the velocity yields  

\[
(v_x,v_y,v_z)  = ( v'\sin\phi'/\gamma_y,v_y, v'\cos\phi'/\gamma_y )    , 
\]

where $\gamma_y = 1/\sqrt{1-v^2_y/c^2}$, $v'$ is the velocity of the electron in the frame $S'$ and the phase angle $\phi' = \phi$ is invariant. This means that, in order to end up in $S$ with a transverse (to the magnetic field direction)  velocity $v_{\perp} = v_0\sin \chi$, one must start in $S'$ with $v' = \gamma_yv_0\sin \chi$. In the ultrarelativistic approximation 
$\gamma_{\perp}^2 = 1/(1-v^2_{\perp}/c^2) \gg 1$, and one finds the simple result $v_0 = v'$, so that a Lorentz boost with non-relativistic velocity $v_y$ leads to a rotation of the particle velocity $\vec{v}_0$ of the angle $\eta = \pi/2 - \chi \simeq v_y/c \ll 1$ (if angle  $\eta$ is small and $v_0 \simeq c$, we would write  $\gamma_y\sin \chi \simeq 1$).  If one transforms the radiation field for a particle in a circular motion in the system $S'$, one obtains the result that the effect of a boost amounts to a rigid rotation of the angular-spectral distribution of the radiation emitted by the electron moving with velocity $v_0$ on a circle that is, once more,  Eq. ( \ref{EF}) \footnote{A covariant approach to analyzing radiation from helical motion is presented in \cite{OS}. It is generally assumed that 
$\vec{x}(t) = \vec{x}(t)_{cov}$, which is why \cite{OS} makes no distinction between the covariant and non-covariant treatments of electron motion along a helix downstream of the kicker setup.}. 

Note: the subscript "$_{\perp}$" used here to indicate the velocity in the transverse to the magnetic field direction should not be confused with the "$_{\perp}$" referring to an acceleration of the electron in the transverse direction in the proceeding sections.

It follows quite naturally that the covariant analysis of radiation from helical motion, as considered above, is grounded in the use of Lorentz transformations. In other words, within the laboratory frame, Lorentz coordinates are inherently applied. It is assumed that, in this Lorentz lab frame, the electron follows a helical trajectory with velocity $v_0$, which serves as an initial condition. In the ultrarelativistic approximation, applying a Lorentz boost along the direction of the magnetic field with a non-relativistic velocity $v_y$ transforms the motion into a circular trajectory, still with velocity $v_0$. As a result, a single boost along the field direction does not alter the radiation properties.

Let us consider synchrotron radiation from a single electron that has received a transverse kick relative to the nominal orbit in the $(x,z)$-plane. As a result of this kick, the electron experiences an acceleration that induces a velocity component $v_y = v\theta_k$ along the $y$-axis and a corresponding change $\Delta v_z = -v\theta_k^2/2$ along the $z$-axis. Limiting the analysis to second-order terms simplifies the calculations significantly.

To describe the motion of the particle downstream of the kicker, we can employ a sequence of two commuting, non-collinear Lorentz boosts.

When the kick is applied, covariant particle tracking predicts a modification to the initial conditions at the entrance of the synchrotron radiation setup. If we retain the Lorentz coordinate system of the lab frame downstream of the kicker, we find that the covariant velocity of the particle on the resulting helical orbit decreases from $v$ to $v_0 = v - v\theta_k^2/2$. Applying a covariant analysis of radiation for a helical trajectory with this reduced velocity leads to a redshift in the critical wavelength.

It is important to note that the single passive Lorentz boost to the reference frame $S'$, discussed above, is merely a mathematical device. One can interpret this transformation as a change of variables, rather than a physical alteration of the system. Under this kinematic transformation, neither the electron’s nor the observer’s motion relative to the fixed stars is affected. True (i.e., physically real) acceleration is a dynamic process and cannot be captured by coordinate transformations alone.

Therefore, we conclude that accelerating an electron with respect to the fixed stars in the lab inertial frame—before it enters the uniform magnetic field—is an absolute acceleration. The effects of this acceleration are intrinsically encoded in the covariant particle trajectory.

\subsubsection{On the Advanced "Paradox" Related to the Coupling Fields and Particles}

We now wish to highlight that two distinct sets of initial conditions can result in the same uniform translation along the magnetic field direction in the Lorentz lab frame.

To illustrate this, we first consider an electron moving along a circular trajectory in the $(x,z)$-plane. Next, we rotate the magnetic field vector $\vec{B}$ within the $(y,z)$-plane by a small angle $\theta_0$, under the assumption that $(\theta_0 \gamma)^2 \ll 1$. In this configuration, the electron moves uniformly along the direction of the magnetic field with a velocity component $v\theta_0$. Importantly, this rotation of the magnetic field does not affect the radiation characteristics of the electron in circular motion. This invariance is intuitive: after rotating the bending magnet, the electron retains its velocity, and due to the Doppler effect, it continues to emit radiation along the direction of motion. The change in the curvature radius $R$ due to the rotation is of order $\theta_0^2$, and thus can be neglected.

We now consider a second scenario. Suppose a kicker is installed in the straight section upstream of the bending magnet, imparting a kick in the $y$-direction with angle $\theta_k = \theta_0$. When such a kick is applied, a redshift in the critical wavelength of the emitted radiation occurs. This is a consequence of the relativistic velocity addition law: the longitudinal component of the electron's velocity is reduced from $v$ to $v - v\theta_0^2/2$ after the kick. The relative redshift of the critical frequency $\omega_c$ is given by: 

\[
\Delta\omega_c/\omega_c = - (3/2)\gamma^2\theta_0^2    . 
\]

Here we observe a second-order correction in $\theta_0$, but it is enhanced by the large factor $\gamma^2$.

In the covariant framework, the outcome depends on the absolute magnitude of the kick angle $\theta_0$. Radiation emitted along the velocity direction experiences a redshift only when the kick angle is nonzero. This result reflects the concept of "absolute" acceleration—acceleration relative to the fixed stars.

The distinction between these two situations—both ultimately resulting in uniform translation along the direction of the magnetic field—is quite intriguing. This difference arises due to the presence of two distinct Lorentz coordinate systems within the lab frame. When we accelerate the electron upstream of the bending magnet, we effectively change its associated Lorentz frame.

To maintain a consistent Lorentz coordinate system in the lab frame after the electron receives a transverse kick, a clock resynchronization must be performed. Consequently, we should expect the electron’s velocity to be modified. This explains the difference between the two experimental setups: when the electron's motion upstream of the bending magnet remains unperturbed (relative to the fixed stars), no clock resynchronization is necessary. However, when we do perturb the electron’s motion, resynchronization is required.

\begin{figure}
	\centering
	\includegraphics[width=0.9\textwidth]{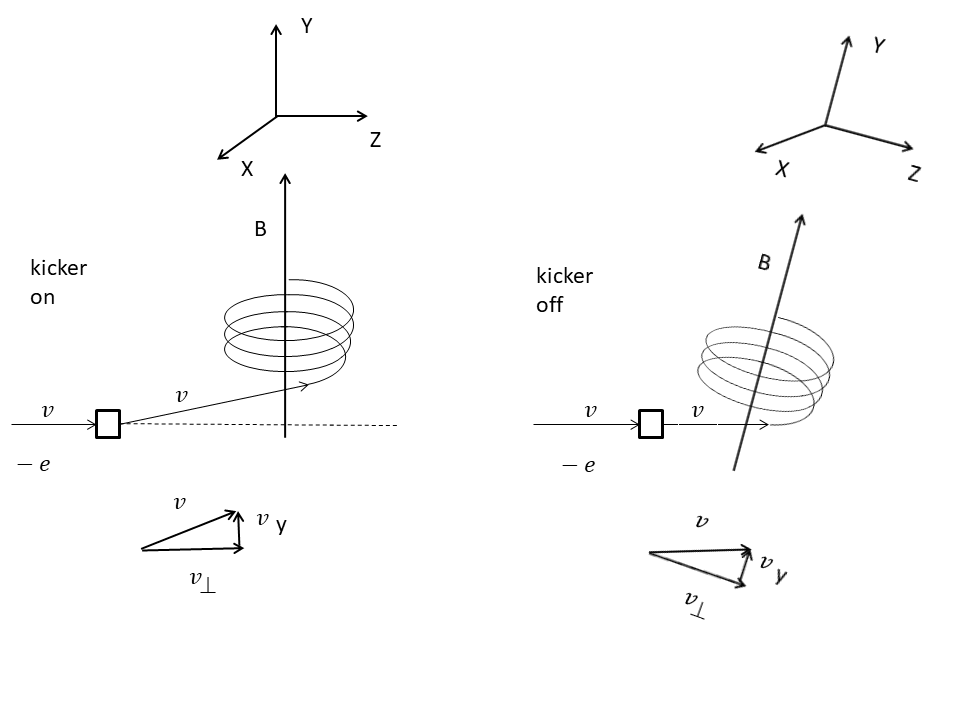}
	\caption{Two sets of initial conditions result in the same uniform motion along the magnetic field direction when using absolute time coordinatization in the lab frame. In both setups, the magnitude of the electron’s velocity and the orientation of the velocity vector relative to the magnetic field vector are identical.}
	\label{B95}
\end{figure}

This leads us to consider an apparent paradox. The argument is as follows: under absolute time coordinatization in the lab frame, the initial conditions at the entrance to the bending magnet appear to be identical in both cases. Specifically, the electron’s velocity magnitude and its orientation with respect to the magnetic field are the same. Yet, when we accelerate the electron upstream of the magnet, this information is seemingly absent from its noncovariant trajectory.

So, where is the information about this acceleration encoded? Since the electron is traditionally considered a structureless particle, the situation seems paradoxical.

However, this paradox hinges on a subtle but critical point: an ultrarelativistic electron is not truly structureless in practice. Consider an electron moving uniformly at a constant velocity. As its speed approaches the speed of light, the electromagnetic fields it generates begin to resemble those of a radiation beam. This virtual radiation beam has a macroscopic transverse extent of order $\lambdabar\gamma$, and its field distribution is described by the Ginzburg-Frank formula (see Appendix A4).

When the electron is subjected to a transverse kick—such as by a kicker magnet—its self-fields are perturbed, now reflecting information about the acceleration. Under the traditional (Galilean) kinematics, the orientation of the virtual radiation phase front remains unchanged. However, Maxwell’s equations are not invariant under Galilean transformations. As discussed throughout this book, adopting Galilean kinematics leads to anisotropic field equations.

As a result, although the phase front of the virtual beam remains planar, the direction of propagation is no longer perpendicular to the phase front. That is, the motion of the virtual radiation beam and the normal to its phase front diverge. Therefore, within the absolute time synchronization convention, electrodynamics predicts that the virtual radiation beam propagates in the kicked direction with a phase front tilt $\theta_k$.

In this way, the information about the electron’s acceleration is embedded in the perturbation of its self-electromagnetic field.

\newpage

\section{Relativity and X-Ray Free Electron Lasers}

\subsection{Introductory Remarks}

In the previous chapter, we examined why the error in radiation theory remained undetected for so long. Within the covariant framework, relativistic kinematic effects in synchrotron radiation emerge in successive orders of approximation. Rather than using the total velocity parameter  $(v/c)$ as in non-relativistic cases, we employ the small transverse velocity parameter  $(v_{\perp}/c)$. 

In our earlier discussion of bending magnet radiation, we found that the motion of a single ultrarelativistic electron in a constant magnetic field, according to relativistic theory, affects kinematic terms only at the second order, i.e., 
 $(v_{\perp}/c)^2$  It has been shown that, due to a combination of the ultrarelativistic (or paraxial) approximation and the specific symmetry of conventional synchrotron radiation setups, second-order relativistic kinematic effects cancel out—except for a non-zero redshift in the critical frequency. This redshift arises when the electron experiences perturbations in the direction of the bending magnetic field. However, because synchrotron radiation from bending magnets spans a broad frequency range, the output intensity is largely insensitive to this redshift. As a result, spontaneous synchrotron radiation shows no observable sensitivity to the differences between covariant and non-covariant particle trajectories.
 
 This situation changes in the 21st century with the advent of X-ray Free Electron Lasers (XFELs). In XFELs, first-order kinematic effects $(v_{\perp}/c)$  play a crucial role in the description of radiation. In such cases, the covariant coupling between fields and particles predicts outcomes that sharply contrast with those of conventional treatments.
 
 In this chapter, we critically reexamine existing XFEL theory, with particular focus on coherent undulator radiation from modulated electron beams. The discussion is primarily intended for readers with limited background in accelerator and XFEL physics.

\begin{figure}
	\centering
	\includegraphics[width=0.9\textwidth]{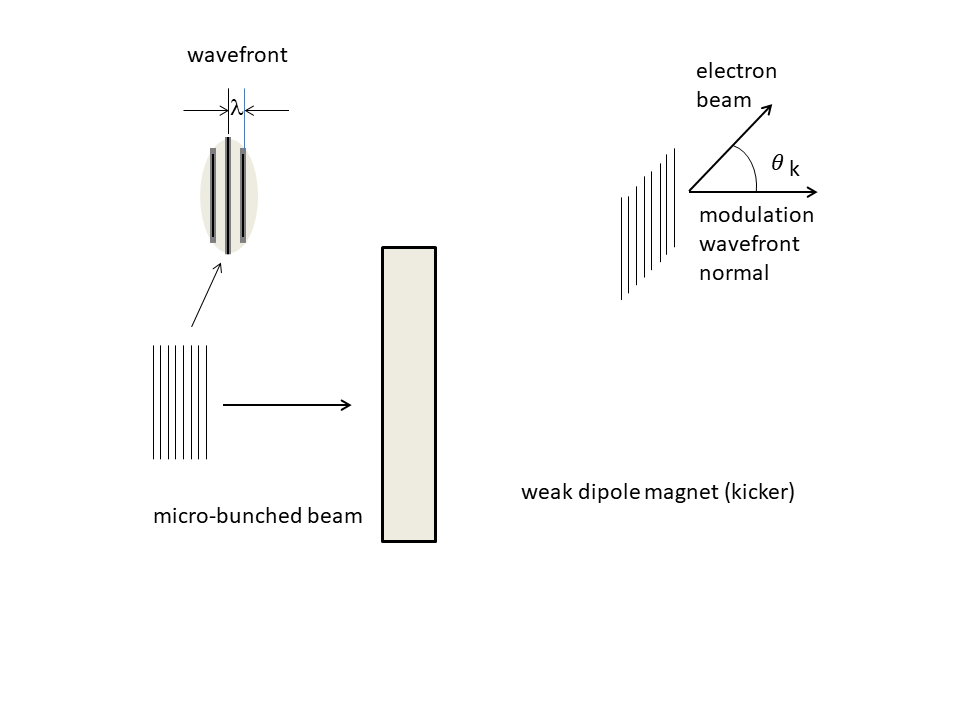}
	\caption{Illustration of a well-known outcome in conventional (non-covariant) particle tracking: a micro-bunched electron beam passes through a weak dipole magnet (kicker), receiving a kick by an angle $\theta_k$. The beam’s propagation axis is deflected, while the orientation of the wavefront remains unchanged. }
	\label{B79}
\end{figure}

The conventional theory of X-ray Free Electron Lasers (XFELs) typically relies on a hybrid approach: Newtonian kinematics—corrected for relativistic mass—is applied to particle dynamics, while Einsteinian principles underpin electrodynamics. In practice, the relativistic treatment of particle motion often reduces to a modified form of Newton’s second law, with velocity composition still grounded in Galilean transformations.

For rectilinear motion of a modulated electron beam, both covariant and non-covariant approaches yield identical particle trajectories. Consequently, Maxwell’s equations remain consistent with conventional particle tracking in such cases. However, the relativity of simultaneity—which involves the intermixing of spatial and temporal coordinates—leads to fundamental differences between covariant and non-covariant descriptions when the beam follows a curved path. According to the theory of relativity, these discrepancies emerge specifically in the presence of acceleration along curved trajectories.

First-order relativistic effects in XFELs can arise under various conditions, particularly when the electron beam experiences a trajectory kick\footnote{Angular kicks are commonly used in XFEL diagnostics and experiments. For instance, standard gain length measurements rely on such kicks. They are also employed in beam-splitting techniques, where polarization components are separated through angular deflection of the modulated beam \cite{L, NUHN}.}.

Among these, one of the most illustrative effects involves the generation of coherent undulator radiation by an ultrarelativistic, modulated electron beam that has been deflected by a weak dipole field prior to entering a downstream undulator. This scenario serves as a critical test for the correct coupling between electromagnetic fields and particle dynamics.

To set the stage, it is helpful to begin with a broad overview of key results. Let us now consider the predictions of the standard XFEL theory in the context of non-collinear electron beam motion. A well-established result from conventional particle tracking states that, following a weak transverse kick, the trajectory of the electron beam changes direction while the modulation wavefront retains its original orientation (see Fig. \ref{B79}). This results in a misalignment between the electron beam direction and the normal to the modulation wavefront—commonly referred to as a wavefront tilt.

In the conventional framework, this wavefront tilt is treated as a physical phenomenon. That is, a transverse kick does not alter the orientation of the modulation wavefront, thereby reducing the efficiency of radiation in the direction of the beam motion  \cite{TKS}.

\begin{figure}
	\centering
	\includegraphics[width=0.9\textwidth]{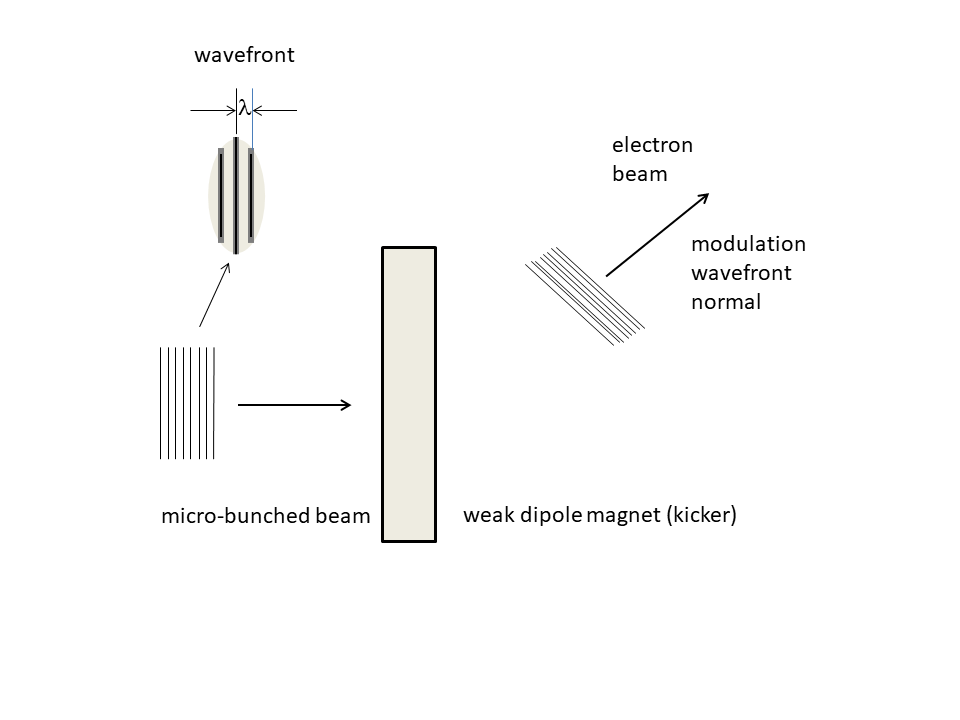}
	\caption{ Result of covariant particle tracking. In the ultrarelativistic limit, the modulation wavefront—or equivalently, the plane of simultaneity—is always oriented perpendicular to the electron beam's velocity when the evolution of the modulated beam is described using Lorentz coordinates.
		According to the theory of relativity, a modulated electron beam in this limit exhibits the same kinematic behavior, in Lorentz coordinates, as a laser beam. Maxwell's equations dictate that the wavefront of a laser beam is always orthogonal to its direction of propagation.}
	\label{B80}
\end{figure}

The covariant approach, applied within the frameworks of both mechanics and electrodynamics, predicts an effect that stands in stark contrast to conventional theory. Specifically, in the ultrarelativistic limit, the modulation wavefront—defined as a plane of simultaneity—is always perpendicular to the velocity of the electron beam (see Fig. \ref{B80}). Consequently, Maxwell's equations predict strong coherent undulator radiation emitted in the direction of the kick imparted to the modulated electron beam.

Experimental results confirm this prediction. Observations from XFEL facilities have demonstrated that even the direction of coherent undulator radiation emission lies outside the predictive scope of the conventional theory.

It is worth noting that the lack of a dynamical explanation for the readjustment of the modulation wavefront under Lorentz coordinatization has been a point of concern for some XFEL experts. However, we suggest that a useful way to conceptualize this readjustment is to view it as a consequence of transforming to a new time variable within the Galilean formulation of electrodynamics.

\subsection{Modulation Wavefront Orientation}

Let us consider a modulated electron beam propagating along the $z$-axis of a Cartesian coordinate system 
$(x,y,z)$ in the laboratory frame. For example, assume that the modulation wavefront is perpendicular to the beam velocity $v_z$. How can we determine this orientation? 

As the electron bunch moves, its position changes over time. A natural way to characterize the orientation of the modulation wavefront is to ask: at what time does each electron cross the $y$-axis of the reference frame?

If a synchronization convention has been adopted—i.e., a method for timing distant events—we can use it to define the orientation of the modulation wavefront. Specifically, if electrons corresponding to the region of maximum density cross the $y$-axis simultaneously at a given position $z$, then the wavefront is perpendicular to the $z$-axis. In other words, the modulation wavefront can be identified as a plane of simultaneous events—where the events are the arrivals of electrons at the point of maximum density. In short, it is a plane of simultaneity.

We now examine the case where the electron beam is accelerated in the lab frame to acquire a small transverse velocity 
component $v_y$ along the $y$-axis.
This raises an important question: how should synchronization be defined in the lab frame after this acceleration?
Prior to the kicker setup, we chose a Lorentz coordinate system for the lab frame.  
After the beam acquires a small transverse velocity  $v_y$,  maintaining the original synchronization leads to complications in the electrodynamics of moving charges. 
As a result of such a boost, the transformation of time and spatial coordinates has the form of a Galilean transformation.

To maintain a Lorentz coordinate system in the lab frame following the acceleration, it is necessary to perform a clock resynchronization. Specifically, this involves a time shift of the form given in Eq.~\ref{GGT3}: 
$t \to t + yv_y/c^2$. This adjustment is valid in the first-order approximation, where $v_y/c$ is so small that $v_y^2/c^2$ can be neglected and one arrives at the coordinate transformation $y \to y + v_yt$, $t \to t + yv_y/c^2$. 
This differs from a pure Galilean transformation by the inclusion of the relativity of simultaneity—the only first-order relativistic correction in $v_y/c$.

This resynchronization introduces a time shift for electrons located at different transverse positions. For instance, electrons at a transverse position $y$ with maximum density will cross the lab frame’s $y$-axis at a later time compared to electrons at  $y =0$. The resulting time shift is given by $\Delta t = yv_y/c^2$.
This time shift results in an effective rotation of the modulation wavefront by an angle: $v_z\Delta t/y = v_zv_y/c^2$ in the first order approximation. 
In ultrarelativistic limits, $v_z \simeq c$, and the modulation wavefront rotates exactly as the velocity vector $\vec{v}$.

What does this readjustment of the wavefront mean in terms of measurements?
In the framework of absolute time coordination, simultaneity between two events is considered absolute. This absolute nature of temporal coincidence stems from the convention of absolute time synchronization. In this classical kinematics view, the modulation wavefront remains unchanged.
However, within the covariant approach, we adopt a criterion for simultaneity based on the invariance of the speed of light. It becomes evident that, due to the motion of electrons along the $y$-axis (i.e., along the plane of simultaneity prior to the boost), with a velocity  $v_y$, the simultaneity of events is no longer absolute. Instead, it becomes dependent on the kick angle $\theta = v_y/c$. This reasoning mirrors the principles of Einstein’s train-embankment thought experiment.

The orientation of the wavefront lacks an exact objective meaning due to the relativity of simultaneity. However, the statement that the wavefront orientation has an objective meaning within a certain accuracy can be illustrated by considering the wavefront in its proper orientation, with the angle’s uncertainty (or “blurring”) given by $\Delta \theta \simeq v_z(v_y/c^2)$. This relationship defines the limits within which nonrelativistic theory remains applicable.

For a very nonrelativistic electron beam, where $v_z^2/c^2$  is negligible, the angle of "blurring" becomes extremely small. In this case, the wavefront tilt angle $\theta = v_y/v_z$ is almost perfectly sharp, with $\Delta \theta/\theta \simeq v_z^2/c^2 \ll 1$. This represents the limiting case of nonrelativistic kinematics. The angle "blurring" is a characteristic feature of relativistic beam motion.

In the ultrarelativistic limit the wavefront tilt has no exact objective meaning. This is because, due to the finite speed of light, no experimental method exists by which we could measure this tilt with certainty.

\subsection{XFEL Radiation Configuration}

One of the most fundamental effects that serve as a crucial test for the proper coupling of fields and particles is the production of coherent undulator radiation by a modulated ultrarelativistic electron beam, which is deflected by a weak dipole field before entering a downstream undulator. Our goal is to study the emission of coherent undulator radiation from such a system.

At the heart of an XFEL source is the undulator, which compels electrons to follow curved, periodic trajectories. There are two primary undulator configurations: helical and planar. To understand the basic principles of undulator operation, let us begin by examining the helical undulator.

The magnetic field along the axis of the helical undulator is given by

\begin{eqnarray}
&& \vec{B}_w = \vec{e}_xB_w\cos(k_wz) -\vec{e}_y B_w\sin(k_wz)  ~ ,
\end{eqnarray}

where $k_w = 2\pi/\lambda_w$ is the undulator wavenumber and $\vec{e}_{x,y}$ are unit vectors directed along the $x$ and $y$ axes. We neglected the transverse variation of the magnetic field. It is necessary to mention that in XFEL engineering we deal with very high-quality undulator systems, which have a sufficiently wide good-field-region,  so that our studies, which refer to a simple model of undulator field nevertheless yield a correct quantitative description in a large variety of practical problems.
The Lorentz force $\vec{F} = -e\vec{v}\times \vec{B}_w/c$ is used to derive the equation of motion      
of electrons with charge $-e$ and mass $m$ in the presence of magnetic field

\begin{eqnarray}
&& m\gamma\frac{dv_x}{dt} = \frac{e}{c}v_z B_y = -\frac{e}{c}v_z B_w \sin(k_wz) ~ ,
\cr &&
m\gamma\frac{dv_y}{dt} = - \frac{e}{c}v_z B_x = -\frac{e}{c}v_z B_w \cos(k_wz) ~ .
\end{eqnarray}

Introducing $\tilde{v} = v_x + iv_y$, $dz = v_zdt$ we obtain  

\begin{eqnarray}
&& m\gamma\frac{d\tilde{v}}{dz} = -i\frac{e}{c}(B_x + iB_y) 
= -i\frac{e}{c}B_w\exp(-ik_wz) ~ .
\end{eqnarray}

Integration of the latter equation gives

\begin{eqnarray}
&& \frac{\tilde{v}}{c} = \theta_w\exp(-k_wz) ~ ,
\end{eqnarray}

where $\theta_w = K/\gamma$ and $K = eB_w/(k_w mc^2)$ is the undulator parameter.
The explicit expression for the electron velocity in the field of the helical undulator has the form 

\begin{eqnarray}
&& \vec{v}_{\perp} = c\theta_w[\vec{e}_x \cos(k_wz) - \vec{e}_y\sin(k_wz)]  ~ ,
\end{eqnarray}

This means that the reference electron in the undulator moves along the constrained helical trajectory parallel to the $z$ axis. As a rule, the electron rotation angle $\theta_w$ is small and the longitudinal electron velocity $v_z$ is close to the velocity of light, 

\[
v_z = \sqrt{v^2 - v^2_{\perp}} \simeq v(1-\theta_w^2/2) \simeq c  .
\]

Let us consider a modulated ultrarelativistic electron beam moving along the $z$-axis within the field of a helical undulator. In this study, we make the following assumptions: First, in the absence of any deflection, the electrons follow constrained helical trajectories that are parallel to the $z$-axis. Second, the electron beam density at the entrance of the undulator is given by

\begin{eqnarray}
&& n = n_0(\vec{r}_{\perp})[1 + a\cos \omega(z/v_z - t)]  ~ ,
\label{BM}
\end{eqnarray}  
 
where $a = \mathrm{const.}$ In other words we consider the case in which there are no variations in amplitude and phase of the density modulation in the transverse plane. Under these assumptions, the transverse current density may be written in the form 

\begin{eqnarray}
&& \vec{j}_{\perp} = -e\vec{v}_{\perp}(z)n_0(\vec{r}_{\perp})[1+a\cos \omega(z/v_z -t)]  ~ .
\end{eqnarray}

Even though the measured quantities are real, it is generally more convenient to use complex representation, starting with real $\vec{j}_{\perp}$, one defines the complex transverse current density: 

\begin{eqnarray}
&&j_x+ij_y = -ec\theta_wn_0(\vec{r}_{\perp})
\exp(-ik_wz)[1+a\cos \omega(z/v_z -t)]  ~ .
\end{eqnarray}

The transverse current density has an angular frequency $\omega$ and two waves traveling in the same direction with variations 

\[
\exp i(\omega z/v_z - k_wz -\omega t) 
\]

and  

\[
\exp - i(\omega z/v_z + k_wz -\omega t) 
\]

will add to give a total current proportional to 

\[
\exp(-ik_wz)[1+a\cos \omega(z/v_z -t)]. 
\]

The factor  

\[
\exp  i(\omega z/v_z - k_wz -\omega t)  
\]

indicates a fast wave, while the factor  

\[
\exp - i(\omega z/v_z + k_wz -\omega t)
\]

indicates a slow wave. The use of the word "fast"  ("slow") here implies a wave with a phase velocity faster (slower) than the beam velocity.

Having defined the sources, we now turn to the electrodynamics problem. Maxwell's equations can be manipulated in various ways to derive forms that are more suitable for specific applications. For instance, from Maxwell's equations (Eq. \ref{CD11}), we can derive an equation that depends solely on the electric field vector $\vec{E}$ (in Gaussian units):

\begin{equation}
c^2 \vec{\nabla}\times(\vec{\nabla}\times{\vec{E}}) = - \partial^2
	\vec{E}/\partial t^2 - 4 \pi \partial \vec{j}/\partial
		t ~. \label{elec}
\end{equation}
With the help of the identity

\begin{equation}
\vec{\nabla}\times(\vec{\nabla}\times{\vec{E}}) =
\vec{\nabla}(\vec{\nabla}\cdot{\vec{E}})-\nabla^2 \vec{E}\label{iden}
\end{equation}
and Poisson equation
\begin{equation}
\vec{\nabla}\cdot\vec{E} = 4 \pi \rho \label{pois}
\end{equation}
we obtain the inhomogeneous wave equation for $\vec{E}$

\begin{eqnarray}
&& c^2 \nabla^2 \vec{E} - \partial^2 \vec{E}/\partial t^2 = 4
\pi c^2 \vec{\nabla} \rho + 4 \pi \partial \vec{j}/\partial t ~ . 
\end{eqnarray}

Once the charge and current densities  $\rho$ and $\vec{j}$ are specified as a function of time and position, this equation allows one to calculate the electric field $\vec{E}$ at each point of space and time. This nonhomogeneous wave equation thus serves as the complete and accurate formulation for describing radiation. However, we aim to apply it to a simplified scenario in which the second term on the right-hand side—associated with the current density—dominates the contribution to the radiation field.

It is important to recall that our focus is on coherent undulator radiation, which exhibits a divergence much smaller than the angle  $1/\gamma$. Under this condition, it can be shown that the gradient term, $4\pi c^2 \vec{\nabla} \rho$,
on the right-hand side of the wave equation becomes negligible. As a result, we simplify the wave equation to the following form:

\begin{eqnarray}
&& c^2 \nabla^2 \vec{E} - \partial^2 \vec{E}/\partial t^2 =   4 \pi \partial \vec{j}_{\perp}/\partial t  ~ . 
\end{eqnarray}

We now consider the case in which the phase velocity of the current wave is close to the speed of light. This condition can be satisfied under the resonance condition 

\[
\omega/c = \omega/v_z - k_w    . 
\]

This is the condition for synchronism between the transverse electromagnetic wave and the fast transverse current wave with the propagation constant $\omega/v_z - k_w$. When the phase velocity of the current wave matches that of the electromagnetic wave, a spatial resonance can occur between the electromagnetic field and the electrons. This resonance enables a cumulative interaction between the modulated electron beam and the transverse electromagnetic wave, even in free space.

Under this synchronism condition—and provided the undulator has a sufficiently large number of periods—the contributions of all other (non-synchronous) waves become negligible. Thus, it is justified to focus solely on the resonant wave in analyzing the beam–field interaction.

Here follows an explanation of the resonance condition which is elementary in the sense that we can see what is happening physically. 
The field of electromagnetic wave has only transverse components, so the energy exchange between the electron and electromagnetic wave is due to a transverse component of the electron velocity. For effective energy exchange between the electron and the wave, the scalar product $-e\vec{v}_{\perp}\cdot\vec{E}$ should be kept nearly constant along the whole undulator length. We see that required synchronism  

\[
k_w + \omega/c - \omega/v_z = 0 
\]

takes place when the wave advances the electron beam by the wavelength at one undulator period 

\[
\lambda_w/v_z = \lambda/(c -v_z), 
\]

where $\lambda = 2\pi/\omega$
is the radiation wavelength. This tells us that the angle between the transverse velocity of the particle $\vec{v}_{\perp}$ and the vector of the electric field $\vec{E}$ remains nearly constant.
Since $v_z \simeq c$ this resonance condition may be written as 

\[
\lambda \simeq \lambda_w/(2\gamma_z^2) = \lambda_w (1 + K^2)/(2\gamma^2).     
\]

We will employ an adiabatic approximation, which is applicable in all practical XFEL scenarios where the modulation wavelength is much shorter than the electron bunch length $\sigma_b$, i.e. $\sigma_b\omega/c \gg 1$. Since our focus is on coherent emission near the modulation wavelength, the theory of coherent undulator radiation is most naturally formulated in the space-frequency domain. This is because, in such cases, one is typically interested in the radiation properties at a fixed modulation frequency.

We first apply a temporal Fourier transformation  to the inhomogeneous wave equation to obtain the inhomogeneous  Helmholtz equation

\begin{eqnarray}
&& c^2 \nabla^2 \vec{\bar{E}} + \omega^2 \vec{\bar{E}} =  - 4 \pi i \omega \vec{\bar{j}}_{\perp} ~ ,
\end{eqnarray}

where 
$\vec{\bar{j}}_{\perp}(\vec{r},\omega)$ is the Fourier transform of the current density
$\vec{j}_{\perp}(\vec{r},t)$.
The solution  can be represented as a weighted superposition of solutions corresponding to a unit point source located at $\vec{r}'$. The Green function for the inhomogeneous Helmholtz equation is given by (for unbounded space and outgoing waves)

\begin{eqnarray}
&& 4\pi G(\vec{r}, \vec{r'}, \omega) =   \frac{1}{|\vec{r} - \vec{r'}|}\exp\left[i\frac{\omega}{c}
|\vec{r} - \vec{r'}|\right] ~ ,
\end{eqnarray}

with 

\[
|\vec{r} - \vec{r'}| = \sqrt{(x' - x )^2 + (y' - y)^2 + (z' - z)}^2   .
\]

With the help of this Green function we can write a formal solution for
the field equation as: 

\begin{eqnarray}
&& \vec{\bar{E}} = \int d\vec{r'} ~G(\vec{r}, \vec{r'}) \left[- 4 \pi i \frac{\omega}{c^2} \vec{\bar{j}_{\perp}}\right]  ~ .
\end{eqnarray}

This is simply a mathematical description of the concept of Huygens' secondary sources and wave propagation, which is, of course, well-known. However, it is worth recalling how this directly follows from Maxwell's equations. We can consider the amplitude of the radiation emitted by the plane of oscillating electrons as the resultant of radiated spherical wavelets. This is because Maxwell's theory exhibits no intrinsic anisotropy. The electrons situated on the plane of simultaneity each generate spherical wavelets, which, according to Huygens' principle, combine to form the resulting radiated wave. If the plane of simultaneity is the $xy$-plane (i.e., the beam modulation wavefront is perpendicular to the $z$-axis), Huygens' construction reveals that plane wavefronts will be emitted along the $z$-axis.

In summary, according to Maxwell's electrodynamics, coherent radiation is always emitted in the direction normal to the modulation wavefront. We have already emphasized that Maxwell's equations are valid only within a Lorentz reference frame, i.e., in an inertial frame where the Einstein synchronization procedure is applied to assign values to the time coordinates. It is crucial to apply Einstein's time order consistently, both in dynamics and electrodynamics. Our previous discussion naturally leads to the conclusion that Maxwell's equations in the lab frame are compatible only with covariant trajectories, $\vec{x}_{cov}(t)$, which are calculated using Lorentz coordinates and, therefore, include relativistic kinematic effects.

Let us revisit the modulated electron beam that was kicked transversely with respect to its direction of motion, as discussed earlier. Conventional particle tracking indicates that, while the direction of the electron beam changes after the kick, the orientation of the modulation wavefront remains unchanged. In other words, the direction of the electron's motion and the normal to the wavefront are not aligned. Therefore, according to the conventional coupling of fields and particles, which we consider incorrect, the coherent undulator radiation produced in the kicked direction downstream in the undulator is expected to be significantly suppressed once the kick angle exceeds the divergence of the output coherent radiation.

To estimate the loss in radiation efficiency in the kicked direction using the conventional coupling of fields and particles, we assume that the spatial profile of the modulation closely follows the electron beam’s transverse distribution, modeled as a Gaussian with standard deviation $\sigma$.

A modulated electron beam in an undulator can be viewed as a sequence of periodically spaced radiating oscillators. These oscillators emit radiation that interferes constructively in the forward direction  ($\theta = 0$) when they are all in phase, leading to strong on-axis emission.

To understand the angular distribution, consider a triangle formed by a radiation path at a small angle $\theta$, 
with altitude $r \simeq \theta z $ and base $z$. The diagonal path length $s$ exceeds the base by an amount $\Delta = s - z\simeq z\theta^2/2$. When $\Delta$  equals one radiation wavelength, destructive interference occurs, as the phase contributions from different oscillators become uniformly distributed over $0$ to $2\pi$.

In the limit of a small electron beam size ($\sigma \to 0 $), constructive interference occurs within an angle

\[
\Delta\theta \backsimeq \sqrt{c/(\omega L_w)} = 1/(\sqrt{4\pi N_w}\gamma_z) \ll 1/\gamma    , 
\]

where $L_w = \lambda_w N_w$ is the undulator length. In the limit for the large size of the electron beam,  the angle of coherence is about $\Delta\theta \backsimeq c/(\omega\sigma)$ instead. The boundary between these two asymptotes is for sizes of about $\sigma_\mathrm{dif} \backsimeq \sqrt{cL_w/\omega}$. The parameter $\omega\sigma^2/(cL_w)$ can be referred to as the electron beam Fresnel number. For XFELs, the transverse beam size  $\sigma$ typically exceeds $\sigma_\mathrm{dif}$, 
indicating a large Fresnel number. Consequently, the angular distribution of radiation in the far zone is approximately Gaussian with standard deviation  $\sigma_\mathrm{\theta} \backsimeq c/(\sqrt{2}\omega\sigma)$. 

However, under the conventional framework, a discrepancy arises after the beam is kicked: the direction of electron motion no longer aligns with the modulation wavefront. As a result, the radiation intensity in the kicked direction is suppressed and can be approximated as $I \backsimeq I_0  \exp[- \theta_k^2/(2\sigma_\mathrm{\theta}^2)]$, where $I_0$ is the on-axis intensity without kick and $\theta_k$ is the kick angle. This exponential suppression reflects the misalignment between the wavefront and the new direction of electron motion.

We presented a study of the very idealized situation to illustrate the difference between the conventional and covariant coupling of fields and particles. We solved the dynamics problem of the motion of relativistic electrons in the prescribed force field of a weak kicker magnet by working only up to the order of $\gamma\theta_k$. This approximation is of particular theoretical interest because it is relatively simple and at the same time forms the basis for understanding relativistic kinematic effects such as relativity of simultaneity.

Let us discuss the region of validity of our small kick angle approximation $\theta_k\gamma \ll 1$. Since in XFELs, the Fresnel number is rather large,  we  can always consider a kick angle that is relatively large compared to the divergence of the output coherent radiation, and, at the same time, it is relatively small compared to the angle $1/\gamma$. In fact, from $\omega\sigma^2/(cL_w) \gg 1$, with some rearranging, we obtain 

\[
\sigma_\theta^2 \simeq c^2/(\omega^2\sigma^2)\ll c/(\omega L_w)    . 
\]

Then we recall that  

\[
\sqrt{c/(\omega L_w)} = 1/(\sqrt{4\pi N_w}\gamma_z) \ll 1/\gamma   .
\]

Therefore, the first-order approximation used to analyze the kicker setup in this chapter is of practical significance for XFEL engineering.

One of the goals of this chapter is to demonstrate the experimental predictions that we expect from our corrected radiation theory. To illustrate the essential physical principles clearly, we worked out a simple case. Surprisingly, the first-order approximation used to analyze the kicker setup in this chapter also has important practical applications.

\begin{figure}
	\centering
	\includegraphics[width=0.9\textwidth]{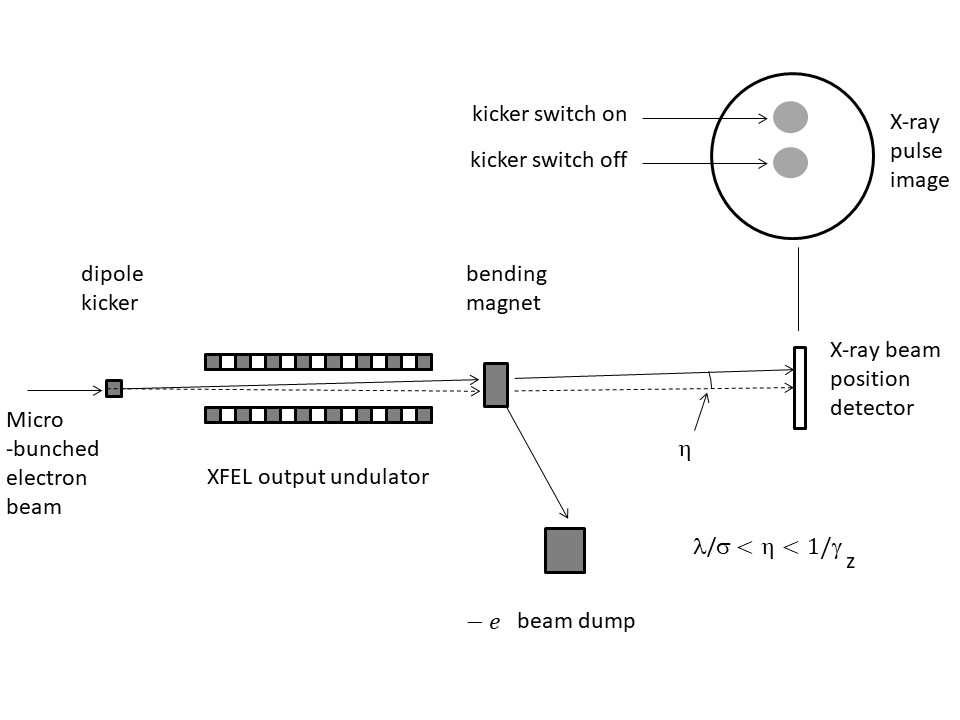}
	\caption{Experimental setup for testing XFEL theory. According to the covariant approach, in the ultrarelativistic limit, the modulation wavefront is always perpendicular to the velocity of the electron beam. As a result, Maxwell's equations predict a strong emission of coherent undulator radiation in the direction of the applied kick. Experimental results confirm that this prediction holds true.}
	\label{B91}
\end{figure}

As shown above, our corrected coupling of fields and particles predicts an effect that is in complete contrast to the conventional treatment. Specifically, in the ultrarelativistic limit, the plane of simultaneity—the wavefront orientation of the modulation—is always perpendicular to the electron beam velocity. As a result, we predict strong emission of coherent undulator radiation from the modulated electron beam in the direction of the kick, as shown in Fig. \ref{B91}.

XFEL experts have indeed observed an apparent wavefront readjustment due to relativistic kinematics effects, though this conclusion was never drawn. In this book, we are the first to consider the idea that the results of the conventional theory of radiation from relativistically moving charges are inconsistent with special relativity.
In previous literature, identification of the trajectories in the source part of the usual Maxwell's equations with the trajectories calculated by conventional particle tracking in the ("single") lab frame has always been taken as self-evident.

\subsection{Modulation Wavefront Tilt and Maxwell's Theory}

In the existing literature, the theoretical analysis of XFELs driven by electron beams with wavefront tilt is typically based on standard simulation codes and conventional applications of Maxwell's equations. However, even when using only a kicker setup—without the inclusion of undulator radiation—we can demonstrate that the traditional coupling between fields and particles in XFEL theory is fundamentally flawed.

This conventional XFEL theory relies on the absolute time convention (i.e., old kinematics) for describing particle dynamics. In this work, we present a straightforward demonstration of the inconsistency between standard particle tracking methods and Maxwell's electrodynamics. Our goal is to show, in a clear and simple manner, that conventional XFEL theory fails to accurately describe the electromagnetic field distribution produced by a fast-moving, modulated electron beam downstream of a kicker.

Under Maxwell's electrodynamics, the fields associated with a modulated electron beam moving at constant velocity exhibit intriguing behavior as the beam velocity approaches the speed of light. Specifically, in the space-time domain, these fields increasingly resemble those of a laser beam (see Appendix A4). For a rapidly moving, modulated electron beam, the electric and magnetic fields are nearly equal in magnitude, transverse, and mutually perpendicular. In the limit $v \to c$, they become virtually indistinguishable from the fields of a laser beam. According to Maxwell's equations, the wavefront of such a beam is always orthogonal to its direction of propagation. This is indeed the case for virtual laser-like radiation beam in the region upstream the kicker.

Let us now examine the impact of a transverse kick on the modulation wavefront of the electron beam. Conventional particle tracking predicts that the kick causes a misalignment between the direction of electron motion and the normal to the modulation wavefront, effectively tilting the wavefront.

This leads to an inherent contradiction: within the conventional "single-frame" approach, the post-kick propagation direction of the radiation beam is no longer perpendicular to its wavefront. In other words, the direction of motion of the virtual radiation beam and the normal to its wavefront diverge. As a result, the beam appears to propagate along the kicked trajectory while maintaining a tilted wavefront—a scenario indistinguishable from real radiation in the ultrarelativistic limit, yet fundamentally inconsistent with Maxwell’s electrodynamics.

The existing literature includes theoretical treatments of XFELs driven by electron beams with tilted modulation wavefronts. These analyses typically rely on Maxwell’s equations and standard simulation tools. However, by isolating the kicker setup—excluding the undulator—we demonstrate that the conventional XFEL framework fails to correctly model the coupling between fields and particles.

This discrepancy underscores a persistent issue in XFEL physics. Initially emerging in the context of coherent undulator radiation from ultrarelativistic, modulated electron beams, the problem has now shifted focus to the tilt of the 
self-electromagnetic fields of the modulated electron beam.

\subsection{Discussion}

Finally, we offer some remarks on an alternative perspective for analyzing our complex problem. While we have already obtained results using Lorentz coordinatization, it is instructive to understand why coherent undulator radiation is still emitted in the kicked direction when described within the framework of absolute time synchronization.

The underlying physics is, in fact, quite simple. If we interpret the emission of coherent undulator radiation from a kicked, modulated electron beam in terms of the aberration of light, we can gain a remarkably intuitive understanding of the phenomena occurring in the undulator. Within the framework of Galilean-transformed electrodynamics, a measurement of the coherent radiation will detect emission only along the kicked direction. This is directly analogous to the classical aberration effect—i.e., the apparent deviation in the direction of energy transport—associated with light emitted by a moving source in an inertial frame (see Fig. \ref{B103}).

There is an intuitively appealing way to grasp the aberration of light in this context. As shown in Fig. \ref{B557}, the aberration effect in an inertial frame can be readily understood within the corpuscular (particle-like) theory of light. This phenomenon arises naturally from the transformation of velocities between reference frames.

Applying this analogy to a transversely kicked electron beam in an undulator, it is reasonable to expect a similar aberration effect. This becomes especially clear when one considers that a pulse of undulator radiation carries a quantifiable amount of electromagnetic energy. Like mass, energy is a conserved quantity, and in many respects, a coherent X-ray pulse behaves analogously to a stream of particles. Accordingly, we expect that the group velocity of undulator radiation pulses transforms according to the same velocity addition rules that apply to particles in an inertial frame. A more rigorous analysis based on the wave theory of light fully supports this expectation.

Let us now demonstrate how Galilean-transformed electrodynamics predicts a deviation in the direction of energy transport for a radiated X-ray pulse.

A modulated electron beam with finite transverse size, when traversing an undulator, effectively acts as an active medium. This medium diffracts the emitted radiation into multiple plane-wave components, each corresponding to a Fourier component of the spatial modulation of the beam. Consider an ultrarelativistic electron beam propagating along the $z$-axis and kicked transversely along the $y$-axis. The transverse electron density is assumed to vary according to
$\rho_{e} = g(K_{\perp})\cos(K_{\perp}y)$, where $K_{\perp}$ is the transverse wavenumber associated with a particular Fourier component.

By applying the Galilean transformation and performing the necessary partial differentiations, one arrives at the modified wave equation, Eq.~(\ref{GGT2}). The additional terms introduced by the Galilean transformation give rise to a predicted Doppler effect. One significant consequence of this effect, in the context of absolute time coordinatization, is an angular frequency dispersion in the light emitted from the kicked, modulated electron beam with finite transverse size.

Specifically, the Doppler shift $\Delta\omega$ of the radiated wave, to first order, is given by $\Delta\omega = K_{\perp}v_y$, where $v_y$ is the transverse velocity component of the kicked beam. This relation implies that a coherent X-ray beam with finite transverse extent propagates along the $y$-axis with a group velocity given by $d\omega/dk_y = v_y$.

Thus, according to the corrected coupling between conventional particle tracking and electrodynamics, the direction of X-ray beam propagation is identified with the direction of energy transport, rather than with the orientation of the wavefront. This distinction arises because, under Galilean transformation, the energy transport direction and the wave normal transform differently.

\subsection{Experimental Test: Electrodynamics - Dynamics Coupling}

The fact that our theory aligns well with experimental observations is clearly illustrated by its agreement with results from an "X-ray beam-splitting" experiment. This technique separates a circularly polarized XFEL pulse from the linearly polarized XFEL background, thereby maximizing the degree of circular polarization.

In this experiment, it was demonstrated that when a modulated electron beam is kicked by an angle significantly larger than the divergence of the XFEL radiation, the modulation wavefront realigns itself along the new direction of beam motion (see Fig. \ref{B91}). This realignment is essential to explain the emission of coherent radiation from a short undulator placed downstream of the kicker and oriented along the kicked direction—see, for example, Fig. 14 in \cite{CONF}.

Although these results were unexpected at the time, they had immediate practical implications. The observed "apparent wavefront readjustment" made it possible to eliminate the unwanted linearly polarized background radiation without any additional hardware—an important breakthrough for XFEL polarization control.

In the existing literature, theoretical treatments of XFELs driven by electron beams with tilted modulation wavefronts have been presented (e.g., \cite{BH, ML}), relying on Maxwell’s equations and implemented through standard simulation codes using Maxwell solvers. However, we assert that this approach is fundamentally flawed. In the ultrarelativistic limit relevant to XFELs, a tilted modulation wavefront is inconsistent with Maxwell's electrodynamics.

Specifically, within the Lorentz lab frame—where Maxwell’s equations are valid and Einstein synchronization is enforced—the concept of a tilted modulation wavefront contradicts the principles of special relativity. When the beam is treated properly in Lorentz coordinates, the modulation wavefront must remain perpendicular to the electron beam velocity. Therefore, the notion of a persistent wavefront tilt under these conditions is not physically meaningful.

It is important to note that the results of the beam-splitting experiment at LCLS support our revised understanding of spontaneous undulator emission \cite{NUHN}. The experiment clearly demonstrated that when a modulated electron beam is kicked by an angle much larger than the divergence of the XFEL radiation, the modulation wavefront realigns itself along the new direction of beam motion. As a result, coherent radiation emitted from an undulator placed downstream of the kicker is observed along the kicked direction, with practically no suppression.

Within the framework of conventional XFEL theory, this leads to a second major puzzle. According to the standard model of undulator radiation, if a modulated electron beam is in perfect resonance before the kick, then after the kick it should remain in resonance along the new velocity direction. However, experimental observations contradict this expectation. Specifically, the data show a redshift in the resonance wavelength following the kick. Maximum radiation power is achieved only when the undulator is detuned to match the reduced longitudinal velocity of the beam after the kick \cite{NUHN}.

It is worth emphasizing that any linear superposition of radiation fields from individual electrons preserves the fundamental characteristics of single-particle emission—such as its parametric dependence on undulator settings and polarization. Consider a modulated beam deflected by a weak dipole field before entering a downstream undulator. The resulting radiation field is simply the coherent sum of emissions from individual electrons. Since the observed coherent undulator radiation exhibits a redshift after the kick, it follows that the radiation from each individual electron must also experience a redshift under the same conditions.

This line of reasoning reinforces the conclusion that the results of the beam-splitting experiment in \cite{NUHN} validate our correction to the conventional theory of spontaneous undulator emission.

\subsection{Wavefront Tilt and Degradation of Electron Beam Modulation }

In conventional XFEL theory, wavefront tilt is often treated as a physically meaningful and measurable quantity. However, a common misconception in accelerator physics concerns the interpretation of this tilt. In the ultrarelativistic regime, wavefront tilt lacks a unique, objective definition. Its value depends on the choice of clock synchronization convention in the laboratory frame and can therefore be assigned arbitrarily within the interval $(0, \theta_k)$.

For example, when the evolution of a modulated electron beam is described using Lorentz coordinates, the orientation of the modulation wavefront is always perpendicular to the electron beam velocity—i.e.,  $\theta_{tilt} = 0$. This illustrates that wavefront tilt is a coordinate-dependent artifact rather than a physical observable. As a general principle, no genuine physical effect can depend on an arbitrary constant or an arbitrary function.

Let us consider a specific example. Some papers (see, e.g., \cite{TKS}) claim that wavefront tilt leads to significant degradation of electron beam modulation in XFELs. To analyze this, suppose the modulation wavefront is initially perpendicular to the beam velocity.

One effect that can influence XFEL performance is the presence of betatron oscillations, which introduce an additional spread in longitudinal velocity. Even for particles with identical energies, differing betatron angles lead to different longitudinal velocities. Thus, beyond the velocity spread caused by energy dispersion, betatron motion introduces another source of longitudinal velocity spread.

To assess the impact of this effect, we consider the total dispersion in longitudinal velocity. The deviation from the nominal longitudinal velocity is given by 

\[
\Delta v_z = v\Delta \gamma/\gamma^3 - v\Delta \theta^2/2  . 
\]

The finite angular spread of the electron beam results in a difference in arrival times at a given longitudinal position, which is the well-known normal debunching effect.

From the standpoint of conventional XFEL theory, this time difference is said to be amplified by the kick angle
$\theta_k$. According to conventional (non-covariant) particle tracking, the wavefront is tilted by an angle  $\theta_{tilt} = \theta_k$. It is commonly believed that this tilt has physical significance, and that the deviation of the longitudinal velocity component—defined as the component perpendicular to the wavefront under Galilean kinematics—is given by 

\[
\Delta v_z =  - v|\Delta \vec{\theta} + \vec{\theta}_k|^2/2     .  
\]

If this description were accurate, the cross term $v\vec{\theta}_k\cdot\Delta \vec{\theta}$ would indeed lead to a significant degradation of the modulation amplitude. This phenomenon, referred to as modulation smearing, is proposed as a separate mechanism from normal debunching (see \cite{TKS}).

However, this interpretation overlooks a critical point. Many experts assume that all forms of debunching are physically meaningful. According to the theory of relativity, while normal debunching is an objective effect, the so-called smearing mechanism is not. The proposed smearing is a coordinate-dependent artifact—it arises solely from the choice of a particular reference frame and synchronization convention in four-dimensional spacetime, and thus lacks physical meaning.

Let us now examine, from a physical perspective, why the smearing mechanism is not valid even within Galilean kinematics. In this classical framework, the cross term  $v\vec{\theta}_k\cdot\Delta \vec{\theta}$ would appear to cause modulation degradation in the forward direction. However, Galilean-transformed electrodynamics tells us that coherent radiation is observable only in the kicked direction. In that direction, the cross term does not appear in the expression for the deviation of the longitudinal velocity component. It follows naturally that the smearing effect is not a real physical phenomenon.

\subsection{Modulation Wavefront Rotation: Correct  and Incorrect Explanations}

The modulation wavefront rotation observed in X-ray free-electron lasers (XFELs) is a relativistic kinematic effect. In the ultrarelativistic limit, when the motion of the electron beam is described in Lorentz coordinates, the modulation wavefront—that is, the plane of simultaneity—is always perpendicular to the beam velocity.

Wavefront rotation accompanies any change in the beam direction along a curvilinear trajectory, for example, during motion through an undulator with small orbit perturbations produced by the focusing system. Since this effect originates entirely from relativistic kinematics, it is not caused by the magnetic forces responsible for changing the beam trajectory.

An alternative explanation of the LCLS observations has been proposed in Ref.~\cite{ML}, where the modulation wavefront rotation is attributed to conventional beam-dynamics effects rather than to relativistic kinematics. In our opinion, this interpretation results from treating the problem within the framework of non-covariant particle tracking, thereby overlooking the role of the relativity of simultaneity.

The analysis presented in Ref.~\cite{ML} starts from the conventional assumption that a transverse kick does not immediately change the orientation of the modulation wavefront. Using non-covariant particle tracking, the authors conclude that the wavefront gradually rotates under the action of quadrupole focusing, as illustrated in Fig.~1 of Ref.~\cite{ML}. According to this interpretation, the resulting rotation improves off-axis lasing and is consistent with both numerical simulations and experimental observations.

To examine this claim, we analyze the argument presented in Ref.~\cite{ML}.
 At LCLS, the corrector dipole magnets are integrated into the quadrupoles of the undulator FODO lattice. The corrector immediately upstream of the Delta undulator deflects the modulated electron beam in the vertical direction.

We choose a coordinate system in which the unperturbed beam propagates along the positive $z$-axis. At $z=0$, the beam receives a vertical kick through an angle $\theta_k$, after which it continues along the kicked trajectory. This configuration corresponds to that shown in Fig.~1 of Ref.~\cite{ML}.

For $z>0$, the transverse position of an individual electron is

\[
y(z)=y_0+\left(\Delta\theta_y+\frac{y_0}{2f}+\theta_k\right)z,
\]

where $f$ is the focal length of the quadrupole and the origin $z=0$ is taken at its center. A positive value of $f$ corresponds to a quadrupole that is defocusing in the vertical plane.

The argument that  modulation wavefront kicked in the defocusing quadrupole experiences rotation  proceeds as follows.
It is most straightforward to calculate the difference in longitudinal position between a reference electron and the dashed  reference trajectory  along the $z'$ axis of Fig. 1.   

According to conventional (non-covariant) particle tracking, the wavefront is tilted by an angle  $\theta_{tilt} = \theta_k$. It is commonly believed that this tilt has physical significance, and that the deviation of the longitudinal velocity component—defined as the component perpendicular to the wavefront under Galilean kinematics—is given by 

\[
\Delta v_z =  - v|\Delta \vec{\theta}  + y_0/(2f) - x_0/(2f) + \vec{\theta}_k|^2/2.  
\]

According to authors of \cite{ML},  the cross term  $v\theta_k y_0/(2f)$  would indeed lead to a modulation wavefront rotation. 
This effect is entirely single particle dynamics in given fields and no FEL gain is needed to explain this wavefront rotation.
The other cross terms  represent modulation smearing and other detrimental correlations.

The next question concerns the evolution of the modulation wavefront along the Delta undulator. The calculations presented in Ref.~\cite{ML} are carried out in the Cartesian coordinate system $(x,y,z)$ shown in Fig.~1. It should be emphasized, however, that the choice of the spatial coordinate system is entirely arbitrary. An exact solution can be obtained almost immediately by introducing a rotated Cartesian coordinate system $(x',y',z')$, whose $z'$-axis is aligned with the electron beam velocity downstream of the kicker.

In this rotated coordinate system, the cross term does not appear in the expression for the longitudinal velocity component along the $z'$-axis. Furthermore, the betatron motion in the $y'$--$z'$ plane is identical to that in the FODO lattice upstream of the kicker. Within the framework of conventional particle tracking, a transverse kick does not alter the orientation of the modulation wavefront. Consequently, the wavefront remains tilted by an angle $\theta_{\mathrm{tilt}}=\theta_k$ with respect to the $z'$-axis.

Likewise, the cross term $v\vec{\theta}_k\cdot\Delta\vec{\theta}$ is absent from the expression for the longitudinal velocity spread $\Delta v_{z'}$. As a result, the beam modulation is not degraded by any additional smearing along the kicked $z'$-direction (see Section~22.7 for a more detailed discussion).

Our analysis begins with the coherent betatron motion in an undulator FODO lattice. To investigate the influence of betatron beating on the modulation wavefront evolution, we performed numerical simulations of the electron beam size along the lattice. The FODO half-cell length was taken to be 3.84~m, approximately equal to the length of a single undulator module. The electron beam energy was set to 4.3~GeV, and the average betatron function was chosen to be 10~m. For a normalized emittance of 0.4~$\mu$m and an rms energy spread of 1.5~MeV, the average rms beam size is $\sigma_0\simeq22~\mu$m.

The evolution of the rms beam size along the undulator exhibits the familiar zigzag behavior characteristic of a FODO lattice. In particular, after a defocusing quadrupole the rms beam size increases approximately linearly from $\sigma_{\min}\simeq18~\mu$m immediately downstream of the quadrupole to $\sigma_{\max}\simeq26~\mu$m at the exit of the undulator.

As the beam size increases, the tilt angle $\alpha$ decreases correspondingly. Using the simple geometric relation

\[
\sigma_{\min}\alpha_{\max} = \sigma(z')\left[\alpha_{\max}+\Delta\alpha(z')\right],
\]

we readily obtain the evolution of the tilt angle $\Delta\alpha(z')$ as a function of the distance from the undulator entrance. In the case of interest, where $\alpha_{\max}=\theta_k$, the change in the tilt angle at the undulator exit is approximately

\[
\Delta\alpha\simeq-0.3~\theta_k.
\]

The resulting analytical expression is exact for a matched beam and incorporates the effects of quadrupole focusing, emittance, and energy spread. Although the same result can be obtained numerically, the analytical solution provides considerably greater physical insight.

There is an intuitive reason why the wavefront rotation can be analyzed entirely within the framework of three-dimensional betatron beating, without resorting to phase-space simulations. The key observation is that electron-beam microbunching is a coherent phenomenon. Consequently, the evolution of the modulation wavefront following a transverse kick is determined solely by the coherent motion of the electron beam. In particular, the wavefront rotation manifests itself as a change in the transverse beam size along the FODO lattice.

In contrast, the numerical calculations reported in Ref.~\cite{ML} are performed in the same coordinate system, $(x,y,z)$, that was used before the kick (see Fig.~1 of Ref.~\cite{ML}). Retaining this coordinate system after the beam has been deflected introduces unnecessary complications. In particular, the particle dynamics acquires the non-standard cross terms $v\theta_k y_0/(2f)$ and $v\vec{\theta}_k\cdot\Delta\vec{\theta}$, which arise solely from the chosen coordinate description.

The predictions of Ref.~\cite{ML} differ substantially from the wavefront evolution obtained in our analysis. One notable example concerns the interpretation of debunching. The authors of Ref.~\cite{ML} state: "The rotation comes at a cost, however. During the microbunch rotation, smearing occurs. ... There is an unavoidable degradation ... This is a consequence of microbunch smearing." We believe that this conclusion results from the use of a coordinate system that is not aligned with the beam trajectory.

After the kick, the electron beam propagates along the $z'$-axis. Accordingly, the deviation of the longitudinal velocity component along the beam direction is

\[
\Delta v_{z'}=-\frac{v}{2}\left|\Delta\vec{\theta}\right|^2,
\]

which is simply the familiar debunching associated with incoherent betatron motion. No additional smearing mechanism is present.

By contrast, Ref.~\cite{ML} evaluates the longitudinal velocity with respect to the original $z$-axis, obtaining

\[
\Delta v_z=-\frac{v}{2}\left|\Delta\vec{\theta}+\vec{\theta}_k\right|^2,
\]

which contains the additional cross terms responsible for the apparent "microbunch smearing." These terms are not intrinsic physical effects but arise from describing the beam in a coordinate system that is no longer aligned with its direction of motion. As a consequence, the expressions derived in Ref.~\cite{ML} for the downstream evolution of the microbunching include coordinate-dependent artifacts that disappear when the natural $(x',y',z')$ coordinate system is employed.

Let us now return to the results obtained within the framework of relativistic dynamics. According to conventional particle tracking, the modulation wavefront undergoes a rotation downstream of the kicker, as illustrated in Fig.~1 of Ref.~\cite{ML}. This rotation is a purely kinematic effect and is not produced by any physical force acting on the microbunching wavefront.

We emphasize, however, that this wavefront rotation should not be regarded as an observer-independent physical quantity. Rather, it reflects the choice of synchronization convention used to define distant simultaneity. To understand this point, it is essential to recognize that the relativity of simultaneity—a fundamental consequence of special relativity—is inseparably connected with extended relativistic objects such as the microbunching wavefront of an ultrarelativistic electron beam.

In standard special relativity, the time coordinate assigned to a distant event is not an intrinsic physical property of that event. Instead, it is a coordinate label determined by the adopted clock-synchronization convention, or equivalently, by the choice of spacelike hypersurfaces used to define simultaneity.

The orientation of the microbunching wavefront is determined by the simultaneity of spatially separated events. These are spacelike-separated events, occurring at different locations outside one another's light cones, so that no signal traveling at or below the speed of light can establish a causal connection between them within the time interval under consideration. Consequently, special relativity does not assign an observer-independent temporal ordering to such events.

When discussing the orientation of the modulation wavefront with respect to the $y$-axis, one must therefore account for the conventional nature of distant simultaneity. The wavefront orientation is not an objectively measurable quantity in the same sense as a particle's path $y(z)$ or momentum. Rather, there exists an intrinsic ambiguity in assigning simultaneity to two events separated by a distance $y$. This ambiguity, of order $y/c$, follows directly from the Minkowski geometry of spacetime. Because information cannot propagate faster than the speed of light, no experimental procedure can determine the simultaneity of such distant events with greater precision.

As discussed above, the rotation of the microbunching wavefront changes the arrival time at a given longitudinal position by an amount satisfying

\[
\Delta t(y)<\frac{y\theta_k}{c},
\]

which is well within the intrinsic uncertainty of order $y/c$ associated with distant simultaneity. We therefore conclude that the rotation of the microbunching wavefront along the undulator does not possess a well-defined operational meaning.

On the other hand, the physical predictions of relativistic dynamics are encoded in convention-invariant quantities, such as the  particle's path  in the magnetic field and,  consequently, its three-momentum. By contrast, the  trajectory $\vec{r}(t)$ in the laboratory frame depends on the synchronization convention adopted for clocks in that frame.

To illustrate this distinction, suppose that two apertures, $A$ and $A'$, are located at the coordinates $(y_1,z)$ and $(y_2,z)$, respectively. 
From the solution of the equations of motion, we conclude that the first particle passes through  aperture $A$ while the second passes through  aperture $A'$ simultaneously. The occurrence of these two events—particles passing through $A$ and $A'$—has an exact objective meaning, making them convention-invariant.  However, whether the two events are regarded as simultaneous depends on the chosen synchronization convention and therefore has no observer-independent meaning.

Although relativistic dynamics employs the concept of a particle trajectory, this coordinate-dependent quantity is not measured directly. Rather, it serves as an intermediate mathematical construct for calculating convention-invariant observables in electrodynamics.

The description of relativistic particle motion in a constant magnetic field used in accelerator physics closely resembles its non-relativistic counterpart in Newtonian mechanics. Consequently, the conventional particle-tracking approach adopted in Ref.~\cite{ML} treats spacetime in a non-relativistic manner, as a $(3+1)$ manifold. Within this framework, the only modification to classical mechanics is the introduction of relativistic momentum (or, historically, relativistic mass), while time remains separate from the three spatial coordinates.

As a result, the particle trajectory obtained from the relativistic form of Newton's second law does not incorporate relativistic kinematic effects. In particular, it is based on the Galilean composition of velocities and therefore neglects the relativity of simultaneity. We emphasize that the velocity of light transforms according to the same kinematic principles as the velocity of a material particle. Consequently, it is not possible to combine classical kinematics for particle dynamics with relativistic kinematics for electrodynamics within a single self-consistent theoretical framework.

It is well known that Maxwell's equations are form-invariant under Lorentz transformations. These transformations, however, mix spatial and temporal coordinates. Consequently, if Maxwell's equations are to retain their standard form in the laboratory frame, the description of particle motion must incorporate the same relativistic kinematics. Equivalently, particle trajectories in the laboratory frame must be obtained as the result of successive infinitesimal Lorentz transformations rather than successive Galilean transformations.

The absence of relativistic kinematic effects is therefore a fundamental limitation of conventional non-covariant particle tracking. In particular, the relativity of simultaneity—the inseparable connection between spatial coordinates and time—is absent from this formulation. As a consequence, coupling Maxwell's equations to particle trajectories computed within this non-covariant framework leads to a description that is not fully consistent with special relativity.

For rectilinear motion of both the modulated electron beam and the emitted radiation, the covariant and non-covariant approaches predict identical particle paths. In this special case, Maxwell's equations can therefore be combined with conventional particle tracking without producing observable inconsistencies. Difficulties arise only when this approach is extended to non-collinear geometries.

The essential feature of the collinear configuration is that the electron beam velocity is perpendicular to the modulation wavefront, that is, to the plane of simultaneity. Consequently, the orientation of the plane of simultaneity is identical in both Lorentz and absolute-time coordinatizations. In this particular geometry, the difference between the two synchronization conventions does not manifest itself in the orientation of the modulation wavefront.

The situation changes fundamentally when the electron beam is deflected by a transverse kick. In this case, the beam is accelerated within the initial plane of simultaneity, and the orientation of the modulation wavefront is no longer an absolute quantity. Instead, it depends on the synchronization convention adopted in the laboratory frame.

This immediately raises the question of how clock synchronization should be assigned after the acceleration. Before the kick, the laboratory frame is described using Lorentz coordinates. If one retains the original synchronization convention after the kick, the electrodynamics of the moving charges becomes considerably more complicated because the transformation between the original and the new beam direction is no longer expressed in the standard Lorentz form. Maxwell's equations preserve their standard form in the laboratory frame only when Lorentz coordinatization is maintained throughout the description.

To preserve Lorentz coordinatization after the kick, it is therefore necessary to introduce a new set of space-time coordinates $(x_L,y_L,z_L,t_L)$, defined by Eq.~(\ref{GGT3}). In this coordinate system, the light emitted by the  kicked electron beam propagating through the  undulator continues to propagate with the invariant speed $c$ in every direction, exactly as required by Maxwell's electrodynamics.

It is important to recognize that Maxwell's equations retain their standard form only in Lorentz coordinates. Consequently, Einstein's concept of simultaneity must be applied consistently in both dynamics and electrodynamics.

According to Maxwell's electrodynamics, the wavefront of a light beam is always perpendicular to its direction of propagation. In the ultrarelativistic limit, a modulated electron beam exhibits the same kinematic behavior as a laser beam when described in Lorentz coordinates. In particular, the modulation wavefront remains orthogonal to the electron beam velocity.

Under a Lorentz boost, the time coordinate transforms according to

\[
t \rightarrow t+\frac{yv}{c^2}.
\]

To first order in $v/c$, this time transformation is equivalent to a rotation of the modulation wavefront in the laboratory frame through an angle approximately equal to $v/c$.

Since electromagnetic radiation is emitted normal to the modulation wavefront, the emitted light beam is likewise rotated by an angle $v/c$. This is simply the familiar phenomenon of the aberration of light.

Before leaving the subject of wavefront rotation in the LCLS experiments, it is worthwhile to discuss a subtle effect that arises in a FODO undulator lattice. Even in the case of rectilinear motion, where the modulated electron beam propagates along the same axis as the emitted radiation, betatron beating produces a slight deformation of the modulation wavefront.

Consider a modulated electron beam that, after the kick, propagates along the $z'$-axis. The clock resynchronization associated with the Lorentz transformation rotates the modulation wavefront by the angle $\theta_k$. Nevertheless, in the new time coordinate there remains a coherent component of the electron motion parallel to the modulation wavefront due to betatron beating. Such coherent motion is an inherent feature of any XFEL FODO lattice, including the undulator section upstream of the kicker.

For the LCLS parameters, the corresponding wavefront deformation angle may be estimated as

\[
\frac{v_{\mathrm{coh}}}{c}
\le
\frac{\Delta\sigma_{\max}}{l}
\simeq
3~\mu\mathrm{rad},
\]

where $v_{\mathrm{coh}}$ is the coherent betatron velocity along the modulation wavefront and $l$ is the FODO half-cell length. This value should be compared with the typical kick angles used at LCLS,

\[
30~\mu\mathrm{rad}\le\theta_k\le50~\mu\mathrm{rad}.
\]

Thus, the deformation of the radiation wavefront produced by coherent betatron motion is negligibly small under the experimental conditions at LCLS.\footnote{It is worth noting that the modulation-wavefront deformation caused by betatron beating is also smaller than the divergence of the coherent undulator radiation, $\sigma_{\mathrm{dif}}\simeq\lambdabar/\sigma_0\simeq10~\mu\mathrm{rad}$ at $\lambda=1.5~\mathrm{nm}$.}

\subsection{Concluding Remarks}

Conventional XFEL theory adopts the absolute time convention to describe particle dynamics. Notably, employing the standard coupling of Maxwell’s equations with the corrected Newtonian equations to compute radiation from an ultrarelativistic electron beam in an undulator does not inherently introduce inaccuracies. In the case of straight-line motion—whether for the electron beam or the emitted undulator radiation—both covariant and non-covariant approaches produce identical trajectories

In standard XFEL operation, the electron beam is guided to remain closely aligned with the undulator axis. Nonetheless, random perturbations in the focusing system can introduce angular deviations or "kicks." The resultant misalignment between beam trajectory and wavefront normal has been discussed in the literature. It is traditionally understood that coherent radiation is emitted along the wavefront normal. Consequently, under the conventional field-particle coupling (which we argue is flawed), the mismatch between these directions diminishes radiation efficiency \cite{TKS}. Studies of trajectory errors in XFEL amplification have shown that undulator magnetic fields must meet stringent tolerances. Interestingly, these tolerances have proven to be more stringent than necessary according to the corrected, four-dimensional (4D) covariant XFEL theory—an insight that helps explain the remarkable success of XFEL technology in recent decades.

\newpage

\section{Appendix}

\subsection{A1. Radiation from Moving Charges}

We start with the solution of
Maxwell's equation in the space-time domain, the well-known
Lienard-Wiechert expression, and we subsequently apply a Fourier
transformation. The Lienard-Wiechert expression for the electric
field of a point charge $(-e)$ reads (see, e.g. \cite{JACK}):

\begin{eqnarray}
\vec{{E}}(\vec{r}_o,t) &=& -e {\vec{n}-\vec{\beta}\over{\gamma^2
		(1-\vec{n}\cdot\vec{\beta})^3 |\vec{r}_o-\vec{r'}|^2}} -{e
	\over{c}} {\vec{n}\times[(\vec{n}-\vec{\beta})\times
	\dot{\vec{\beta}}]\over{(1-\vec{n}\cdot\vec{\beta})^3|\vec{r}_o-\vec{r'}|}}
. \label{LWt}
\end{eqnarray}
$R=|\vec{r}_o-\vec{r'}(t')|$ denotes the displacement
vector from the retarded position of the charge to the point where
the fields are calculated. Moreover, $\vec{\beta}=\vec{v}/c$,
$\vec{\dot{\beta}}=\vec{\dot{v}}/c$, while $\vec{v}$ and
$\vec{\dot{v}}$ denote the retarded velocity and acceleration of
the electron. Finally, the observation time $t$ is linked with the
retarded time $t'$ by the retardation condition $R =  c(t-t')$. As
is well-known, Eq. (\ref{LWt}) serves as a basis for the
decomposition of the electric field into a sum of two quantities.
The first term on the right-hand side of Eq. (\ref{LWt}) is
independent of acceleration, while the second term linearly
depends on it. For this reason, the first term is called "velocity
field", and the second "acceleration field" \cite{JACK}. The
velocity field differs from the acceleration field in several
respects, one of which is the behavior in the limit for a very
large distance from the electron. There one finds that the
velocity field decreases like $R^{-2}$, while the acceleration
field only decreases as $R^{-1}$. Let us apply a Fourier
transformation:

\begin{eqnarray}
\vec{\bar{E}}(\vec{r}_o,\omega) &=& -e\int_{-\infty}^{\infty} dt'
{\vec{n}-\vec{\beta}\over{\gamma^2 (1-\vec{n}\cdot\vec{\beta})^2
		|\vec{r}_o-\vec{r'}|^2}} \exp
\left[i\omega\left(t'+{|\vec{r}_o-\vec{r'}(t')|\over{c}}\right)\right]
\cr &&-{e \over{c}}\int_{-\infty}^{\infty}dt'
{\vec{n}\times[(\vec{n}-\vec{\beta})\times
	\dot{\vec{\beta}}]\over{(1-\vec{n}\cdot\vec{\beta})^2|\vec{r}_o-\vec{r'}|}}
\exp
\left[i\omega\left(t'+{|\vec{r}_o-\vec{r'}(t')|\over{c}}\right)\right]
. \label{LW}
\end{eqnarray}
As in Eq. (\ref{LWt}) one may formally recognize a velocity and an
acceleration term in Eq. (\ref{LW}) as well. Since Eq. (\ref{LW})
follows directly from Eq. (\ref{LWt}), that is valid in the time
domain, the magnitude of the velocity and acceleration parts in
Eq. (\ref{LW}), that include terms in $1/R^2$ and $1/R$
respectively, do not depend on the wavelength $\lambda$. It is instructive to take advantage of
integration by parts. With the help of

\begin{eqnarray}
{1\over{c}}{d\over{dt'}}|\vec{r_o}-\vec{r'}(t')| =
-\vec{n}\cdot\vec{\beta}~~~~~\mathrm{and}~~~~{d\vec{n}\over{dt'}}
= {c\over{|\vec{r}_o-\vec{r'}(t')|}}
\left[-\vec{\beta}+\vec{n}\left(\vec{n}\cdot\vec{\beta}\right)\right]~,
\label{usefulrel}
\end{eqnarray}
Eq. (\ref{LW}) can be written as

\begin{eqnarray}
\vec{\bar{E}}&(&\vec{r}_o, \omega) = -{ e} \int_{-\infty}^{\infty}
dt'{\vec{n}\over{|\vec{r}_o-\vec{r'}(t')|^2}}  \exp
\left[i\omega\left(t'+{|\vec{r}_o-\vec{r'}(t')|\over{c}}\right)\right]
\cr &&+ {e\over{c}} \int_{-\infty}^{\infty} dt'{d\over{dt'}}\left[
{\vec{\beta}-\vec{n}\over{({1-\vec{n}\cdot\vec{\beta}})|\vec{r}_o-\vec{r'}(t')|}}\right]
\exp
\left[i\omega\left(t'+{|\vec{r}_o-\vec{r'}(t')|\over{c}}\right)\right].\cr
&& \label{revtrasfbiss}
\end{eqnarray}
Eq. (\ref{revtrasfbiss}) may now be integrated by parts. When edge
terms can be dropped one obtains \cite{GE}

\begin{eqnarray}
\vec{\bar{E}}(\vec{r_o},\omega) &=& -{i\omega e\over{c}}
\int_{-\infty}^{\infty} dt'
\left[{\vec{\beta}-\vec{n}\over{|\vec{r_o}-\vec{r'}(t')|}}-{ic\over{\omega}}{\vec{n}\over{
		|\vec{r_o}-\vec{r'}(t')|^2}}\right]\cr &&\times
\exp\left\{i\omega\left(t'+{|\vec{r_o}-\vec{r'}(t')|\over{c}}\right)\right\}~.
\label{revtrasf}
\end{eqnarray}

The only assumption made going from
Eq. (\ref{LW}) to Eq. (\ref{revtrasf}) is that edge terms in the
integration by parts can be dropped. This assumption can be
justified by means of physical arguments in the most general
situation accounting for the fact that the integral in $dt'$ has
to be performed over the entire history of the particle and that
at $t'=-\infty$ and $t'=+\infty$, the electron does not contribute
to the field anymore. Let us give a concrete example for an
ultra-relativistic electron. Imagine that bending magnets are
placed at the beginning and at the end of a given setup, such that
they deflect the electron trajectory of an angle much larger than
the maximal observation angle of interest for radiation from a
bending magnet. This means that the magnets would be longer than
the formation length associated with the bends, i.e.
$L_\mathrm{fb} \sim (c \rho^2/\omega)^{1/3}$, where $\rho$ is the
bending radius. In this way, intuitively, the magnets act like
switches: the first magnet switches the radiation on, and the second
switches it off. Then, what precedes the upstream bend and what
follows the downstream bend does not contribute to the field
detected at the screen position. With this caveat Eq.
(\ref{revtrasf}) is completely equivalent to Eq. (\ref{LW}).

The derivation of Eq. (\ref{revtrasf}) is particularly instructive
because shows that each term in Eq. (\ref{revtrasf}) is due to a
combination of velocity and acceleration terms in Eq. (\ref{LW}).
In other words the terms in $1/R$ and in $1/R^2$ in Eq.
(\ref{revtrasf}) appear as a combination of the terms in $1/R$
(acceleration term) and $1/R^2$ (velocity term) in Eq. (\ref{LW}).
As a result, one can say that there exist contributions to the
radiation from the velocity part in Eq. (\ref{LW}). The
presentation in Eq. (\ref{revtrasf})  is
more interesting with respect to that in Eq. (\ref{LW}) (although
equivalent to it) because of the magnitude of the $1/R^2$-term in Eq.
(\ref{revtrasf}) can directly be compared with the magnitude of
the $1/R$-term inside the integral sign.

The bottom line is that physical sense can be ascribed only to the
integral in Eq. (\ref{LW}) or Eq. (\ref{revtrasf}). The integrand
is, in fact, an artificial construction. In this regard, it is
interesting to note that the integration by parts giving Eq.
(\ref{revtrasf}) is not unique. 
First, we find that \cite{GE}

\begin{eqnarray}
{\vec{n}\times[(\vec{n}-\vec{\beta})\times\vec{\dot{\beta}}]
	\over{|\vec{r_o}-\vec{r'}|(1-\vec{n}\cdot\vec{\beta})^2}} &=&
{1\over{|\vec{r_o}-\vec{r'}|}} {d\over{dt'}}
\left[{\vec{n}\times(\vec{n}\times\vec{\beta})\over{(1-\vec{n}\cdot\vec{\beta})}}\right]
\cr &&
-\left[{\vec{\dot{n}}(\vec{n}\cdot\vec{\beta})+\vec{n}(\vec{\dot{n}}\cdot\vec{\beta})
	-
	\vec{\dot{n}}(\vec{n}\cdot\vec{\beta})^2-\vec{\beta}(\vec{\dot{n}}\cdot\vec{\beta})
	\over{|\vec{r_o}-\vec{r'}|(1-\vec{n}\cdot\vec{\beta})^2}}\right].
\label{luccioparti}
\end{eqnarray}
Note that Eq. (\ref{luccioparti}) accounts for the fact that
$\vec{n} = (\vec{r_o}-\vec{r'}(t'))/|\vec{r_o}-\vec{r'}(t')|$ is
not a constant in time. Using Eq. (\ref{luccioparti}) in the
integration by parts, we obtain

\begin{eqnarray}
\vec{\bar{E}}(\vec{r_o},\omega) &=& -{i \omega e\over{c}}
\int_{-\infty}^{\infty} dt'
\left[-{\vec{n}\times(\vec{n}\times{\vec{\beta}})\over{|\vec{r_o}-\vec{r'}(t')|}}
+ {i {c}\over{\omega}}
{\vec{\beta}-\vec{n}-2\vec{n}(\vec{n}\cdot\vec{\beta})\over{|\vec{r_o}-\vec{r'}(t')|^2}}\right]
\cr && \times
\exp\left\{i\omega\left(t'+{|\vec{r_o}-\vec{r'}(t')|\over{c}}\right)\right\}
~. \label{revluccio}
\end{eqnarray}
Similarly as before,  the edge terms have been dropped. Eq.
(\ref{LW}), Eq. (\ref{revtrasf}) and Eq. (\ref{revluccio}) are
equivalent but include different integrands. This is no mistake,
as different integrands can lead to the same integral.

If the position of the observer is far away enough from the charge, one can make the expansion $\left| \vec{r}_0-\vec{r}(t) \right|= r_0 - \vec{n} \cdot \vec{r}(t)$.
Using Eq. (\ref{revluccio}), we obtain Eq. (\ref{revwied}).

\newpage

\subsection{A2. Undulator Radiation in the Far Zone}

Calculations pertaining undulator radiation are well established  see e.g. \cite{ONUKI}. 
In all generality, the field in Eq. (\ref{undurad_ap1}) can
be written as

\begin{eqnarray}
&&{\vec{\widetilde{E}}}= \exp\left[i\frac{\omega \theta^2
	z_0}{2c}\right] \frac{i \omega e}{c^2 z_0} \cr && \times \int_{-L/2}^{L/2}
dz'\left\{\frac{K}{2 i \gamma}\left[\exp\left(2 i k_w
z'\right)-1\right]\vec{e}_x +\vec{\theta}\exp\left(i k_w
z'\right)\right\} \cr &&\times \exp\left[i \left(C + {\omega
	\theta^2 \over{2 c }}\right) z' -
{K\theta_x\over{\gamma}}{\omega\over{k_w c}}\cos(k_w z')  -
{K^2\over{8\gamma^2}} {\omega\over{k_w c}} \sin(2 k_w z') \right]
~. \cr &&\label{undurad2_ap1}
\end{eqnarray}
Here $\omega = \omega_r + \Delta \omega$, $C =  k_w \Delta\omega/\omega_r$ and

\begin{eqnarray}
\omega_r = 2 k_w c \bar{\gamma}_z^2~, \label{res_ap1}
\end{eqnarray}
is the fundamental resonance frequency.

Using the Anger-Jacobi expansion:

\begin{equation}
\exp\left[i a \sin(\psi)\right] = \sum_{p=-\infty}^{\infty} J_p(a)
\exp\left[ip\psi\right]~, \label{alfeq}
\end{equation}
where $J_p(\cdot)$ indicates the Bessel function of the first kind
of order $p$, to write the integral in Eq. (\ref{undurad2_ap1}) in a
different way:

\begin{eqnarray}
&&{\vec{\widetilde{E}}}= \exp\left[i\frac{\omega \theta^2
	z_0}{2c}\right] \frac{i \omega e}{c^2 z_0} \sum_{m,n=-\infty}^\infty
J_m(u) J_n(v) \exp\left[\frac{i \pi n}{2}\right] \cr && \times
\int_{-L/2}^{L/2} dz'\exp\left[i \left(C + {\omega \theta^2
	\over{2 c }}\right) z'\right] \cr &&\times \left\{\frac{K}{2 i \gamma}
\left[\exp\left(2 i k_w z'\right)-1\right]\vec{e}_x
+\vec{\theta}\exp\left(i k_w z'\right)\right\}
\exp\left[i (n+2m) k_w z'\right] ~,\cr &&\label{undurad3_ap1}
\end{eqnarray}
where\footnote{Here the parameter $v$ should not be confused with the velocity.}

\begin{equation}
u = - \frac{K^2 \omega}{8 \gamma^2 k_w c}~~~~\mathrm{and}~~~v = -
\frac{K \theta_x \omega}{\gamma k_w c}~. \label{uv}
\end{equation}
Up to now we just re-wrote Eq. (\ref{undurad_ap1}) in a different way.
Eq. (\ref{undurad_ap1}) and Eq. (\ref{undurad3_ap1}) are equivalent. Of
course, definition of $C$ is suited to
investigate frequencies around the fundamental harmonic but no
approximation is taken besides the paraxial approximation.

Whenever

\begin{equation}
C  + \frac{\omega \theta^2}{{2 c}} \ll k_w \label{eqq_ap1} ~,
\end{equation}
the first phase term in $z'$ under the integral sign in Eq.
(\ref{undurad3_ap1}) is varying slowly on the scale of the undulator
period $\lambda_w$. As a result, simplifications arise when $N_w
\gg 1$, because fast oscillating terms in powers of  $\exp[i k_w
z']$ effectively average to zero. When these simplifications are
taken,  resonance approximation is applied, in the sense that one
exploits the large parameter $N_w \gg 1$. This is possible under
condition (\ref{eqq_ap1}). Note that (\ref{eqq_ap1}) restricts the range
of frequencies for positive values of $C$ independently of the
observation angle ${\theta}$, but for any value $C<0$ (i.e. for
wavelengths longer than $\lambdabar_r = c/\omega_r$) there is
always some range of $\theta$ such that Eq. (\ref{eqq_ap1}) can be
applied. Altogether, application of the resonance approximation is
possible for frequencies around $\omega_r$ and lower than
$\omega_r$. Once any frequency is fixed, (\ref{eqq_ap1}) poses
constraints on the observation region where the resonance
approximation applies. Similar reasonings can be done for
frequencies around higher harmonics with a more convenient
definition of the detuning parameter $C$.

Within the resonance approximation we further select frequencies
such that $|\Delta \omega|/\omega_r \ll 1$ (i.e. $|C|
\ll k_w $).
Note that this condition on frequencies automatically selects
observation angles of interest $\theta^2 \ll 1/\gamma_z^2$. In
fact, if one considers observation angles outside the range
$\theta^2 \ll 1/\gamma_z^2$, condition (\ref{eqq_ap1}) is not
fulfilled, and the integrand in Eq. (\ref{undurad3_ap1}) exhibits fast
oscillations on the integration scale $L$. As a result, one
obtains zero transverse field, $\vec{\widetilde{E}} = 0$,
with accuracy $1/N_w$. Under the constraint imposed by $|\Delta \omega|/\omega_r \ll 1$
, independently of the value of $K$ and for
observation angles of interest $\theta^2 \ll 1/\gamma_z^2$, we
have

\begin{equation}
|v|={K|\theta_x|\over{\gamma}}{\omega\over{k_w c}} =
\left(1+\frac{\Delta \omega}{\omega_r}\right) \frac{2 \sqrt{2}
	K}{\sqrt{2+K^2}} \bar{\gamma}_z |\theta_x| \lesssim
\bar{\gamma}_z |\theta_x| \ll 1~. \label{drop}
\end{equation}
This means that, independently of $K$, $|v| \ll 1$ and we may
expand $J_n(v)$ in Eq. (\ref{undurad3_ap1}) according to $J_n(v)
\simeq [2^{-n}/\Gamma(1+n)]~v^n$, $\Gamma(\cdot)$ being the Euler
gamma function  $\Gamma(z) = \int_0^\infty dt~t^{z-1} \exp[-t]$

Similar reasonings can be done for frequencies around higher
harmonics with a different definition of the detuning parameter
$C$. However, around odd harmonics, the before-mentioned
expansion, together with the application of the resonance
approximation for $N_w \gg 1$ (fast oscillating terms in powers of
$\exp[i k_w z']$ effectively average to zero), yields
extra-simplifications.

Here we are dealing specifically with the first harmonic.
Therefore, these extra-simplifications apply. We
neglect both the  term in $\cos(k_w z')$ in the phase of Eq. (\ref{undurad2_ap1})
and the term in $\vec{\theta}$ in Eq. (\ref{undurad2_ap1}). First,
non-negligible terms in the expansion of $J_n(v)$ are those for
small values of $n$, since $J_n(v) \sim v^n$, with $|v|\ll 1$. The
value $n=0$ gives a non-negligible contribution $J_0(v) \sim 1$.
Then, since the integration in $d z'$ is performed over a large
number of undulator periods $N_w\gg 1$, all terms of the expansion
in Eq. (\ref{undurad3_ap1}) but those for $m=-1$ and $m=0$ average to
zero due to resonance approximation. Note that surviving
contributions are proportional to $K/\gamma$, and can be traced
back to the term in $\vec{e}_x$ only, while the
term in $\vec{\theta}$ in Eq. (\ref{undurad3_ap1}) averages to zero
for $n=0$. Values $n=\pm 1$ already give negligible contributions.
In fact, $J_{\pm 1}(v) \sim v$. Then, the term in $\vec{e}_x$ in
Eq. (\ref{undurad3_ap1}) is $v$ times the term with $n=0$ and is
immediately negligible, regardless of the values of $m$. The
term in $\vec{\theta}$ would survive averaging when $n=1,
~m=-1$ and when $n=-1, ~m=0$. However, it scales as $\vec{\theta}
v$. Now, using condition $|\Delta \omega|/\omega_r \ll 1$ we see that, for
observation angles of interest $\theta^2 \ll 1/\gamma_z^2$,
$|\vec{\theta}|~ |v| \sim (\sqrt{2}~ K~/\sqrt{2+K^2}~)
~\bar{\gamma}_z \theta^2 \ll K/\gamma$. Therefore, the
term in $\vec{\theta}$ is negligible with respect to the term in $\vec{e}_x$
for $n=0$, that scales as $K/\gamma$. All terms corresponding to
larger values of $|n|$ are negligible.

Summing up, all terms of the expansion in Eq. (\ref{alfeq}) but
those for $n=0$ and $m=-1$ or $m=0$ give negligible contribution.
After definition of

\begin{eqnarray}
A_{JJ} = J_0\left(\frac{\omega K^2}{8 k_w c \gamma^2}\right) -
J_1\left(\frac{\omega K^2}{8 k_w c \gamma^2}\right) ~,\label{AJJ}
\end{eqnarray}
that can be calculated at $\omega = \omega_r$ since $|C| \ll k_w$,
we have

\begin{eqnarray}
&&{\vec{\widetilde{E}}}= - \frac{K \omega e }{2
	c^2 z_0 \gamma} A_{JJ} \exp\left[i\frac{\omega \theta^2 z_0}{2c}\right]
\int_{-L/2}^{L/2} dz' \exp\left[i \left(C + {\omega \theta^2
	\over{2  c }}\right) z' \right] \vec{e}_x~.\cr &&\label{undurad5finale}
\end{eqnarray}

\newpage

\subsection{A3. Approximating the Electron Path}

Let us now discuss the case of the radiation from a single electron with an arbitrary angular deflection $\vec{\eta}$ and an arbitrary offset $\vec{l}$ with respect to a reference orbit defined as the path through the origin of the coordinate system, that is $x(0) = y(0) = 0$.

If the magnetic field in the setup does not depend on the transverse coordinates, i.e. $B = B(z)$, an initial offset $x(0) = l_x$, $y(0) = l_y$ shifts the path of an electron of $\vec{l}$. Similarly, an angular deflection $\vec{\eta} = (\eta_x, \eta_y)$ at $z=0$ tilts the path without modifying it. Cases when the magnetic field of SR sources include focusing elements (or the natural focusing of insertion devices) are out of the scope of this paper. Assuming further that $\eta_x \ll 1$ and $\eta_y \ll 1$, which is typically justified for ultrarelativistic electron beams, one obtains the following approximation for the electron path:

\begin{eqnarray}
&& x(z) = x_r(z) + \eta_x z + l_x ~,  y(z) = y_r(z) + \eta_y z + l_y ~,  \cr &&
\label{shiftilt}
\end{eqnarray}
where the subscript `r' refers to the reference path. This  gives a parametric description of the path of a single electron with offset $\vec{l}$ and deflection $\vec{\eta}$. The curvilinear abscissa on the path can then be written as

\begin{eqnarray}
&& s(z) = \int_0^z dz' \left[1+\left(\frac{dx}{dz'} \right)^2+\left(\frac{dy}{dz'} \right)^2\right]^{1/2} \cr &&
\simeq \int_0^z dz' \left[1 + \frac{1}{2} \left(\frac{dx_r}{dz'} \right)^2+\frac{1}{2}\left(\frac{dy_r}{dz'}\right)^2 + \frac{1}{2} \left(\eta_x^2 + \eta_y^2\right) + \eta_x \frac{d x_r}{dz'} + \eta_y \frac{d y_r}{dz'}\right] \cr &&= s_r(z) + \frac{\eta^2 z}{2} + x_r(z)\eta_x + y_r(z)\eta_y ~,
\label{curvabs}
\end{eqnarray}
where we expanded the square root around unity in the first passage, we made use of Eq. (\ref{shiftilt}), and of the fact that the curvilinear abscissa along the reference path is $s_r(z) \simeq z + \int_0^z dz'[(dx_r/dz')^2/2 +(dy_r/dz')^2/2]$.

We now substitute Eq. (\ref{shiftilt}) and Eq. (\ref{curvabs}) into Eq. (\ref{generalfin}) to obtain:

\begin{eqnarray}
&& \vec{\widetilde{{E}}}(z_0, \vec{r}_0,\omega) = -{i
	\omega e\over{c^2}z_0} \int_{-\infty}^{\infty} dz' {\exp{\left[i
		\Phi_T\right]}} \cr && \times  \left[\left({v_x(z')\over{c}}
-(\theta_x-\eta_x)\right){\vec{e_x}}
+\left({v_y(z')\over{c}}-(\theta_y-\eta_y)\right){\vec{e_y}}\right]
~,\cr && \label{generalfin2}
\end{eqnarray}
where the total phase $\Phi_T$ is

\begin{eqnarray}
&&\Phi_T = \omega \left[\frac{s_r(z')}{v} +\frac{\eta^2 z'}{2 v} + \frac{1}{v}[x_r(z')\eta_x +y_r(z')\eta_y] -{z'\over{c}}\right] \cr &&
+ \frac{\omega}{2c}\left[z_0 (\theta_x^2+\theta_y^2) - 2 \theta_x x_r(z')  - 2 \theta_x \eta_x z' - 2 \theta_x l_x \right. \cr &&  \left. - 2 \theta_y y(z') - 2 \theta_y \eta_y z' - 2 \theta_y l_y  + z'(\theta_x^2+\theta_y^2)\right]~,
\label{totph2}
\end{eqnarray}
which can be rearranged as

\begin{eqnarray}
&&\Phi_T \simeq
\omega \left[{s_r(z')\over{v}}-{z'\over{c}}\right]  - \frac{\omega}{c} (\theta_x l_x + \theta_y l_y) \cr &&
+ \frac{\omega}{2c}\left[z_0 (\theta_x^2+\theta_y^2) - 2 (\theta_x-\eta_x) x_r(z') \right. \cr && \left. - 2 (\theta_y-\eta_y) y_r(z') + z'\left((\theta_x-\eta_x)^2+(\theta_y-\eta_y)^2\right)\right] ~.\cr &&
\label{totph3}
\end{eqnarray}

\newpage

\subsection{A4. Self-Fields of a Modulated Electron Beam}

In general, the electrodynamical theory is based on the exploitation, for the ultra-relativistic particles, of the small parameter $1/\gamma^2$. By this, Maxwell's equations are reduced to much simpler equations with the help of paraxial approximation.

Whatever the method used to present results, one needs to solve
Maxwell's equations in unbounded space. We introduce a cartesian
coordinate system, where a point in space is identified by a
longitudinal coordinate $z$ and transverse position $\vec{r}_{\perp}$.
Accounting for electromagnetic sources, i.e. in a region of space
where current and charge densities are present, the following
equation for the field in the space-frequency domain holds in all
generality:

\begin{equation}
c^2 \nabla^2 \vec{\bar{E}} + \omega^2 \vec{\bar{E}} = 4 \pi c^2
\vec{\nabla} \bar{\rho} - 4 \pi i \omega \vec{\bar{j}}~,
\label{trdisoo}
\end{equation}
where $\bar{\rho}(z,\vec{r}_{\perp},\omega)$ and
$\vec{\bar{j}}(z,\vec{r}_{\perp},\omega)$ are the Fourier
transforms of the charge density $\rho(z,\vec{r}_{\perp},t)$ and of the current
density $\vec{j}(z,\vec{r}_{\perp},t)$. Eq. (\ref{trdisoo}) is the
well-known Helmholtz equation. Here $\vec{\bar{E}}$ indicates the
Fourier transform of the electric field in the space-time domain.

Eq. (\ref{trdisoo}) can be solved with the help of an appropriate
Green's function $G(z-z',\vec{r}_{\bot }-\vec{r'}_\bot)$
yielding

\begin{eqnarray}
\vec{{\bar{E}}}(z, \vec{r}_{\bot },\omega) &=& - 4\pi
\int_{-\infty}^{\infty} dz' \int d \vec{r'}_{\bot}
\left(\frac{i\omega}{c^2} \vec{\bar{j}} - \nabla' \bar{\rho}\right)  G(z-z',\vec{r}_{\bot
}-\vec{r'}_\bot) ~, \label{sol1}
\end{eqnarray}
the integration in $d \vec{r'}_{\bot}$ being performed over the
entire transverse plane. An explicit expression for the Green's
function to be used in Eq. (\ref{sol1}) is given by

\begin{eqnarray}
G(z-z',\vec{r}_{\bot }-\vec{r'}_\bot) &=& -\frac{\exp\left\{ i
	(\omega/c) \left[\left|\vec{r}_{\bot
	}-\vec{r'}_\bot\right|^2+\left(z-z'\right)^2\right]^{1/2}\right\}}{4\pi
	\left[\left|\vec{r}_{\bot
	}-\vec{r'}_\bot\right|^2+\left(z-z'\right)^2\right]^{1/2}}
\label{greenhyp}~,
\end{eqnarray}
that automatically includes the proper boundary conditions at
infinity.

The transverse field $\vec{\bar{E}}_\bot$ can be treated in terms
of paraxial Maxwell's  equations in the space-frequency domain
(see e.g. \cite{GE}). From the paraxial approximation
follows that the electric field envelope $\vec{\widetilde{E}}_\bot
= \vec{\bar{E}}_\bot \exp{[-i\omega z/c]}$ does not vary much
along $z$ on the scale of the reduced wavelength
$\lambdabar=\lambda/(2\pi)$. As a result, the following field
equation holds:

\begin{eqnarray} \mathcal{D}
\left[\vec{\widetilde{E}}_\bot(z,\vec{r}_\bot,\omega)\right] =
\vec{g}(z, \vec{r}_\bot,\omega) ~,\label{field1}
\end{eqnarray}
where the differential operator $\mathcal{D}$ is defined by

\begin{eqnarray}
\mathcal{D} \equiv \left({\nabla_\bot}^2 + {2 i \omega \over{c}}
{\partial\over{\partial z}}\right) ~,\label{Oop}
\end{eqnarray}
${\nabla_\bot}^2$ being the Laplacian operator over transverse
cartesian coordinates. Eq. (\ref{field1}) is Maxwell's equation in
paraxial approximation. 

Eq. (\ref{trdisoo}),
which is an elliptic partial differential equation, has thus been
transformed into Eq. (\ref{field1}), which is of parabolic type.
Note that the applicability of the paraxial approximation depends
on the ultra-relativistic assumption $\gamma^2 \gg 1$ but not on
the choice of the $z$ axis. If, for a certain choice of the
longitudinal $z$ direction, part of the trajectory is such that
$\gamma_z^2 \sim 1$, the formation length is very short ($L_f \sim
\lambdabar$), and the radiated field is practically zero. As a
result, Eq. (\ref{trdisoo}) can always be applied, i.e. the
paraxial approximation can always be applied, whenever $\gamma^2
\gg 1$.

Complementarily, it should also be remarked here that the status
of the paraxial equation Eq. (\ref{field1}) in Synchrotron
Radiation theory is different from that of the paraxial equation
in Physical Optics. In the latter case, the paraxial approximation
is satisfied only by small observation angles. For example, one
may think of a setup where a thermal source is studied by an
observer positioned at a long distance from the source and behind
a limiting aperture. Only if a small-angle acceptance is
considered the paraxial approximation can be applied. On the
contrary, due to the ultra-relativistic nature of the emitting
electrons, contributions to the SR field from parts of the
trajectory with formation length $L_f \gg \lambdabar$ (the only
non-negligible) are highly collimated. As a result, the paraxial
equation can be applied at any angle of interest because it
practically returns zero field at angles where it should not be
applied.

The source-term vector $\vec{g}(z,
\vec{r}_\bot)$ is specified by the trajectory of the source
electrons, and can be written in terms of the Fourier transform of
the transverse current density, $\vec{\bar{j}}_\bot
(z,\vec{r}_\bot,\omega)$, and of the charge density,
$\bar{\rho}(z,\vec{r}_\bot,\omega)$, as

\begin{eqnarray}
\vec{g} = && - {4 \pi} \exp\left[-\frac{i \omega z}{c}\right]
\left(\frac{i\omega}{c^2}\vec{\bar{j}}_\bot -\vec{\nabla}_\bot
\bar{\rho}\right)  ~. \label{fv}
\end{eqnarray}
$\vec{\bar{j}}_\bot$ and $\bar{\rho}$ are regarded as given data. We will treat $\vec{\bar{j}}_\bot$ and $\bar{\rho}$
as macroscopic quantities, without investigating individual
electron contributions.  In the time domain, we may write the charge density
$\rho(\vec{r},t)$ and the current density $\vec{j}(\vec{r},t)$ as

\begin{eqnarray}
\rho(\vec{r},t) =  \frac{1}{v}
\rho_\bot(\vec{r}_\bot)
f\left(t-\frac{z}{v}\right) \label{charge}
\end{eqnarray}
and

\begin{eqnarray}
\vec{j}(\vec{r},t) &=& \frac{1}{v} \vec{v}
\rho_\bot(\vec{r}_\bot)f\left(t-\frac{z}{v}\right)
~, \cr&& \label{curr}
\end{eqnarray}

where $v$ denote the velocity of the electron.
The quantity $\rho_\bot$ has the meaning of transverse electron
beam distribution, while $f$ is the longitudinal charge density
distribution.

In the space-frequency domain, Eq. (\ref{charge}) and Eq.
(\ref{curr}) transform to:

\begin{equation}
\bar{\rho}(\vec{r}_\bot,z,\omega) =
\rho_\bot\left(\vec{r}_\bot\right)
\bar{f}(\omega) \exp\left[{i \omega z)/v}\right] ~,
\label{charge2tr}
\end{equation}
and

\begin{equation}
\vec{\bar{j}}(\vec{r}_\bot,z,\omega) =  \vec{v}
\rho_\bot\left(\vec{r}_\bot\right)
\bar{f}(\omega)\exp\left[{i \omega z/v}\right]~.
\label{curr2tr}
\end{equation}

It should be remarked that $\bar{\rho}$ and $\vec{\bar{j}} =
\bar{\rho} \vec{v}$ satisfy the continuity equation. In other
words, one can find $\vec{\nabla} \cdot \vec{\bar{j}} = i\omega
\bar{\rho}$.

We find an exact solution of Eq. (\ref{field1}) without any other
assumption about the parameters of the problem. A Green's function
for Eq. (\ref{field1}), namely the solution corresponding to the unit
point source can be written as (see e.g. \cite{GE}):

\begin{eqnarray}
G(z-z';\vec{r_{\bot}}-\vec{r'_\bot}) &=& -{1\over{4\pi (z-z')}}
\exp\left\{ i\omega{\mid \vec{r_{\bot}}
	-\vec{r'_\bot}\mid^2\over{2c (z-z')}}\right\}\label{green}~,
\end{eqnarray}
assuming $z-z' > 0$. When $z-z' < 0$ the paraxial approximation
does not hold, and the paraxial wave equation Eq. (\ref{field1})
should be substituted, in the space-frequency domain, by a more
general Helmholtz equation. Yet, the radiation formation length
for $z - z'<0$ is very short with respect to the case $z - z' >0$,
i.e. we can neglect contributions from sources located at $z-z'
<0$.

Since it is assumed that electrons are moving along the $z$-axis, we have  $\vec{\bar{j}}_{\perp} = 0$.
Thus, after integration by parts, we obtain the solution

\begin{eqnarray}
\vec{\widetilde{E}}_{\bot}(z, \vec{r}_{\bot}) &=& - \frac{i \omega
}{c}\bar{f}(\omega)  \int_{0}^{z} dz'   \int d \vec{r'}_{\bot} \exp\left\{i
\omega\left[\frac{\mid \vec{r}_{\bot }-\vec{r'}_\bot \mid^2}{2c
	(z-z')}\right]+  i \frac{ \omega z'}{2 c
	\gamma^2} \right\} \cr&&\times \frac{1}{z-z'}
{\rho_\bot}\left(\vec{r'}_\bot\right)
\left(
\frac{\vec{r}_{\bot}-\vec{r'}_{\bot}}{z-z'}\right)~. \cr &&
\label{ggeneralfin}
\end{eqnarray}
Eq. (\ref{ggeneralfin}) describes the field at any position $z$.

First, we make a change in the integration variable
from $z'$ to $\xi \equiv z-z'$. In the limit for $z
\longrightarrow \infty$, corresponding to the condition
$z \gg \gamma^2 \lambdabar$, we can
write for the transverse field

\begin{eqnarray}
\vec{\widetilde{E}}_{\bot}(z, \vec{r}_{\bot}) &=& -\frac{i \omega
	\bar{f}(\omega)}{c} \int d \vec{r'}_{\bot}
{\rho_\bot}\left(\vec{r'}_\bot\right) \exp\left[ \frac{i\omega z}{2 c
	\gamma^2}\right] \Bigg\{
\left[\frac{ic}{\omega}
\frac{\vec{r}_{\bot}-\vec{r'}_{\bot}} {\mid\vec{r}_{\bot}
	-\vec{r'}_{\bot}\mid}\cdot\frac{d}{d\left[\mid\vec{r}_{\bot}-\vec{r'}_{\bot}\mid\right]}\right]\cr
&& ~~~~~~~~~~~~~~~~~~~~~~~~~~~~~~~\times\int_{0}^{\infty} \frac{
	d\xi}{\xi} \exp\left[+i \omega \frac{\mid
	\vec{r}_{\bot}-\vec{r'}_\bot \mid^2}{2c \xi}- \frac{i\omega \xi}{2
	c \gamma^2}\right] \Bigg\}\label{effp2}
\end{eqnarray}

We now use the fact that, for any real number $\alpha>0$:

\begin{eqnarray}
\int_{0}^{\infty} {d \xi}
\exp\left[i\left(-\xi+{\alpha}/{\xi}\right)\right]/{\xi} = 2
K_0\left(2~
\sqrt{\alpha}\right) ~ ,
\label{rele}
\end{eqnarray}

where $K_0$ is the zero order modified Bessel function of the
second kind. Using Eq. (\ref{rele}) 
we can write Eq. (\ref{effp2}) as

\begin{eqnarray}
&& \vec{\widetilde{E}}_{\bot}(z, \vec{r}_{\bot}) = \frac{i \omega
	\bar{f}(\omega)}{c} \exp\left[ \frac{i\omega z}{2 c
	\gamma^2}\right] \int d \vec{r'}_{\bot}
{\rho_\bot}\left(\vec{r'}_\bot\right) \cr && \times \Bigg\{\left[\frac{ic}{\omega}\frac{\vec{r}_{\bot}-\vec{r'}_{\bot}}{\mid\vec{r}_{\bot}-\vec{r'}_{\bot}\mid}\right]
\frac{2}{\bar{\gamma}_z \lambdabar} K_1\left( \frac{\mid
	\vec{r}_{\bot}-\vec{r'}_\bot \mid}{\gamma \lambdabar}
\right) \Bigg\} ~, \label{Teffp3}
\end{eqnarray}

where $K_1(\cdot)$ is the modified Bessel function of the first order.

Let us assume a Gaussian transverse charge density distribution of the electron bunch with rms size $\sigma$  i.e. $\rho_\bot = (2\pi\sigma^2)^{-1}\exp[-r_\bot^2/(2\sigma^2)]$.
Within the deep asymptotic region when the transverse size of the modulated electron beam $\sigma \ll \lambdabar\gamma$ the Ginzburg-Frank formula can be applied \cite{GIF}

\begin{eqnarray}
\vec{\widetilde{E}}_{\bot}(z, \vec{r}_{\bot}) =
-\frac{2 \omega e }{c^2 \gamma} \exp\left[ \frac{i\omega z}{2 c
	\gamma^2}\right] \frac{\vec{r}_{\bot}}{r_{\bot}} K_1\left(\frac{\omega r_{\bot}}{c\gamma} \right) ~ .\label{virpm05}
\end{eqnarray}

Analysis of Eq.(\ref{virpm05}) shows a typical scale related to the transverse field distribution of order $ \lambdabar\gamma$ in dimensional units.  
Here $\lambda$ is the modulation wavelength.  In this asymptotic region radiation can be considered as virtual radiation from a filament electron beam (with no transverse dimensions).  

However, in XFEL practice we only deal with the deep asymptotic region where $\sigma \gg  \lambdabar\gamma$. Then, it can be seen that the field distribution in the space-time domain is essentially a convolution in the space domain between the transverse charge distribution of the electron beam and the field spread function described by the Ginzburg-Frank formula. Assuming a Gaussian (azimuthally-symmetric) transverse density distribution of the electron beam we obtain the
radially polarized virtual radiation beam.

\end{document}